\documentclass[a4paper,fleqn]{book}

\usepackage[utf8x]{inputenc} 	      
\usepackage[T1]{fontenc}            
\usepackage{csquotes}
\usepackage{enumerate}
\usepackage{enumitem}
\usepackage[title,titletoc,page]{appendix} 
\usepackage{pdfpages}
\usepackage{graphicx}         
\usepackage{tikz}
\usepackage{caption,subcaption}
\usepackage{epigraph}


\usepackage{booktabs}
\usepackage{array}
\newcolumntype{C}[1]{>{\centering\let\newline\\\arraybackslash\hspace{0pt}}m{#1}}
\usepackage{xcolor,comment}           
\usepackage{amsmath,amssymb,amscd,amsfonts,amsthm,esint,mathrsfs,stmaryrd,bbding,bm,epic,eepic,epsfig,latexsym}
\usepackage[nohints]{minitoc} 
\usepackage{framed}
\usepackage{bookmark}
\usepackage{BOONDOX-calo}
\usepackage{diagbox}
\usepackage{textcomp}
\usepackage{url}

\usepackage[margin=2cm]{geometry}
\usepackage{tocloft}
\renewcommand{\cftpartfont}{\bf \large }
\renewcommand{\cftchapfont}{Chapter }
\cftsetindents{part}{0em}{2em}
\cftsetindents{chapter}{1em}{3em}
\cftsetindents{section}{2em}{3em}
\renewcommand{\cftpartleader}{\cftdotfill{\cftdotsep}}
\renewcommand{\cftchapleader}{\cftdotfill{\cftdotsep}}

\usepackage{titletoc}

\renewcommand{\contentsname}{Table of Contents}
\mtcsettitle{parttoc}{}
\mtcsetfeature{parttoc}{before}{\vspace{0.6cm}}
\mtcsetfeature{parttoc}{open}{\vspace{0.3cm}}
\mtcsetfeature{parttoc}{close}{\vspace{0.2cm}}

\renewcommand\thechapter{\arabic{chapter}}

\usepackage{fancyhdr}

\newtheorem{theorem}{Theorem}[section]
\newtheorem{lemma}[theorem]{Lemma}
\newtheorem{corollary}[theorem]{Corollary}
\newtheorem{proposition}[theorem]{Proposition}
\newtheorem{definition}[theorem]{Definition}
\newtheorem{remark}[theorem]{Remark}
\newtheorem{defi/prop}[theorem]{Definition/Proposition}
\newtheorem{example}[theorem]{Example}

\newtheorem{fact}[theorem]{Fact}

\newcommand{\N}{\mathbf{N}}

\newcommand{\Q}{\mathbf{Q}}
\newcommand{\R}{\mathbf{R}}
\newcommand{\C}{\mathbf{C}}
\renewcommand{\P}{\mathbf{P}}
\DeclareMathOperator{\E}{\mathbf{E}}

\newcommand{\e}{\varepsilon}
\renewcommand{\leq}{\leqslant}
\renewcommand{\geq}{\geqslant}

\newcommand{\st}{\  : \ }

\renewcommand{\H}{\mathrm{H}}
\newcommand{\A}{\mathrm{A}}
\newcommand{\B}{\mathrm{B}}
\newcommand{\rE}{\mathrm{E}}

\newcommand{\Id}{\mathrm{Id}}
\newcommand{\id}{\mathrm{id}}
\newcommand{\cH}{\mathcal{H}}
\newcommand{\cK}{\mathcal{K}}
\newcommand{\cU}{\mathcal{U}}
\newcommand{\cF}{\mathcal{F}}
\newcommand{\cD}{\mathcal{D}}
\newcommand{\cS}{\mathcal{S}}
\newcommand{\cP}{\mathcal{P}}
\newcommand{\cX}{\mathcal{X}}
\newcommand{\cY}{\mathcal{Y}}
\newcommand{\cZ}{\mathcal{Z}}
\newcommand{\cA}{\mathcal{A}}
\newcommand{\cB}{\mathcal{B}}
\newcommand{\cN}{\mathcal{N}}
\newcommand{\cM}{\mathcal{M}}
\newcommand{\cL}{\mathcal{L}}

\newcommand{\iy}{\infty}

\DeclareMathOperator{\vol}{vol}
\DeclareMathOperator{\vrad}{vrad}
\DeclareMathOperator{\conv}{conv}

\DeclareMathOperator{\spec}{spec}

\DeclareMathOperator{\diag}{diag}
\DeclareMathOperator{\im}{Im}
\DeclareMathOperator{\tr}{Tr}
\DeclareMathOperator{\Sym}{Sym}
\DeclareMathOperator{\Asym}{Asym}
\DeclareMathOperator{\Span}{Span}

\newcommand{\NS}{{\text{NS}}}
\newcommand{\SNOS}{{\text{SNOS}}}

\DeclareFontFamily{U}{mathx}{\hyphenchar\font45}
\DeclareFontShape{U}{mathx}{m}{n}{
      <5> <6> <7> <8> <9> <10>
      <10.95> <12> <14.4> <17.28> <20.74> <24.88>
      mathx10
      }{}
\DeclareSymbolFont{mathx}{U}{mathx}{m}{n}
\DeclareMathSymbol{\bigtimes}{1}{mathx}{"91}

\newcommand{\scalar}[2]{\langle #1 , #2\rangle}
\newcommand{\braket}[2]{\langle #1 | #2\rangle}
\newcommand{\ketbra}[2]{| #1 \rangle\!\langle #2 |}
\newcommand{\bra}[1]{\langle #1 |}
\newcommand{\ket}[1]{| #1 \rangle}
\newcommand{\proj}[1]{| #1 \rangle\!\langle #1 |}

\begin{document}


\pagenumbering{arabic}
\thispagestyle{empty}
$\,$

\vfil
\vfil


\begin{center}
\begin{Large}
Universit\'{e} Claude Bernard Lyon 1 \& Universitat Aut\`{o}noma de Barcelona
\end{Large}
\end{center}

\vfil

\begin{center}
\textbf{\LARGE
{\scalebox{1}[1.4]{High dimension and symmetries} \\
\vskip7pt
\scalebox{1}[1.4]{in quantum information theory}
}}
\end{center}

\vfil

\begin{figure}[!h]
\begin{center}
\includegraphics[scale=0.4]{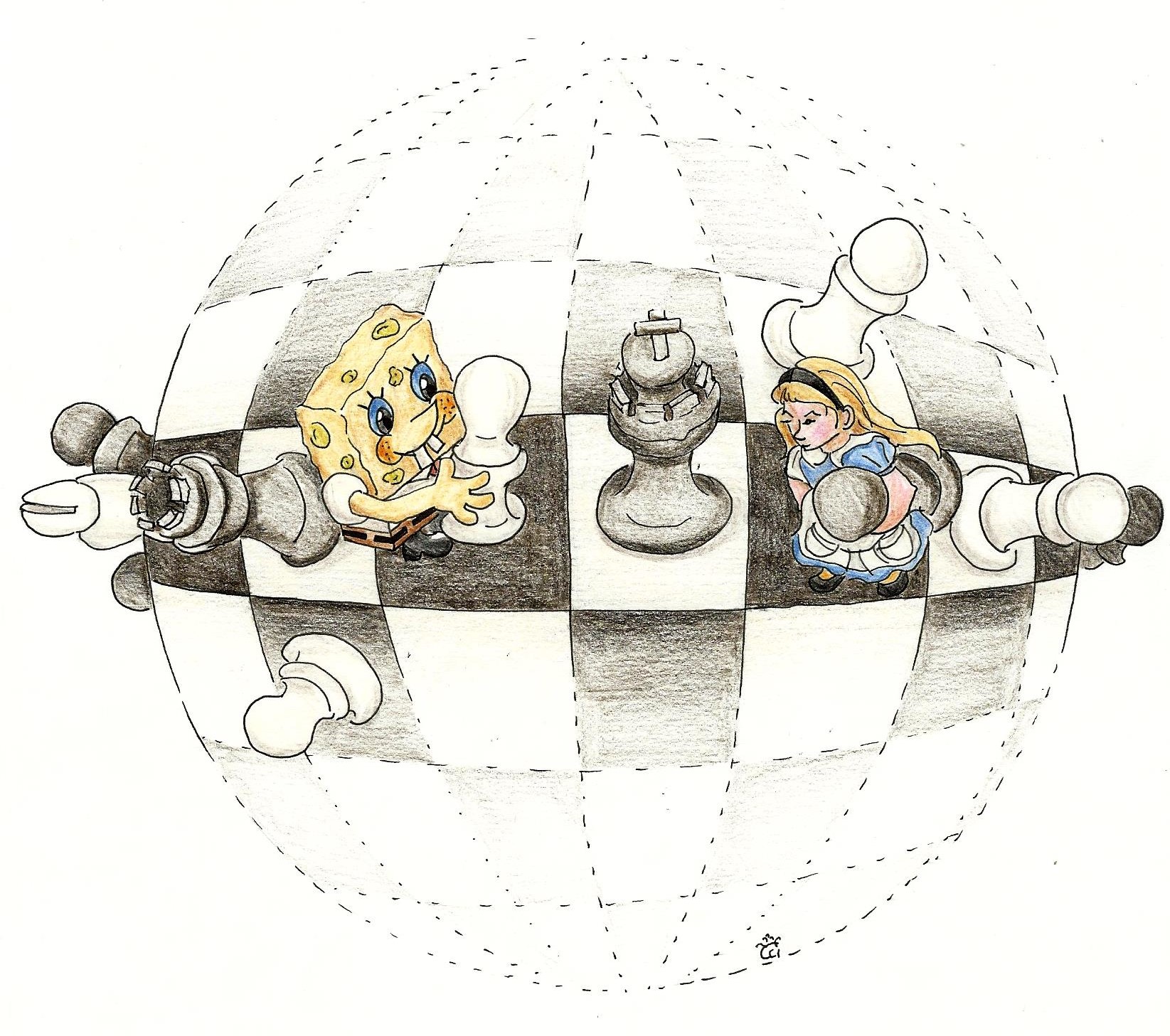}\\
Alice and Bob living on a planet whose mass concentrates around the equator.\\
Drawing in coloured pencil and ink by Aur\'{e}lie Garnier.
\end{center}	
\end{figure}

\vfil

\begin{center}
\begin{LARGE}
\textbf{PhD Thesis}
\end{LARGE}
\end{center}

\vfil

\begin{center}
{\Large
Publicly defended on June 9th 2016 by}
\end{center}

\vfil

\begin{center}
  \begin{Large}
\textbf{C\'{e}cilia Lancien}
  \end{Large}
\end{center}

\vfil

\begin{center}
{\Large in front of the committee composed of:}

\vfil
\begin{large}
\begin{tabular}{lll}
Mr St\'{e}phane Attal & \quad Universit\'{e} Claude Bernard Lyon 1  & \qquad {\small Examiner}\\[3pt]
Mr Guillaume Aubrun &\quad Universit\'{e} Claude Bernard Lyon 1 &\qquad {\small PhD advisor}\\[3pt]
Mr Florent Benaych-Georges & \quad Universit\'{e} Descartes Paris 5  & \qquad {\small Examiner}\\[3pt]
Mr John Calsamiglia &\quad Universitat Aut\`{o}noma de Barcelona &\qquad {\small Examiner}\\[3pt]
Mr Olivier Gu\'{e}don & \quad Universit\'{e} Paris-Est Marne-La-Vall\'{e}e  & \qquad {\small Examiner}\\[3pt]
Mr Marius Junge & \quad University of Illinois at Urbana-Champaign & \qquad {\small Reviewer}\\[3pt]
Mrs Stephanie Wehner & \quad Technische Universiteit Delft & \qquad {\small Reviewer}\\[3pt]
Mr Andreas Winter &\quad Universitat Aut\`{o}noma de Barcelona &\qquad {\small PhD advisor}\\[3pt]
\end{tabular}
\end{large}
\end{center}

\vfil


\newpage
\thispagestyle{empty}
$\,$
\newgeometry{lmargin=2cm,rmargin=2cm,tmargin=2cm,bmargin=2cm}

\newpage
\thispagestyle{empty}
\vspace*{2cm}

\setlength{\epigraphwidth}{230pt}
\epigraph{\textit{Pourquoi fait-on des math\'{e}matiques? Parce que c'est int\'{e}ressant, parce qu'on est curieux, sans doute, mais surtout parce qu'elles sont utiles.}\\
\vskip3pt
Gilles Godefroy, Les math\'{e}matiques mode d'emploi}


\newpage

\doparttoc

\pagestyle{fancy}
\renewcommand{\chaptermark}[1]{%
\markboth{\MakeUppercase{%
\chaptername}\ \thechapter.%
\ #1}{}}
\fancyhead{}
\fancyfoot{}
\lhead[\thepage]{\leftmark}
\rhead[]{\thepage}

\chapter*{Abstract / R\'{e}sum\'{e} / Resumen}

\section*{Abstract}

If a one-phrase summary of the subject of this thesis were required, it would be something like: miscellaneous large (but finite) dimensional phenomena in quantum information theory. That said, it could nonetheless be helpful to briefly elaborate. Starting from the observation that quantum physics unavoidably has to deal with high dimensional objects, basically two routes can be taken: either try and reduce their study to that of lower dimensional ones, or try and understand what kind of universal properties might precisely emerge in this regime. We actually do not choose which of these two attitudes to follow here, and rather oscillate between one and the other.

\smallskip

In the first part of this manuscript (Chapters \ref{chap:channel-compression} and \ref{chap:zonoids}), our aim is to reduce as much as possible the complexity of certain quantum processes, while of course still preserving their essential characteristics. The two types of processes we are interested in are quantum channels and quantum measurements. In both cases, complexity of a transformation is measured by the number of operators needed to describe its action, and proximity of the approximating transformation towards the original one is defined in terms of closeness between the two outputs, whatever the input. We propose universal ways of achieving our quantum channel compression and quantum measurement sparsification goals (based on random constructions) and prove their optimality.

\smallskip

Oppositely, the second part of this manuscript (Chapters \ref{chap:data-hiding}, \ref{chap:SDrelaxations} and \ref{chap:k-extendibility}) is specifically dedicated to the analysis of high dimensional quantum systems and some of their typical features. Stress is put on multipartite systems and on entanglement-related properties of theirs. We essentially establish the following: as the dimensions of the underlying spaces grow, being barely distinguishable by local observers is a generic trait of multipartite quantum states, and being very rough approximations of separability itself is a generic trait of separability relaxations. On the technical side, these statements stem mainly from average estimates for suprema of Gaussian processes, combined with the concentration of measure phenomenon.

\smallskip

In the third part of this manuscript (Chapters \ref{chap:deFinetti} and \ref{chap:SNOS}), we eventually come back to a more dimensionality reduction state of mind. This time though, the strategy is to make use of the symmetries inherent to each particular situation we are looking at in order to derive a problem-dependent simplification. By quantitatively relating permutation-symmetry and independence, we are able to show the multiplicative behaviour of several quantities showing up in quantum information theory (such as support functions of sets of states, winning probabilities in multi-player non-local games etc.). The main tool we develop for that purpose is an adaptable de Finetti type result.

\newpage
\section*{R\'{e}sum\'{e}}

S'il fallait r\'{e}sumer le sujet de cette th\`{e}se en une expression, cela pourrait \^{e}tre quelque chose comme: ph\'{e}nom\`{e}nes de grande dimension (mais n\'{e}anmoins finie) en th\'{e}orie quantique de l'information. Cela \'{e}tant dit, essayons toutefois de d\'{e}velopper bri\`{e}vement. La physique quantique a in\'{e}luctablement affaire \`{a} des objets de grande dimension. Partant de cette observation, il y a, en gros, deux strat\'{e}gies qui peuvent \^{e}tre adopt\'{e}es: ou bien essayer de ramener leur \'{e}tude \`{a} celle de situations de plus petite dimension, ou bien essayer de comprendre quels sont les comportements universels pr\'{e}cis\'{e}ment susceptibles d'\'{e}merger dans ce r\'{e}gime. Nous ne donnons ici notre pr\'{e}f\'{e}rence \`{a} aucune de ces deux attitudes, mais au contraire oscillons constamment entre l'une et l'autre.

\smallskip

Notre but dans la premi\`{e}re partie de ce manuscrit (Chapitres \ref{chap:channel-compression} et \ref{chap:zonoids}) est de r\'{e}duire autant que possible la complexit\'{e} de certains processus quantiques, tout en pr\'{e}servant, \'{e}videmment, leurs caract\'{e}ristiques essentielles. Les deux types de processus auxquels nous nous int\'{e}ressons sont les canaux quantiques et les mesures quantiques. Dans les deux cas, la complexit\'{e} d'une transformation est mesur\'{e}e par le nombre d'op\'{e}rateurs n\'{e}cessaires pour d\'{e}crire son action, tandis que la proximit\'{e} entre la transformation d'origine et son approximation est d\'{e}finie par le fait que, quel que soit l'\'{e}tat d'entr\'{e}e, les deux \'{e}tats de sortie doivent \^{e}tre proches l'un de l'autre. Nous proposons des solutions universelles (bas\'{e}es sur des constructions al\'{e}atoires) \`{a} ces probl\`{e}mes de compression de canaux quantiques et d'amenuisement de mesures quantiques, et nous prouvons leur optimalit\'{e}.

\smallskip

La deuxi\`{e}me partie de ce manuscrit (Chapitres \ref{chap:data-hiding}, \ref{chap:SDrelaxations} et \ref{chap:k-extendibility}) est, au contraire, sp\'{e}cifiquement d\'{e}di\'{e}e \`{a} l'analyse de syst\`{e}mes quantiques de grande dimension et certains de leurs traits typiques. L'accent est mis sur les syst\`{e}mes multi-partites et leurs propri\'{e}t\'{e}s ayant un lien avec l'intrication. Les principaux r\'{e}sultats auxquels nous aboutissons peuvent se r\'{e}sumer de la fa\c{c}on suivante: lorsque les dimensions des espaces sous-jacents augmentent, il est g\'{e}n\'{e}rique pour les \'{e}tats quantiques multi-partites d'\^{e}tre \`{a} peine distinguables par des observateurs locaux, et il est g\'{e}n\'{e}rique pour les relaxations de la notion de s\'{e}parabilit\'{e} d'en \^{e}tre des approximations tr\`{e}s grossi\`{e}res. Sur le plan technique, ces assertions sont \'{e}tablies gr\^{a}ce \`{a} des estimations moyennes de suprema de processus gaussiens, combin\'{e}es avec le ph\'{e}nom\`{e}ne de concentration de la mesure.

\smallskip

Dans la troisi\`{e}me partie de ce manuscrit (Chapitres \ref{chap:deFinetti} et \ref{chap:SNOS}), nous revenons pour finir \`{a} notre \'{e}tat d'esprit de r\'{e}duction de dimensionnalit\'{e}. Cette fois pourtant, la strat\'{e}gie est plut\^{o}t: pour chaque situation donn\'{e}e, tenter d'utiliser au maximum les sym\'{e}tries qui lui sont inh\'{e}rentes afin d'obtenir une simplification qui lui soit propre. En reliant de mani\`{e}re quantitative sym\'{e}trie par permutation et ind\'{e}pendance, nous nous retrouvons en mesure de montrer le comportement multiplicatif de plusieurs quantit\'{e}s apparaissant en th\'{e}orie quantique de l'information (fonctions de support d'ensembles d'\'{e}tats, probabilit\'{e}s de succ\`{e}s dans des jeux multi-joueurs non locaux etc.). L'outil principal que nous d\'{e}veloppons dans cette optique est un r\'{e}sultat de type de Finetti particuli\`{e}rement mall\'{e}able.

\newpage
\section*{Resumen}

En unas palabras, el tema de esta tesis se podr\'{\i}a resumir como: fen\'{o}menos varios en alta (pero finita) dimensi\'{o}n en teor\'{\i}a cu\'{a}ntica de la informaci\'{o}n. Dicho esto, sin embargo podemos dar algunos detalles de m\'{a}s. Empezando con la observaci\'{o}n que la f\'{\i}sica cu\'{a}ntica ineludiblemente tiene que tratar con objetos de alta dimensi\'{o}n, se pueden seguir esencialmente dos caminos: o intentar reducir su estudio al de otros que tienen dimensi\'{o}n m\'{a}s baja, o intentar comprender qu\'{e} tipo de comportamiento universal surge precisamente en este r\'{e}gimen. Aqu\'{\i} no elegimos cu\'{a}l de estas dos posturas hay que adoptar, sino que oscilamos constantemente entre una y la otra.

\smallskip

En la primera parte de este manuscrito (Cap\'{\i}tulos \ref{chap:channel-compression} y \ref{chap:zonoids}), nuestro objetivo es reducir al m\'{\i}nimo posible la complejidad de ciertos procesos cu\'{a}nticos, preservando sus caracter\'{\i}sticas esenciales. Los dos tipos de procesos que nos interesan son canales cu\'{a}nticos y medidas cu\'{a}nticas. En ambos casos, la complejidad de una transformaci\'{o}n se cuantifica con el n\'{u}mero de operadores necesarios para describir su acci\'{o}n, y la proximidad entre la transformaci\'{o}n de origen y su aproximaci\'{o}n se define por el hecho de que, cualquiera que sea el estado de entrada, los respectivos estados de salida deben ser suficientemente similares. Proponemos maneras universales de alcanzar nuestras metas de compresi\'{o}n de canales cu\'{a}nticos y rarefacci\'{o}n de medidas cu\'{a}nticas (basadas en construcciones aleatorias) y demostramos su optimalidad.

\smallskip

En contrapartida, la segunda parte de este manuscrito (Cap\'{\i}tulos \ref{chap:data-hiding}, \ref{chap:SDrelaxations} y \ref{chap:k-extendibility}) se dedica espec\'{\i}ficamente al an\'{a}lisis de sistemos cu\'{a}nticos de alta dimensi\'{o}n y sus rasgos t\'{\i}picos. El \'{e}nfasis se pone sobre sistemos multipartidos y sus propiedades de entrelazamiento. En resumen, establecemos principalmente lo siguiente: cuando las dimensiones de los espacios subyacentes aumentan, es gen\'{e}rico para estados cu\'{a}nticos multipartidos ser pr\'{a}cticamente indistinguible mediante observaciones locales, y es gen\'{e}rico para relajaciones de la noci\'{o}n de separabilidad ser burdas aproximaciones de ella. Desde un punto de vista t\'{e}cnico, estos resultados se derivan de estimaciones de promedio para supremosa de procesos gaussianos, combinadas con el fen\'{o}meno de concentraci\'{o}n de la medida.

\smallskip

En la tercera parte de este manuscrito (Cap\'{\i}tulos \ref{chap:deFinetti} y \ref{chap:SNOS}), finalmente volvemos a una filsof\'{\i}a de reducci\'{o}n de dimensionalidad. Pero esta vez, nuestra estrategia es utilizar las simetr\'{\i}as inherentes a cada situaci\'{o}n particular que consideramos para derivar una simplificaci\'{o}n adecuada. Vinculamos de manera cuantitativa simetr\'{\i}a por permutaci\'{o}n y independencia, lo que nos permite establecer el comportamiento multiplicativo de varias cuantidades que ocurren en teor\'{\i}a cu\'{a}ntica de la informaci\'{o}n (funciones de soporte de conjuntos de estados, probabilidad de \'{e}xito en juegos multi-jugadores no locales etc.). La principal herramienta t\'{e}cnica que desarrollamos con este fin es un resultado de tipo de Finetti muy adaptable.

\chapter*{Acknowledgements / Remerciements}

Life is a succession of coincidences, in my scientific case, most lucky ones I must say. When I arrived in Bristol in 2012 to do a first Master's research project with Andreas, the only thing I knew about him was that he was a bad boy wearing sunglasses. When I arrived in Lyon in 2013 to do a second Master's research project with Guillaume, the only thing I knew about him was that he was the successful inventor of a board game. So how on earth could I have guessed from this meagre prior knowledge that it would then be with both of them, science-wise and person-wise, such love at first sight? That being so, I had the inestimable luxury of launching into my PhD already eager and confident that I was in the best possible hands! There are so many things I would like to thank Andreas and Guillaume for that I will unavoidably have to content myself with a random selection here. I am indebted (and immensely grateful) to Andreas for teaching me that solving a problem with basic tools is not shameful (quite the reverse), for inculcating in me (or at least trying to) that if you are not convinced yourself of the interest of what you are doing there is no way the person in front is, for replying at any weird time to any weird math questioning of mine... and also for guiding me up the highest peak in the Pyrenees. I am beholden (and extremely thankful) to Guillaume for exemplifying from every viewpoint the concept of ``few words but perfectly chosen words'' (which I am obviously still totally unable to follow myself), for never letting pass the slightest bold (hence frequently erroneous) statement of mine, for replying at any non-weird time to any weird math questioning of mine... and also for sharing delightful running discussions. One summarizing word, and then I promise I move on: I thank my two scientific daddies of course as a priority for their availability and dedication during these past few years, for the skilful balance they achieved between guidance and ``manage by yourself'', for getting me going keenly on fascinating problems, but also for the not so anecdotal fact that they are as madly workaholic and perfectionist as me, so that, interacting with them, I almost felt like a sane person!

\smallskip

Merci \`{a} Aur\'{e}lie qui ne m'a elle non plus pas (trop) prise pour une folle quand je suis venue lui expliquer ce que je souhaitais comme dessin de couverture, et s'est ensuite laiss\'{e} s\'{e}duire par les charmes d'Alice in Wonderland et de Sponge Bob!

\smallskip

Having Marius Junge and Stephanie Wehner as referees of this manuscript represents something special for me. My very first contact with quantum information theory, and at the same time my very first researcher's experience, consisted in generalizing a result of Stephanie's. After one week, I was feeling so passionately about the topic that it was already clear to me what I wanted to dedicate my professional life to! Still before even starting my PhD was my quite memorable first encounter with Marius: me giving a talk at the end of which he casually asked ``but isn't what you're doing approximating zonoids by zonotopes?''. Needless to say this turned out to be true (even though I had absolutely no clue at that time what the hell these objects could be), and gave the questions I was investigated a new dimension! Hence, neither the irreconcilable footballistic disagreement with Stephanie nor the teasing of Marius on my mathematical antecedents alter the great deal of respect I have for them. I am therefore extremely happy to have them both reviewing my PhD manuscript today.

\smallskip

I am also grateful to St\'{e}phane Attal and John Calsamiglia for accepting to be the two ``locals'' in my PhD committee. I think it is fair to say that St\'{e}phane is the one from whom this whole story originates, as it was thanks to him that I ended up doing my first Master's research project with Andreas. I therefore owe it to him in a sense to be today where I am, all the more given that, since then, he has been continuously taking on a crucial mentor duty for me. As for John, I have rarely met such a scientifically curious person. From interacting with him, even useless mathematicians like me get the feeling that what they are doing is actually interesting and exciting! Seemingly of secondary importance perhaps (but this is also what builds a relationship going far beyond strictly professional), during each of my stays in Barcelona, I could share with him early morning office discussions on basically everything (including hopeless Franco-Spanish administration or politics).

\smallskip

Florent Benaych-Georges et Olivier Gu\'{e}don me font l'honneur de compl\'{e}ter mon jury, et je les en remercie. Mon tout premier contact avec l'analyse fonctionnelle asymptotique a \'{e}t\'{e} via un cours de Master donn\'{e} par Olivier. Au vu de mes centres d'int\'{e}r\^{e}t math\'{e}matiques actuels, il semble inutile de pr\'{e}ciser que cette introduction au sujet a \'{e}t\'{e} loin d'\^{e}tre rebutante. Quant \`{a} Florent, je me suis retrouv\'{e} l'an dernier \`{a} \'{e}tudier plusieurs de ses papiers dont je pensais pouvoir utiliser les r\'{e}sultats. Le projet s'est sold\'{e} par un \'{e}chec, mais ce n'est pas de sa faute, et cela me fait dans tous les cas tr\`{e}s plaisir de pouvoir enfin le rencontrer pour de vrai.

\smallskip

Merci aux probabilistes de l'ICJ et de l'UMPA pour leur immense ouverture d'esprit: ils m'ont gentillement accueillie parmi eux, et m'ont m\^{e}me fait m'y sentir \`{a} ma place, en d\'{e}pit du fait que je sois bien loin d'\^{e}tre une v\'{e}ritable probabiliste! La qualit\'{e} de mes s\'{e}jours \`{a} Lyon doit aussi beaucoup \`{a} l'ambiance au sein du groupe de th\'{e}sards. Que vous soyez 11h30-iste, 12h30-iste ou non-affilié, \`{a} tous un grand merci donc. Avec une petite d\'{e}dicace particuli\`{e}re \`{a} l'intention d'Adriane, Benjamin et Xavier, qui m'ont l\'{e}g\`{e}rement pr\'{e}c\'{e}d\'{e}e (et avec qui je pouvais partager \'{e}tats d'\^{a}me linguistiques, m\'{e}taphysiques et math\'{e}matiques, respectivement), ainsi que de Christian, Hugo, Luigia, Simon A., Simon B., Sylvain et Tom\'{a}s, que j'ai l\'{e}g\`{e}rement devanc\'{e}s. Enfin, petite d\'{e}dicace super particuli\`{e}re \`{a} l'intention d'Ivan, ``ing\'{e}nieur'' comme moi, et de Nadja, ``matheuse ultime'' quant \`{a} elle, mes deux compagnons de tout pendant ces trois ann\'{e}es, du scientifique au presque philosophique, en passant par le plus festif: ma vie braconnienne, et tellement au-del\`{a}, aurait \'{e}t\'{e} bien fade sans vous! La quasi-simultan\'{e}it\'{e} de nos soutenances (la petite tortue que je suis \'{e}tant logiquement la derni\`{e}re \`{a} passer) est une fort belle conclusion \`{a} notre ``threesome d'atypiques''. Une pens\'{e}e enfin \`{a} l'\'{e}gard de Dario et Thomas (malgr\'{e} les cauchemars qu'ils m'ont fait faire sur ma soutenance) pour les sympathiques soir\'{e}es sportives pass\'{e}es ensemble. Et parfaite transition avec le paragraphe suivant: merci \`{a} Daniel, le Catalan-presque-Lyonnais, gr\^{a}ce \`{a} qui j'ai eu un toit \`{a} Barcelone.

\smallskip

Many thanks to all GIQitos for the emotional ``so nice that you're back'' or ``so sad that you're gone'' each time I was coming to or leaving from Barcelona, for the enthusiasm in front of each new cake on average or smelly French cheese of mine, for the passionate Catalunya vs Rest of the World discussions, and for so much more! A specific thought for the Graciosos, Christoph, Gianni, Kk, Ludovico, Mohammad and Sara, who granted me the status of emeritus member of theirs, having at least spiritual, if not physical anymore, presence in this awesome group. Another special thought for the Austrian Dream Team, aka Claude, Marcus and Pauli, for nothing being ever a problem (especially not hosting me anytime anywhere), for everything being always super geil (from French underground music to creepy maths theorems), well basically for being so refreshingly generous and excited! And a last personalised thought for Alex and Milan, my two functional analysis companions in Barcelona. Amongst other, I owe to Alex a now more or less correct Spanish summary of my thesis, and I owe to Milan a few unforgettable hikes! Finally, before closing these Barcelona thanks, let us not forget about expressing our gratitude to some emblematic host institutions there: LIQUID and MAFIA for their ``always welcome'' motto, Chivuo's, Gasterea, Quimet et al. for their gastro-social federating power!

\smallskip

There are many people without whom my scientific life up to now would not have been the same, precisely because what they brought me goes way beyond the purely scientific field. There is no good manner of ordering them in my acknowledgements so I will just go for chronological order of first encounter. Toby Cubitt was enthusiastically paying me a beer on the day where my first paper was released on the arXiv, and two days later a guy I did not know at this baby mathematician stage, Ashley Montanaro, had already carefully read the paper in question and was sending me detailed comments about it! Since then, both of them have shown (or at least well-pretended to show) constant interest in what I was doing, even enquiring more often than not about the most repelling details of certain proofs. Getting such kind of positive feedback is simply invaluable as a youngster. I met Fernando Brandao and Aram Harrow shortly later. Realising that they were not so much older than me, while the works of theirs that I knew had made me expect old respectful professors, was naturally a bit depressing at first. But them as well have had from then on such a supportive and interactive attitude towards me that I do not hold it against them! David Reeb and Alexander M\"{u}ller-Hermes are two other scientists I met in my early mathematical youth, and with whom I had constant interchanges ever after. It is indeed a quite pleasant feeling to realise that there are at least two other persons in the world who care about the same insignificant (but annoyingly open) questions as you! The Fall 2013 semester that I spent participating to the thematic programme ``Mathematical challenges in quantum information'' at the Isaac Newton Institute in Cambridge, at the very beginning of my PhD, was the occasion to make several influential encounters. The one with Nilanjana Datta and Debbie Leung was for sure determining. I would even go as far as seeing it as me projecting myself onto how I would dream to evolve in my future scientific behaviour! The ones with Matthias Christandl and Michael Walter were definitely shaping as well. The two of them patiently filled my group representation gaps and cheerfully kept suggesting new directions when absolutely everything we were working on was collapsing! Subsequent collaborations with Matthias have always been of that same joyful, hence most pleasant, vein.


\smallskip

Je suis extr\^{e}mement reconnaissante envers Ion Nechita pour n'avoir jamais failli \`{a} son r\^{o}le de mentor. Il m'a en effet d\`{e}s mes d\'{e}buts chapeaut\'{e}e, et ce \`{a} tous points de vue: passant sans probl\`{e}me des journ\'{e}es enti\`{e}res \`{a} r\'{e}pondre \`{a} mes questions math\'{e}matiques, me mettant sur le devant de la sc\`{e}ne lors d'\`{a} peu pr\`{e}s chacune des conf\'{e}rences qu'il a organis\'{e}es, et me faisant m\^{e}me d\'{e}couvrir la folle vie nocturne roumaine! Je profite d'ailleurs de cette occasion pour remercier aussi Maria Anastasia Jivulescu, organisatrice en chef (et surtout la plus attentionn\'{e}e qu'on puisse concevoir) de cette m\'{e}morable conf\'{e}rence en Roumanie! Toujours par association d'id\'{e}es, merci aux autres StoQistes toulousains, Cl\'{e}ment Pellegrini et Tristan Benoist, pour leur disponibilit\'{e} (intra ou extra scientifique) absolument sans faille. Enfin, Benoit Collins et Staszek, mon grand-père mathématique, ont eux aussi \'{e}t\'{e} des interlocuteurs constamment disponibles, avec qui j'appr\'{e}cie toujours \'{e}changer des id\'{e}es et aupr\`{e}s desquels j'ai \'{e}norm\'{e}ment appris.

\smallskip

The MathQI group in Madrid should be my next home, and I am pretty much looking forward to it, for the very simple reason that I actually already feel at home there. From my first encounter with Angelo, Carlos G.G., Carlos P.C., David P.G., Ignacio, and friends, during the ``Intensive month on operator algebras and quantum information'' that they were organizing at the ICMAT in Summer 2013, I already knew there were the best conceivable vibes between us, surpassing by far the professional collaboration standpoint!

\smallskip

Ces trois ann\'{e}es de th\`{e}se (entre bien d'autres choses) n'auraient pas eu la m\^{e}me saveur sans le groupe de grimpeurs de la promo X2009 et ses pi\`{e}ces rapport\'{e}es (que nous appellerons par la suite Plastik par souci de concision). Je remercie donc Plastik du fond du c{\oe}ur pour avoir accept\'{e} qu'on peut taper une preuve en TeX \`{a} 6h du matin et \^{e}tre n\'{e}anmoins quelqu'un de fr\'{e}quentable. Plastik, c'est un peu ma deuxi\`{e}me famille, dont je resterai (d'un point de vue scientifique en tout cas) toujours la Maman, puisque premi\`{e}re de la troupe \`{a} acc\'{e}der au grade de Docteur. Votre soutien pr\'{e}-soutenance m'a particuli\`{e}rement touch\'{e}e: vos suggestions sur quoi faire le jour J et comment se d\'{e}tendre avant n'ont pas forc\'{e}ment toutes \'{e}t\'{e} mises en {\oe}uvre, mais ont \'{e}t\'{e} appr\'{e}ci\'{e}es pour leur inventivit\'{e}! A Guillaume et Harmonie, un immense merci pour votre relecture orthogonalement suppl\'{e}mentaire de ce manuscrit: Harmonie \'{e}tant probablement herm\'{e}tique \`{a} la beaut\'{e} math\'{e}matique de mes \'{e}quations (bien que fonci\`{e}rement convaincue de leur g\'{e}nialitude) mais rep\'{e}rant les moindres fautes d'anglais, Guillaume ne me corrigeant absolument rien mais s'enthousiasmant d'absolument tout!

\smallskip

Enfin, je tiens \`{a} dire un grand merci \`{a} ma famille, qui a toujours eu la curiosit\'{e} d'essayer de comprendre ce que je pouvais bien passer mes journ\'{e}es \`{a} faire. Merci en particulier \`{a} Grounchy et Schnoky pour leur soutien ind\'{e}fectible: ``Alors cet ordi quantique, \c{c}a avance?'', ``Ah ta preuve marche, c'est cool! On sait tr\`{e}s bien que demain elle marchera plus, mais t'en fais pas, apr\`{e}s-demain elle re-marchera...''. And last but not least, merci \`{a} mes parents pour m'avoir appris tr\`{e}s t\^{o}t \`{a} faire la diff\'{e}rence entre un Banach et un canard. Ils ont bien tent\'{e} de me sugg\'{e}rer que j'aurais probablement beaucoup mieux \`{a} faire de ma vie que de prouver (voire le plus souvent m\^{e}me pas) des th\'{e}or\`{e}mes dont trois personnes au monde se soucient. Mais il faut croire que l'image qu'ils m'ont renvoy\'{e} du chercheur en maths n'a pas \'{e}t\'{e} suffisamment d\'{e}courageante pour me convaincre...

\cleardoublepage
\tableofcontents
\addtocontents{toc}{~\hfill\textbf{Page}\par}
\addtocontents{toc}{\protect\mbox{}\protect\hrulefill\par}

\part{Introduction and background}

\newpage
\textbf{\LARGE{Part I -- Table of contents}}
\parttoc

\chapter{What is this thesis about and how is this manuscript organized?}
\label{chap:motivations}

\section{Motivations and context of the thesis}

A one-particle quantum system in a pure state is described by a unit vector in some Hilbert space (or to be precise, by the projection onto the corresponding line). For multi-particle systems, the associated Hilbert space is simply given by the tensor product of the individual ones. Its dimension thus grows exponentially with the number of subsystems, making the classical modelling of quantum systems practically unfeasible as soon as more than a few particles are involved.
\smallskip

One way around this curse of dimensionality is to make use of any extra information which may be held, a priori, on the state of the system under consideration. One could know, for instance, that the latter has certain symmetries, and exploit the reduction in the effective number of degrees of freedom that this fact implies. The archetypical example is the following: Assume that a system is composed of $n$ indistinguishable particles (meaning that their labelling is irrelevant), with associated finite-dimensional Hilbert space $\mathrm{H}$. Then, a pure state of such system actually lives in the symmetric subspace $\mathrm{Sym}^n(\mathrm{H})$ of $\mathrm{H}^{\otimes n}$, whose dimension is only polynomial, rather than exponential, in $n$.
\smallskip

On the other hand, one should not conclude too quickly that having to deal with high dimensional objects is necessarily a misfortune. Indeed, it may also be the case that, as the dimension grows, certain universal behaviours emerge. This is precisely what asymptotic geometric analysis and random matrix theory teach us. In quantum information theory, these subfields of mathematics can serve at least two main purposes. First of all, they may be used as a tool to determine what are the properties that big quantum systems, subject to whatever relevant restrictions, are expected to exhibit. More concretely, one could be interested in knowing what are the typical characteristics of either quantum states or quantum transformations, under several types of constraints such as locality, noise, energy etc. Second of all, they may help in proving the existence of objects having a given property. In this context, the paradigmatic idea is that constructing the latter explicitly might be harder than asserting that a suitably chosen random one will do with overwhelming probability. And with both aims in view, high dimension of the underlying space is an asset, what makes average features become generic.
\smallskip

The purpose of this thesis is therefore clearly two-sided: in some places the focus is on how to reduce the study of large (or even infinite) dimensional situations to that of lower dimensional ones, while in some others the goal is precisely to understand the typical aspects which may arise as the dimension of the studied object grows. The latter objective is clearly the underlying motivation of Part \ref{part:entanglement}. As for the former objective, two different routes can be taken to achieve it. The first approach, which is the one followed in Part \ref{part:complexity}, consists in trying to compress the initial data set as much as possible while still preserving the essential information it contains. The second approach consists in exploiting potential additional knowledge on the initial data to argue that they already belong, effectively, to a much smaller set. This is the spirit of Part \ref{part:symmetry}.
\smallskip

All the questions posed here initially arise from quantum information theory. But in order to solve them, several fields of pure mathematics had to get heavily involved. For a start, random matrix theory plays undeniably a prominent role in all our work. Nevertheless, we are not so much preoccupied with the asymptotic study of random matrices, as (free) probabilists usually are, but rather with the non-asymptotic one. Indeed, for the applications we have in mind, knowing that a certain behaviour emerges almost surely in the regime where the size of the matrix goes to infinity is not so useful: we need instead to understand, for a large but finite size, what is the probability that the properties of the matrix do not deviate too much from their limiting ones. This is where concentration of measure enters the picture, allowing us to assert that, if the function we are looking at is regular enough, then it should be close to its expected value with overwhelming probability as the size of its random input grows. We exploit this key phenomenon under multiple forms throughout these pages. It however appears clear that, before even trying to bound the deviation probability of a given function (and more often than not, also as an end in itself), we have to be able to estimate its average value. For that purpose, we usually call in geometric and probabilistic techniques in Banach space theory. Besides, taking advantage of the symmetries of our problem, is an idea that we try to apply whenever possible. With that goal in view, information theory tools, both classical and quantum, are crucial to us. They are in fact what makes it for instance possible to relate in a quantitative way exchangeability and independence.

\section{Structure of the present manuscript}

Each chapter revolves around a piece of work which either already led to a paper (sometimes published yet, sometimes in preprint form still) or should lead to one shortly. In each case though, modifications have been done on the original version. Occasionally it is just a matter of notation or presentation, to simply fit better with the rest of the manuscript (unfortunately, it is likely that neither complete non-redundancy nor absolute style unity have been achieved between the chapters). But here and there further developments, that were investigated only afterwards, have also been added.
\smallskip

The remainder of this introductory part is dedicated to setting the common ground and tools on which subsequent parts develop. If it achieves the goal it was designed for, Chapter \ref{chap:QIT} could be renamed ``Everything you need to know about quantum physics if you are a mathematician willing to read this manuscript''. As for Chapters \ref{chap:symmetries} and \ref{chap:toolbox}, they can be seen as the two toolboxes in which we will regularly dig: in Chapter \ref{chap:symmetries} for all the basics about permutation-symmetry, and in Chapter \ref{chap:toolbox} for all the basics about high dimensional convex geometry and deviation inequalities.
\smallskip

As already mentioned, the core scientific material in this thesis is then divided into three parts. Let us try to summarize in a few words the content of each of them, even though the needed definitions in order to do so in a satisfactory way have not been introduced yet (see Chapter \ref{chap:QIT} for most of them). Precisely for that matter, it is only at the beginning of each part that a proper account is made of its objectives, its main achievements (and techniques to reach them), its non-ignorable difficulties etc.
\smallskip

The main theme of Part \ref{part:complexity} is that of approximating complex processes by simpler ones. In Chapter \ref{chap:channel-compression} the question which is investigated is: given a quantum channel, to how much can its number of Kraus operators be brought down, with the constraint that each input state is still sent close to its original output state? In some sense, Chapter \ref{chap:zonoids} deals with a similar issue. Nevertheless, the channels which are now considered are quantum-classical channels (aka quantum measurements), and the approximation requirement is completely different: namely that, for each input state, outcome statistics close to its original ones are obtained. Our objective in both cases is to exhibit universal schemes which attain the maximum doable compression.
\smallskip

The goal of Part \ref{part:entanglement} is to study several properties of quantum states (more specifically, entanglement-related properties of multipartite quantum states), and see how (un)common they become as the size of the quantum system grows. Chapter \ref{chap:data-hiding} is in line with Chapter \ref{chap:zonoids} since it also focusses on the issue of distinguishing quantum states from their measurement outcome statistics. But this time the question is: how well can local observers, having access only to their own subsystem, typically perform in this task? Then, in both Chapters \ref{chap:SDrelaxations} and \ref{chap:k-extendibility}, the aim is to quantify the average strength of certain semidefinite relaxations of separability. In Chapter \ref{chap:SDrelaxations}, stress is put on exporting entanglement detection tools from the bipartite setting, relying on positive maps, to certify genuinely multipartite entanglement. Oppositely, the interest in Chapter \ref{chap:k-extendibility} is in a purely bipartite and not positive map based separability criterion (more precisely, a complete hierarchy of necessary conditions for bi-separability build on symmetric extensions).
\smallskip

This opens naturally the route to Part \ref{part:symmetry}, where symmetries are exploited to their maximum, with the purpose of reducing the understanding of permutation-invariant scenarios to that of i.i.d.~ones. Hence, a major difference with Part \ref{part:complexity} is that we are now making extensive use of the specificities of the situation we have at hand in order to get a highly problem-dependent simplification. Chapter \ref{chap:deFinetti} is dedicated to setting a very general and adaptable framework in which one can indeed do so. As a consequence, several quantities arising in quantum information theory can be shown to exhibit a multiplicative behaviour. Chapter \ref{chap:SNOS} is then entirely devoted to seeing through one important application of the previously developed machinery, namely to the parallel repetition problem for multi-player non-local games.
\smallskip

Perhaps now is the time to fleetingly elaborate on the choice of the thesis title. There are two distinct ways in which we are principally dealing with ``high dimension'': either because we are looking at one system whose underlying dimension is large, or because we are looking at many copies of a system. Yet, in both situations ``symmetries'' play a central role. To attack the first problem, it is usually tools from asymptotic geometric analysis that we call on for help. This field of mathematics can be seen as a middle ground between geometry (which traditionally deals with small-dimensional objects) and functional analysis (which traditionally deals with infinite-dimensional objects). And its whole purpose is to identify and exploit approximate symmetries that escaped both the too rigid area of geometry and the too qualitative area of functional analysis. As for the second issue, it is usually tackled via information theory techniques. And it is more often than not making use of the permutation-symmetry of the multi-copy scenario which allows relating asymptotic performances to single-shot ones, and hence get an understanding of them.
\bigskip

This manuscript essentially puts together the material appearing in the following publications or preprints:
\begin{itemize}[topsep=0cm,itemsep=-1em,parsep=0cm,leftmargin=*]
\item C.~Lancien. Quantum channel compression. \\
\textit{In preparation}. (cf.~Chapter \ref{chap:channel-compression})\\
\item G.~Aubrun and C.~Lancien. Zonoids and sparsification of quantum measurements. \\
\textit{Positivity}, 20(1):1--23, 2016. arXiv[quant-ph]:1309.6003 (cf.~Chapter \ref{chap:zonoids})\\
\item G.~Aubrun and C.~Lancien. Locally restricted measurements on a multipartite quantum system: data hiding is generic. \\
\textit{Quant. Inf. Comput.}, 15(5--6):512--540, 2014. arXiv[quant-ph]:1406.1959. (cf.~Chapter \ref{chap:data-hiding})\\
\item O.~G\"{u}hne, M.~Huber, C.~Lancien and R.~Sengupta. Relaxations of separability in multipartite systems:
semidefinite programs, witnesses and volumes. \\
\textit{J. Phys. A: Math. Theor.}, 48(505302), 2015. arXiv[quant-ph]:1504.01029. (cf.~Chapter \ref{chap:SDrelaxations})\\
\item C.~Lancien. k-extendibility of high-dimensional bipartite quantum states. \\
\textit{Preprint}. arXiv[quant-ph]:1504.06459. (cf.~Chapter \ref{chap:k-extendibility})\\
\item C.~Lancien and A.~Winter. Flexible constrained de Finetti reductions and applications. \\
\textit{Preprint}. arXiv[quant-ph]:1605.09013. (cf.~Chapter \ref{chap:deFinetti})\\
\item C.~Lancien and A.~Winter. Parallel repetition and concentration for (sub-)no-signalling games via a flexible constrained de Finetti reduction. \\
\textit{Preprint}. arXiv[quant-ph]:1506.07002. (cf.~Chapter \ref{chap:SNOS})\\
\end{itemize}

Other publications or preprints on which this manuscript does not focus:
\begin{itemize}[topsep=0cm,itemsep=-1em,parsep=0cm,leftmargin=*]
\item C.~Lancien and A.~Winter. Distinguishing multi-partite states by local measurements. \\
\textit{Commun. Math. Phys.}, 323:555--573, 2013. arXiv[quant-ph]:1206.2884.\\
\item G.~Adesso, S.~Di Martino, M.~Huber, C.~Lancien, M.~Piani and A.~Winter. Should entanglement measures be monogamous or faithful? \\
\textit{Preprint}. arXiv[quant-ph]:1604.02189.\\
\end{itemize}

\section{Notation and conventions}
\label{sec:notation}

Here is a list of notation that shall be used repeatedly throughout the whole manuscript (it might happen though that some are recalled as the text goes). Oppositely, notation which are needed more sporadically will be introduced only in due time.
\smallskip

Given a complex Hilbert space $\mathrm{H}$, we denote by $\mathcal{L}(\mathrm{H})$ the space of all linear operators on $\mathrm{H}$. When $\mathrm{H}\equiv\C^d$ is finite-dimensional (which will almost always be the case we will be dealing with), we identify $\mathcal{L}(\mathrm{H})$ with the space of $d\times d$ complex matrices. Important subsets of $\mathcal{L}(\mathrm{H})$ include: the group of unitary operators on $\mathrm{H}$, which we denote by $\mathcal{U}(\mathrm{H})$, the space of Hermitian operators on $\mathrm{H}$, which we denote by $\mathcal{H}(\mathrm{H})$, the cone of positive semidefinite operators on $\mathrm{H}$, which we denote by $\mathcal{H}_+(\mathrm{H})$. The notation $\Id$ (or $\Id_{\mathrm{H}}$ if there is a risk of confusion) stands for the identity operator on $\mathrm{H}$, while the notation $\mathcal{I\!d}$ (or $\mathcal{I\!d}_{\mathrm{H}}$) is used for the identity (super-)operator on $\mathcal{L}(\mathrm{H})$. Also, for any $X\in\cL(\mathrm{H})$, we denote by $X^{\dagger}$ its adjoint (i.e.~transpose conjugate) and by $|X|=\sqrt{X^{\dagger}X}$ its absolute value.
\smallskip

For each $p\in[1,+\infty]$, we define $\|\cdot\|_p$ as the Schatten $p$-norm on $\mathcal{L}(\mathrm{H})$, i.e.~\[ \forall\ X\in\mathcal{L}(\mathrm{H}),\ \|X\|_p=\left(\tr |X|^p\right)^{1/p}\ \text{if}\ p\in[1,+\infty[\ \text{and}\ \|X\|_{\infty}=\lim_{p\rightarrow+\infty}\|X\|_p. \]
Particular instances of interest are the trace class norm $\|\cdot\|_1$, the Hilbert--Schmidt norm $\|\cdot\|_2$ (which is nothing else than the Euclidean norm arising from the Hilbert--Schmidt inner product $\langle X,Y\rangle \mapsto \tr \left(X^{\dagger}Y\right)$ on $\mathcal{L}(\mathrm{H})$), and the operator norm $\|\cdot\|_{\infty}$. Most of the time, we will look at Schatten $p$-norms not on the full space $\cL(\H)$ but in restriction to its subspace $\cH(\H)$. We denote by $B_p(\H)$, resp.~$S_p(\H)$, the corresponding unit ball, resp.~ sphere, in $\cH(\H)$. In the special case $p=2$, we may also use the notation $B_{HS}(\H)$ and $S_{HS}(\H)$ instead.
\smallskip

We denote by $\|\cdot\|_p$ as well the $p$-norms on $\R^d$ or $\C^d$. One exception is again the Euclidean norm $\|\cdot\|_2$, for which we shall usually simply use the notation $\|\cdot\|$, while $S^{d-1}$ and $S_{\C^d}$ stand for the corresponding Euclidean unit spheres in $\R^d$ and $\C^d$, respectively.
\smallskip

The notation $|\cdot|$ stands both for the cardinality when applied to a finite set and for the dimension when applied to a finite-dimensional Hilbert space.
\smallskip

Given $n\in\N$, we may use the shorthand notation $[n]$ for $\{1,\ldots,n\}$, and we denote by $\mathfrak{S}(n)$, resp.~$\mathfrak{P}(n)$, the set of permutations, resp.~partitions, of $[n]$.
\smallskip

In the remainder of this manuscript, we are often interested in the asymptotic regime, when the dimensions of the underlying finite-dimensional Hilbert spaces tend to infinity. In that setting, the letters $C,c,c_0,\ldots$ will always denote (non-negative) numerical constants, independent from any other parameters such as the dimension. The value of these constants may change from occurrence to occurrence. Similarly $c(\e)$ denotes a constant depending only on the parameter $\e$. Also, given functions $f,g$ of the underlying dimension $d$, we will write $f=O(g)$, resp.~$f=\Omega(g)$, if there exists $C>0$ such that, for all $d\in\N$, $f(d)\leq Cg(d)$, resp.~$f(d)\geq g(d)/C$, and we will write $f=\Theta(g)$ if both $f=O(g)$ and $f=\Omega(g)$ hold.
\smallskip

When working with a random variable $X$, we will use the notation $\P(\mathcal{E}(X))$ to denote the probability of the event $\mathcal{E}(X)$, and the notation $\E\left(f(X)\right)$ to denote the expectation of the function $f(X)$.
\smallskip

More quantum information orientated notation include (see Chapter \ref{chap:QIT}, Sections \ref{sec:states-POVMs} and \ref{sec:sep-ent}, for the corresponding definitions): $\mathcal{D}(\mathrm{H})$ for the set of quantum states on the Hilbert space $\mathrm{H}$, $\mathcal{S}(\mathrm{H}_1{:}\cdots{:}\mathrm{H}_k)$ for the set of separable quantum states on the tensor product Hilbert space $\mathrm{H}_1\otimes\cdots\otimes\mathrm{H}_k$ (across the $k$-partite cut $\mathrm{H}_1{:}\cdots{:}\mathrm{H}_k$).

\chapter{The mathematics of quantum information theory in a nutshell}
\chaptermark{The mathematics of quantum information theory in a nutshell}
\label{chap:QIT}

The purpose of this chapter is to explain why certain mathematical notions enter the theory of quantum mechanics, by giving an idea of their physical interpretation. There is no attempt to being exhaustive, far from that, the accent being put only on concepts that will then play a central role throughout this manuscript. The reader is for instance referred to the lecture notes \cite{Wolf}, Chapter 1, 2 and 3, for a much more comprehensive justification of the correspondence between mathematical objects and physical situations in the quantum mechanical framework. The book \cite{ASbook}, Chapters 2 and 3, would be another recommendation for going way deeper into that topic.

\section{Quantum states and observables: preparation and measurement}
\label{sec:states-POVMs}

In quantum mechanics, the state of a system is described by a unit vector $\psi$ in a complex separable Hilbert space $(\mathrm{H},\braket{\cdot}{\cdot})$. The physical picture is then the following: There exists a distinguished orthonormal basis $\{e_i,\ i\in I\}$ of $\mathrm{H}$, where $I$ is countable (finite or infinite) and labels the possible ``levels'' of the system. If $\psi=e_{i}$ for some $i\in I$, then (as one would expect) the system is said to be in level $i$. But in general, $\psi$ is in a superposition of the basis vectors, i.e.~\[ \psi=\sum_{i\in I}\braket{e_i}{\psi}e_i, \]
And the only thing that one can tell is that the probability of obtaining outcome $i\in I$ when measuring its level is given by $|\braket{e_i}{\psi}|^2$. We already see with this description that, for any $\alpha\in\C$ such that $|\alpha|^2=1$, states $\psi$ and $\alpha\psi$ cannot be distinguished. It is therefore more accurate to say that the state of the quantum system is characterized by the rank-$1$ projection $\ketbra{\psi}{\psi}$ on $\mathrm{H}$, which reads
\[ \ketbra{\psi}{\psi} = \sum_{i,j\in I}\braket{e_i}{\psi}\braket{\psi}{e_j}\ketbra{e_i}{e_j}. \]
And hence, simply rewriting what we just explained, the probability of obtaining outcome $i\in I$ when measuring its level is given by $\tr\left(\ketbra{e_i}{e_i}\ketbra{\psi}{\psi}\right)$.
\smallskip

This describes well the situation where the physical system under consideration can be perfectly prepared and kept in any given target state $\psi$. Such scenario is of course completely idealistic, and in a more realistic framework noise has to be taken into account. The latter may arise either from imprecisions in the preparation procedure or from interactions of the system of interest with its environment. In both cases, one can only assign probabilities $\{p_x,\ x\in X\}$ to the system being in one amongst the so-called \textit{pure states} $\{\ketbra{\psi_x}{\psi_x},\ x\in X\}$. And the true state of the system is the so-called \textit{mixed state} $\rho$ defined as
\[ \rho=\sum_{x\in X}p_x\ketbra{\psi_x}{\psi_x}. \]
Here again, the formula for the probability of getting outcome $i\in I$ when measuring the system's level reads $\tr\left(\ketbra{e_i}{e_i}\rho\right)$.
\smallskip

Hence, probabilities appear at two very distinct stages in quantum mechanics: Just as in classical mechanics, they model our ignorance of the precise state in which the system is. But they also occur, more fundamentally, because even having perfect knowledge of the system's state, we cannot (in general) predict with certainty which value a measurement performed on it will yield. That is why the measurement process, as a transition from possibilities to facts, plays such a central role in the quantum theory.
\smallskip

Up to now, we mentioned only one specific measurement, namely the one performed in the canonical basis of $\mathrm{H}$, which determines the level of the system. Its action on a state $\rho$ consists in sending it on one amongst the pure states $\{\ketbra{e_i}{e_i},\ i\in I\}$, with respective probabilities $\left\{ \tr\left(\ketbra{e_i}{e_i}\rho\right),\ i\in I \right\}$. Still looking at this particular measurement, one could however consider the case where it cannot be performed with such perfect accuracy. For instance, it could be that levels which are too close to one another are simply seen as being the same, or that certain outcomes are sometimes mistaken for one another. The resulting ``blurred'' measurement gives outcomes labelled by $j\in J$, and its action on a state $\rho$ consists in sending it on one amongst the mixed states $\{ M_j/\tr M_j,\ j\in J \}$, with respective probabilities $\{\tr(M_j\rho),\ j\in J\}$, where for each $j\in J$, there exists a probability distribution $\{p_{j,i},\ i\in I\}$ such that $M_j=\sum_{i\in I} p_{j,i}\ketbra{e_i}{e_i}$.
\smallskip

Let us summarize and formalize the above discussion, focussing for simplicity on the finite-dimensional case, which is the one we shall almost exclusively consider in the sequel. The Hilbert space associated to a $d$-level quantum system can simply be identified with $\C^d$, equipped with its canonical inner product. The state of such quantum system is entirely characterized by a \textit{density operator} $\rho$ on $\C^d$, which is nothing else than a convex combination (or mixture) of rank-$1$ projectors on $\C^d$ (or pure states on $\C^d$). Equivalently, $\rho$ has to satisfy the two properties of being positive semidefinite and having trace $1$. Denoting by $\mathcal{H}(\C^d)$ the set of Hermitian operators on $\C^d$, the set of density operators on $\C^d$ is thus defined as
\[ \mathcal{D}(\C^d) := \conv\{\ketbra{\psi}{\psi},\ \psi\in\C^d,\ |\braket{\psi}{\psi}|^2=1\} =\{ \rho\in\mathcal{H}(\C^d),\ \rho\geq 0,\ \tr\rho=1\}. \]
Besides, a measurement on such quantum system is described by a \textit{Positive Operator-Valued Measure (POVM)}, which is a resolution of the identity on $\C^d$, i.e.~a set of positive semidefinite operators summing to the identity on $\C^d$.
\smallskip

Understanding the geometry of the set $\mathcal{D}(\C^d)$ and of certain of its subsets is a wide topic, upon which we shall touch in several places of this manuscript (see \cite{BZ} or \cite{ASbook}, Chapters 2 and 9, for a comprehensive exposition). For now, let us just state some immediate facts. $\mathcal{D}(\C^d)$ is a compact convex set whose extreme points are exactly the pure states, i.e.~the rank-$1$ states. It is included in $\{M\in\mathcal{H}(\C^d),\ \tr M=1\}$, hyperplane of $\mathcal{H}(\C^d)$, where it has non-empty interior. $\mathcal{D}(\C^d)$ is therefore a convex body of real dimension $d^2-1$ (recall that $\mathcal{H}(\C^d)$ has real dimension $d^2$). Its center of mass is the so-called \textit{maximally mixed state} $\Id/d$.

\section{Composite quantum systems: separability vs entanglement}
\label{sec:sep-ent}

Assume now that the quantum system under consideration is composed of $k$ subsystems, with associated Hilbert spaces $\mathrm{H}_1,\ldots,\mathrm{H}_k$. Then, the Hilbert space associated to the global $k$-partite system is simply the tensor product Hilbert space $\mathrm{H}=\mathrm{H}_1\otimes\cdots\otimes\mathrm{H}_k$. 
A state $\rho$ on $\mathrm{H}$ is called \textit{separable} if it can be written as a convex combination of product states, i.e.~states of the form $\rho_1\otimes\cdots\otimes\rho_k$ where $\rho_1,\ldots,\rho_k$ are states on $\mathrm{H}_1,\ldots,\mathrm{H}_k$ respectively. Otherwise, it is called \textit{entangled}.
\smallskip

As we shall later see on several occasions, the dichotomy between separability and entanglement is fundamental in quantum information theory (see in particular Chapters \ref{chap:data-hiding}, \ref{chap:SDrelaxations}, \ref{chap:k-extendibility} and \ref{chap:deFinetti}). So let us start here with a few basic observations. If a pure state is separable, then it is necessarily product, which means that its local subsystems are completely uncorrelated. In the general mixed case, a separable state may have local subsystems which exhibit correlations, but classical ones only (in the sense that such state can always be prepared by local parties sharing just common randomness).
\smallskip

In the finite-dimensional case, the set of separable states on $\C^{d_1}\otimes\cdots\otimes\C^{d_k}\equiv\C^d$ (i.e.~$d=d_1\times\cdots\times d_k$) is thus defined as
\[ \mathcal{S}(\C^{d_1}{:}\cdots{:}\C^{d_k}) := \conv\left\{\rho_1\otimes\cdots\otimes\rho_k,\ \forall\ 1\leq i\leq k,\ \rho_i\in\mathcal{D}(\C^{d_i}) \right\}. \]
And the extreme points of $\mathcal{S}(\C^{d_1}{:}\cdots{:}\C^{d_k})$ are actually easy to characterize: these are precisely the pure separable states, i.e.~the pure product states. Just as $\mathcal{D}(\C^d)$, $\mathcal{S}(\C^{d_1}{:}\cdots{:}\C^{d_k})$ has real dimension $d^2-1$ and its center of mass is the maximally mixed state $\Id/d$ (here again, see \cite{BZ} or \cite{ASbook}, Chapters 2 and 9, for a much more exhaustive presentation).

\smallskip

For the sake of clarity, let us focus for now on the bipartite case $\mathrm{H}=\mathrm{H}_1\otimes\mathrm{H}_2$. Given a state $\rho$ on $\mathrm{H}$, we can define its \textit{reduced state} on the first subsystem $\rho_1=\tr_2 \rho$, which is the state on $\mathrm{H}_1$ characterized by the property that, for any operator $M_1$ on $\mathrm{H}_1$, $\tr(\rho_1M_1)=\tr(\rho\, M_1\otimes\Id_2)$. Conversely, given a state $\rho_1$ on $\mathrm{H}_1$, we say that a state $\rho$ on $\mathrm{H}$ is an \textit{extension} of $\rho_1$ if $\tr_2\rho=\rho_1$, and that it is more precisely a \textit{purification} of $\rho_1$ if it is additionally pure. For any state $\rho_2$ on $\mathrm{H}_2$, the state $\rho_1\otimes\rho_2$ is obviously an extension of $\rho_1$, and when the latter is pure its extensions are in fact necessarily of this product form (see e.g.~\cite{ChNi}, Chapter 2). This means in other words that a system in a pure state cannot share any kind of correlations (not even classical ones) with another system.

\section{Evolution of a quantum system}
\label{sec:channels}

When talking about the time evolution of a physical system (classical or quantum), two cases have to be distinguished: that of \textit{closed systems} and that of \textit{open systems}. A closed system is one which is considered perfectly isolated from the outside world, and thus undergoes reversible dynamics only. On the contrary, when studying an open system, one takes into account that it may interact with its surrounding, so that irreversibility may arise from this coupling (it is only the system of interest accompanied by its environment which is seen as a whole as closed).
\smallskip

Let us see how these ideas are mathematically formalized when looking at a quantum system, with associated Hilbert space $\mathrm{H}$. If the system is considered closed, the evolutions it may undergo are unitary transformations. This means that there exists a unitary operator $U$ on $\mathrm{H}$ such that, if it is in the initial state $\rho$, then it is brought in the final state $U\rho U^{\dagger}$. If this time the system is considered open, coupled with some ancilla Hilbert space $\mathrm{K}$, it is the global system $\mathrm{H}\otimes\mathrm{K}$ which is subject to a unitary transformation. Hence, there exist a unitary operator $U$ on $\mathrm{H}\otimes\mathrm{K}$ and a pure state $\ketbra{\psi}{\psi}$ on $\mathrm{K}$ such that, at the level of the system of interest, if it is in the initial state $\rho$, then it is brought in the final state $\tr_{\mathrm{K}}\left(U\rho\otimes\ketbra{\psi}{\psi} U^{\dagger}\right)$.
\smallskip

A crucial observation is that these two situations can be encompassed into one common description, by saying that the evolution of a quantum system is, in full generality, characterized by a so-called \textit{completely positive and trace preserving} (CPTP) map \cite{Stinespring}. Let us explain what this means. A linear map $\mathcal{N}$ from operators on $\mathrm{H}$ to operators on $\mathrm{H}'$ is \textit{positive} (P) if, for any positive semidefinite operator $X$ on $\mathrm{H}$, $\mathcal{N}(X)$ is a positive semidefinite operator on $\mathrm{H}'$. And it is \textit{completely positive} (CP) if, for any $\mathrm{K}$, the linear map $\mathcal{N}\otimes\mathcal{I\!d}$ from operators on $\mathrm{H}\otimes\mathrm{K}$ to operators on $\mathrm{H}'\otimes\mathrm{K}$ is positive. Besides, it is called \textit{trace preserving} (TP) if, as the name suggests, for any trace-class operator $X$ on $\mathrm{H}$, $\tr \mathcal{N}(X) =\tr X$. A CPTP map $\mathcal{N}$ is therefore usually referred to as a \textit{quantum channel}, transforming the input state $\rho$ into the output state $\mathcal{N}(\rho)$.
\smallskip

These definitions will be crucial to us in Chapter \ref{chap:channel-compression}, and to some extent also in Chapter \ref{chap:deFinetti}. Note that the measurement procedure described before falls into this category: a POVM is a particular type of quantum channel, which maps quantum states to probability distributions, and is thus sometimes called a quantum-classical (QC) channel. These QC channels will be thoroughly studied in Chapters \ref{chap:zonoids} and \ref{chap:data-hiding}. Oppositely, a preparation procedure, which consists in mapping a probability distribution to a mixture of quantum states, is sometimes called a classical-quantum (CQ) channel. And the fully classical analogue of these notions is simply that of a stochastic map (aka conditional probability distribution), which will play a central role in Chapter \ref{chap:SNOS}.

\section{Separability criteria on multipartite quantum systems}
\label{sec:sep-multi}

Already in the bipartite case, deciding whether a given state is entangled or separable is known to be, in general, a hard task, both from a mathematical and computational point of view. In fact, even the weak membership problem for separability (i.e.~deciding if the state under consideration is far away from or close to the set of separable states) is a NP-hard problem \cite{Gurvits,Gharibian}. There exist necessary conditions for separability, though, which are much easier to check. A whole family of them is based on the following easy observation: if $\rho$ is a separable state on $\mathrm{H}_1\otimes\mathrm{H}_2$, then for any positive map $\mathcal{N}_1$ from operators on $\mathrm{H}_1$ to operators on $\mathrm{H}'_1$, $\mathcal{N}_1\otimes\mathcal{I\!d}_2(\rho)$ is a positive semidefinite operator on $\mathrm{H}_1\otimes\mathrm{H}_2$.
\smallskip

Widely studied and used examples of positive (yet not completely positive) maps, giving rise to a corresponding (non trivial) necessary condition for separability, include:
\begin{itemize}[topsep=0cm,itemsep=-1em,parsep=0cm,leftmargin=*]
\item The transposition map $\mathcal{T}:X\in\mathcal{L}(\mathrm{H})\mapsto X^T\in\mathcal{L}(\mathrm{H})$ \cite{Peres,HHH}.\\
\item The reduction map $\mathcal{R}:X\in\mathcal{L}(\mathrm{H})\mapsto(\tr X)\Id-X\in\mathcal{L}(\mathrm{H})$ \cite{HH}.\\
\item The Choi map $\mathcal{C}:X\in\mathcal{L}(\mathrm{H})\mapsto -X + (|\mathrm{H}|-1)\Delta(X) + \Delta'(X)\in\mathcal{L}(\mathrm{H})$, where the action of $\Delta$ and $\Delta'$ on any $X\in\mathcal{L}(\mathrm{H})$ is defined by, for each $1\leq i,j\leq |\mathrm{H}|$, $\Delta(X_{i,j})=\delta_{i,j}X_{i,i}$ and $\Delta'(X_{i,j})=\delta_{i,j}X_{i-1,i-1}$ \cite{CL}.
\end{itemize}
\smallskip

Taken altogether, these separability tests based on positive maps actually define a necessary and sufficient condition for separability: a state $\rho$ on $\mathrm{H}_1\otimes\mathrm{H}_2$ is separable if and only if, for all positive maps $\mathcal{N}_1$ from operators on $\mathrm{H}_1$ to operators on $\mathrm{H}'_1$, $\mathcal{N}_1\otimes\mathcal{I\!d}_2(\rho)$ is a positive semidefinite operator on $\mathrm{H}_1'\otimes\mathrm{H}_2$ \cite{Choi}. This result is a fundamental one in quantum information theory. It translates in the language of so-called \textit{entanglement witnesses} as follows: If a state $\rho$ on $\mathrm{H}_1\otimes\mathrm{H}_2$ is entangled, then there exists a block-positive operator $M$ on $\mathrm{H}_1\otimes\mathrm{H}_2$ such that $\tr(\rho M) <0$ \cite{HHH}. $M$ is thus said to witness the entanglement of $\rho$, since for any separable state $\sigma$ on $\mathrm{H}_1\otimes\mathrm{H}_2$, we have $\tr(\sigma M) \geq 0$. In other words, the operator $M$ defines a hyperplane in $\mathcal{H}( \mathrm{H}_1\otimes\mathrm{H}_2)$ which separates the convex body $\mathcal{S}( \mathrm{H}_1{:}\mathrm{H}_2)$ from the point $\rho$. One given entanglement witness can only detect the entanglement in a small fraction of states (those on the same side as $\rho$ of the hyperplane it defines). However, entanglement witnesses carry the advantage of being experimentally easily accessible.
\smallskip

\begin{figure}[h] \caption{Several degrees of separability on a tripartite system}
\label{fig:separable}
\begin{center}
\includegraphics[width=9cm]{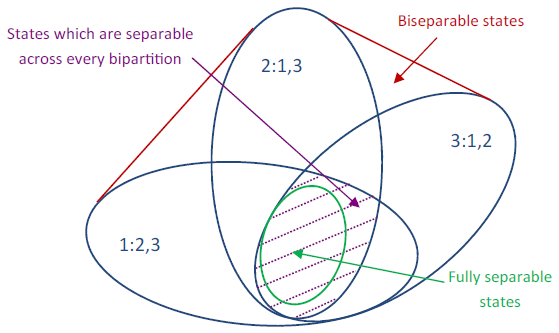}
\end{center}
\end{figure}

In the multipartite case, the picture becomes even more complex. Indeed, checking separability of a given state across all bipartite splitting of the subsystems is not enough to guarantee that it is fully separable. Motivated by this easy observation, one can define a hierarchy of relaxations of the latter notion: a state on $\mathrm{H}_1\otimes\cdots\otimes\mathrm{H}_k$ is $\ell$-separable, where $2\leq \ell\leq k$, if it is a convex combination of states, each of which is separable across a given splitting of the $k$ subsystems into $\ell$ groups (note that, in general, there is no splitting of the $k$ subsystems into $\ell$ groups across which an $\ell$-separable state is separable). $k$-separability is then just full separability, while a state which is not $2$-separable is called \textit{genuinely multipartite entangled}. These various notions of separability are illustrated in the tripartite case $\mathrm{H}=\mathrm{H}_1\otimes\mathrm{H}_2\otimes\mathrm{H}_3$ in Figure \ref{fig:separable} (where $x\,{:}\,y,z$ denotes the set of states on $\mathrm{H}$ which are separable across the bipartite cut $\mathrm{H}_x{:}\mathrm{H}_y\otimes\mathrm{H}_z$).

All these notions concerning entanglement detection will be crucial in several places of this manuscript, most notably in Chapters \ref{chap:SDrelaxations} and \ref{chap:k-extendibility}, but also to a lesser extent in Chapters \ref{chap:data-hiding} and \ref{chap:deFinetti}.

\section{A few notions from quantum Shannon theory}
\label{sec:Shannon}

Classical information theory is concerned with quantifying the amount of information and correlations present in probability distributions. By analogy, quantum information theory is concerned with doing the same for density operators. We gather here some basic definitions and facts from the quantum Shannon theory that we shall later need here and there. The reader is for instance referred to the book \cite{Wilde}, Chapter 11, for a complete expounding of these notions.
\smallskip

The \textit{von Neumann entropy} of a quantum state $\rho$ is defined as
\[ S(\rho):=-\tr(\rho\log\rho). \]
It is always non-negative, and equal to $0$ if and only if $\rho$ is a pure state. In the case where the underlying Hilbert space $\mathrm{H}$ is finite-dimensional, it is upper-bounded by $\log|\mathrm{H}|$, with equality if and only if $\rho$ is the maximally mixed state $\Id/|\mathrm{H}|$.
\smallskip

The \textit{mutual information} of a state $\rho$, on a bipartite Hilbert space $\mathrm{H}_1\otimes\mathrm{H}_2$, is defined as
\[ I(\mathrm{H}_1{:}\mathrm{H}_2)(\rho) := S(\rho_{1})+S(\rho_{2})-S(\rho). \]
By sub-additivity of the quantum entropy \cite{ArLi}, it is always non-negative, and equal to $0$ if and only if $\rho=\rho_{1}\otimes\rho_{2}$, i.e.~ if and only if there is no correlation (not even classical ones) between the two subsystems.
\smallskip

The \textit{conditional mutual information} of a state $\rho$, on a tripartite Hilbert space $\mathrm{H}_1\otimes\mathrm{H}_2\otimes\mathrm{H}_3$, is defined as
\[ I(\mathrm{H}_1{:}\mathrm{H}_2|\mathrm{H}_3)(\rho) := S(\rho_{13})+S(\rho_{23})-S(\rho_{3})-S(\rho). \]
By strong sub-additivity of the quantum entropy \cite{LR}, it is always non-negative.
\smallskip

Finally, the \textit{relative entropy} between states $\rho$ and $\sigma$ is defined as
\[ D(\rho\|\sigma) := \tr(\rho(\log\rho-\log\sigma)). \]
It is always non-negative, and infinite whenever the support of $\rho$ and the orthogonal of the support of $\sigma$ have a non-zero intersection. For a state $\rho$ on a bipartite Hilbert space $\mathrm{H}_1\otimes\mathrm{H}_2$, we have the notable property that $D(\rho\|\rho_1\otimes\rho_2) = I(\mathrm{H}_1{:}\mathrm{H}_2)(\rho)$.
\smallskip

The main reason why we shall later be interested in these entropic quantities is because several measures of the amount of entanglement present in multipartite quantum systems can be built from them. We will need all these definitions chiefly in Chapters \ref{chap:channel-compression} and \ref{chap:deFinetti}.

\chapter{Permutation-symmetry and de Finetti type theorems}
\chaptermark{Permutation-symmetry and de Finetti type theorems}
\label{chap:symmetries}

This chapter collects together some standard definitions and results revolving around permutation-symmetry in quantum information theory. The reader is referred to the review \cite{Harrow} or to the lecture notes \cite{Watrous}, Chapter 22, for a much more satisfying treatment of this material. Here, we focus only on what we will make use of in this manuscript (most notably in Chapters \ref{chap:k-extendibility}, \ref{chap:deFinetti} and \ref{chap:SNOS}, but also more punctually in Chapter \ref{chap:data-hiding}, and at a technical rather than fundamental level, in Chapters \ref{chap:channel-compression} and \ref{chap:zonoids}).

\section{The symmetric subspace: a brief review of some basic notions and facts}
\label{sec:sym}

Let $\mathrm{H}$ be a Hilbert space and $\{\ket{i},\ 1\leq i\leq |\mathrm{H}|\}$ be an orthonormal basis of $\mathrm{H}$ (with the convention that $\{1,\ldots,|\mathrm{H}|\}=\N$ whenever $\mathrm{H}$ is infinite-dimensional). For any $n\in\N$ and any permutation $\pi\in\mathfrak{S}(n)$, denote by $U(\pi)\in\mathcal{U}(\mathrm{H}^{\otimes n})$ the associated permutation unitary on $\mathrm{H}^{\otimes n}$, characterized by
\[ \forall\ 1\leq i_1,\ldots,i_n\leq |\mathrm{H}|,\ U(\pi)\ket{i_{1}}\otimes\cdots\otimes\ket{i_{n}} = \ket{i_{\pi(1)}}\otimes\cdots\otimes\ket{i_{\pi(n)}}. \]
Note that this definition is actually independent of the basis. The $n$-symmetric subspace of $\mathrm{H}^{\otimes n}$ can then be defined as the simultaneous $+1$-eigenspace of all $U(\pi)$'s,
\begin{align*}
\Sym^n\left(\mathrm{H}\right) := & \left\{ \ket{\psi}\in\mathrm{H}^{\otimes n} \st \forall\ \pi\in\mathfrak{S}(n),\ U(\pi)\ket{\psi}=\ket{\psi} \right\}\\
= & \Span\left\{ \ket{v_{i_1,\ldots,i_n}}=\sum_{\pi\in\mathfrak{S}(n)} \ket{i_{\pi(1)}}\otimes\cdots\otimes\ket{i_{\pi(n)}} \st 1\leq i_1\leq\cdots\leq i_n\leq|\mathrm{H}| \right\}.
\end{align*}
The orthogonal projector onto $\Sym^n\left(\mathrm{H}\right)$ may thus be written as
\[ P_{\Sym^n(\mathrm{H})} = \frac{1}{n!}\sum_{\pi\in\mathfrak{S}(n)}U(\pi) = \sum_{1\leq i_1\leq\cdots\leq i_n\leq |\mathrm{H}|} \ket{\psi_{i_1,\ldots,i_n}}\bra{\psi_{i_1,\ldots,i_n}}, \]
where for each $1\leq i_1\leq\cdots\leq i_n\leq |\mathrm{H}|$, $\ket{\psi_{i_1,\ldots,i_n}}$ denotes the unit vector having same direction as $\ket{v_{i_1,\ldots,i_n}}$. In the case where $\mathrm{H}$ is finite-dimensional, this can also be re-written as
\[ P_{\Sym^n(\mathrm{H})} = {n+|\mathrm{H}|-1 \choose n} \int_{\ket{\psi}\in S_{\mathrm{H}}} \ket{\psi}\bra{\psi}^{\otimes n}\mathrm{d}\psi, \]
where $\mathrm{d}\psi$ stands for the uniform probability measure on the unit sphere $S_{\mathrm{H}}$ of $\mathrm{H}$. This is due to Schur's Lemma, since $\Sym^n\left(\mathrm{H}\right)$ is an irreducible representation of the commutant action of $\{U(\pi),\ \pi\in \mathfrak{S}(n)\}$, which is that of the local unitaries $\{U^{\otimes n},\ U\in \mathcal{U}(\mathrm{H})\}$ (see e.g.~\cite{Simon} for more group representation background).
\smallskip

A state $\rho$ on $\mathrm{H}^{\otimes n}$ is said to be \textit{permutation-invariant} if it satisfies $U(\pi)\rho U(\pi)^{\dagger} = \rho$ for all $\pi\in\mathfrak{S}(n)$. This condition is actually equivalent to the existence of a unit vector $\ket{\psi}\in\Sym^n\left(\mathrm{H}\otimes\mathrm{H}'\right)$, where $\mathrm{H}'\equiv\mathrm{H}$, such that $\rho=\tr_{\mathrm{H}'}\ketbra{\psi}{\psi}$. That is why permutation-invariant states are also sometimes simply referred to as \textit{symmetric} states.

\section{Permutation-invariant states in the bipartite finite-dimensional case}
\label{sec:Werner}

The permutation-invariant states on $(\C^d)^{\otimes 2}$ are the so-called \textit{Werner states}. These are mixtures of the (fully) symmetric state $\pi_s$ and the (fully) antisymmetric state $\pi_a$ on $(\C^d)^{\otimes 2}$ (which are defined as the renormalized projectors onto the symmetric and antisymmetric subspaces of $(\C^d)^{\otimes 2}$, respectively). Concretely, set
\begin{align*}
& \Sym^2(\C^d) := \Span\left\{ \ket{i_1,i_2}+\ket{i_2,i_1} \st 1\leq i_1\leq i_2\leq d \right\},\\
& \Asym^2(\C^d) := \Span\left\{ \ket{i_1,i_2}-\ket{i_2,i_1} \st 1\leq i_1< i_2\leq d \right\}.
\end{align*}
We then have the orthogonal direct sum decomposition
\[ \left(\C^d\right)^{\otimes 2}=\Sym^2(\C^d)\overset{\perp}{\oplus}\Asym^2(\C^d),\ \text{with}\
\begin{cases} \dim(\Sym^2(\C^d))=d(d+1)/2 \\ \dim(\Asym^2(\C^d))=d(d-1)/2 \end{cases}. \]
The two states $\pi_s$ and $\pi_a$ on $(\C^d)^{\otimes 2}$ are therefore defined as
\[ \pi_s:=\frac{2}{d(d+1)}P_{\Sym^2(\C^d)}\ \text{and}\ \pi_a:=\frac{2}{d(d-1)}P_{\Asym^2(\C^d)}. \]
And we define next, for each $0\leq\lambda\leq 1$, the Werner state $\rho_{\lambda}:=\lambda\pi_s+(1-\lambda)\pi_a$ on $(\C^d)^{\otimes 2}$.
\smallskip

This one-parameter family of states was introduced in \cite{Werner} in order to understand the relation between entanglement and violation of Bell inequalities. It has received considerable interest since then. Indeed, due to their symmetry property, quantities which may be hard to compute in general become much easier to analyze for Werner states. And still, permutation-invariance is an assumption which is more often than not natural to make, so that they encompass a wide enough range of situations. For instance, separability vs entanglement is simple to characterize for Werner states, because it coincides with their being positive under partial transposition or not (see Chapter \ref{chap:QIT}, Section \ref{sec:sep-multi}, for definitions). The result is that $\rho_{\lambda}$ is separable for $1/2\leq\lambda\leq 1$ and entangled for $0\leq\lambda<1/2$ \cite{Werner}.

\section{Classical and quantum de Finetti type theorems}
\label{sec:deFinetti}

The motivation behind all de Finetti type theorems is to reduce the study of permutation-invariant scenarios to that of i.i.d.~ones.
This is precisely what the seminal classical finite de Finetti theorem, proved by Diaconis and Freedman in \cite{DF}, enables. Indeed, it tells us that the marginal probability distribution (in a few random variables) of an exchangeable probability distribution (in a lot of random variables) is well-approximated by a convex combination of product probability distributions. More precisely, we have the quantitative error-bound provided by Theorem \ref{th:dFclassical} below.

\begin{theorem}[Classical finite de Finetti theorem, \cite{DF}] \label{th:dFclassical}
Let $P^{(n)}$ be an exchangeable probability distribution in $n$ random variables, meaning that, for any $\pi\in\mathfrak{S}(n)$, $P^{(n)}\circ\pi=P^{(n)}$. For any $k\leq n$, denote by $P^{(k)}$ the marginal probability distribution of $P^{(n)}$ in $k$ random variables.
Then, there exists a probability distribution $\mu$ on the set of  probability distributions in $1$ random variable such that
\[ \left\|P^{(k)}-\int_{Q}Q^{\otimes k}\mathrm{d}\mu(Q)\right\|_1\leq \frac{k^2}{n}. \]
\end{theorem}

\smallskip

The first quantum analogue of this result was established by Christandl, K\"{o}nig, Mitchison and Renner in \cite{CKMR}. The statement is of similar spirit: the reduced state (on a few subsystems) of a permutation-invariant state (on a lot of subsystems) is well-approximated by a convex combination of product states. Again, the worst-case error can be quantitatively upper-bounded, which is the content of Theorem \ref{th:dFquantum} below.

\begin{theorem}[Quantum finite de Finetti theorem, \cite{CKMR}] \label{th:dFquantum}
Let $\rho^{(n)}$ be a permutation-invariant state on $(\C^d)^{\otimes n}$, meaning that, for any $\pi\in\mathfrak{S}(n)$, $U(\pi)\rho^{(n)}U(\pi)^{\dagger}=\rho^{(n)}$. For any $k\leq n$, denote by $\rho^{(k)}=\tr_{(\C^d)^{\otimes n-k}}\rho^{(n)}$ the reduced state of $\rho^{(n)}$ on $(\C^d)^{\otimes k}$. Then, there exists a probability distribution $\mu$ on the set of states on $\C^d$ such that
\[ \left\|\rho^{(k)}-\int_{\sigma}\sigma^{\otimes k}\mathrm{d}\mu(\sigma)\right\|_1\leq \frac{2kd^2}{n}. \]
\end{theorem}

\smallskip

Note that the main difference between these two theorems is that the classical one applies to probability distributions on an infinite size alphabet while the quantum one only applies to quantum states on a finite-dimensional Hilbert space. And it is known that the appearance of the dimension in the upper bound is not an artefact of the proof techniques: there actually exist permutation-invariant states $\rho^{(n)}$ on $(\C^d)^{\otimes n}$ which are such that the distance of $\rho^{(k)}$ to the set of separable states on $(\C^d)^{\otimes k}$ scales as $kd/n$ (see \cite{CKMR}, Section C, or \cite{NOP1}, Section III, for such examples).
\smallskip

In Chapter \ref{chap:k-extendibility}, a hierarchy of necessary conditions for separability, based on symmetric extensions, is studied in extensive depth. And it is Theorem \ref{th:dFquantum} under this precise form which allows to prove that this hierarchy converges to separability. However, in many applications, one does not actually need such a strong approximation result: knowing only that the considered permutation-invariant state can be upper bounded by a convex combination of tensor power states (up to some multiplicative factor $C$) is enough. Indeed, assume that you have such an operator-ordering, and that you know that a given order-preserving linear form $f$ satisfies $f\leq\epsilon$, for some $0<\epsilon<1$, on $1$-particle states. Then, you can conclude that $f^{\otimes n}\leq C\epsilon^n$ on permutation-invariant $n$-particle states, which decays exponentially to $0$ with $n$. These de Finetti results of different kind are usually referred to as \textit{de Finetti reductions} or \textit{post-selection lemmas}. Establishing and applying appropriate variants of them is at the heart of Chapters \ref{chap:deFinetti} and \ref{chap:SNOS}.

\chapter{Asymptotic geometric analysis toolbox}
\label{chap:toolbox}

This chapter gathers several notions from asymptotic geometric analysis that will be used in many places of this manuscript. Section \ref{ap:convex-geometry} introduces standard definitions, notation and results from classical convex geometry, in particular related to estimating the size of a convex body, which is something that we will have to do at various occasions. Section \ref{ap:deviations} summarizes two basic incarnations of the concentration of measure phenomenon (taking the view point of either functions on a sphere or sums of independent random variables), on which we shall build later more elaborate deviation estimates, tailored to our specific needs. Additional functional analytic tools (e.g.~from random matrix theory or from the local theory of Banach spaces), which play a more sporadic role in this manuscript, will be introduced only in due time.

\section{Classical convex geometry}
\label{ap:convex-geometry}

The standard convex geometry concepts expounded in this section will be crucial tools in (at least part of) Chapters \ref{chap:zonoids}, \ref{chap:data-hiding}, \ref{chap:SDrelaxations} and \ref{chap:k-extendibility}. The reader is e.g.~referred to the lecture notes \cite{Ball} or \cite{Vershynin} for a detailed and accessible presentation.

\subsection{Some vocabulary}

We work in the Euclidean space $\R^n$, where we denote by $\|\cdot\|$ the Euclidean norm, and by $B^n$, resp.~$S^{n-1}$, the associated unit ball, resp.~sphere. We denote by $\vol_n(\cdot)$ or simply $\vol(\cdot)$ the
$n$-dimensional Lebesgue measure. A {\em convex body} $K \subset \R^n$ is a convex compact set with non-empty interior.
A convex body $K$ is {\em symmetric} if $K=-K$.

\smallskip

The \textit{gauge} (or \textit{Minkowski functional}) associated to
a convex body $K$ in $\R^n$ is the function $\|\cdot\|_K$ defined for $x \in \R^n$ by
\[ \|x\|_K := \inf \{ t \geq 0 \st x \in tK \}. \]
This is a norm if and only if $K$ is symmetric. In such case, this actually defines a one-to-one correspondence between norms on $\R^n$ and symmetric convex bodies in $\R^n$: $K$ is simply the unit ball for $\|\cdot\|_K$. This identification has the following elementary properties: $K\subset L$ if and only if $\|\cdot\|_K\geq\|\cdot\|_L$, and for all $t>0$, $\|\cdot\|_{tK}=\|\cdot\|_K/t$.

\smallskip

If $K \subset \R^n$ is a convex body with origin in its interior,
the {\em polar} of $K$ is the convex body $K^\circ$ defined as
\[ K^\circ := \{ y \in \R^n \st \forall\ x \in K,\ \langle x,y \rangle \leq 1 \} .\]
In the symmetric case, the norms $\|\cdot\|_K$ and $\|\cdot\|_{K^\circ}$ are dual to each other, meaning that
\[ \forall\ x\in\R^n,\ \|x\|_{K^{\circ}} = \sup\{ \langle x,y \rangle \st \|y\|_K\leq 1 \} = \sup\{ \langle x,y \rangle \st y\in K \}. \]
If $x\in S^{n-1}$, we shall sometimes call the quantity defined above the {\em support
function} of $K$ in the direction $x$, and use the notation
\[ h_K(x):= \|x\|_{K^\circ}. \]
Note that $h_K(x)$ is the
distance from the origin to the hyperplane tangent to $K$ in the
direction $x$.

\smallskip

Two global invariants associated to a convex body $K \subset \R^n$, the \textit{volume radius} and the \textit{mean width}, will play an important role in many of our proofs.

\begin{definition}
\label{def:vrad}
The volume
radius of a convex body $K \subset \R^n$ is defined as
\[ \vrad (K) := \left( \frac{\vol K}{\vol B^n} \right)^{1/n}.\]
In words, $\vrad(K)$ is the radius of the Euclidean ball with same volume as $K$.
\end{definition}

\begin{figure}[h] \caption{Width of the convex body $K$ in the direction $u$}
\label{fig:width-direction}
\begin{center}
\includegraphics[width=5cm]{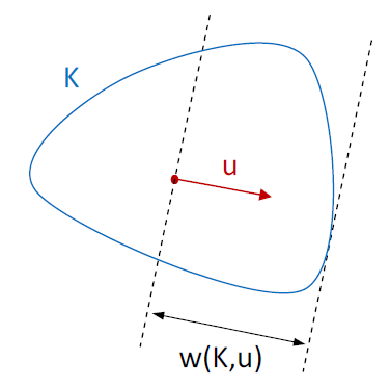}
\end{center}
\end{figure}

\begin{definition}
\label{def:w}
The mean width of a subset $K \subset \R^n$ is defined as
\[ w (K):=\int _{S^{n-1}} \max_{x\in K} \langle u,x \rangle \,\mathrm{d}\sigma(u) ,\]
where $\sigma$ is the uniform probability measure on $S^{n-1}$.
In words, $w(K)$ is the average, for $u$ uniformly distributed on $S^{n-1}$, of the (half-)width of $K$ in the direction $u$ (see Figure \ref{fig:width-direction}).
\end{definition}
If $K$ is a convex body, we have
\[ w (K)=\int _{S^{n-1}}h_K(u)\,\mathrm{d}\sigma(u)=\int _{S^{n-1}} \|u\|_{K^\circ} \, \mathrm{d}\sigma(u). \]
Moreover, one can check that the mean width satisfies the additivity property $w(K+L)=w(K)+w(L)$, for any $K,L\subset \R^n$ bounded.

\smallskip

The inequality below (see e.g.~\cite{Pisier}, Corollary 1.4, for a proof) is a
fundamental result which compares the volume radius and the mean
width. It asserts that, among sets of given
volume, the mean width is minimized for Euclidean balls.

\begin{theorem}[Urysohn inequality]
\label{theorem:urysohn}
For any convex body $K \subset \R^n$, we have
\[ \vrad(K) \leq w(K) .\]
\end{theorem}

\smallskip

It is convenient to compute the mean width using Gaussian rather than spherical integration. Let $G$ be a standard Gaussian vector in $\R^n$,
i.e.~such that its
coordinates, in any orthonormal basis, are independent and following a Gaussian distribution with mean $0$ and variance $1$. Denoting $\gamma_n = \E \|G\| \sim_{n\rightarrow+\infty} \sqrt{n}$, we have,
for any compact set
$K \subset \R^n$,
\[ w_G(K) := \E \max_{x \in K} \langle G,x \rangle = \gamma_n w(K) .\]
The Gaussian mean width has the advantage over the spherical one of being independent from the ambient
dimension. Indeed, if $K\subset E$ is a compact set in a subspace $E$ of $\R^n$, the value
of $w_G(K)$ does not depend on whether it is computed in $E$ or in $\R^n$, contrary to $w(K)$. It is also usually easier to compute. For example, it allows to compute the mean width of a segment: if $u \in S^{n-1}$ is a
unit vector, then
\[ \alpha_n := w (\conv \{ \pm u\}) = \frac{1}{\gamma_n} \sqrt{\frac{2}{\pi}} \underset{n\rightarrow+\infty}{\sim} \sqrt{\frac{2}{\pi n}}. \]
It also shows how to control the mean width of an intersection or a projection. Let $K \subset \R^n$ be a compact set, and $E \subset \R^n$ be a $k$-dimensional subspace.
Denoting by $P_E$ the orthogonal projection onto $E$, we have
$w_G(P_E K) \leq w_G(K)$, and therefore
\begin{equation} \label{eq:width-projection}
w(K \cap E) \leq w(P_E K) \leq \frac{\gamma_n}{\gamma_k} w(K).
\end{equation}

\smallskip

We will also need the two following lemmas, which are incarnations of the familiar ``union bound''. Lemma \ref{lemma:mean-width-polytope} appears for example in \cite{LT}, Chapter 3, under the equivalent formulation via suprema of Gaussian processes. Lemma \ref{lemma:mean-width-union-bound} on the other hand, seems to appear nowhere in the literature, even under a different phrasing. Since it is a result that we will use at several occasions (in Chapters \ref{chap:data-hiding} and \ref{chap:SDrelaxations}) we write a short proof of it for completeness. The route we follow is the ``usual'' one in that setting, detailed e.g. in \cite{LT}, Chapter 3, to which the reader is referred for the details that we may pass over.

\begin{lemma}[Bounding the mean width of a polytope] \label{lemma:mean-width-polytope}
Let $v_1,\ldots,v_N$ be points in $\R^n$ such that $v_i\in\lambda B^n$ for every index $1\leq i\leq N$, for some $\lambda\geq 0$. Then,
\[ w\left(\conv(v_1,\ldots,v_N)\right) \leq \lambda \sqrt{\frac{2\ln N}{n}}. \]
\end{lemma}

\begin{lemma}[Bounding the mean width of a union] \label{lemma:mean-width-union-bound}
Let $K_1,\dots,K_N$ be convex sets in $\R^n$ such that $K_i \subset \lambda B^n$ for every index $1\leq i\leq N$, for some $\lambda\geq 0$. Then,
\[ w \left( \conv \left( \bigcup_{i=1}^N K_i \right) \right) \leq 2\left(\max_{1 \leq i \leq N} w(K_i) +  \lambda \sqrt{\frac{2\ln N}{n}} \right).\]
\end{lemma}

\begin{proof}
By homogeneity, it is enough to prove Lemma \ref{lemma:mean-width-union-bound} in the case $\lambda=1$. Now, for $G$ a standard Gaussian in $\R^n$, we have that, for any $\delta\geq 0$ (to be fixed later),
\[ \E \underset{1\leq i\leq N}{\max}\big| h_{K_i}(G)-w_G(K_i)\big| \leq \delta + \sum_{i=1}^N\int_{\delta}^{+\infty} \P\big( \big| h_{K_i}(G)-w_G(K_i)\big| >t \big)\mathrm{d}t. \]
Yet, by the Gaussian concentration inequality (see e.g. \cite{LT}, Chapter 1, and also Section \ref{ap:deviations} below for a more detailed exposition of closely related results), we know that, for each $1\leq i\leq N$,
\[ \forall\ t>0,\ \P\big( \big| h_{K_i}(G)-w_G(K_i)\big| >t \big) \leq e^{-t^2/2}. \]
This is because, by the assumption $K_i\subset B^n$, $h_{K_i}$ is a $1$-Lipschitz function:
\[ \forall\ G,H\in\R^n,\ \big| h_{K_i}(G) -  h_{K_i}(H) \big| \leq  h_{K_i}(G-H) \leq \|G-H\|. \]
We therefore have in the end
\[ \E \underset{1\leq i\leq N}{\max}\big| h_{K_i}(G)-w_G(K_i)\big| \leq \delta + N\int_{\delta}^{+\infty} e^{-t^2/2}\mathrm{d}t \leq \delta + N\sqrt{\frac{\pi}{2}}e^{-\delta^2/2}. \]
Choosing $\delta=\sqrt{2\ln N}$ in the above inequality eventually yields
\[ \E \underset{1\leq i\leq N}{\max} h_{K_i}(G) \leq 2\underset{1\leq i\leq N}{\max}w_G(K_i) + \sqrt{2\ln N} + \sqrt{\frac{\pi}{2}} \leq 2\left(\underset{1\leq i\leq N}{\max}w_G(K_i) + \sqrt{2\ln N}\right). \]
And going back to spherical rather than Gaussian variables gives precisely the advertised result.
\end{proof}

\subsection{Some volume inequalities}

We will use several times (in Chapters \ref{chap:data-hiding}, \ref{chap:SDrelaxations} and \ref{chap:k-extendibility}) the following result, established by Milman and Pajor.

\begin{theorem}[Milman--Pajor inequality, \cite{MP}, Corollary 3]
\label{theorem:Milman-Pajor}
If $K,L$ are convex bodies in $\R^n$ with the same center of mass, then
\[ \mathrm{vrad}(K\cap L)\mathrm{vrad}(K-L)\geq\mathrm{vrad}(K)\mathrm{vrad}(L),\]
where $K-L=\left\{x-y \st x\in K,\ y\in L\right\}$ stands for the Minkowski sum of the convex bodies $K$ and $-L$.
\end{theorem}

Choosing $K=-L$ in Theorem \ref{theorem:Milman-Pajor} yields the following corollary.

\begin{corollary}
\label{corollary:Milman-Pajor}
If $K$ is a convex body in $\R^n$ with center of mass at the origin, then
\[\mathrm{vrad}(K\cap -K)\geq\frac{1}{2}\mathrm{vrad}(K),\]
and more generally for any orthogonal transformation $\theta$,
\[\mathrm{vrad}(K\cap\theta(K))\geq\frac{1}{2}\frac{\mathrm{vrad}(K)^2}{w(K)}.\]
\end{corollary}

The latter inequality is simply because, on the one hand, $\mathrm{vrad}(\theta(K))=\mathrm{vrad}(K)$, and on the other hand, by Theorem \ref{theorem:urysohn}, $\mathrm{vrad}(K-\theta(K))\leq w(K-\theta(K)) = w(K)+w(\theta(K)) = 2w(K)$.

\smallskip

We will typically use Corollary \ref{corollary:Milman-Pajor} in the following way: if $K$ is a convex body with center of
mass at the origin which satisfies a ``reverse'' Urysohn inequality,
i.e.~$\mathrm{vrad}(K) \geq \alpha w(K)$ for some constant $0<\alpha<1$, we can conclude that the volume radius of $K \cap \theta(K)$
is comparable to the volume radius of $K$.

\smallskip

Another volume inequality which will be useful to us (in Chapter \ref{chap:data-hiding}) is the one below, due to Rogers and Shepard.

\begin{theorem}[Rogers--Shephard inequality, \cite{RS}] \label{theorem:rogers-shephard}
Let $u\in S^{n-1}$, $h>0$, and consider the affine hyperplane
\[ H = \{ x \in \R^n \st \scalar{x}{u} = h \} .\]
Let $K$ be a convex body inside $H$ and $L = \conv (K,-K)$. Then,
\[  2h \vol_{n-1}(K) \leq \vol_n(L) \leq 2h \vol_{n-1}(K) \frac{2^{n-1}}{n}. \]
Consequently,
\begin{equation} \label{eq:RS} \mathrm{vrad}(L) \simeq h^{1/n} \mathrm{vrad}(K)^{1-1/n}. \end{equation}
\end{theorem}

We can infer from equation \eqref{eq:RS} that for sets $K$ with ``reasonable'' volume (which will be the case of all the sets that we will consider)
$\mathrm{vrad}(K)$ and $\mathrm{vrad}(L)$ are comparable.

\section{Concentration of measure and deviation inequalities}
\label{ap:deviations}

\subsection{From individual to global concentration estimates}

Levy's Lemma guarantees that, in high dimension, regular enough functions typically do not deviate much from their average behaviour. We will use repeatedly variations and refinements of this general paradigm (in Chapter \ref{chap:data-hiding} a multi-variate version, in Chapter \ref{chap:SDrelaxations} a Gaussian rather than spherical version, in Chapter \ref{chap:k-extendibility} a local version). So let us recall precisely the seminal formulation here.

\begin{lemma}[Levy's Lemma for Lipschitz functions on the sphere, \cite{Levy}] \label{lemma:levy}
Let $n\in\N$. For any $L$-Lipschitz function $f:S^{n-1}\rightarrow\R$ and any $t>0$, if $x$ is uniformly distributed on $S^{n-1}$, then
\[ \P( | f(x) - \E f| > t ) \leq e^{-cn t^2/L^2}, \]
where $c>0$ is a universal constant.
\end{lemma}

Combining such type of individual deviation probability estimates and a discretization with \textit{nets} of reasonable size, one can then usually derive global deviation probability estimates via the union bound. Let us specify what we mean.

\begin{definition} Fix $\varepsilon>0$, and let $\|\cdot\|_{\sharp},\|\cdot\|_{\flat}$ be two norms in $\R^n$, with associated unit balls $B_{\sharp},B_{\flat}$. $\mathcal{A}$ is an $\varepsilon$-net for $\|\cdot\|_{\flat}$ within $B_{\sharp}$ if $\mathcal{A}\subset B_{\sharp}$ and, for all $x\in B_{\sharp}$, there exists $x'\in\mathcal{A}$ such that $\|x-x'\|_{\flat}\leq\varepsilon$.
\end{definition}

Usually, one is interested in having such a discretized version of a given unit ball with as few elements as possible. Now, just observing that an $\varepsilon$-separated set with maximal cardinality actually forms an $\varepsilon$-net, it is easy to get the following cardinality upper bound by simply comparing volumes.

\begin{lemma}[Bounding the cardinality of nets via a volumetric argument, \cite{Pisier}, Lemmas 4.16 and 4.10] \label{lemma:nets}
Fix $\varepsilon>0$, and let $\|\cdot\|_{\sharp},\|\cdot\|_{\flat}$ be two norms in $\R^n$, with associated unit balls $B_{\sharp},B_{\flat}$. Then, there exists an $\varepsilon$-net $\mathcal{A}$ for $\|\cdot\|_{\flat}$ within $B_{\sharp}$ satisfying
\[ |\mathcal{A}| \leq \frac{\vol\left((2/\varepsilon)B_{\sharp}+B_{\flat}\right)}{\vol\left(B_{\flat}\right)}. \]
So in particular, there exists an $\varepsilon$-net $\mathcal{A}$ for $\|\cdot\|_{\sharp}$ within $B_{\sharp}$ such that $|\mathcal{A}|\leq \left(1+2/\varepsilon\right)^n$.
\end{lemma}

Subsequently, making use of both Lemmas \ref{lemma:levy} and \ref{lemma:nets} in a careful way, one gets the celebrated Dvoretzky's Theorem, which is quoted below. This is precisely the kind of strategy we adopt in Chapters \ref{chap:channel-compression} and \ref{chap:zonoids} to go from individual to global concentration phenomenon.

\begin{lemma}[Dvoretzky's Theorem for Lipschitz functions on the sphere, \cite{Milman,Gordon,Sche}] \label{lemma:dvo}
Let $n\in\N$. For any symmetric $L$-Lipschitz function $f:S^{n-1}\rightarrow\R$ and any $t>0$, if $H$ is a uniformly distributed $Cnt^2/L^2$-dimensional subspace of $\R^n$, with $C>0$ a universal constant, then
\[ \P\left(\exists\ x\in H\cap S^{n-1}:\ | f(x) - \E f| > t \right) \leq e^{-cn t^2/L^2}, \]
where $c>0$ is a universal constant.
\end{lemma}

This tangible version of Dvoretzky's theorem is essentially due to Milman \cite{Milman}, but with a dependence on $t$ in the maximal dimension of $H$ which scales as $t^2/\log(1/t)$. This extra logarithmic factor was later removed by Gordon in \cite{Gordon} and by Schechtman in \cite{Sche} (the latter proof is based on concentration of measure, like the original one, while the former proof uses instead comparison inequalities for Gaussian processes).

\subsection{$\psi_{\alpha}$ random variables}

The reader is for instance referred to the review \cite{CGLP}, Chapter 1, for a complete presentation of the theory of Orlicz spaces. A particular instance of these are the so-called $L_{\psi_{\alpha}}$ spaces, of which we gather here only the few properties that we will need to exploit.
\smallskip

For any $\alpha\geq 1$, a random variable $X$ is called a $\psi_{\alpha}$ random variable if its $\psi_{\alpha}$-norm $\|X\|_{\psi_{\alpha}}$ is finite. The latter may be defined in several equivalent ways, the most standard one being through the Orlicz function $x \mapsto \exp(x^{\alpha})-1$. This characterization is not the one that will be most practical for us though, and we shall rather use the one via the growth of absolute moments, which leads to the equivalent norm (see e.g.~\cite{CGLP}, Corollary 1.1.6)
\[ \|X\|_{\psi_{\alpha}} = \sup_{p \in \N}\frac{\big(\E |X|^p\big)^{1/p}}{p^{1/\alpha}} .\]
A $\psi_1$, resp.~$\psi_2$, random variable is also referred to as as sub-exponential, resp.~sub-gaussian. Indeed, the $\psi_1$-norm, resp.~ $\psi_2$-norm, of a random variable quantifies the exponential, resp.~gaussian, decay of its tail.
\smallskip

As a crucial tool in several of our coming reasonings (in Chapters \ref{chap:channel-compression} and \ref{chap:zonoids}), we will need the Bernstein-type deviation inequality for a sum of independent $\psi_1$ random variables which is quoted below (see e.g.~\cite{CGLP}, Theorem 1.2.5, for a proof).

\begin{theorem}[Bernstein's inequality]
\label{th:Bernstein}
Let $X_1,\ldots,X_N$ be $N$ independent centered $\psi_1$ random variables.
Setting $M=\max_{1\leq i\leq N}\|X_i\|_{\psi_1}$ and $\sigma^2=\sum_{1\leq i\leq N}\big(\|X_i\|_{\psi_1}\big)^2/N$, we have
\[ \forall\ t>0,\ \P\left(\left|\frac{1}{N}\sum_{i=1}^N X_i \right| >t \right)\leq 2\exp\left(-c N\min\left(\frac{t^2}{\sigma^2},\frac{t}{M}\right) \right),\]
where $c>0$ is a universal constant.
\end{theorem}

We see in Theorem \ref{th:Bernstein} above that, for a sum of independent $\psi_1$ random variables, there are two distinct regimes which enter the picture in the behaviour of the tail: sub-gaussian for moderate deviations (when the central limit phenomenon dominates) and sub-exponential for large deviations (when the prominent role is played by the tails of the individual variables).


\part{Complexity reduction in quantum information theory}
\label{part:complexity}

A central issue in asymptotic geometric analysis is that of quantifying how close a given, potentially complex, convex body is to one which is easier to describe. There are several reasons why a convex body may be considered simple: because there is an efficient way of deciding membership for it (algorithmic point of view), because it is the unit ball for a Euclidean norm or because it has few vertices/faces (geometric point of view), because it can be covered with few Euclidean balls (entropic point of view) etc. There are also several notions of distance between convex bodies that one might consider depending on the context. In any case, two sample (dual) take-home messages from asymptotic geometric analysis are: given a high dimensional convex body, firstly most of its information is already encoded in its projection onto a lower dimensional subspace, and secondly most of its complexity disappears when looking at its section by a lower dimensional subspace. Mathematically, these two results are known, respectively, as the Johnson--Lindenstrauss lemma and as the Dvoretzky theorem. From an information theory standpoint, they have a clear data-compression interpretation, with an achievability side (preservation of almost all information when projecting on a subspace of dimension above a certain value) and a converse side (loss of almost all information when intersecting with a subspace of dimension below a certain value).

\smallskip

Similar considerations appear in many contexts of quantum information theory. Phrased very generally, a natural wonder would usually be: given an ideal process, with many (even potentially infinitely many) degrees of freedom, is it possible to approximate it by a more realistic one, i.e.~one which can be described with few parameters? Or in other words: by just allowing some small error, can we execute a task which potentially requires a lot of resources with much less resources? In this part, we take a closer look at two such problems: the question of compressing a quantum channel into one with a small environment is treated in Chapter \ref{chap:channel-compression}, and the question of sparsifying a quantum measurement into one with few outcomes is treated in Chapter \ref{chap:zonoids}. Here is a summary of the main results in each of them.

\smallskip

Chapter \ref{chap:channel-compression} deals with the following problem: given a quantum channel $\cN$, find a quantum channel $\widehat{\cN}$ with environment as small as possible such that, for any input state $\rho$, the output states $\cN(\rho)$ and $\widehat{\cN}(\rho)$ are close to one another. We investigate different notions of closeness: standard ones quantified in terms of Schatten $p$-norm distance, but also stronger ones defined in terms of operator ordering. Those are especially well-suited for then deriving closeness results involving, for instance, entropic output quantities. In brief, our main result is that any channel $\cN$ with input and output dimensions $d$ can be approximated, say in $(1{\rightarrow}1)$-norm distance, by a channel $\widehat{\cN}$ with environment dimension $O(d\log d)$, hence much smaller than the a priori environment dimension $d^2$ of $\cN$. In the case where $\cN$ is sufficiently noisy (meaning that all its output states are sufficiently mixed), this result can be improved to $O(d)$. Such dimensional dependence is shown to be optimal, contrary to the general one for which we do not know whether or not the $\log d$ factor can be removed. On the technical side, all our statements stem, as a first crucial step, from large deviation inequalities for sums of independent sub-exponential random variables.

\smallskip

In Chapter \ref{chap:zonoids}, attention is focussed on a particular kind of quantum channels, namely quantum-classical channels (i.e.~POVMs). However, the notion of approximation which is investigated there is completely different from that of Chapter \ref{chap:channel-compression}. Let us specify what we mean. One can associate to any POVM $\mathrm{M}$ its so-called \textit{distinguishability norm} $\|\cdot\|_{\mathrm{M}}$, which quantifies how well it performs in the task of discriminating two quantum states. Given a POVM $\mathrm{M}$, with potentially many (or even infinitely many) outcomes, we are interested in constructing a POVM $\mathrm{M}'$, with as few outcomes as possible, behaving almost as the POVM $\mathrm{M}$ in terms of distinguishability norm (in the sense that $\|\cdot\|_{\mathrm{M}}\simeq\|\cdot\|_{\mathrm{M}'}$, up to some small error). This notion of closeness between POVMs thus has an operational significance: two POVMs are comparable if they yield comparable biases on any pair of states to be discriminated. Our first realization is that the unit ball for a POVM's distinguishability norm is a very particular object: precisely, its polar is, in general, a \textit{zonoid}, and in the case of a discrete POVM, a \textit{zonotope}. What kind of convex body are these? A zonotope is the Minkowski sum of a finite number of segments, while a zonoid is the limit of a sequence of zonotopes (in Hausdorff distance). So our problem can almost be rephrased as: to approximate a zonoid by a zonotope, how many segments are needed? And this turns out to be a well-studied topic in classical convex geometry. Exporting these ideas from the local theory of Banach spaces, we are able to prove the following main results: on $\C^d$, a POVM which is symmetric enough (such as e.g.~the uniform POVM, the most symmetric one) can be sparsified by a POVM with $O(d^2)$ outcomes, and any POVM can be sparsified by a sub-POVM with $O(d^2\log d)$ outcomes. The dimensional dependence in the former result is optimal, and we leave it open whether or not the latter result can be improved. What is more, we can also deal with the multipartite setting, after considering the appropriate notion of tensor product for zonoids. We establish the, desirable but not a priori obvious, fact that local POVMs can be sparsified locally.

\smallskip

Looking at it more closely afterwards, there are analogies between Chapters \ref{chap:channel-compression} and \ref{chap:zonoids} which go beyond their common complexity reduction motivation. To begin with, we propose in both cases a universal random construction which has the property of working optimally in very balanced situations (such as the fully randomizing channel or the uniform POVM), but maybe not as well in very unbalanced ones. Hence the question of coming up with more adaptive schemes in one instance or the other. Note that this issue is far from being specific to the two problems we are dealing with. On the contrary, it is for instance a recurrent one when trying to embed a Banach space into another one almost isometrically: how can their geometry be taken into account in a clever way? Furthermore, the routes we follow in both chapters in order to establish our main theorems have the exact same spirit (which is, admittedly, a very standard one): first we prove concentration for one fixed data point (which follows from certain random variables having a nice sub-exponential behaviour), and then we derive concentration for all data points by discretizing our data set with a well-chosen net. The reader is referred to Chapter \ref{chap:toolbox}, Section \ref{ap:deviations}, for the archetypal example of this strategy, namely the derivation of Dvoretzky's theorem from Levy's lemma.

\newpage
\textbf{\LARGE{Part II -- Table of contents}}
\parttoc

\chapter{Quantum channel compression}
\chaptermark{Quantum channel compression}
\label{chap:channel-compression}

\textsf{Based on ``Quantum channel compression'' \cite{Lancien-et-al}.}

\bigskip

We study the problem of approximating a quantum channel by one with as few Kraus operators as possible (in the sense that, for any input state, the output states of the two channels should be close to one another). Our main result is that any quantum channel mapping states on some input Hilbert space $\A$ to states on some output Hilbert space $\B$ can be compressed into one with order $d\log d$ Kraus operators, where $d=\max(|\A|,|\B|)$, hence much less than $|\A||\B|$. In the case where the channel's outputs are all very mixed, this can be improved to order $d$. We discuss the optimality of this result as well as some consequences.

\section{Introduction}
\label{sec:channel-intro}

Quantum channels are the most general framework in which the transformations that a quantum system may undergo are described. These are defined as completely positive and trace preserving (CPTP) maps from the set of bounded operators on some input Hilbert space $\A$ to the set of bounded operators on some output Hilbert space $\B$. Indeed, to be a physically valid evolution in the open system setting, a linear map $\cN$ has to preserve quantum states (i.e.~positive semi-definiteness and unit-trace conditions) even when tensorized with the identity map $\mathcal{I\!d}$ on an auxiliary system. The reader is referred to Chapter \ref{chap:QIT}, Section \ref{sec:channels}, for a more developed exposition.

In the remainder of this chapter, we shall use the general notation introduced in Chapter \ref{chap:motivations}, Section \ref{sec:notation} (in particular concerning operators and operator-norms on Hilbert spaces). Also, given a finite-dimensional Hilbert space $\mathrm{H}$ (which will be the case of all the Hilbert spaces we will deal with in the sequel) we shall denote by $|\mathrm{H}|$ its dimension.

So assume from now on that the Hilbert spaces $\A$ and $\B$ are finite-dimensional. Then, we know by Choi's representation theorem \cite{Choi} that a CPTP map $\mathcal{N}:\cL(\A)\rightarrow\cL(\B)$ can always be written as
\begin{equation} \label{eq:Kraus} \mathcal{N}: X\in\cL(\A) \mapsto \sum_{i=1}^s K_i X K_i^{\dagger} \in\cL(\B), \end{equation}
where the operators $K_i:\A\rightarrow\B$, $1\leq i\leq s$, are called the Kraus operators of $\mathcal{N}$ and satisfy the normalization relation $\sum_{i=1}^s K_i^{\dagger}K_i = \Id_{\A}$. The minimal $s\in\N$ such that $\mathcal{N}$ can be decomposed in the Kraus form \eqref{eq:Kraus} is called the Kraus rank of $\mathcal{N}$, which we shall denote by $r_K(\mathcal{N})$. By Stinespring's dilatation theorem \cite{Stinespring}, another alternative way of characterizing a CPTP map $\mathcal{N}:\cL(\A)\rightarrow\cL(\B)$ is as follows
\begin{equation} \label{eq:Stinespring} \mathcal{N}: X\in\cL(\A) \mapsto \tr_{\rE}\left(VXV^{\dagger}\right) \in\cL(\B), \end{equation}
for some environment Hilbert space $\rE$ and some isometry $V:\A\hookrightarrow\B\otimes\rE$ (i.e.~$V^{\dagger}V=\Id_{\A}$). In such picture, $r_K(\mathcal{N})$ is then nothing else than the minimal environment dimension $|\rE|\in\N$ such that $\mathcal{N}$ may be expressed in the Stinespring form \eqref{eq:Stinespring}. It may be worth pointing out that there is a lot of freedom in representation \eqref{eq:Kraus}: two sets of Kraus operators $\{K_i,\ 1\leq i\leq s\}$ and $\{L_i,\ 1\leq i\leq s\}$ give rise to the same quantum channel as soon as there exists a unitary $U$ on $\C^s$ such that, for all $1\leq i\leq s$, $L_i=\sum_{j=1}^sU_{ij}K_j$. On the contrary, representation \eqref{eq:Stinespring} is essentially unique, up to the (usually irrelevant) transformation $V\mapsto(\Id\otimes U) V$, for $U$ a unitary on $\rE$. That is why we will often prefer working with the latter than with the former.

Yet another way of viewing the Kraus rank of a CPTP map $\mathcal{N}:\cL(\A)\rightarrow\cL(\B)$ is as the rank of its associated Choi-Jamiolkowski state. Denoting by $\psi$ a maximally entangled state on $\A\otimes\A$, i.e. $\ket{\psi}=\sum_{i=1}^{|\A|}\ket{ii}/\sqrt{|\A|}$ for $\{\ket{i},\ 1\leq i\leq|\A|\}$ an orthonormal basis of $\A$, the latter is defined as the state $\tau(\cN) = \mathcal{I\!d}\otimes\cN(\psi)$ on $\A\otimes\B$.
Consequently, it holds that any quantum channel from $\A$ to $\B$ has Kraus rank at most $|\A||\B|$. And the extremal such quantum channels have Kraus rank less than $|\A|$. In particular, the case $r_K(\cN)=1$ corresponds to $\cN$ being a unitary, hence reversible, evolution, whereas whenever $r_K(\cN)>1$, one can view $\cN$ as a noisy summary of a unitary evolution on a larger system. The Kraus rank of a quantum channel can thus legitimately be seen as a measure of its ``complexity'': it quantifies the minimal amount of ancillary resources needed to implement it (or equivalently the amount of degrees of freedom in it that one is ignorant of). A natural question in this context would therefore be: given any quantum channel, is it possible to reduce its complexity while not affecting too much its action, or in other words to find a channel with much smaller Kraus rank which approximately simulates it?

One last definition we shall need concerning CP maps is the following: the conjugate (or dual) of a CP map $\cN:\cL(\A)\rightarrow\cL(\B)$ is the CP map $\cN^*:\cL(\B)\rightarrow\cL(\A)$ defined by
\[ \forall\ X\in\cL(\A),\ \forall\ Y\in\cL(\B),\ \tr(\cN(X)Y) = \tr(X\cN^*(Y)). \]
It is characterized as well by saying that $\{K_i,\ 1\leq i\leq s\}$ is a set of Kraus operators for $\cN$ if and only if $\{L_i=K_i^{\dagger},\ 1\leq i\leq s\}$ is a set of Kraus operators for $\cN^*$. Hence obviously, $\cN$ and $\cN^*$ have same Kraus rank, while the trace-preservingness condition $\sum_{i=1}^sK_i^{\dagger}K_i=\Id$ for $\cN$ is equivalent to the unitality condition $\sum_{i=1}^sL_iL_i^{\dagger}=\Id$ for $\cN^*$.

\section{Quantum channel approximation: definitions and already known facts}

Before going any further, we need to specify a bit what we mean by ``approximating a quantum channel'', since indeed, several definitions of approximation may be considered. In our setting, the most natural one is probably that of approximation in $(1{\rightarrow}1)$-norm: given CPTP maps $\mathcal{N},\widehat{\mathcal{N}}:\cL(\A)\rightarrow\cL(\B)$, we will say that $\widehat{\cN}$ is an $\varepsilon$-approximation of $\cN$ in $(1{\rightarrow}1)$-norm, where $\varepsilon>0$ is some fixed parameter, if
\begin{equation} \label{eq:def_1->1} \forall\ \rho\in\cD(\A),\ \left\|\widehat{\cN}(\rho)-\cN(\rho)\right\|_1 \leq \varepsilon. \end{equation}
One could think at first sight that an even more natural error quantification in such context would be in terms of completely-bounded $(1{\rightarrow}1)$-norm (aka diamond norm). That is, in order to call $\widehat{\cN}$ an $\varepsilon$-approximation of $\cN$, we would require that, for any Hilbert space $\A'$,
\begin{equation} \label{eq:def_cb-1->1} \forall\ \rho\in\cD(\A\otimes\A'),\ \left\|\widehat{\cN}\otimes\mathcal{I\!d}(\rho)-\cN\otimes\mathcal{I\!d}(\rho)\right\|_1 \leq \varepsilon. \end{equation}
Nevertheless, this notion of approximation is too strong for our purposes. Indeed, if $\mathcal{N}$ and $\widehat{\mathcal{N}}$ satisfy equation \eqref{eq:def_cb-1->1}, it implies in particular that their associated Choi-Jamiolkowski states have to be $\varepsilon$-close in trace-norm distance. And this, in general, is possible only if $\mathcal{N}$ and $\widehat{\mathcal{N}}$ have a number of Kraus operators which scale the same: for instance, if $\tau(\cN)=P/r_K(\cN)$ with $P$ a projector on a $r_K(\cN)$-dimensional subspace of $\A\otimes\B$, then $\|\tau(\widehat{\cN})-\tau(\cN)\|_1\leq\varepsilon$ can hold only if $r_K(\widehat{\cN})\geq (1-\varepsilon/2)\,r_K(\cN)$. So no environment dimensionality reduction can be achieved in that sense.

The question of quantum channel compression has already been studied in one specific case, which is the one of the fully randomizing (or depolarizing) channel. Let us recall what is known there. The fully randomizing channel $\mathcal{R}:\cL(\A)\rightarrow\cL(\A)$ is the CPTP map with same input and output spaces defined by
\[ \mathcal{R}:X\in\cL(\A)\mapsto (\tr X)\frac{\Id}{|\A|} \in\cL(\A), \]
so that, in particular, all input states $\rho\in\cD(\A)$ are sent to the maximally mixed state $\Id/|\A|\in\cD(\A)$. $\mathcal{R}$ has maximal Kraus rank $|\A|^2$ (because $\tau(\mathcal{R})$ is simply $\Id/|\A|^2$, and hence has rank $|\A|^2$). This was of course to be expected, if adhering to the intuitive idea that the bigger is the Kraus rank of channel, the noisier is the channel. One possible minimal Kraus decomposition for $\mathcal{R}$ is
\[ \mathcal{R}:X\in\cL(\A)\mapsto  \frac{1}{|\A|^2}\sum_{i,j=1}^{|\A|} V_{ij}X V_{ij}^{\dagger} \in\cL(\A), \]
where for each $1\leq i,j\leq |\A|$, $V_{ij}=\Sigma_x^j\Sigma_z^k$ with $\Sigma_x,\Sigma_z$ the generalized Pauli operators on $\A$.
It was initially established in \cite{HLSW} and later improved in \cite{Aubrun2} that there exist almost randomizing channels with drastically smaller Kraus rank. More specifically, the following was proved: for any $0<\varepsilon<1$, the CPTP map $\mathcal{R}$ can be $\varepsilon$-approximated in $(1{\rightarrow}1)$-norm by a CPTP map $\widehat{\mathcal{R}}$ with Kraus rank at most $C|\A|/\varepsilon^2$, where $C>0$ is a universal constant. Actually, something stronger was established, namely
\[ \forall\ \rho\in\cD(\A),\ \left\|\widehat{\mathcal{R}}(\rho)-\mathcal{R}(\rho)\right\|_{\infty} \leq \frac{\varepsilon}{|\A|}, \]
which obviously implies that, for any $1\leq p\leq\infty$, $\widehat{\mathcal{R}}$ is an $\varepsilon$-approximation of $\mathcal{R}$ in $(1{\rightarrow}p)$-norm, in the sense that
\begin{equation} \label{eq:def_1->p} \forall\ \rho\in\cD(\A),\ \left\|\widehat{\mathcal{R}}(\rho)-\mathcal{R}(\rho)\right\|_p \leq \frac{\varepsilon}{|\A|^{1-1/p}}. \end{equation}

The question we investigate here is whether such kind of statement actually holds true for any channel. Note however that, for a channel which is not the fully randomizing one, the notion of approximation in Schatten $p$-norm appearing in equation \eqref{eq:def_1->p} is maybe not what we would expect as being the ``correct'' one. In fact, it would seem more accurate to quantify closeness in terms of relative error. Hence, given a CPTP map $\mathcal{N}:\cL(\A)\rightarrow\cL(\B)$, we would rather be interested in finding a CPTP map $\widehat{\mathcal{N}}:\cL(\A)\rightarrow\cL(\B)$ with Kraus rank as small as possible, and such that
\[ \forall\ \rho\in\cD(\A),\ \left\|\widehat{\cN}(\rho)-\cN(\rho)\right\|_p \leq \varepsilon\left\|\cN(\rho)\right\|_p. \]

\section{Statement of the main results}

\begin{theorem} \label{th:main}
Fix $0<\varepsilon<1$ and let $\cN:\cL(\A)\rightarrow\cL(\B)$ be a CPTP map with Kraus rank $|\rE|\geq |\A|,|\B|$. Then, there exists a CP map $\widehat{\cN}:\cL(\A)\rightarrow\cL(\B)$ with Kraus rank at most $C\max(|\A|,|\B|)\log(|\rE|/\varepsilon)/\varepsilon^2$ (where $C>0$ is a universal constant) and such that
\begin{equation} \label{eq:approxmain} \forall\ \rho\in\cD(\A),\ -\varepsilon\left(\cN(\rho) +\frac{\Id}{|\B|}\right) \leq \widehat{\cN}(\rho)-\cN(\rho)\leq \varepsilon\left(\cN(\rho) +\frac{\Id}{|\B|}\right). \end{equation}
\end{theorem}

\begin{remark} \label{remark:schatten}
Note that if $\widehat{\cN}$ satisfies equation \eqref{eq:approxmain}, then it especially implies that it approximates $\cN$ in any Schatten-norm in the following sense
\[ \forall\ p\in\N,\ \forall\ \rho\in\cD(\A),\ \left\|\widehat{\cN}(\rho)-\cN(\rho)\right\|_p \leq \varepsilon\left(\left\|\cN(\rho)\right\|_p+\frac{1}{|\B|^{1-1/p}}\right). \]
In particular, we have the worth pointing out $(1{\rightarrow}1)$-norm approximation of $\cN$ by $\widehat{\cN}$
\[ \forall\ \rho\in\cD(\A),\ \left\|\widehat{\cN}(\rho)-\cN(\rho)\right\|_1 \leq 2\varepsilon,  \]
in which we can further impose that $\widehat{\cN}$ is strictly, and not just up to an error $2\varepsilon$, trace preserving (cf.~the proof of Theorem \ref{th:main}).
\end{remark}

One important question at that point is the one of optimality in Theorem \ref{th:main}. A first obvious observation to make in order to answer it is the following: if a CP map has Kraus-rank $s$, then it necessarily sends rank $1$ inputs on rank at most $s$ outputs. This is of course informative only if $s$ is smaller than the output space dimension. But as we shall see, having this is mind will be useful to prove that certain channels cannot be compressed further than as guaranteed by Theorem \ref{th:main}.

Our constructions will be based on the existence of so-called tight normalized frames. Namely, for any $N,d\in\N$ with $N\geq d$, there exist unit vectors $\ket{\psi_1},\ldots,\ket{\psi_N}$ in $\C^d$ such that
\[ \frac{1}{N}\sum_{k=1}^N\ketbra{\psi_k}{\psi_k}=\frac{\Id}{d}. \]
Denoting by $\{\ket{j},\ 1\leq j\leq d\}$ an orthonormal basis of $\C^d$, a possible way of constructing such vectors is e.g.~to make the choice
\begin{equation} \label{eq:frame'} \forall\ 1\leq k\leq N,\ \ket{\psi_k}=\frac{1}{\sqrt{d}}\sum_{j=1}^d e^{2i\pi jk/N}\ket{j}. \end{equation}
Note that if this so, then any basis vector $\ket{j}$, $1\leq j\leq d$, is such that, for each $1\leq k\leq N$, $|\braket{\psi_k}{j}|^2=1/d$.

Let us now come back to our objective. What we want to exhibit here are CPTP maps $\cN:\cL(\A)\rightarrow\cL(\B)$ with either one or the other of the following two properties: if a CP map $\widehat{\cN}:\cL(\A)\rightarrow\cL(\B)$ satisfies
\begin{equation} \label{eq:approx1} \forall\ R\in\mathcal{H}_+(\A),\
(1-\varepsilon)\cN(R)-\varepsilon(\tr R)\frac{\Id}{|\B|} \leq \widehat{\cN}(R) \leq (1+\varepsilon)\cN(R)+\varepsilon(\tr R)\frac{\Id}{|\B|}, \end{equation}
then it necessarily has to be such that either $r_K(\widehat{\cN})\geq |\A|$ or $r_K(\widehat{\cN})\geq |\B|$. Besides, note that the CP maps $\cN,\widehat{\cN}$ fulfilling condition \eqref{eq:approx1} above is equivalent to the conjugate CP maps $\cN^*,\widehat{\cN}^*$ fulfilling condition \eqref{eq:approx2} below
\begin{equation} \label{eq:approx2} \forall\ R\in\mathcal{H}_+(\B),\
(1-\varepsilon)\cN^*(R)-\varepsilon(\tr R)\frac{\Id}{|\B|} \leq \widehat{\cN}^*(R) \leq (1+\varepsilon)\cN^*(R)+\varepsilon(\tr R)\frac{\Id}{|\B|}. \end{equation}
Depending on what we want to establish, it will be more convenient to work with either one or the other of these requirements.


Assume first of all that $|\B|\geq |\A|$, and consider $\cM:\cL(\A)\rightarrow\cL(\B)$ a so-called quantum-classical channel (aka measurement). More specifically, define the CPTP map
\begin{equation} \label{eq:qc} \mathcal{M}:X\in\cL(\A)\mapsto\frac{|\A|}{|\B|}\sum_{i=1}^{|\B|}\bra{\psi_i}X\ket{\psi_i}\ketbra{x_i}{x_i} \in\cL(\B), \end{equation}
where $\{\ket{x_i},\ 1\leq i\leq|\B|\}$ is an orthonormal basis of $\B$ and $\ket{\psi_1},\ldots,\ket{\psi_{|\B|}}$ are unit vectors of $\A$, defined in terms of an orthonormal basis $\{\ket{j},\ 1\leq j\leq|\A|\}$ of $\A$ as by equation \eqref{eq:frame'}. Just to connect with the considerations in Chapter \ref{chap:zonoids}, note that this tight normalized frame assumption implies that $\left\{(|\A|/|\B|)\ketbra{\psi_i}{\psi_i}\right\}_{1\leq i\leq|\B|}$ forms a rank-$1$ POVM on $\A$ (hence a posteriori the justification of the denomination for $\cM$). Setting, for each $1\leq i\leq|\B|$, $K_i=\sqrt{|\A|/|\B|}\,\ketbra{x_i}{\psi_i}$, we can clearly re-write $\mathcal{M}:X\in\cL(\A)\mapsto \sum_{i=1}^{|\B|}K_iXK_i^{\dagger} \in\cL(\B)$, so $r_K(\cM)\leq |\B|$. And what we actually want to show is that it is even impossible to approximate $\cM$ in the sense of Theorem \ref{th:main} with strictly less than $|\B|$ Kraus operators. Observe that by construction, say, $\ket{1}$ is such that, for each $1\leq i\leq |\B|$, $|\braket{\psi_i}{1}|^2=1/|\A|$, so that $\cM(\ketbra{1}{1})= \Id/|\B|$. Yet, assume that $\widehat{\cM}:\cL(\A)\rightarrow\cL(\B)$ is a CPTP map such that $\cM,\widehat{\cM}$ fulfill equation \eqref{eq:approx1} for some $0<\varepsilon<1/2$. Then, the l.h.s.~of equation \eqref{eq:approx1} yields in particular, $\widehat{\cM}(\ketbra{1}{1})\geq (1-2\varepsilon)\,\Id/|\B|$, so that $\widehat{\cM}(\ketbra{1}{1})$ has to have full rank. And therefore, it cannot be that $r_K(\widehat{\mathcal{M}})<|\B|$.

Assume now that $|\A|\geq |\B|$, and consider $\cN:\cL(\A)\rightarrow\cL(\B)$ a so-called classical-quantum channel. More specifically, define the CPTP map
\begin{equation} \label{eq:cq} \mathcal{N}:X\in\cL(\A)\mapsto\sum_{i=1}^{|\A|}\bra{x_i}X\ket{x_i}\ketbra{\psi_i}{\psi_i} \in\cL(\B), \end{equation}
where $\{\ket{x_i},\ 1\leq i\leq|\A|\}$ is an orthonormal basis of $\A$ and $\ket{\psi_1},\ldots,\ket{\psi_{|\A|}}$ are unit vectors in $\B$. Setting, for each $1\leq i\leq|\A|$, $K_i=\ketbra{\psi_i}{x_i}$, we can clearly re-write $\mathcal{N}:X\in\cL(\A)\mapsto \sum_{i=1}^{|\A|}K_iXK_i^{\dagger} \in\cL(\B)$, so $r_K(\cN)\leq |\A|$. Now, we want to show that, at least for certain choices of $\ket{\psi_1},\ldots,\ket{\psi_{|\A|}}$, it is even impossible to approximate $\cN$ in the sense of Theorem \ref{th:main} with strictly less than $|\A|$ Kraus operators. For that, we impose that they are defined in terms of an orthonormal basis $\{\ket{j},\ 1\leq j\leq|\B|\}$ of $\B$ as by equation \eqref{eq:frame'}. Since the conjugate of $\cN$ is the CP unital map
\[ \mathcal{N}^*:X\in\cL(\B)\mapsto\sum_{i=1}^{|\A|}\bra{\psi_i}X\ket{\psi_i}\ketbra{x_i}{x_i} \in\cL(\A), \]
we have in this case that $\mathcal{M}=(|\B|/|\A|)\cN^*$ is precisely of the form \eqref{eq:qc} (with the roles of $\A$ and $\B$ switched). Hence, as we already showed, if $\widehat{\cM}:\cL(\B)\rightarrow\cL(\A)$ is a CPTP map such that $\cM,\widehat{\cM}$ fulfill equation \eqref{eq:approx1} (with the roles of $\A$ and $\B$ switched) for some $0<\varepsilon<1/2$, then it cannot be that $r_K(\widehat{\mathcal{M}})<|\A|$. This means equivalently that if $\widehat{\cN}^*:\cL(\B)\rightarrow\cL(\A)$ is a CP map such that $\cN^*,\widehat{\cN}^*$ fulfill equation \eqref{eq:approx2} for some $0<\varepsilon<1/2$, then it cannot be that $r_K(\widehat{\mathcal{N}}) = r_K(\widehat{\mathcal{N}}^*)<|\A|$.


Summarizing, we just established that $n\geq \max (|\A|,|\B|)$ is for sure necessary in Theorem \ref{th:main}. But it is not clear whether or not the $\log |\rE|$ factor can be removed. In the case of ``well-behaved'' channels, whose range is only composed of sufficiently mixed states, we can answer affirmatively, which is the content of Theorem \ref{th:main'} below. However, we leave the question open in general.

\begin{theorem} \label{th:main'}
Fix $0<\varepsilon<1$ and let $\cN:\cL(\A)\rightarrow\cL(\B)$ be a CPTP map with Kraus rank $|\rE|\geq |\A|,|\B|$. Then, there exists a CP map $\widehat{\cN}:\cL(\A)\rightarrow\cL(\B)$ with Kraus rank at most $C\max(|\A|,|\B|)/\varepsilon^2$ (where $C>0$ is a universal constant) and such that
\[ \label{eq:approx} \sup_{\rho\in\cD(\A)} \left\| \widehat{\cN}(\rho)-\cN(\rho) \right\|_{\infty} \leq \varepsilon\sup_{\rho\in\cD(\A)} \left\|\cN(\rho)\right\|_{\infty}. \]
\end{theorem}

\section{Proof of the main results}

As a crucial step in establishing Theorems \ref{th:main} and \ref{th:main'}, we will need a large deviation inequality for sums of independent $\psi_1$ (aka sub-exponential) random variables, known as Bernstein's inequality. All needed definitions and results concerning $\psi_1$ random variables are gathered in Chapter \ref{chap:toolbox}, Section \ref{ap:deviations}. Our application of Bernstein's inequality (recalled as Theorem \ref{th:Bernstein} in Chapter \ref{chap:toolbox}, Section \ref{ap:deviations}) to a suitably chosen sum of independent $\psi_1$ random variables will yield Proposition \ref{prop:fixed} below. Note that in the latter, as well as in several other places in the remainder of this chapter, we shall use the following shorthand notation, whenever no confusion is at risk: given a unit vector $\phi$ in $\C^n$, we also denote by $\phi$ the corresponding pure state $\ketbra{\phi}{\phi}$ on $\C^n$.

\begin{proposition} \label{prop:fixed}
Let $\cN:\cL(\A)\rightarrow\cL(\B)$ be a CPTP map with Kraus rank $|\rE|$, defined by
\begin{equation} \label{eq:N} \forall\ \rho\in\cD(\A),\ \cN(\rho)=\tr_{\rE}\left[V\rho V^{\dagger}\right], \end{equation}
for some isometry $V:\A\hookrightarrow\B\otimes\rE$.

For any given unit vector $\varphi$ in $\rE$ define next the CP map $\cN_{\varphi}:\cL(\A)\rightarrow\cL(\B)$ by
\begin{equation} \label{eq:N_phi} \forall\ \rho\in\cD(\A),\ \cN_{\varphi}(\rho)= |\rE|\, \tr_{\rE}\left[\left(\Id\otimes\varphi\right) V\rho V^{\dagger} \left(\Id\otimes\varphi\right) \right]. \end{equation}

Now, fix unit vectors $x$ in $\A$, $y$ in $\B$, and pick random unit vectors $\varphi_1,\ldots,\varphi_n$ in $\rE$, independently and uniformly. Then,
\[ \forall\ 0<\varepsilon<1,\ \P\left( \left|\frac{1}{n}\sum_{i=1}^n \bra{y}\cN_{\varphi_i}\left(x\right)\ket{y} - \bra{y}\cN\left(x\right)\ket{y} \right| > \varepsilon\bra{y}\cN\left(x\right)\ket{y} \right) \leq e^{-cn\varepsilon^2}, \]
where $c>0$ is a universal constant.
\end{proposition}

In order to derive this concentration result, we will need first of all an estimate on the $\psi_1$-norm of a certain random variable appearing in our construction. This is the content of Lemma \ref{lemma:psi_1} below.

\begin{lemma} \label{lemma:psi_1}
Fix $d,s\in\N$. Let $\sigma$ be a state on $\C^{d}\otimes\C^s$ and $y$ be a unit vector in $\C^{d}$. Next, for $\varphi$ a uniformly distributed unit vector in $\C^s$ define the random variable
\[ X_{\varphi}(\sigma,y) = \tr\left[ y\otimes\varphi\,\sigma\right]. \]
Then, $X_{\varphi}(\sigma,y)$ is a $\psi_1$ random variable with mean and $\psi_1$-norm satisfying
\begin{equation} \label{eq:mean-psi_1} \E X_{\varphi}(\sigma,y) = \frac{1}{s}\tr\left[ y\otimes\Id\,\sigma\right]\ \ \text{and}\ \ \|X_{\varphi}(\sigma,y)\|_{\psi_1} \leq \frac{1}{s}\tr\left[ y\otimes\Id\,\sigma\right]. \end{equation}
\end{lemma}

\begin{proof}
To begin with, recall that, for any $p\in\N$, we have, for $\varphi$ a uniformly distributed unit vector in $\C^s$,
\[ \E\varphi^{\otimes p} = \frac{1}{{s+p-1 \choose p}} P_{\Sym^{p}(\C^s)}, \]
where $P_{\Sym^{p}(\C^s)}$ denotes the orthogonal projector onto the completely symmetric subspace of $(\C^s)^{\otimes p}$ (see \cite{Harrow} and Chapter \ref{chap:symmetries}, Section \ref{sec:sym}, of this manuscript for further details).

Now, setting $\sigma_y= \tr_{\C^{d}}\left[y\otimes\Id\, \sigma\right]$, positive operator on $\C^s$, we see that $X_{\varphi}(\sigma,y) = \tr\left[\varphi\,\sigma_y\right]$. Hence, we clearly have for a start the first statement in equation \eqref{eq:mean-psi_1}, namely
\[ \E X_{\varphi}(\sigma,y) = \frac{1}{s} \tr \left[\Id\,\sigma_y\right] =\frac{1}{s} \tr\left[y\otimes\Id\,\sigma\right]. \]
What is more, for any $p\in\N$, $\left|X_{\varphi}(\sigma,y)\right|^p = \left(\tr \left[ \varphi\,\sigma_y\right]\right)^p = \tr \left[ \varphi^{\otimes p}\,\sigma_y^{\otimes p}\right]$. And therefore,
\[ \E \left|X_{\varphi}(\sigma,y)\right|^p = \frac{1}{{s+p-1 \choose p}} \tr\left[P_{\Sym^{p}(\C^s)}\sigma_y^{\otimes p}\right] \leq \frac{1}{{s+p-1 \choose p}} \tr\left[\sigma_y^{\otimes p}\right] \leq \left(\frac{p}{s} \tr\left[\sigma_y\right]\right)^p, \]
where the last inequality is simply by the rough bounds $p!\leq p^p$ and $(s+p-1)!/(s-1)!\geq s^p$.
So in the end, we get as wanted the second statement in equation \eqref{eq:mean-psi_1}, namely
\[ \|X_{\varphi}(\sigma,y)\|_{\psi_1} =\sup_{p\in\N} \frac{\left(\E \left|X_{\varphi}(\sigma,y)\right|^p\right)^{1/p}}{p} \leq \frac{1}{s}\tr\left[y\otimes\Id\,\sigma\right]. \]
This concludes the proof of Lemma \ref{lemma:psi_1}.
\end{proof}

\begin{proof}[Proof of Proposition \ref{prop:fixed}] Note first of all that we can obviously re-write
\[ \bra{y}\cN(x)\ket{y} = \tr\left[y\otimes\Id\, VxV^{\dagger}\right]\ \text{and}\ \forall\ \varphi\in S_{\rE},\ \bra{y}\cN_{\varphi}(x)\ket{y} = |\rE|\,\tr\left[y\otimes\varphi\, VxV^{\dagger}\right]. \]
Next, for each $1\leq i\leq n$, define the random variable $Y_i=\bra{y}\cN_{\varphi_i}(x)\ket{y}$. By Lemma \ref{lemma:psi_1}, combined with the observation just made above, we know that these are independent $\psi_1$ random variables with mean $\bra{y}\cN(x)\ket{y}$ and $\psi_1$-norm upper bounded by $\bra{y}\cN(x)\ket{y}$. So by Bernstein's inequality, recalled as Theorem \ref{th:Bernstein} in Chapter \ref{chap:toolbox}, Section \ref{ap:deviations}, we get that
\[ \forall\ t>0,\ \P\left( \left|\frac{1}{n}\sum_{i=1}^n Y_i - \bra{y}\cN(x)\ket{y} \right| > t \right) \leq \exp\left(-c_0n\min\left(\frac{t^2}{\bra{y}\cN(x)\ket{y}^2},\frac{t}{\bra{y}\cN(x)\ket{y}}\right) \right), \]
where $c_0>0$ is a universal constant. And hence,
\[ \forall\ 0<\varepsilon<1,\ \P\left( \left|\frac{1}{n}\sum_{i=1}^n Y_i - \bra{y}\cN(x)\ket{y} \right| > \varepsilon \bra{y}\cN(x)\ket{y} \right) \leq e^{-c_0n\varepsilon^2}, \]
which is precisely the result announced in Proposition \ref{prop:fixed}.
\end{proof}

Having at hand the ``fixed $x,y$'' concentration inequality of Proposition \ref{prop:fixed}, we can now get its ``for all $x,y$'' counterparts by a standard net-argument. It appears as the following Propositions \ref{prop:forall} and \ref{prop:forall'}. Note that the approach is very similar to the one leading to the derivation of Dvoretzky's theorem from Levy's lemma, recalled in Chapter \ref{chap:toolbox}, Section \ref{ap:deviations}.

\begin{proposition} \label{prop:forall}
Let $\cN:\cL(\A)\rightarrow\cL(\B)$ be a CPTP map, as characterized by equation \eqref{eq:N}, and for each unit vector $\varphi$ in $\rE$ define the CP map $\cN_{\varphi}:\cL(\A)\rightarrow\cL(\B)$ as in equation \eqref{eq:N_phi}. Next, for $\varphi_1,\ldots,\varphi_n$ independent uniformly distributed unit vectors in $\rE$, set $\cN_{\varphi^{(n)}}=\big(\sum_{i=1}^n\cN_{\varphi_i}\big)/n$. Then, for any $0<\varepsilon<1$,
\[ \P\left( \forall\ x\in S_{\A},y\in S_{\B},\ \left|\bra{y}\cN_{\varphi^{(n)}}(x)-\cN(x)\ket{y} \right| \leq \varepsilon\bra{y}\cN(x)\ket{y} +\frac{\varepsilon}{|\B|} \right) \geq 1 - \left(\frac{24|\rE||\B|}{\varepsilon}\right)^{2(|\A|+|\B|)} e^{-cn\varepsilon^2}, \]
where $c>0$ is a universal constant.
\end{proposition}

\begin{proof}
Fix $0<\alpha,\beta<1$ and consider $\mathcal{A}_{\alpha},\mathcal{B}_{\beta}$ minimal $\alpha,\beta$-nets within the unit spheres of $\A,\B$, so that by a standard volumetric argument $\left|\mathcal{A}_{\alpha}\right|\leq (3/\alpha)^{2|\A|},\left|\mathcal{B}_{\beta}\right|\leq (3/\beta)^{2|\B|}$ (see Lemma \ref{lemma:nets} in Chapter \ref{chap:toolbox}, Section \ref{ap:deviations}). Then, by Proposition \ref{prop:fixed} and the union bound, we get that, for any $\varepsilon>0$,
\begin{equation} \label{eq:M_delta} \P\left( \forall\ x\in\mathcal{A}_{\alpha},y\in\mathcal{B}_{\beta},\ \left| \bra{y} \cN_{\varphi^{(n)}}(x)-\cN(x) \ket{y} \right| \leq \varepsilon\bra{y}\cN\left(x\right)\ket{y} \right) \geq 1 - \left(\frac{3}{\alpha}\right)^{2|\A|}\left(\frac{3}{\beta}\right)^{2|\B|}e^{-cn\varepsilon^2}. \end{equation}
Now, fix $\varepsilon>0$ and suppose that $\mathcal{E}:\cL(\A)\rightarrow\cL(\B)$ is a Hermiticity-preserving map which is such that
\begin{equation} \label{eq:true-net} \forall\ x\in\mathcal{A}_{\alpha},\ \forall\ y\in\mathcal{B}_{\beta},\ \left|\bra{y}\mathcal{E}(x)\ket{y}\right| \leq \varepsilon\bra{y}\cN\left(x\right)\ket{y}. \end{equation}
Assume that $\mathcal{E}$ additionally satisfies the boundedness property
\begin{equation} \label{eq:boundedness} \forall\ x\in S_{\A},\ \forall\ y\in S_{\B},\ \left|\bra{y}\mathcal{E}(x)\ket{y}\right| \leq |\rE|.\end{equation}
Note that if $\mathcal{E}$ is Hermicity-preserving, then for any $x,y$, $\sup_{v,v'}|\bra{y}\mathcal{E}(\ketbra{v}{v'})\ket{y}|= \sup_{v}|\bra{y}\mathcal{E}(\ketbra{v}{v})\ket{y}|$ and $\sup_{w,w'}|\bra{w}\mathcal{E}(\ketbra{x}{x})\ket{w'}|= \sup_{w}|\bra{w}\mathcal{E}(\ketbra{x}{x})\ket{w}|$ (this is because for any $X$, $\mathcal{E}(X^{\dagger})=\mathcal{E}(X)^{\dagger}$). Hence, it will be useful to us later on to keep in mind that assumption \eqref{eq:boundedness} is actually equivalent to
\[ \forall\ x,x'\in S_{\A},\ \forall\ y,y'\in S_{\B},\ \begin{cases} \left|\bra{y}\mathcal{E}(\ketbra{x}{x'})\ket{y}\right| \leq |\rE| \\ \left|\bra{y}\mathcal{E}(\ketbra{x}{x})\ket{y'}\right| \leq |\rE| \end{cases}. \]
Then, for any unit vectors $x\in S_{\A}$, $y\in S_{\B}$, we know by definition that there exist $\tilde{x}\in\mathcal{A}_{\alpha}$, $\tilde{y}\in\mathcal{B}_{\beta}$ such that $\|x-\tilde{x}\|\leq\alpha$, $\|y-\tilde{y}\|\leq\beta$. Hence, first of all
\begin{align*} \left|\bra{y}\mathcal{E}(\proj{x})\ket{y}\right| & \leq \left|\bra{\tilde{y}}\mathcal{E}(\proj{x})\ket{\tilde{y}}\right| + \left|\bra{y-\tilde{y}}\mathcal{E}(\proj{x})\ket{\tilde{y}}\right| + \left|\bra{y}\mathcal{E}(\proj{x})\ket{y-\tilde{y}}\right|\\
& \leq \left|\bra{\tilde{y}}\mathcal{E}(\proj{x})\ket{\tilde{y}}\right| + 2\beta|\rE|, \end{align*}
where the second inequality follows from the boundedness property \eqref{eq:boundedness} of $\mathcal{E}$, combined with the fact that $\|y-\tilde{y}\|\leq\beta$. Then similarly, because $\|x-\tilde{x}\|\leq\alpha$,
\begin{align*} \left|\bra{\tilde{y}}\mathcal{E}(\proj{x})\ket{\tilde{y}}\right| & \leq \left|\bra{\tilde{y}}\mathcal{E}(\proj{\tilde{x}})\ket{\tilde{y}}\right| + \left|\bra{\tilde{y}}\mathcal{E}(\ketbra{x-\tilde{x}}{\tilde{x}})\ket{\tilde{y}}\right| + \left|\bra{\tilde{y}}\mathcal{E}(\ketbra{x}{x-\tilde{x}})\ket{\tilde{y}}\right| \\
& \leq \left|\bra{\tilde{y}}\mathcal{E}(\proj{\tilde{x}})\ket{\tilde{y}}\right| + 2\alpha|\rE|. \end{align*}
Putting together the two previous upper bounds, we see that we actually have
\[ \left|\bra{y}\mathcal{E}(\proj{x})\ket{y}\right| \leq \left|\bra{\tilde{y}}\mathcal{E}(\proj{\tilde{x}})\ket{\tilde{y}}\right| + 2|\rE|(\alpha+\beta) \leq \varepsilon \bra{\tilde{y}}\mathcal{N}(\proj{\tilde{x}})\ket{\tilde{y}} + 2|\rE|(\alpha+\beta), \]
where the second inequality is by assumption \eqref{eq:true-net} on $\mathcal{E}$. Now, arguing just as before (using this time that $\mathcal{N}$ satisfies the boundedness property $\left|\bra{y}\mathcal{N}(\ketbra{x}{x'})\ket{y'}\right|\leq 1$ for any $x,x'\in S_{\A}$ and $y,y'\in S_{\B}$), we get
\[ \bra{\tilde{y}}\mathcal{N}(\proj{\tilde{x}})\ket{\tilde{y}} \leq \bra{y}\mathcal{N}(\proj{x})\ket{y}+2(\alpha+\beta) . \]
So eventually, what we obtain is
\[ \left|\bra{y}\mathcal{E}(x)\ket{y}\right| \leq \varepsilon\big(\bra{y}\cN(x)\ket{y} + 2(\alpha+\beta)\big) + 2|\rE|(\alpha+\beta) \leq \varepsilon\bra{y}\cN(x)\ket{y} + 4|\rE|(\alpha+\beta). \]
Therefore, choosing $\alpha=\beta=\varepsilon/(8|\rE||\B|)$ (and observing that, by the way $\cN_{\varphi^{(n)}}$ is constructed, $\cN_{\varphi^{(n)}}-\cN$ fulfills condition \eqref{eq:boundedness}), it follows from equation \eqref{eq:M_delta} that, for any $\varepsilon>0$,
\[ \P\left( \forall\ x\in S_{\A},y\in S_{\B},\ \left| \bra{y}\cN_{\varphi^{(n)}}(x)-\cN(x)\ket{y} \right| \leq \varepsilon\bra{y}\cN\left(x\right)\ket{y} +\frac{\varepsilon}{|\B|} \right) \geq 1 - \left(\frac{24|\rE||\B|}{\varepsilon}\right)^{2(|\A|+|\B|)}e^{-cn\varepsilon^2}, \]
which is exactly what we wanted to show.
\end{proof}

\begin{proposition} \label{prop:forall'}
Let $\cN:\cL(\A)\rightarrow\cL(\B)$ be a CPTP map, as characterized by equation \eqref{eq:N}, and for each unit vector $\varphi$ in $\rE$ define the CP map $\cN_{\varphi}:\cL(\A)\rightarrow\cL(\B)$ as in equation \eqref{eq:N_phi}. Next, for $\varphi_1,\ldots,\varphi_n$ independent uniformly distributed unit vectors in $\rE$, set $\cN_{\varphi^{(n)}}=\big(\sum_{i=1}^n\cN_{\varphi_i}\big)/n$. Then, for any $0<\varepsilon<1$,
\[ \P\left( \sup_{x\in S_{\A},y\in S_{\B}} \left| \bra{y}\cN_{\varphi^{(n)}}(x)-\cN(x)\ket{y} \right| \leq \varepsilon \sup_{x\in S_{\A},y\in S_{\B}} \bra{y}\cN(x)\ket{y} \right) \geq 1 - 225^{|\A|+|\B|} e^{-cn\varepsilon^2}, \]
where $c>0$ is a universal constant.
\end{proposition}

\begin{proof}
We will argue in a way very similar to what was done in the proof of Proposition \ref{prop:forall}, and hence skip some of the details here. Again, fix $0<\alpha,\beta< 1/4$ and consider $\mathcal{A}_{\alpha},\mathcal{B}_{\beta}$ minimal $\alpha,\beta$-nets within the unit spheres of $\A,\B$.
Now, fix $\varepsilon>0$ and suppose that $\mathcal{E}:\cL(\A)\rightarrow\cL(\B)$ is a Hermiticity-preserving map which is such that,
\[ \forall\ x\in\mathcal{A}_{\alpha},\ \forall\ y\in\mathcal{B}_{\beta},\ \left|\bra{y}\mathcal{E}(x)\ket{y}\right| \leq \varepsilon\bra{y}\cN\left(x\right)\ket{y}. \]
Then, for any unit vectors $x\in S_{\A}$, $y\in S_{\B}$,
\begin{align*}
\left|\bra{y}\mathcal{E}(\proj{x})\ket{y}\right| \leq\, & \varepsilon \big( \bra{y}\cN(\proj{x})\ket{y} + 2\alpha\,{\sup}_{v,v'}\bra{y}\cN(\ketbra{v}{v'})\ket{y} + 2\beta\,{\sup}_{w,w'} \bra{w}\cN(\proj{\tilde{x}})\ket{w'} \big) \\
& + 2\alpha\,{\sup}_{v,v'}\left|\bra{\tilde{y}}\mathcal{E}(\ketbra{v}{v'})\ket{\tilde{y}}\right| + 2\beta\,{\sup}_{w,w'}\left|\bra{w}\mathcal{E}(\proj{x})\ket{w'}\right|,
\end{align*}
where $\tilde{x}\in\mathcal{A}_{\alpha}$, $\tilde{y}\in\mathcal{B}_{\beta}$ are such that $\|x-\tilde{x}\|\leq\alpha$, $\|y-\tilde{y}\|\leq\beta$. And consequently, taking supremum over unit vectors $x\in S_{\A}$, $y\in S_{\B}$, we get
\[ {\sup}_{x,y}\left|\bra{y}\mathcal{E}(x)\ket{y}\right| \leq\,  \varepsilon(1+2(\alpha+\beta))\,{\sup}_{x,y} \bra{y}\cN(x)\ket{y} + 2(\alpha+\beta) \,{\sup}_{x,y}\left|\bra{y}\mathcal{E}(x)\ket{y}\right|, \]
that is equivalently,
\[ {\sup}_{x,y}\left|\bra{y}\mathcal{E}(x)\ket{y}\right| \leq \varepsilon\,\frac{1+2(\alpha+\beta)}{1-2(\alpha+\beta)}\,{\sup}_{x,y} \bra{y}\cN(x)\ket{y}. \]
Therefore, choosing $\alpha=\beta=1/5$, so that $(1+2(\alpha+\beta))/(1-2(\alpha+\beta))=9$ and $3/\alpha=3/\beta=15$, we eventually obtain that, for any $0<\varepsilon<1$,
\[ \P\left( \sup_{x\in S_{\A},y\in S_{\B}} \left| \bra{y}\cN_{\varphi^{(n)}}(x)-\cN(x)\ket{y} \right| \leq 9\varepsilon \sup_{x\in S_{\A},y\in S_{\B}} \bra{y}\cN(x)\ket{y} \right) \geq 1 - 15^{2(|\A|+|\B|)}e^{-cn\varepsilon^2}, \]
which, after relabelling $9\varepsilon$ in $\varepsilon$, implies precisely the result announced in Proposition \ref{prop:forall'}.
\end{proof}

\begin{proof}[Proof of Theorem \ref{th:main}]
Because operator-ordering is preserved by convex combinations, it follows from Proposition \ref{prop:forall} that there exists a universal constant $c>0$ such that, for any $\varepsilon>0$,
\[ \P\left( \forall\ \rho\in\cD(\A),\ \left| \cN_{\varphi^{(n)}}(\rho) -\cN(\rho) \right| \leq \varepsilon\left(\cN(\rho) +\frac{\Id}{|\B|}\right) \right) \geq 1 - \left(\frac{24|\rE||\B|}{\varepsilon}\right)^{2(|\A|+|\B|)}e^{-cn\varepsilon^2}. \]
The r.h.s.~of the latter inequality becomes larger than, say, $1/2$ as soon as $n\geq C\max(|\A|,|\B|)\log (|\rE|/\varepsilon)/\varepsilon^2$, for $C>0$ some universal constant.

Recapitulating, what we have shown sofar is that there exists a completely positive map $\cN^{(n)}$ with Kraus rank $n\leq C\max(|\A|,|\B|)\log (|\rE|/\varepsilon)/\varepsilon^2$, for $C>0$ some universal constant, such that,
\begin{equation} \label{eq:CP} \forall\ \rho\in\cD(\A),\ -\varepsilon\left(\cN(\rho) +\frac{\Id}{|\B|}\right) \leq \cN^{(n)}(\rho)-\cN(\rho)\leq \varepsilon\left(\cN(\rho) +\frac{\Id}{|\B|}\right). \end{equation}
In particular, equation \eqref{eq:CP} implies that, for any $\rho\in\cD(\A)$, $\left|\tr\left(\cN^{(n)}(\rho)\right) -1\right|\leq 2\varepsilon$, so that $\cN^{(n)}$ is almost trace preserving, up to an error $2\varepsilon$. As a consequence of equation \eqref{eq:CP}, we also have
\begin{equation} \label{eq:1-norm} \forall\ \rho\in\cD(\A),\ \left\|\cN^{(n)}(\rho)-\cN(\rho)\right\|_1 \leq 2\varepsilon, \end{equation}
and to get only such trace-norm approximation, it is actually possible to impose that $\cN^{(n)}$ is strictly trace preserving. Indeed, denote by $\{K_1,\ldots,K_n\}$ a set of Kraus operators for $\cN^{(n)}$, and set $S=\sum_{i=1}^nK_i^{\dagger}K_i$. Equation \eqref{eq:1-norm} guarantees that $\|S-\Id\|_{\infty}\leq 2\varepsilon$, so that $S$ is in particular invertible, as soon as $\varepsilon<1/2$. Hence, assume in the sequel that, in fact, $\varepsilon<1/4$, and consider the completely positive map $\widehat{\cN}^{(n)}$ having $\{K_1S^{-1/2},\ldots,K_nS^{-1/2}\}$ as a set of Kraus operators, which means that $\widehat{\cN}^{(n)}(\cdot)=\cN^{(n)}(S^{-1/2}\,\cdot\, S^{-1/2})$. The latter is trace preserving by construction, and such that
\begin{equation} \label{eq:TP-approx} \forall\ \rho\in\cD(\A),\ \left\|\widehat{\cN}^{(n)}(\rho)-\cN(\rho)\right\|_1\leq \left\|\widehat{\cN}^{(n)}(\rho)-\cN^{(n)}(\rho)\right\|_1 + \left\|\cN^{(n)}(\rho)-\cN(\rho)\right\|_1 \leq 5\varepsilon. \end{equation}
Indeed, for any $\rho\in\cD(\A)$, we have the chain of inequalities
\[ \|S^{-1/2}\rho S^{-1/2}-\rho\|_1\leq \left(\|S^{-1/2}\|_{\infty}+\|\Id\|_{\infty}\right) \|\rho\|_1 \|S^{-1/2}-\Id\|_{\infty} \leq (1+2\varepsilon+1)\,2\varepsilon\leq 3\varepsilon, \]
where the first inequality follows from the triangle and H\"{o}lder inequalities (after simply noticing that, setting $\Delta=\Id-S^{-1/2}$, we can rewrite $S^{-1/2}\rho S^{-1/2}-\rho$ as $\Delta\rho\Id + S^{-1/2}\rho\Delta$), while the second inequality is because, for any $0<x<1/4$, $(1+2x)^{-1/2}\geq 1-x$ and $(1-2x)^{-1/2}\leq 1+2x$, so that $\|S^{-1/2}\|_{\infty}\leq 1+2\varepsilon$ and $\|S^{-1/2}-\Id\|_{\infty}\leq 2\varepsilon$. This implies that, for any $\rho\in\cD(\A)$,
\[ \left\|\widehat{\cN}^{(n)}(\rho)-\cN^{(n)}(\rho)\right\|_1 = \left\|\cN^{(n)}\left(S^{-1/2}\rho S^{-1/2}-\rho\right)\right\|_1 \leq 3\varepsilon, \]
which, combined with \eqref{eq:1-norm}, justifies the last inequality in \eqref{eq:TP-approx}.

This concludes the proof of Theorem \ref{th:main} and of Remark \ref{remark:schatten} following it.
\end{proof}

\begin{proof}[Proof of Theorem \ref{th:main'}]
By extremality of pure states amongst all states, it follows from Proposition \ref{prop:forall'} that there exists a universal constant $c>0$ such that, for any $\varepsilon>0$,
\[ \P\left( \sup_{\rho\in\cD(\A)} \left\| \cN_{\varphi^{(n)}}(\rho) -\cN(\rho) \right\|_{\infty} \leq \varepsilon \sup_{\rho\in\cD(\A)} \left\|\cN(\rho)\right\|_{\infty} \right) \geq 1 - 144^{|\A|+|\B|}e^{-cn\varepsilon^2}. \]
The r.h.s.~of the latter inequality becomes larger than, say, $1/2$ as soon as $n\geq C\max(|\A|,|\B|)/\varepsilon^2$, for $C>0$ some universal constant. And the proof of Theorem \ref{th:main'} is thus complete.
\end{proof}

\section{Consequences and applications}
\label{sec:applications}

\subsection{Approximation in terms of output entropies or fidelities}

This section gathers some (more or less straightforward) corollaries of Theorem \ref{th:main} concerning approximation of quantum channels in other distance measures than the $(1{\rightarrow}1)$-norm distance mostly studied up to now.

Given a state $\rho$ on some Hilbert space $\H$, we define, for any $p\in]1,\infty[$, its R\'{e}nyi entropy of order $p$ as
\[ S_p(\rho)=-\frac{p}{p-1}\log\|\rho\|_p, \]
and the latter definition is extended by continuity to $p\in\{1,\infty\}$ as
\[ S_1(\rho)=S(\rho)=-\tr(\rho\log\rho)\ \text{and}\ S_{\infty}(\rho)=-\log\lambda_{\max}(\rho). \]
R\'{e}nyi $p$-entropies thus measure the amount of information present in a quantum state, generalizing the case $p=1$ of the von Neumann entropy introduced in Chapter \ref{chap:QIT}, Section \ref{sec:Shannon}. Besides, given states $\rho,\sigma$ on some Hilbert space $\H$, their fidelity is defined as $F(\rho,\sigma) = \|\sqrt{\rho}\sqrt{\sigma}\|_1$.

Now, given a channel $\cN$, from some input Hilbert space $\A$ to some output Hilbert space $\B$, it is quite important to understand quantities such as its minimum output R\'{e}nyi $p$-entropy, i.e.~$S_p^{\min}(\cN)=\min_{\rho\in\cD(\A)} S_p\big(\cN(\rho)\big)$, or its maximum output fidelity with a fixed state $\sigma$ on $\B$, i.e.~$F^{\max}(\cN,\sigma)=\max_{\rho\in\cD(\A)} F\big(\cN(\rho),\sigma\big)$. Hence the interest of having a channel $\widehat{\cN}$ which is simpler than $\cN$ but nevertheless shares approximately the same $S_p^{\min}$ and $F^{\max}(\cdot,\sigma)$.

\begin{proposition} \label{prop:S_p}
Let $\cN:\cL(\A)\rightarrow\cL(\B)$ be a CPTP map, and assume that the CP map $\widehat{\cN}:\cL(\A)\rightarrow\cL(\B)$ satisfies
\begin{equation} \label{eq:order} \forall\ \rho\in\cD(\A),\ (1-\varepsilon)\cN(\rho) -\varepsilon\frac{\Id}{|\B|} \leq \widehat{\cN}(\rho)\leq (1+\varepsilon)\cN(\rho) +\varepsilon\frac{\Id}{|\B|}, \end{equation}
for some $0<\varepsilon<1/2$. Then, for any $p\in]1,\infty]$, we have
\[ \forall\ \rho\in\cD(\A),\ S_p\big(\cN(\rho)\big) - \frac{p}{p-1}2\varepsilon \leq S_p\big(\widehat{\cN}(\rho)\big) \leq S_p\big(\cN(\rho)\big) + \frac{p}{p-1}4\varepsilon. \]
Hence, for any $p\in]1,\infty]$, $\widehat{\cN}$ is close to $\cN$ in terms of output $p$-entropies, in the sense that
\[ \forall\ \rho\in\cD(\A),\ \left| S_p\big(\widehat{\cN}(\rho)\big)-S_p\big(\cN(\rho)\big) \right| \leq \frac{p}{p-1}4\varepsilon. \]
\end{proposition}

\begin{proof}
Setting $\sigma=\cN(\rho)$, $\widehat{\sigma}=\widehat{\cN}(\rho)$ and $\tau=\Id/|\B|$, we can re-write equation \eqref{eq:order} as the two inequalities
\[ \widehat{\sigma}\leq (1+\varepsilon)\sigma +\varepsilon\tau\ \text{and}\ \sigma\leq \frac{1}{1-\varepsilon}\widehat{\sigma} +\frac{\varepsilon}{1-\varepsilon}\tau \leq (1+2\varepsilon)\widehat{\sigma} +2\varepsilon\tau. \]
By operator monotonicity and the triangle inequality for $\|\cdot\|_p$, these imply the two estimates
\begin{equation} \label{eq:p} \|\widehat{\sigma}\|_p \leq (1+\varepsilon)\|\sigma\|_p +\varepsilon\|\tau\|_p\ \text{and}\ \|\sigma\|_p \leq (1+2\varepsilon)\|\widehat{\sigma}\|_p +2\varepsilon\|\tau\|_p. \end{equation}
Now, from the first inequality in equation \eqref{eq:p}, we get
\[ \log\|\widehat{\sigma}\|_p \leq \log\left((1+\varepsilon)\|\sigma\|_p +\varepsilon\|\tau\|_p\right) \leq \log(1+\varepsilon) + \log\|\sigma\|_p + \frac{\varepsilon}{1+\varepsilon}\frac{\|\tau\|_p}{\|\sigma\|_p} \leq \log\|\sigma\|_p + 2\varepsilon, \]
where we used first that $\log$ is non-decreasing, then twice that $\log(1+x)\leq x$, and finally that $\|\sigma\|_p\leq \|\tau\|_p$. Similarly, we derive from the second inequality in equation \eqref{eq:p} that
\[ \log\|\sigma\|_p  \leq \log\|\widehat{\sigma}\|_p + 4\varepsilon. \]
Multiplying the two previous inequalities by $-p/(p-1)<0$, we eventually obtain
\[ S_p(\widehat{\sigma}) \geq S_p(\sigma) -\frac{p}{p-1}2\varepsilon\ \text{and}\ S_p(\sigma) \geq S_p(\widehat{\sigma}) -\frac{p}{p-1}4\varepsilon. \]
And all the conclusions of Proposition \ref{prop:S_p} follow.
\end{proof}

\begin{proposition} \label{prop:S}
Let $\cN:\cL(\A)\rightarrow\cL(\B)$ be a CPTP map, and assume that the CP map $\widehat{\cN}:\cL(\A)\rightarrow\cL(\B)$ satisfies
\begin{equation} \label{eq:approx} \forall\ \rho\in\cD(\A),\ \left\|\widehat{\cN}(\rho)-\cN(\rho)\right\|_1 \leq \frac{2\varepsilon}{\log|\B|}, \end{equation}
for some $0<\varepsilon<1/2$. Then, we have
\[ \forall\ \rho\in\cD(\A),\ \left| S\big(\widehat{\cN}(\rho)\big) - S\big(\cN(\rho)\big) \right| \leq \varepsilon + \frac{2\varepsilon}{\log|\B|} + \sqrt{\frac{\varepsilon}{\log|\B|}}. \]
Hence, $\widehat{\cN}$ is close to $\cN$ in terms of output entropies, in the sense that
\[ \forall\ \rho\in\cD(\A),\ \left| S\big(\widehat{\cN}(\rho)\big)-S\big(\cN(\rho)\big) \right| \leq 4\sqrt{\varepsilon}. \]
\end{proposition}

\begin{proof}
By Fannes-Audenaert inequality \cite{Audenaert}, equation \eqref{eq:approx} implies that
\[ \left| S\big(\widehat{\cN}(\rho)\big) - S\big(\cN(\rho)\big) \right| \leq \varepsilon - \frac{\varepsilon}{\log|\B|}\log \left(\frac{\varepsilon}{\log|\B|}\right) - \left(1-\frac{\varepsilon}{\log|\B|}\right)\log\left(1-\frac{\varepsilon}{\log|\B|}\right). \]
Now, for any $0<x<1/2$, we have on the one hand $x\log (1/x)\leq \sqrt{x}$, while we have on the other hand $\log(1/(1-x))\leq\log(1+2x)\leq 2x$ so that $(1-x)\log(1/(1-x))\leq 2x$. All the conclusions of Proposition \ref{prop:S} then follow.
\end{proof}

\begin{theorem} \label{th:entropies}
Fix $0<\varepsilon<1$ and let $\cN:\cL(\A)\rightarrow\cL(\B)$ be a CPTP map with Kraus rank $|\rE|\geq |\A|,|\B|$. Then, there exists a CP map $\widehat{\cN}:\cL(\A)\rightarrow\cL(\B)$ with Kraus rank at most $C\max(|\A|,|\B|)\log(|\rE|/\varepsilon)/\varepsilon^2$ (where $C>0$ is a universal constant) and such that
\[ \forall\ p\in]1,\infty],\ \forall\ \rho\in\cD(\A),\  \left| S_p\big(\widehat{\cN}(\rho)\big)-S_p\big(\cN(\rho)\big) \right| \leq \frac{p}{p-1}\varepsilon. \]
Besides, there exists a CPTP map $\widehat{\cN}':\cL(\A)\rightarrow\cL(\B)$ with Kraus rank at most $C\max(|\A|,|\B|)\log^5(|\rE|/\varepsilon^2)/\varepsilon^4$ (where $C>0$ is a universal constant) and such that
\begin{equation} \label{eq:approx-S} \forall\ \rho\in\cD(\A),\ \left| S\big(\widehat{\cN}'(\rho)\big)-S\big(\cN(\rho)\big) \right| \leq \varepsilon. \end{equation}
\end{theorem}

\begin{proof}
This is a direct consequence of Theorem \ref{th:main}, combined with Propositions \ref{prop:S_p} and \ref{prop:S}.
\end{proof}

We already argued about optimality in Theorem \ref{th:main}, showing that there indeed exist some CPTP maps $\cN:\cL(\A)\rightarrow\cL(\B)$ for which at least $|\A|$ or $|\B|$ Kraus operators are needed to approximate them in the sense of equation \eqref{eq:approxmain}. We will now establish that, even to get the weaker notion of approximation of equation \eqref{eq:approx-S}, a Kraus-rank of at least $|\A|$ or $|\B|$ might, in some cases, still be necessary.

Let $\cN:X\in\cL(\A)\mapsto\tr_{\mathrm{E}}(VX V^{\dagger})\in\cL(\B)$ be a CPTP map with isometry $V:\A\hookrightarrow\B\otimes\rE$. Given $\rho\in\cD(\A)$, we consider its input entropy $S(\rho)$, its output entropy $S\left(\cN(\rho)\right)$, and its entropy exchange $S\left(\rho,\cN\right)$. The latter quantity is defined as follows: let $\varphi_{A'A}$ be an extension of $\rho_A$, $\widetilde{\varphi}_{A'BE}=\left(\Id_{A'}\otimes V_{A\rightarrow BE}\right)\varphi_{A'A}$, and set
\[ S\left(\rho_A,\cN_{A\rightarrow B}\right)= S\left(\tr_{\rE}\widetilde{\varphi}_{A'BE}\right)= S\left(\tr_{\A'\B}\widetilde{\varphi}_{A'BE}\right). \]
By non-negativity of the loss and the noise of a quantum channel, we then have (see \cite{GH}, Section 4.5)
\[ \forall\ \rho\in\cD(\A),\ \left|S(\rho)-S(\cN(\rho))\right| \leq S(\rho,\cN). \]
Yet, for any $\rho\in\cD(\A)$, obviously $S(\rho,\cN)\leq\log|\rE|$. And hence as a consequence,
\[ \log|\rE| \geq \max\left\{ \left|S(\rho)-S(\cN(\rho))\right| \st \rho\in\cD(\A) \right\}. \]
In particular, we may derive the two following lower bounds on $|\rE|$, for certain CPTP maps $\cN$,
\begin{equation} \label{eq:S-exchange} \exists\ \psi_{\A}:\ \cN\left(\psi_A\right)=\frac{\Id_B}{|\B|}\ \Rightarrow\ |\rE|\geq|\B|\ \text{and}\ \exists\ \psi_B:\ \cN\left(\frac{\Id_A}{|\A|}\right)=\psi_B\ \Rightarrow\ |\rE|\geq|\A|\ . \end{equation}
And this remains approximately true for an approximation of $\cN$. Concretely, let $\widehat{\cN}:\cL(\A)\rightarrow\cL(\B)$ be a CPTP map such that
\[ \forall\ \rho\in\cD(\A),\ \left| S\big(\widehat{\cN}(\rho)\big)-S\big(\cN(\rho)\big) \right| \leq \varepsilon. \]
If $\cN$ satisfies the first condition in equation \eqref{eq:S-exchange}, then
\[ S\big(\widehat{\cN}\left(\psi_A\right)\big) \geq S\big(\cN\left(\psi_A\right)\big) -\varepsilon = \log|\B| - \varepsilon,\ \text{so that}\ \log r_K\big(\widehat{\cN}\big) \geq \log|\B|-\varepsilon,\ \text{i.e.}\ r_K\big(\widehat{\cN}\big) \geq e^{-\varepsilon}|\B|.\]
And if $\cN$ satisfies the second condition in equation \eqref{eq:S-exchange}, then
\[ S\left(\widehat{\cN}\left(\frac{\Id_A}{|\A|}\right)\right) \leq S\left(\cN\left(\frac{\Id_A}{|\A|}\right)\right) +\varepsilon = \varepsilon,\ \text{so that}\ \log r_K\big(\widehat{\cN}\big) \geq \log|\A|-\varepsilon,\ \text{i.e.}\ r_K\big(\widehat{\cN}\big) \geq e^{-\varepsilon}|\A|.\]
Therefore, the conclusion of this study is that, in Theorem \ref{th:entropies}, $r_K\big(\widehat{\cN}\big)\geq(1-\varepsilon)\max(|\A|,|\B|)$ is for sure necessary, in general, to have the entropy approximation \eqref{eq:approx-S}. It additionally tells us that there is a channel-dependent lower bound on $r_K\big(\widehat{\cN}\big)$ so that the latter holds, namely
\[ r_K\big(\widehat{\cN}\big) \geq (1-\varepsilon) \max\left\{ \left|S(\rho)-S(\cN(\rho))\right| \st \rho\in\cD(\A) \right\}. \]

\begin{proposition} \label{prop:F}
Let $\cN:\cL(\A)\rightarrow\cL(\B)$ be a CPTP map, and assume that the CP map $\widehat{\cN}:\cL(\A)\rightarrow\cL(\B)$ satisfies
\begin{equation} \label{eq:order'} \forall\ \rho\in\cD(\A),\ (1-\varepsilon)\cN(\rho) -\varepsilon\frac{\Id}{|\B|} \leq \widehat{\cN}(\rho)\leq (1+\varepsilon)\cN(\rho) +\varepsilon\frac{\Id}{|\B|},  \end{equation}
for some $0<\varepsilon<1/2$. Then, $\widehat{\cN}$ is close to $\cN$ in terms of output fidelities, in the sense that
\[ \forall\ \rho\in\cD(\A),\forall\ \omega\in\cD(\B),\ \left| F\big(\widehat{\cN}(\rho),\omega\big) - F\big(\cN(\rho),\omega\big) \right| \leq \frac{3}{\sqrt{2}}\sqrt{\varepsilon}. \]
\end{proposition}

\begin{proof}
As noted in the proof of Proposition \ref{prop:S_p}, setting $\sigma=\cN(\rho)$, $\widehat{\sigma}=\widehat{\cN}(\rho)$ and $\tau=\Id/|\B|$, we can re-write equation \eqref{eq:order'} as the two inequalities $\widehat{\sigma}\leq (1+\varepsilon)\sigma +\varepsilon\tau$ and $\sigma\leq (1+2\varepsilon)\widehat{\sigma} +2\varepsilon\tau$.
By operator monotonicity of $F(\cdot,\omega)$, and the fact that it is upper bounded by $1$, these imply the two estimates
\[ F(\widehat{\sigma},\omega) \leq \sqrt{1+\varepsilon}\,F(\sigma,\omega) +\sqrt{\varepsilon}\,F(\tau,\omega) \leq  F(\sigma,\omega) + \frac{\varepsilon}{2} + \sqrt{\varepsilon}, \]
\[ F(\sigma,\omega) \leq \sqrt{1+2\varepsilon}\,F(\widehat{\sigma},\omega) +\sqrt{2\varepsilon}\,F\big(\tau,\omega) \leq  F(\widehat{\sigma},\omega) + \varepsilon + \sqrt{2\varepsilon}. \]
Finally, just observing that $\varepsilon\leq \sqrt{\varepsilon/2}$ for $0<\varepsilon<1/2$, the conclusion of Proposition \ref{prop:F} directly follows.
\end{proof}

\begin{theorem} \label{th:fidelitiesies}
Fix $0<\varepsilon<1$ and let $\cN:\cL(\A)\rightarrow\cL(\B)$ be a CPTP map with Kraus rank $|\rE|\geq |\A|,|\B|$. Then, there exists a CP map $\widehat{\cN}:\cL(\A)\rightarrow\cL(\B)$ with Kraus rank at most $C\max(|\A|,|\B|)\log(|\rE|/\varepsilon)/\varepsilon^4$ (where $C>0$ is a universal constant) and such that
\[ \forall\ \rho\in\cD(\A),\forall\ \omega\in\cD(\B),\ \left| F\big(\widehat{\cN}(\rho),\omega\big) - F\big(\cN(\rho),\omega\big) \right| \leq \varepsilon. \]
\end{theorem}

\begin{proof}
This is a direct consequence of Theorem \ref{th:main}, combined with Proposition \ref{prop:F}.
\end{proof}

\subsection{Destruction of correlations with few resources}

It was observed in \cite{HLSW}, Section 3, that an $\varepsilon$-randomizing channel (i.e.~a channel which is an $\varepsilon$-approximation of the fully randomizing channel) approximately destroys the correlations between the system it acts on and any system the latter might be coupled to, in the following two senses: First of all, a state which is initially just classically correlated becomes almost uncorrelated (or in other words any separable state is sent close to a product state, in $1$-norm distance). And second of all, whatever the initial state, the correlations present in it become almost invisible to local observers (or in other words any state is sent close to a product state, in one-way-LOCC-norm distance). Hence, having an $\varepsilon$-randomizing channel with few Kraus operators can be seen as having an efficient way to decouple a system of interest from its environment. Thanks to Theorem \ref{th:main}, we can generalize these results into Theorem \ref{th:correlations} below. The reader is referred to Chapter \ref{chap:data-hiding} for a precise definition of locally restricted measurement norms (such as the one-way-LOCC-norm) and much more on the topic of data-locking and data-hiding.

\begin{theorem} \label{th:correlations}
Let $\A,\B,\mathrm{C}$ be Hilbert spaces, and assume that $d=\max(|\A|,|\B|)<+\infty$. For any $0<\varepsilon<1$ and $\sigma_B^*\in\mathcal{D}(\B)$, there exists a CPTP map $\widehat{\mathcal{N}}:\mathcal{L}(\A)\rightarrow\mathcal{L}(\B)$ with Kraus rank at most $Cd\log(d/\varepsilon)/\varepsilon^2$ (where $C>0$ is a universal constant) and such that
\begin{equation} \label{eq:correlations1} \forall\ \rho_{AC}\in\mathcal{S}(\A{:}\mathrm{C}),\ \left\|\widehat{\mathcal{N}}\otimes\mathcal{I\!d}(\rho_{AC})-\sigma_B^*\otimes\rho_C\right\|_1 \leq \varepsilon, \end{equation}
\begin{equation} \label{eq:correlations2} \forall\ \rho_{AC}\in\mathcal{D}(\A\otimes\mathrm{C}),\ \left\|\widehat{\mathcal{N}}\otimes\mathcal{I\!d}(\rho_{AC})-\sigma_B^*\otimes\rho_C\right\|_{\mathbf{LOCC^{\rightarrow}(\B{:}\mathrm{C})}} \leq \varepsilon. \end{equation}
\end{theorem}

\begin{proof}
Define the completely forgetful CPTP map $\mathcal{N}:X_A\in\mathcal{L}(\A)\mapsto(\tr X_A)\sigma_B^*\in\mathcal{L}(\B)$ (i.e.~$\mathcal{N}$ sends every input state on the output state $\sigma_B^*$). By Theorem \ref{th:main}, there exists a CPTP map $\widehat{\mathcal{N}}:\mathcal{L}(\A)\rightarrow\mathcal{L}(\B)$ with Kraus rank at most $Cd\log(d/\varepsilon)/\varepsilon^2$ such that
\[ \forall\ \rho_A\in\mathcal{D}(\A),\ \left\|\widehat{\mathcal{N}}(\rho_A)-\mathcal{N}(\rho_A)\right\|_1\leq\varepsilon\ \text{i.e.}\ \left\|\widehat{\mathcal{N}}(\rho_A)-\sigma_B^*\right\|_1\leq\varepsilon. \]
Now, following the exact same route as in the proofs of Lemmas III.1 and III.2 in \cite{HLSW}, we get that this implies precisely equations \eqref{eq:correlations1} and \eqref{eq:correlations2}, respectively. We will therefore only briefly recall the arguments here.

Concerning equation \eqref{eq:correlations1}, let $\rho_{AC}\in\mathcal{S}(\A{:}\mathrm{C})$, i.e.~ $\rho_{AC}=\sum_xp_x\rho_A^{(x)}\otimes\rho_C^{(x)}$. Then,
\[ \left\|\widehat{\mathcal{N}}\otimes\mathcal{I\!d}(\rho_{AC})-\sigma_B^*\otimes\rho_C\right\|_1 = \left\|\sum_x p_x \left(\widehat{\mathcal{N}}\left(\rho_{A}^{(x)}\right) -\sigma_B^*\right)\otimes\rho_C^{(x)} \right\|_1
\leq \sum_x p_x \left\|\widehat{\mathcal{N}}\left(\rho_A^{(x)}\right)-\sigma_B^*\right\|_1 \leq \varepsilon, \]
where the last inequality is because, by assumption, for each $x$, $\big\|\widehat{\mathcal{N}}\big(\rho_A^{(x)}\big)-\sigma_B^*\big\|_1 \leq \varepsilon$, and $\sum_x p_x=1$.

As for equation \eqref{eq:correlations2}, let $\mathrm{M}=\big(M_B^{(x)}\otimes M_C^{(x)}\big)_x\in\mathbf{LOCC^{\rightarrow}(\B{:}\mathrm{C})}$, i.e.~ for each $x$, $0\leq M_B^{(x)},M_C^{(x)}\leq \Id$, and $\sum_x M_B^{(x)}=\Id$. Then, for any $\rho_{AC}\in\mathcal{D}(\A\otimes\mathrm{C})$,
\begin{align*} \left\| \widehat{\mathcal{N}}\otimes\mathcal{I\!d}(\rho_{AC})-\mathcal{N}\otimes\mathcal{I\!d}(\rho_{AC})\right\|_{\mathrm{M}} = & \sum_x \left| \tr\left[ M_B^{(x)}\otimes M_C^{(x)} \left( \widehat{\mathcal{N}}\otimes\mathcal{I\!d}(\rho_{AC})- \mathcal{N}\otimes\mathcal{I\!d}(\rho_{AC}) \right) \right] \right| \\
= & \sum_x \left| \tr\left[ \left(\widehat{\mathcal{N}}^*\left(M_B^{(x)}\right)- \mathcal{N}^*\left(M_B^{(x)}\right)\right)\otimes M_C^{(x)} \rho_{AC}\right] \right| \\
\leq & \sum_x \left\| \widehat{\mathcal{N}}^*\left(M_B^{(x)}\right)- \mathcal{N}^*\left(M_B^{(x)}\right) \right\|_{\infty} \\
\leq & \,\varepsilon,
\end{align*}
where the next-to-last inequality is because, $\|\rho_{AC}\|_1\leq 1$ and for each $x$, $\big\|M_C^{(x)}\big\|_{\infty}\leq 1$, while the last inequality is because, by assumption, for each $x$, $\big\| \widehat{\mathcal{N}}^*\big(M_B^{(x)}\big)- \mathcal{N}^*\big(M_B^{(x)}\big) \big\|_{\infty} \leq \varepsilon \tr M_B^{(x)}/|\B|$, and $\sum_x \tr M_B^{(x)}=|\B|$.
\end{proof}

\subsection{The case of Werner channels}

An interesting case to which Theorem \ref{th:main'} applies is that of the so-called Werner channels. These are defined as the family of CPTP maps
\[ \mathcal{W}_{\lambda}:X\in\mathcal{L}(\A)\mapsto\frac{1}{|\A|+2\lambda-1}\left[(\tr X)\Id +(2\lambda-1)X^T\right]\in\mathcal{L}(\A),\ 0\leq\lambda\leq 1. \]
Denoting by $\pi_s$ and $\pi_a$ the symmetric and anti-symmetric states on $\A\otimes\A$, it is easy to check that, for each $0\leq\lambda\leq 1$, the Choi-Jamiolkowski state $\tau(\mathcal{W}_{\lambda})$ associated to $\mathcal{W}_{\lambda}$ is nothing else than the Werner state $\rho_{\lambda}=\lambda\pi_s+(1-\lambda)\pi_a$ (see Chapter \ref{chap:symmetries}, Section \ref{sec:Werner}, for all definitions if need be). Hence, $\mathcal{W}_{\lambda}$ has Kraus rank $|\A|^2$ whenever $0<\lambda<1$, and $|\A|(|\A|+1)/2$, resp.~$|\A|(|\A|-1)/2$, when $\lambda=1$, resp.~ $\lambda=0$, i.e.~in any case full or almost full Kraus rank. These channels are thus typically of the kind that we would like to compress into more economical ones. What is more, they have the property of having only very mixed output states. Indeed,
\[ \max_{\rho\in\mathcal{D}(\A)}\left\|\mathcal{W}_{\lambda}(\rho)\right\|_{\infty} = \begin{cases} 2\lambda/(|\A|+2\lambda-1)\ \text{if}\ \lambda\geq 1/2\\ 1/(|\A|+2\lambda-1)\ \text{if}\ \lambda< 1/2 \end{cases} \leq \frac{2}{|\A|}. \]
So by Theorem \ref{th:main'}, we get that, for each $0\leq\lambda\leq 1$, given $0<\varepsilon<1$, there exists a CP map $\widehat{\mathcal{W}}_{\lambda}:\mathcal{L}(\A)\rightarrow\mathcal{L}(\A)$ with Kraus rank at most $C|\A|/\varepsilon^2$ (where $C>0$ is a universal constant) such that
\[ \forall\ \rho\in\mathcal{D}(\A),\ \left\| \widehat{\mathcal{W}}_{\lambda}(\rho)-\mathcal{W}_{\lambda}(\rho)\right\|_{\infty} \leq \frac{\varepsilon}{|\A|}. \]
In words, this means that the Werner CPTP maps can be $(\varepsilon/|\A|)$-approximated in $(1{\rightarrow}\infty)$-norm distance (hence in particular $\varepsilon$-approximated in $(1{\rightarrow}1)$-norm distance) by CP maps having Kraus rank $C|\A|/\varepsilon^2\ll |\A|^2$.

\section{Discussion}

We have generalized in several senses the result established in \cite{HLSW} and \cite{Aubrun2}. First, we have shown that it holds for all quantum channels and not only for the fully randomizing one: any CPTP map from $\mathcal{L}(\mathrm{A})$ to $\mathcal{L}(\mathrm{B})$ can be $\varepsilon$-approximated in $(1{\rightarrow}1)$-norm distance by a CPTP map with Kraus rank of order $d\log(d/\varepsilon)/\varepsilon^2$, where $d=\max(|\A|,|\B|)$. Second, we have established that a stronger notion of approximation can actually be proven, namely an $\varepsilon$-ordering of the two CP maps, which allows to derive approximation results in terms of various output quantities (that are tighter than those induced by the rougher norm distance closeness). In the case where the channel under consideration is, as the fully randomizing channel, very noisy (meaning that all output states are very mixed), the extra $\log(d/\varepsilon)$ factor in our result can be removed. However, we do not know if this is true in general. On a related note, our study of optimality shows that there exist channels which cannot be compressed below order $d$ Kraus operators (even to achieve the weakest notions of approximation). But what about channel-dependent lower bounds? For a given channel, would there be a more clever construction than ours (i.e.~a non-universal one) that would enable its compression to a number of Kraus operators whose $\log$ would be, for instance, of order its maximum input-output entropy difference?

\smallskip

Finally, full or partial derandomization of our construction would be desirable. Here again the main difficulty is that most of the techniques which apply to very noisy channels may fail in general. Let us specify a bit what we mean. In \cite{Aubrun2}, two approximation schemes were proposed for the fully randomizing channel $\mathcal{R}:\cL(\C^d)\rightarrow\cL(\C^d)$. They consisted in taking as Kraus operators $\{U_i/\sqrt{n},\ 1\leq i\leq n\}$ with $U_1,\ldots,U_n$ sampled either from the Haar measure on $\cU(\C^d)$ or from any other isotropic (aka unitary $1$-design) measure on $\cU(\C^d)$. It was then shown that, in order to approximate $\mathcal{R}$ up to error $\varepsilon/d$ in $(1{\rightarrow}\infty)$-norm, $n$ of order $d/\varepsilon^2$ was enough in the Haar-distributed case and $n$ of order $d\log^6d/\varepsilon^2$ was enough in the, more general, isotropically-distributed case. The advantage of the second result compared to the first one is that there exist isotropic measures which are much simpler than the Haar measure on $\cU(\C^d)$, in particular discrete ones (e.g.~the uniform measure over any unitary orthogonal basis of $\cL(\C^d)$). Hence, from a practical point of view, generating such a measure is arguably more realistic than generating the Haar measure (the reader is e.g. referred to \cite{BHH} for a more precise formulation of the claim that implementing a Haar distributed unitary is hard and an extensive discussion on how to approximate such a unitary by a more easily implementable one). Now, if $\cN:\cL(\C^d)\rightarrow\cL(\C^d)$ is a channel, with environment $\C^s$, such that $\sup_{\rho\in\cD(\C^d)}\|\cN(\rho)\|_{\infty} \leq C/d$, then arguments of the same type apply to our construction: to approximate $\mathcal{N}$ up to error $\varepsilon/d$ in $(1{\rightarrow}\infty)$-norm by sampling unit vectors in $\C^s$, order $d/\varepsilon^2$ of them is enough if they are Haar-distributed (which is the content of Theorem \ref{th:main}) and order $d\log^6d/\varepsilon^2$ of them is enough if they are only assumed to be isotropically-distributed. Here as well, the gain in terms of needed amount of randomness is obvious: there exist isotropic measures which are much simpler to sample from than the Haar-measure on $S_{\C^s}$ (e.g.~the uniform measure on any orthonormal basis of $\C^s$). Unfortunately, this whole reasoning (based on Dudley's upper bounding of Bernoulli averages by covering number integrals and on a sharp entropy estimate for the suprema of empirical processes in Banach spaces) fails completely for channels that have some of their outputs which are too pure.

\chapter{Zonoids and sparsification of quantum measurements}
\chaptermark{Zonoids and sparsification of quantum measurements}
\label{chap:zonoids}

\textsf{Based on ``Zonoids and sparsification of quantum measurements'', in collaboration with G. Aubrun \cite{AL2}.}

\bigskip

In this chapter, we establish a connection between zonoids (a concept from classical convex geometry) and the distinguishability norms associated to quantum measurements or POVMs (Positive Operator-Valued Measures), recently introduced in quantum information theory.

This correspondence allows us to state and prove the POVM version of classical results from the local theory of Banach spaces about the approximation of zonoids by zonotopes. We show that on $\C^d$, the uniform POVM (the most symmetric POVM) can be sparsified, i.e.~approximated by a discrete POVM, the latter having only $O(d^2)$ outcomes. We also show that similar (but weaker) approximation results actually hold for any POVM on $\C^d$.

By considering an appropriate notion of tensor product for zonoids, we extend our results to the multipartite setting: we show,
roughly speaking, that local POVMs may be sparsified locally. In particular, the local uniform POVM on $\C^{d_1}\otimes\cdots\otimes\C^{d_k}$
can be approximated by a discrete POVM which is local and has $O(d_1^2 \times \cdots \times d_k^2)$ outcomes.

\section{Introduction}

A classical result by Lyapounov (\cite{RudinFA}, Theorem 5.5) asserts that the range of a non-atomic $\R^n$-valued vector measure is closed and convex.
Convex sets in $\R^n$ obtained in this way are called zonoids. Zonoids are equivalently characterized as
convex sets which can be approximated by finite sums of segments.

Here we consider a special class of vector measures: Positive Operator-Valued Measures (POVMs). In the formalism of quantum mechanics, POVMs
represent the most general form of a quantum measurement. Recently, Matthews, Wehner and Winter \cite{MWW} introduced
the distinguishability norm associated to a POVM. This norm has an operational interpretation as the bias of the POVM for the state
discrimination problem (a basic task in quantum information theory) and is closely related to the zonoid arising from Lyapounov's theorem.

A well-studied question in high-dimensional convexity is the approximation of zonoids by zonotopes. The series of papers
\cite{FLM,Schechtman,BLM,Talagrand}
culminates in the following result: any zonoid in $\R^n$ can be approximated by the sum of $O(n \log n)$ segments. The aforementioned connection
between POVMs and zonoids allows us to state and prove approximation results for POVMs, which improve on previously known bounds. Precise statements
appear as Theorem \ref{theorem:approximation-of-U} and \ref{theorem:approximation-any}.

This chapter is organized as follows. Section \ref{section:POVM} introduces POVMs and their associated distinguishability norms.
Section \ref{section:POVMs-zonoids} connects POVMs with zonoids. Section \ref{section:tensorizing} introduces a notion of tensor product
for POVMs, and the corresponding notion for zonoids.
Section \ref{section:sparsification} pushes forward this connection to state the POVM
version of approximation results for zonoids, which are proved in Sections \ref{sec:uniform-POVM} and \ref{sec:approximation-any}. Section
\ref{section:sparsification-multipartite} provides sparsification results for local POVMs on multipartite systems.

The reader may have a look at Table \ref{table:analogies}, which summarizes analogies between zonoids and POVMs.

\subsection*{Notation}


Let us recall a few standard concepts from classical convex geometry that we will need throughout our proofs. Much more on that matter is gathered in Chapter \ref{chap:toolbox}, Section \ref{ap:convex-geometry}. The support function $h_K$ of a convex compact set $K \subset \R^n$
is the function defined for $x \in \R^n$ by $ h_K(x) = \sup \{ \langle x,y \rangle \st y \in K \}$.
Moreover, for a pair $K,L$ of convex compact sets, the inclusion $K \subset L$ is equivalent to the inequality $h_K \leq h_L$.
The polar of a convex set $K \subset \R^n$ is $K^\circ = \{ x \in \R^n \st \langle x,y \rangle \leq 1 \textnormal{ whenever } y \in K \}$.
The bipolar theorem (see e.g.~\cite{Barvinok}) states that $(K^\circ)^\circ$ is the closed convex hull of $K$ and $\{0\}$.
A convex body is a convex compact set with non-empty interior.

Whenever we apply tools from convex geometry in the (real) space $\cH(\C^d)$ of Hermitian operators on $\C^d$ (e.g.~polars or support functions), we use the Hilbert--Schmidt inner product $\langle A,B\rangle \mapsto \tr (AB)$ to define the Euclidean structure. In that setting, in addition to the general notation specified in Chapter \ref{chap:motivations}, Section \ref{sec:notation}, we introduce the following one: $[-\Id,\Id]$ stands for the set of $A\in\mathcal{H}(\C^d)$ such that $-\Id \leq A \leq \Id$. In
other words $[-\Id,\Id]$ is the Hermitian part of the unit ball for $\|\cdot\|_{\infty}$ on $\mathcal{L}(\C^d)$.

We also use the following convention: whenever a formula is given for the dimension of a (sub)space, it is tacitly understood that one should take the integer part.

\section{POVMs and distinguishability norms} \label{section:POVM}

In quantum mechanics, the state of a $d$-dimensional system is described by a positive operator on $\C^d$ with trace $1$.
The most general form of a
measurement that may be performed on such a quantum system is encompassed by the formalism of Positive Operator-Valued Measures (POVMs).
Given a set $\Omega$ equipped with a $\sigma$-algebra $\mathcal{F}$, a POVM on $\C^d$
is a map $\mathrm{M} :\cF \to \cH_+(\C^d)$  which is $\sigma$-additive and such that $\mathrm{M}(\Omega) = \Id$.
In this definition the space $(\Omega,\cF)$ could potentially be infinite, so that the POVMs defined on it would be continuous. However, we often restrict ourselves to the subclass
of discrete POVMs, and a main point of this chapter is to substantiate this ``continuous to discrete'' transition. The reader is referred to Chapter \ref{chap:QIT}, Section \ref{sec:states-POVMs}, for the physical motivations behind this mathematical formalism.

A discrete POVM is a POVM in which the underlying $\sigma$-algebra $\cF$ is required to be finite. In that case there is a finite
partition $\Omega = A_1 \cup \cdots \cup A_n$ generating $\cF$. The positive operators $M_i = \mathrm{M}(A_i)$ are often referred to as the elements of
the POVM, and they satisfy the condition $M_1 + \dots + M_n = \Id$. We usually identify a discrete POVM with the set of its elements by writing
$\mathrm{M} = (M_i)_{1 \leq i \leq n}$. The index set $\{1,\dots,n\}$ labels the outcomes of the measurement.
The integer $n$ is thus the number of outcomes of $\mathrm{M}$ and can be seen as a crude way to measure the
complexity of $\mathrm{M}$.

What happens when measuring with a POVM $\mathrm{M}$ a quantum system in a state $\rho$?
In the case of a discrete POVM $\mathrm{M}=(M_i)_{1 \leq i \leq n}$, we know from Born's rule that the outcome $i$ is output with probability
$\tr (\rho M_i)$.
This simple formula can be used to quantify the efficiency of a POVM to perform the task of state discrimination. State discrimination can be
described as follows: a quantum system is prepared in an unknown state which is either $\rho$ or $\sigma$ (both hypotheses being a priori
equally likely), and we have to guess the unknown state. After measuring it with the discrete POVM $\mathrm{M} =
(M_i)_{1 \leq i \leq n}$, the optimal strategy,
based on the maximum likelihood probability, leads to a probability of wrong guess equal to \cite{Holevo,Helstrom}
\[ \P_{error} = \frac{1}{2}\left( 1 - \frac{1}{2} \sum_{i=1}^n \left| \tr (\rho M_i) - \tr(\sigma M_i) \right| \right) . \]
In this context, the quantity $\left(\sum_{i=1}^n \left| \tr (\rho M_i) - \tr(\sigma M_i) \right|\right)/2$
is therefore called the bias of the POVM $\mathrm{M}$ on the state pair $(\rho,\sigma)$.

Following \cite{MWW}, we introduce a norm on $\cH(\C^d)$, called the distinguishability norm associated to $\mathrm{M}$,
and defined for $\Delta \in \cH(\C^d)$ by
\begin{equation} \label{eq:definition-norm-discrete} \|\Delta\|_{\mathrm{M}} = \sum_{i=1}^n \left| \tr (\Delta M_i) \right|. \end{equation}
It is such that $\P_{error} = \left(1 - \| \rho - \sigma \|_{\mathrm{M}}/2\right)/2$, and thus quantifies how powerful the POVM
$\mathrm{M}$ is in discriminating one state from another with the smallest probability of error.

The terminology ``norm'' is slightly abusive since one may have $\|\Delta\|_{\mathrm{M}}=0$
for a nonzero $\Delta \in \cH(\C^d)$. The functional $\|\cdot\|_{\mathrm{M}}$ is however always a semi-norm, and
it is easy to check that $\|\cdot\|_{\mathrm{M}}$ is a norm if and only if the POVM elements $(M_i)_{1 \leq i \leq n}$ span $\cH(\C^d)$ as a vector
space. Such POVMs
are called informationally complete in the quantum information literature.

Similarly, the distinguishability norm associated to a general POVM $\mathrm{M}$, defined on a set $\Omega$ equipped with a $\sigma$-algebra
$\mathcal{F}$, is described for $\Delta \in \cH(\C^d)$
by
\begin{equation}
\label{eq:definition-norm-continuous}
\| \Delta \|_{\mathrm{M}} = \| \tr(\Delta \mathrm{M}(\cdot)) \|_{\mathrm{TV}} = \sup_{A \in \cF} \big[ \tr(\Delta \mathrm{M}(A)) -
\tr(\Delta \mathrm{M}(\Omega \setminus A)) \big]  = \sup_{M \in \mathrm{M}(\mathcal{F})} \tr (\Delta(2M-\Id)).\end{equation}
Here $\|\mu\|_{\mathrm{TV}}$ denotes the total variation of a measure $\mu$. When $\mathrm{M}$ is discrete, formulae
\eqref{eq:definition-norm-discrete} and
\eqref{eq:definition-norm-continuous} coincide.
Note also that the inequality $\|\cdot\|_{\mathrm{M}} \leq \|\cdot\|_1$ holds for any POVM $\mathrm{M}$, with equality on $\cH_+(\C^d)$.

Given a POVM $\mathrm{M}$, we denote by $B_{\mathrm{M}} = \{ \|\cdot \|_{\mathrm{M}} \leq 1 \}$ the unit ball for the distinguishability norm,
and $K_{\mathrm{M}}= (B_{\mathrm{M}})^\circ$ its polar, i.e.~\[ K_{\mathrm{M}} = \{ A \in \cH(\C^d) \st \tr (AB) \leq 1 \textnormal{ whenever } \|B\|_{\mathrm{M}} \leq 1 \}. \]
The set $K_{\mathrm{M}}$ is a compact convex set. Moreover $K_{\mathrm{M}}$ has nonempty interior if and only if the POVM $\mathrm{M}$
is informationally complete. It follows from the inequality $\|\cdot\|_{\mathrm{M}} \leq \|\cdot\|_1$ that $K_{\mathrm{M}}$ is always included in the
operator interval $[-\Id,\Id]$.

On the other hand, it follows from \eqref{eq:definition-norm-continuous} that $B_{\mathrm{M}} = ( 2 \mathrm{M}(\mathcal{F}) - \Id)^{\circ}$,
and the bipolar theorem implies that
\begin{equation} \label{eq:K_M-polar} K_{\mathrm{M}} = 2 \conv ( \mathrm{M}(\mathcal{F})) - \Id . \end{equation}
By Lyapounov's theorem,  the convex hull operation is not needed when $\mathrm{M}$ is non-atomic.
For a discrete POVM $\mathrm{M} = ( M_i )_{1 \leq i \leq n}$, equation \eqref{eq:K_M-polar} may be rewritten in the form
\begin{equation} \label{eq:K_M-zonotope} K_{\mathrm{M}} = \conv \{\pm M_1 \} + \cdots + \conv \{ \pm M_n \}, \end{equation}
where the addition of convex sets should be understood as the Minkowski sum: $A+B = \{a+b \st a \in A,\ b \in B \}$.

We are going to show that POVMs can be sparsified, i.e approximated by discrete POVMs with few outcomes.
The terminology ``approximation'' here refers to the associated distinguishability norms: a POVM $\mathrm{M}$ is considered to be ``close'' to a
POVM $\mathrm{M}'$ when
their distinguishability norms satisfy inequalities of the form
\[ (1-\e)\|\cdot\|_{\mathrm{M}'} \leq \|\cdot\|_{\mathrm{M}} \leq (1+\e)\|\cdot\|_{\mathrm{M}'}. \]
This notion of approximation has an operational significance: two POVMs are comparable when both lead to comparable biases when used for any state
discrimination task. Let us perhaps stress that point: if one has additional information on the states to be discriminated,
it may of course be used to design a POVM specifically efficient for those (one could for instance be interested in the problem
of distinguishing pairs of low-rank states \cite{Sen,AE}).

In this chapter, we study the distinguishability norms from a functional-analytic point of view. We are mostly
interested in the asymptotic regime,
when the dimension $d$ of the underlying Hilbert space is large.

\begin{table}[h] \caption{A ``dictionary'' between zonoids and POVMs}
\label{table:analogies}
\renewcommand{\arraystretch}{1.5}
\small{
\begin{tabular}{| C{8.7cm} | C{7.3cm} |}
\hline
Zonotope which is the Minkowski sum of $N$ segments & Discrete POVM with $N$ outcomes \\
\hline
Zonoid = limit of zonotopes  & General POVM = limit of discrete POVMs \\
\hline
Tensor product of zonoids  & Local POVM on a multipartite system \\
\hline
Euclidean unit ball $B^n$ & Uniform POVM $\mathrm{U}_d$ \\
= most symmetric zonoid in $\R^n$ & = most symmetric POVM on $\C^d$ \\
\hline
``4th moment method'' (explicit \cite{Rudin60}): $c B^n \subset Z \subset C B^n$,  & ``Approximate $4$-design POVM'' \cite{AE}: \\
with $Z$ a zonotope which is the sum of $O(n^2)$ segments. & explicit sparsification of $\mathrm{U}_d$ with $O(d^4)$ outcomes. \\
\hline
Measure concentration (non-explicit \cite{FLM}): $(1-\e) B^n \subset Z \subset (1+\e) B^n$, & Theorem \ref{theorem:approximation-of-U}:
a randomly chosen POVM  \\
with $Z$ a zonotope which is the sum of $O_\e(n)$ segments. & with $O(d^2)$ outcomes is a sparsification of $\mathrm{U}_d$. \\
\hline
Derandomization \cite{GW,LS,IS} & ? \\
\hline
Any zonoid in $\R^n$ can be approximated by a zonotope & Theorem \ref{theorem:approximation-any}: any POVM on $\C^d$ can be sparsified \\
which is the sum of $O(n \log n)$ segments \cite{Talagrand}. & into a sub-POVM with $O(d^2 \log d)$ outcomes. \\
\hline
\end{tabular}
}
\end{table}

\section{POVMs and zonoids}

\label{section:POVMs-zonoids}

\subsection{POVMs as probability measures on states}

The original definition of a POVM involves an abstract measure space, and the specification of this measure space is irrelevant when considering the
distinguishability norms. The following proposition, which is probably well-known, gives a more concrete look at POVMs as probability measures
on the set $\mathcal{D}(\C^d)$ of states on $\C^d$.

\begin{proposition} \label{proposition:POVM-states}
Let $\mathrm{M}$ be a POVM on $\C^d$. There is a unique Borel probability measure $\mu$ on $\mathcal{D}(\C^d)$ with barycenter equal to $\Id/d$ and such
that, for any $\Delta \in \cH(\C^d)$,
\begin{equation} \label{eq:support-function-POVM} \|\Delta\|_{\mathrm{M}} = d \int_{\mathcal{D}(\C^d)} \left| \tr(\Delta \rho)\right| \,
\mathrm{d} \mu(\rho). \end{equation}
Conversely, given a Borel probability measure $\mu$ with barycenter equal to $\Id/d$, there is a POVM $\mathrm{M}$ such that
\eqref{eq:support-function-POVM} is satisfied.
\end{proposition}

\begin{proof}
We use the polar decomposition for vector measures, which follows from applying the Radon--Nikodym theorem to vector measures (see \cite{RudinRCA}, Theorem 6.12): a vector measure $\mu$ defined on a $\sigma$-algebra
$\mathcal{F}$ on $\Omega$ and
taking values in a normed space $(\R^n,\|\cdot\|)$ satisfies $d\mu = h d|\mu|$ for some measurable function $h : \Omega\to \R^n$. Moreover, one
has $\|h\| = 1$ $|\mu|$-a.e. Here $|\mu|$ denotes the total variation measure of $\mu$.

Let $\mathrm{M}$ be a POVM on $\C^d$, defined on a $\sigma$-algebra $\mathcal{F}$ on $\Omega$. We equip $\cH(\C^d)$ with the trace norm, so that we simply
have $|\mathrm{M}| = \tr \mathrm{M}$ and $|\mathrm{M}|(\Omega)=d$. The polar decomposition yields a measurable function $h : \Omega \to \cH(\C^d)$
such that $\|h\|_1 = 1$ $|\mathrm{M}|$-a.e. Moreover, the fact that $\mathrm{M}(\mathcal{F})\subset\cH_+(\C^d)$ implies that $h\in\cH_+(\C^d)$ $|\mathrm{M}|$-a.e. Let $\mu$ be the push forward of $|\mathrm{M}|/d$
under the map $h$. We have
\[ \Id = \mathrm{M}(\Omega) = \int_{\Omega} h \, \mathrm{d}|\mathrm{M}| = d \int_{\cH(\C^d)} \rho \, \mathrm{d}\mu(\rho).\]
And since $h\in \mathcal{D}(\C^d)$ a.e., $\mu$ is indeed
a Borel probability measure on $\mathcal{D}(\C^d)$, with barycenter equal to $\Id/d$. Finally, for any $\Delta \in \mathcal{H}(\C^d)$,
\[ \|\Delta\|_{\mathrm{M}} = \int_\Omega |\tr (\Delta h)| \, \mathrm{d} |\mathrm{M}| = d \int_{\mathcal{D}(\C^d)} |\tr (\Delta \rho) |  \, \mathrm{d} \mu(\rho) .\]
We postpone the proof of uniqueness to the next subsection (see after Proposition \ref{proposition:POVM}).

Conversely, given a Borel probability measure $\mu$ on $\mathcal{D}(\C^d)$ with barycenter at $\Id/d$, consider the vector measure
$\mathrm{M}:\mathcal{B} \to \cH(\C^d)$, where $\mathcal{B}$ is the Borel $\sigma$-algebra on $\mathcal{D}(\C^d)$, defined by
\[ \mathrm{M}(A) = d \int_A \rho \, \mathrm{d} \mu (\rho) .\]
It is easily checked that $\mathrm{M}$ is a POVM and that formula \eqref{eq:support-function-POVM} is satisfied.
\end{proof}

Note that in the case of a discrete POVM $\mathrm{M} = (M_i)_{1 \leq i \leq n}$, the corresponding probability measure is
\[ \mu = \frac{1}{d} \sum_{i=1}^n \left( \tr M_i \right) \, \delta_{M_i/\tr M_i} .\]

\begin{corollary} \label{corollary:approximation}
Given a POVM $\mathrm{M}$ on $\C^d$, there is a sequence $(\mathrm{M}_n)_n$ of discrete POVMs such that $K_\mathrm{M_n}$ converges to $K_\mathrm{M}$
in Hausdorff distance. Moreover, if $\mu$ (resp.~$\mu_n$) denotes the probability measure on $\mathcal{D}(\C^d)$ associated to $\mathrm{M}$
(resp.~to $\mathrm{M}_n$)
as in \eqref{eq:support-function-POVM}, we can guarantee that the support of $\mu_n$ is contained into the support of $\mu$.
\end{corollary}

\begin{proof}
Let $\mu$ be the probability measure associated to $\mathrm{M}$.
Given $n$, let $(Q_k)_k$ be a finite partition of $\mathcal{D}(\C^d)$ into sets of diameter at most $1/n$ with respect to the trace norm.
Let $\rho_k \in \mathcal{D}(\C^d)$ be the barycenter of the restriction of $\mu$ to $Q_k$
(only defined when $\mu(Q_k)>0$). The probability measure
\[ \mu_n = \sum_{k} \mu(Q_k) \delta_{\rho_k} \]
has the same barycenter as $\mu$, and the associated POVM $\mathrm{M}_n$ satisfies
\[ \left| h_{K_\mathrm{M}} (\Delta) - h_{K_{\mathrm{M}_n}} (\Delta) \right| \leq d \frac{\|\Delta\|_{\infty}}{n} ,\]
and therefore $K_\mathrm{M_n}$ converges to $K_\mathrm{M}$.

The condition on the supports can be enforced by changing slightly the definition
of $\mu_n$. For each $k$ we can write $\rho_k = \sum \lambda_{k,j} \rho_{k,j}$, where $(\lambda_{k,j})_j$ is a convex
combination and $(\rho_{k,j})_j$ belong to the support of $\mu$ restricted to $Q_k$. The measure
\[ \mu'_n = \sum_{k} \mu(Q_k) \sum_j \lambda_{k,j} \delta_{\rho_{k,j}} \]
satisfies the same properties as $\mu_n$, and its support is contained into the support of $\mu$.
\end{proof}

\subsection{POVMs and zonoids}

\begin{figure}[h] \caption{A symmetric zonotope in $\R^2$}
\label{fig:zonotope}
\begin{center}
\includegraphics[width=8cm]{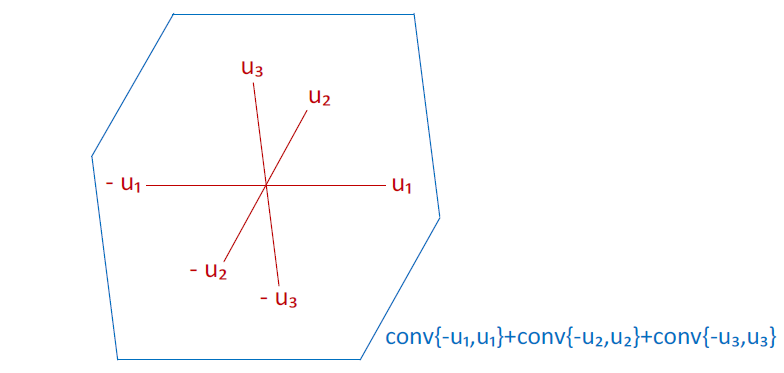}
\end{center}
\end{figure}

We connect here POVMs with zonoids, which form an important family of convex bodies (see \cite{Bolker,SW,GW} for surveys on zonoids
to which we refer for all the material presented here).
A zonotope $Z \subset \R^n$ is a closed convex set which can be written as the Minkowski sum of finitely many segments,
i.e.~such that there exist finite sets of vectors
$(u_i)_{1 \leq i \leq N}$ and $(v_i)_{1 \leq i \leq N}$ in $\R^n$ such that
\begin{equation} \label{eq:def-zonotope} Z = \conv \{ u_1,v_1 \} + \cdots + \conv \{ u_N,v_N \} .\end{equation}
A zonoid is a closed convex set which can be approximated by zonotopes (with respect to the Hausdorff distance).
Every zonoid has a center of symmetry, and therefore can be translated into a (centrally) symmetric zonoid.
Note that for a centrally symmetric zonotope, we can choose $v_i=-u_i$ in \eqref{eq:def-zonotope}. An example of symmetric zonotope in $\R^2$ appears in Figure \ref{fig:zonotope}.

Here are equivalent characterizations of zonoids.

\begin{proposition} \label{proposition:zonoids}
Let $K \subset \R^n$ be a symmetric closed convex set. The following are equivalent.
\begin{enumerate}
\item[(i)] $K$ is a zonoid.
\item[(ii)] There is a Borel positive measure $\nu$ on the Euclidean unit sphere $S^{n-1}$ which is even (i.e.~such that $\nu(A)=\nu(-A)$ for any Borel set
$A \subset S^{n-1}$) and such that, for every $x\in\R^n$,
\begin{equation} \label{eq:support-function-zonoid} h_K(x) = \int_{S^{n-1}} | \langle x,\theta \rangle | \, \mathrm{d}\nu(\theta) . \end{equation}
\item[(iii)] There is a vector measure $\mu : (\Omega,\cF) \to \R^n$ such that $K = \mu(\cF)$.
\end{enumerate}
Moreover, when these conditions are satisfied, the measure $\nu$ is unique.
\end{proposition}

\begin{remark}
Having the measure $\nu$ supported on the sphere and be even is only a matter of
normalization and a way to enforce uniqueness: if $\nu$ is a Borel measure on $\R^n$ for which linear forms are integrable,
there is a symmetric zonoid $K \subset \R^n$ such that
 \[ h_K(x) = \int_{\R^n} | \langle x, y \rangle | \, \mathrm{d}\nu(y) . \]
\end{remark}

As an immediate consequence, we characterize which subsets of $[-\Id,\Id]$ arise as $K_\mathrm{M}$ for some POVM $\mathrm{M}$.

\begin{proposition} \label{proposition:POVM}
Let $K \subset \cH(\C^d)$ be a symmetric closed convex set. Then the following are equivalent.
\begin{enumerate}
\item[(i)] $K$ is a zonoid such that $K \subset [-\Id,\Id]$ and $\pm \Id \in K$.
\item[(ii)] There exists a POVM $\mathrm{M}$ on $\C^d$ such that $K=K_{\mathrm{M}}$.
\end{enumerate}
Moreover, $K$ is a zonotope only if the POVM $\mathrm{M}$  can be chosen to be discrete.
\end{proposition}

\begin{proof}
Let $K$ be a zonoid such that $\pm \Id \in K \subset [-\Id,\Id]$. From Proposition \ref{proposition:zonoids}, there is a vector measure
$\mu$ defined on a $\sigma$-algebra $\mathcal{F}$ on a set $\Omega$, whose range is $K$. Let $A \in \mathcal{F}$ such that $\mu(A) = -\Id$.
The vector measure $\mathrm{M}$ defined for $B \in \mathcal{F}$ by
\[ \mathrm{M}(B) = \frac{1}{2} \left( \mu(B \setminus A) - \mu( B \cap A) \right)= \frac{1}{2}\left( \mu( B \Delta A) + \Id \right) \]
is a POVM. Indeed, its range, which equals $( K +\Id)/2$, lies inside the positive semidefinite cone, and contains $\Id$. We get from
\eqref{eq:K_M-polar} that $K_\mathrm{M}=K$.

Conversely, for any POVM $\mathrm{M}$, formula \eqref{eq:K_M-polar} implies that $\pm \Id \in K \subset [-\Id,\Id]$.
The fact that $K$ is a zonoid follows, using
the general fact that the convex hull of the range of a vector measure is a zonoid (see \cite{Bolker}, Theorem 1.6).

In the case of zonotopes and discrete POVMs, these arguments have more elementary analogues which we do not repeat.
\end{proof}

We can now argue about the uniqueness part in Proposition \ref{proposition:POVM-states}. This is indeed a consequence
of the uniqueness of the measure associated to a zonoid in Proposition \ref{proposition:zonoids}: after rescaling and symmetrization, a measure
$\mu$ on $\mathcal{D}(\C^d)$ satisfying
\eqref{eq:support-function-POVM} naturally induces a measure $\nu$ on the Hilbert--Schmidt sphere satisfying \eqref{eq:support-function-zonoid}
for $K=K_{\mathrm{M}}$.

Another characterization of zonoids involves the Banach space $L^1=L^1([0,1])$. A symmetric convex body $K$
is a zonoid if and only if the normed space $(\R^n,h_K)$ embeds isometrically into $L^1$. Therefore, Proposition \ref{proposition:POVM} can be restated
as a characterization of distinguishability norms on $\mathcal{H}(\C^d)$.

\begin{corollary}
Let $\|\cdot\|$ be a norm on $\mathcal{H}(\C^d)$. The following are equivalent
\begin{enumerate}
 \item[(i)] There is POVM $\mathrm{M}$ on $\C^d$ such that $\|\cdot\|=\|\cdot\|_{\mathrm{M}}$.
 \item[(i)] The normed space $(\mathcal{H}(\C^d),\|\cdot\|)$ is isometric to a subspace of $L^1$, and the following inequality is satisfied for any
 $\Delta \in \mathcal{H}(\C^d)$
 \[ | \tr \Delta | \leq \|\Delta\| \leq \tr | \Delta | .\]
\end{enumerate}
\end{corollary}

\section{Local POVMs and tensor products of zonoids}

\label{section:tensorizing}

\subsection{Tensor products for zonoids}

There is a natural notion of tensor product for subspaces of $L^1$ which appeared in the Banach space literature (see e.g.~\cite{FJ}).

\begin{definition} \label{definition:1-tensor}
Let $X,Y$ be two Banach spaces which can be embedded isometrically into $L^1$, i.e.~such that there exist linear
norm-preserving maps $i : X \to L^1(\mu)$ and $j : Y \to L^1(\nu)$. Then, the $1$-tensor product of $X$ and $Y$ is defined as the completion of the
algebraic
tensor product $X \otimes Y$ for the norm
\[ \left\| \sum_k x_k \otimes y_k \right\|_{X \otimes^1 Y} = \int \int \left| \sum_k i(x_k)(s)j(y_k)(t) \right| \, \mathrm{d}\mu(s)
\mathrm{d}\nu(t) .\]
\end{definition}

It can be checked that the norm above is well-defined and does not depend on the particular choice of the embeddings $i,j$ (see e.g.~\cite{FJ} or Lemma 2 in \cite{RosSza}).

In the finite-dimensional case, subspaces of $L^1$ are connected to zonoids. Therefore, Definition \ref{definition:1-tensor}
leads naturally to a notion of tensor product for (symmetric) zonoids.

\begin{definition} \label{def:otimes_Z}
Let $K \subset\R^m$ and $L \subset\R^n$ be two symmetric zonoids, and suppose that $\nu_K$ and $\nu_L$ are Borel measures on $S^{m-1}$ and $S^{n-1}$
respectively, such that for any $x \in \R^m$ and $y \in \R^n$,
\[ h_K(x) = \int_{S^{m-1}} | \langle x,\theta \rangle | \, \mathrm{d}\nu_K(\theta)\ \text{ and }\ h_L(y) = \int_{S^{n-1}}
| \langle y,\phi \rangle | \, \mathrm{d}\nu_L(\phi). \]
The zonoid tensor product of $K$ and $L$ is defined as the zonoid $K \otimes^Z L \subset \R^n \otimes \R^m$ whose support function satisfies
\begin{equation} \label{eq:tensorizing-zonoids} h_{K \otimes^Z L}(z) =\int_{S^{m-1}}\int_{S^{n-1}}
| \langle z,\theta\otimes\phi \rangle | \, \mathrm{d}\nu_K(\theta)\mathrm{d}\nu_L(\phi) \end{equation}
for any $z \in \R^m \otimes \R^n$.
\end{definition}

As in Definition \ref{definition:1-tensor}, this construction does not depend on the choice of the measures $\nu_K$ and $\nu_L$. This can be seen
directly: given $z \in \R^m \otimes \R^n$ and $\phi\in S^{n-1}$, set $\widetilde{z}(\phi)=\left(\Id\otimes\bra{\phi}\right)(z)$. We have
\begin{equation} \label{eq:ztilde-phi} h_{K \otimes^Z L}(z) =
\int_{S^{n-1}}h_K(\widetilde{z}(\phi))\, \mathrm{d}\nu_L(\phi),
\end{equation}
and therefore $K \otimes^Z L$ does not depend on $\nu_K$. The same argument applies for $\nu_L$.

In the case
of zonotopes, the zonoid tensor product takes a simpler form :
\[ \left( \sum_i \conv \{ \pm v_i \} \right) \otimes^Z \left( \sum_j \conv \{ \pm w_j \} \right)
= \sum_{i} \sum_{j} \conv \{ \pm v_i \otimes w_j \} .\]

Here is a first simple property of the zonoid tensor product.

\begin{lemma}
Given symmetric zonoids $K,L$ and linear maps $S,T$, we have
\[ S(K) \otimes^Z T(L) = (S \otimes T) ( K \otimes^Z L) \]
\end{lemma}

Additionally, and crucially for the applications we have in mind, the zonoid tensor product is compatible with inclusions.

\begin{lemma}
\label{lemma:tensorizing-zonoids}
Let $K,K'$ be two symmetric zonoids in $\R^m$ with $K\subset K'$, and let $L,L'$ be two symmetric zonoids in $\R^n$ with $L\subset L'$. Then
\[ K \otimes^Z L \subset K' \otimes^Z L'. \]
\end{lemma}

\begin{proof}
This is a special case of Lemma 2 in \cite{RosSza}. Here is a proof in the language of zonoids.
We may assume that $L=L'$, the general case following then by arguing that
$K \otimes^Z L \subset K' \otimes^Z L \subset K' \otimes^Z L'$.

In terms of support functions, we are thus reduced to showing that the inequality $h_K \leq h_{K'}$ implies the inequality
$h_{K \otimes^Z L} \leq h_{K' \otimes^Z L}$,
which is an easy consequence of \eqref{eq:ztilde-phi}.
\end{proof}

Suppose that $(X,\|\cdot\|_X)$ and $(Y,\|\cdot\|_Y)$ are Banach spaces with Euclidean norms, i.e.~induced by some inner
products $\scalar{\cdot}{\cdot}_X$ and $\scalar{\cdot}{\cdot}_Y$. Their Euclidean tensor product $X \otimes^2 Y$ is defined (after completion)
by the norm induced by the inner product on the algebraic tensor product which satisfies
\[ \scalar{x \otimes y}{x' \otimes y'} = \scalar{x}{x'}_X \scalar{y}{y'}_Y. \]
It turns out that, for Euclidean norms, the tensor norms $\otimes^1$ and $\otimes^2$ are equivalent.

\begin{proposition}[see \cite{RosSza,Bennett}] \label{proposition:2-vs-1}
If $X$ and $Y$ are two Banach spaces equipped with Euclidean norms, then
\[ \sqrt{\frac{2}{\pi}} \| \cdot \|_{X \otimes^2 Y} \leq \| \cdot \|_{X \otimes^1 Y}
\leq \| \cdot \|_{X \otimes^2 Y} .\]
\end{proposition}

\subsection{Local POVMs}

In quantum mechanics, when a system is shared by several parties, the underlying global Hilbert space is the tensor product of the local Hilbert spaces
corresponding to each of the subsystems. A physically relevant class of POVMs on such a multipartite system is the one of local POVMs,
describing the situation where each party is only able to perform measurements on his own subsystem (cf.~Chapter \ref{chap:data-hiding}, which is entirely dedicated to this topic of POVMs satisfying locality constraints).

\begin{definition}
For $i=1,2$, let $\mathrm{M}_i$ denote a POVM on $\C^{d_i}$, defined on a $\sigma$-algebra $\mathcal{F}_i$ on a set $\Omega_i$. The tensor POVM
$\mathrm{M}_1 \otimes \mathrm{M}_2$ is the unique map defined on the product $\sigma$-algebra $\mathcal{F}_1 \otimes \mathcal{F}_2$ on $\Omega_1
\times \Omega_2$, and such that
\[ (\mathrm{M}_1 \otimes \mathrm{M}_2) ( A_1 \times A_2) = \mathrm{M}_1(A_1) \otimes \mathrm{M}_2(A_2) \]
for every $A_1 \in \mathcal{F}_1, A_2 \in \mathcal{F}_2$. By construction, $\mathrm{M}_1 \otimes \mathrm{M}_2$ is a POVM on $\C^{d_1} \otimes \C^{d_2}$.
\end{definition}

In the discrete case, this definition becomes more transparent: if $\mathrm{M}=(M_i)_{1\leq i\leq m}$ and
$\mathrm{N}=(N_j)_{1\leq j\leq n}$ are discrete POVMs, then $\mathrm{M}\otimes\mathrm{N}$ is also discrete, and
\[ \mathrm{M}\otimes\mathrm{N} = (M_i\otimes N_j)_{1\leq i\leq m,1\leq j\leq n} .\]

POVMs on $\C^{d_1} \otimes \C^{d_2}$ which can be decomposed as tensor product of two POVMs are called local POVMs.
If we identify the POVMs $\mathrm{M}_1$ and $\mathrm{M}_2$ with measures $\mu_1$ and $\mu_2$ as in Proposition \ref{proposition:POVM-states},
then the measure corresponding to $\mathrm{M}_1 \otimes \mathrm{M}_2$ is the image of the product measure $\mu_1 \times \mu_2$ under the map
$(\rho,\sigma) \mapsto \rho \otimes \sigma$. It thus follows that

\begin{proposition} \label{proposition:tensor-POVM}
If $\mathrm{M}$ and $\mathrm{N}$ are two POVMs, then
$ \|\cdot\|_{\mathrm{M} \otimes \mathrm{N}} = \|\cdot\|_{\mathrm{M}} \otimes^1 \|\cdot\|_{\mathrm{N}} $ and
$ K_{\mathrm{M} \otimes \mathrm{N}} = K_{\mathrm{M}} \otimes^Z K_{\mathrm{N}} $.
\end{proposition}

These definitions and statements are given here only in the bipartite case for the sake of clarity, but
can be extended to the situation where a system is shared between any number $k$ of parties.

\section{Sparsifying POVMs}

\label{section:sparsification}

\subsection{The uniform POVM}

It has been proved in \cite{MWW} that, in several senses, the ``most efficient'' POVM on $\C^d$ is the ``most symmetric'' one, i.e.~the uniform POVM $\mathrm{U}_d$, which corresponds to the uniform measure on the set of pure states in the representation
\eqref{eq:support-function-POVM} from Proposition \ref{proposition:POVM}.

The corresponding norm is
\begin{equation} \label{eq:def-normU} \|\Delta\|_{\mathrm{U}_d} = d \E | \langle \psi | \Delta | \psi \rangle |, \end{equation}
where $\psi$ is a random Haar-distributed unit vector on $\C^d$.

An important property is that the norm $\|\cdot\|_{\mathrm{U}_d}$ is equivalent
to a ``modified'' Hilbert--Schmidt norm.
\begin{proposition}[\cite{HMS,LW1}] \label{proposition:norm-equivalence}
For every $\Delta \in \cH(\C^d)$, we have
\begin{equation} \label{eq:U-vs-modifiedHS} \frac{1}{\sqrt{18}}\|\Delta\|_{2(1)}\leq\|\Delta\|_{\mathrm{U}_d} \leq\|\Delta\|_{2(1)}, \end{equation}
where the norm $\|\cdot\|_{2(1)}$ is defined as
\begin{equation} \label{eq:modifiedHS} \|\Delta\|_{2(1)}=\sqrt{\mathrm{Tr}(\Delta^2)+(\mathrm{Tr}\Delta)^2}. \end{equation}
\end{proposition}

One can check that $\|\Delta\|_{2(1)}$ equals the $L^2$ norm of the random variable $\langle g | \Delta | g \rangle$, where $g$ is a standard Gaussian
vector in $\C^d$, while the $L^1$ norm of this random variable is nothing else than $\|\Delta\|_{\mathrm{U}_d}$. Therefore
Proposition \ref{proposition:norm-equivalence} can be seen as a reverse H\"older inequality, and an interesting problem would be to find the optimal constant in that inequality (the factor $\sqrt{18}$ is presumably far from optimal).

This dimension-free lower bound on the
distinguishing power of the uniform POVM is of interest in quantum information theory.
One could cite as one of its applications the possibility to establish lower bounds on the dimensionality reduction of quantum states \cite{HMS}. However,
from a computational or algorithmic point of view, this statement involving a continuous POVM is of no practical use.
There has been interest therefore in the question of sparsifying $\mathrm{U}_d$, i.e.~of finding a discrete POVM, with as few outcomes as possible,
which would be equivalent to $\mathrm{U}_d$ in terms of discriminating efficiency. Examples of such constructions arise from the theory
of projective $4$-designs.

Given an integer $t \geq 1$, an (exact) $t$-design is a finitely supported probability measure $\mu$ on $S_{\C^d}$ such that
\[ \int_{S_{\C^d}} \ketbra{\psi}{\psi}^{\otimes t} \, \mathrm{d}\mu(\psi) =
\int_{S_{\C^d}} \ketbra{\psi}{\psi}^{\otimes t} \, \mathrm{d}\sigma(\psi) = \binom{d+t-1}{t}^{-1} P_{\Sym^t(\C^d)} .\]
Here, $\sigma$ denotes the Haar probability measure on $S_{\C^d}$, and $P_{\Sym^t(\C^d)}$ denotes the orthogonal projection onto the symmetric subspace $\Sym^t(\C^d) \subset (\C^d)^{\otimes t}$ (see Chapter \ref{chap:symmetries}, Section \ref{sec:sym}, for further details).

Note that a $t$-design is also a $t'$-design for any $t' \leq t$. Let $\mu$ be a $1$-design. The map $\psi \mapsto \ketbra{\psi}{\psi}$
pushes forward $\mu$ into a measure $\tilde{\mu}$ on the set of (pure) states, with barycenter equal to $\Id/d$. By Proposition \ref{proposition:POVM},
this measure corresponds to a POVM, and in the following we identify $t$-designs with the associated POVMs. For example the uniform POVM
$\mathrm{U}_d$ is a $t$-design for any $t$.

This notion can be relaxed: define an $\e$-approximate $t$-design to be a finitely supported measure $\mu$ on $S_{\C^d}$ such that
\[ (1-\e) \int_{S_{\C^d}} \ketbra{\psi}{\psi}^{\otimes t} \, \mathrm{d}\sigma(\psi) \leq
\int_{S_{\C^d}} \ketbra{\psi}{\psi}^{\otimes t} \, \mathrm{d}\mu(\psi) \leq
(1+\e) \int_{S_{\C^d}} \ketbra{\psi}{\psi}^{\otimes t} \, \mathrm{d} \sigma(\psi) .\]

It has been proved in \cite{AE} that a $4$-design (exact or approximate) supported on $N$ points yields a POVM $\mathrm{M}$ with $N$ outcomes such that
\begin{equation} \label{eq-AE} C^{-1} \|\cdot\|_{\mathrm{U}_d} \leq \|\cdot\|_{\mathrm{M}} \leq C \|\cdot\|_{\mathrm{U}_d} \end{equation}
for some constant $C$. The proof is based on the fourth moment method, which is used to control the first absolute moment of a random variable
by its second and fourth moments.

Now, what is the minimal cardinality of a $4$-design? The support of any exact or $\e$-approximate (provided $\e<1$) $4$-design must
contain at least $\dim (\Sym^4(\C^d)) = \binom{d+3}{4} = \Omega(d^4)$ points. Conversely, an argument
based on Carath\'eodory's theorem shows that there exist exact $4$-designs with $O(d^8)$ points. Starting from such an exact $4$-design,
the sparsification procedure from \cite{BSS} gives a deterministic and efficient algorithm which outputs an $\e$-approximate $4$-design supported
by $O(d^4/\e^2)$ points.

However, this approach has two drawbacks: the constant $C$ from \eqref{eq-AE} cannot be taken close to $1$, and the number of outcomes
has to be $\Omega(d^4)$. We are going to remove both inconveniences in our Theorem \ref{theorem:approximation-of-U}.

\subsection{Euclidean subspaces}

How do these ideas translate into the framework of zonoids? The analogue of $\mathrm{U_d}$ is the most symmetric zonoid, namely the Euclidean
ball $B^n \subset \R^n$. To connect with literature from functional analysis, it is worth emphasizing that approximating $B^n$ by a zonotope
which is the sum of $N$ segments is equivalent to embedding the space $\ell_2^n=(\R^n,\|\cdot\|_2)$ into the space $\ell_1^N=(\R^N,\|\cdot\|_1)$.
Indeed, assume that $x_1,\dots,x_N$ are points in $\R^n$ such that, for some constants $c,C$,
\[ c Z \subset B^n \subset C Z ,\]
where $Z = \conv \{\pm x_1 \} + \dots + \conv \{ \pm x_N \}$. Then the map $u: \R^n \to \R^N$ defined by
\[ u(x) = \Big( \scalar{x}{x_1}, \cdots, \scalar{x}{x_N} \Big) \]
satisfies $c \|u(x)\|_1 \leq \|x\|_2 \leq C\|u(x)\|_1$ for any $x \in \R^n$. In this context,
the ratio $C/c$ is often called the distortion of the embedding.

An early result by Rudin \cite{Rudin60} shows an explicit embedding of $\ell_2^n$ into $\ell_1^{O(n^2)}$ with distortion $\sqrt{3}$. This is proved by
the fourth moment method and can be seen as the analogue of the constructions based on 4-designs.
The following theorem (a variation on Dvoretzky's theorem, recalled as Lemma \ref{lemma:dvo} in Chapter \ref{chap:toolbox}, Section \ref{ap:deviations})
has been a major improvement on Rudin's result, showing that $\ell_1^N$ has almost Euclidean sections of proportional dimension.

\begin{theorem}[\cite{FLM}] \label{theorem-dvoretzky}
For every $0<\e<1$, there exists a subspace $E \subset \R^N$ of dimension $n=c(\e)N$ such that for any $x \in E$,
\begin{equation} \label{eq:dvoretzky} (1-\e) M \|x\|_2 \leq \|x\|_1 \leq (1+\e) M\|x\|_2, \end{equation}
where $M$ denotes the average of the $1$-norm over the Euclidean unit sphere $S^{N-1}$.
\end{theorem}

Theorem \ref{theorem-dvoretzky} was first proved in \cite{FLM}. As explained in greater depth in Chapter \ref{chap:toolbox}, Section \ref{ap:deviations}, the reasoning makes a seminal use of individual measure concentration in the form of L\'evy's lemma (see Lemma \ref{lemma:levy}) and a discretization via nets (see Lemma \ref{lemma:nets}) to yield global concentration of measure of Dvoretzky-type (see Lemma \ref{lemma:dvo}). The argument
shows that a generic subspace $E$ (i.e.~picked uniformly at random amongst all $c(\e)N$-dimensional subspaces of $\R^N$) satisfies the conclusion of the
theorem with high probability whenever $c(\e)=O\left(\e^2 |\log \e|^{-1}\right)$.
This was later improved in \cite{Gordon} to $c(\e)=O\left(\e^2\right)$.

\subsection{Sparsification of the uniform POVM}

Translated in the language of zonotopes, Theorem \ref{theorem-dvoretzky} states that the sum of $O(n)$ randomly chosen segments in $\R^n$
is close to the Euclidean ball $B^n$. More precisely, for any $0<\e<1$, if $N=c(\e)^{-1}n$ and $x_1,\ldots,x_N$ are randomly chosen points in $\R^n$, the zonotope $Z=\conv\{\pm x_1\}+\cdots+\conv\{\pm x_N\}$ is $\e$-close to the Euclidean ball $B^n$, in the sense
that $(1-\e)Z\subset B^n\subset(1+\e)Z$.

By analogy, we expect a POVM constructed from $O(d^2)$ randomly chosen elements to be close to the uniform POVM.
This random construction can be achieved as follows:
let $(\psi_i)_{1\leq i \leq n}$ be independent random vectors, uniformly chosen on the unit sphere of $\C^d$. Set
$P_i = \ketbra{\psi_i}{\psi_i}$, $1 \leq i \leq n$, and $S = P_1 + \dots + P_n$. When $n \geq d$, $S$ is almost surely invertible,
and we may consider the random POVM
\begin{equation} \label{eq:randomPOVM} \mathrm{M} = (S^{-1/2} P_i S^{-1/2})_{1 \leq i \leq n } .\end{equation}

\begin{theorem} \label{theorem:approximation-of-U}
Let $\mathrm{M}$ be a random POVM on $\C^d$ with $n$ outcomes, defined as in \eqref{eq:randomPOVM}, and let $0<\e<1$.
If $n \geq C\e^{-2} |\log \e| d^2$, then with high probability
the POVM $\mathrm{M}$ satisfies the inequalities
\[ (1-\e) \|\Delta\|_{\mathrm{U}_d} \leq \|\Delta\|_{\mathrm{M}} \leq (1+\e) \|\Delta\|_{\mathrm{U}_d} \]
for every $\Delta \in \cH(\C^d)$.
\end{theorem}

By ``with high probability'' we mean that the probability that the conclusion fails is less than $\exp(-c(\e)d)$ for some constant $c(\e)$.
Theorem \ref{theorem:approximation-of-U} is proved in Section \ref{sec:uniform-POVM}, the proof being based on a careful use of $\e$-nets and deviation
inequalities. It does not seem possible to deduce formally Theorem \ref{theorem:approximation-of-U} from the existing Banach space literature.

Theorem \ref{theorem:approximation-of-U} shows that the uniform POVM on $\C^d$ can be $\e$-approximated
(in the sense of closeness of distinguishability norms)
by a
POVM with $n=O(\e^{-2}|\log\e|d^2)$ outcomes. Note that the dependence of $n$ with respect to $d$ is optimal: since a POVM on $\C^d$ must have at least
$d^2$ outcomes to be informationally complete, one cannot hope for a tighter dimensional dependence. The dependence with respect to $\e$ is less clear:
the factor $|\log \e|$ can probably be removed but we do not pursue this direction.

Our construction is random and a natural question is whether deterministic constructions yielding comparable properties exist. A lot of effort has
been put in derandomizing Theorem \ref{theorem-dvoretzky}. We refer to \cite{IS} for bibliography and mention two of the latest results.
Given any $0<\gamma<1$, it is shown in \cite{IS} how to construct, from $cn^{\gamma}$ random bits (i.e.~an amount of randomness sub-linear in $n$) a
subspace of $\ell_1^N$ satisfying
\eqref{eq:dvoretzky} with $N\leq (\gamma\e)^{-C\gamma}n$. A completely explicit construction appears in \cite{Indyk}, with
$N\leq n2^{C(\e)(\log\log n)^2}=n^{1+C(\e)o(n)}$. It is not obvious how to adapt these constructions to obtain sparsifications of the uniform POVM
using few or no randomness.

\subsection{Sparsification of any POVM}

Theorem \ref{theorem-dvoretzky}
initiated intensive research in the late 80's \cite{Schechtman,BLM,Talagrand} on the theme of ``approximation of zonoids by zonotopes'',
trying to extend the result for
the Euclidean ball (the most symmetric zonoid) to an arbitrary zonoid. This culminated in Talagrand's proof \cite{Talagrand} that for any zonoid
$Y \subset \R^n$ and any $0<\e<1$, there exists a zonotope $Z \subset \R^n$ which is the sum of $O(\e^{-2}n \log n)$ segments and such that
$(1-\e)Y \subset Z \subset (1+\e)Y$. A more precise version is stated in Section \ref{sec:approximation-any}. Whether the $\log n$ factor
can be removed is still an open problem.

This result easily implies a similar result for POVMs, provided we consider the larger class of sub-POVMs.
A discrete sub-POVM with $n$ outcomes is a
finite family $\mathrm{M} = (M_i)_{1\leq i\leq n}$ of $n$ positive operators such that $S = \sum_{i=1}^n M_i \leq \Id$.
As for POVMs, the norm associated to a sub-POVM $\mathrm{M}$ is defined for $\Delta \in \cH(\C^d)$ by
\[ \| \Delta \|_{\mathrm{M}} = \sum_{i=1}^n |\tr (\Delta M_i)|. \]
We prove the following result in Section \ref{sec:approximation-any}.

\begin{theorem}
\label{theorem:approximation-any}
Given any POVM $\mathrm{M}$ on $\C^d$ and any $0<\e<1$, there is a sub-POVM $\mathrm{M}' = (M'_i)_{1\leq i\leq n}$, with $n \leq C \e^{-2}d^2 \log(d)$
such that, for any $\Delta \in \cH(\C^d)$,
\[ (1-\e) \| \Delta \|_{\mathrm{M}} \leq \|\Delta\|_{\mathrm{M}'} \leq \|\Delta\|_{\mathrm{M}}. \]
Moreover, we can guarantee that the states $M'_i/\tr (M'_i)$ belong to the support of the measure $\mu$ associated to $\mathrm{M}$.
\end{theorem}

We do not know whether Theorem \ref{theorem:approximation-any} still holds if we want $\mathrm{M}'$ to be a POVM.
Given a sub-POVM $(M_i)_{1\leq i\leq n}$, there are at least
two natural ways to modify it into a POVM. A solution is to add an extra outcome corresponding to the operator $\Id-S$,
and another one is to substitute $S^{-1/2}M_iS^{-1/2}$
in place of $M_i$, as we proceeded in \eqref{eq:randomPOVM}. However for a general POVM, the error terms arising from this renormalization step may
exceed the quantity to be approximated.

\section{Sparsifying local POVMs}
\label{section:sparsification-multipartite}

Proposition \ref{prop:tensorizing-sparsifications} below is an immediate corollary of Lemma \ref{lemma:tensorizing-zonoids} and Proposition
\ref{proposition:tensor-POVM}. In words, it shows that, on a
multipartite system, a local POVM can be sparsified by tensorizing sparsifications of each of its factors.

\begin{proposition}
\label{prop:tensorizing-sparsifications}
Let $0<\e<1$. Let $\mathrm{M}_1,\ldots,\mathrm{M}_k$ be POVMs and $\mathrm{M}_1',\ldots,\mathrm{M}_k'$ be (sub-)POVMs, on $\C^{d_1},\ldots,\C^{d_k}$
respectively, satisfying, for all $1\leq i\leq k$, and for all $\Delta\in\cH(\C^{d_i})$,
\[ (1-\e)\|\Delta\|_{\mathrm{M}_i}\leq\|\Delta\|_{\mathrm{M}_i'}\leq(1+\e)\|\Delta\|_{\mathrm{M}_i}. \]
Then, for any $\Delta\in\cH(\C^{d_1}\otimes\cdots\otimes\C^{d_k})$,
\[ (1-\e)^k\|\Delta\|_{\mathrm{M}_1\otimes\cdots\otimes\mathrm{M}_k} \leq\|\Delta\|_{\mathrm{M}_1'\otimes\cdots\otimes\mathrm{M}_k'}
\leq(1+\e)^k\|\Delta\|_{\mathrm{M}_1\otimes\cdots\otimes\mathrm{M}_k}. \]
\end{proposition}


Let us give a concrete application of Proposition \ref{prop:tensorizing-sparsifications}. We consider $k$ finite-dimensional Hilbert spaces
$\C^{d_1},\ldots,\C^{d_k}$ and define the local uniform POVM on the $k$-partite Hilbert space $\C^{d_1}\otimes\cdots\otimes\C^{d_k}$ as the tensor
product of the $k$ uniform POVMs $\mathrm{U}_{d_1},\ldots,\mathrm{U}_{d_k}$. We will denote it by $\mathrm{LU}$. The
corresponding distinguishability norm can be described, for any $\Delta\in\cH(\C^{d_1}\otimes\cdots\otimes\C^{d_k})$, as
\[ \|\Delta\|_{\mathrm{LU}} = d \E \left| \langle \psi_1 \otimes \cdots \otimes \psi_k | \Delta | \psi_1 \otimes \cdots \otimes \psi_k
\rangle \right|, \]
where $d=d_1\times\cdots\times d_k$ is the dimension of the global Hilbert space, and where the random unit vectors $\psi_1, \dots ,\psi_k$ are
independent and Haar-distributed in $\C^{d_1},\ldots,\C^{d_k}$ respectively.

The following multipartite generalization of Proposition \ref{proposition:norm-equivalence} shows that the norm $\|\cdot\|_{\mathrm{LU}}$,
in analogy to the norm $\|\cdot\|_{\mathrm{U}}$, is equivalent to a ``modified'' Hilbert--Schmidt norm.

\begin{proposition} [\cite{LW1}] \label{proposition:norm-equivalence-multipartite}
For every $\Delta\in\cH(\C^{d_1}\otimes\cdots\otimes\C^{d_k})$, we have
\begin{equation} \label{eq:U-vs-modifiedHS-multi} \frac{1}{18^{k/2}}\|\Delta\|_{2(k)}\leq\|\Delta\|_{\mathrm{LU}} \leq\|\Delta\|_{2(k)}, \end{equation}
where the norm $\|\cdot\|_{2(k)}$ is defined as
\begin{equation} \label{eq:modifiedHS-multi} \|\Delta\|_{2(k)}=\sqrt{\sum_{I\subset \{1,\dots,k\}} \tr \left[\big(\tr_{I}\Delta\big)^2\right]}.
\end{equation}
Here $\tr_I$ denotes the partial trace over all parties $I \subset \{1,\dots,k\}$.
\end{proposition}

\begin{proof}[Proof of Proposition \ref{proposition:norm-equivalence-multipartite}]
A direct proof appears in \cite{LW1}, but we find interesting to show that it can be deduced (with a worst constant) from Proposition
\ref{proposition:norm-equivalence}.
If we denote by $\scalar{\cdot}{\cdot}_H$ the inner product inducing a Euclidean norm $\|\cdot\|_H$, we have
\[ \langle A_1\otimes\cdots\otimes A_k, B_1\otimes\cdots\otimes B_k \rangle_{2(k)} =\langle A_1, B_1 \rangle_{2(1)}\times\cdots\times\langle A_k,
B_k \rangle_{2(1)} \]
which is equivalent to saying that
\[ \|\cdot\|_{2(k)} = \|\cdot\|_{2(1)} \otimes^2 \cdots \otimes^2 \|\cdot\|_{2(1)} .\]
We thus get by Proposition \ref{proposition:2-vs-1} that
\[ c_0^{k-1}\|\cdot\|_{2(k)} \leq \|\cdot\|_{2(1)} \otimes^1 \cdots
\otimes^1  \|\cdot\|_{2(1)} \leq \|\cdot\|_{2(k)} \]
with $c_0=\sqrt{2/\pi}$.
Now, we just have to observe that, we know by Proposition \ref{proposition:tensor-POVM} that, on $\cH(\C^{d_1}\otimes\cdots\otimes\C^{d_k})$, $\|\cdot\|_{\mathrm{LU}}=
\|\cdot\|_{\mathrm{U}_{d_1}}\otimes^1\cdots\otimes^1\|\cdot\|_{\mathrm{U}_{d_k}}$, and by Proposition \ref{proposition:norm-equivalence}
that $c\|\cdot\|_{2(1)}\leq\|\cdot\|_{\mathrm{U}_d}\leq \|\cdot\|_{2(1)}$ for some constant $c$ (e.g.~$c=1/\sqrt{18}$ works). So by Lemma
\ref{lemma:tensorizing-zonoids},
\[ c^k \, \|\cdot\|_{2(1)} \otimes^1 \cdots \otimes^1 \|\cdot\|_{2(1)}
\leq \|\cdot\|_{\mathrm{LU}} \leq  \|\cdot\|_{2(1)} \otimes^1 \cdots \otimes^1 \|\cdot\|_{2(1)}, \]
and therefore
\[ c_0^{k-1} c^k \|\cdot\|_{2(k)} \leq \|\cdot\|_{\mathrm{LU}} \leq \|\cdot\|_{2(k)}.  \qedhere\]
\end{proof}

Remarkably, local dimensions do not appear in equation \eqref{eq:U-vs-modifiedHS-multi}.
This striking fact that local POVMs can have asymptotically non-vanishing distinguishing power can be used to construct an algorithm that solves the Weak Membership Problem for separability in quasi-polynomial time (see \cite{BCY} for a description of the latter algorithm in the bipartite case, which is based on symmetric extension search, the central topic of Chapter \ref{chap:k-extendibility}). Hence the importance of being able to sparsify the local uniform POVM by a POVM for which the locality property is preserved and which has a number of outcomes that optimally scales as the square of the global dimension. We state the corresponding multipartite version of Theorem \ref{theorem:approximation-of-U}, which is straightforwardly obtained by combining the unipartite version with Proposition \ref{prop:tensorizing-sparsifications}.

\begin{theorem} \label{theorem:approximation-LU}
Let $0<\e<1$. For all $1\leq i\leq k$, let $\mathrm{M}_i$ be a random POVM on $\C^{d_i}$ with $n_i\geq C\e^{-2}|\log\e|d_i^2$ outcomes, defined as in
\eqref{eq:randomPOVM}. Then, with high probability, the local POVM $\mathrm{M}_1\otimes\cdots\otimes\mathrm{M}_k$ on
$\C^{d_1}\otimes\cdots\otimes\C^{d_k}$ is such that, for any $\Delta\in\cH(\C^{d_1}\otimes\cdots\otimes\C^{d_k})$,
\[ (1-\e)^k\|\Delta\|_{\mathrm{LU}} \leq\|\Delta\|_{\mathrm{M}_1\otimes\cdots\otimes\mathrm{M}_k} \leq(1+\e)^k\|\Delta\|_{\mathrm{LU}}. \]
\end{theorem}

Let us put in words the content of Theorem \ref{theorem:approximation-LU}: the local uniform POVM on $\C^{d_1}\otimes\cdots\otimes\C^{d_k}$ can be
$k\e$-approximated (in terms of distinguishability norms) by a POVM which is also local and has a total number of outcomes
$n=O(C^k\e^{-2k}|\log\e|^kd^2)$, where $d=d_1\times\cdots\times d_k$. Note that the dimensional dependence of $n$ is optimal. On the contrary,
the dependence of $n$ on $\e$ deteriorates as $k$ grows. The high-dimensional situation our result applies to is thus really
the one of a ``small'' number of ``large'' subsystems (i.e.~$k$ fixed and $d_1,\ldots,d_k\rightarrow+\infty$), and not of
a ``large'' number of ``small'' subsystems.

\section{Proof of the main theorem concerning the sparsification of the uniform POVM}
\label{sec:uniform-POVM}

\subsection{Proof of Theorem \ref{theorem:approximation-of-U}}


In this subsection we prove Theorem \ref{theorem:approximation-of-U}.
Let $n\in\N$ and $(\psi_i)_{1 \leq i \leq n}$ be independent random unit vectors,
uniformly distributed on the unit sphere of $\C^d$. Our main technical estimates are a couple of probabilistic inequalities. Proposition
\ref{proposition:Wishart} is an immediate consequence of Theorem 1 in \cite{Aubrun2}. Proposition \ref{proposition:large-deviations} is a consequence of a Bernstein-type inequality, recalled as Theorem \ref{th:Bernstein} in Chapter \ref{chap:toolbox}, Section \ref{ap:deviations}. However, its proof requires some careful estimates which we postpone to Subsection \ref{section:bernstein}.

\begin{proposition} \label{proposition:Wishart}
If $(\psi_i)_{1 \leq i \leq n}$ are independent random vectors, uniformly distributed on the unit sphere of $\C^d$, then for every $0<\eta<1$
\[ \P \left( (1-\eta) \frac{\Id}{d} \leq \frac{1}{n} \sum_{i=1}^n \ketbra{\psi_i}{\psi_i} \leq (1+\eta) \frac{\Id}{d} \right) \geq 1 -
C^d \exp(-c n \eta^2). \]
\end{proposition}

\begin{proposition} \label{proposition:large-deviations}
Fix $\Delta \in \cH(\C^d)$, and let $(\psi_i)_{1 \leq i \leq n}$ be independent random vectors, uniformly distributed on the unit sphere of $\C^d$.
For each $1 \leq i \leq n$, consider next the random variables $X_i = d| \bra{\psi_i} \Delta \ket{\psi_i} |$ and $Y_i=X_i-\E X_i=X_i - \|\Delta\|_{\mathrm{U}_d}$.
Then, for any $t > 0$,
\[ \P \left( \left| \frac{1}{n} \sum_{i=1}^n Y_i \right| \geq t \| \Delta \|_{\mathrm{U}_d} \right) \leq 2 \exp ( - c'_0 n \min(t,t^2)) .\]
\end{proposition}

We now show how to derive Theorem \ref{theorem:approximation-of-U} from the estimates in Propositions \ref{proposition:Wishart} and \ref{proposition:large-deviations}.
For each $1\leq i\leq n$, set $P_i=\ketbra{\psi_i}{\psi_i}$, and introduce the (random) norm defined for any $\Delta\in\cH(\C^d)$ as
\[ |||\Delta||| = \frac{d}{n} \sum_{i=1}^n |\tr (\Delta P_i)|. \]
We will now prove that $|||\cdot|||$ is, with probability close to $1$, a good approximation to $\|\cdot\|_{\mathrm{U}_d}$.
First, using Proposition \ref{proposition:large-deviations}, we obtain that for any $0 < \e < 1$ and any $\Delta \in \cH(\C^d)$
\begin{equation} \label{eq:single-Delta}
\P \left( (1-\e) \|\Delta\|_{\mathrm{U}_d} \leq |||\Delta||| \leq (1+\e) \|\Delta\|_{\mathrm{U}_d} \right) \geq 1- 2\exp(-c'_0 n \e^2).
\end{equation}

We next use a net argument. Fix $0<\e<1/3$ and a $\e$-net $\mathcal{A}$ inside the unit ball for the norm $\|\cdot\|_{\mathrm{U}_d}$,
with respect to the distance induced by
$\|\cdot\|_{\mathrm{U}_d}$. A standard volumetric argument (see Lemma \ref{lemma:nets} in Chapter \ref{chap:toolbox}, Section \ref{ap:deviations}) shows that we may assume
$|\mathcal{A}| \leq (1 + 2/\e)^{d^2} \leq (3/\e)^{d^2}$.
Introduce the quantities
\[ A := \sup \{ ||| \Delta ||| \st \| \Delta \|_{\mathrm{U}_d} \leq 1 \}, \]
\[ A' := \sup \{ ||| \Delta ||| \st \Delta \in \mathcal{A} \}. \]
Given $\Delta$ such that $\|\Delta\|_{\mathrm{U}_d} \leq 1$, there is $\Delta_0 \in \mathcal{A}$ with $\|\Delta-\Delta_0\|_{\mathrm{U}_d} \leq \e$.
By the triangle inequality,
we have $|||\Delta||| \leq A' + |||\Delta - \Delta_0||| \leq A'+ \e A$. Taking supremum over $\Delta$ yields $A \leq A'+\e A$ i.e.~$A \leq
A'/(1-\e)$.

If we introduce $B := \inf \{ ||| \Delta ||| \st \| \Delta \|_{\mathrm{U}_d} = 1 \}$ and $B' := \inf \{ ||| \Delta ||| \st \Delta \in \mathcal{A} \}$,
a similar argument shows that $B \geq B' - \e A$, so that in fact $B \geq B' - \e A'/(1-\e)$.
We therefore have the implications
\begin{equation} \label{eq:implication} 1-\e \leq B' \leq A' \leq 1+\e \ \ \Rightarrow \ \  1-\e-\frac{\e(1+\e)}{1-\e} \leq  B \leq A \leq
\frac{1+\e}{1-\e}
\ \ \Rightarrow \ \ 1-3\e \leq B \leq A \leq 1+ 3\e. \end{equation}
By the union bound, we get from \eqref{eq:single-Delta} that $\P(1-\e \leq B' \leq A' \leq 1+\e) \geq 1-2 | \mathcal{A}| \exp(-c'_0n \e^2)$.
Combined with (\ref{eq:implication}), and using homogeneity of norms, this yields
\begin{equation} \label{eq:global-Delta} \P \Big( (1-3\e) \|\cdot \|_{\mathrm{U}_d} \leq |||\cdot||| \leq (1+3\e) \|\cdot\|_{\mathrm{U}_d} \Big) \geq 1- 2 \left(\frac{3}{\e}\right)^{d^2}
\exp(-c'_0n \e^2). \end{equation}
This probability estimate is non-trivial, and can be made close to $1$, provided $n \gtrsim d^2 \e^{-2} | \log \e |$.

Whenever $n \geq d$, the
vectors $(\psi_i)_{1 \leq i \leq n}$ generically span $\C^d$, and therefore the operator $S = P_1 + \cdots + P_n$ is invertible. We may
then define $\widetilde{P}_i = S^{-1/2} P_i S^{-1/2}$ so that $\mathrm{M} = (\widetilde{P}_i)_{1 \leq i \leq n}$ is a POVM. The norm associated to
$\mathrm{M}$ is, for any $\Delta\in\cH(\C^d)$,
\[ \|\Delta\|_{\mathrm{M}} = \sum_{i=1}^n |\tr (\Delta\widetilde{P}_i)|. \]
We now argue that the norms $|||\cdot|||$ and
$\|\cdot\|_{\mathrm{M}}$ are similar enough (modulo normalization), because the modified operators $\widetilde{P}_i$ are close
enough to the initial ones $P_i$. This is achieved by showing that $T:=\sqrt{n/d}\,S^{-1/2}$ is close to $\Id$ (in operator-norm distance).
We use Proposition \ref{proposition:Wishart} for $\eta = \e \|\Delta\|_{\mathrm{U}_d}/\|\Delta\|_1$.
By Proposition \ref{proposition:norm-equivalence}, we have $\eta \geq \e / \sqrt{18d}$.
Proposition \ref{proposition:Wishart} implies that
\begin{equation} \label{eq:boundZ} \P( \|T-\Id\|_{\infty} \geq \eta ) \leq \P( \|T^{-2}-\Id\|_{\infty} \geq \eta ) \leq C^d \exp(-c' n \e^2 /d).
\end{equation}
This upper bound is much smaller than $1$ provided $n \geq C_1 \e^{-2}d^2$. Also, note that the event $\|T-\Id\|_{\infty} \leq \eta$
implies that
\[ \|\Delta-T \Delta T\|_{\mathrm{M}} \leq \|\Delta - T\Delta T\|_{1} \leq  \|\Delta\|_1 \|\Id-T\|_{\infty}
\left( 1 + \|T\|_{\infty} \right) \leq  2 \eta \|\Delta\|_1 = 2 \e \|\Delta\|_{\mathrm{U}_d}. \]

Using the cyclic property of the trace, we check that $\|T\Delta T\|_{\mathrm{M}} = |||\Delta|||$.
Now, choose $n$ larger than both $C_0 \e^{-2} |\log \e| d^2$ and $C_1 \e^{-2}d^2$.
With high probability, the events from equations \eqref{eq:global-Delta} and \eqref{eq:boundZ} both hold.
We then obtain for every $\Delta \in \cH(\C^d)$,
\[
\|\Delta\|_{\mathrm{M}} \leq \| T\Delta T\|_{\mathrm{M}} + \|\Delta-T\Delta T\|_{\mathrm{M}} \leq |||\Delta||| + 2\e \|\Delta\|_{\mathrm{U}_d} \leq
(1+ 5\e) \|\Delta\|_{\mathrm{U}_d}
\]
and similarly $\|\Delta\|_{\mathrm{M}} \geq (1- 5\e) \|\Delta\|_{\mathrm{U}_d}$.
This is precisely the result from Theorem \ref{theorem:approximation-of-U} with $5\e$ instead of $\e$, which of course can be absorbed by renaming
the constants appropriately.

\subsection{Proof of Proposition \ref{proposition:large-deviations}}
\label{section:bernstein}

The proof is a direct application of a large deviation inequality for sums of independent $\psi_1$ (aka sub-exponential) random variables. More details on that topic are gathered in Chapter \ref{chap:toolbox}, Section \ref{ap:deviations}.



For $\Delta \in \cH(\C^d)$, consider for each $1\leq i\leq n$ the random variables $X_i = d|\tr (\Delta P_i)|$ with $P_i=\ketbra{\psi_i}{\psi_i}$,
and $Y_i=X_i - \E X_i =d|\tr (\Delta P_i)| - \|\Delta\|_{\mathrm{U}_d}$. The random variables
$Y_i$, $1\leq i\leq n$, are independent and have mean zero. The key lemma is a bound on their $\psi_1$-norm.

\begin{lemma} \label{lemma:psi1}
Let $\Delta\in \cH(\C^d)$ and consider the random variable $X:=d|\tr (\Delta P) |$, where $P=\ketbra{\psi}{\psi}$
with $\psi$ uniformly distributed on the unit sphere of $\C^d$.
Then, first of all $\|X\|_{\psi_1} \leq \|\Delta\|_{2(1)}$, and as a consequence $\|X - \E X\|_{\psi_1} \leq 3 \|\Delta\|_{2(1)} \leq 3\sqrt{18} \|\Delta\|_{\mathrm{U}_d}$.
\end{lemma}

Therefore, we may apply Bernstein's inequality, recalled as Theorem \ref{th:Bernstein} in Chapter \ref{chap:toolbox}, Section \ref{ap:deviations}, with $M= \sigma \leq 3\sqrt{18} \| \Delta \|_{\mathrm{U}_d} $, yielding Proposition \ref{proposition:large-deviations}.

\begin{proof}[Proof of Lemma \ref{lemma:psi1}]
For each integer $q$, we compute
\[ \E \left[\tr (\Delta P)\right]^{2q}   = \E \tr\left(\Delta^{\otimes 2q}P^{\otimes 2q}\right) = \tr \left( \Delta^{\otimes 2q}
\left[\E P^{\otimes 2q}\right] \right) .\]
In order to go further, we simply observe that
\[ \E P^{\otimes 2q} = \frac{(2q)!}{(d+2q-1)\times\cdots\times d}
P_{\Sym^{2q}(\C^d)} =\frac{1}{(d+2q-1)\times\cdots\times d} \sum_{\pi\in\mathfrak{S}(2q)} U(\pi) ,\]
where $P_{\Sym^{2q}(\C^d)}$ denotes the orthogonal projection
onto the symmetric subspace $\Sym^{2q}(\C^d)\subset(\C^d)^{\otimes 2q}$, and for each permutation $\pi\in\mathfrak{S}(2q)$, $U(\pi)$ denotes the associated permutation unitary on $(\C^d)^{\otimes 2q}$ (see e.g.~\cite{Harrow} and Chapter \ref{chap:symmetries}, Section \ref{sec:sym}, of this manuscript for much more on that matter). This yields
\[ \E \left[\tr (\Delta P)\right]^{2q} = \frac{1}{(d+2q-1)\times\cdots\times d}\sum_{\pi\in\mathfrak{S}(2q)}
\tr \left(\Delta^{\otimes 2q}U(\pi)\right) .\]
If $\ell_1,\dots,\ell_k$ denote the lengths of the cycles appearing in the cycle decomposition of a permutation $\pi \in \mathfrak{S}(2q)$, we have
$\ell_1 + \cdots + \ell_k = 2q$ and
\[ \tr \left(\Delta^{\otimes 2q}U(\pi)\right) = \prod_{i=1}^k \tr (\Delta^{\ell_i}). \]
Now, for any integer $\ell \geq 2$, we have $| \tr (\Delta^\ell) | \leq [\tr (\Delta^2)]^{\ell/2} \leq \|\Delta\|_{2(1)}^\ell$. The inequality
$| \tr (\Delta^\ell) | \leq \|\Delta\|_{2(1)}^\ell$ is also (trivially) true for $\ell=1$. Therefore
$\left|\tr \left(\Delta^{\otimes 2q}U(\pi)\right) \right| \leq \|\Delta\|_{2(1)}^{2q}$. It follows that
\[ \E \left[\tr (\Delta P)\right]^{2q} \leq \frac{(2q)!}{d^{2q}} \|\Delta\|_{2(1)}^{2q} \leq \left( \frac{2q \|\Delta\|_{2(1)}}{d} \right)^{2q}, \]
so that $\left( \E X^{2q} \right)^{1/2q} \leq 2q \|\Delta\|_{2(1)}$, and thus $\|X\|_{\psi_1} \leq \|\Delta\|_{2(1)}$.
The last part of the Lemma follows from the triangle
inequality, since $\| \E X \|_{\psi_1}=|\E X| \leq 2 \| X \|_{\psi_1}$, and from the equivalence \eqref{eq:U-vs-modifiedHS} between the norms
$\|\cdot\|_{\mathrm{U}_d}$
and $\|\cdot\|_{2(1)}$.
\end{proof}

\section{Proof of the main theorem concerning the sparsification of any POVM}
\label{sec:approximation-any}

In this section we prove Theorem \ref{theorem:approximation-any}. Here is a version of Talagrand's theorem which is suitable for our purposes.

\begin{theorem}[\cite{Talagrand}] \label{theorem:talagrand}
Let $Z \subset \R^n$ be a symmetric zonotope, with
\[ Z = \sum_{i \in I} \conv \{ \pm u_i \} \]
for a finite family of vectors $(u_i)_{i \in I}$. Then for every $\e >0$ there exists a subset $J \subset I$
with $|J| \leq C n \log n/\e^2$, and positive numbers $(\lambda_i)_{i \in J}$ such that the zonotope
\[ Z' = \sum_{i \in J} \conv \{ \pm \lambda_i u_i \} \]
satisfies $Z' \subset Z \subset (1+\e) Z'$.
\end{theorem}

Theorem \ref{theorem:approximation-any} is a very simple consequence of Theorem \ref{theorem:talagrand}. Let $\mathrm{M}$ be a POVM to
be sparsified. Using Corollary
\ref{corollary:approximation}, we may assume that $\mathrm{M} = (M_i)_{i \in I}$ is discrete. Applying Theorem \ref{theorem:talagrand} to the
zonotope $K_\mathrm{M}=\sum_{i \in I} \conv \{ \pm M_i \}$ (which lives in a $d^2$-dimensional space), we obtain a zonotope $Z' = \sum_{i \in J} \conv \{ \pm \lambda_i M_i \}$
with $|J| \leq C d^2 \log d /\e^2$ such that $Z' \subset K_{\mathrm{M}} \subset (1+\e) Z'$.
It remains to show that $\mathrm{M}'=(\lambda_i M_i)_{i \in J}$ is a sub-POVM. We know that $h_{Z'} \leq h_{K_\mathrm{M}}$.
Therefore, given a unit vector
$x \in \C^d$, the inequality $h_{Z'}(\Delta) \leq h_{K_\mathrm{M}}(\Delta)$ applied with $\Delta=\ketbra{x}{x}$ shows that
\[  \sum_{i \in J} \lambda_i \left| \langle x | M_i | x \rangle \right| \leq \| \ketbra{x}{x} \|_{\mathrm{M}} \leq \|
\ketbra{x}{x} \|_1 =1, \]
and therefore $\sum_{i \in J} \lambda_iM_i \leq \Id$, as required. Since the inclusions $Z' \subset K_{\mathrm{M}} \subset (1+\e) Z'$ are
equivalent to the inequalities $\|\cdot\|_{\mathrm{M}'} \leq \|\cdot\|_{\mathrm{M}} \leq (1+\e) \|\cdot\|_{\mathrm{M}'}$, Theorem
\ref{theorem:approximation-any} follows.

\section{What about approximating distinguishability norms only up to multiplicative factors?}

In this chapter, the focus has been from the start on how to sparsify a POVM into one with fewer outcomes, so that their distinguishability norms are equivalent with dominating constants $1-\varepsilon,1+\varepsilon$. However, one could also be interested in approximation only up to multiplicative factors $c,C$. Both questions are slightly different, even though they have a similar flavour, so we do aim at making a thorough analysis of the latter here. Nevertheless, let us briefly explain what kind of results can be obtained, by looking at the simplest example, namely that of the uniform POVM.

\begin{theorem} \label{th:U-multiplicative}
For any $\eta>0$, there exists $c_{\eta}>0$ such that, for any $d\in\N$, there is a POVM $\mathrm{M}$ on $\C^d$ having $(1+\eta)\,d^2$ outcomes and satisfying for all traceless Hermitian $\Delta$ on $\C^d$,
\begin{equation} \label{eq:U-multiplicative} c_{\eta}\,\|\Delta\|_{\mathrm{U}_d}\leq \|\Delta\|_{\mathrm{M}} \leq C\,\|\Delta\|_{\mathrm{U}_d}, \end{equation}
where one can take $C=2\sqrt{3}$ as universal constant on the r.h.s.~of equation \eqref{eq:U-multiplicative}.
\end{theorem}

In words, Theorem \ref{th:U-multiplicative} tells us the following: at the cost of having the constant on the l.h.s.~of equation \eqref{eq:U-multiplicative} becoming smaller and smaller, it is possible to sparsify the uniform POVM on $\C^d$ by one having a number of outcomes getting arbitrarily close to $d^2$.

\begin{proof} As a starting point we observe that, in restriction to traceless Hermitians, the uniform norm is dimension-independently equivalent to the Hilbert--Schmidt norm. This is a consequence of Proposition \ref{proposition:norm-equivalence} (just noticing that the modified $2$-norm of a traceless Hermitian reduces to its usual $2$-norm). The best known dominating constants in this norm equivalence are those appearing in \cite{AE}: for any traceless Hermitian $\Delta$ on $\C^d$,
\begin{equation} \label{eq:U-2} \frac{1}{\sqrt{3}}\|\Delta\|_2 \leq \|\Delta\|_{\mathrm{U}_d} \leq \|\Delta\|_2. \end{equation}
Hence, our problem actually boils down to finding a POVM with few outcomes whose distinguishability norm is dimension-independently equivalent to the Hilbert--Schmidt norm.

The question of approximating the Euclidean ball by a zonotope which is the sum of few segments, up to factors $c,C$ rather than $1-\varepsilon,1+\varepsilon$, has also been studied. The seminal results from \cite{Kas} and the follow-up ones from \cite{Sza} translate in our context into: for any $\eta>0$, there exists $c_{\eta}>0$ such that, for any $d\in\N$, there exist Hermitians $X_1,\ldots,X_N$ on $\C^d$, where $N=(1+\eta)\,d^2$, such that the zonotope $Z=\conv\{\pm X_1\}+\cdot+\conv\{\pm X_N\}$ satisfies
\[ c_{\eta}\,B_2(\C^d) \subset Z \subset B_2(\C^d). \]
Now, set for each $1\leq i\leq N$, $M_i=(\Id+X_i)/2$. $\mathrm{M}=(M_i)_{1\leq i\leq N}$ is by construction a sub-POVM on $\C^d$ which is such that $Z=K_{\mathrm{M}}$ (the fact that, indeed, $M_i\leq 0$ for each $1\leq i\leq N$ and $M_1+\cdots+M_N\leq \Id$ follows from the fact that $Z\subset B_2(\C^d)\subset B_{\infty}(\C^d)$). So the result above is equivalent to: there exists a sub-POVM $\mathrm{M}$ on $\C^d$ with $N=(1+\eta)\,d^2$ outcomes such that $c_{\eta}\,\|\cdot\|_2\leq\|\cdot\|_{\mathrm{M}}\leq \|\cdot\|_2$. Combining this with equation \eqref{eq:U-2} yields that, for any traceless Hermitian $\Delta$ on $\C^d$,
\[ c_{\eta}\,\|\Delta\|_{\mathrm{U}_d}\leq \|\Delta\|_{\mathrm{M}} \leq \sqrt{3}\,\|\Delta\|_{\mathrm{U}_d}. \]
To conclude, one just has to consider, instead of the sub-POVM $\mathrm{M}$, the POVM $\mathrm{M}'=(M'_i)_{1\leq i\leq N+1}$ defined by $M'_i=M_i$ for $1\leq i\leq N$ and $M'_{N+1}=\Id-(M_1+\cdots+M_N)$. The latter is in fact such that, for any traceless Hermitian $\Delta$, $\|\Delta\|_{\mathrm{M}}\leq \|\Delta\|_{\mathrm{M}'}\leq 2\,\|\Delta\|_{\mathrm{M}}$, and the proof of Theorem \ref{th:U-multiplicative} is thus complete.
\end{proof}


\part{Some aspects of generic entanglement: data-hiding and separability relaxations in high dimensions}
\label{part:entanglement}

When talking about entanglement in multipartite quantum systems, the picture changes drastically depending on the system's size. In small dimensions, relaxing the notion of separability to one which is easier to handle is usually a quite fruitful approach. Oppositely, as the dimensions grow, any too simple necessary condition for separability is doomed to be very rough. These very general and hand-waving assertions actually apply to many more specific settings and can be made much more precise by using tools from high-dimensional convex geometry. Such line of study was arguably initiated by Hayden, Leung and Winter in \cite{HLW}. In this seminal paper, the typical value of various correlation measures was estimated for random high-dimensional multipartite states. Among others, it was already observed there that, in bipartite quantum systems, features such as large entanglement of formation and small distillable entanglement are the rule rather than the exception when the dimensions of the two subsystems are large. It may have appeared surprising at first that having such bound-entangled like properties is in fact a common trait. Indeed, exhibiting explicit examples of states being so is usually hard, because they may be rare (or even nonexisting) in small dimensions while computations rapidly become intractable as dimensions grow.

\smallskip

Chapter \ref{chap:data-hiding} is in keeping with Chapter \ref{chap:zonoids} in the previous part, where the functional-analytical study of distinguishability norms associated to POVMs was initiated. And actually, the first question we look at could just as well have been put in Part \ref{part:complexity}, since it consists in looking at finite sub-families of the infinite family $\mathbf{ALL}$ of all POVMs, and asking: how many POVMs do they have to contain to achieve near to the maximum discrimination efficiency attained by $\mathbf{ALL}$? However, the focus turns next to multipartite systems, where POVMs satisfying certain locality constraints are considered. The goal is then to try and quantify how such restrictions might affect the ability to distinguish global states. More concretely, we look at several classes of locally restricted POVMs on multipartite quantum systems, and ask the following: as the dimensions of the underlying local spaces increase, does there exist an unbounded gap between them, and if so, is it an exceptional or a typical feature? Let us briefly summarize the obtained results, concentrating on the bipartite case for the sake of clarity. In the hierarchy between the families of local POVMs $\mathbf{LO}$, local POVMs with one-way classical communication $\mathbf{LOCC^{\rightarrow}}$, local POVMs with two-way classical communication $\mathbf{LOCC}$, separable POVMs $\mathbf{SEP}$, positive under partial transposition POVMs $\mathbf{PPT}$ and all POVMs $\mathbf{ALL}$, there exist unbounded gaps between the corresponding distinguishability norms at each step. Furthermore, this unbounded gap is generic in $\mathbf{SEP}$ vs $\mathbf{PPT}$, but not generic in $\mathbf{LOCC^{\rightarrow}}$ vs $\mathbf{LOCC}$, $\mathbf{LOCC}$ vs $\mathbf{SEP}$, and $\mathbf{PPT}$ vs $\mathbf{ALL}$ (concerning the unbounded gap $\mathbf{LO}$ vs $\mathbf{LOCC^{\rightarrow}}$, we are unable to conclude about its typicality). These results translate nicely in terms of data-hiding considerations: this phenomenon, consisting of two multipartite states which are very different (hence very well distinguishable by some global measurement) but nevertheless look almost the same to observers that can only perform measurements on their subsystem and communicate classically, is in fact exhibited by most multipartite states. Let us also say just one word concerning the main ingredients in the proofs. In order to estimate what is the value of $\|\cdot\|_{\mathbf{M}}$ to be expected, for $\mathbf{M}$ being one of the above-mentioned families of POVMs, two steps are required: one must first determine the average value of $\|\cdot\|_{\mathbf{M}}$ by estimating certain size parameters of (the polar of) its unit ball, and second make use of random matrix theory and concentration of measure to argue that this average behavior occurs with overwhelming probability in high dimension.

\smallskip

Chapter \ref{chap:SDrelaxations} is dedicated to the study of separability and entanglement in multipartite quantum systems. Certifying that a bipartite state is not separable can always be done by exhibiting an entanglement witness constructed from a positive map. In the multipartite case, the picture becomes more intricate. Indeed, even asserting that a state is not biseparable (i.e.~a convex combination of states which are separable across a given bipartition) may be a delicate task. We show however that it is always possible to construct such a genuine multipartite entanglement witness by lifting entanglement witnesses which only reveal bipartite entanglement. In small dimensions, this approach is quite versatile since it allows for a formulation of the problem as a semidefinite program, and can therefore be solved efficiently. Nevertheless, as one could have expected, any state-independent construction is condemned to become weaker and weaker as the dimensions grow. We back up this affirmation by focussing on one specific positive map relaxation of separability, namely positivity under partial transposition. We thus prove that, on high dimensional multipartite systems, the set of states which have a positive partial transpose across every cut is much bigger than the set of biseparable states. We additionally construct a whole class of random multipartite states which have the property of being, with overwhelming probability as the dimensions increase, fully positive under partial transposition and nevertheless (robustly) genuinely multipartite entangled. These two arguments substantiate that our universal schemes, even though efficient and effective to detect genuine multipartite entanglement in, for instance, three-qutrit systems, will miss most genuinely multipartite entangled states on higher dimensional systems.

\smallskip

Finally, in Chapter \ref{chap:k-extendibility}, we take a closer look at one particular hierarchy of separability tests, known as the hierarchy of $k$-extendibility tests, indexed by $k\in\N$. It consists of a sequence of increasingly constraining necessary conditions for separability (expressible as semidefinite programs of increasing size) which is asymptotically also sufficient. In real life however, only a finite number of checks can be performed, so it makes sense to ask, for a fixed $k\in\N$, what is the strength of the $k^{\text{th}}$ test. Now, it is known that there exist states which are far from separable (in either standard or operational distance measures) even though highly extendible. But these worst case scenarios do not exclude the possibility of making stronger statements about average or typical behaviors. This is precisely the question we tackle in Chapter \ref{chap:k-extendibility}, being especially interested in the high-dimensional regime. There are at least two distinct ways of answering it in a quantitative manner. The first approach consists in estimating a certain size parameter of the set of $k$-extendible states, and compare the obtained value with the known corresponding estimate for the set of separable states to see how the sizes of these two sets of states scale with one another. The second approach consists in looking at \textit{random-induced states} (i.e.~random mixed states which are obtained by partial tracing over an ancilla space a uniformly distributed pure state) and characterizing when these are with high probability $k$-extendible or not, so that again, comparing the obtained result with the known one for separability provides some information on how powerful the $k$-extendibility test typically is to detect entanglement. These two routes lead to the same conclusion, namely that if $k\in\N$ is a fixed parameter, then $k$-extendibility becomes a very loose relaxation of separability as the dimensions of the underlying local spaces grow. Nevertheless, whatever other well-studied separability criterion is defeated by the $k$-extendibility criterion above a certain dimension independent value of $k$ (from either one or the other point of view). Furthermore, it is possible to partially extend these results to the case were $k$ is not fixed but instead grows with the local dimensions as well. We are thus able to see that, for some growth rate of $k$, the set of $k$-extendible states lies strictly in-between the set of separable states and the set of all states.

\smallskip

Similarly in Chapters \ref{chap:data-hiding}, \ref{chap:SDrelaxations} and \ref{chap:k-extendibility}, the tools that we use to estimate the sizes of the convex bodies under consideration (either sets of POVMs or sets of states) are both geometric and probabilistic. Indeed, there are two main size parameters that we consider, the \textit{volume radius} and the \textit{mean width}. The volume radius, as the name indicates, is a volumetric quantity. Understanding how it behaves under intersection, Minkowski sum, symmetrization, tensor product etc. is the very essence of classical convex geometry. The mean width on the other hand, again in line with the denomination, is a probabilistic quantity: estimating it can be phrased as estimating the average of the supremum of a certain Gaussian process. Powerful union-bound type arguments exist for that (yielding sharp bounds at their highest level of sophistication and, for our purposes, already pretty good ones in their simpler forms).
Then, again in all three chapters, making statements about typical behaviours for random states requires, in addition to a mean value estimate, one extra ingredient, namely concentration of Lipschitz functions (in spherical or Gaussian random variables) around their average in high dimension. Note that in order to get, as a starting point, the average case behaviour, there are basically two routes that we may follow: either we relate it to some (already computed) mean width (appealing to the heuristic that unitary invariant random matrix ensembles having comparable spectra have comparable norms), or we compute it directly using standard tools from random matrix theory (e.g.~moment method).

\newpage
\textbf{\LARGE{Part III -- Table of contents}}
\parttoc

\chapter{Locally restricted measurements on multipartite quantum systems}
\chaptermark{Locally restricted measurements on multipartite quantum systems}
\label{chap:data-hiding}

\textsf{Based on ``Locally restricted measurements on a multipartite quantum system: data hiding is generic'', in collaboration with G. Aubrun \cite{AL1}.}

\bigskip

We study the distinguishability norms associated to families of locally restricted POVMs on multipartite systems.
These norms (introduced by Matthews, Wehner and Winter) quantify
how quantum measurements, subject to locality constraints, perform in the task of discriminating two multipartite quantum states.
We mainly address the following question regarding the behaviour of these
distinguishability norms in the high-dimensional regime: On a bipartite space, what are the relative strengths of standard classes of
locally restricted measurements? We show that the class of PPT measurements typically
performs almost as well as the class of all measurements whereas restricting to local measurements and classical communication,
or even just to separable measurements, implies a substantial loss. We also provide examples of state pairs which can be perfectly
distinguished by local measurements if (one-way) classical communication is allowed between the parties, but very poorly without it.
Finally, we study how many POVMs are needed to distinguish almost perfectly any pair of states on $\C^d$,
showing that the answer is $\exp(\Theta(d^2))$.

\section{Introduction}

How quantum measurements can help us make decisions? We consider a basic problem, the task of distinguishing two quantum states,
where this question has a neat answer. Given a POVM (Positive Operator-Valued Measure) $\mathrm{M}$ on $\C^d$, Matthews, Wehner and Winter \cite{MWW}
introduced its distinguishability norm $\|\cdot\|_\mathrm{M}$, which has the property that given a pair $(\rho,\sigma)$ of quantum states,
$\|\rho-\sigma\|_\mathrm{M}$ is the bias observed when the POVM $\mathrm{M}$ is used optimally to distinguish $\rho$ from $\sigma$ (the larger
is the norm, the more efficient is the POVM).
More generally, we can associate to a family of POVMs $\mathbf{M}$ the norm $\|\cdot\|_{\mathbf{M}} = \sup \{ \|\cdot\|_{\mathrm{M}} \st
\mathrm{M} \in \mathbf{M} \}$ which corresponds to the bias achieved by the best POVM from the family.

Here, in the line of Chapter \ref{chap:zonoids}, we study these norms from a functional-analytic point of view and
are mostly interested in the asymptotic regime, when the dimension of the underlying Hilbert space tends to infinity.

\subsection*{How many essentially distinct POVMs are there?}

The (infinite) family $\mathbf{ALL}$ of all POVMs on $\C^d$ achieves maximal efficiency in the distinguishability task, and in some sense gives us perfect information. It was indeed one of
the seminal observations by Holevo \cite{Holevo} and Helstrom \cite{Helstrom} that $\|\cdot\|_{\mathbf{ALL}} = \|\cdot\|_1$, so that two orthogonal quantum states could be perfectly
distinguished (i.e.~with a zero probability of error) by a suitable measurement.
But how ``complex'' is the class $\mathbf{ALL}$? What about finite subfamilies? How many POVMs are needed to obtain near-to-optimal efficiency?
We show (Theorem \ref{theorem:approximation-of-ALL})
that $\exp(\Theta(d^2))$ different POVMs are necessary (and sufficient) to obtain approximation within a constant factor. The concept of mean width (from convex geometry) plays an important
role in our proof,
which is detailed in Section \ref{sec:all-POVMs}.

\subsection*{Locally restricted POVMs on a multipartite quantum system}

On a multipartite quantum system, experimenters usually cannot implement any global observable. For instance, they may be only able to perform
quantum measurements on their own subsystem
(and then perhaps to communicate the results classically). A natural question in such situation is thus to quantify the relative strengths of
several classes of measurements, restricted by these locality constraints, such as LOCC,
separable or PPT measurements (precise definitions appear in Section \ref{sec:local-POVM}).

Let us summarize the main result in this chapter (restricting here to the bipartite case for the sake of clarity). We consider typical discrimination tasks, in the following sense.
Let $\rho$ and $\sigma$ be states chosen independently and uniformly at random within
the set of all states on $\C^d\otimes\C^d$. We show that our ability to distinguish $\rho$ from $\sigma$ depends in an essential way on the class
of the allowed measurements.
Indeed, with high probability, $\|\rho-\sigma\|_{\mathbf{PPT}}$ is of order $1$ (as $\|\rho-\sigma\|_{\mathbf{ALL}}$) while
$\|\rho-\sigma\|_{\mathbf{SEP}}$, $\|\rho-\sigma\|_{\mathbf{LOCC}}$ and $\|\rho-\sigma\|_{\mathbf{LOCC^{\rightarrow}}}$ are of order $1/\sqrt{d}$.
This shows that data hiding is generic: typically, high-dimensional quantum states cannot be distinguished locally even
though they look different globally. These results appear as Theorem \ref{theorem:typical-states} in Section \ref{sec:local-POVM-statement}. The proofs are detailed in Section \ref{sec:local-POVM-proofs}. They rely, as a first essential step, on estimates
on the volume radius and the mean width of the (polar of) the unit balls associated to the norms
$\|\cdot\|_{\mathbf{PPT}}$, $\|\cdot\|_{\mathbf{SEP}}$ and $\|\cdot\|_{\mathbf{LOCC}}$ (Theorem \ref{theorem:vrad-w-PPT-SEP}). The use of concentration of measure and random matrix theory
(Proposition \ref{proposition:Delta-vs-difference-of-states}) then allows to pass from these global estimates to the estimates in a typical
direction quoted above. In Section \ref{sec:data-hiding} corollaries on quantum data hiding are derived and detailed, both in the bipartite
and in the generalized multipartite case.

We also provide examples of random bipartite states $\rho,\sigma$ on $\C^d\otimes\C^d$ which are such that $\|\rho-\sigma\|_{\mathbf{LOCC^{\rightarrow}}}=2$ while, with high probability, $\|\rho-\sigma\|_{\mathbf{LO}}$ is of order $1/\sqrt{d}$. The precise result appears in Theorem \ref{theorem:LO-vs-LOCC} and is proved in Section \ref{section:LO-vs-LOCC}. Following the same proof technique, one can then construct, more generally, random bipartite states $\rho,\sigma$ such that, either $\|\rho-\sigma\|_{\mathbf{LOCC^{\rightarrow}}}\gg\|\rho-\sigma\|_{\mathbf{LO}}$ or $\|\rho-\sigma\|_{\mathbf{LOCC}}\gg\|\rho-\sigma\|_{\mathbf{LOCC^{\rightarrow}}}$.

Table \ref{table:summary} summarizes our various conclusions.

\begin{table}[h] \caption{Unbounded gaps between locally restricted distinguishability norms}
\label{table:summary}
\begin{center}
\renewcommand{\arraystretch}{1.5}
\begin{tabular}{| C{1.70cm} | C{1.165cm} C{0.275cm} C{1.615cm} C{0.275cm} C{1.615cm} C{0.275cm} C{1.615cm} C{0.275cm} C{1.615cm} C{0.275cm} C{1.165cm} |}
\hline
Norm hierarchy & $\ \|\cdot\|_{\mathbf{LO}}$ & $\leq$ & $\|\cdot\|_{\mathbf{LOCC^{\rightarrow}}}$ & $\leq$ & $\|\cdot\|_{\mathbf{LOCC}}$ & $\leq$ & $\|\cdot\|_{\mathbf{SEP}}$ & $\leq$ & $\|\cdot\|_{\mathbf{PPT}}$ & $\leq$ & $\|\cdot\|_{\mathbf{ALL}}\ $ \\
\hline
Existing unbounded gap? & & yes & & yes & & $\ \text{?}\ $ & & yes & & yes & \\
\hline
Generic unbounded gap? & & $\ \text{?}\ $ & & no & & no & & yes & & no & \\
\hline
\end{tabular}
\end{center}
\end{table}

\subsection*{Notation}

In addition to the general notation specified in Chapter \ref{chap:motivations}, Section \ref{sec:notation}, we introduce the following one: When $A,B$ are Hermitian matrices, we denote by $[A,B]$ the order interval, i.e.~the set of Hermitian
matrices $C$ such that both $C-A$ and $B-C$ are positive semidefinite matrices. In particular, $[-\Id,\Id]$ is the Hermitian part of the unit ball for $\|\cdot\|_{\infty}$ on the set of all complex matrices.

When $A$ and $B$ are quantities depending on the dimension, the notation $A \lesssim B$ means that there is a constant $C$ such that
$A \leq CB$. The notation $A \simeq B$ means both $A \lesssim B$ and $B \lesssim A$, and $A \sim B$ means that the ratio $A/B$ tends to $1$ when the dimension tends to infinity.

Extra notation, concepts and results from convex geometry, which are needed throughout our proofs, are gathered in Chapter \ref{chap:toolbox}, Section \ref{ap:convex-geometry}.

\section{Distinguishing quantum states: survey of our results}
\label{sec:POVM-geometry}

\subsection{General setting}
\label{sec:measurement-norm}

In this section, we gather some basic information about norms associated to POVMs, and refer e.g.~to \cite{MWW} for more details and proofs.
A POVM (Positive Operator-Valued Measure) on $\C^d$ is a finite family $\mathrm{M}=(M_i)_{i\in I}$ of positive operators on $\C^d$ such that
\[ \underset{i\in I}{\sum} M_i = \Id . \]
One could consider also continuous POVMs, where the finite sum is replaced by an integral. However this is not necessary, since continuous POVMs appear as limit
cases of discrete POVMs which we consider here (see Chapter \ref{chap:zonoids}).

Given a POVM $\mathrm{M}=(M_i)_{i\in I}$ on $\C^d$, and denoting by $\{|i\rangle,\ i\in I\}$ an orthonormal basis of $\C^{|I|}$,
we may associate to
$\mathrm{M}$ the CPTP (Completely Positive and Trace-Preserving) map
\[ \mathcal{M}:\Delta\in\mathcal{H}(\C^d)\mapsto\sum_{i\in I} \tr \big(M_i\Delta\big)\ketbra{i}{i}\in\mathcal{H}\big(\C^{|I|}\big). \]
The reader is referred to Chapter \ref{chap:QIT}, Section \ref{sec:channels}, for further comments.
The measurement (semi-)norm associated to $\mathrm{M}$ is then defined, for any $\Delta\in\mathcal{H}(\C^d)$, as
\[ \|\Delta\|_{\mathrm{M}}:=\|\mathcal{M}(\Delta)\|_1=\sum_{i\in I}|\tr \big(M_i\Delta\big)| . \]
Note that for any $\Delta\in\cH(\C^d)$, $\|\Delta\|_{\mathrm{M}}\leq\|\Delta\|_1$, with equality if $\Delta\in\cH_+(\C^d)$.

In general, $\|\cdot\|_{\mathrm{M}}$ is a semi-norm, and may vanish on non-zero Hermitians. A necessary and sufficient condition for $\|\cdot\|_{\mathrm{M}}$ to be
a norm is that the POVM $\mathrm{M}=(M_i)_{i \in I}$ is informationally complete, i.e.~that the family of operators $(M_i)_{i\in I}$ spans $\mathcal{H}(\C^d)$ as a linear space.
This especially implies that $\mathrm{M}$ has a total number of outcomes satisfying $|I| \geq d^2=\mathrm{dim}\ \mathcal{H}(\C^d)$.

We denote by $B_{\|\cdot\|_{\mathrm{M}}}$ the unit ball associated to $\|\cdot\|_{\mathrm{M}}$, and by $K_{\mathrm{M}}$ the polar of $B_{\|\cdot\|_{\mathrm{M}}}$ (i.e.~the
unit ball associated to the norm dual to $\|\cdot\|_{\mathrm{M}}$). In
other words, this means that the support function of $K_{\mathrm{M}}$ is defined, for any $\Delta\in\mathcal{H}(\C^d)$, as
\begin{equation} \label{eq:definition-K} h_{K_{\mathrm{M}}}(\Delta) = \|\Delta\|_{\mathrm{M}}. \end{equation}
Precise definitions of these concepts are given in Chapter \ref{chap:toolbox}, Section \ref{ap:convex-geometry}.

More generally, one can define the ``measurement'' or ``distinguishability'' norm associated to a whole set $\mathbf{M}$ of POVMs on $\C^d$ as
\[ \|\cdot\|_{\mathbf{M}}:=\underset{\mathrm{M}\in\mathbf{M}}{\sup}\|\cdot\|_{\mathrm{M}}. \]
The corresponding unit ball, and its polar, are
\[ B_{\|\cdot\|_{\mathbf{M}}} = \bigcap_{\mathrm{M} \in \mathbf{M}} B_{\|\cdot\|_{\mathrm{M}}} ,\]
\[ K_{\mathbf{M}} = \conv \left( \bigcup_{\mathrm{M} \in \mathbf{M}} K_{\mathrm{M}} \right) .\]

As mentioned earlier on in the introduction, these measurement norms are related to the task of distinguishing quantum states.
Let us consider the situation where a system (with associated Hilbert space $\C^d$) can be either in state $\rho$ or in state $\sigma$, with equal prior
probabilities $1/2$. It is known \cite{Holevo, Helstrom} that a decision process based on the maximum likelihood rule after performing the POVM $\mathrm{M}$
on the system yields a probability of error
\[ \P_{error}  =\frac{1}{2}\left(1-\left\|\frac{1}{2} \rho-\frac{1}{2}\sigma\right\|_{\mathrm{M}}\right). \]
In this context, the operational interpretation of the quantity $\|\rho-\sigma\|_{\mathrm{M}}$ is thus clear (and actually justifies the terminology of ``distinguishability norm''): up to a factor $1/2$, it is nothing else than the bias of the POVM $\mathrm{M}$ on the state pair $(\rho,\sigma)$.

Something that is worth pointing at is that, for any set $\mathbf{M}$ of POVMs on $\C^d$, there exists a set $\mathbf{\widetilde{M}}$ of $2$-outcome POVMs on
$\C^d$ which is such that $\|\cdot\|_{\mathbf{M}}=\|\cdot\|_{\mathbf{\widetilde{M}}}$. It may be explicitly defined as
\[ \mathbf{\widetilde{M}}:=\left\{\big(M,\Id-M\big),\ \exists\ (M_i)_{i\in I}\in\mathbf{M},\ \exists\ \widetilde{I}\subset I:\
M=\underset{i\in\widetilde{I}}{\sum}M_i\right\} .\]
Note then that
\[ K_{\mathbf{M}}=\conv \big\{2M-\Id,\ (M,\Id-M)\in\mathbf{\widetilde{M}}\big\}. \]

\subsection{On the complexity of the class of all POVMs}

Denote by $\mathbf{ALL}$ the family of all POVMs on $\C^d$. As we already noticed, $\|\cdot\|_{\mathbf{ALL}} = \|\cdot\|_1$ and therefore
$K_{\mathbf{ALL}}$ equals $[-\Id,\Id]$, which is the unit ball in $\cH(\C^d)$ for the operator norm.

The family $\mathbf{ALL}$ is obviously infinite. Since real-life situations can involve only finitely many apparatuses,
it makes sense to ask what must be the cardinality of
a finite family of POVMs $\mathbf{M}$ which achieves close to perfect discrimination, i.e.~such that the inequality $\|\cdot\|_{\mathbf{M}} \geq \lambda \|\cdot\|_{\mathbf{ALL}}$ holds for some $0<\lambda<1$. We show that the answer
is exponential in $d^2$. More precisely, we have the theorem below.

\begin{theorem} \label{theorem:approximation-of-ALL}
There are universal constants $c,C>0$ such that the following holds:
\begin{enumerate}
 \item[(i)] For any dimension $d$ and any $0<\e<1$, there is a family $\mathbf{M}$ consisting of at most  $\exp(C |\log\e| d^2)$ POVMs on
 $\C^d$ such that
 $\|\cdot\|_{\mathbf{M}} \geq (1-\e) \|\cdot\|_{\mathbf{ALL}}$.
 \item[(ii)] For any $\e>C/\sqrt{d}$, any family $\mathbf{M}$ of POVMs on $\C^d$ such that $\|\cdot\|_{\mathbf{M}} \geq \e \|\cdot\|_{\mathbf{ALL}}$
 contains at least $\exp(c \e^2 d^2)$ POVMs.
 \end{enumerate}
\end{theorem}

Theorem \ref{theorem:approximation-of-ALL} is proved in Section \ref{sec:all-POVMs}.
It is clear that the conclusion of (ii) fails for $\e \lesssim 1/\sqrt{d}$, since a single POVM
$\mathrm{M}$ (e.g.~the uniform POVM, see \cite{MWW} or Chapter \ref{chap:zonoids} of the present manuscript) may satisfy $\|\cdot\|_{\mathrm{M}} \gtrsim \|\cdot\|_1/\sqrt{d}$.

\subsection{Locally restricted measurements on a bipartite quantum system}
\label{sec:local-POVM}

We now study the class of locally restricted POVMs. We assume that the underlying global Hilbert space is the tensor product of several local
Hilbert spaces. However, for simplicity,
we focus on the case of a bipartite system
in which both parts play the same role and consider the Hilbert space $\mathrm{H} = \C^d \otimes \C^d$.
Several classes of POVMs can be defined on $\mathrm{H}$ due to various levels of locality restrictions (consult \cite{MWW} or \cite{LW1} for further information).

The most restricted class of POVMs on $\mathrm{H}$ is the one of local measurements, whose elements are tensor products of measurements on each of the
sub-systems:
\[\mathbf{LO}:=\left\{\left(M_{i} \otimes N_{j} \right)_{i\in I,j \in J} \st \ M_i \geq 0, \ N_j \geq 0, \ \sum_{i \in I} M_i = \Id_{\C^d}, \ \sum_{j \in J} N_j = \Id_{\C^d} \right\}. \]
This corresponds to the situation where parties are not allowed to communicate.

Then, we consider the class of separable measurements, whose elements are the measurements on $\mathrm{H}$ made of tensor operators
\[ \mathbf{SEP}:=\left\{\left(M_j \otimes N_j\right)_{j\in J} \st \ M_j \geq 0, \ N_j \geq 0, \ \sum_{j\in J}M_j \otimes N_j =\Id_{\C^d \otimes \C^d} \right\}. \]

An important subclass of $\mathbf{SEP}$ is the class $\mathbf{LOCC}$ (Local Operations and Classical Communication)
of measurements that can be implemented by a finite sequence of local operations on the sub-systems
followed by classical communication between the parties. This class can be described recursively as the smallest subclass of $\mathbf{SEP}$ which
contains $\mathbf{LO}$ and is stable under the following operation: given a POVM $\mathrm{M}=(M_i)_{i \in I}$ on $\C^d$, and for each $i \in I$ a
$\mathbf{LOCC}$ POVM $\big(R(i)_j \otimes S(i)_j\big)_{j \in J_i}$, the POVMs
\[ \left( M_i^{1/2} R(i)_j M_i^{1/2} \otimes S(i)_j \right)_{i \in I, j \in J_i} \ \ \textnormal{and}
\ \ \left( R(i)_j \otimes M_i^{1/2}  S(i)_j M_i^{1/2} \right)_{i \in I, j \in J_i} \]
are in $\mathbf{LOCC}$. A subclass of $\mathbf{LOCC}$ is the class $\mathbf{LOCC^\rightarrow}$ of one-way LOCC POVMs, which has a simpler description
\[\mathbf{LOCC^\rightarrow}:=\left\{\left(M_{i} \otimes N_{i,j} \right)_{i\in I,j \in J_i} \st \ M_i \geq 0, \ N_{i,j} \geq 0,
\ \sum_{i \in I} M_i = \Id_{\C^d}, \ \sum_{j \in J_i} N_{i,j} = \Id_{\C^d} \right\}. \]

Finally, we consider the class of positive under partial transpose (PPT) measurements, whose elements are the measurements on $\mathrm{H}$
made of operators that
remain positive when partially transposed on one sub-system:
\[ \mathbf{PPT}:=\left\{(M_j)_{j\in J} \st M_j \geq 0,\ M_j^{\Gamma}\geq\mathrm{0},\ \sum_{j\in J}M_j=\Id_{\C^d \otimes \C^d} \right\}. \]
The partial transposition $\Gamma$ is defined by its action on tensor
operators on $\mathrm{H}$: $(M\otimes N)^{\Gamma}:= M^T \otimes N$, $M^T$ denoting the usual transpose of $M$.
Let us point out that, even though the expression of a matrix transpose depends on the chosen basis, its eigenvalues on the contrary are intrinsic.
Therefore the PPT notion is basis-independent.

It is clear from the definitions that we have the chain of inclusions
\[ \mathbf{LO}\subset\mathbf{LOCC^\rightarrow}\subset \mathbf{LOCC} \subset\mathbf{SEP}\subset\mathbf{PPT}\subset \mathbf{ALL} \]
and consequently the chain of norm inequalities
\begin{equation}\label{eq:hierarchy} \|\cdot\|_{\mathbf{LO}} \leq \|\cdot\|_{\mathbf{LOCC^{\rightarrow}}} \leq \|\cdot\|_{\mathbf{LOCC}}
\leq \|\cdot\|_{ \mathbf{SEP}} \leq
\|\cdot\|_{\mathbf{PPT}} \leq \|\cdot\|_{\mathbf{ALL}}. \end{equation}

All the inequalities in \eqref{eq:hierarchy} are known to be strict provided $d >2$. Note though that the difference between the norms
$\|\cdot\|_{\mathbf{LOCC^{\rightarrow}}}$ and $\|\cdot\|_{\mathbf{LOCC}}$, as well as between $\|\cdot\|_{\mathbf{LOCC}}$ and $\|\cdot\|_{\mathbf{SEP}}$,
has been established only very recently \cite{CH}.

Here, we are interested in the high-dimensional behaviour of these norms, and the general question we investigate is whether or not the various gaps in the
hierarchy are bounded (independently of the dimension of the subsystems). It is already known that the gap between $\mathbf{PPT}$ and $\mathbf{ALL}$ is unbounded. An important example is provided by the symmetric state $\pi_s$ and the antisymmetric state $\pi_a$ on $\C^d\otimes\C^d$ (see Chapter \ref{chap:symmetries}, Section \ref{sec:Werner}, for precise definitions and further comments), which satisfy (see e.g.~\cite{DVLT2})
\[ \| \pi_s - \pi_a \|_{\mathbf{ALL}} =2  \ \ \textnormal{ while } \ \ \| \pi_s - \pi_a \|_{\mathbf{PPT}} =
\frac{4}{d+1}.\]
We show however (see Theorem \ref{theorem:typical-states}) that such feature is not generic. This is in contrast with the gap between $\mathbf{SEP}$ and
$\mathbf{PPT}$ which we prove to be generically unbounded (see Theorem \ref{theorem:typical-states}). We also provide examples of unbounded gap between $\mathbf{LO}$
and $\mathbf{LOCC^{\rightarrow}}$ (see Theorems \ref{theorem:LO-vs-LOCC} and \ref{th:LO-LOCC1}), as well as between $\mathbf{LOCC^{\rightarrow}}$ and $\mathbf{LOCC}$ (see Theorem \ref{th:LOCC1-LOCC2}). But we do not know if this situation is typical. Regarding the gap between $\mathbf{LOCC}$ and $\mathbf{SEP}$, determining whether it is bounded is still an open problem.

Note also that for states of low rank, the gaps between these norms remain bounded. It follows from the results of \cite{LW1} that, for $\Delta \in \cH(\C^d \otimes \C^d)$
of rank $r$,
we have
\[ \|\Delta\|_{\mathrm{LO}} \geq \frac{1}{18\sqrt{r}} \|\Delta\|_{\mathbf{ALL}}. \]

\subsection{Discriminating power of the different classes of locally restricted measurements}
\label{sec:local-POVM-statement}

Our main result compares the efficiency of the classes $\mathbf{LOCC^{\rightarrow}}$, $\mathbf{LOCC}$, $\mathbf{SEP}$, $\mathbf{PPT}$ and $\mathbf{ALL}$ to perform a typical discrimination task.
Here ``typical'' means the following: we consider the problem of distinguishing $\rho$ from $\sigma$, where $\rho$ and $\sigma$ are random states, chosen independently at random with respect to the
uniform measure (i.e.~the
Lebesgue measure induced by the Hilbert--Schmidt distance) on the set of all states.
It turns out that the PPT constraint on the allowed measurements is not very restrictive, affecting typically the performance by only a constant
factor, while the separability one implies
a more substantial loss.
This shows that generic bipartite states are data hiding: separable measurements (and even more so local measurements followed by classical communication) can poorly distinguish them (see \cite{HLSW} for another
instance of this phenomenon and Section \ref{sec:data-hiding} for a more detailed discussion on that topic).

\begin{theorem} \label{theorem:typical-states}
There are universal constants $C,c>0$ such that the following holds. Given a dimension $d$,
let $\rho$ and $\sigma$ be random states, independent and uniformly distributed on the set of states on $\C^d \otimes \C^d$. Then, with high probability,
\[ c \leq \| \rho - \sigma \|_{\mathbf{PPT}} \leq \| \rho - \sigma \|_{\mathbf{ALL}} \leq C, \]
\[ \frac{c}{\sqrt{d}} \leq \| \rho - \sigma \|_{\mathbf{LOCC^{\rightarrow}}} \leq \| \rho - \sigma \|_{\mathbf{LOCC}} \leq \| \rho - \sigma \|_{\mathbf{SEP}} \leq \frac{C}{\sqrt{d}}. \]
Here, ``with high probability'' means that the probability that one of the conclusions fails is less than $\exp(-c_0 d^3)$ for some constant $c_0>0$.
\end{theorem}

An immediate consequence of the high probability estimates is that one can find in $\C^d \otimes \C^d$ exponentially many states which are pairwise data hiding.

\begin{corollary}
There are constants $C,c>0$ such that, if $\mathcal{A}$ denotes a set of $\exp(cd^3)$ independent random states uniformly distributed on the set
of states on $\C^d \otimes \C^d$, with high probability any pair of distinct states $\rho,\sigma \in \mathcal{A}$ satisfies the conclusions of
Theorem \ref{theorem:typical-states}.
\end{corollary}

We deduce Theorem \ref{theorem:typical-states} from estimates on the mean width and the volume of the unit balls $K_{\mathbf{LOCC^{\rightarrow}}}$,
$K_{\mathbf{SEP}}$ and $
K_{\mathbf{PPT}}$. The use of concentration
of measure allows to pass from these global estimates to the estimates in a typical direction that appear in Theorem \ref{theorem:typical-states}.
We include all this material
in Section \ref{sec:local-POVM-proofs}.

We also show that even the smallest amount of communication has a huge influence:
we give examples of states which are perfectly distinguishable under local measurements and one-way classical communication
but very poorly distinguishable under local measurements with no communication between the parties.

\begin{theorem} \label{theorem:LO-vs-LOCC}
There is a universal constant $C>0$ such that the following holds: for any $d$, there exists states $\rho$ and $\sigma$ on $\C^d \otimes \C^d$
such that
\[ \|\rho-\sigma\|_{\mathbf{LOCC^{\rightarrow}}} = 2, \]
and
\begin{equation} \label{equation:small-LO} \|\rho-\sigma\|_{\mathbf{LO}} \leq \frac{C}{\sqrt{d}}. \end{equation}
\end{theorem}

These states are constructed as follows: assuming without loss of generality that $d$ is even, let
$E$ be a fixed $d/2$-dimensional subspace of $\C^d$, let $U_1,\ldots,U_d$ be random independent Haar-distributed unitaries on $\C^d$, and
define the random states $\rho_i=U_iP_EU_i^\dagger/(d/2)$ and $\sigma_i=U_iP_{E^{\perp}}U_i^\dagger/(d/2)$, $1\leq i\leq d$, on $\C^d$ (where $P_E$ and $P_{E^{\perp}}$ denote the orthogonal projections onto $E$ and $E^{\perp}$ respectively).
Then, denoting by $\{\ket{1},\ldots,\ket{d}\}$ an orthonormal basis of $\C^d$, define
\[ \rho=\frac{1}{d}\sum_{i=1}^d\ketbra{i}{i}\otimes\rho_i\ \ \text{and}\ \ \sigma=\frac{1}{d}\sum_{i=1}^d\ketbra{i}{i}\otimes\sigma_i. \]
The pair $(\rho,\sigma)$ satisfies \eqref{equation:small-LO} with high probability.

Theorem \ref{theorem:LO-vs-LOCC} is proved in Section \ref{section:LO-vs-LOCC}. It is built on the idea that, typically, a single POVM cannot succeed
simultaneously in several ``sufficiently different'' discrimination tasks. The above construction is also generalized there to show the existence of an unbounded gap between $\mathbf{LOCC^{\rightarrow}}$ and $\mathbf{LOCC}$.

\section{On the complexity of the class of all POVMs}
\label{sec:all-POVMs}

In this section, we determine how many distinct POVMs a set $\mathbf{M}$ of POVMs on $\C^d$ must contain in order
to approximate the set $\mathbf{ALL}$ of all POVMs on $\C^d$ (in the sense that
$\lambda \|\cdot\|_{\mathbf{ALL}}\leq\|\cdot\|_{\mathbf{M}}\leq\|\cdot\|_{\mathbf{ALL}}$ for some $0 < \lambda < 1$).

The reason for the $\exp(d^2)$ scaling in the first part of Theorem \ref{theorem:approximation-of-ALL} is that these POVMs should be able to discriminate any two states within the family of
states
$\{P_E/\dim E\}$, where $E$ varies among all subspaces of $\C^d$, and $P_E$ denotes the orthogonal projection onto $E$.
The set of $k$-dimensional subspaces of $\C^d$
has dimension $k(d-k)$, which is of order $d^2$ when $k$ is proportional to $d$.

The second part of Theorem \ref{theorem:approximation-of-ALL} requires an extra ingredient, since
a single POVM may be able to discriminate exponentially many pairs of subspaces. The concept of mean width (see Chapter \ref{chap:toolbox}, Section \ref{ap:convex-geometry}) provides a neat answer to this problem.

\medskip

To begin with, we prove the first part of Theorem \ref{theorem:approximation-of-ALL}.
Note that the condition $\|\cdot\|_{\mathbf{M}} \geq (1-\e) \|\cdot\|_{\mathbf{ALL}}$
is equivalent to $K_{\mathbf{M}} \supset (1-\e) [-\Id, \Id]$, the set $K_{\mathrm{M}}$ being defined in \eqref{eq:definition-K}.
We thus only have to make use of the well-known lemma below.

\begin{lemma}[Approximation of convex bodies by polytopes] \label{lemma:approximation-polytope}
Given a symmetric convex body $K \subset \R^n$ and $0<\e<1$, there is a finite family $(x_i)_{i \in I}$ such that $|I| \leq (3/\e)^n$ and
\[ (1-\e) K \subset \conv \{ \pm x_i \st i \in I \} \subset K .\]
\end{lemma}

\begin{proof}
Let $\mathcal{A}$ be an $\e$-net in $K$, with respect to $\|\cdot\|_K$ (the gauge of $K$, as defined in Chapter \ref{chap:toolbox}, Section \ref{ap:convex-geometry}).
A standard volumetric argument (see Lemma \ref{lemma:nets} in Chapter \ref{chap:toolbox}, Section \ref{ap:deviations}) shows that we may ensure that $|\mathcal{A}|
\leq  (3/\e)^n$. Let $P := \conv ( \pm \mathcal{A}) \subset K$.
Given any $x \in K$, there exists $x' \in \mathcal{A}$ such that $\|x-x'\|_K \leq \e$. Therefore
\[ \|x\|_P \leq \|x'\|_P+\|x-x'\|_P \leq 1 + \e A, \]
where $A := \sup \{ \|y\|_P \st y \in K\}$. Taking supremum over $x \in K$, we obtain $A \leq 1 +\e A$ and therefore ($A$ is easily seen to be finite)
$A \leq (1-\e)^{-1}$. We thus proved the inequality $\|\cdot\|_P \leq (1-\e)^{-1} \|\cdot\|_K$, which is equivalent to the inclusion $(1-\e)K \subset P$.
\end{proof}

When applied to the $d^2$-dimensional convex body $K_{\mathbf{ALL}} = [-\Id,\Id]$, Lemma \ref{lemma:approximation-polytope} implies that there is a
finite family
$(A_i)_{i \in I} \subset [-\Id,\Id]$ with $|I| \leq (3/\e)^{d^2}$ and $\conv \{ \pm A_i \st i \in I \} \supset (1-\e) [-\Id,\Id]$.
For every $i \in I$, we may consider the POVM
\[ \mathrm{M}_i := \left( \frac{\Id+A_i}{2} , \frac{\Id-A_i}{2} \right) .\]
If we denote $\mathbf{M} := \{ \mathrm{M}_i \st i \in I \}$, then for any $i \in I$, $\pm A_i \in K_{\mathrm{M}_i}$ and therefore
$(1-\e) [-\Id,\Id] \subset K_{\mathbf{M}}$, which is precisely what we wanted to prove.

\medskip

We now show the second part of Theorem \ref{theorem:approximation-of-ALL}. The key observation is the following lemma, where
we denote by $\alpha_n$ the mean width of a segment $[-x,x]$ for $x$ a unit vector in $\R^n$, so that $\alpha_n \sim \sqrt{2/\pi n}$ (see Chapter \ref{chap:toolbox}, Section \ref{ap:convex-geometry}).

\begin{lemma} \label{lemma:mean-width-POVM}
Let $\mathrm{M}$ be a POVM on $\C^d$. Then the mean width of the set $K_{\mathrm{M}}$, defined by equation \eqref{eq:definition-K}, satisfies
$w(K_{\mathrm{M}}) \leq d \alpha_{d^2}$,
with equality if  $\mathrm{M}$ is a rank-$1$ POVM (note that $d \alpha_{d^2}$ is of order $1$).
\end{lemma}

It may be pointed out that the assertion of Lemma \ref{lemma:mean-width-POVM} implies that, as far as the mean width is concerned, all rank-$1$
POVMs are comparable.

\begin{proof}
Given any POVM $\mathrm{M}$, there is a rank-$1$ POVM $\mathrm{M}'$ such that $K_{\mathrm{M}} \subset K_{\mathrm{M'}}$
(this is easily seen by splitting the POVM
elements from $\mathrm{M}$ as a sum of rank-$1$ operators). Therefore, it suffices to show that $w(K_{\mathrm{M}})=d \alpha_{d^2}$ for any rank-$1$ POVM.
Let $\mathrm{M}=\left(p_i\ketbra{\psi_i}{\psi_i}\right)_{i\in I}$ be a rank-$1$ POVM, where $(p_i)_{i\in I}$ are positive numbers and
$(\psi_i)_{i\in I}$ are unit vectors such that
\[ \sum_{i\in I} p_i\ketbra{\psi_i}{\psi_i} =\Id . \]
By taking the trace, we check that the total mass of $\{ p_i \st i\in I\}$ equals $d$. We then have, for any $\Delta \in \cH(\C^d)$,
\[ h_{K_{\mathrm{M}}}(\Delta) = \sum_{i\in I} p_i |\langle \psi_i |\Delta| \psi_i \rangle|. \]
Hence, denoting by $S_{HS}(\C^d)$ the Hilbert--Schmidt unit sphere of $\cH(\C^d)$ (which has dimension $d^2-1$) equipped with the uniform
measure $\sigma$, the mean width of
$K_{\mathrm{M}}$ can be computed as
\[ w(K_{\mathrm{M}}) = \int_{S_{HS}(\C^d)} h_{K_{\mathrm{M}}}(\Delta) \, \mathrm{d} \sigma(\Delta) =
\sum_{i\in I} p_i \left( \int_{S_{HS}(\C^d)} |\langle \psi_i |\Delta| \psi_i \rangle|  \, \mathrm{d} \sigma(\Delta)  \right)
 = \sum_{i\in I} p_i \alpha_{d^2} = d \alpha_{d^2} . \qedhere \]
\end{proof}

Assume that $\mathbf{M}$ is a family of $N$ POVMs such that
$\|\Delta\|_{\mathbf{M}} \geq \e \|\Delta\|_1$ for any $\Delta \in \cH(\C^d)$. This implies that $K_{\mathbf{M}} \supset \e [-\Id,\Id]$ and therefore
that
\begin{equation} \label{wKM-lower-bound} w(K_{\mathbf{M}}) \geq \e w([-\Id,\Id]) \simeq \e \sqrt{d},
\end{equation}
where we used last the estimate on the mean width of $[-\Id,\Id]$ from Theorem \ref{theorem:operator-norm} in Appendix \ref{ap:standard}.
On the other hand,  we have
\begin{equation} \label{eq:convex-hull} K_{\mathbf{M}} = \conv \left( \bigcup_{\mathrm{M} \in \mathbf{M}} K_{\mathrm{M}} \right) ,\end{equation}
so that $K_{\mathbf{M}}$ is
the convex hull of $N$ sets, each of them of mean width bounded by an absolute constant (by Lemma \ref{lemma:mean-width-POVM}).
We may apply Lemma \ref{lemma:mean-width-union-bound} in Chapter \ref{chap:toolbox}, Section \ref{ap:convex-geometry}, with $\lambda = \sqrt{d}$
since $[-\Id,\Id]$ is contained in the Hilbert--Schmidt ball of radius $\sqrt{d}$. Recalling that the ambient dimension is $n=d^2$, we get
\begin{equation} \label{wKM-upper-bound} w(K_\mathbf{M}) \leq C \left( 1 + \frac{\sqrt{\log N}}{\sqrt{d}} \right) .\end{equation}
A comparison of the bounds \eqref{wKM-lower-bound} and \eqref{wKM-upper-bound} immediately yields $\log N \gtrsim \e^2 d^2$, as required.

\section{Existing unbounded gap between $\mathbf{LO}$ and $\mathbf{LOCC}$}
\label{section:LO-vs-LOCC}

\subsection{Unbounded gap $\mathbf{LO}/\mathbf{LOCC^{\rightarrow}}$}

In this section we give a proof of Theorem \ref{theorem:LO-vs-LOCC}.
Let $\{\ket{1},\ldots,\ket{d}\}$ be an orthonormal basis of $\C^d$. For $d$ even, we consider a fixed $d/2$-dimensional subspace $E \subset \C^d$,
and denote $\Delta_0 = 2P_E - \Id$. We then pick $U_1,\ldots,U_d$ random independent Haar-distributed unitaries on $\C^d$, and for $1 \leq i \leq d$ we
consider the random operators $\Delta_i=U_i\Delta_0U_i^\dagger$. We finally introduce
\begin{equation} \label{eq:def-Delta} \Delta = \sum_{i=1}^d \ketbra{i}{i}\otimes\Delta_i. \end{equation}

For each $1\leq i\leq d$, let $\mathrm{M}_i=(M_i,\Id-M_i)$ be a POVM on $\C^d$ such that $\|\Delta_i\|_{\mathrm{M}_i} = \|\Delta_i\|_1$. Then,
\[ \mathrm{M} = \left( \ketbra{i}{i} \otimes M_i, \ketbra{i}{i} \otimes (\Id-M_i) \right)_{1 \leq i \leq d} \]
is a POVM on $\C^d\otimes\C^d$ which is in $\mathbf{LOCC^{\rightarrow}}$. Therefore,
\[ \|\Delta\|_{\mathbf{LOCC^{\rightarrow}}} \geq \|\Delta\|_{\mathrm{M}}  = \sum_{i=1}^d  \|\Delta_i\|_1 = d^2, \]
and there is actually equality in the inequality above since we also have
\[ \|\Delta\|_{\mathbf{LOCC^{\rightarrow}}} \leq \|\Delta\|_1 = \sum_{i=1}^d  \|\Delta_i\|_1 = d^2. \]
Theorem \ref{theorem:LO-vs-LOCC} will follow (with $\rho$ and $\sigma$ being the positive and negative parts of $\Delta$,
after renormalization) if we prove that $\|\Delta\|_{\mathbf{LO}} \lesssim Cd^{3/2}$ with high probability.

\begin{proposition} \label{prop:discretization-LO}
For $\Delta \in \cH(\C^d \otimes \C^d)$ defined as in \eqref{eq:def-Delta}, we have
\begin{equation} \label{eq:LO-Delta} \|\Delta\|_{\mathbf{LO}} =
\sup\left\{ \sum_{i=1}^d  \|\Delta_i\|_{\mathrm{N}} \st \mathrm{N}\ \text{POVM on}\ \C^d \right\}. \end{equation}
This quantity can be upper bounded as follows, where $\mathcal{A}$ denotes a $1/16$-net in $S_{\C^d}$
\begin{eqnarray}
\label{eq:bound-LO}
\|\Delta\|_{\mathbf{LO}} & \leq & d \sup_{x \in S_{\C^d}} \sum_{i=1}^d \left| \bra{x} \Delta_i \ket{x} \right| \\
\label{eq:bound-LO-2}
& \leq & 2d \sup_{x \in \mathcal{A}} \sum_{i=1}^d \left| \bra{x} \Delta_i \ket{x} \right|.
\end{eqnarray}
\end{proposition}

\begin{proof}
The inequality $\geq$ in \eqref{eq:LO-Delta} follows by considering the $\mathbf{LO}$ POVM $(\ketbra{i}{i})_{1 \leq i \leq d} \otimes \mathrm{N}$.
Conversely, given
POVMs $\mathrm{M}=(M_j)_{j\in J}$ and $\mathrm{N}=(N_k)_{k\in K}$ on $\C^d$, we have
\begin{align*} \|\Delta\|_{\mathrm{M}\otimes\mathrm{N}} = & \sum_{j\in J, k\in K} \left|\sum_{i=1}^d \tr\left(\left(\ketbra{i}{i}\otimes\Delta_i\right) \left(M_j\otimes N_k\right)\right)\right|\\
\leq & \sum_{i=1}^d \left(\sum_{j\in J}\left|\bra{i}M_j\ket{i}\right|\right)
\left(\sum_{k\in K}\left|\tr\left(\Delta_i N_k\right)\right|\right)\\
\leq & \sum_{i=1}^d  \|\Delta_i\|_{\mathrm{N}},
\end{align*}
the last inequality being because, for each $1\leq i\leq d$, $\sum_{j\in J}\left|\bra{i}M_j\ket{i}\right| =\sum_{j\in J}\bra{i}M_j\ket{i}
=\braket{i}{i} =1$. Taking the supremum over $\mathrm{M}$ and $\mathrm{N}$ gives the inequality $\leq$ in  \eqref{eq:LO-Delta}.

The supremum in \eqref{eq:LO-Delta} is unchanged when restricting to the supremum on POVMs whose elements have rank $1$, since splitting the POVM elements
as sum of rank $1$ operators does not decrease the distinguishability norm. If $\mathrm{N}$ is such a POVM, its elements can be written as
$(\alpha_k \ketbra{x_k}{x_k})_{k \in K}$, where $(x_k)_{k \in K}$ are unit vectors and $(\alpha_k)_{k \in K}$ positive numbers satisfying $\sum_{k \in K} \alpha_k=d$.
We thus have in that case
\[ \sum_{i=1}^d \|\Delta_i\|_{\mathrm{N}} = \sum_{i=1}^d \sum_{k \in K} |\tr(\Delta_i \cdot \alpha_k \ketbra{x_k}{x_k} )|
\leq d \sup_{x \in S_{\C^d}} \sum_{i=1}^d \left| \bra{x} \Delta_i \ket{x} \right|, \]
proving \eqref{eq:bound-LO}.

To prove \eqref{eq:bound-LO-2}, we introduce the function $g$ defined for $x,y \in \C^d$ by $g(x,y) = \sum_{i=1}^d \left|\bra{x}\Delta_i \ket{y}\right|$, and the function $f$ defined for $x \in \C^d$ by $f(x)=g(x,x)$. Denote by $G$ the supremum of $g$ over $S_{\C^d} \times S_{\C^d}$, by
$F$ the supremum of $f$ over $S_{\C^d}$ and by $F'$ the supremum of $f$ over a $\delta$-net $\mathcal{A}$.
For any $x,y\in\C^d$, we have by the polarisation identity
\[\bra{x}\Delta_i\ket{y} = \frac{1}{4}\left(\bra{x+y}\Delta_i\ket{x+y}
+i\bra{x+iy}\Delta_i\ket{x+iy} -\bra{x-y}\Delta_i\ket{x-y} -i\bra{x-iy}\Delta_i\ket{x-iy}\right), \]
so that $g(x,y) \leq \left(f(x+y)+f(x+iy)+f(x-y)+f(x-iy)\right)/4$ and therefore $G\leq 4F$.

Given $x\in S_{\C^d}$, there exists $x'\in\mathcal{A}$ such that $\|x-x'\|_2 \leq\delta$, and by the triangle inequality
\[ \left|\bra{x}\Delta_i\ket{x}\right| \leq \left|\bra{x}\Delta_i\ket{x-x'}\right| + \left|\bra{x-x'}\Delta_i\ket{x'}\right| +
\left|\bra{x'}\Delta_i\ket{x'}\right| .\]
Summing over $i$ and taking supremum over $x \in S_{\C^d}$ gives
\[ F \leq 2 \delta G + F' \leq 8 \delta F +F' .\]
For $\delta=1/16$, we obtain $F \leq 2F'$, and therefore \eqref{eq:bound-LO-2} follows from \eqref{eq:bound-LO}.
\end{proof}

To bound $\|\Delta\|_{\mathbf{LO}}$, we combine Proposition \ref{prop:discretization-LO} with the following result.

\begin{proposition}
\label{prop:individual}
Let $x$ be a fixed unit vector in $\C^d$, $E$ be a fixed $d/2$-dimensional subspace of $\C^d$ and $\Delta_0 = 2P_E -\Id$, $(U_i)_{1 \leq i \leq n}$ be Haar-distributed independent random unitaries on $\C^d$, and for each $1\leq i\leq n$, set
$\Delta_i=U_i\Delta_0U_i^\dagger$. Then, for any $t>1$,
\[ \P\left(\sum_{i=1}^n \left|\bra{x}\Delta_i\ket{x}\right|\geq (1+t) n \E |\bra{x} \Delta_1 \ket{x}|\right)\leq e^{-c_0 nt},\]
$c_0>0$ being a universal constant.
\end{proposition}

\begin{proof}
Proposition \ref{prop:individual} is a consequence of Proposition \ref{proposition:large-deviations} in Chapter \ref{chap:zonoids}, which is itself a variation on Bernstein inequalities (recalled as Theorem \ref{th:Bernstein} in Chapter \ref{chap:toolbox}, Section \ref{ap:deviations}).
The quantity $\E |\bra{x} \Delta_1 \ket{x}|$ is equal to the so-called ``uniform norm'' of $\Delta_1$ (see \cite{MWW} and Chapter \ref{chap:zonoids} of the present manuscript) and we use the bound from \cite{LW1}
\[ \E |\bra{x} \Delta_1 \ket{x}| \leq \frac{1}{d} \|\Delta_1\|_{2} = \frac{1}{\sqrt{d}}. \qedhere \]
\end{proof}

We now complete the proof of Theorem \ref{theorem:LO-vs-LOCC}.
Let $\mathcal{A}$ be a minimal $1/16$-net in $S_{\C^d}$, so that $|\mathcal{A}|\leq 48^{2d}$ (see Lemma \ref{lemma:nets} in Chapter \ref{chap:toolbox}, Section \ref{ap:deviations}).
Using Propositions \ref{prop:discretization-LO} and \ref{prop:individual} (for $n=d$), and the union bound, we obtain that for any $t>1$
\[ \P \left( \|\Delta\|_{\mathbf{LO}} \geq 2(1+t) d^{3/2} \right) \leq
\P\left( \exists \ x\in \mathcal{A} \st \sum_{i=1}^d \left|\bra{x}\Delta_i\ket{x}\right| \geq (1+t)\sqrt{d}\right) \leq
48^{2d} e^{-c_0dt} . \]

This estimate is less than $1$ when $t$ is larger than some number $t_0$. This shows that
$\|\Delta\|_{\mathbf{LO}} \leq 2(1+t_0)d^{3/2}$ with high probability while $\|\Delta\|_{\mathbf{LOCC}^{\rightarrow}} = d^2$, and Theorem \ref{theorem:LO-vs-LOCC} follows.

\begin{remark}
The operator $\Delta$ defined by equation \eqref{eq:def-Delta} can be rewritten as $\Delta = d^2(\rho' -
\mathrm{Id}/d^2)$, with
\[ \rho' = \frac{2}{d^2} \sum_{i=1}^d \ketbra{i}{i}
\otimes U_iP_{E}U_i^{\dagger}. \]
Hence by Theorem \ref{theorem:LO-vs-LOCC}, $\| \rho' - {\mathrm{Id}}/d^2
\|_{\mathbf{LO}}\leq C/\sqrt{d}$ with high probability, while
$\| \rho' - {\mathrm{Id}}/d^2\|_{\mathbf{LOCC}^{\rightarrow}}=1$. This property is
characteristic of data locking states. These are states whose accessible
mutual information (i.e.~the maximum classical mutual information that can
be achieved by local measurements) drastically underestimates their
quantum mutual information (see \cite{DVHLST} for the original description
of this phenomenon, and Chapter \ref{chap:QIT}, Section \ref{sec:Shannon}, in this manuscript for a reminder of the definition of mutual information). Now, following \cite{DFHL} and \cite{FHS}, data locking
may also be defined in terms of distinguishability from the maximally
mixed state by local measurements: informally, a state $\rho$ on
${\mathbf{C}}^d\otimes{\mathbf{C}}^d$ which is such that $\| \rho -
{\mathrm{Id}}/d^2\|_{\mathbf{LO}} \ll \|\rho-
{\mathrm{Id}}/d^2\|_{\mathbf{LOCC}^{\rightarrow}}$ may be used
for information locking.
\end{remark}

\subsection{Generalization: unbounded gaps $\mathbf{LO}/\mathbf{LOCC^{\rightarrow}}$ and $\mathbf{LOCC^{\rightarrow}}/\mathbf{LOCC}$}

For each $r\in\N$, we define $\mathbf{LOCC^{(r)}}$ as the class of local POVMs with $r$ rounds of classical communication between the parties. In particular, making the link with the notation that were previously introduced, we have $\mathbf{LOCC^{\rightarrow}}=\mathbf{LOCC^{(1)}}$ while $\mathbf{LOCC}=\lim_{r\rightarrow+\infty}\mathbf{LOCC^{(r)}}$.
As before, we let $E$ be a $d/2$-dimensional subspace of $\C^d$, and denote by $P$ the orthogonal projector onto $E$, $P^{\perp}$ the orthogonal projector onto $E^{\perp}$. Then, define the two orthogonal states $\tau_+,\tau_-$ on $\C^d$ by
\[ \tau_+=\frac{P}{d/2}\ \ \text{and}\ \ \tau_-=\frac{P^{\perp}}{d/2}. \]

Given $m\in\N$, let $\{\ket{1},\ldots,\ket{m}\}$ be an orthonormal basis of $\C^m$, and $U_1,\ldots,U_m$ be unitaries on $\C^d$. Then, define the two bipartite states $\rho_+^{(1)},\rho_-^{(1)}$ on $\mathrm{A}\otimes\mathrm{B}$, with $\mathrm{A}\equiv\C^m$, $\mathrm{B}\equiv\C^d$, as
\begin{equation} \label{eq:rho1}
\left[\rho_{+/-}^{(1)}\right]_{\A\B}=\frac{1}{m}\sum_{i=1}^{m}\ketbra{i}{i}_\A\otimes \left[U_i\tau_{+/-} U_i^{\dagger}\right]_\B.
\end{equation}

Given $n\in\N$, let $\{\ket{1},\ldots,\ket{n}\}$ be an orthonormal basis of $\C^n$, and $V_1,\ldots,V_n$ be unitaries on $\C^m$. Then, define the two bipartite states $\rho_+^{(2)},\rho_-^{(2)}$ on $\mathrm{A}\otimes\mathrm{B}$, with $\mathrm{A}\equiv\C^n\otimes\C^d$, $\mathrm{B}\equiv\C^m$, as
\begin{equation} \label{eq:rho2}
\left[\rho_{+/-}^{(2)}\right]_{\A\B}=\frac{1}{mn}\sum_{i=1}^{m}\sum_{j=1}^{n}\left[\ketbra{j}{j}\otimes U_i\tau_{+/-} U_i^{\dagger}\right]_\A\otimes \left[V_j\ketbra{i}{i}V_j^{\dagger}\right]_\B.
\end{equation}

\begin{theorem} \label{th:LO-LOCC1}
Let $0\leq\delta<1/2$. Consider the states $\rho_+^{(1)}$ and $\rho_-^{(1)}$ as defined by equation \eqref{eq:rho1} and the Hermitian $\Delta^{(1)}=\rho_+^{(1)}-\rho_-^{(1)}$ on $\mathrm{A}\otimes\mathrm{B}\equiv\C^m\otimes\C^d$. For $m\simeq d^{1-\delta}$, there exist $U_1,\ldots,U_m$ such that
\[ \|\Delta^{(1)}\|_{\mathbf{LOCC^{(1)}}}=2\ \ \text{and}\ \ \|\Delta^{(1)}\|_{\mathbf{LO}}\lesssim\frac{1}{d^{1/2-\delta}}. \]
\end{theorem}

In words: if the hypotheses of Theorem \ref{th:LO-LOCC1} are satisfied, the states $\rho_+^{(1)}$ and $\rho_-^{(1)}$ are perfectly distinguishable by local POVMs and one round of classical communication, but very poorly by local POVMs without any classical communication.

\begin{proof}
The $\mathbf{LOCC^{(1)}}$ POVM which enables perfect discrimination of $\rho_+^{(1)}$ and $\rho_-^{(1)}$ is
\[ \left(\ketbra{i}{i}_\A \otimes \left[U_iPU_i^{\dagger}\right]_\B , \ketbra{i}{i}_\A \otimes \left[U_iP^{\perp}U_i^{\dagger}\right]_\B\right)_{1\leq i\leq m}. \]
So the first equality is clear.

The second inequality is just a slight generalization of Theorem \ref{theorem:LO-vs-LOCC}. Indeed, it was just observed in its proof that, setting $\Delta_i=U_i\tau_+U_i^{\dagger}-U_i\tau_-U_i^{\dagger}$ for each $1\leq i\leq m$, we have
\[ \|\Delta^{(1)}\|_{\mathbf{LO}}= \sup\left\{\frac{1}{m}\sum_{i=1}^{m}\|\Delta_i\|_{\mathrm{N}} \st \mathrm{N}\ \text{POVM on}\ \C^d\right\}. \]
It was then established that, for $m=Cd^{1-\delta}$ and $U_1,\ldots,U_m$ independent Haar distributed unitaries on $\C^d$, the latter quantity is, with probability greater than $1/2$, smaller than $C'/d^{1/2-\delta}$ (where $C,C'>0$ are universal constants).
\end{proof}

\begin{theorem} \label{th:LOCC1-LOCC2}
Let $0\leq\delta<1/2$. Consider the states $\rho_+^{(2)}$ and $\rho_-^{(2)}$ as defined by equation \eqref{eq:rho2} and the Hermitian $\Delta^{(2)}=\rho_+^{(2)}-\rho_-^{(2)}$ on $\mathrm{A}\otimes\mathrm{B}\equiv\left(\C^n\otimes\C^d\right)\otimes\C^m$. For $n\simeq m\simeq d^{1-\delta}$, there exist $U_1,\ldots,U_m$ and $V_1,\ldots,V_n$ such that
\[ \|\Delta^{(2)}\|_{\mathbf{LOCC^{(2)}}}=2\ \ \text{and}\ \ \|\Delta^{(2)}\|_{\mathbf{LOCC^{(1)}}}\lesssim\frac{1}{d^{1/2-\delta}}. \]
\end{theorem}

In words: if the hypotheses of Theorem \ref{th:LOCC1-LOCC2} are satisfied, the states $\rho_+^{(2)}$ and $\rho_-^{(2)}$ are perfectly distinguishable by local POVMs and two rounds of classical communication, but very poorly by local POVMs and one round of classical communication only.

\begin{proof}
The $\mathbf{LOCC^{(2)}}$ POVM which enables perfect discrimination of $\rho_+^{(2)}$ and $\rho_-^{(2)}$ is
\[ \left( \ketbra{j}{j}_{\A_1}\otimes \left[U_iPU_i^{\dagger}\right]_{\A_2}\otimes\left[V_j\ketbra{i}{i}V_j^{\dagger}\right]_\B , \ketbra{j}{j}_{\A_1}\otimes \left[U_iP^{\perp}U_i^{\dagger}\right]_{\A_2}\otimes\left[V_j\ketbra{i}{i}V_j^{\dagger}\right]_\B \right)_{1\leq i\leq m,\ 1\leq j\leq n}. \]
So the first equality is clear.

Let us turn to showing the second inequality. Setting $\Delta_i=U_i\tau_+U_i^{\dagger}-U_i\tau_-U_i^{\dagger}$ for each $1\leq i\leq m$, we have by definition
\begin{align*} \|\Delta^{(2)}\|_{\mathbf{LOCC^{(1)}}} =\, \sup & \, \sum_{\alpha}\frac{1}{mn} \sum_{i=1}^m\sum_{j=1}^n\|\Delta_i\otimes\ketbra{j}{j}\|_{\mathrm{N}_{\alpha}} \tr\left(V_j\ketbra{i}{i}V_j^{\dagger}M_{\alpha}\right)\\
& \mathrm{s.t.}\ \mathrm{M}=(M_{\alpha})_{\alpha}\ \text{POVM on}\ \C^m\ \text{and}\ \mathrm{N}_{\alpha}\ \text{POVMs on}\ \C^d\otimes\C^n.
\end{align*}
Yet, if $\mathrm{N}=(N_{\beta})_{\beta}$ is a POVM on $\C^d\otimes\C^n$, then for each $1\leq j\leq n$, $\mathrm{N}(j)=\left(\tr_n\left[\Id_d\otimes\ketbra{j}{j}_n N_{\beta}\right]\right)_{\beta}$ is a POVM on $\C^d$ which is such that, for any Hermitian $\Delta$ on $\C^d$, $\|\Delta\otimes\ketbra{j}{j}\|_{\mathrm{N}}=\|\Delta\|_{\mathrm{N}(j)}$. Therefore,
\begin{align*}
\|\Delta^{(2)}\|_{\mathbf{LOCC^{(1)}}} \leq\, \sup & \, \sum_{\alpha}\frac{1}{m}\sum_{i=1}^{m} \left(\frac{1}{n}\sum_{j=1}^{n}\tr\left(V_j\ketbra{i}{i}V_j^{\dagger}M_{\alpha}\right)\right) \|\Delta_i\|_{\mathrm{N}_{\alpha}}\\
& \mathrm{s.t.}\ \mathrm{M}=(M_{\alpha})_{\alpha}\ \text{POVM on}\ \C^m\ \text{and}\ \mathrm{N}_{\alpha}\ \text{POVMs on}\ \C^d.
\end{align*}
Now choose unitaries $V_1,\ldots,V_n$ on $\C^m$ which are $1$-randomizing, i.e.~such that, for any unit vector $x\in\C^m$,
\begin{equation} \label{eq:randomizing}
\left\| \frac{1}{n}\sum_{j=1}^{n}V_j\ketbra{x}{x}V_j^{\dagger} -\frac{\Id}{m}\right\|_{\infty} \leq \frac{1}{m}.
\end{equation}
This is achieved with probability greater than $1/2$ by independent Haar distributed unitaries on $\C^m$ if $n=Cm$, where $C>0$ is a universal constant (see \cite{HLSW}). In that case, for any POVM $\mathrm{M}=(M_{\alpha})_{\alpha}$ on $\C^m$, we have that for each $\alpha$ and $1\leq i\leq m$,
\[ \frac{1}{n}\sum_{j=1}^{n}\tr\left(V_j\ketbra{i}{i}V_j^{\dagger}M_{\alpha}\right) \leq \left\| \frac{1}{n}\sum_{j=1}^{n}V_j\ketbra{x}{x}V_j^{\dagger}\right\|_{\infty}\left\|M_{\alpha}\right\|_1 \leq \frac{2}{m}\tr M_{\alpha}, \]
the last inequality being by assumption \eqref{eq:randomizing}, and because $M_{\alpha}\geq 0$. Since additionally $\sum_{\alpha}M_{\alpha}=\Id$, so that $\sum_{\alpha}\tr M_{\alpha}/m=1$, we finally get
\begin{align*} \|\Delta^{(2)}\|_{\mathbf{LOCC^{(1)}}} \leq & \sup\left\{\frac{2}{m}\sum_{i=1}^{m} \sum_{\alpha}\lambda_{\alpha}\|\Delta_i\|_{\mathrm{N}_{\alpha}} \st \lambda=(\lambda_{\alpha})_{\alpha}\ \text{p.d.},\ \mathrm{N}_{\alpha}\ \text{POVMs on}\ \C^d\right\}\\
= & \sup\left\{\frac{2}{m}\sum_{i=1}^{m} \|\Delta_i\|_{\mathrm{N}} \st \mathrm{N}\ \text{POVM on}\ \C^d\right\}.
\end{align*}
As explained in the proof of Theorem \ref{th:LO-LOCC1}, we know that there exist universal constants $C,C'>0$ such that, for $m=Cd^{1-\delta}$ and $U_1,\ldots,U_m$ independent Haar distributed unitaries on $\C^d$, the latter quantity is, with probability greater than $1/2$, smaller than $C'/d^{1/2-\delta}$.
\end{proof}

\section{Generic unbounded gap between $\mathbf{SEP}$ and $\mathbf{PPT}$}
\label{sec:local-POVM-proofs}

\subsection{Volume and mean width estimates}

The first step towards Theorem \ref{theorem:typical-states} is to estimate globally the size of the (dual) unit balls $K_{\mathbf{PPT}}$, $K_{\mathbf{SEP}}$ and $K_{\mathbf{LOCC^{\rightarrow}}}$ associated to the
measurement norms $\|\cdot\|_{\mathbf{PPT}}$, $\|\cdot\|_{\mathbf{SEP}}$ and $\|\cdot\|_{\mathbf{LOCC^{\rightarrow}}}$. Classical invariants which are generally useful to quantify the size of convex
bodies include the volume radius and
the mean width, which are defined in Chapter \ref{chap:toolbox}, Section \ref{ap:convex-geometry}.

Recall that whenever we use tools from convex geometry in the space $\mathcal{H}(\C^d \otimes \C^d)$ (which has real dimension $d^4$) it is tacitly
understood that we use the Euclidean structure
induced by the
Hilbert--Schmidt inner product.
The definitions of the volume radius and the mean width of $K_{\mathbf{M}}$ thus become
\[ \vrad (K_{\mathbf{M}}) = \left( \frac{\vol K_{\mathbf{M}}}{\vol B_{HS}(\C^d\otimes\C^d)} \right)^{1/d^4}\ \text{and}\ w (K_{\mathbf{M}}) = \int_{S_{HS}(\C^d\otimes\C^d)} \| \Delta \|_{\mathbf{M}} \, \mathrm{d} \sigma (\Delta), \]
where $B_{HS}(\C^d\otimes\C^d)$ denotes the Hilbert--Schmidt unit ball of $\mathcal{H}(\C^d \otimes \C^d)$ and $S_{HS}(\C^d\otimes\C^d)$ its Hilbert--Schmidt unit sphere equipped
with the uniform measure $\sigma$.
Here are the estimates on the volume radius and the mean width of $K_{\mathbf{PPT}}$, $K_{\mathbf{SEP}}$ and $K_{\mathbf{LOCC^{\rightarrow}}}$. As a reference, recall that
(on $\C^d \otimes \C^d$)
\[ \vrad(K_{\mathbf{ALL}}) \simeq w(K_{\mathbf{ALL}}) \simeq d .\]
This follows from Theorem \ref{theorem:operator-norm} in Appendix \ref{ap:standard} once we have in mind that $K_{\mathbf{ALL}} = [-\Id,\Id]$.

\begin{theorem}
\label{theorem:vrad-w-PPT-SEP}
In $\C^d \otimes \C^d$, one has
\[ \mathrm{vrad}\left(K_{\mathbf{PPT}}\right)\simeq w\left(K_{\mathbf{PPT}}\right)\simeq d\ \]
and
\[ \mathrm{vrad}\left(K_{\mathbf{LOCC^{\rightarrow}}}\right) \simeq w \left(K_{\mathbf{LOCC^{\rightarrow}}}\right)\simeq
\mathrm{vrad}\left(K_{\mathbf{LOCC}}\right) \simeq w \left(K_{\mathbf{LOCC}}\right)\simeq
\mathrm{vrad}\left(K_{\mathbf{SEP}}\right) \simeq w \left(K_{\mathbf{SEP}}\right)\simeq \sqrt{d}. \]
\end{theorem}

To prove these results, we will make essential use of the Urysohn inequality (Theorem \ref{theorem:urysohn} in Chapter \ref{chap:toolbox}, Section \ref{ap:convex-geometry}): for any convex body
$K \subset \R^n$, we have $\mathrm{vrad}(K) \leq w(K)$. In particular, Theorem \ref{theorem:vrad-w-PPT-SEP} follows from the following four
inequalities: (a) $w(K_{\mathbf{PPT}})\lesssim d$, (b) $\vrad(K_{\mathbf{PPT}}) \gtrsim d$, (c) $w(K_{\mathbf{SEP}})\lesssim \sqrt{d}$, and
(d) $\vrad(K_{\mathbf{LOCC^{\rightarrow}}}) \gtrsim \sqrt{d}$.

\subsection*{(a) Proof that $w(K_{\mathbf{PPT}})\lesssim d$} This follows from the inclusion $K_{\mathbf{PPT}} \subset [-\Id,\Id]$, together with the estimate on the mean width of $[-\Id,\Id]$
 from Theorem \ref{theorem:operator-norm} in Appendix \ref{ap:standard}.

\subsection*{(b) Proof that $\vrad(K_{\mathbf{PPT}})\gtrsim d$} We start by noticing that
\[K_{\mathbf{PPT}}=[-\Id,\Id]\cap [-\Id,\Id]^{\Gamma}.\]
We apply the Milman--Pajor inequality (Corollary \ref{corollary:Milman-Pajor} in Chapter \ref{chap:toolbox}, Section \ref{ap:convex-geometry}) to the convex body $[-\Id,\Id]$
(which indeed has the origin as center of mass) and to the orthogonal transformation $\Gamma$ (the partial transposition). This yields
\[\mathrm{vrad}\left(K_{\mathbf{PPT}}\right)\geq \frac{1}{2}\frac{\mathrm{vrad}\left([-\Id,\Id]\right)^2}{w\left([-\Id,\Id]\right)}\simeq d,\]
where we used the estimates on the volume radius and the mean width of $[-\Id,\Id]$ from Theorem \ref{theorem:operator-norm} in Appendix \ref{ap:standard}.

\subsection*{(c) Proof that $w(K_{\mathbf{SEP}})\lesssim \sqrt{d}$} We are going to relate $K_{\mathbf{SEP}}$ with the set $\mathcal{S}$ of
separable states on $\C^d \otimes \C^d$. In fact, denoting the cone with base $\mathcal{S}$ by
\[ \R^+\mathcal{S} := \{ \lambda \rho \st \lambda \in \R^+,\ \rho \in \mathcal{S} \}, \]
we have $K_{\mathbf{SEP}} = L \cap (-L)$, where
\[ L := 2\left(\R^+\mathcal{S}\cap[0,\Id]\right)-\Id .\]
This gives immediately an upper bound on the mean width of $K_{\mathbf{SEP}}$
\[ w(K_{\mathbf{SEP}}) \leq w(L) \leq 2 w(\R^+\mathcal{S}\cap[0,\Id]) \leq 2 w( \{ \lambda \rho \st \lambda \in [0,d^2],\ \rho \in \mathcal{S} \}  )
=2d^2 w(\conv( \{0\}, \mathcal{S})).\]
Now, if $K,K'$ are two convex sets such that $K \cap K' \neq \emptyset$, then $w(\conv(K,K')) \leq w(K) + w(K')$. So, denoting  by
$\alpha_n$ the mean width of a segment $[-x,x]$ for $x$ a unit vector in $\R^n$, we have
\[ w(\conv(\{0\},\mathcal{S})) \leq w(\conv \{0,\Id/d^2 \}) + w (\mathcal{S}) \lesssim \frac{\alpha_{d^4}}{d} + \frac{1}{d^{3/2}} \lesssim
\frac{1}{d^{3/2}}, \]
where we used the estimate $w(\mathcal{S}) \simeq d^{-3/2}$ from Theorem \ref{theorem:separable-states} in Appendix \ref{ap:standard}, and the fact that $\alpha_n\simeq n^{-1/2}$
(see Chapter \ref{chap:toolbox}, Section \ref{ap:convex-geometry}).

\subsection*{(d) Proof that $\vrad(K_{\mathbf{LOCC^{\rightarrow}}})\gtrsim \sqrt{d}$} We consider the following set of states on $\C^d \otimes \C^d$
\[ T = \conv \left\{ \ketbra{\psi}{\psi} \otimes \sigma \st \psi \in S_{\C^d}, \ \sigma
\text{ a state on }\C^d \text{ such that } \|\sigma\|_{\iy} \leq 3/d \right\}. \]
A connection between $T$ and $\mathbf{LOCC^{\rightarrow}}$ is given by the following lemma.

\begin{lemma} \label{lemma:LOCC}
Let $\rho$ and $\rho'$ be operators in $T$ such that $\rho+\rho'=2 \Id/d^2$. Then, the operators
$d^2\rho/6 $ and $d^2\rho'/6$ belong to $K_{\mathbf{LOCC^{\rightarrow}}}$.
\end{lemma}

\begin{proof}
There exist convex combinations $(\alpha_i)_{i \in I}, (\alpha'_j)_{j \in J}$, unit vectors $(\psi_i)_{i \in I}, (\psi'_j)_{j \in J}$
and states $(\sigma_i)_{i \in I},(\sigma'_j)_{j \in J}$ satisfying $\|\sigma_i\|_{\iy} \leq 3/d$, $\|\sigma'_j\|_{\iy} \leq 3/d$, such that
\[ \rho = \sum_{i \in I} \alpha_i \ketbra{\psi_i}{\psi_i} \otimes \sigma_i \ \ \textnormal{ and } \ \ \rho' = \sum_{j \in J} \alpha'_j
\ketbra{\psi'_j}{\psi'_j} \otimes \sigma'_j .\]
Define states $(\tau_i)_{i \in I}$ and $(\tau'_j)_{j \in J}$ by the relations $\sigma_i + 2\tau_i = \sigma'_j + 2\tau'_j= 3\Id/d$. It can then be checked that the following POVM is in
$\mathbf{LOCC^{\rightarrow}}$
\[ \mathrm{M} = \left( \frac{d^2}{6} \alpha_i \ketbra{\psi_i}{\psi_i} \otimes \sigma_i,\, \frac{d^2}{6} \alpha_i \ketbra{\psi_i}{\psi_i} \otimes 2\tau_i,\, \frac{d^2}{6} \alpha'_j \ketbra{\psi'_j}{\psi'_j} \otimes \sigma'_j,\, \frac{d^2}{6} \alpha'_j \ketbra{\psi'_j}{\psi'_j} \otimes 2\tau'_j \right)_{i\in I,\, j\in J}. \]
Hence, the operators $d^2\rho/6$ and $d^2\rho'/6$ belong to $K_{\mathrm{M}}$ and therefore to $K_{\mathbf{LOCC^{\rightarrow}}}$.
\end{proof}

Let $\widetilde{T}$ be the symmetrization of $T$ defined as $\widetilde{T} = T \cap \left\{ 2 \Id/d^2 - T \right\}$.
By Lemma \ref{lemma:LOCC} and the fact that $K_{\mathbf{LOCC}^\rightarrow}$ is centrally symmetric, we have
\[ \frac{d^2}{6} \conv ( \widetilde{T}, -\widetilde{T} ) \subset K_{\mathbf{LOCC}^\rightarrow} .\]

We are going to give a lower
bound on the volume radius of $\widetilde{T}$.
The center of mass of the set $T$ equals the maximally mixed state $\Id/d^2$ (indeed, the center of mass commutes with local unitaries). By the Milman--Pajor inequality (Corollary \ref{corollary:Milman-Pajor} in Chapter \ref{chap:toolbox}, Section \ref{ap:convex-geometry}) this implies that $\vrad( \widetilde{T} ) \geq \vrad(T)/2$. On the other hand, one has (see definitions in Appendix \ref{ap:standard})
\begin{equation} \label{eq:convT} \conv(T, - T) \supset \frac{1}{d} \cdot B_1(\C^d) \hat{\otimes} B_{\iy}(\C^d) .\end{equation}
Let us check \eqref{eq:convT}. An extreme point of $\frac{1}{d} \cdot B_1(\C^d) \hat{\otimes} B_{\iy}(\C^d)$ has the form $\pm \ketbra{\psi}{\psi} \otimes A$
for $\psi \in S_{\C^d}$ and $A \in \cH(\C^d)$ such that $\|A\|_{\iy} \leq 1/d$. Let $\e = 2 - \|A\|_{1} \geq 1$ and let $A^+,A^-$ be the positive and
negative parts of $A$. Set $\lambda^{\pm} = \e/4 + \tr A^\pm/2$ (so that $\lambda^++\lambda^-=1$), and consider the states
$\rho^\pm = ( \e/4 \cdot \Id/d + A^\pm/2)/\lambda^\pm$. We have
\[ \| \rho^{\pm} \|_{\iy} \leq \frac{{\e}/{4d}+ {1}/{2d}}{{\e}/{4}} \leq \frac{3}{d}\]
and therefore $\rho^\pm \in T$. Since $A = \lambda^+ \rho^+ - \lambda^- \rho^-$, this shows \eqref{eq:convT}. Using Theorem
\ref{theorem:volume-S1-Sinfini} in Appendix \ref{ap:standard}, it follows that
\[ \vrad( \conv(T,-T) ) \gtrsim d^{-3/2} .\]
And therefore,
\[ \vrad( \conv ( \widetilde{T}, -\widetilde{T} ) ) \gtrsim \vrad ( \widetilde{T}) \gtrsim \vrad(T) \gtrsim \vrad( \conv(T,-T) ) \gtrsim d^{-3/2} , \]
the first and third inequalities being due to the Rogers--Shephard inequality (Theorem \ref{theorem:rogers-shephard} in Chapter \ref{chap:toolbox}, Section \ref{ap:convex-geometry}).
We eventually get
\[ \vrad( K_{\mathbf{LOCC^{\rightarrow}}} ) \gtrsim \sqrt{d} .\]

\subsection{Discriminating between two generic states}

Let $\mathbf{M}$ be a family of POVMs on $\C^d$ (possibly reduced to a single POVM). We relate the mean width $w(K_{\mathbf{M}})$ to the typical
performance of $\mathbf{M}$ for
discriminating two random states, chosen independently and uniformly from the set $\mathcal{D}(\C^d)$ of all states on $\C^d$.

\begin{proposition} \label{proposition:Delta-vs-difference-of-states}
Let $\mathbf{M}$ be a family of POVMs on $\C^d$, and denote $\omega := w(P_{H_0} K_{\mathbf{M}})$, where $P_{H_0}$ stands for the orthogonal
projection onto
the hyperplane $H_0 \subset \cH(\C^d)$ of trace $0$ Hermitian operators on $\C^d$.
Let $\rho$ and $\sigma$ be two random states, chosen independently with respect to the uniform measure on $\mathcal{D}(\C^d)$.
Then,
\begin{equation}
\label{eq:expectation}
\E := \E \| \rho - \sigma \|_{\mathbf{M}} \simeq \frac{\omega}{\sqrt{d}} .
\end{equation}
Moreover, we have the concentration estimate
\begin{equation}
\label{eq:concentration}
\forall\ t>0,\ \P \left( \left| \| \rho - \sigma \|_{\mathbf{M}} - \E \right| > t \right) \leq 2 \exp (-cd^2t^2), \end{equation}
$c>0$ being a universal constant.
\end{proposition}

We first deduce Theorem \ref{theorem:typical-states} from Theorem \ref{theorem:vrad-w-PPT-SEP} and
Proposition \ref{proposition:Delta-vs-difference-of-states} (we warn the reader that we apply the latter on the space $\C^d \otimes \C^d$,
and therefore the ambient dimension is $d^2$ instead of $d$).

\begin{proof}[Proof of Theorem \ref{theorem:typical-states}]
Let $\mathbf{M} \in \{ \mathbf{LOCC}, \mathbf{LOCC^{\rightarrow}},\mathbf{SEP},\mathbf{PPT} \}$.
While we computed $w(K_{\mathbf{M}})$ in Theorem \ref{theorem:vrad-w-PPT-SEP}, the relevant quantity here
is $w(P_{H_0} K_{\mathbf{M}})$. We show that both are comparable. We first have the upper bound (see \eqref{eq:width-projection} from Chapter \ref{chap:toolbox}, Section \ref{ap:convex-geometry})
\[ w(P_{H_0}K_{\mathbf{M}}) \lesssim w(K_{\mathbf{M}}). \]
To get the reverse bound, we consider the volume radius rather than the mean width. If we denote more generally by $H_t$ the hyperplane of
trace $t$ operators on $\C^d$, we have by
Fubini's theorem
\[ \vol_{d^4}(K_{\mathbf{M}}) = \frac{1}{d} \int_{-d^2}^{d^2} \vol_{d^4-1}( K_{\mathbf{M}} \cap H_t) \,\mathrm{d}t .\]
By the Brunn--Minkowski inequality, the function under the integral is maximal when $t=0$, and therefore
\[ \vol_{d^4}(K_{\mathbf{M}}) \leq 2d \vol_{d^4-1}(K_{\mathbf{M}} \cap H_0). \]
It follows easily that $w(P_{H_0}K_{\mathbf{M}})\geq\vrad(P_{H_0} K_{\mathbf{M}}) \geq \vrad ( K_{\mathbf{M}} \cap H_0 ) \gtrsim
\vrad(K_{\mathbf{M}}) \simeq w(K_{\mathbf{M}})$,
the first inequality being the Urysohn inequality (Theorem \ref{theorem:urysohn} in Chapter \ref{chap:toolbox}, Section \ref{ap:convex-geometry}) and the last estimate being by Theorem
\ref{theorem:vrad-w-PPT-SEP}.
Once this is known, Theorem \ref{theorem:typical-states}
is immediate from Proposition \ref{proposition:Delta-vs-difference-of-states}.
\end{proof}

\begin{proof}[Proof of Proposition \ref{proposition:Delta-vs-difference-of-states}]
We first show the concentration estimate \eqref{eq:concentration}, using the following representation due to \.Zyczkowski and
Sommers \cite{ZS}: $\rho$ has the same distribution as
$MM^\dagger$, where $M$ is uniformly distributed on the Hilbert--Schmidt unit sphere of $d \times d$ complex matrices.
We estimate the Lipschitz
constant of the function $(M,N) \mapsto \|MM^\dagger-NN^\dagger\|_{\mathbf{M}}$, defined on the product of two such unit spheres, as follows:
\begin{align*}
\| M_1M_1^\dagger-N_1N_1^\dagger\|_{\mathbf{M}} - \| M_2M_2^\dagger-N_2N_2^\dagger\|_{\mathbf{M}} \leq & \, \| M_1M_1^\dagger-M_2M_2^\dagger\|_{\mathbf{M}} +
\| N_1N_1^\dagger-N_2N_2^\dagger\|_{\mathbf{M}} \\
\leq & \, \| M_1M_1^\dagger-M_2M_2^\dagger\|_{1} + \| N_1N_1^\dagger-N_2N_2^\dagger\|_{1} \\
\leq & \, 2\left(\|M_1-M_2\|_{2} + \|N_1-N_2\|_{2} \right).
\end{align*}
The second inequality is simply because $\|\cdot\|_{\mathbf{M}}\leq\|\cdot\|_1$, while the third inequality follows from Cauchy--Schwarz inequality (and the fact that $M_1,M_2,N_1,N_2$ have unit Hilbert--Schmidt norm), after noticing that $||AA^\dagger-BB^\dagger||_{1} \leq
||(A-B)B^\dagger||_{1}+||A(A-B)^\dagger||_{1}$. We obtain as a consequence of Lemma \ref{lemma:levy-two-spheres} below, a variation on L\'evy's lemma (recalled as Lemma \ref{lemma:levy} in Chapter \ref{chap:toolbox}, Section \ref{ap:deviations}),
the desired estimate
\[ \P \left( \left| \|\rho-\sigma\|_{\mathbf{M}} -\E \right| > t \right) \leq 2 \exp(-cd^2t^2) .\]
In our application of Lemma \ref{lemma:levy-two-spheres}, we identify the set of complex $d \times d$ matrices with $\R^n$ for $n=2d^2$, and use $L=2$.

\begin{lemma}
\label{lemma:levy-two-spheres}
Let $S^{n-1}$ be the Euclidean unit sphere in $\R^n$, and equip $S^{n-1} \times S^{n-1}$ with the measure $\mu \otimes \mu$,
where $\mu$ is the
uniform probability measure on $S^{n-1}$, and with the metric $d((x,y),(x',y')):=\|x-x'\|+\|y-y'\|$. For any $L$-Lipschitz function $f : S^{n-1} \times S^{n-1} \to \R$ and any $t>0$,
\[ \P( | f - \E f| > t ) \leq 2 \exp(-cn t^2/L^2), \]
$c>0$ being a universal constant.
\end{lemma}

Lemma \ref{lemma:levy-two-spheres} can be deduced quickly from the usual L\'evy lemma (see Lemma \ref{lemma:levy} in Chapter \ref{chap:toolbox}, Section \ref{ap:deviations}) which quantifies the phenomenon of
concentration of measure on the sphere.
If we define, for any fixed $x\in S^{n-1}$, $E_x := \int_{S^{n-1}} f(x,y) \,\mathrm{d}\mu(y)$, we may apply L\'evy's lemma
to show that the function $y\in S^{n-1} \mapsto f(x,y)$ concentrates around its expectation $E_x$, and again L\'evy's lemma to show that the function
$x\in S^{n-1} \mapsto E_x$ (which is $L$-Lipschitz,
as an average of $L$-Lipschitz functions) is also well-concentrated.

\medskip

We now prove the first part of Proposition \ref{proposition:Delta-vs-difference-of-states}.
Let $\Delta$ be a random matrix uniformly chosen from the Hilbert--Schmidt sphere in the hyperplane $H_0$, and $\rho,\sigma$ be independent
random states with uniform distribution.
We claim that, from a very rough perspective,
the spectra of $\rho-\sigma$ and $\Delta/\sqrt{d}$ look similar. More precisely, we have the following lemma.

\begin{lemma} \label{lemma:global-spectra}
Let $\rho,\sigma$ be independent random states uniformly chosen from $\mathcal{D}(\C^d)$, and $\Delta$ be a random
matrix uniformly chosen from the Hilbert--Schmidt sphere in the hyperplane $H_0$. Then with large probability, on the one hand
\[ \|\Delta\|_1 \simeq \sqrt{d},\ \|\Delta\|_{2} = 1\ \text{and}\ \|\Delta\|_{\infty} \simeq 1/\sqrt{d}, \]
while on the other hand
\[ \| \rho - \sigma \|_1 \simeq 1,\ \|\rho-\sigma\|_{2} \simeq 1/\sqrt{d}\ \text{and}\ \|\rho-\sigma\|_{\infty} \simeq 1/d. \]
Moreover these statements hold in expectation: for example $\E \| \Delta \|_{\infty}\simeq 1/\sqrt{d}$ and $\E \|\rho-\sigma\|_{\infty} \simeq 1/d$.
\end{lemma}

In order to compare $\rho-\sigma$ with $\Delta$, we rely on the following lemma. For $x=(x_1,\dots,x_n) \in \R^n$, we denote
$\|x\|_{\iy} = \max \{ |x_i| \st 1 \leq i \leq n \}$ and $\|x\|_1= \sum_{i=1}^n |x_i|$.

\begin{lemma} \label{lemma:comparison-of-norms} Let $E = \{ x \in \R^n \st \sum_{i=1}^n x_i =0 \}$ and let
$|||\cdot|||$ be a norm on $E$ which is invariant under permutation of
coordinates.
Then, for any nonzero vectors $x,y \in E$, we have
\begin{equation}
\label{eq:equiv}
 |||x|||  \leq  2n\frac{\|x\|_{\infty}}{\|y\|_1} |||y|||.
\end{equation}
\end{lemma}

Assuming both lemmas, we now complete the proof of Proposition \ref{proposition:Delta-vs-difference-of-states}. On the hyperplane $E \subset \R^d$
of vectors whose sum of coordinates is zero,
we define a norm by
\[ |||x||| := \int_{\mathcal{U}(d)} \| U \diag(x) U^\dagger \|_{\mathbf{M}} \, \mathrm{d}U , \]
where the integral is taken with respect to the Haar measure on the unitary group, and $\diag(x)$ denotes the diagonal matrix on $\C^d$ with diagonal
elements equal to
the coordinates of $x$. Note that $|||\cdot|||$ is obviously invariant under permutation of coordinates. Also, $\Delta$ has the same distribution as
$U \diag\left(\spec(\Delta)\right) U^\dagger$, where $U$ is a Haar-distributed unitary
matrix independent from $\Delta$ and $\spec(A) \in \R^d$ denotes the spectrum of $A \in \cH(\C^d)$ (the ordering of eigenvalues being irrelevant).
The same holds for $\rho-\sigma$ instead of $\Delta$, and it follows that
\[ \E |||\spec(\Delta)||| = \E ||\Delta||_{\mathbf{M}}  \ \ \ \textnormal{and} \ \ \ \E |||\spec(\rho-\sigma)||| = \E ||\rho - \sigma||_{\mathbf{M}}. \]

Let us show that
\begin{equation} \label{eq:half} \E ||\rho-\sigma||_{\mathbf{M}} \simeq \E \frac{1}{\sqrt{d}} ||\Delta||_{\mathbf{M}}. \end{equation}

We first prove the inequality $\lesssim$. Say that a vector $y \in E$ satisfies the condition $(\star)$ if
$\|y\|_{1} \geq c\sqrt{d}$, where we may choose the constant $c$ such that the random vector $\spec(\Delta)$ satisfies the condition $(\star)$
with probability larger than $1/2$ (this is possible, as we check using Lemma \ref{lemma:global-spectra}).
Now, by Lemma \ref{lemma:comparison-of-norms}, for any $y \in E$ satisfying
condition $(\star)$ and any $x \in E$, we have
\[ |||x||| \lesssim \sqrt{d} \|x\|_{\infty} \cdot |||y||| .\]
We apply this inequality with $x=\spec(\rho-\sigma)$ and take expectation.
This gives (using the statement about expectations in Lemma \ref{lemma:global-spectra})
\[ \E |||\rho-\sigma|||_{\mathbf{M}} \lesssim \frac{1}{\sqrt{d}} |||y||| .\]
This inequality is true for any $y \in E$ satisfying condition $(\star)$. Therefore,
\[ \E ||\Delta||_{\mathbf{M}} = \E |||\spec(\Delta)||| \gtrsim \sqrt{d} \cdot \P\big(\spec(\Delta)
\textnormal{ satisfies condition }(\star) \big) \E \|\rho-\sigma\|_{\mathbf{M}}
\simeq \sqrt{d} \E \|\rho-\sigma\|_{\mathbf{M}},\]
as needed. This proves one half of \eqref{eq:half}, and the reverse inequality is proved along the exact same lines.
Finally, we note that
\[ \E ||\Delta||_{\mathbf{M}} = w\left(P_{H_0} K_{\mathbf{M}} \right),\]
which, together with \eqref{eq:half}, shows \eqref{eq:expectation}, and concludes the proof.
\end{proof}

\begin{proof}[Proof of Lemma \ref{lemma:global-spectra}]
This is folklore in random matrix theory, in fact much more precise results are known
(for example, $\simeq$ can be replaced with $\sim$, with specific constants implicit
in that notation). However, most of the literature focusses on slightly different
random setups. Accordingly, we sketch an essentially self-contained elementary argument for completeness.

First of all, we observe that it is enough to prove the upper estimate for
$\|\cdot\|_{\infty}$ and the lower estimate for $\|\cdot\|_{2}$. Indeed, the
remaining upper estimates and the lower estimate for $\|\cdot\|_{\infty}$
follow then from the generally valid inequalities
$\|\cdot\|_{1} \leq \sqrt{d} \|\cdot\|_{2} \leq d \|\cdot\|_{\infty}$, while
the lower bound for $ \|\cdot\|_{1}$ follows from $\|\cdot\|_{2} \leq \|\cdot\|_{1}^{1/2} \|\cdot\|_{\infty}^{1/2}$.

The upper bound on $\|\cdot\|_{\infty}$ can be proved by a standard net argument. The lower bound on $\|\Delta\|_{2}$ is trivial,
while for $\|\rho -\sigma\|_{2}$ we may proceed as follows. First, using concentration of measure in the form of Lemma \ref{lemma:levy-two-spheres},
$\E \| \rho - \sigma \|_{2}$ is comparable to $\left( \E \| \rho - \sigma \|^2_{2} \right)^{1/2}$. Next, by Jensen inequality,
\[ \E \| \rho - \sigma \|^2_{2} \geq \E \| \rho - \Id/d \|^2_{2}. \]
Recalling that $\rho$ can be represented as $MM^\dagger$, with $M$ uniformly distributed on the Hilbert--Schmidt unit sphere of $d\times d$ complex matrices, the last
quantity can be expanded as
\[ \E \left\| \rho - \frac{\Id}{d} \right\|_{2}^2 = \E \tr |M|^4 - \frac{1}{d}\]
and it can be checked by moments expansion that $\E \tr |M|^4 \sim 2/d$.
\end{proof}

\begin{proof}[Proof of Lemma \ref{lemma:comparison-of-norms}]
Define $\alpha = 2n \|x\|_{\iy}/\|y\|_1$. By elementary properties of majorization (see e.g.~Chapter II in \cite{Bhatia})
it is enough to show that $x$ is majorized by $\alpha y$, i.e.~that for every $1 \leq k \leq n$,
\[ \sum_{i=1}^k x_i^\downarrow \leq \alpha \sum_{i=1}^k y_i^\downarrow ,\]
where $(x_i^\downarrow)_{1 \leq i \leq n}, (y_i^\downarrow)_{1 \leq i \leq n}$ denote the non-increasing rearrangement of $x,y$. This follows from
the inequalities
\begin{equation} \label{eq:majorization}
\frac{1}{\|x\|_{\iy}} \sum_{i=1}^k x_i^\downarrow \leq \min (k,n-k) \leq \frac{2n}{\|y\|_1} \sum_{i=1}^k y_i^\downarrow.
\end{equation}
The left-hand inequality in \eqref{eq:majorization} follows from the triangle inequality, once we have in mind that
\[ x_1^\downarrow + \cdots + x_k^\downarrow = -(x_{k+1}^\downarrow + \cdots + x_n^\downarrow). \]
To prove the right-hand inequality in \eqref{eq:majorization}, note that
the sum of positive coordinates of $y$ and the sum of negative coordinates of $y$ both equal $\|y\|_1/2$. Let $\ell$ be the number of positive coordinates
of $y$. If $k \leq \ell$, then
\[ y_1^\downarrow+\cdots+y_k^\downarrow \geq \frac{k}{\ell} \frac{\|y\|_1}{2} \geq \frac{k}{2n} \|y\|_1, \]
while if $k > \ell$, then
\[ y_1^\downarrow+\cdots+y_k^\downarrow = -(y_{k+1}^\downarrow + \cdots + y_n^\downarrow) \geq \frac{n-k}{n-\ell} \frac{\|y\|_1}{2} \geq \frac{n-k}{2n} \|y\|_1. \]
And the proof of Lemma \ref{lemma:comparison-of-norms} is thus complete.
\end{proof}

\section{Applications to quantum data hiding}
\label{sec:data-hiding}

\subsection{Bipartite data hiding}

As already mentioned, what Theorem \ref{theorem:typical-states} establishes is that generic bipartite states are data hiding for separable measurements but not for PPT measurements. This fact somehow counterbalances the usually cited constructions of data hiding schemes using Werner states (see e.g.~\cite{DVLT1,DVLT2,EW} and \cite{MWW,LW1}). Werner states are indeed data hiding in the exact same way for both separable and PPT measurements.
\smallskip

Besides, results in the same vein as those from Theorem \ref{theorem:typical-states} but more specifically orientated towards applications to quantum data hiding may be quite directly written down. In fact, one often thinks of data hiding states as being orthogonal states, hence perfectly distinguishable by the suitable global measurement, that are nevertheless barely distinguishable by any local measurement. The following theorem provides a statement in that direction.

\begin{theorem}
\label{theorem:data-hiding}
There are universal constants $C,c>0$ such that the following holds. Given a dimension $d$,
let $E$ be a $d^2/2$-dimensional subspace of $\C^d\otimes\C^d$ (we assume without loss of generality that $d$ is even). Let also
$\rho=UP_EU^\dagger/(d^2/2)$ and $\sigma=UP_{E^{\perp}}U^\dagger/(d^2/2)$, where $U$ is a Haar-distributed random unitary on $\C^d\otimes\C^d$. Then,
\[ \| \rho - \sigma \|_{\mathbf{ALL}}=2, \]
whereas with high probability,
\[ c \leq \| \rho - \sigma \|_{\mathbf{PPT}} \leq C\ \text{and}\ \frac{c}{\sqrt{d}} \leq \| \rho - \sigma \|_{\mathbf{SEP}} \leq \frac{C}{\sqrt{d}}. \]
Here, ``with high probability'' means that the probability that one of the conclusions fails is less than $\exp(-c_0 d^3)$ for some constant $c_0>0$.
\end{theorem}

\begin{proof}
The first part of Theorem \ref{theorem:data-hiding} is clear: the random states $\rho$ and $\sigma$ are orthogonal by construction, so that $\| \rho - \sigma \|_{\mathbf{ALL}}=\| \rho - \sigma \|_1=2$.

To prove the second part of Theorem \ref{theorem:data-hiding}, the only thing we have to show is that Proposition \ref{proposition:Delta-vs-difference-of-states} also holds for the random states $\rho$ and $\sigma$ considered here.

Now, for any family $\mathbf{M}$ of POVMs on $\C^d\otimes\C^d$, the function $f : U\in\mathcal{U}(\C^d\otimes\C^d)\mapsto \left\|U(P_E-P_{E^{\perp}})U^\dagger/(d^2/2)\right\|_{\mathbf{M}}$ is $8/d$-Lipschitz. Indeed, by the same arguments as in the proof of \eqref{eq:concentration},
\begin{align*}
f(U_1)-f(U_2) \leq & \, \frac{2}{d^2}\left(\|U_1P_EU_1^\dagger-U_2P_EU_2^\dagger\|_{\mathbf{M}} +\|U_1P_{E^{\perp}}U_1^\dagger-U_2P_{E^{\perp}}U_2^\dagger\|_{\mathbf{M}} \right)\\
\leq & \, \frac{2}{d} \left(\|U_1P_EU_1^\dagger-U_2P_EU_2^\dagger\|_2 +\|U_1P_{E^{\perp}}U_1^\dagger-U_2P_{E^{\perp}}U_2^\dagger\|_2 \right)\\
\leq & \, \frac{4}{d} \left(\|U_1P_E-U_2P_E\|_2 +\|U_1P_{E^{\perp}}-U_2P_{E^{\perp}}\|_2 \right)\\
\leq & \, \frac{8}{d} \|U_1-U_2\|_2.
\end{align*}
And any $L$-Lipschitz function $g : \mathcal{U}(\C^n)\rightarrow\R$ satisfies the concentration estimate (see the Appendix in \cite{MM})
\[ \forall\ t>0,\ \P( | g - \E g| > t ) \leq 2 \exp(-cn t^2/L^2),\ \ c\ \text{being a universal constant}. \]
The function $f$ thus satisfies $\P(|f-\E f|>t)\leq 2\exp(-cd^4t^2)$. So the concentration estimate \eqref{eq:concentration} in Proposition \ref{proposition:Delta-vs-difference-of-states} is in fact still true for the random states under consideration.

What is more, the results from Lemma \ref{lemma:global-spectra} remain valid too because we here even have the equalities
\[ \left\|\frac{2}{d^2}U(P_E-P_{E^{\perp}})U^\dagger\right\|_1=2,\ \ \left\|\frac{2}{d^2}U(P_E-P_{E^{\perp}})U^\dagger\right\|_2=\frac{2}{d}\ \ \text{and}\ \ \left\|\frac{2}{d^2}U(P_E-P_{E^{\perp}})U^\dagger\right\|_{\infty}=\frac{2}{d^2}. \]
So since the random Hermitians $U(P_E-P_{E^{\perp}})U^\dagger$ and $V\diag\left(\spec\left(U(P_E-P_{E^{\perp}})U^\dagger\right)\right)V^\dagger$ have the same distribution, for $U,V\in\mathcal{U}(\C^d\otimes\C^d)$ independent and Haar-distributed, one may apply Lemma \ref{lemma:comparison-of-norms} to conclude that the expectation estimate \eqref{eq:expectation} in Proposition \ref{proposition:Delta-vs-difference-of-states} is in fact still true too for the random states under consideration.
\end{proof}

In words, Theorem \ref{theorem:data-hiding} stipulates the following. Picking a subspace $E$ uniformly at random from the set of $d^2/2$-dimensional subspaces of $\C^d\otimes\C^d$, and then considering the states $\rho=P_E/(d^2/2)$ and $\sigma=P_{E^{\perp}}/(d^2/2)$, one gets examples of states which are perfectly distinguishable by some global measurement and which are with high probability data-hiding for separable measurements but not data-hiding for PPT measurements.

\begin{remark} Let us come back on the example of the symmetric state $\pi_s$ and the antisymmetric state $\pi_a$ on $\C^d\otimes\C^d$. They satisfy (see e.g.~\cite{DVLT2})
\begin{equation}
\label{eq:sym-anti}
\|\pi_s-\pi_a\|_{\mathbf{SEP}} = \|\pi_s-\pi_a\|_{\mathbf{PPT}} = \frac{4}{d+1} = \frac{2}{d+1}\|\pi_s-\pi_a\|_{\mathbf{ALL}}.
\end{equation}
They are consequently ``exceptional'' data hiding states for two reasons. First, as mentioned before, because they are equally PPT and SEP data hiding. And second because they are ``more'' data hiding than generic states: their SEP norm is of order $1/d\ll 1/\sqrt{d}$, hence almost reaching the known lower-bound valid for any states $\rho,\sigma$ on $\C^d\otimes\C^d$ (see e.g.~\cite{MWW}) namely $\|\rho-\sigma\|_{\mathbf{SEP}}\geq 2\|\rho-\sigma\|_{\mathbf{ALL}}/d$.
\end{remark}

\subsection{Multipartite vs bipartite data hiding}

In Theorem \ref{theorem:vrad-w-PPT-SEP}, we focussed on the bipartite case $\mathrm{H}=(\C^d)^{\otimes 2}$ for the sake of clarity. However, generalizations to the
general $k$-partite case $\mathrm{H}=(\C^d)^{\otimes k}$ are quite straightforward, at least in the situation where the high-dimensional composite system of interest is made of a
``small'' number of ``large'' subsystems (i.e.~$k$ is fixed and $d$ tends to infinity).

Let us denote by $\mathbf{PPT}_{d,k}$ and $\mathbf{SEP}_{d,k}$ the sets of respectively $k$-PPT and $k$-separable POVMs on $(\C^d)^{\otimes k}$.
On the one hand, an iteration of the
Milman--Pajor inequality (Corollary \ref{corollary:Milman-Pajor} in Chapter \ref{chap:toolbox}, Section \ref{ap:convex-geometry}) leads to the estimate
\[ c^{2^k}d^{k/2} \leq \mathrm{vrad}(K_{\mathbf{PPT}_{d,k}})\leq w(K_{\mathbf{PPT}_{d,k}})\leq Cd^{k/2}, \]
for some constants $c,C>0$ depending neither on $k$ nor on $d$.

On the other hand, the generalization of Theorem \ref{theorem:separable-states} in Appendix \ref{ap:standard} to the set $\mathcal{S}_{d,k}$ of $k$-separable states on $(\C^d)^{\otimes k}$
is known, namely (see \cite{AS1}, Theorem 1)
\[ \frac{c^{k}}{d^{k-1/2}}\leq \mathrm{vrad}(\mathcal{S}_{d,k})\leq w(\mathcal{S}_{d,k})\leq C\frac{\sqrt{k\log k}}{d^{k-1/2}}, \]
and implies that
\[ c^{k}d^{1/2}\leq \mathrm{vrad}(K_{\mathbf{SEP}_{d,k}})\leq w(K_{\mathbf{SEP}_{d,k}})\leq C\sqrt{k\log k}d^{1/2}, \]
for some constants $c,C>0$ depending neither on $k$ nor on $d$.

A multipartite analogue of Theorem \ref{theorem:typical-states} can then be derived, following the exact same lines of proof.

\begin{theorem}
\label{th:multiparty}
There exist constants $c_k,C_k>0$, depending only on $k$, such that the following holds. Given a dimension $d$, let $\rho$ and $\sigma$ be random states, independent and uniformly distributed on the set of states on $(\C^d)^{\otimes k}$. Then, with high probability,
\[ c_k \leq \| \rho - \sigma \|_{\mathbf{PPT}_{d,k}} \leq \| \rho - \sigma \|_{\mathbf{ALL}} \leq C_k, \]
\[ \frac{c_k}{\sqrt{d^{k-1}}} \leq \| \rho - \sigma \|_{\mathbf{SEP}_{d,k}} \leq \frac{C_k}{\sqrt{d^{k-1}}}. \]
\end{theorem}

This means that, forgetting about the dependence on $k$ and only focussing on the one on $d$, for ``typical'' states $\rho,\sigma$
on $(\C^d)^{\otimes k}$,
$\|\rho-\sigma\|_{\mathbf{PPT}_{d,k}}$ is of order $1$, like $\|\rho-\sigma\|_{\mathbf{ALL}}$, while $\|\rho-\sigma\|_{\mathbf{SEP}_{d,k}}$
is of order $1/\sqrt{d^{k-1}}$.

\medskip

In this multipartite setting, another quite natural question is the one of finding states that local observers can poorly distinguish if they remain alone but that they can distinguish substantially better though by gathering into any possible two groups. This type of problem was especially studied in \cite{EW}. Here is another result in that direction.

Define $\mathbf{bi{-}SEP}_{d,k}$ as the set of POVMs on $(\C^d)^{\otimes k}$ which are biseparable across any bipartition of $(\C^d)^{\otimes k}$. It may then be shown that for random states $\rho,\sigma$, independent and uniformly distributed on the set of states on $(\C^d)^{\otimes k}$, with high probability,
$\| \rho - \sigma \|_{\mathbf{bi{-}SEP}_{d,k}} \simeq d^{-k/4}$ (whereas $\| \rho - \sigma \|_{\mathbf{SEP}_{d,k}} \simeq d^{-(k-1)/2}$ by Theorem \ref{th:multiparty}). This means that on $(\C^d)^{\otimes k}$, with $k>2$ fixed, restricting to POVMs which are biseparable across every bipartition
is roughly the same as restricting to POVMs which are biseparable across one bipartition, whereas imposing $k$-separability is a much tougher constraint that implies a dimensional loss in the distinguishing ability. The reader is referred to Chapter \ref{chap:SDrelaxations}, Section \ref{section:volumes}, for related analyses.

\begin{remark} This result might not be as strong as one could hope for. It only shows that $\|\cdot\|_{\mathbf{bi{-}SEP}_{d,k}}$ typically vanishes
slower than $\|\cdot\|_{\mathbf{SEP}_{d,k}}$ when the local dimension $d$ grows, but it does not provide examples of states
$\rho,\sigma$ on $(\C^d)^{\otimes k}$ for which $\|\rho-\sigma\|_{\mathbf{bi{-}SEP}_{d,k}}$ would be of order $1$ while
$\|\rho-\sigma\|_{\mathbf{SEP}_{d,k}}$ would tend to zero.
\end{remark}

\section{Miscellaneous remarks and questions}

\subsection{On the complexity of the classes of separable and PPT POVMs}
Having at hand the estimates on the mean width of $K_{\mathbf{SEP}}$ (or $K_{\mathbf{LOCC}}$) and $K_{\mathbf{PPT}}$ provided by Theorem \ref{theorem:vrad-w-PPT-SEP}, one may follow the exact same lines as in the proof of Theorem \ref{theorem:approximation-of-ALL} to show that on $\C^d\otimes\C^d$, $\exp(\Theta(d^4))$ different POVMs are necessary and sufficient to approximate the class $\mathbf{PPT}$. For the class $\mathbf{SEP}$ (or $\mathbf{LOCC}$), we lack a complete answer since the same arguments show that the minimal number of POVMs is between $\exp(\Omega(d^3))$ and $\exp(O(d^4))$.

\subsection{What is the typical performance of the class $\mathbf{LO}$?}
\label{sec:LO-LOCC-SEP}

While Theorem \ref{theorem:LO-vs-LOCC} shows that the gap between the classes $\mathbf{LO}$ and $\mathbf{LOCC}$ may be unbounded,
we do not know if this situation is typical or not. Note that asking whether the norms $\|\cdot\|_{\mathbf{LO}}$ and $\|\cdot\|_{\mathbf{LOCC}}$ are comparable in a typical direction is more or less equivalent to asking whether the ratio $\vrad(K_{\mathbf{LOCC}}) / \vrad(K_{\mathbf{LO}})$ is bounded as the dimension increases. It is thus a pure convex geometry question, which now only requires to understand how the convex body $K_{\mathbf{LO}}$ looks like.

\subsection{Can the gap $\mathbf{LOCC}/\mathbf{SEP}$ be unbounded?}
Or conversely, does there exist an absolute
constant $c$ such that the inequality $\|\cdot\|_{\mathbf{LOCC}} \geq c \|\cdot\|_{\mathbf{SEP}}$ holds for any dimension? Also, we showed the existence of unbounded gaps in the two steps of the hierarchy $\mathbf{LOCC^{(0)}}/\mathbf{LOCC^{(1)}}/\mathbf{LOCC^{(2)}}$. More generally, is it possible to find, for any $r\in\N$, states which can be distinguished very poorly by $r$ rounds of local measurements and classical communication but very well if one extra round is allowed? The main difficulty in generalizing Theorems \ref{th:LO-LOCC1} and \ref{th:LOCC1-LOCC2} to the case $r\geq 2$ lies in understanding the disturbance induced on the states to be distinguished by all successive rounds of measurements.

\subsection{Typical value of other locally restricted distance measures}

We solved the issue of determining, for several classes of measurements $\mathbf{M}$, what is the typical value of the measured trace distance $\|\rho-\sigma\|_{\mathbf{M}}$ between two states $\rho,\sigma$. Several other ``filtered through measurements'' distances between $\rho$ and $\sigma$ can be defined in a completely analogous way, such as e.g.~the measured fidelity distance $F_{\mathbf{M}}(\rho,\sigma)$ or the measured relative entropy distance $D_{\mathbf{M}}(\rho\|\sigma)$ (see e.g.~\cite{Piani}). These quantities are all closely related to one another by well-known inequalities. Our statements can thus be straightforwardly translated into statements on the typical value of $F_{\mathbf{M}}(\rho,\sigma)$ or $D_{\mathbf{M}}(\rho\|\sigma)$. Besides, such restricted distance measures have already found a tremendous amount of applications in quantum information theory (see e.g.~\cite{BCHW} or \cite{LS} for two very recent ones, in two quite different topics, and also Chapter \ref{chap:deFinetti} of the present manuscript). Understanding better what is their generic scaling (and ultimately the one of their regularised versions) is therefore of prime interest, amongst other, to assess how optimal are the bounds where they appear, what is the efficiency of the quantum information processing protocols where they are involved etc.

\subsection{Locally restricted measurements on a multipartite quantum system}
There are at least two ways for a multipartite system such as $(\C^d)^{\otimes k}$ to be high-dimensional: either with $k$ fixed and $d$ large
(few large subsystems) or $k$ large and $d$ fixed (many small subsystems). Theorem \ref{th:multiparty} tells us what is the
typical discriminating power of $k$-PPT and $k$-separable POVMs, but in the first setting only. The extension to the case of many small subsystems
seems a challenging problem.

\section{Appendix: Volume estimates for some Schatten norm unit balls and related convex bodies}
\label{ap:standard}

In this appendix we gather estimates on the mean width and the volume radius of ``standard'' sets, which are used in our proofs. These concepts from classical convex geometry (and several others which we alude to here) were introduced in Chapter \ref{chap:toolbox}, Section \ref{ap:convex-geometry}, to which the reader is referred for further details. We also recall the two following notation specified in Chapter \ref{chap:motivations}, Section \ref{sec:notation}, for the unit balls associated to Schatten norms in the space of self-adjoint operators on $\C^d$:
\[ B_1(\C^d) = \{ A \in \mathcal{H}(\C^d) \st \|A\|_1 \leq 1 \}, \]
\[ B_{\iy}(\C^d) = \{ A \in \mathcal{H}(\C^d) \st \|A\|_\iy \leq 1 \} = [ -\Id , \Id]. \]

Moreover, given symmetric convex bodies $K \subset \R^n$ and $K' \subset \R^{n'}$, their projective tensor product is the convex body in $\R^n\otimes\R^{n'}$ defined as
\[ K \hat{\otimes} K' = \conv \{ x \otimes x' \st x \in K, x' \in K' \} \subset \R^n \otimes \R^{n'} \]

\begin{theorem}
\label{theorem:operator-norm}
We have
\[\mathrm{vrad}(B_{\iy}(\C^d)) \simeq  w(B_{\iy}(\C^d)) \simeq \sqrt{d}.\]
\[\mathrm{vrad}(B_1(\C^d)) \simeq  w(B_1(\C^d)) \simeq \frac{1}{\sqrt{d}}.\]
\end{theorem}

\begin{proof}
The estimates on the mean width follow from the semicircle law. Indeed,
the standard Gaussian vector in the space of self-adjoint operators on $\C^d$ is exactly a $d\times d$ GUE matrix $G$, and therefore (see \cite{AGZ}, Chapter 2, for a proof of these asymptotic estimates)
\begin{align*}
& w_G(B_{\iy}(\C^d)) = \E \|G\|_1 = d^{3/2} \left(\int_{-2}^2 |x| \frac{\sqrt{4-x^2}}{2\pi} \,\mathrm{d}x + o(1)\right) = d^{3/2} \left(\frac{8}{3\pi} + o(1)\right),\\
& w_G(B_1(\C^d)) = \E \|G\|_{\infty} = d^{1/2}(2+o(1)).
\end{align*}
Hence, setting $\gamma(d)=\E\|G\|_2$, which is known to satisfy $\gamma(d)\sim d$ (see again \cite{AGZ}, Chapter 2, for a proof of this asymptotic estimate), we get
\begin{align*}
& w(B_{\iy}(\C^d)) = \frac{1}{\gamma(d)} w_G(B_{\iy}(\C^d)) \sim \frac{8\sqrt{d}}{3\pi}, \\
& w(B_1(\C^d))= \frac{1}{\gamma(d)} w_G(B_1(\C^d)) \sim \frac{2}{\sqrt{d}}.
\end{align*}
Since $B_1(\C^d)$ and $B_{\iy}(\C^d)$ are polar to each other and symmetric, the Santal\'{o} and reverse Santal\'o inequalities (see \cite{Santalo} and \cite{BM}) yield
\[ \vrad(B_{\iy}(\C^d)) \vrad(B_1(\C^d)) \simeq 1.\]
If we then use the Urysohn inequality (see Theorem \ref{theorem:urysohn} in Chapter \ref{chap:toolbox}, Section \ref{ap:convex-geometry}), we obtain
\[ 1\simeq \vrad(B_{\iy}(\C^d)) \vrad(B_1(\C^d)) \leq w(B_{\iy}(\C^d))w(B_1(\C^d)) \simeq \sqrt{d}\, \frac{1}{\sqrt{d}} \simeq 1 ,\]
and therefore all these inequalities are sharp up to a multiplicative constant.
\end{proof}

We also need volume estimates on projective tensor products of Schatten norm unit balls.

\begin{theorem} \label{theorem:volume-S1-Sinfini}
We have the following estimates
\[ \mathrm{vrad}\big(B_1(\C^d) \hat{\otimes} B_{\iy}(\C^d)\big) \simeq w\big(B_1(\C^d) \hat{\otimes} B_{\iy}(\C^d)\big) \simeq \frac{1}{\sqrt{d}}.\]
\end{theorem}

A very similar proof shows that the estimates of Theorem \ref{theorem:volume-S1-Sinfini} are also valid when we consider the full complex Schatten classes, without the self-adjoint constraint. The question of estimating the volume radius of projective
tensor product of Schatten classes has been considered in \cite{DM}, where the question is answered (in a general setting) only up to a factor $\log d$.

\begin{proof}
An upper bound on the mean width can be obtained by a discretization argument, which we just sketch since only the lower bound will we used.
There is a polytope $P$ with $\exp(Cd)$ vertices such that $B_1(\C^d) \subset P \subset 2B_1(\C^d)$, and a polytope $Q$ with $\exp(Cd^2)$ vertices
such that $B_{\iy}(\C^d) \subset Q \subset 2B_{\iy}(\C^d)$. The polytope $P \hat{\otimes} Q$ satisfies
\[ B_1(\C^d) \hat{\otimes} B_{\iy}(\C^d) \subset P \hat{\otimes} Q \subset 4 B_1(\C^d) \hat{\otimes} B_{\iy}(\C^d) .\]
The polytope $P \hat{\otimes} Q$ is the convex hull of $\exp(C'd^2)$ points with Hilbert--Schmidt norm at most $4\sqrt{d}$. Using standard bounds for the mean width of polytopes (see Lemma \ref{lemma:mean-width-polytope} in Chapter \ref{chap:toolbox}, Section \ref{ap:convex-geometry}) gives the desired estimate $w\big(B_1(\C^d) \hat{\otimes} B_{\iy}(\C^d)\big) \lesssim 1/\sqrt{d}$.

We now give a lower bound on the volume radius.
We denote by $B_1^n \subset \R^n$ the unit ball for $\|\cdot\|_1$ in $\R^n$. We have the following formula.

\begin{lemma} \label{lemma:volume-ell_1-projective}
Let $m,n$ be integers and $K \subset \R^m$ be a symmetric convex body. Then
\[ \vol ( B_1^n \hat{\otimes} K ) = \frac{(m!)^n}{(mn)!} \vol(K)^n .\]
Consequently,
\[ \vrad (B_1^n \hat{\otimes} K ) \simeq \frac{1}{\sqrt{n}} \vrad(K) .\]
\end{lemma}

\begin{proof}
If $\{e_1,\dots,e_n\}$ denotes the canonical basis of $\R^n$, we have, for any $x_1,\dots,x_n \in \R^m$,
\[ \left\| \sum_{i=1}^n e_i \otimes x_i \right\|_{B_1^n \hat{\otimes} K} = \sum_{i=1}^n \|x_i\|_K .\]
So Lemma \ref{lemma:volume-ell_1-projective} follows easily from the formula below, valid for any integer $p$ and any symmetric convex body $L\subset\R^p$,
\begin{equation}\label{eq:volume-convex-body} \vol(L) = \frac{1}{p!} \int_{\R^p} \exp ( - \|x\|_L) \, \mathrm{d}x .\end{equation}
Equation \eqref{eq:volume-convex-body} itself may be obtained by the following chain of equalities
\[ \int_{\R^p} e^{- \|x\|_L} \, \mathrm{d}x = \int_{\R^p}\int_{\|x\|_L}^{+\infty} e^{-t}  \, \mathrm{d}t  \, \mathrm{d}x = \int_{0}^{+\infty}\int_{\{\|x\|_L<t\}} e^{-t}  \, \mathrm{d}x  \, \mathrm{d}t = \int_{0}^{+\infty} e^{-t} \vol(tL)  \, \mathrm{d}t = \vol(L)p!, \]
the last equality being because $\int_{0}^{+\infty} t^pe^{-t}\, \mathrm{d}t = p!$.
\end{proof}

Denote by $\{\ket{j} \st 1 \leq j \leq d\}$ an orthonormal basis of $\C^d$. The family
\[ \left\{ \ketbra{j}{j} \st 1 \leq j \leq d \right\} \cup \left\{ \frac{1}{\sqrt{2}} ( \ketbra{j}{k} + \ketbra{k}{j} ) \st 1 \leq j < k \leq d \right\}
 \cup \left\{ \frac{i}{\sqrt{2}} ( \ketbra{j}{k} - \ketbra{k}{j} ) \st 1 \leq j < k \leq d \right\}
\]
is an orthonormal basis of $\cH(\C^d)$ whose elements live in $\sqrt{2} B_1(\C^d)$. It follows that
\[ \vrad\big(B_1(\C^d) \hat{\otimes} B_{\iy}(\C^d)\big) \geq \frac{1}{\sqrt{2}} \vrad\big(B_1^{d^2} \hat{\otimes} B_{\iy}(\C^d)\big) \gtrsim \frac{1}{d} \vrad(B_{\iy}(\C^d)) ,\]
the last estimate being a consequence of Lemma \ref{lemma:volume-ell_1-projective}.

Using Theorem \ref{theorem:operator-norm} one may thus conclude that
$\vrad\big(B_1(\C^d) \hat{\otimes} B_{\iy}(\C^d)\big) \gtrsim 1/\sqrt{d}$.
\end{proof}

We also need a result on the volume radius and the mean width of the set of separable states, which is taken from \cite{AS1}, Theorem 1.

\begin{theorem}
\label{theorem:separable-states}
On $\C^d\otimes\C^d$, denoting by $\mathcal{S}$ the set of separable states, we have
\[d^{-3/2}\simeq \mathrm{vrad}(\mathcal{S})\leq w(\mathcal{S})\simeq d^{-3/2}.\]
\end{theorem}

\chapter{Relaxations of separability in multipartite quantum systems}
\chaptermark{Relaxations of separability in multipartite quantum systems}
\label{chap:SDrelaxations}

\textsf{Based on ``Relaxations of separability in multipartite systems: semidefinite programs, witnesses and volumes'', in collaboration with O. G\"{u}hne, M. Huber and R. Sengupta \cite{GHLS}.}

\bigskip

While entanglement is believed to be an important ingredient in
understanding quantum many-body physics, the complexity of its
characterization scales very unfavorably with the size of the
system. Finding super-sets of the set of separable states that
admit a simpler description has proven to be a fruitful
approach in the bipartite setting. In this chapter we discuss a
systematic way of characterizing multiparticle entanglement
via various relaxations. We furthermore describe an
operational witness construction arising from such relaxations
that is capable of detecting every entangled state. Finally,
we also derive an analytic upper bound on the volume of biseparable
states and show that the volume of the states with a positive partial
transpose for any split rapidly outgrows this volume.
This proves that simple semidefinite relaxations in the
multiparticle case cannot be an equally good approximation
for any scenario.

\section{Introduction}

Without a doubt entanglement can be considered one of the most important
concepts in quantum physics, clearly distinguishing quantum systems from
classical ones. It can be harnessed to enable novel ways of processing
quantum information in numerous ways, from communication to computation.
Many of these operational tasks require an operational detection or even
quantification of this indispensable resource. While in bipartite systems
of low dimensions this can be achieved in an efficient way, the complexity
of the characterization of entangled states makes a complete and computable
framework of entanglement detection impossible in high dimensions and thus
also for multipartite systems \cite{HHHH,GT}.

A possible way for deriving statements on the presence of entanglement
is to discard the actual complex structure of the border
between separable and entangled states and try to find good approximations
that admit a more amenable description. In the bipartite
case positive maps play a central role in such approximations: all separable
states remain positive semidefinite under application of a positive, yet not
completely positive map to one of its subsystems. The most well-known example of
such a map is the partial transposition: this map generally changes the
eigenvalues of a matrix, but separable states have a positive partial transpose. The approach of positive maps allows for a
characterization of super-sets of the set of separable states using techniques
from semidefinite programming, and it nevertheless captures the whole structure:
a state remains positive under {\it all} positive maps if and only if the state
is indeed separable \cite{HHHH,GT}. The reader is referred to Chapter \ref{chap:QIT}, Section \ref{sec:sep-multi}, for extra information.

In order to gauge the efficiency of certain maps for characterizing
entanglement, one of the most relevant issues is how the volume of
states that remain positive under the map in question compares to the
volume of separable states \cite{HLSZ}. The sobering and non-surprising answer from
bipartite systems can be gained from convex geometry considerations and
shows that for all known maps the states that remain positive are most
likely entangled in high dimensions \cite{BS}. For small dimensions,
however, a given positive map can detect a large fraction of entangled states.
This is, for instance, true for the partial transposition, which delivers
a necessary and sufficient criterion for $2 \times 2$ and $2 \times 3$ systems.

In multipartite systems the characterization of entanglement constitutes
an even greater challenge. Since partial separability of multipartite
states can no longer be defined as a purely bipartite concept, the application
of positive maps to subsystems alone can reveal little more than entanglement
across a fixed partition of the multipartite state. Nevertheless, recently
several works succeeded in defining suitable mixtures of positive maps, which
can be used to develop strong criteria for genuine multiparticle entanglement
\cite{GJM,HS}.

In this chapter we first develop a framework that allows for the semidefinite
relaxation of partially separable states, opening the possibility for harnessing
well developed techniques based on positive maps to detect genuine multipartite
entanglement (an approach which has already been shown to yield useful
results with different relaxations in \cite{CEGH,BV1,BV2}).
We achieve this goal by first formally defining semidefinite
relaxations of partially separable states using positive maps.

Due to the formulation as semidefinite programs these constructions yield versatile
criteria for detecting multipartite entanglement in low dimensional systems. To unlock
these powerful techniques for more complex quantum states we proceed to discuss a recently
introduced program of lifting bipartite witnesses \cite{HS}. We prove that it is always
possible to exploit witnesses that only reveal bipartite entanglement in order to construct
witnesses for genuine multipartite entanglement. This facilitates this notoriously hard problem, and
we showcase this technique with some exemplary multipartite entangled states.

In a second step, we ask which fraction of genuinely multipartite entangled states can be detected with such relaxation methods. We prove an upper bound on the volume of the set of biseparable states, and a lower bound on the volume of a set
of states that can never be detected with relaxation methods based on the partial
transposition. For large dimensions, both values deviate significantly. This shows
that while the relaxation approaches are strong for small systems, they
fail to deliver a good approximation in the general case.

\section{Characterizing relaxations of separability with semidefinite programs}
\label{section:SD relaxations}

The most straightforward relaxation of separability in multipartite systems
is again given by positive maps. Trying to justify this assertion is the
object of the current section.

Let us start with some basic definitions and notation (see e.g.~\cite{GT,ES} for a general review on these notions, and Chapter \ref{chap:QIT}, Section \ref{sec:sep-multi}, of this manuscript for a brief recap). On a multipartite
system, given a bipartition $\{I|I^c\}$ of the
subsystems, we denote by $\mathcal{S}_I$ the set of states which are biseparable
across this cut, i.e.~\[ \rho\in\mathcal{S}_I\ \text{if}\ \rho=\sum_x q_x|\phi_I^{(x)}\rangle\!\langle\phi_I^{(x)}|\otimes|\phi_{I^c}^{(x)}\rangle\!\langle\phi_{I^c}^{(x)}|,
\]
where $\{q_x\}_x$ is a convex combination, and for each $x$, $|\phi_I^{(x)}\rangle,|\phi_{I^c}^{(x)}\rangle$ are pure states on the subsystems in $I,I^c$ respectively.
We then define the set $\mathcal{S}_{(2)}$ of biseparable states as being the convex hull of $\{\mathcal{S}_I\}_I$, i.e.~\[ \rho\in\mathcal{S}_{(2)}\ \text{if}\ \rho=\sum_{I}p_I\sigma_I, \]
where $\{p_I\}_I$ is a convex combination, and for each $I$, $\sigma_I\in\mathcal{S}_I$. If a state is not biseparable it is called \emph{genuinely multipartite entangled} (GME).

This definition admits a simple relaxation with a positive semidefinite characterization. Given, for each bipartition $\{I|I^c\}$, a positive map $\mathcal{N}_I$, acting on the substystems in $I$, we define the set $\mathcal{R}_{\{\mathcal{N}_I\}_I}$ of $\{\mathcal{N}_I\}_I$-relaxation of $\mathcal{S}_{(2)}$ by
\begin{equation} \label{eq:relaxation} \rho\in\mathcal{R}_{\{\mathcal{N}_I\}_I}\ \text{if}\ \rho=\sum_{I}p_I\sigma_{\mathcal{N}_I}, \end{equation}
where $\{p_I\}_I$ is a convex combination and for each $I$, $\mathcal{N}_I\otimes\mathcal{I\!d}_{I^c}\left(\sigma_{\mathcal{N}_I}\right)\geq 0$.

Such a relaxation carries the operational advantage that it
can be approached via semidefinite programming (SDP).
Besides, this definition can easily be extended to $\ell$-separable
states by applying different maps to the induced partitions.
As we are however mainly interested in characterizing the
strongest form of multipartite entanglement we focus here on the
distinction between biseparable states and genuinely
multipartite entangled ones.

For instance, these relaxations can be particularly useful when optimizing convex functions over the set of biseparable states. Indeed, for any function $f$, we trivially have that, for any set of positive maps $\{\mathcal{N}_I\}_I$,
\begin{equation} \label{eq:optimize} \min_{\sigma\in\mathcal{S}_{(2)}}f(\sigma)\geq\min_{\sigma\in\mathcal{R}_{\{\mathcal{N}_I\}_I}}f(\sigma).
\end{equation}
In particular, for any convex function $f$, equation~\eqref{eq:optimize} provides a relaxation of the optimization of $f$ over $\mathcal{S}_{(2)}$ which can be cast as an SDP. One of the most straightforward applications of such a strategy is to testing whether a given density matrix $\rho$ is indeed biseparable, i.e.~applying it e.g.~to $f:\sigma\mapsto\left(\mathrm{Tr}\left[(\rho-\sigma)^2\right]\right)^{1/2}$. It yields in that case the equivalence
\[ \forall\ \{\mathcal{N}_I\}_I,\ \min_{\sigma\in\mathcal{R}_{\{\mathcal{N}_I\}_I}}\mathrm{Tr}\big[\sigma(\sigma-2\rho)\big] \leq \mathrm{Tr}\left[\rho^2\right] \Leftrightarrow\ \rho\in\mathcal{S}_{(2)}. \]

While this defines, in theory, a necessary and sufficient program for deciding whether a given state is biseparable, checking all possible sets of positive maps is of course not feasible. However even making one particular choice, such as the transposition map for instance, has already proven to yield very strong witnesses, and through the dual of the program one can additionally often extract analytical constructions for important classes of states \cite{GJM}.

Let us specify a bit what we mean in the simplest tripartite case. Given a tripartite state $\rho$ and a positive map $\mathcal{N}$ acting on one subsystem, the condition $\rho\in\mathcal{R}_{\{\mathcal{N}_1,\mathcal{N}_2,\mathcal{N}_3\}}$ is equivalent to the following SDP yielding a non-negative value:
\[ \max s\ \text{s.t.}\ \sigma_1,\sigma_2,\sigma_3\geq s\Id,\ \rho=\sigma_1+\sigma_2+\sigma_3,\ \mathcal{N}_1\otimes\mathcal{I\!d}_{23}(\sigma_1),\mathcal{N}_2\otimes\mathcal{I\!d}_{13}(\sigma_2),\mathcal{N}_3\otimes\mathcal{I\!d}_{12}(\sigma_3) \geq s\Id. \]
While not being quantitative as the above program it can be easily implemented in MATLAB, using the packages YALMIP \cite{YALMIP} and the SEDUMI solver \cite{SEDUMI}. What is more, if an $s\geq0$ is found, the program also returns a feasible $\sigma_1,\sigma_2,\sigma_3$ achieving the maximum, and can thus be used to find actual $\mathcal{N}$-positive decompositions of $\rho$.

To test the further prospects of such SDP approach, we have first programmed it choosing either the transposition map $\mathcal{T}$ (initially introduced in \cite{Peres} and \cite{HHH}) or the Choi map $\mathcal{C}$ (initially introduced in \cite{CL}) as positive map (see Chapter \ref{chap:QIT}, Section \ref{sec:sep-multi}, for precise definitions). We have applied them to the family of states $\rho(\{\lambda_1,\lambda_2,\lambda_3\})$ introduced in \cite{HS}, Example 2. It was shown there that the state $\rho(\lambda):=\rho(\{\lambda,\lambda,\lambda\})$ is GME for $0<\lambda<1/3$. Testing the two programs on this family of states returns a positive under partial transposition map (PPT) or positive under partial Choi map (PPC) decomposition for $1/3\leq\lambda<1$, thus showing that on the set of PPT or PPC mixtures the witness presented in \cite{HS} is weakly optimal. It furthermore proves that for $1/3\leq\lambda<1$ the state can be decomposed into states which are either PPT or PPC.

We have then extended the program to demand simultaneous positivity under the transposition and the Choi maps. That is formally, we looked for
\[ \max s\ \text{s.t.}\  \sigma_1,\sigma_2,\sigma_3\geq s\Id,\ \rho=\sigma_1+\sigma_2+\sigma_3,\
\begin{cases} \mathcal{C}_1\otimes\mathcal{I\!d}_{23}(\sigma_1), \mathcal{C}_2\otimes\mathcal{I\!d}_{13}(\sigma_2), \mathcal{C}_3\otimes\mathcal{I\!d}_{12}(\sigma_3) \geq s\Id \\ \mathcal{T}_1\otimes\mathcal{I\!d}_{23}(\sigma_1),\mathcal{T}_2\otimes\mathcal{I\!d}_{13}(\sigma_2),\mathcal{T}_3\otimes\mathcal{I\!d}_{12}(\sigma_3) \geq s\Id \end{cases}. \]
With this program, one can check that, in fact, for every $0<\lambda<1$, $\rho(\lambda)$ is GME. This is because the program shows that, even in the range $1/3\leq\lambda\leq 1$, $\rho(\lambda)$ cannot be decomposed into a mixture of states which are both PPT and PPC.

This example showcases the versatility of this approach in small dimensions. Indeed, relaxing being biseparable to just being a mixture of states which are positive under partial application of two positive maps already proves quite fruitful. Let us mention though that, in high dimensions, making the SDP so slightly more constraining, by imposing positivity under two maps instead of one, is not expected to drastically improve its efficiency as in this particular case (see \cite{AS2} for a mathematically precise formulation of this assertion).

\section{Constructing multiparticle witnesses from bipartite witnesses}
\label{sec:witnesses}

While the semidefinite program presented before technically gives sufficient criteria for deciding biseparability, it becomes quickly intractable beyond a few qubits. There are, however, frequent situations in which one can use some additional knowledge to facilitate witness constructions.
Genuinely multipartite entangled states of course also have to be entangled across every bipartition of the system. And since the construction of bipartite entanglement witnesses can be a rather straightforward affair (e.g.~through positive maps) one can ask whether there is a possibility to construct multipartite entanglement witnesses from a collection of bipartite operators.

\subsection{A systematic construction for lifting bipartite witnesses}

In \cite{HS} a general
witness construction method was introduced, which enables the construction
of multipartite entanglement witnesses from a set of
bipartite witnesses across every possible bipartition. Such
a problem can be formalized as follows.

Given, for each bipartition $\{I|I^c\}$, a witness $W_I$, i.e.~a self-adjoint operator such
that $\mathrm{Tr}(W_I\sigma_I)\geq0$ for all $\sigma_I\in\mathcal{S}_I$, we are looking for a self-adjoint
operator $W_{GME}$ with the following property
\begin{equation} \forall\ I,\ W_{GME}\geq W_I. \label{general} \end{equation}

The construction of \cite{HS} is one particular instance of the following general method. Let $Q$ be some operator, and set, for each $I$,
\begin{equation}
\label{eq:W_b}
T_I=Q-W_I.
\end{equation}
Define next $W_{GME}$ as
\begin{equation}
\label{eq:W_GME}
W_{GME}=Q+\sum_I\big[T_I\big]_+,
\end{equation}
where, for any self-adjoint operator $A$, we denote by $[A]_+$ the projection of $A$ onto the positive semidefinite cone.
It is then easy to see that condition (\ref{general}) holds for every $I$. So of course, the crucial issue here is first of all the one of the
generality of such a construction: given a set of bipartite entanglement witnesses $\{W_I\}_I$ for some genuinely multipartite entangles state $\rho_{GME}$, does there always exist a $Q$ such that $W_{GME}$ as defined by equation \eqref{eq:W_GME} is a genuinely multipartite entanglement witness for $\rho_{GME}$? Second, one can also ask the question of the optimal choice of $Q$ (in a sense to be defined).

Using a special choice for $Q$, it was shown in \cite{HS} that there exist
genuinely multipartite entangled states and a set of
bipartite witnesses for which the expectation value of
$W_{GME}$ is negative, proving that this construction can indeed
succeed in generating witnesses for multipartite
entanglement. Here we begin with proving the following generality result.

\begin{theorem}
\label{th:generality}
For every genuinely multipartite entangled state $\rho_{GME}$, there
exists a set of weakly optimal bipartite entanglement witnesses
$\{W_I\}_I$ such that
\[ \mathrm{Tr}\left(\rho_{GME}\left(Q+\sum_I\big[T_I\big]_+\right)\right)<0, \]
where the operators $Q$ and $\{T_I\}_I$ are defined by equation \eqref{eq:W_b}.
\end{theorem}

\begin{proof}
Let $Q$ be a genuine multipartite entanglement witness for $\rho_{GME}$, meaning that $\mathrm{Tr}(Q\sigma)\geq 0$ for any $\sigma\in\mathcal{S}_{(2)}$, while $\mathrm{Tr}(Q\rho_{GME}) < 0$. From the former assumption it follows that, for each bipartition $\{I|I^c\}$, $\alpha_I:=\min_{\sigma_{I}\in\mathcal{S}_I}\mathrm{Tr}(Q\sigma_{I})\geq 0$. As this minimization is convex in the space of states it is clear that the minimal overlap with $Q$ can also be reached by an optimal pure state $|\varphi_{I}\rangle\!\langle\varphi_{I}|\in\mathcal{S}_I$. Now we can choose the following bipartite witness: $W_{I}=Q-\alpha_{I}|\varphi_I\rangle\!\langle\varphi_I|$. We can then verify that, for each bipartition $\{I|I^c\}$, on the one hand
\[ \forall\ \sigma_I\in\mathcal{S}_I,\ \mathrm{Tr}(W_{I}\sigma_{I}) \geq \mathrm{Tr}(Q\sigma_{I}) -\alpha_I \geq 0,\ \text{and}\  \mathrm{Tr}(W_{I}|\varphi_{I}\rangle\!\langle\varphi_{I}|)=0, \]
while on the other hand
\[ \mathrm{Tr}\left(W_{I}\rho_{GME}\right) \leq \mathrm{Tr}\left(Q\rho_{GME}\right) <0. \]
So indeed, for each bipartition $\{I|I^c\}$, $W_I$ is a weakly optimal witness detecting that $\rho_{GME}\notin\mathcal{S}_I$.

Inserting this set of optimal bipartite
witnesses into the construction from above using $[-\alpha_{I}|\varphi_I\rangle\!\langle\varphi_I|]_+=0$ yields $W_{GME}=Q$, proving that each genuine multipartite entanglement witness can be gained from the construction using weakly optimal bipartite entanglement witnesses.
\end{proof}

As a direct consequence of Theorem \ref{th:generality} we have, as wanted, that for any set of bipartite entanglement witnesses $\{W_I\}_I$ for $\rho_{GME}$, there exists a $Q$ such that $W_{GME}$ as defined by equation \eqref{eq:W_GME} is a genuinely multipartite entanglement witness for $\rho_{GME}$.

\begin{corollary}
Every multipartite entanglement witness $W_{GME}$ can be constructed
using the framework summarized by equations \eqref{eq:W_b} and \eqref{eq:W_GME}. Furthermore, it is possible to impose that, for every bipartition $\{I|I^c\}$,
$W_I=\mathcal{N}^*_{I}\otimes\mathcal{I\!d}_{I^c}\left(|\psi_I\rangle\!\langle\psi_I|\right)$ for some choice of positive map $\mathcal{N}_{I}$, acting on the subsystems in $I$, and some choice of pure state $|\psi_I\rangle$.
\end{corollary}

\begin{proof}
First it is important to notice that, for every state $\rho$ which is
entangled across a specific bipartition $\{I|I^c\}$,
there exists a positive map $\mathcal{N}_I$, acting on the subsystems in $I$, such that
$\mathcal{N}_I\otimes\mathcal{I\!d}_{I^c}(\rho)$ is not positive semidefinite,
i.e.~it is detected to be entangled by this positive map
(see \cite{HHHH, dePillis, Jamiolkowski}). This implies in turn that there exists a
unit vector $|\psi_I\rangle$ such that $\langle\psi_I|\mathcal{N}_I\otimes\mathcal{I\!d}_{I^c}(\rho)|\psi_I\rangle< 0$, i.e.~ $\mathrm{Tr}\left(\rho\mathcal{N}^*_{I}\otimes\mathcal{I\!d}_{I^c}\left(|\psi_I\rangle\!\langle\psi_I|\right)\right)<0$.
Through continuity this implies that, for every extremal biseparable state $\ketbra{\phi}{\phi}$, there exists a weakly optimal witness of the form $\widetilde{W}_I=\mathcal{N}^*_{I}\otimes\mathcal{I\!d}_{I^c}\left(|\psi_I\rangle\!\langle\psi_I|\right)$ such that $\mathrm{Tr}\big(\widetilde{W}_I|\phi\rangle\!\langle\phi|\big)=0$. Now, invoking the Choi-Jamiolkowski isomorphism, we can conclude that every possible hyperplane intersecting the biseparable set corresponds to a witness derived from a positive map. This implies that amongst all the $\{\widetilde{W}_I\}_I$, there is at least one of them $\widetilde{W}_{I_0}$ which satisfies the following: for each bipartition $\{I|I^c\}$, $W_I=\widetilde{W}_{I_0}-\beta_I\Id$, for some $\beta_I$, is an optimal bipartite witness for $\rho$ (and $\beta_{I_0}=0$). This concludes the proof: the $\{W_I\}_I$ are all obtained from shifting the same $\widetilde{W}_{I_0}=\mathcal{N}^*_{I_0}\otimes\mathcal{I\!d}_{I_0^c}\left(|\psi_{I_0}\rangle\!\langle\psi_{I_0}|\right)$.
\end{proof}

\subsection{Illustration on one example}

To showcase the strength of the above constructions let us illustrate it with a peculiar example. First of all let us make a specific choice for $Q$ that has already proven to work well in \cite{HS}. Given a set of witnesses for every bipartition $\{W_I\}_I$, we will construct $Q$ by finding element wise the matrix of largest common negative matrix entries
\[ N=\sum_{j,j'=1}^d|j\rangle\!\langle j'|\min\left[0,\max_{I}\big[\Re e\bra{j}W_I\ket{j'}\big]\right]\,, \]
and the matrix of smallest common positive matrix entries
\[ P=\sum_{j,j'=1}^d|j\rangle\!\langle j'|\max\left[0,\min_{I}\big[\Re e\bra{j}W_I\ket{j'}\big]\right]\,, \]
where $d$ denotes the global dimension. With these two matrices we can then define
\[ Q:=N+P\,. \]
Now for the purpose of elucidating how this construction works in practice let us follow it through step by step in an exemplary three-qutrit case. In order to write down our target state, we first define the vector $|\psi_0\rangle=|111\rangle+|222\rangle+|333\rangle$. We next fix $a_1,a_2,a_3>0$ and define the following vectors
\begin{align*}
& |\psi_1\rangle=\sqrt{a_1} |112\rangle+\sqrt{\frac{1}{a_1}}|221\rangle,\ \ |\psi_2\rangle=\sqrt{a_1} |121\rangle+\sqrt{\frac{1}{a_1}}|212\rangle,\ \ |\psi_3\rangle=\sqrt{a_1} |211\rangle+\sqrt{\frac{1}{a_1}}|122\rangle, \\
& |\psi_4\rangle=\sqrt{a_2} |223\rangle+\sqrt{\frac{1}{a_2}}|332\rangle,\ \ |\psi_5\rangle=\sqrt{a_2} |232\rangle+\sqrt{\frac{1}{a_2}}|323\rangle,\ \ |\psi_6\rangle=\sqrt{a_2} |322\rangle+\sqrt{\frac{1}{a_2}}|233\rangle, \\
& |\psi_7\rangle=\sqrt{a_3} |331\rangle+\sqrt{\frac{1}{a_3}}|113\rangle,\ \ |\psi_8\rangle=\sqrt{a_3} |313\rangle+\sqrt{\frac{1}{a_3}}|131\rangle,\ \ |\psi_9\rangle=\sqrt{a_3} |133\rangle+\sqrt{\frac{1}{a_3}}|311\rangle.
\end{align*}
We can then construct a three-qutrit mixed state $\rho$ as follows
\[ \rho:=\frac{\tilde{\rho}}{\text{Tr}\,\tilde{\rho}},\ \text{where}\ \tilde{\rho}:=\sum_{i=0}^{9}|\psi_i\rangle\!\langle\psi_i|+p(|112\rangle\!\langle 112|+\Id). \]
By construction this state is PPT across all three cuts. If we choose $a_1=10^{-6}$, $a_2=300$ and $a_3=12\times10^{-3}$, then for $p>0.0003$ it is also PPC across all three cuts. As it is both fully PPT and PPC it is fair to say that if it is entangled it is only very weakly so (from a standard detection tool point of view). The system's size is still small enough to apply the positive map mixer introduced in the previous section, and to reveal that it is nonetheless indeed genuinely multipartite entangled for various values of $p>0.0003$.

Hence, this is a good opportunity to demonstrate the power of witness liftings. We can use the indecomposable map introduced in \cite{Osaka}, choosing $c_1=1$, $c_2=10^{-3}$ and $c_3=10^3$, revealing entanglement across all three cuts. In fact, even without ever calculating any eigenvalues, we can just apply the map on the three subsystems of $|\psi_{0}\rangle$ only. This means taking
\[ W_1=\mathcal{N}^{c_1,c_2,c_3}_1\otimes\mathcal{I\!d}_{23}\left(|\psi_{0}\rangle\!\langle\psi_{0}|\right),\ W_2=\mathcal{N}^{c_1,c_2,c_3}_2\otimes\mathcal{I\!d}_{13}\left(|\psi_{0}\rangle\!\langle\psi_{0}|\right),\ W_3=\mathcal{N}^{c_1,c_2,c_3}_3\otimes\mathcal{I\!d}_{12}\left(|\psi_{0}\rangle\!\langle\psi_{0}|\right). \]
Plugging the three resulting witnesses in our construction for $Q$ we get a GME-witness that is able to reveal genuine multipartite entanglement in the state for a range of $0\leq p\leq 0.00069$. This example illustrates that even bound entangled states which are positive with respect to paradigmatic maps can exhibit multipartite entanglement. As the state itself is full rank and not symmetric, there is no other known method that could have revealed it to be GME, clearly demonstrating the power of positive map mixers and witness liftings.

One question at this point is that of how exceptional the fact of being fully PPT and PPC, but nevertheless GME, is. Finding explicit examples of such states is not so easy. Indeed, in small dimensions things are quite contrived and there is not much room to move (as just exemplified), while as the dimensions grow computations quickly become untractable. However, as we shall see in Section \ref{section:volumes}, this feature is actually generic in large dimensions.

\subsection{Finding a multiparticle witness with semidefinite
programming}

In \cite{HS} and in the previous example, the multipartite witness $W_{GME}$ was constructed
by starting with bipartite witnesses $W_I$ for each bipartition
$\{I|I^c\}$, and then one possible choice for the operator
$Q$ was explicitly  constructed. While the presented choice
works well for many examples, as just illustrated,
it is not clear why it would be the best one. We leave open this optimality question at this level of generality. Let us make two easy observations though. On the one hand, note that the choice $Q=0$ yields $W_{GME}=\sum_I[W_I]_+\geq 0$, which detects absolutely no GME state. On the other hand, in the case where there exists an $I_0$ such that, for each $I$, $W_I=W_{I_0}-\alpha_I\Id$ with $\alpha_I\geq 0$, then the choice $Q=W_{I_0}$ yields $W_{GME}=W_{I_0}$, which detects all the GME states whose bipartite entanglement is detected by the $\{W_I\}_I$.

Let us also mention that, for a given state $\rho$, the optimal
multipartite witness $W_{GME}$ can directly be computed as a semidefinite program that is easier to run than the map-mixers.

For that, consider the following constrained optimization problem
\[ \mbox{ \text{minimize:} } \quad  \mathrm{Tr}\big(\rho W_{GME}\big)\ \mbox{ \text{subject to:} } \quad \forall\ I,\ W_{GME} \geq W_I. \]
This is a semidefinite program, which can be easily and efficiently solved
using standard numerical techniques.

There is, in addition, a variation of the presented method for obtaining
a multiparticle witness from a set of bipartite witnesses. The condition
imposed by equation~(\ref{general}) guarantees that, for each $I$,
$\mathrm{Tr}(\sigma W_{GME}) \geq \mathrm{Tr}(\sigma W_{I})$ for any
state $\sigma$, so that $W_{GME}$ is indeed an entanglement witness.
However, for being an entanglement witness it suffices that, for
each $I$, $\mathrm{Tr}(\sigma_I W_{GME}) \geq \mathrm{Tr}(\sigma_I W_{I})$
for any state $\sigma_I$ which is separable for the bipartition
$\{I|I^c\}$. And this is already guaranteed if, for all
$I$, there exists a positive map $\mathcal{N}_I$, acting on subsystems
in $I$, such that
\[ \mathcal{N}_I \otimes \mathcal{I\!d}_{I^c} \left(W_{GME}\right) \geq \mathcal{N}_I \otimes \mathcal{I\!d}_{I^c} \left(W_{I}\right). \]
So, for computing a multipartite witness
for a given state one can also consider the semidefinite program
\[ \mbox{ \text{minimize:} } \quad  \mathrm{Tr}\big(\rho W_{GME}\big)\ \mbox{ \text{subject to:} } \quad \forall\ I,\ \mathcal{N}_I \otimes \mathcal{I\!d}_{I^c} \left(W_{GME}\right) \geq \mathcal{N}_I \otimes \mathcal{I\!d}_{I^c} \left(W_I\right). \]
These two semidefinite programs can be used to construct the best witness for a given
state systematically. They might be useful, if the analytical method from \cite{HS}
does not work.

It should be noted, however, that the presented formulations are not necessarily
the best way to construct a multiparticle witness from  bipartite witnesses, if
one wishes to use semidefinite programming. The reason is the following: For each
bipartite witness one can construct via the Choi-Jamiolkowski isomorphism a positive, but not
completely positive map. This map detects more states than the original witness.
Given these maps one can then evaluate the corresponding map relaxations, as described
in Section \ref{section:SD relaxations}. This criterion will be stronger than the semidefinite programs presented
above. For that reason, we do not discuss detailed examples here.

\section{Relaxations of separability beyond positive maps}
\label{section:CCNR}

So far, we considered only the approximation of the biseparable set by
super-sets which are associated to positive maps. One can, however, also
use other bipartite separability criteria, such as the computable cross norm (CCNR), aka
realignment, criterion \cite{Rudolph,CW} or the symmetric extension, aka $k$-extendibility, criterion
\cite{DPS} (for the latter, see also Chapter \ref{chap:k-extendibility} of this manuscript). In the following we explain how the CCNR criterion can be used in the multipartite setting.

\subsection{Description of the method}

Let us start by explaining the CCNR criterion. Any quantum
state $\rho$ on the bipartite Hilbert space $\mathrm{H}_1\otimes\mathrm{H}_2$ can be expressed (via the Schmidt decomposition in the space of linear operators) as
\[ \rho = \sum_i \lambda_i G_1^{(i)} \otimes G_2^{(i)}, \]
where the $\lambda_i$ are positive coefficients and the $G_1^{(i)}$, $G_2^{(i)}$
are orthogonal observables on $\mathrm{H}_1$, $\mathrm{H}_2$, i.e.~they fulfill $\mathrm{Tr}\big(G_1^{(i)}G_1^{(j)}\big)= \mathrm{Tr}\big(G_2^{(i)}G_2^{(j)}\big)= \delta_{ij}$. With this representation one can easily prove that the
following holds:
\[ \rho\in\mathcal{S}(\mathrm{H}_1{:}\mathrm{H}_2)\  \Rightarrow\  \sum_i \lambda_i \leq 1. \]
And this necessary condition for separability is known as the CCNR criterion.
The criterion has the advantage that it detects entanglement in many states
where the PPT criterion fails. On the other hand, not all two-qubit entangled
states can be detected by this test.

From this structure, one can easily write down entanglement witnesses.
Namely, any operator of the form
\[ W = \Id - \sum_i G_1^{(i)} \otimes G_2^{(i)} \]
is an entanglement witness, as it is positive on all states with $\sum_i \lambda_i \leq 1$.

This structure can be used for constructing witnesses for
genuine multipartite entanglement as follows. Consider an operator $W_{GME}$ on the $k$-partite Hilbert space $\mathrm{H}_1\otimes\cdots\otimes\mathrm{H}_k$ which is such that,
for any bipartition $\{I|I^c\}$ of the $k$ subsystems,
\begin{equation}
W_{GME} =  P_I + \Id - \sum_i G_I^{(i)} \otimes G_{I^c}^{(i)},
\label{ccnrcond}
\end{equation}
where the $G_{I}^{(i)}$, $G_{I^c}^{(i)}$ are orthogonal observables on
$\mathrm{H}_I$, $\mathrm{H}_{I^c}$, and $P_I\geq 0$ is positive semidefinite.
Clearly, if a state obeys the CCNR criterion for some bipartition,
the mean value of the witness $W_{GME}$ will be non-negative.
Consequently, the witness is also non-negative on all biseparable
states.

\subsection{Example: the three-qubit GHZ state}

The witnesses from the CCNR criterion are more difficult to handle
than the witnesses from positive maps. The reason is that no approach
via semidefinite programming is possible. Moreover, the condition from
equation~(\ref{ccnrcond}) is more difficult to check than the condition in
equation~(\ref{general}). Nevertheless, we will present an example where
known optimal entanglement witnesses have this structure.

Consider first the three-qubit Greenberger-Horne-Zeilinger (GHZ)
state $\ket{GHZ}= (\ket{111}+\ket{222})/\sqrt{2}.$
The typical witness for this state  is $W= \Id/2 - \ketbra{GHZ}{GHZ}$.
Now, the GHZ state can be expressed in terms of its stabilizers as
\[ \ketbra{GHZ}{GHZ} = \frac{1}{8}(111 + ZZ1 + Z1Z + 1ZZ + XXX - XYY - YXY - YYX), \]

where $1,X,Y,Z$ represent the Pauli matrices $\Id, \sigma_x, \sigma_y, \sigma_z$,
and tensor product signs have been omitted. After a change of the normalization
this can be used to write the witness as
\[ W = {\Id} - 2 \ketbra{GHZ}{GHZ} =  \Id - \left[\frac{Z}{\sqrt{2}} \frac{Z1+1Z}{\sqrt{8}} +
\frac{X}{\sqrt{2}} \frac{XX-YY}{\sqrt{8}} +
\frac{-Y}{\sqrt{2}} \frac{YX+XY}{\sqrt{8}} +
\frac{1}{\sqrt{2}} \frac{11+ZZ}{\sqrt{8}}\right]. \]
From this representation, it is clear that $W$ is a witness as in
equation~(\ref{ccnrcond}) for the bipartition $\{\{1\}|\{2,3\}\}$, with $P_1=0.$
Due to the symmetry, this works for all bipartitions.

\section{Estimating the performance of PPT relaxations in high dimensions}
\label{section:volumes}

In this final section we want to discuss the overall performance of such relaxations in multipartite systems, using the paradigmatic partial transpose map. In order to estimate the performance of using PPT relaxations to detect randomly chosen multipartite entangled states we derive lower bounds on the fraction of multipartite entangled states, among states which are positive under partial transposition across every cut. The latter condition is strictly stronger than the relaxation employed in equation~\eqref{eq:relaxation}, thus providing an upper bound on the fraction of states in $\mathcal{R}_{\{\mathcal{T}_I\}_I}$ (where, as before, $\mathcal{T}_I$ stands for the transposition map on the subsystems in $I$) that are also in $\mathcal{S}_{(2)}$.

Our main result can be summarized as follows: for a fixed number of parties, the ratio between the size of fully PPT states and the size of  biseparable states (as measured by either the volume radius or the mean width) scales at least as $\sqrt{d}$, where $d$ is the local dimension.

In order to precisely formulate this result, we need to introduce first some of the basic notions and definitions that will be employed in the derivations.

\subsection{Notation and preliminary technical remarks}


All notation, concepts and results from classical convex
geometry, which are required throughout our proofs, are
gathered in Chapter \ref{chap:toolbox}, Section \ref{ap:convex-geometry} (see also Chapter \ref{chap:motivations}, Section \ref{sec:notation}, for general notation that we use in the remainder of this section).

It may, however, be worth mentioning that whenever we use
tools from convex geometry in the space $\cH(\C^n)$ of Hermitian operators on $\C^n$ (which
has real dimension $n^2$) it is tacitly understood that we
use the Euclidean structure induced by the Hilbert--Schmidt
inner product $\langle A,B \rangle = \tr(AB)$. For instance,
Definition \ref{def:vrad} of the volume radius of a convex
body $K\subset\cH(\C^n)$ becomes, denoting by $B_{HS}(\C^n)$ the
Hilbert--Schmidt unit ball of $\cH(\C^n)$,
\[ \vrad (K) = \left( \frac{\vol K}{\vol B_{HS}(\C^n)}
\right)^{1/n^2}. \]
While Definition \ref{def:w} of its mean width is, denoting
by $S_{HS}(\C^n)$ the Hilbert--Schmidt unit sphere of $\cH(\C^n)$
equipped with the uniform probability measure $\sigma$,
\[ w (K) = \int_{S_{HS}(\C^n)} \underset{M\in K}{\max}\tr(X M) \, \mathrm{d} \sigma (X).  \]
As also mentioned in Chapter \ref{chap:toolbox}, Section \ref{ap:convex-geometry}, the
latter quantity can be re-expressed via Gaussian variables,
which yields here
\[ w(K) = \frac{1}{\gamma(n)} \E \left(\underset{M\in K}{\max}\tr(G M)\right), \]
where $G$ is a matrix from the Gaussian Unitary Ensemble (GUE) on $\C^n$ and $\gamma(n)=\E\|G\|_2\sim_{n\rightarrow+\iy}n$ (see e.g.~\cite{AGZ}, Chapter 2, for a proof).

To be fully rigorous, let us make one last comment. All the
convex bodies of $\cH(\C^n)$ that we shall consider will
actually be included in the set $\cD(\C^n)$ of density
operators on $\C^n$ (i.e.~the set of positive and trace $1$
operators on $\C^n$). So we will in fact be working in an
ambient space of real dimension $n^2-1$ (namely the
hyperplane of $\cH(\C^n)$ composed of trace $1$ elements).
This subtlety will not be an issue though, since we will be
mostly interested in the asymptotic regime
$n\rightarrow+\iy$. In this setting, the operator that will
play for us the role of the origin will naturally be the
center of mass of $\cD(\C^n)$, i.e.~the maximally mixed
state $\Id/n$.


\begin{theorem} \label{th:vrad-w-states}
On $\C^n$, the volume radius and the mean width of the set $\cD(\C^n)$ of all quantum states satisfy the asymptotic estimates,
\begin{equation} \label{eq:vrad-states} \vrad\big(\cD(\C^n)\big) \underset{n\rightarrow+\iy}{\sim}\frac{e^{-1/4}}{\sqrt{n}}, \end{equation}
\begin{equation} \label{eq:w-states} w\big(\cD(\C^n)\big) \underset{n\rightarrow+\iy}{\sim}\frac{2}{\sqrt{n}}. \end{equation}
\end{theorem}

\begin{proof}
Equation~\eqref{eq:vrad-states} was established in \cite{SZ}.

Equation~\eqref{eq:w-states} is a direct consequence of Wigner's semicircle law (see e.g.~\cite{AGZ}, Chapter 2, for a proof). Indeed, we have by definition
\[ w\big(\cD(\C^n)\big) = \frac{1}{\gamma(n)}\E \left(\underset{\rho\in\cD(\C^n)}{\max}\tr\left[G\left(\rho-\frac{\Id}{n}\right)\right]\right)
= \frac{1}{\gamma(n)}\E \left(\underset{\rho\in\cD(\C^n)}{\max}\tr[G\rho]\right)
= \frac{1}{\gamma(n)} \E \left(\lambda_{max}(G)\right), \]
where $G$ is a GUE matrix on $\C^n$ and we denoted by $\lambda_{max}(G)$ its largest eigenvalue (the second equality being because $\E G=0$). The claimed result then follows from $\gamma(n)\sim_{n\rightarrow+\infty}n$ and $\E \left(\lambda_{max}(G)\right)\sim_{n\rightarrow+\infty}2\sqrt{n}$.
\end{proof}

\subsection{Volume estimates}

In the sequel, we shall consider the multipartite system $(\C^d)^{\otimes k}$, and slightly adapt and generalize the notation introduced in Section \ref{section:SD relaxations}. We shall denote by $\cS$ and $\cP$ the sets of states on $(\C^d)^{\otimes k}$ which are, respectively, separable and PPT across any bi-partition, and by $\cS_{(2)}$ and $\cP_{(2)}$ the sets of states on $(\C^d)^{\otimes k}$ which are, respectively, bi-separable and bi-PPT. These sets may be more precisely defined in the following way. There are $N_k=2^{k-1}-1$ different bi-partitions of the $k$ subsystems $\C^d$. Denoting by $\left\{\cS^1,\ldots,\cS^{N_k}\right\}$ and by $\left\{\cP^1,\ldots,\cP^{N_k}\right\}$ the sets of states which are, respectively, bi-separable and bi-PPT across one of these, we have
\begin{align*} & \cS=\bigcap_{i=1}^{N_k}\cS^i\ \ \text{and} \ \ \cP=\bigcap_{i=1}^{N_k}\cP^i,\\
& \cS_{(2)}=\conv\left(\bigcup_{i=1}^{N_k}\cS^i\right)\ \ \text{and} \ \ \cP_{(2)}=\conv\left(\bigcup_{i=1}^{N_k}\cP^i\right). \end{align*}

\begin{theorem} \label{th:fully-ppt} There exist positive constants $c_d\rightarrow_{d\rightarrow+\infty}1$ such that, on $(\C^d)^{\otimes k}$, the volume radius and the mean width of the set of states which are PPT across any bi-partition satisfy
\begin{equation} \label{eq:vrad-w-ppt} w(\cP)\geq\vrad(\cP)\geq c_d\, \frac{c^{2^k}}{d^{k/2}}, \end{equation}
where one may choose $c=e^{-1/4}/4$.
\end{theorem}

\begin{proof}
The first inequality in equation~\eqref{eq:vrad-w-ppt} is just by the Urysohn inequality (see Theorem \ref{theorem:urysohn} in Chapter \ref{chap:toolbox}, Section \ref{ap:convex-geometry}).

To show the second inequality in equation~\eqref{eq:vrad-w-ppt}, we will use repeatedly the Milman--Pajor inequality (see Theorem \ref{theorem:Milman-Pajor}) and more specifically its Corollary \ref{corollary:Milman-Pajor} (both of them in Chapter \ref{chap:toolbox}, Section \ref{ap:convex-geometry}). We will in fact show more precisely that there exist $c_d\rightarrow_{d\rightarrow+\infty}1$ such that
\begin{equation}\label{eq:vrad-ppt} \vrad(\cP)\geq c_d\,\frac{c^{N_k}\,e^{-1/4}}{d^{k/2}}. \end{equation}
The first thing to note is that, denoting by $\Gamma_1,\ldots,\Gamma_{N_k}$ the partial transpositions across the $N_k$ different bi-partitions of the $k$ subsystems $\C^d$, we have
\[ \cP = \cD\cap\cD^{\Gamma_1}\cap\cdots\cap\cD^{\Gamma_{N_k}}. \]
Now, by Corollary \ref{corollary:Milman-Pajor} applied to the convex body $\cD\subset\cH((\C^d)^{\otimes k})$ (which indeed has the origin $\Id/d^k$ as center of mass) and to the isometry $\Gamma_1$, we get
\[ \vrad\left(\cD\cap\cD^{\Gamma_1}\right)\geq\frac{1}{2}\frac{\vrad(\cD)^2}{w(\cD)} \underset{d\rightarrow+\iy}{\sim}c\times\frac{e^{-1/4}}{d^{k/2}}, \]
the last equivalence being by Theorem \ref{th:vrad-w-states}. We may then conclude recursively that equation~\eqref{eq:vrad-ppt} actually holds. \end{proof}

\begin{theorem} \label{th:bi-sep} On $(\C^d)^{\otimes k}$, the volume radius and the mean width of the set of bi-separable states satisfy
\begin{equation} \label{eq:vrad-w-sep2} \vrad(\cS_{(2)})\leq w(\cS_{(2)})\leq \frac{C+C_{d,k}}{d^{(k+1)/2}}, \end{equation}
where one may choose $C=\min\left\{6\sqrt{\ln(1+2/\delta)}/(1-2\delta^2)^2 \st 1/10<\delta<1/4 \right\}$ and $C_{d,k}=\sqrt{8\ln(2)/d^{k-1}}$, so that $C\leq 11$ and $C_{d,k}\rightarrow_{d\rightarrow+\infty}0$.
\end{theorem}

\begin{proof}
The first inequality in equation~\eqref{eq:vrad-w-sep2} is just by the Urysohn inequality (see Theorem \ref{theorem:urysohn} in Chapter \ref{chap:toolbox}, Section \ref{ap:convex-geometry}).

To show the second inequality in equation~\eqref{eq:vrad-w-sep2}, we start from this observation: for each $1\leq i\leq N_k$,
\begin{equation}\label{eq:w(S^i)} w(\cS^i) \leq \frac{C/2}{d^{(k+1)/2}}, \end{equation}
where $C=\min\big\{6\sqrt{\ln(1+2/\delta)}/(1-2\delta^2)^2 \st 1/10<\delta<1/4 \big\}$. It relies on the already known fact that there exists a universal constant $\widetilde{C}$ such that, for any $m,n\in\N$ with $m\leq n$, the mean width of the set $\cS$ of separable states on $\C^m\otimes\C^n$ is upper bounded by $\widetilde{C}/m\sqrt{n}$. In that way, the upper bound appearing in equation \eqref{eq:w(S^i)} is simply the upper bound obtained for one of the $k$ sets of states on $(\C^d)^{\otimes k}$ which are separable across a given bipartite cut $\C^d{:}(\C^d)^{\otimes k-1}$ (the largest of all upper bounds for sets of states on $(\C^d)^{\otimes k}$ which are separable across some bipartite cut). The former result was basically proved in \cite{AS1}, Theorem 1, but since specifically stated there in the balanced case $m=n$ only, for $\vrad(\cS)$ rather than $w(\cS)$ and without specifying that one may choose $\widetilde{C}=C/2$, we briefly recall the argument here.

Let $1/10<\delta<1/4$ and consider $\mathcal{A}_{\delta}$, $\mathcal{B}_{\delta}$ $\delta$-nets for $\|\cdot\|$ within the Euclidean unit spheres of $\C^m$ and $\C^n$ respectively. Imposing that $\mathcal{A}_{\delta}$, $\mathcal{B}_{\delta}$ have minimal cardinality, we know by volumetric arguments (see Lemma \ref{lemma:nets} in Chapter \ref{chap:toolbox}, Section \ref{ap:deviations}) that $\left|\mathcal{A}_{\delta}\right|\leq\left(1+2/\delta\right)^{2m}$ and $\left|\mathcal{B}_{\delta}\right|\leq\left(1+2/\delta\right)^{2n}$. Then, it may be checked that
\[ \conv\left(\cS\cup-\cS\right)\subset\frac{1}{(1-2\delta^2)^2}\conv\big\{\pm\ketbra{x}{x}\otimes\ketbra{y}{y} \st \ket{x}\in\mathcal{A}_{\delta},\ \ket{y}\in\mathcal{B}_{\delta}\big\}. \]
So by Lemma \ref{lemma:mean-width-polytope} in Chapter \ref{chap:toolbox}, Section \ref{ap:convex-geometry}, we get
\begin{align*} w(\cS) \leq\, & w\left(\conv\left(\cS\cup-\cS\right)\right)\\
\leq\, & \frac{1}{(1-2\delta^2)^2}\sqrt{\frac{2\ln\left(2(1+2/\delta)^{2m}(1+2/\delta)^{2n}\right)}{(mn)^2}}\\
=\, & \frac{1}{(1-2\delta^2)^2}\frac{\sqrt{4(m+n)\ln\left(1+2/\delta\right)+\ln(4)}}{mn}\\
\leq\, & \frac{3\sqrt{\ln\left(1+2/\delta\right)}}{(1-2\delta^2)^2m\sqrt{n}}, \end{align*}
which is precisely the content of equation~\eqref{eq:w(S^i)}.

Now, we also have that, for each $1\leq i\leq N_k$, $\cS^i\subset B_{1}\subset B_{HS}$. Hence, by Lemma \ref{lemma:mean-width-union-bound} in Chapter \ref{chap:toolbox}, Section \ref{ap:convex-geometry}, we get
\[ w(\cS_{(2)}) \leq 2\left(\underset{1\leq i\leq N_k}{\max} w(\cS^i) + \sqrt{\frac{2\ln(N_k)}{(d^k)^2}}\right) \leq \frac{C}{d^{(k+1)/2}} + \frac{\sqrt{8\ln(2)k}}{d^k}=\frac{C+C_{d,k}}{d^{(k+1)/2}}, \]
where $C_{d,k}=\sqrt{8\ln(2)/d^{k-1}}$.
\end{proof}

The conclusion of Theorems \ref{th:fully-ppt} and \ref{th:bi-sep} may be phrased as follows. On a multipartite system which is composed of a small number of big subsystems ($k$ fixed and $d\rightarrow+\infty$), imposing that a state is PPT across any bi-partition (i.e.~the strongest notion of PPT) is still, on average, a much less restrictive constraint than imposing that it is bi-separable (i.e.~the weakest notion of separability). Indeed, the ``sizes'' of these two sets of states (measured by either their volume radii or their mean widths) scale completely differently: the ``size'' of the former is at least of order $1/d^{k/2}$ while the ``size'' of the latter is at most of order $1/d^{(k+1)/2}$, hence differing by a factor of order at least $\sqrt{d}$.

\subsection{A class of fully PPT and GME states}

In Section \ref{sec:witnesses} an explicit class of GME states which are PPT across all cuts was presented. In small dimensions it is a hard task to find such examples, but the results from the previous section suggest that at least in high dimensions being GME should be a generic feature of fully PPT states. To emphasize this fact we present a construction of random states which are with high probability PPT across all cuts and GME.

Consider the following random state model on $(\C^d)^{\otimes k}$: fix some parameter $0<\alpha<1/4$ (independent of $d$), pick $G$ a traceless GUE matrix on $(\C^d)^{\otimes k}$, and define the ``maximally mixed + gaussian noise'' state on $(\C^d)^{\otimes k}$
\begin{equation}\label{eq:rho_G} \rho_G=\frac{1}{d^k}\left(\Id+\frac{\alpha}{d^{k/2}}G\right). \end{equation}
Then, typically (i.e.~with probability going to $1$ as $d$ grows) $\rho_G$ is fully PPT and nevertheless GME.

More quantitatively, we will show that the following result holds.

\begin{theorem} \label{th:fullyPPT-GME}
Let $G$ be a traceless GUE matrix on $(\C^d)^{\otimes k}$. Then, the state $\rho_G$ on $(\C^d)^{\otimes k}$, as defined by equation~\eqref{eq:rho_G}, is fully PPT and not bi-separable with probability greater than $1-\exp(-cd^{k-1})$, for some universal constant $c>0$.
\end{theorem}

Theorem \ref{th:fullyPPT-GME} is a straightforward consequence of Propositions \ref{prop:fullyPPT} and \ref{prop:GME} below. Before stating and proving them, let us elude once and for all a slight issue: a GUE matrix on $\C^n$ is the standard Gaussian vector in $\cH(\C^n)$, while a traceless GUE matrix on $\C^n$ is the standard Gaussian vector in the hyperplane of $\cH(\C^n)$ composed of trace $0$ elements. So in the asymptotic regime $n\rightarrow+\infty$, all the known results on $n\times n$ GUE matrices that we shall use also hold for traceless $n\times n$ GUE matrices (because the ambient spaces of these two Gaussian vectors have equivalent dimensions in this limit).

\begin{proposition} \label{prop:fullyPPT}
Let $G$ be a traceless GUE matrix on $(\C^d)^{\otimes k}$. Then, the state $\rho_G$ on $(\C^d)^{\otimes k}$, as defined by equation~\eqref{eq:rho_G}, satisfies
\[ \P\left( \rho_G\notin\cP\right) \leq N_ke^{-cd^k}, \]
where $c>0$ is a universal constant.
\end{proposition}

\begin{proof} In \cite{Aubrun3}, a deviation inequality is proved for the smallest eigenvalue of a GUE matrix, namely: Let $G$ be a GUE matrix on $\C^n$ and denote by $\lambda_{min}(G)$ its smallest eigenvalue. Then, for any $\e>0$,
\begin{align} \P\left(\lambda_{min}(G)<-(2+\e)\sqrt{n}\right) \leq e^{-c\e^{3/2}n}, \end{align}
where $c>0$ is a universal constant.

Now, observe that $G$ as well as all its partial transpositions $G^{\Gamma_i}$, $1\leq i\leq N_k$, are GUE matrices on $(\C^d)^{\otimes k}$. Hence, Proposition \ref{prop:fullyPPT} follows directly, by choosing for instance $\e=1$. Indeed, by assumption on $\alpha$, we have $3\alpha<3/4<1$, so the probability that $\rho_G$ or any $\rho_G^{\Gamma_i}$, $1\leq i\leq N_k$, is not positive is less than $e^{-cd^k}$. The advertised result follows by the union bound.
\end{proof}

\begin{proposition} \label{prop:GME}
Let $G$ be a traceless GUE matrix on $(\C^d)^{\otimes k}$. Then, the state $\rho_G$ on $(\C^d)^{\otimes k}$, as defined by equation~\eqref{eq:rho_G}, satisfies
\[ \P\left( \rho_G\in\cS_{(2)}\right) \leq e^{-cd^{k-1}}, \]
where $c>0$ is a universal constant.
\end{proposition}

\begin{proof} Our strategy to show Proposition \ref{prop:GME} is to exhibit a Hermitian $M$ on $(\C^d)^{\otimes k}$ which is with probability greater than $1-\exp(-cd^{k-1})$ a GME witness for the state $\rho_G$ (i.e.~such that $\tr(\rho_G M)<0$ while $\tr(\rho M)>0$ for any bi-separable state $\rho$).

Note first of all that, on the one hand,
\begin{equation} \label{eq:E1} \E\tr(\rho_GG) = \frac{\alpha}{d^{3k/2}}\E\tr(G^2) \underset{d\rightarrow +\infty}{\sim} \frac{\alpha}{d^{3k/2}}d^{2k} = \alpha d^{k/2}, \end{equation}
while on the other hand, by Theorem \ref{th:bi-sep},
\begin{equation} \label{eq:E2} \E\underset{\rho\in\cS_{(2)}}{\sup}\tr(\rho G) \leq \frac{C+C_{d,k}}{d^{(k+1)/2}} \E[\tr(G^2)]^{1/2} \underset{d\rightarrow +\infty}{\sim} \frac{C}{d^{(k+1)/2}}d^k= Cd^{(k-1)/2}, \end{equation}
where we used that for $G$ a GUE matrix on $\C^n$, $\E\tr(G^2)\sim_{n\rightarrow+\infty} n^2$ and $\E[\tr(G^2)]^{1/2}\sim_{n\rightarrow+\infty} n$ (see e.g.~\cite{AGZ}, Chapter 2, for a proof).

Let us now show that the functions $G\mapsto\tr(\rho_GG)$ and $G\mapsto\sup_{\rho\in\cS_{(2)}}\tr(\rho G)$ concentrate around their respective average values. In that aim, we shall make use of the following Gaussian deviation inequality (see e.g.~\cite{Pisier}, Chapter 2, for a proof), which is the analogue of Levy's spherical version appearing as Lemma \ref{lemma:levy} in Chapter \ref{chap:toolbox}, Section \ref{ap:deviations}: Assume that $f$ is a function satisfying, for any Gaussian random variables $G,H$, $|f(G)-f(H)|\leq \sigma_{G,H}\|G-H\|_2$ for some $\sigma_{G,H}$ such that $\E\sigma_{G,H}\leq L$. Then, for any $\e>0$,
\begin{equation} \label{eq:levy-gaussian} \P\left( \left|f-\E f\right| > \e \right) \leq e^{-c_0\e^2/L^2}, \end{equation}
where $c_0>0$ is a universal constant.

Define $f:G\in \mathcal{H}(\C^n)\mapsto\tr(G^2)$, and $f_{\Sigma}:G\in \mathcal{H}(\C^n)\mapsto\sup_{\rho\in\Sigma}\tr(\rho G)$, for any given set of states $\Sigma$ on $\C^n$. We have first,
\[ |f(G)-f(H)| = |\tr(GG^{\dagger})-\tr(HH^{\dagger})| \leq \|GG^{\dagger}-HH^{\dagger}\|_1 \leq (\|G\|_2+\|H\|_2)\|G-H\|_2, \]
where the last inequality is by the triangle inequality, the Cauchy--Schwarz inequality, and the invariance of $\|\cdot\|_2$ under conjugate transposition, after noticing that $GG^{\dagger}-HH^{\dagger}=G(G^{\dagger}-H^{\dagger})+(G-H)H^{\dagger}$. And second,
\[ \left|f_{\Sigma}(G)-f_{\Sigma}(H)\right| = \left|\underset{\rho\in\Sigma}{\sup}\tr(\rho G) - \underset{\rho\in\Sigma}{\sup}\tr(\rho H)\right|
\leq \underset{\rho\in\Sigma}{\sup}\left|\tr(\rho[G-H])\right| \leq \|G-H\|_{\infty} \leq \|G-H\|_2, \]
where the next to last inequality is because $\Sigma$ is a subset of the $1$-norm unit ball of Hermitians on $\C^n$.

Hence, the functions $f$ and $f_{\Sigma}$ both satisfy the hypotheses of the Gaussian deviation inequality \eqref{eq:levy-gaussian}, with $L=2\gamma(d^k)\sim_{d\rightarrow+\infty}2d^k$ and $L=1$ respectively. So by the mean estimates \eqref{eq:E1} and \eqref{eq:E2}, we have that for any $0<\e<1$,
\begin{equation} \label{eq:F1} \P\left( \tr\left(\rho_GG\right) < (1-\e)\alpha d^{k/2} \right) \leq \exp\left(-c_0\left(\e\alpha d^{2k}\right)^2/\left(2d^k\right)^2\right) = \exp\left(-c'_0\e^2d^{2k}\right), \end{equation}
\begin{equation} \label{eq:F2} \P\left( \underset{\rho\in\cS_{(2)}}{\sup}\tr(\rho G) > (1+\e)Cd^{(k-1)/2} \right) \leq \exp\left(-c_0\left(\e Cd^{(k-1)/2}\right)^2\right) = \exp\left(-c'_0\e^2d^{k-1}\right). \end{equation}

As a consequence, we have that for any $\beta_d$ satisfying $1/2Cd^{(k-1)/2}\leq \beta_d\leq 3/2\alpha d^{k/2}$, the Hermitian $M=\Id-\beta_dG$ on $(\C^d)^{\otimes k}$ is a GME witness for $\rho_G$ with probability greater than $1-\exp(-cd^{k-1})$, where $c>0$ is a universal constant. Indeed, choosing $\e=1/6$ in equation~\eqref{eq:F1} and $\e=1/2$ in equation~\eqref{eq:F2}, we get
\[ \P\left( \tr\left(\rho_GM\right) > -\frac{1}{4} \right) \leq e^{-cd^{2k}}\ \text{and}\ \P\left( \underset{\rho\in\cS_{(2)}}{\sup}\tr(\rho M) < \frac{1}{4} \right) \leq e^{-cd^{k-1}}, \]
which concludes the proof.
\end{proof}

We may actually say even more on the random state $\rho_G$ defined by equation \eqref{eq:rho_G}. Indeed, define for all $0<\e<1$ the state $\widetilde{\rho}_G(\e)$ on $(\C^d)^{\otimes k}$ by
\[ \widetilde{\rho}_G(\e)= \e\rho_G+(1-\e)\frac{\Id}{d^k} = \frac{1}{d^k}\left(\Id+\e\frac{\alpha}{d^{k/2}}G\right). \]
Then, what the proof of Proposition \ref{prop:GME} additionally tells us is that, as long as $\e\geq \gamma/\sqrt{d}$, for some constant $\gamma>0$, $\widetilde{\rho}_G(\e)$ is with high probability not bi-separable. This means that $\rho_G$ is typically a fully PPT state on $(\C^d)^{\otimes k}$ which is not bi-separable, and whose random robustness of genuinely multipartite entanglement (as defined in \cite{TV}) additionally grows at least as $\sqrt{d}$ when $d\rightarrow+\infty$.

\section{Conclusion}

The problem of characterizing genuine multipartite entanglement and biseparability
is difficult. Therefore, a natural approach lies in the relaxation of the definition of
biseparability: instead of considering states which are separable with respect to some
bipartition, one replaces this set by an appropriate superset, e.g.~defined by the PPT
condition or some other positivity under partial application of a positive, yet not completely positive, map.

\smallskip

In this chapter we investigated this angle of attack from several perspectives. First, we
established how this relaxation approach with positive maps can be evaluated with semidefinite
programming and how it can be used to construct entanglement witnesses for this problem. Then, we
showed that, in principle, also other relaxations, besides those obtained from positive maps (e.g.~based on
the CCNR criterion), are possible. Finally, we studied the accurateness of the relaxation approach.
We proved rigorous bounds on the volume of the set of biseparable states as well as on the volume of the set of
states which are PPT for any cut. In this way, we showed that in the limit of large dimensional multipartite systems, the relaxation approach detects only a small fraction of the
multiparticle entangled states. It must be stressed, however, that this does not mean
that the relaxation method is not fruitful. Indeed, it is a well known fact from the theory
of two-particle entanglement that, already in such case, simple entanglement criteria miss most of the states if
the dimension of the local spaces increases \cite{BS,AS1,AS2}. However, from a practical point of view,
relaxation techniques are clearly the best tools for characterizing multiparticle
entanglement available at the moment \cite{HS, GJM}.

\smallskip

For future research, there are many open questions to address. First, a more systematic
analysis for the various positive maps besides the transposition would be desirable.
Then, an approach for characterizing separability classes besides biseparability
(e.g.~$\ell$-separability, for any $\ell$) would be useful. Finally, methods to certify the Schmidt-rank or the dimensionality of entanglement \cite{HdV} in high-dimensional systems are needed for current experiments. Investigating the generic scaling of these quantities could also be of interest.


\chapter{$k$-extendibility of high-dimensional bipartite quantum states}
\label{chap:k-extendibility}

\textsf{Based on ``$k$-extendibility of high-dimensional bipartite quantum states'' \cite{Lancien}.}

\bigskip

The idea of detecting the entanglement of a given bipartite state by searching for symmetric extensions of this state was first proposed by Doherty, Parrilo and Spedialeri. The complete family of separability tests it generates, often referred to as the hierarchy of \textit{$k$-extendibility tests}, has already proved to be most promising. The goal of this chapter is to try and quantify the efficiency of this separability criterion in typical scenarios. For that, we essentially take two approaches. First, we compute the average width of the set of $k$-extendible states, in order to see how it scales with the one of separable states. And second, we characterize when random-induced states are, depending on the ancilla dimension, with high probability violating or not the $k$-extendibility test, and compare the obtained result with the corresponding one for entanglement vs separability. The main results can be precisely phrased as follows: on $\C^d\otimes\C^d$, when $d$ grows, the average width of the set of $k$-extendible states is equivalent to $(2/\sqrt{k})/d$, while random states obtained as partial traces over an environment $\C^s$ of uniformly distributed pure states are violating the $k$-extendibility test with probability going to $1$ if $s<((k-1)^2/4k)d^2$. Both statements converge to the conclusion that, if $k$ is fixed, $k$-extendibility is asymptotically a weak approximation of separability, even though any of the other well-studied separability relaxations is outperformed by $k$-extendibility as soon as $k$ is above a certain (dimension independent) value.

\section{Introduction}

Deciding whether a given bipartite quantum state is entangled or separable (or even just close to separable) is known to be a computationally hard task (see \cite{Gurvits} and \cite{Gharibian}). Several much more easily checkable necessary conditions for separability do exist though, the most famous and widely used ones being perhaps the positivity of partial transpose criterion \cite{Peres}, the realignment criterion \cite{CW} or the $k$-extendibility criterion \cite{DPS}. All of them have in common that verifying if a given state fulfils them or not may be cast as a Semi-Definite Program (SDP) and hence be efficiently solved (see Chapter \ref{chap:SDrelaxations} of this manuscript, and the quite extensive review \cite{Doherty}, for much more on that topic).

We focus here on a relaxation of the notion of separability of quite different kind: the so-called $k$-extendibility criterion for separability, which was introduced in \cite{DPS}. It is especially appealing because it provides a hierarchy of increasingly powerful separability tests (expressible as SDPs of increasing dimension), which is additionally complete, meaning that any entangled state is guaranteed to fail a test after some finite number of steps in the hierarchy. Let us be more precise.

\begin{definition}
Let $k\in\N$. A state $\rho_{\A\B}$ on a bipartite Hilbert space $\mathrm{A}\otimes\mathrm{B}$ is $k$-extendible with respect to $\mathrm{B}$ if there exists a state $\rho_{\A\B^k}$ on $\mathrm{A}\otimes\mathrm{B}^{\otimes k}$ which is invariant under any permutation of the $\mathrm{B}$ subsystems and such that $\rho_{\A\B}=\tr_{\B^{k-1}}\rho_{\A\B^k}$.
\end{definition}

\begin{theorem}[The complete family of $k$-extendibility criteria for separability, \cite{DPS}] \label{th:k-extendibility-criterion}
A state on a bipartite Hilbert space $\mathrm{A}\otimes\mathrm{B}$ is separable if and only if it is $k$-extendible with respect to $\mathrm{B}$ for all $k\in\N$.
\end{theorem}

Note that one direction in Theorem \ref{th:k-extendibility-criterion} is obvious, namely that a separable state on some bipartite system is necessarily $k$-extendible for all $k\in\N$ (with respect to both subsystems). Indeed, if $\rho=\sum_{x}p_x\sigma_x\otimes\tau_x$ is separable, then $\sum_{x}p_x\sigma_x^{\otimes k}\otimes\tau_x$ and $\sum_{x}p_x\sigma_x\otimes\tau_x^{\otimes k}$ are symmetric extensions of $\rho$ to $k$ copies of the first and second subsystems respectively. The other direction in Theorem \ref{th:k-extendibility-criterion} follows from the quantum finite De Finetti theorem (see e.g.~\cite{KR,CKMR} and Chapter \ref{chap:symmetries}, Section \ref{sec:deFinetti}, of this manuscript). The latter establishes, roughly speaking, that starting from a permutation-invariant state on some tensor power system and tracing out all except a few of the subsystems, one gets a state that may be well-approximated by a convex combination of tensor power states (with a vanishing error as the initial number of subsystems increases).

\begin{figure}[h] \caption{The nested and converging sequence of $k$-extendibility relaxations of separability, $k\in\N$}
\label{fig:k-ext}
\begin{center}
\includegraphics[width=6cm]{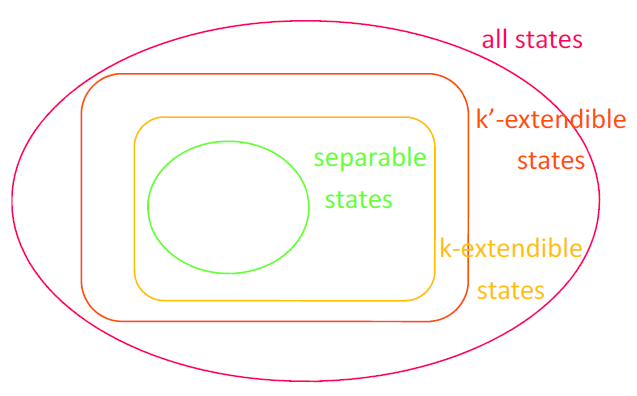}
\end{center}
\end{figure}

It is easy to see that if a state is $k$-extendible for some $k\in\N$, then it is automatically $k'$-extendible for all $k'\leq k$. Hence, the necessary and sufficient condition for separability provided by Theorem \ref{th:k-extendibility-criterion} actually decomposes into a series of increasingly constraining necessary conditions for separability, which are only asymptotically also sufficient (see Figure \ref{fig:k-ext}). In real life however, checks can only be done up to a finite level in this hierarchy. It thus makes sense to ask, given a finite $k\in\N$, how ``powerful'' the $k$-extendibility test is to detect entanglement.

Actually, various more quantitative versions of Theorem \ref{th:k-extendibility-criterion} do exist, that put bounds on how far a $k$-extendible state can be from separable. Let us mention two quite different statements in that direction. The original result, appearing in \cite{CKMR}, establishes that a state on $\mathrm{A}\otimes\mathrm{B}$ which is $k$-extendible with respect to $\mathrm{B}$ is at distance at most $2d_{\B}^2/k$, in $1$-norm, from the set of separable states. It is a direct consequence of one of the quantitative versions of the quantum finite de Finetti theorem. A more recent result, proved essentially in \cite{BCY} and improved in \cite{BH}, stipulates that such a state is at distance at most $\sqrt{2\ln d_{\A}/k}$, in $\mathbf{LOCC}^{\rightarrow}$-norm, from the set of separable states (see Chapter \ref{chap:data-hiding}, Section \ref{sec:POVM-geometry}, for a precise definition of the operational one-way-LOCC norm). It relies on the observation that a $k$-extendible state has a small squashed entanglement, and therefore cannot be distinguished well from a separable state by local observers (see Chapter \ref{chap:deFinetti}, Section \ref{sec:sep2}, for a related discussion). The main problem of such estimates is that they become non-trivial only when $k\gg d_{\B}$ or $k\gg \ln d_{\A}$. So in the case where $d_{\A},d_{\B}$ are ``big'', can anything interesting still be said for a ``not too big'' $k$? On the other hand, these bounds valid for any $k$-extendible state are known to be close from optimal (there are examples of $k$-extendible states whose closest separable state is at distance of order $d_{\B}/k$ in $1$-norm or of order $\sqrt{\ln d_{\A}/k}$ in $\mathbf{LOCC}^{\rightarrow}$-norm). Consequently, one may only hope to make stronger statements about average behaviours.

This is precisely the general question we address here, being especially interested in the case of high-dimensional bipartite quantum systems. We try and quantify in two distinct ways the typical efficiency of the $k$-extendibility criterion for separability in this asymptotic regime.

The first approach consists in estimating a specific size parameter (known as the \textit{mean width}) of the set of $k$-extendible states when the dimension of the underlying Hilbert space goes to infinity. Comparing the obtained value with the known asymptotic estimate for the mean width of the set of separable states then tells us how the sizes of these two sets of states scale with one another. The computation is carried out in Section \ref{section:w(k-ext)} (where all needed notions related to high-dimensional convex geometry are properly defined as well) and ends with the concluding Theorem \ref{th:w(k-ext)}, some technical parts being relegated to Appendix \ref{appendix:gaussian}. In Section \ref{section:w-comparison}, the result is commented and comparisons are made between the mean-widths of, on the one hand, $k$-extendible states, and on the other, separable or PPT states. Besides, a smaller upper bound is derived, in Section \ref{section:w(k-ext-ppt]} and its companion technical Appendix \ref{appendix:gaussian-gamma}, on the mean width of the set of $k$-extendible states whose extension is required to be PPT (precise definitions and motivations to look at this set of states appear there).

The second approach consists in looking at random mixed states which are obtained by partial tracing over an ancilla space a uniformly distributed pure state, and characterizing when these are, with overwhelming probability as the dimension of the system grows, $k$-extendible or not. Again, comparing the obtained result with the known one for separability provides some information on how powerful the $k$-extendibility test is to detect entanglement. Section \ref{section:random-states} introduces all required material regarding the considered model of random-induced states and one possible way of detecting their non-$k$-extendibility. The adopted strategy is next seen through in Section \ref{section:non-k-ext-witness}, relying on technical statements put in Appendix \ref{appendix:wishart}, and concludes as Theorem \ref{th:not-kext}. The determined environment dimension below which random-induced states are with high probability violating the $k$-extendibility criterion is then compared, in Section \ref{section:threshold-comparison}, with the previously established ones for violating other separability criteria, and for actually not being separable.

Finally, generalizations to the unbalanced case are stated in Section \ref{section:unbalanced} (Theorems \ref{th:unbalanced} and \ref{th:unbalanced'}), while Section \ref{section:miscellaneous} exposes miscellaneous concluding remarks and loose ends.

The reader may have a look at Table \ref{table:comparison} for a sample corollary of this study.

\begin{table}[h] \caption{Comparison of the average and typical case performance of the $k$-extendibility criterion with that of the PPT and realignment criteria}
\label{table:comparison}
\begin{center}
\renewcommand{\arraystretch}{1.7}
\begin{tabular}{| C{8cm} | C{2.5cm} | C{2.5cm} |}
\hline
\diagbox[width=8cm]{$k$-extendibility ``beats''}{from the point of view of} & mean width of the set & entanglement detection of random states\\
\hline
PPT & for $k\geq 11$ & for $k\geq 17$ \\
\hline
realignment & ? & for $k\geq 5$ \\
\hline
\end{tabular}
\end{center}
\end{table}

Appendix \ref{appendix:combinatorics} gathers a bunch of standard definitions and facts about the combinatorics of permutations and partitions which are necessary for our purposes. All employed notation on that matter are also introduced there. In Appendix \ref{appendix:wick}, the connection is made between computing moments of GUE or Wishart matrices and counting permutations having a certain genus. These general observations play a key role in the moments' derivations of Appendices \ref{appendix:gaussian}, \ref{appendix:gaussian-gamma} and \ref{appendix:wishart}, which are, as for them, specifically the ones that we need to obtain our various statements. To get tractable expressions, though, a formula relating the number of cycles in some specific permutations is additionally required, whose proof is detailed in Appendix \ref{appendix:technical}. Appendix \ref{appendix:geodesics}, finally, is devoted to establishing the last crucial ingredient in most of our reasonings, namely bounding the number of non-geodesic permutations (in terms of the number of geodesic ones) in some particular instances which are of interest to us. Aside, Appendix \ref{appendix:convergence} is dedicated to proving more precise results than the ones which are strictly needed on the convergence of the studied random matrix ensembles, and in generalizing the developed method to establish the asymptotic freeness of certain gaussian random matrices.

\subsection*{Notation} Beside the general notation specified in Chapter \ref{chap:motivations}, Section \ref{sec:notation} (Hermitian and positive semidefinite operators on a Hilbert space, Schatten $p$-norm etc.). there are a few more specific ones that we will use repeatedly in the remainder of this chapter.
Here, we will in fact always consider the case where the state space $\mathrm{H}$ of interest is finite-dimensional and bipartite, i.e.~ $\mathrm{H}=\mathrm{A}\otimes\mathrm{B}$, with $\mathrm{A},\mathrm{B}$ finite-dimensional Hilbert spaces. And we introduce the additional notation: $\mathcal{D}(\mathrm{A}\otimes\mathrm{B})$ for the set of all states on $\mathrm{H}$, $\mathcal{S}(\A{:}\B)$ for the set of separable states on $\mathrm{H}$, $\mathcal{P}(\A{:}\B)$ for the set of PPT states on $\mathrm{H}$ (in both cases in the cut $\A{:}\B$), and for each $k\in\N$, $\mathcal{E}_k(\A{:}\B)$ for the set of $k$-extendible states on $\mathrm{H}$ (in the cut $\A{:}\B$ and with respect to $\mathrm{B}$). What is more, we will actually almost exclusively deal with the balanced case, i.e.~when $\mathrm{A}\equiv\mathrm{B}\equiv\C^d$.

\subsection*{Preliminary technical lemma}

It will be essential for us in the sequel to express in a more tractable way the quantity $\sup_{\sigma\in\mathcal{E}_k(\A{:}\B)}\mathrm{Tr}(M\sigma)$, for any given Hermitian $M$ on $\A\otimes\B$. Such amenable expression is provided by Lemma \ref{lemma:sup-k-ext} below.

\begin{lemma} \label{lemma:sup-k-ext}
Let $k\in\N$. For any $M_{\A\B}\in\mathcal{H}(\mathrm{A}\otimes \mathrm{B})$, we have
\[ \underset{\sigma_{\A\B}\in\mathcal{E}_k(\A{:}\B)}{\sup}\ \tr\big(M_{\A\B}\sigma_{\A\B}\big) = \left\|\frac{1}{k}\underset{j=1}{\overset{k}{\sum}}\widetilde{M}_{\A\B^k}(j)\right\|_{\infty}, \]
where for each $1\leq j\leq k$, denoting by $\Id_{\widehat{\B}^k_j}$ the identity on $\B_1\otimes\cdots\otimes \B_{j-1}\otimes \B_{j+1}\otimes\cdots \otimes \B_k$, we defined $\widetilde{M}_{\A\B^k}(j)=M_{\A\B_j}\otimes\Id_{\widehat{\B}^k_j}$.
\end{lemma}

Before proving Lemma \ref{lemma:sup-k-ext}, let us introduce once and for all the following notation, which we shall later use on several occasions: for any $M_{\A\B^k}\in\mathcal{H}(\mathrm{A}\otimes \mathrm{B}^{\otimes k})$, we define its symmetrisation with respect to $\mathrm{B}^{\otimes k}$ as
\[ \Sym_{\A{:}\B^k}(M_{\A\B^k}) = \frac{1}{k!}\underset{\pi\in\mathfrak{S}(k)}{\sum}\big(\Id_\A\otimes U(\pi)_{\B^k}\big) M_{\A\B^k}\big(\Id_\A\otimes U(\pi)_{\B^k}\big)^{\dagger}, \]
where for each permutation $\pi\in\mathfrak{S}(k)$, $U(\pi)_{\B^k}$ denotes the associated permutation unitary on $\mathrm{B}^{\otimes k}$ (see e.g.~\cite{Harrow} and Chapter \ref{chap:symmetries}, Section \ref{sec:sym}, of this manuscript for further details).

\begin{proof}
By definition, the condition $\sigma_{\A\B}\in\mathcal{E}_k(\A{:}\B)$ is equivalent to the condition $\sigma_{\A\B}=\tr_{\B^{k-1}}\Sym_{\A{:}\B^k}(\sigma_{\A\B^k})$ for some $\sigma_{\A\B^k}\in\mathcal{D}(\mathrm{A}\otimes \mathrm{B}^{\otimes k})$. Hence, for any $M_{\A\B}\in\mathcal{H}(\mathrm{A}\otimes \mathrm{B})$, we have
\begin{align*}
\underset{\sigma_{\A\B}\in\mathcal{E}_k(\A{:}\B)}{\sup}\ \tr_{\A\B}\big[M_{\A\B}\sigma_{\A\B}\big] &=\underset{\sigma_{\A\B^k}\in\mathcal{D}(\mathrm{A}\otimes\mathrm{B}^{\otimes k})}{\sup}\ \tr_{\A\B} \big[M_{\A\B} \tr_{\B^{k-1}}\Sym_{\A{:}\B^k}(\sigma_{\A\B^k})\big]\\
&= \underset{\sigma_{\A\B^k}\in\mathcal{D}(\mathrm{A}\otimes\mathrm{B}^{\otimes k})}{\sup}\ \tr_{\A\B^{k}} \big[\big(M_{\A\B}\otimes\Id_{\B^{k-1}}\big) \Sym_{\A{:}\B^k}(\sigma_{\A\B^k})\big]\\
&= \underset{\sigma_{\A\B^k}\in\mathcal{D}(\mathrm{A}\otimes\mathrm{B}^{\otimes k})}{\sup}\ \tr_{\A\B^{k}} \big[ \Sym_{\A{:}\B^k}\big(M_{\A\B}\otimes\Id_{\B^{k-1}}\big)\sigma_{\A\B^k}\big]\\
&= \left\|\Sym_{\A{:}\B^k} \big(M_{\A\B}\otimes\Id_{\B^{k-1}}\big)\right\|_{\infty}.
\end{align*}
Now, for each $\pi\in\mathfrak{S}(k)$, $\big(\Id_\A\otimes U(\pi)_{\B^k}\big) \big(M_{\A\B}\otimes\Id_{\B^{k-1}}\big) \big(\Id_\A\otimes U(\pi)_{\B^k}\big)^{\dagger}= M_{\A\B_{\pi(1)}}\otimes \Id_{\widehat{\B}^k_{\pi(1)}}$. Therefore, grouping together, for each $1\leq j\leq k$, the permutations $\pi\in\mathfrak{S}(k)$ such that $\pi(1)=j$, we get
\[ \Sym_{\A{:}\B^k} \big(M_{\A\B}\otimes\Id_{\B^{k-1}}\big)=\frac{1}{k}\sum_{j=1}^k M_{\A\B_j}\otimes\Id_{\widehat{\B}^k_j}, \]
and hence the advertised result.
\end{proof}

\section{Mean width of the set of $k$-extendible states for ``small'' $k$}
\label{section:w(k-ext)}

\subsection{Preliminaries on convex geometry}

Let us introduce a few notions coming from classical convex geometry which we shall need in the sequel. Much more details on that matter appear in Chapter \ref{chap:toolbox}, Section \ref{ap:convex-geometry}. For any $K\subset\mathcal{H}(\C^n)$ and any $M\in\mathcal{H}(\C^n)$ having unit Hilbert--Schmidt norm, we define the width of $K$ in the direction $M$ as
\[ w(K,M) = \underset{\Delta\in K}{\sup}\mathrm{Tr}(M\Delta) . \]
The mean width of $K$ is then defined as the average of $w(K,\cdot)$ over the whole Hilbert--Schmidt unit sphere $S_{HS}(\C^n)$ of $\mathcal{H}(\C^n)$ (equipped with the Haar measure $\sigma$) i.e.~\[ w(K)= \int_{M\in S_{HS}(\C^n)}w(K,M)\mathrm{d}\sigma(M) = \int_{M\in S_{HS}(\C^n)} \left[\underset{\Delta\in K}{\sup}\mathrm{Tr}(M\Delta)\right]\mathrm{d}\sigma(M). \]
This average width $w$ is an interesting size parameter, on its own, but also because it is related to other important geometric quantities, such as e.g.~the volume radius $\mathrm{vrad}$, which is defined as the radius of the Euclidean ball having same volume (i.e.~Lebesgue measure). For instance, we have for any convex body $K$ the Urysohn inequality $w(K)\geq \mathrm{vrad}(K)$, and for most of the convex bodies $K$ we shall be considering a ``reverse'' Urysohn inequality $w(K)\leq \mu\,\mathrm{vrad}(K)$ for some $\mu\geq 1$. These connections and the precise formulation of these convex geometry results are exemplified in Section \ref{section:w-comparison}.

In order to compute the quantity $w(K)$, it is often convenient to re-express it as a Gaussian rather than spherical averaging. We thus denote by $GUE(n)$ the Gaussian Unitary Ensemble on $\C^n$, which is the standard Gaussian vector in $\mathcal{H}(\C^n)$ (equivalently, $G\sim GUE(n)$ if $G=(H+H^{\dagger})/\sqrt{2}$ with $H$ a $n\times n$ matrix having independent complex normal entries). And we define the Gaussian mean width of $K$ as
\[ w_G(K) = \E_{G\sim GUE(n)}\left[\underset{\Delta\in K}{\sup}\mathrm{Tr}(G\Delta)\right] .\]
Just observing that for $G\sim GUE(n)$, $G/\|G\|_2$ is uniformly distributed over $S_{HS}(\C^n)$, and $G/\|G\|_2$, $\|G\|_2$ are independent random variables, we get that the link between both quantities is, setting $\gamma(n)=\E_{G\sim GUE(n)}\|G\|_2$, which is known to satisfy $\gamma(n)\sim_{n\rightarrow +\infty}n$ (see e.g.~\cite{AGZ}, Chapter 2, for a proof),
\begin{equation} \label{eq:w-wG} w(K)= \frac{1}{\gamma(n)}w_G(K). \end{equation}

\begin{remark} All the sets $K$ that we will consider in the sequel will actually be subsets of $\mathcal{D}(\C^n)$, hence living in the hyperplane of $\mathcal{H}(\C^n)$ composed of trace $1$ elements, i.e.~in a space of real dimension $n^2-1$, rather than $n^2$. It would thus seem more natural to define their mean width $w(K)$ as an average width over a $n^2-2$, rather than $n^2-1$, dimensional Euclidean unit sphere. The Gaussian mean width $w_G(K)$, on the other hand, is an intrinsic notion that does not depend on the ambient dimension (because marginals of standard Gaussian vectors are themselves standard Gaussian vectors). As a consequence, we see from equation \eqref{eq:w-wG} that computing the mean width of $K$ as if it was a $n^2$ dimensional set is asymptotically equivalent to computing it taking into account that it is in fact a $n^2-1$ dimensional set. We may therefore serenely forget about this issue.
\end{remark}


Our aim is now to estimate, for any fixed $k\in\N$, the mean width of the set of $k$-extendible states on $\A\otimes\B$ when $\A\equiv\B\equiv\C^d$ and $d\rightarrow+\infty$. By the definitions above, we have
\[ w\big(\mathcal{E}_k(\A{:}\B)\big) 
= \frac{1}{\gamma(d^2)} \mathbf{E}_{G_{\A\B}\sim GUE(d^2)} \left[ \underset{\sigma_{\A\B}\in \mathcal{E}_k(\A{:}\B)}{\sup}\mathrm{Tr}\big(G_{\A\B}\sigma_{\A\B}\big) \right]. \]
Using the result of Lemma \ref{lemma:sup-k-ext}, and the notation introduced there, we thus get,
\begin{equation} \label{eq:w(k-ext)-definition} w\big(\mathcal{E}_k(\A{:}\B)\big) = \frac{1}{\gamma(d^2)} \mathbf{E}_{G_{\A\B}\sim GUE(d^2)} \left\| \frac{1}{k} \underset{j=1}{\overset{k}{\sum}}\widetilde{G}_{\A\B^k}(j) \right\|_{\infty}. \end{equation}

\subsection{An operator-norm estimate}

As justified above, to obtain the mean width of the set of $k$-extendible states on $\C^d\otimes\C^d$, what we need is to compute the average operator-norm $\mathbf{E} \left\|\sum_{j=1}^k\widetilde{G}_{\A\B^k}(j)\right\|_{\infty}$, for $G$ a GUE matrix on $\C^d\otimes\C^d$. We will show that the following asymptotic estimate holds.

\begin{proposition} \label{prop:gaussian-infty}
Fix $k\in\N$. Then,
\[ \mathbf{E}_{G_{\A\B}\sim GUE(d^2)} \left\| \underset{j=1}{\overset{k}{\sum}}\widetilde{G}_{\A\B^k}(j) \right\|_{\infty} \underset{d\rightarrow+\infty}{\sim} 2\sqrt{k}d .\]
\end{proposition}

As a preliminary step towards estimating the sup-norm $\mathbf{E} \left\|\sum_{j=1}^k\widetilde{G}_{\A\B^k}(j)\right\|_{\infty}$, we will look at the $2p$-order moments $\mathbf{E}\, \mathrm{Tr}\left[\left(\sum_{j=1}^k\widetilde{G}_{\A\B^k}(j)\right)^{2p}\right]$, $p\in\mathbf{N}$, and show that they can be expressed in terms of the $2p$-order moments of a centered semicircular distribution of appropriate parameter.

So let us recall first a few required definitions. For any $\sigma>0$, we shall denote by $\mu_{SC(\sigma^2)}$ the centered semicircular distribution of variance parameter $\sigma^2$, whose density is given by
\[ \mathrm{d}\mu_{SC(\sigma^2)}(x)=\frac{1}{2\pi\sigma^2}\sqrt{4\sigma^2-x^2} \mathbf{1}_{[-2\sigma,2\sigma]}(x)\mathrm{d}x .\]
We shall also denote, for each $p\in\N$, by $\mathrm{M}_{SC(\sigma^2)}^{(p)}$ its $p$-order moment, i.e.~ $\mathrm{M}_{SC(\sigma^2)}^{(p)}=\int_{-\infty}^{+\infty}x^{p}\mathrm{d}\mu_{SC(\sigma^2)}(x)$. It is well-known that
\[ \forall\ p\in\N,\ \mathrm{M}_{SC(\sigma^2)}^{(2p-1)}=0\ \text{and}\ \mathrm{M}_{SC(\sigma^2)}^{(2p)}=\sigma^{2p}\,\mathrm{Cat}_p, \]
where $\mathrm{Cat}_p$ is the $p^{\text{th}}$ Catalan number defined in Lemma \ref{lemma:catalan}.

\begin{proposition} \label{prop:gaussian-p} Fix $k\in\N$. Then, when $d\rightarrow+\infty$, the random matrix $\left(\sum_{j=1}^k\widetilde{G}_{\A\B^k}(j)\right)/d$ converges in moments towards a centered semicircular distribution of parameter $k$. Equivalently, this means that, for any $p\in\N$,
\begin{align*}
& \mathbf{E}_{G_{\A\B}\sim GUE(d^2)}\, \mathrm{Tr}\left[ \left( \underset{j=1}{\overset{k}{\sum}}\widetilde{G}_{\A\B^k}(j) \right)^{2p-1} \right]=0,\\
& \mathbf{E}_{G_{\A\B}\sim GUE(d^2)}\, \mathrm{Tr}\left[ \left( \underset{j=1}{\overset{k}{\sum}}\widetilde{G}_{\A\B^k}(j) \right)^{2p} \right] \underset{d\rightarrow+\infty}{\sim} \mathrm{M}_{SC(k)}^{(2p)}d^{2p+k+1}.
\end{align*}
\end{proposition}

\begin{remark}
Stronger convergence results than the one established in Proposition \ref{prop:gaussian-p} may in fact be proved, as discussed in Appendix \ref{appendix:convergence}.
\end{remark}

\begin{proof} [Proof of Proposition \ref{prop:gaussian-p}]
Let $p\in\N$. Computing the value of the $2p$-order moment $\mathbf{E}\, \mathrm{Tr}\left[\left(\sum_{j=1}^k\widetilde{G}_{\A\B^k}(j)\right)^{2p}\right]$ may be done using the Gaussian Wick formula (see Lemma \ref{lemma:wick} for the statement and Appendix \ref{appendix:wick-gaussian} for a succinct summary of how to derive moments of GUE matrices from it). In our case, what we get by the computations carried out in Appendix \ref{appendix:gaussian} and summarized in Proposition \ref{prop:gaussian-p-preliminary} is that, for any $d\in\N$, denoting by $\sharp(\cdot)$ the number of cycles in a permutation,
\begin{align}
\mathbf{E}\, \mathrm{Tr}\left[\left(\sum_{j=1}^k\widetilde{G}_{\A\B^k}(j)\right)^{2p}\right] & =  \sum_{f:[2p]\rightarrow[k]} \mathbf{E}\, \mathrm{Tr}\left[\underset{i=1}{\overset{2p}{\overrightarrow{\prod}}}\widetilde{G}_{\A\B^k}(f(i))\right]\\
& =  \underset{f:[2p]\rightarrow[k]}{\sum} \underset{\lambda\in\mathfrak{P}^{(2)}(2p)}{\sum} d^{\sharp(\gamma^{-1}\lambda)+\sharp(\gamma_f^{-1}\lambda) + k - |\im(f)|},
\label{eq:p-moment-gaussian} \end{align}
where we defined on $\{1,\ldots,2p\}$, $\mathfrak{P}^{(2)}(2p)$ as the set of pair partitions, $\gamma=(2p\,\ldots\,1)$ as the canonical full cycle, and for each $f:[2p]\rightarrow[k]$, $\gamma_f=\gamma_{f=1}\cdots\gamma_{f=k}$ as the product of the canonical full cycles on each of the level sets of $f$.

We now have to understand which $\lambda\in\mathfrak{P}^{(2)}(2p)$ and $f:[2p]\rightarrow[k]$ contribute to the dominating term in the moment expansion \eqref{eq:p-moment-gaussian}, i.e.~are such that the quantity $\sharp(\gamma^{-1}\lambda)+\sharp(\gamma_f^{-1}\lambda) + k - |\im(f)|$ is maximal.

First of all, for any $\lambda\in\mathfrak{P}^{(2)}(2p)$, we have
\begin{equation} \label{eq:lambda-lambda_f'}
\sharp(\lambda)+\sharp(\gamma^{-1}\lambda) = 4p - \left(|\lambda|+|\gamma^{-1}\lambda|\right) \leq 4p-|\gamma^{-1}| = 4p-(2p-1)=2p+1,
\end{equation}
where the first equality is by Lemma \ref{lemma:cycles-transpositions}, while the second inequality is by equation \eqref{eq:geodesic'} in Lemma \ref{lemma:distance} and is an equality if and only if the pair-partition $\lambda$ is non-crossing. Next, for any $\lambda\in\mathfrak{P}^{(2)}(2p)$ and $f:[2p]\rightarrow[k]$, we have
\begin{equation} \label{eq:lambda-lambda_f''}
\sharp(\lambda)+\sharp(\gamma_f^{-1}\lambda) = 4p-\left(|\lambda|+|\gamma_f^{-1}\lambda|\right) \leq 4p - |\gamma_f^{-1}|= 4p - \left(2p-|\im(f)|\right) =2p+|\im(f)|,
\end{equation}
where the first equality is again by Lemma \ref{lemma:cycles-transpositions}, while the second inequality is by equation \eqref{eq:geodesic''} in Lemma \ref{lemma:distance} and is an equality if and only if the pair-partition $\lambda$ is non-crossing and is finer than the partition of $\{1,\ldots,2p\}$ induced by $\gamma_f$ (i.e.~$f$ takes the same value on elements belonging to the same pair-block of $\lambda$).

Putting equations \eqref{eq:lambda-lambda_f'} and \eqref{eq:lambda-lambda_f''} together, we get that for any $\lambda\in\mathfrak{P}^{(2)}(2p)$ and $f:[2p]\rightarrow[k]$ (just keeping in mind that necessarily $\sharp(\lambda)=p$),
\begin{equation} \label{eq:lambda-lambda_f} \sharp(\gamma^{-1}\lambda)+\sharp(\gamma_f^{-1}\lambda) + k - |\im(f)|\leq 2p+k+1, \end{equation}
with equality if and only if $\lambda\in NC^{(2)}(2p)$ and $f\circ\lambda=f$. Since it is well-known that there are $\mathrm{Cat}_p$ elements in $NC^{(2)}(2p)$, and for each of these there are $k^p$ functions which are constant on each of its $p$ pair-blocks, we indeed get the asymptotic estimate announced in Proposition \ref{prop:gaussian-p}, namely
\[ \mathbf{E}\, \mathrm{Tr}\left[ \left( \underset{j=1}{\overset{k}{\sum}}\widetilde{G}_{\A\B^k}(j) \right)^{2p} \right] \underset{d\rightarrow+\infty}{\sim} k^p\,\mathrm{Cat}_p\,d^{2p+k+1} =\mathrm{M}_{SC(k)}^{(2p)}d^{2p+k+1} . \qedhere \]
\end{proof}

\begin{proof} [Proof of Proposition \ref{prop:gaussian-infty}]
The convergence in moments stated in Proposition \ref{prop:gaussian-p} implies 
that, asymptotically, the matrix $\left(\sum_{j=1}^k\widetilde{G}_{\A\B^k}(j)\right)/d$ has a smallest eigenvalue and a largest eigenvalue which are, on average, at most the lower-edge and at least the upper-edge of the support of $\mu_{SC(k)}$, i.e.~$-2\sqrt{k}$ and $2\sqrt{k}$. Indeed, convergence of all moments of the empirical spectral distribution of $\left(\sum_{j=1}^k\widetilde{G}_{\A\B^k}(j)\right)/d$ entails convergence of all polynomial functions, and therefore of all continuous functions with bounded support, when integrated against it. And this in turn entails (when applied to continuous functions with support strictly included in $[-2\sqrt{k},2\sqrt{k}]$) that the extreme eigenvalues of $\left(\sum_{j=1}^k\widetilde{G}_{\A\B^k}(j)\right)/d$ cannot be, on average, strictly bigger than $-2\sqrt{k}$ or smaller $2\sqrt{k}$. The reader is referred to \cite{AGZ}, Chapter 2, for all the technical details of the argument. Hence in other words, Proposition \ref{prop:gaussian-p} guarantees that there exist positive constants $c_d\rightarrow_{d\rightarrow+\infty}1$ such that
\begin{equation} \label{eq:lower-bound-gaussian} \mathbf{E} \left\| \underset{j=1}{\overset{k}{\sum}}\widetilde{G}_{\A\B^k}(j) \right\|_{\infty} 
\geq c_d\, 2\sqrt{k}d .\end{equation}

In the opposite direction, Proposition \ref{prop:gaussian-p} only guarantees that the matrix $\left(\sum_{j=1}^k\widetilde{G}_{\A\B^k}(j)\right)/d$ asymptotically has, on average, no strictly positive fraction of eigenvalues strictly below $-2\sqrt{k}$ or above $2\sqrt{k}$. So to show that the reverse inequality to \eqref{eq:lower-bound-gaussian} holds too, a little more care is required. Indeed, to say it roughly, we have to make sure that in the moment's expression \eqref{eq:p-moment-gaussian}, the permutations contributing to the non-dominating terms (in $d$) are not too numerous.

For $d\in\N$ fixed, it holds thanks to Jensen's inequality and monotonicity of Schatten norms that
\begin{equation} \label{eq:infty-p-gaussian} \forall\ p\in\N,\ \mathbf{E} \left\|\sum_{j=1}^k\widetilde{G}_{\A\B^k}(j)\right\|_{\infty} \leq \left( \mathbf{E} \left\|\sum_{j=1}^k\widetilde{G}_{\A\B^k}(j)\right\|_{\infty}^{2p}\right)^{1/2p} \leq \left(\mathbf{E}\, \mathrm{Tr}\left[\left(\sum_{j=1}^k\widetilde{G}_{\A\B^k}(j)\right)^{2p}\right] \right)^{1/2p} .\end{equation}
So let us fix $d\in\N$ and $p\in\N$, and rewrite \eqref{eq:p-moment-gaussian} explicitly as an expansion in powers of $d$, keeping in the sum the permutations not saturating equation \eqref{eq:lambda-lambda_f}. Being cautious only with the permutations not saturating equation \eqref{eq:lambda-lambda_f''}, and not with those not saturating equation \eqref{eq:lambda-lambda_f'}, we get
\begin{equation} \label{eq:momentsG-defect}
\mathbf{E}\, \mathrm{Tr}\left[\left(\sum_{j=1}^k\widetilde{G}_{\A\B^k}(j)\right)^{2p}\right] \leq \left(\underset{f:[2p]\rightarrow[k]}{\sum} \sum_{\delta=0}^{\lfloor(p+k)/2\rfloor} \left|\mathfrak{P}^{(2)}_{f,\delta}(2p)\right| d^{-2\delta}\right) d^{2p+k+1},
\end{equation}
where we defined, for each $f:[2p]\rightarrow[k]$ and each $0\leq\delta\leq \lfloor(p+k)/2\rfloor$,
\[ \mathfrak{P}^{(2)}_{f,\delta}(2p)=\left\{ \lambda\in\mathfrak{P}^{(2)}(2p) \st \sharp(\gamma_f^{-1}\lambda) =p+|\im(f)|-2\delta \right\}. \]
In words, $\mathfrak{P}^{(2)}_{f,\delta}(2p)$ is nothing else than the set of permutations which have a defect $2\delta$ from lying on the geodesics between the identity and the product of the canonical full cycles on each of the level sets of $f$. This justifies in particular \textit{a posteriori} why the summation in \eqref{eq:momentsG-defect} is only over even defects (see the parity argument in Lemma \ref{lemma:distance-general}).

Now, by Lemma \ref{lemma:number-functions,pairings-defect}, we know that, if $0\leq\delta\leq\lfloor p/2\rfloor$, then
\[ \left|\left\{ (f,\lambda) \st \lambda\in\mathfrak{P}^{(2)}_{f,\delta}(2p) \right\}\right| \leq k^p\,\mathrm{Cat}_p \times\left(\frac{kp^2}{2}\right)^{2\delta}. \]
And if $\lceil p/2\rceil\leq\delta\leq \lfloor (p+k)/2\rfloor$, then trivially
\[ \left|\left\{ (f,\lambda) \st \lambda\in\mathfrak{P}^{(2)}_{f,\delta}(2p) \right\}\right| \leq k^{2p}\,\frac{(2p)!}{2^pp!} \leq k^p\,\mathrm{Cat}_p\times \left(\frac{kp^2}{2}\right)^p. \]
Putting everything together, we therefore get,
\[ \mathbf{E}\, \mathrm{Tr}\left[\left(\sum_{j=1}^k\widetilde{G}_{\A\B^k}(j)\right)^{2p}\right] \leq k^p\,\mathrm{Cat}_p \left(1+\sum_{\delta=1}^{\lfloor p/2\rfloor}\left(\frac{kp^2}{2d}\right)^{2\delta}+\frac{k}{2}\left(\frac{kp^2}{2d}\right)^p\right)d^{2p+k+1}. \]
Yet, $\max\left\{\left(kp^2/2d\right)^{2\delta} \st 1\leq\delta\leq \lceil p/2\rceil\right\}$ is attained for $\delta=1$, provided $p\leq(2d/k)^{1/2}$. So if such is the case,
\[ \sum_{\delta=1}^{\lfloor p/2\rfloor}\left(\frac{kp^2}{2d}\right)^{2\delta}+\frac{k}{2}\left(\frac{kp^2}{2d}\right)^p \leq \frac{p+k}{2} \frac{k^2p^4}{4d^2} \leq \frac{k^2p^5}{4d^2}, \]
where the last inequality holds as long as $p\geq k$. And hence, under all the previous assumptions,
\[ \mathbf{E}\, \mathrm{Tr}\left[\left(\sum_{j=1}^k\widetilde{G}_{\A\B^k}(j)\right)^{2p}\right] \leq\, \mathrm{M}_{SC(k)}^{(2p)}\left(1 +\frac{k^2p^5}{4d^2}\right) d^{2p+k+1}. \]

So set $p_d=(2d/k)^{(2-\epsilon)/5}$ for some $0<\epsilon<1$ (which is indeed smaller than $(2d/k)^{1/2}$ and bigger than $k$ for $d$ big enough, in particular bigger than $k^{7/2}/2$). And using inequality \eqref{eq:infty-p-gaussian} in the special case $p=p_d$, we eventually get
\begin{equation} \label{eq:upper-bound-gaussian} \mathbf{E} \left\|\sum_{j=1}^k\widetilde{G}_{\A\B^k}(j)\right\|_{\infty}  \leq \left(\mathrm{M}_{SC(k)}^{(2p_d)} \left(1 + \frac{k^2p_d^5}{4d^2}\right)\right)^{1/2p_d} d^{1+(k+1)/2p_d} \underset{d\rightarrow+\infty}{\sim} 2\sqrt{k}d. \end{equation}
Combining the lower bound in equation \eqref{eq:lower-bound-gaussian} and the upper bound in equation \eqref{eq:upper-bound-gaussian} yields Proposition \ref{prop:gaussian-infty}.
\end{proof}

\subsection{Conclusion}

Combining Proposition \ref{prop:gaussian-infty} with equation \eqref{eq:w(k-ext)-definition}, we straightforwardly obtain the estimate we were looking for, which is stated in Theorem \ref{th:w(k-ext)} below.

\begin{theorem} \label{th:w(k-ext)}
Let $k\in\N$. The mean width of the set of $k$-extendible states on $\C^d\otimes\C^d$ satisfies
\[ w\big(\mathcal{E}_k(\C^d{:}\C^d)\big) \underset{d\rightarrow +\infty}{\sim} \frac{2}{\sqrt{k}}\frac{1}{d} .\]
\end{theorem}

\section{Discussion and comparison with the mean width of the set of PPT states}
\label{section:w-comparison}

It was shown in \cite{AS1} that the mean width of the set of separable states on $\C^d\otimes\C^d$ is of order $1/d^{3/2}$. And we just showed in Theorem \ref{th:w(k-ext)} that, for $k\in\N$ fixed, the mean width of the set of $k$-extendible states on $\C^d\otimes\C^d$ is of order $1/d$, so that, for $d$ large,
\[ w\big(\mathcal{S}(\C^d{:}\C^d)\big) \ll w\big(\mathcal{E}_k(\C^d{:}\C^d)\big). \]
This result is not surprising: it just means that, when $d$ grows, if $k$ does not grow in some way too, then the set of $k$-extendible states becomes an increasingly poor approximation of the set of separable states on $\C^d\otimes\C^d$. There had been several evidences, already, in that direction, with examples of highly-extendible, though entangled, states (see e.g.~\cite{BCY} and \cite{NOP1}).

It is well-known that the exact same feature is actually exhibited by the set of PPT states on $\C^d\otimes\C^d$, whose mean width is of order $1/d$ too. Let us be more precise.

\begin{proposition} \label{prop:w(ppt)}
There exist positive constants $c_d,C_d\rightarrow_{d\rightarrow+\infty}1$ such that the mean width of the set of PPT states on $\C^d\otimes\C^d$ satisfies
\[ c_d\,\frac{e^{-1/2}}{d} \leq w\big(\mathcal{P}(\C^d{:}\C^d)\big) \leq C_d\,\frac{2}{d}. \]
\end{proposition}

\begin{proof} Proposition \ref{prop:w(ppt)} was basically established in \cite{AS1}, but not stated in this exact way and with these exact constants, so we briefly recall the argument here for the sake of completeness (see also Chapter \ref{chap:SDrelaxations}, Section \ref{section:volumes}, of the present manuscript for a similar discussion).

To get the asymptotic upper bound, we just use
\[ w\big(\mathcal{P}(\C^d{:}\C^d)\big) \leq w\big(\mathcal{D}(\C^d\otimes\C^d)\big) \underset{d\rightarrow+\infty}{\sim} \frac{2}{d}. \]
The last equivalence is a consequence of Wigner's semicircle law (see e.g.~\cite{AGZ}, Chapter 2, for a proof) from which it follows that
\[ w_G\big(\mathcal{D}(\C^d\otimes\C^d)\big) = \E_{G\sim GUE(d^2)} \underset{\sigma\in\mathcal{D}(\C^d\otimes\C^d)}{\sup}\mathrm{Tr}(G\sigma) = \E_{G\sim GUE(d^2)} \|G\|_{\infty} \underset{d\rightarrow+\infty}{\sim}2d. \]

To get the asymptotic lower bound, we will make use of two results from classical convex geometry: Urysohn and Milman--Pajor inequalities (see Theorems \ref{theorem:urysohn} and \ref{theorem:Milman-Pajor}, respectively, in Chapter \ref{chap:toolbox}, Section \ref{ap:convex-geometry}).  But before that, we need one more definition (also appearing in Chapter \ref{chap:toolbox}, Section \ref{ap:convex-geometry}): For any convex body $K$, we denote by $\vrad(K)$ its volume radius, which is defined as the radius of the Euclidean ball having the same volume (i.e.~Lebesgue measure) as $K$. In our case, denoting by $\Gamma$ the partial transposition, we have $\mathcal{P}(\C^d{:}\C^d)=\mathcal{D}(\C^d\otimes\C^d)\cap\mathcal{D}(\C^d\otimes\C^d)^{\Gamma}$, with $\mathcal{D}(\C^d\otimes\C^d)$ having the maximally mixed state $\Id/d^2$ as center of gravity and $\Gamma$ being a linear isometry. Hence,
\[ w\big(\mathcal{P}(\C^d{:}\C^d)\big) \geq \vrad\big(\mathcal{P}(\C^d{:}\C^d)\big) \geq \frac{\vrad\big(\mathcal{D}(\C^d\otimes\C^d)\big)^2}{2w\big(\mathcal{D}(\C^d\otimes\C^d)\big)}, \]
the first inequality being by the Urysohn inequality, and the second being by the consequence of Milman--Pajor inequality stated as Corollary \ref{corollary:Milman-Pajor} in Chapter \ref{chap:toolbox}, Section \ref{ap:convex-geometry}.
Now, we just argued that $w\big(\mathcal{D}(\C^d\otimes\C^d)\big)\sim_{d\rightarrow+\infty} 2/d$, while it was shown in \cite{SZ} that $\vrad\big(\mathcal{D}(\C^d\otimes\C^d)\big)\sim_{d\rightarrow+\infty} e^{-1/4}/d$. Therefore,
\[ w\big(\mathcal{P}(\C^d{:}\C^d)\big) \geq \frac{\vrad\big(\mathcal{D}(\C^d\otimes\C^d)\big)^2}{2w\big(\mathcal{D}(\C^d\otimes\C^d)\big)} \underset{d\rightarrow+\infty}{\sim} \frac{e^{-1/2}}{d}. \qedhere \]
\end{proof}

As a straightforward consequence of Theorem \ref{th:w(k-ext)} and Proposition \ref{prop:w(ppt)}, we have, roughly speaking, that for $k\geq 11$, the set of $k$-extendible states becomes asymptotically a ``better'' approximation of the set of separable states than the set of PPT states, on average. Indeed, if $k\geq 11$, then $2/\sqrt{k}<e^{-1/2}$, so that for $d$ large enough
\[ w\big(\mathcal{E}_k(\C^d{:}\C^d)\big) < w\big(\mathcal{P}(\C^d{:}\C^d)\big). \]

\section{Adding the PPT constraint on the extension}
\label{section:w(k-ext-ppt]}

The hierarchy of SDPs originally proposed in \cite{DPS} to detect entanglement was in fact slightly different from the one that would be derived from Theorem \ref{th:k-extendibility-criterion}. Indeed, for a given bipartite state $\rho_{\A\B}$, the $k^{\text{th}}$ test would here consist in looking for a symmetric extension $\rho_{\A\B^k}$ of $\rho_{\A\B}$, while in \cite{DPS} it was additionally imposed that this extension had to be PPT in any cut of the $k+1$ subsystems. This of course increased quite considerably the size of the SDP to be solved at each step, but with the hope that it would at the same time decrease dramatically the number of steps an entangled state would pass.

Another hierarchy of SDPs was later proposed in \cite{NOP1} and \cite{NOP2}, built on the exact same ideas as those in \cite{DPS}. It was noticed there that only demanding that the (Bose) symmetric extension of the state be PPT in one fixed (even) cut of the $k+1$ subsystems already implied a noticeable speed-up in the convergence of the algorithm. It therefore seems worth taking a closer look at the set of states arising from these constraints. The latter is properly defined as follows.

\begin{definition}
Let $k\in\N$. A state $\rho_{\A\B}$ on a bipartite Hilbert space $\mathrm{A}\otimes\mathrm{B}$ is $k$-PPT-extendible with respect to $\mathrm{B}$ if there exists a state $\rho_{\A\B^k}$ on $\mathrm{A}\otimes\mathrm{B}^{\otimes k}$ which is PPT in the cut $\mathrm{A}\otimes\mathrm{B}^{\otimes \lfloor k/2\rfloor}:\mathrm{B}^{\otimes \lceil k/2\rceil}$, invariant under any permutation of the $\mathrm{B}$ subsystems and such that $\rho_{\A\B}=\tr_{\B^{k-1}}\rho_{\A\B^k}$. We denote by $\mathcal{E}_k^{PPT}(\A{:}\B)$ the set of $k$-PPT-extendible states on $\mathrm{A}\otimes\mathrm{B}$ (in the cut $\A{:}\B$ and with respect to $\mathrm{B}$).
\end{definition}

\begin{theorem} \label{th:w(k-PPT-ext)}
Let $k\in\N$. There exist positive constants $C_d\rightarrow_{d\rightarrow+\infty}1$ such that the mean width of the set of $k$-PPT-extendible states on $\C^d\otimes\C^d$ satisfies
\[ w\big(\mathcal{E}_k^{PPT}(\C^d{:}\C^d)\big) \leq C_d\,\frac{\sqrt{2}}{\sqrt{k}}\frac{1}{d} .\]
\end{theorem}

\begin{proof}
Using the notation introduced in Lemma \ref{lemma:sup-k-ext}, we start from the simple observation that, for any $M_{\A\B}\in\mathcal{H}(\mathrm{A}\otimes \mathrm{B})$,
\begin{align*}
\underset{\sigma_{\A\B}\in\mathcal{E}_k^{PPT}(\A{:}\B)}{\sup}\ \tr\big[M_{\A\B}\sigma_{\A\B}\big]
&= \underset{\sigma_{\A\B^k}\in\mathcal{P}(\mathrm{A}\mathrm{B}^{\lfloor k/2\rfloor}:\mathrm{B}^{\lceil k/2\rceil})}{\sup}\ \tr \big[ \Sym_{\A{:}\B^k}\big(M_{\A\B}\otimes\Id_{\B^{k-1}}\big)\sigma_{\A\B^k}\big]\\
&\leq \min\left( \big\|\Sym_{\A{:}\B^k} \big(M_{\A\B}\otimes\Id_{\B^{k-1}}\big)\big\|_{\infty}, \big\|\left[\Sym_{\A{:}\B^k} \big(M_{\A\B}\otimes\Id_{\B^{k-1}}\big)\right]^{\Gamma}\big\|_{\infty} \right),
\end{align*}
where $\Gamma$ stands here for the partial transposition over the $\lceil k/2\rceil$ last $\mathrm{B}$ subsystems, so that in fact
\[ \left[\Sym_{\A{:}\B^k} \big(M_{\A\B}\otimes\Id_{\B^{k-1}}\big)\right]^{\Gamma} = \frac{1}{k} \left( \sum_{j=1}^{\lfloor k/2\rfloor} M_{\A\B_j}\otimes\Id_{\widehat{\B}^k_j} + \sum_{j=\lfloor k/2\rfloor +1}^k M_{\A\B_j}^{\Gamma}\otimes\Id_{\widehat{\B}^k_j} \right), \]
where $\Gamma$ now stands for the partial transposition over $\mathrm{B}$.

The upper bound in Theorem \ref{th:w(k-PPT-ext)} will thus be a direct consequence of the sup-norm estimate
\[ \mathbf{E}_{G_{\A\B}\sim GUE(d^2)} \left\| \sum_{j=1}^{k}\widetilde{G}_{\A\B^k}(j)^{\Gamma} \right\|_{\infty} \underset{d\rightarrow+\infty}{\sim} \sqrt{2}\sqrt{k}d .\]
The latter is proved in the exact same way as Proposition \ref{prop:gaussian-infty}, i.e.~by first showing that for any $p\in\N$,
\begin{equation} \label{eq:moment-Gamma} \mathbf{E}_{G_{\A\B}\sim GUE(d^2)}\, \mathrm{Tr}\left[ \left( \sum_{j=1}^{k}\widetilde{G}_{\A\B^k}(j)^{\Gamma} \right)^{2p} \right] \underset{d\rightarrow+\infty}{\sim} 2\mathrm{M}_{SC(k/2)}^{(2p)}d^{2p+k/2+1} ,\end{equation}
and second arguing that also $\mathbf{E}\left\| \sum_{j=1}^{k} \widetilde{G}_{\A\B^k}(j)^{\Gamma}\right\|_{\infty}\sim_{d\rightarrow+\infty}\lim_{p\rightarrow+\infty} \left( \mathbf{E}\, \mathrm{Tr}\left[ \left( \sum_{j=1}^{k} \widetilde{G}_{\A\B^k}(j)^{\Gamma}\right)^{2p} \right] \right)^{1/2p}$. This last step will be omitted here since the argument is very similar to the one appearing in the proof of Proposition \ref{prop:gaussian-infty}. Concerning the moment estimate \eqref{eq:moment-Gamma}, it is first of all proved in Appendix \ref{appendix:gaussian-gamma} that
\[ \E\, \mathrm{Tr}\left[\left(\sum_{j=1}^k\widetilde{G}_{\A\B^k}(j)^{\Gamma}\right)^{2p}\right]
 \underset{d\rightarrow+\infty}{\sim} \underset{\lambda\in\mathfrak{P}^{(2)}(2p)}{\sum}\, \underset{f:[2p]\rightarrow\left[\lfloor k/2\rfloor\right]\,\text{or}\,\left[\lceil k/2\rceil\right]}{\sum} d^{\sharp(\gamma^{-1}\lambda)+\sharp(\gamma_f^{-1}\lambda)+k-|\im(f)|}. \]
And by the same arguments as the in the proof of Proposition \ref{prop:gaussian-p}, we can then identify which $\lambda$ and $f$ actually contribute to the dominant order in the latter expression, yielding
\begin{align*}
\E \,\mathrm{Tr}\left[\left(\sum_{j=1}^k\widetilde{G}_{\A\B^k}(j)^{\Gamma}\right)^{2p}\right]
& \underset{d\rightarrow+\infty}{\sim} \underset{\lambda\in NC^{(2)}(2p)}{\sum} \left( \underset{\underset{f\circ\lambda=f}{f:[2p]\rightarrow\left[\lfloor k/2\rfloor\right]}}{\sum} d^{2p+\lfloor k/2\rfloor+1} + \underset{\underset{f\circ\lambda=f}{f:[2p]\rightarrow\left[\lceil k/2\rceil\right]}}{\sum} d^{2p+\lceil k/2\rceil+1} \right) \\
& \underset{d\rightarrow+\infty}{\sim} \mathrm{Cat}_p\left(\lfloor k/2\rfloor^pd^{2p+\lfloor k/2\rfloor+1} + \lceil k/2\rceil^pd^{2p+\lceil k/2\rceil+1}\right),
\end{align*}
which is the announced moment estimate \eqref{eq:moment-Gamma}.
\end{proof}

Comparing Theorem \ref{th:w(k-PPT-ext)} to Theorem \ref{th:w(k-ext)}, we see that the asymptotic mean width of the set of $k$-PPT-extendible states is at least $\sqrt{2}$ smaller than the asymptotic mean width of the set of $k$-extendible states. For instance, the set of $2$-PPT-extendible states is, on average, asymptotically smaller than the set of $4$-extendible states.
This however does not really shed light on why adding the constraint, at each step in the sequence of tests, that the symmetric extension is PPT across one fixed (even) cut would make the entanglement detection notably faster.

\section{Preliminaries on random-induced states and witnesses}
\label{section:random-states}

We will employ the notation $\rho\sim\mu_{n,s}$ to mean that $\rho=\tr_{\C^s}\ketbra{\psi}{\psi}$ with $\ket{\psi}$ a random Haar-distributed pure state on $\C^n\otimes\C^s$ (i.e.~$\rho$ describes an $n$-dimensional system which is obtained by partial-tracing over an $s$-dimensional ancilla space a uniformly distributed pure state on the global ``system+ancilla'' space). An equivalent mathematical characterization of such random state model is $\rho=W/\mathrm{Tr}W$ with $W\sim\mathcal{W}_{n,s}$ an $(n,s)$-Wishart matrix, i.e.~$W=GG^{\dagger}$ with $G$ a $n\times s$ matrix having independent complex normal entries (see e.g.~\cite{ZS}).

Let $K\subset\mathcal{D}(\C^n)$ be a convex body. For any $\rho\in\mathcal{D}(\C^n)$, a standard way of showing that $\rho\notin K$ is to produce a ``not belonging to $K$ witness'', i.e.~some $M\in\mathcal{H}_+(\C^n)$ which is such that
\[ \underset{\sigma\in K}{\sup}\mathrm{Tr}(M\sigma)<\mathrm{Tr}(M\rho). \]
By testing $\rho$ itself as possible such ``not belonging to $K$ witness'', we have
\begin{equation} \label{eq:witness} \underset{\sigma\in K}{\sup}\mathrm{Tr}(\rho\sigma)<\mathrm{Tr}(\rho^2)\ \Rightarrow\ \rho\notin K. \end{equation}

Crucially for the applications we have in mind, the functions $\rho\mapsto\mathrm{Tr}(\rho^2)$ and $\rho\mapsto\sup_{\sigma\in K}\mathrm{Tr}(\rho\sigma)$ both have nice concentration properties around their average. More precisely, we have the two following results.

\begin{proposition} \label{prop:concentration1}
Let $n,s\in\N$. Then, there exist universal constants $c,c'>0$ such that, for any $\eta>0$, first of all
\[ \mathbf{P}_{\rho\sim\mu_{n,s}}\left(\left|\mathrm{Tr}(\rho^2)- \mathbf{E}_{\tau\sim\mu_{n,s}}\big[\mathrm{Tr}(\tau^2)\big]\right|\geq\eta\right) \leq e^{-cs} + e^{-c'n^3s\eta^2}, \]
and second of all, for any convex body $K\subset\mathcal{D}(\C^n)$,
\[ \mathbf{P}_{\rho\sim\mu_{n,s}}\left(\left|\underset{\sigma\in K}{\sup}\mathrm{Tr}(\rho\sigma)-\mathbf{E}_{\tau\sim\mu_{n,s}}\left[\underset{\sigma\in K}{\sup}\mathrm{Tr}(\tau\sigma)\right]\right|\geq\eta\right) \leq e^{-cs} + e^{-c'n^2s\eta^2}. \]
\end{proposition}

\begin{proof} To show Proposition \ref{prop:concentration1}, we will make essential use of a local version of Levy's lemma. The usual Levy lemma is recalled as Lemma \ref{lemma:levy} in Chapter \ref{chap:toolbox}, Section \ref{ap:deviations}. Here we will rely on the following refinement (see \cite{ASY}, Lemma 3.4, for a proof): Let $\Omega\subset S^{m-1}$ be a subset of the Euclidean unit sphere of $\R^m$ satisfying $\P(\Omega)\geq 7/8$. Let also $f:S^{m-1}\rightarrow\R$ be a function whose restriction to $\Omega$ is $L$-Lipschitz and $M$ be a central value for $f$ (i.e.~$\P(\{f\geq M\})\geq 1/4$ and $\P(\{f\leq M\})\geq 1/4$). Then, for any $\eta>0$,
\[ \P\left(\left\{|f-M|\geq\eta\right\}\right) \leq \P\left(S^{m-1}\setminus\Omega\right) + e^{-c_0m\eta^2/L^2}, \]
where $c_0>0$ is a universal constant.

It is well-known (see e.g.~\cite{ZS} for a proof) that $\rho\sim\mu_{n,s}$ is equivalent to $\rho=XX^{\dagger}$ with $X$ uniformly distributed over the Hilbert--Schmidt unit sphere of $n\times s$ complex matrices, and the latter can be identified with the real Euclidean unit sphere $S^{2ns-1}$. Therefore, one may apply Levy's lemma above with $\Omega=\left\{X\in S^{2ns-1} \st \|X\|_{\infty}\leq 3/\sqrt{n}\right\}$, which is such that $\P(S^{2ns-1}\setminus\Omega)\leq e^{-cs}$ for some universal constant $c>0$ (see e.g.~\cite{ASW}, Lemma 6 and Appendix B, for a proof).

Consider first $f:X\in S^{2ns-1}\mapsto\mathrm{Tr}((XX^{\dagger})^2)$, which is $36/n$-Lipschitz on $\Omega$.  Indeed, for any $X,Y\in\Omega$,
\begin{align*}
\left|f(X)-f(Y)\right| & \leq \left\|\left(XX^{\dagger}\right)^2-\left(YY^{\dagger}\right)^2\right\|_1\\
& \leq \left(\|XX^{\dagger}\|_{\infty}+\|YY^{\dagger}\|_{\infty}\right)\left(\|X\|_{2}+\|Y\|_{2}\right)\|X-Y\|_2 \\
& \leq \frac{36}{n} \|X-Y\|_2.
\end{align*}
The second inequality is just by H\"{o}lder's inequality (more specifically $\|ABC\|_1\leq\|\A\|_{\infty}\|B\|_2\|C\|_2$) and the triangle inequality, after noticing that $\left(XX^{\dagger}\right)^2-\left(YY^{\dagger}\right)^2 = XX^{\dagger}\Delta+\Delta YY^{\dagger}$ with $\Delta=X(X^{\dagger}-Y^{\dagger})+(X-Y)Y^{\dagger}$. And the third inequality is because, by assumption, for any $Z\in\Omega$, $\|Z\|_2=1$ and $\|ZZ^{\dagger}\|_{\infty}=\|Z\|_{\infty}^2\leq 9/n$.

Now, the fact that $\P(S^{2ns-1}\setminus\Omega)\leq e^{-cs}$, combined with the fact that $|f|$ is bounded by $1$ on $S^{2ns-1}$, implies that the average of $|f|$ on $S^{2ns-1}\setminus\Omega$ is bounded by $e^{-cs}$, which tends to $0$ when $s$ tends to infinity. While the Lipschitz estimate for $f$ on $\Omega$ implies that the average of $f$ on $\Omega$ differs from its median by at most $C/n^{3/2}s^{1/2}$, which also tends to $0$ when $n,s$ tend to infinity. We can therefore conclude that the average of $f$ is a central value of $f$ for $n,s$ big enough. Hence, taking $M=\E f$ as central value for $f$, we get the concentration estimate
\[ \mathbf{P}_X\left(\left|\mathrm{Tr}\left(\left(XX^{\dagger}\right)^2\right)- \mathbf{E}_Y\mathrm{Tr}\left(\left(YY^{\dagger}\right)^2\right)\right|\geq\eta\right) \leq e^{-cs} + e^{-c'n^3s\eta^2}. \]

Take next $f:X\in S^{2ns-1}\mapsto\sup_{\sigma\in K}\mathrm{Tr}(XX^{\dagger}\sigma)$, which is $6/\sqrt{n}$-Lipschitz on $\Omega$. Indeed, for any $X,Y\in\Omega$,
\begin{align*}
\left|f(X)-f(Y)\right| & \leq \left| \sup_{\sigma\in K}\mathrm{Tr}\left(\left(XX^{\dagger}-YY^{\dagger}\right)\sigma\right)\right|\\
& \leq \left\|XX^{\dagger}-YY^{\dagger}\right\|_{\infty}\\
& \leq \left(\|X\|_{\infty}+\|Y\|_{\infty}\right) \left\|X-Y\right\|_{\infty}\\
& \leq \frac{6}{\sqrt{n}}\|X-Y\|_2.
\end{align*}
The second inequality is just by duality, since $K$ is contained in the unit ball for the $1$-norm. The third inequality is by the triangle inequality, after noticing that $XX^{\dagger}-YY^{\dagger}=(X-Y)X^{\dagger}+Y(X^{\dagger}-Y^{\dagger})$. And the fourth inequality is by the norm inequality $\|\cdot\|_{\infty}\leq\|\cdot\|_2$ and because, by assumption, for any $Z\in\Omega$, $\|Z\|_{\infty}\leq 3/\sqrt{n}$.

Arguing as before, we see that the average of $f$ is a central value of $f$ for $n,s$ big enough (this time, the average of $|f|$ on $S^{2ns-1}\setminus\Omega$ is bounded by $e^{-cs}$ while the average of $f$ on $\Omega$ differs from its median by at most $C/ns^{1/2}$). Hence, taking $M=\E f$ as central value for $f$, we get the concentration estimate
\[ \P_X\left(\left\{\left|\underset{\sigma\in K}{\sup}\mathrm{Tr}(XX^{\dagger}\sigma)-\E_Y\underset{\sigma\in K}{\sup}\mathrm{Tr}(YY^{\dagger}\sigma)\right|\geq\eta\right\}\right) \leq e^{-cs} + e^{-c'n^2s\eta^2}. \]

Hence, we indeed have the two announced deviation probability bounds.
\end{proof}

Combining the two statements in Proposition \ref{prop:concentration1}, together with equation \eqref{eq:witness}, we get as a consequence: Let $K\subset\mathcal{D}(\C^n)$ be a convex body. Then, for any $\eta>0$,
\begin{equation} \label{eq:proba-notinK'} \mathbf{E}_{\rho\sim\mu_{n,s}}\big[\mathrm{Tr}(\rho^2)\big] - \mathbf{E}_{\rho\sim\mu_{n,s}}\left[\underset{\sigma\in K}{\sup}\mathrm{Tr}(\rho\sigma)\right] > \eta
\ \Rightarrow\ \mathbf{P}_{\rho\sim\mu_{n,s}}\big(\rho\notin K\big)\geq 1-e^{-cs\min(1,n^2\eta^2)}, \end{equation}
where $c>0$ is a universal constant.

From now on, we will in fact consider random-induced states on the bipartite space $\C^d\otimes\C^d$. So let $K\subset\mathcal{D}(\C^d\otimes\C^d)$ (such as e.g.~$\mathcal{P}(\C^d{:}\C^d)$ or $\mathcal{E}_k(\C^d{:}\C^d)$, $k\in\N$). It follows from equation \eqref{eq:proba-notinK'} that, for any $\eta>0$,
\begin{equation} \label{eq:proba-notinK} \mathbf{E}_{\rho\sim\mu_{d^2,s}}\big[\mathrm{Tr}(\rho^2)\big] - \mathbf{E}_{\rho\sim\mu_{d^2,s}}\left[\underset{\sigma\in K}{\sup}\mathrm{Tr}(\rho\sigma)\right] > \eta
\ \Rightarrow\ \mathbf{P}_{\rho\sim\mu_{d^2,s}}\big(\rho\notin K\big)\geq 1-e^{-cs\min(1,d^4\eta^2)}, \end{equation}
where $c>0$ is a universal constant.

\section{Non $k$-extendibility of random-induced states for ``small'' $k$}
\label{section:non-k-ext-witness}

\subsection{Strategy}

Our goal in the sequel will be to identify a range of environment size $s$ for which random-induced states on $\C^d\otimes\C^d$ are, with high-probability, not $k$-extendible. In view of equation \eqref{eq:proba-notinK}, this may be done by characterizing
\[ \left\{ s\in\N \st  \mathbf{E}_{\rho\sim\mu_{d^2,s}}\left[\underset{\sigma\in\mathcal{E}_k(\C^d{:}\C^d)}{\sup}\mathrm{Tr}(\rho\sigma)\right] < \mathbf{E}_{\rho\sim\mu_{d^2,s}}\big[\mathrm{Tr}(\rho^2)\big] \right\}. \]

Yet by Lemma \ref{lemma:sup-k-ext}, and using the notation introduced there, we have that for any state $\rho_{\A\B}$ on $\mathrm{A}\otimes\mathrm{B}$,
\[ \underset{\sigma_{\A\B}\in\mathcal{E}_k(\A{:}\B)}{\sup}\mathrm{Tr}\big(\rho_{\A\B}\sigma_{\A\B}\big) =\left\|\underset{j=1}{\overset{k}{\sum}}\widetilde{\rho}_{\A\B^k}(j)\right\|_{\infty} .\]

\subsection{An operator-norm estimate}

As explained above, to know when random-induced states on $\A\otimes\B$ are not $k$-extendible, what we need first is to compute the average operator-norm $\mathbf{E} \left\|\sum_{j=1}^k\widetilde{W}_{\A\B^k}(j)\right\|_{\infty}$, for $W$ a $(d^2,s)$-Wishart matrix. We will proceed in a very similar way to what was done in Section \ref{section:w(k-ext)}, and establish what can be seen as the analogues of Propositions \ref{prop:gaussian-infty} and \ref{prop:gaussian-p} but for Wishart instead of GUE matrices.

\begin{proposition} \label{prop:wishart-infty} Fix $k\in\N$ and $c>0$. Then,
\[ \mathbf{E}_{W_{\A\B}\sim\mathcal{W}_{d^2,cd^2}} \left\|\underset{j=1}{\overset{k}{\sum}}\widetilde{W}_{\A\B^k}(j)\right\|_{\infty} \underset{d\rightarrow+\infty}{\sim} (\sqrt{ck}+1)^2d^2. \]
\end{proposition}

As a preliminary step towards estimating the sup-norm $\mathbf{E} \left\|\sum_{j=1}^k\widetilde{W}_{\A\B^k}(j)\right\|_{\infty}$, we will look at the $p$-order moments $\mathbf{E}\, \mathrm{Tr}\left[\left(\sum_{j=1}^k\widetilde{W}_{\A\B^k}(j)\right)^p\right]$, $p\in\mathbf{N}$, and show that they can be expressed in terms of the $p$-order moments of a Mar\v{c}enko-Pastur distribution of appropriate parameter.

So let us recall first a few required definitions. For any $\lambda>0$, we shall denote by $\mu_{MP(\lambda)}$ the Mar\v{c}enko-Pastur distribution of parameter $\lambda$, whose density is given by
\[ \mathrm{d}\mu_{MP(\lambda)}(x) = \begin{cases} f_{\lambda}(x)\mathrm{d}x\ \text{if}\ \lambda>1 \\
\left(1-\lambda\right)\delta_0 + \lambda f_{\lambda}(x)\mathrm{d}x\ \text{if}\ \lambda\leq 1 \end{cases} ,\]
where, setting $\lambda_{\pm}=(\sqrt{\lambda}\pm 1)^2$, we defined the function $f_{\lambda}$ by
\[ f_{\lambda}(x)=\frac{\sqrt{(\lambda_+-x)(x-\lambda_-)}}{2\pi\lambda x}\mathbf{1}_{[\lambda_-,\lambda_+]}(x). \]
We shall also denote, for each $p\in\N$, by $\mathrm{M}_{MP(\lambda)}^{(p)}$ its $p$-order moment, i.e.~ $\mathrm{M}_{MP(\lambda)}^{(p)}=\int_{-\infty}^{+\infty}x^p\mathrm{d}\mu_{MP(\lambda)}(x)$. It is well-known that
\[ \forall\ p\in\N,\ \mathrm{M}_{MP(\lambda)}^{(p)}=\sum_{m=1}^p\lambda^m\mathrm{Nar}_p^m, \]
where $\mathrm{Nar}_p^m$ is the $(p,m)^{th}$ Narayana number defined in Lemma \ref{lemma:catalan}. In particular,  $\mathrm{M}_{MP(1)}^{(p)}=\mathrm{Cat}_p$, the $p^{\text{th}}$ Catalan number defined in Lemma \ref{lemma:catalan} as well.

\begin{proposition}\label{prop:wishart-p} Fix $k\in\N$ and $c>0$. Then, when $d\rightarrow+\infty$, the random matrix $\left(\sum_{j=1}^k\widetilde{W}_{\A\B^k}(j)\right)/d^2$ converges in moments towards a Mar\v{c}enko-Pastur distribution of parameter $ck$. Equivalently, this means that, for any $p\in\N$,
\[ \mathbf{E}_{W_{\A\B}\sim\mathcal{W}_{d^2,cd^2}}\, \mathrm{Tr}\left[\left(\underset{j=1}{\overset{k}{\sum}}\widetilde{W}_{\A\B^k}(j)\right)^p\right] \underset{d\rightarrow+\infty}{\sim} \mathrm{M}_{MP(ck)}^{(p)}d^{2p+k+1}. \]
\end{proposition}

\begin{remark}
Stronger convergence results than the one established in Proposition \ref{prop:wishart-p} may in fact be proved, as discussed in Appendix \ref{appendix:convergence}.
\end{remark}

\begin{proof} [Proof of Proposition \ref{prop:wishart-p}]
Let $p\in\N$. Computing the value of the $p$-order moment $\mathbf{E}\, \mathrm{Tr}\left[\left(\sum_{j=1}^k\widetilde{W}_{\A\B^k}(j)\right)^p\right]$ may be done using the Gaussian Wick formula (see Lemma \ref{lemma:wick} for the statement and Appendix \ref{appendix:wick-wishart} for a succinct summary of how to derive moments of Wishart matrices from it). In our case, we get by the computations carried out in Appendix \ref{appendix:wishart} and summarized in Proposition \ref{prop:wishart-p-preliminary} that, for any $d,s\in\N$, denoting by $\sharp(\cdot)$ the number of cycles in a permutation,
\begin{align*} \mathbf{E}_{W_{\A\B}\sim\mathcal{W}_{d^2,s}}\, \mathrm{Tr}\left[\left(\sum_{j=1}^k\widetilde{W}_{\A\B^k}(j)\right)^p\right] = & \sum_{f:[p]\rightarrow[k]}\mathbf{E}_{W_{\A\B}\sim\mathcal{W}_{d^2,s}}\, \mathrm{Tr}\left[\underset{i=1}{\overset{p}{\overrightarrow{\prod}}}\widetilde{W}_{\A\B^k}(f(i))\right] \\
= & \underset{f:[p]\rightarrow[k]}{\sum}\underset{\alpha\in\mathfrak{S}(p)}{\sum} d^{\sharp(\gamma^{-1}\alpha)+\sharp(\gamma_f^{-1}\alpha)+k-|\im(f)|}s^{\sharp(\alpha)}, \end{align*}
where we defined on $\{1,\ldots,p\}$, $\mathfrak{S}(p)$ as the set of permutations, $\gamma=(p\,\ldots\,1)$ as the canonical full cycle, and for each $f:[p]\rightarrow[k]$, $\gamma_f=\gamma_{f=1}\cdots\gamma_{f=k}$ as the product of the canonical full cycles on each of the level sets of $f$.

Hence, in the case where $s=cd^2$, for some constant $c>0$, we have
\begin{equation}
\label{eq:p-moment}
\mathbf{E}_{W_{\A\B}\sim\mathcal{W}_{d^2,cd^2}}\, \mathrm{Tr}\left[\left(\sum_{j=1}^k\widetilde{W}_{\A\B^k}(j)\right)^p\right] = \sum_{f:[p]\rightarrow[k]} \sum_{\alpha\in\mathfrak{S}(p)} c^{\sharp(\alpha)} d^{2\sharp(\alpha)+\sharp(\gamma^{-1}\alpha)+\sharp(\gamma_f^{-1}\alpha)+k-|\im(f)|}.
\end{equation}

We now have to understand which $\alpha\in\mathfrak{S}(p)$ and $f:[p]\rightarrow[k]$ contribute to the dominating term in the moment expansion \eqref{eq:p-moment}, i.e.~are such that the quantity $2\sharp(\alpha)+\sharp(\gamma^{-1}\alpha)+\sharp(\gamma_f^{-1}\alpha)+k-|\im(f)|$ is maximal.

First of all, for any $\alpha\in\mathfrak{S}(p)$, we have
\begin{equation} \label{eq:alpha} \sharp(\alpha)+\sharp(\gamma^{-1}\alpha)= 2p-(|\alpha|+|\gamma^{-1}\alpha|)\leq 2p-|\gamma| =p+\sharp(\gamma) =p+1, \end{equation}
where the first equality is by Lemma \ref{lemma:cycles-transpositions}, whereas the second inequality is by equation \eqref{eq:geodesic'} in Lemma \ref{lemma:distance} and is an equality if and only if $\alpha\in NC(p)$. Next, for any $\alpha\in\mathfrak{S}(p)$ and $f:[p]\rightarrow[k]$, we have
\begin{equation} \label{eq:alpha_f} \sharp(\alpha)+\sharp(\gamma_f^{-1}\alpha)= 2p-(|\alpha|+|\gamma_f^{-1}\alpha|)\leq 2p-|\gamma_f| =p+\sharp(\gamma_f) =p+|\im(f)|, \end{equation}
where the first equality is once more by Lemma \ref{lemma:cycles-transpositions}, whereas the second inequality is by equation \eqref{eq:geodesic''} in Lemma \ref{lemma:distance} and is an equality if and only if $\alpha\in NC(p)$ and $f\circ\alpha=f$. So equations \eqref{eq:alpha} and \eqref{eq:alpha_f} together yield that, for any $\alpha\in\mathfrak{S}(p)$ and $f:[p]\rightarrow[k]$,
\begin{equation} \label{eq:alpha-alpha_f} 2\sharp(\alpha)+\sharp(\gamma^{-1}\alpha)+\sharp(\gamma_f^{-1}\alpha)+k-|\im(f)|\leq 2p+k+1, \end{equation}
with equality if and only if $\alpha\in NC(p)$ and $f\circ\alpha=f$.

We thus get the asymptotic estimate
\[ \mathbf{E}\, \mathrm{Tr}\left[\left(\sum_{j=1}^k\widetilde{W}_{\A\B^k}(j)\right)^p\right] \underset{d\rightarrow+\infty}{\sim} \left(\sum_{\alpha\in NC(p)} \sum_{\underset{f\circ\alpha=f}{f:[p]\rightarrow[k]}}c^{\sharp(\alpha)}\right) d^{2p+k+1}. \]
Yet, a function $f$ satisfying $f\circ\alpha=f$ is fully characterized by its value on each of the $\sharp(\alpha)$ cycles of $\alpha$. So there are $k^{\sharp(\alpha)}$ such functions. Hence in the end, the asymptotic estimate
\[ \mathbf{E}\, \mathrm{Tr}\left[\left(\sum_{j=1}^k\widetilde{W}_{\A\B^k}(j)\right)^p\right] \underset{d\rightarrow+\infty}{\sim} \left(\sum_{\alpha\in NC(p)}(ck)^{\sharp(\alpha)}\right) d^{2p+k+1} = \mathrm{M}_{MP(ck)}^{(p)} d^{2p+k+1}, \]
the last equality being because, for any $\lambda>0$, $\sum_{\alpha\in NC(p)}\lambda^{\sharp(\alpha)} = \sum_{m=1}^p\lambda^m\mathrm{Nar}_p^m = \mathrm{M}_{MP(\lambda)}^{(p)}$.
\end{proof}

\begin{proof} [Proof of Proposition \ref{prop:wishart-infty}]
The argument will follow the exact same lines as the one used to derive Proposition \ref{prop:gaussian-infty} from Proposition \ref{prop:gaussian-p}.

As pointed out there, showing the inequality ``$\geq$'' in Proposition \ref{prop:wishart-infty} is easy. Indeed, the convergence in moments established in Proposition \ref{prop:wishart-p} implies 
that, asymptotically, the matrix $\left(\sum_{j=1}^k\widetilde{W}_{\A\B^k}(j)\right)/d^2$ has a largest eigenvalue which is, on average, at least the upper-edge of the support of $\mu_{MP(ck)}$, i.e.~$(\sqrt{ck}+1)^2$. In other words, it guarantees that there exist positive constants $c_d\rightarrow_{d\rightarrow+\infty}1$ such that
\begin{equation} \label{eq:lower-bound} \mathbf{E} \left\|\sum_{j=1}^k\widetilde{W}_{\A\B^k}(j)\right\|_{\infty} 
\geq c_d\,(\sqrt{ck}+1)^2d^2. \end{equation}

Let us now turn to the more tricky part, which is showing the inequality ``$\leq$'' in Proposition \ref{prop:wishart-infty}. For $d\in\N$ fixed, it holds thanks to Jensen's inequality and monotonicity of Schatten norms that
\begin{equation} \label{eq:infty-p-wishart} \forall\ p\in\N,\ \mathbf{E} \left\|\sum_{j=1}^k\widetilde{W}_{\A\B^k}(j)\right\|_{\infty} \leq \left( \mathbf{E} \left\|\sum_{j=1}^k\widetilde{W}_{\A\B^k}(j)\right\|_{\infty}^p \right)^{1/p} \leq \left(\mathbf{E}\, \mathrm{Tr}\left[\left(\sum_{j=1}^k\widetilde{W}_{\A\B^k}(j)\right)^p\right] \right)^{1/p} .\end{equation}
So let us fix $d\in\N$ and $p\in\N$, and rewrite \eqref{eq:p-moment} explicitly as an expansion in powers of $d$, keeping in the sum the permutations not saturating equation \eqref{eq:alpha-alpha_f}. Being cautious only regarding the permutations not saturating equation \eqref{eq:alpha_f}, and not regarding those not saturating equation \eqref{eq:alpha}, we thus get the upper bound
\begin{equation} \label{eq:momentsW-defect}
\mathbf{E}\, \mathrm{Tr}\left[\left(\sum_{j=1}^k\widetilde{W}_{\A\B^k}(j)\right)^p\right]\leq \left(\underset{f:[p]\rightarrow[k]}{\sum} \sum_{\delta=0}^{\lfloor(p+k)/2\rfloor}\sum_{m=1}^p \left|\mathfrak{S}_{f,\delta,m}(p)\right|c^m d^{-2\delta}\right) d^{2p+k+1},
\end{equation}
where we defined, for each $f:[p]\rightarrow[k]$, each $1\leq m\leq p$ and each $0\leq\delta\leq \lfloor(p+k)/2\rfloor$,
\[ \mathfrak{S}_{f,\delta,m}(p)=\left\{ \alpha\in\mathfrak{S}(p) \st \sharp(\alpha)=m\ \text{and}\ \sharp(\alpha)+\sharp(\gamma_f^{-1}\alpha) =p+|\im(f)|-2\delta \right\}. \]
$\mathfrak{S}_{f,\delta,m}(p)$ is thus nothing else than the set of permutations which are composed of $m$ cycles and have a defect $2\delta$ from lying on the geodesics between the identity and the product of the canonical full cycles on each of the level sets of $f$. This justifies in particular \textit{a posteriori} why the summation in \eqref{eq:momentsW-defect} is only over even defects (see the parity argument in Lemma \ref{lemma:distance-general}). Note that the definition of $\mathfrak{S}_{f,\delta,m}(p)$ can actually be extended to all $m\in\N$, with $\left|\mathfrak{S}_{f,\delta,m}(p)\right|=0$ if $m\geq p+|\im(f)|-2\delta$, which we shall do in what follows for writing convenience.

Now, by Lemma \ref{lemma:number-functions,permutations-defect}, we know that, if $0\leq\delta\leq\lfloor p/2\rfloor$, then for any $1\leq m\leq p$,
\[ \left|\big\{ (f,\alpha) \st \alpha\in\mathfrak{S}_{f,\delta,m}(p) \big\}\right| \leq \left(\sum_{\epsilon=0}^{2\delta}k^{m-\epsilon}\mathrm{Nar}_p^{m-\epsilon}\right)\times\left(2k^2p^2\right)^{2\delta}. \]
And if $\lceil p/2\rceil\leq\delta\leq \lfloor (p+k)/2\rfloor$, then trivially for any $1\leq m\leq p$,
\[ \left|\big\{ (f,\alpha) \st \alpha\in\mathfrak{S}_{f,\delta,m}(p) \big\}\right| \leq k^p\,p! \leq \left(\sum_{\epsilon=0}^{p}k^{m-\epsilon}\mathrm{Nar}_p^{m-\epsilon}\right)\times\left(2k^2p^2\right)^{p}. \]
What is more, for a given $0\leq\delta\leq\lceil p/2\rceil$, we have, making the change of summation index $m\mapsto m-\epsilon$,
\begin{align*}
\sum_{m=1}^pc^m\left(\sum_{\epsilon=0}^{2\delta}k^{m-\epsilon}\mathrm{Nar}_p^{m-\epsilon}\right) & = \left(\sum_{\epsilon=0}^{2\delta}c^{-\epsilon}\right) \left(\sum_{m=1}^p(ck)^{m}\mathrm{Nar}_p^{m}\right) \\
& \leq (1+c)^{2\delta} \sum_{m=1}^p(ck)^{m}\mathrm{Nar}_p^{m}\\
& = (1+c)^{2\delta} \mathrm{M}_{MP(ck)}^{(p)}.
\end{align*}
Putting everything together, we therefore get,
\[ \mathbf{E}\, \mathrm{Tr}\left[\left(\sum_{j=1}^k\widetilde{W}_{\A\B^k}(j)\right)^{p}\right] \leq \mathrm{M}_{MP(ck)}^{(p)} \left(1 + \sum_{\delta=1}^{\lfloor p/2\rfloor}\left(\frac{2(1+c)k^2p^2}{d}\right)^{2\delta} +\frac{k}{2}\left(\frac{2(1+c)k^2p^2}{d}\right)^p\right) d^{2p+k+1}.\]
Yet, $\max\left\{\left(2(1+c)k^2p^2/d\right)^{2\delta} \st 1\leq\delta\leq \lceil p/2\rceil\right\}$ is attained for $\delta=1$, provided $p\leq(d/2(1+c)k^2)^{1/2}$. So if such is the case,
\[ \sum_{\delta=1}^{\lfloor p/2\rfloor}\left(\frac{2(1+c)k^2p^2}{d}\right)^{2\delta}+\frac{k}{2}\left(\frac{2(1+c)k^2p^2}{d}\right)^p \leq \frac{p+k}{2} \frac{4(1+c)^2k^4p^4}{d^2} \leq \frac{4(1+c)^2k^4p^5}{d^2}, \]
where the last inequality holds as long as $p\geq k$. And hence, under all the previous assumptions,
\[ \mathbf{E}\, \mathrm{Tr}\left[\left(\sum_{j=1}^k\widetilde{W}_{\A\B^k}(j)\right)^{p}\right] \leq \mathrm{M}_{MP(ck)}^{(p)} \left(1 +\frac{4(1+c)^2k^4p^5}{d^2}\right) d^{2p+k+1}. \]

So set $p_d=(d/2(1+c)k^2)^{(2-\epsilon)/5}$ for some $0<\epsilon<1$ (which is indeed smaller than $(d/2(1+c)k^2)^{1/2}$ and bigger than $k$ for $d$ big enough, in particular bigger than $2(1+c)k^{9/2}$). And using inequality \eqref{eq:infty-p-wishart} in the special case $p=p_d$, we eventually get
\begin{equation} \label{eq:upper-bound} \mathbf{E} \left\|\sum_{j=1}^k\widetilde{W}_{\A\B^k}(j)\right\|_{\infty} \leq \left(\mathrm{M}_{MP(ck)}^{(p_d)} \left(1 + \frac{4(1+c)^2p_d^4}{d^2}\right)\right)^{1/p_d} d^{2+(k+1)/p_d}
\underset{d\rightarrow+\infty}{\sim} (\sqrt{ck}+1)^2d^2.
\end{equation}

Combining the lower bound in equation \eqref{eq:lower-bound} and the upper bound in equation \eqref{eq:upper-bound} yields Proposition \ref{prop:wishart-infty}.
\end{proof}

\subsection{Conclusion} Having at hand the operator-norm estimate from Proposition \ref{prop:wishart-infty}, we can now easily answer our initial question. It is the content of Theorem \ref{th:not-kext} below.

\begin{theorem}
\label{th:not-kext}
Let $k\in\N$, and for any $0<\epsilon<1/2$ define $c_{\epsilon}(k)=\frac{(k-1)^2}{4k}(1-\epsilon)$. Then, there exists a constant $C_{k,\epsilon}>0$ such that
\[ \mathbf{P}_{\rho\sim\mu_{d^2,c_{\epsilon}(k)d^2}} \big(\rho\notin\mathcal{E}_k(\C^d{:}\C^d)\big) \geq 1-e^{-C_{k,\epsilon}d^2}. \]
One can take $C_{k,\epsilon}=C\epsilon^2/k$ for some universal constant $C>0$.
\end{theorem}

\begin{proof}
As a direct consequence of Proposition \ref{prop:wishart-infty}, we have
\[ \mathbf{E}_{W_{\A\B}\sim\mathcal{W}_{d^2,cd^2}} \left[\underset{\sigma_{\A\B}\in\mathcal{E}_k(\A{:}\B)}{\sup}\ \tr\big(W_{\A\B}\sigma_{\A\B}\big)\right] \underset{d\rightarrow+\infty}{\sim} \frac{(\sqrt{ck}+1)^2}{k}d^2. \]
And since $\mathbf{E}_{W\sim\mathcal{W}_{d^2,s}}\mathrm{Tr}W\sim_{d,s\rightarrow +\infty} d^2s$ (see e.g.~\cite{CN} or Appendix \ref{appendix:wick-wishart}), the result we eventually come to after renormalizing by $\mathrm{Tr}W$ is
\begin{equation}
\label{eq:k-ext}
\mathbf{E}_{\rho\sim\mu_{d^2,cd^2}} \left[\underset{\sigma\in\mathcal{E}_k(\C^d{:}\C^d)}{\sup}\ \mathrm{Tr}(\rho\sigma)\right] \underset{d\rightarrow+\infty}{\sim} \frac{1}{d^2cd^2}\frac{(\sqrt{ck}+1)^2}{k}d^2 = \frac{(\sqrt{ck}+1)^2}{ck}\frac{1}{d^2}.
\end{equation}
On the other hand, $\mathbf{E}_{W\sim\mathcal{W}_{d^2,s}}\mathrm{Tr}(W^2)\sim_{d,s\rightarrow +\infty} d^2s^2 + (d^2)^2s$ (see e.g.~\cite{CN} or Appendix \ref{appendix:wick-wishart}), so we also have
\[ \mathbf{E}_{\rho\sim\mu_{d^2,cd^2}} \big[\mathrm{Tr}(\rho^2)\big] \underset{d\rightarrow+\infty}{\sim} \frac{d^2(cd^2)^2+(d^2)^2cd^2}{(d^2cd^2)^2} = \left(1+\frac{1}{c}\right)\frac{1}{d^2}. \]
Now, if $c=(1-\epsilon)(k-1)^2/4k$ for some $0<\epsilon<1/2$, then
\[ \frac{(\sqrt{ck}+1)^2}{ck}-\left(1+\frac{1}{c}\right)=\frac{4\epsilon}{(k-1)(1-\epsilon)} < \frac{8\epsilon}{k-1}. \]
So by equation \eqref{eq:proba-notinK}, we have in such case
\[ \mathbf{P}_{\rho\sim\mu_{d^2,cd^2}} \big(\rho\notin\mathcal{E}_k(\C^d{:}\C^d)\big) \geq 1-\exp\left(-Ckd^2d^4(\epsilon/kd^2)^2\right)= 1 -\exp\left(-Cd^2\epsilon^2/k\right), \]
for some universal constant $C>0$.
\end{proof}

\section{Discussion and comparison with other separability criteria}
\label{section:threshold-comparison}

For each $k\in\N$, define $c_{k-ext}$ as the smallest constant $c$ such that a random state $\rho$ on $\C^d\otimes\C^d$ induced by an environment of dimension $cd^2$ is not $k$-extendible with high probability when $d$ is large. That is,
\[ c_{k-ext}= \inf\left\{ c \st \P_{\rho\sim\mu_{d^2,cd^2}}\left(\rho\notin\mathcal{E}_k(\C^d{:}\C^d)\right) \underset{d\rightarrow+\infty}{\rightarrow} 1 \right\}. \]
What we established in Theorem \ref{th:not-kext} is that $c_{k-ext}\leq(k-1)^2/4k$.

Yet, we know from \cite{ASY} that for $c>0$ fixed, $\rho\sim\mu_{d^2,cd^2}$ is with high probability entangled when $d\rightarrow+\infty$: the threshold for $\rho\sim\mu_{d^2,s(d)}$ being with high probability either entangled or separable occurs for some $s(d)=s_0(d)$ with $d^3\lesssim s_0(d)\lesssim d^3\log^2 d$. So what we proved is that if $c<(k-1)^2/4k$, i.e.~if $k>2c+2\sqrt{c(c+1)}+1$, then this generic entanglement will be generically detected by the $k$-extendibility test.

Furthermore, it is well-known (see e.g.~\cite{ZS}) that $\rho\sim\mu_{d^2,d^2}$ is equivalent to $\rho$ being uniformly distributed on the set of mixed states on $\C^d\otimes\C^d$ (for the Haar measure induced by the Hilbert--Schmidt distance). As just mentioned, when $d\rightarrow+\infty$, such states are typically not separable. Now, for $k\geq 6$, $(k-1)^2/4k>1$, so such states are also typically not $k$-extendible. Hence, entanglement of uniformly distributed mixed states on $\C^d\otimes\C^d$ is typically detected by the $k$-extendibility test for $k\geq 6$.

Let us define, in a similar way to what was done for the $k$-extendibility criterion, $c_{ppt}$, resp.~$c_{ra}$, as the smallest constant $c$ such that a random state $\rho$ on $\C^d\otimes\C^d$ induced by an environment of dimension $cd^2$ is, with probability tending to one when $d$ tends to infinity, not satisfying the PPT, resp.~realignment (see Chapter \ref{chap:SDrelaxations}, Section \ref{section:CCNR}), criterion.
We know from \cite{Aubrun1} that $c_{ppt}=4$, whereas we know from \cite{AN} that $c_{ra}=\left(8/3\pi\right)^2$. Now, for $k\geq 17$, $(k-1)^2/4k>4$, and for $k\geq 5$, $(k-1)^2/4k>\left(8/3\pi\right)^2$. So roughly speaking, this means that the $k$-extendibility criterion for separability becomes ``better'' than the PPT one at most for $k\geq 17$, and ``better'' than the realignment one at most for $k\geq 5$. This is to be taken in the following sense: if $k\geq 17$, resp.~$k\geq 5$, then there is a range of environment dimensions for which random-induced states have a generic entanglement which is generically detected by the $k$-extendibility test but not detected by the PPT, resp.~realignment, test.

Note also that for the reduction criterion \cite{HH}, it was established in \cite{JLN} that the threshold for a random-induced state on $\C^d\otimes\C^d$ either passing or failing it with high probability occurs at an environment dimension $d$, hence much smaller than for all previously mentioned criteria.

\section{The unbalanced case}
\label{section:unbalanced}

For the sake of simplicity, we previously focussed on the case where $\mathrm{H}=\mathrm{A}\otimes\mathrm{B}$ is a balanced bipartite Hilbert space. One may now wonder what happens, more generally, when $\mathrm{A}\equiv\C^{d_\A}$ and $\mathrm{B}\equiv\C^{d_\B}$ with $d_\A$ and $d_\B$ being possibly different. It is easy to see that the results from Theorems \ref{th:not-kext} and \ref{th:w(k-ext)} straightforwardly generalize to the case where $d_\A$ and $d_\B$ both tend to infinity (but possibly at different rates). The corresponding statements appear in Theorem \ref{th:unbalanced} below.

\begin{theorem} \label{th:unbalanced}
Let $k\in\N$ and let $d_\A,d_\B\in\N$. The mean width of the set of $k$-extendible states on $\C^{d_\A}\otimes\C^{d_\B}$ (with respect to $\C^{d_\B}$) satisfies
\[ w\big(\mathcal{E}_k(\C^{d_\A}{:}\C^{d_\B})\big) \underset{d_\A,d_\B\rightarrow+\infty}{\sim} \frac{2}{\sqrt{k}}\frac{1}{\sqrt{d_\A d_\B}} .\]
Also, when $d_\A,d_\B\rightarrow+\infty$, a random state on $\C^{d_\A}\otimes\C^{d_\B}$ which is sampled from $\mu_{d_\A d_\B,c d_\A d_\B}$, with $c<(k-1)^2/4k$, is with high probability not $k$-extendible (with respect to $\C^{d_\B}$).
\end{theorem}

Oppositely, when one of the two subsystems has a fixed dimension and the other one only has an increasing dimension, the sets of $k$-extendible states with respect to either the smaller or the bigger subsystem exhibit different size scalings. This is made precise in Theorem \ref{th:unbalanced'} below.

\begin{theorem} \label{th:unbalanced'}
Let $k\in\N$ and let $d_\A,d_\B\in\N$. If $d_\A$ is fixed, the mean width of the set of $k$-extendible states on $\C^{d_\A}\otimes\C^{d_\B}$ (with respect to $\C^{d_B}$) satisfies
\[ w\big(\mathcal{E}_k(\C^{d_\A}{:}\C^{d_\B})\big) \underset{d_\B\rightarrow+\infty}{\sim} \frac{2}{\sqrt{k}}\frac{1}{\sqrt{d_\A d_\B}} .\]
Whereas if $d_\B$ is fixed, the mean width of the set of $k$-extendible states on $\C^{d_\A}\otimes\C^{d_\B}$ (with respect to $\C^{d_\B}$) satisfies
\[ w\big(\mathcal{E}_k(\C^{d_\A}{:}\C^{d_\B})\big) \underset{d_\A\rightarrow+\infty}{\sim} \frac{2\,C(d_\B,k)}{\sqrt{k}}\frac{1}{\sqrt{d_\A d_\B}} ,\]
with $C(d_\B,k)\geq \left(1+(k-1)/d_\B^2\right)^{1/4}$.
\end{theorem}

\begin{proof}
Using the same notation as in the proof of Proposition \ref{prop:gaussian-p}, we start in both cases from the exact expression for the $2p$-order moment (slightly generalizing Proposition \ref{prop:gaussian-p-preliminary})
\begin{equation} \label{eq:p-dAdB} \mathbf{E}_{G_{\A\B}\sim GUE(d_\A d_\B)}\, \mathrm{Tr}\left[\left(\sum_{j=1}^k\widetilde{G}_{\A\B^k}(j)\right)^{2p}\right] =\underset{f:[2p]\rightarrow[k]}{\sum} \underset{\lambda\in\mathfrak{P}^{(2)}(2p)}{\sum} d_\A^{\sharp(\gamma^{-1}\lambda)}d_\B^{\sharp(\gamma_f^{-1}\lambda)+k-|\im(f)|}. \end{equation}

First, fix $d_\A$. The argument then follows the exact same lines as in the proof of Proposition \ref{prop:gaussian-p}. Indeed, the pair partitions $\lambda\in\mathfrak{P}^{(2)}(2p)$ contributing to the dominant order in $d_B$ in the expansion \eqref{eq:p-dAdB} are the $\mathrm{Cat}_p$ non-crossing pair partitions $\lambda\in NC^{(2)}(2p)$, for which $\sharp(\gamma^{-1}\lambda)=p+1$. Moreover, for each of these $\lambda$, the functions $f:[2p]\rightarrow[k]$ contributing to the dominant order in $d_B$ in the expansion \eqref{eq:p-dAdB} are the $k^p$ functions which are such that $f\circ\lambda=f$, for which $\sharp(\gamma_f^{-1}\lambda)+k-|\im(f)|=p+k$. So we eventually get
\[ \mathbf{E}_{G_{\A\B}\sim GUE(d_\A d_\B)}\, \mathrm{Tr}\left[\left(\sum_{j=1}^k\widetilde{G}_{\A\B^k}(j)\right)^{2p}\right] \underset{d_\B\rightarrow+\infty}{\sim} \mathrm{Cat}_p\,k^pd_\A^{p+1}d_\B^{p+k}. \]

Now, fix $d_\B$. Again, the pair partitions $\lambda\in\mathfrak{P}^{(2)}(2p)$ contributing to the dominant order in $d_\A$ in the expansion \eqref{eq:p-dAdB} are the $\mathrm{Cat}_p$ non-crossing pair partitions $\lambda\in NC^{(2)}(2p)$, for which $\sharp(\gamma^{-1}\lambda)=p+1$.
So consider one of these $\lambda$. Observe that, for any $0\leq\delta\leq \lfloor(p+k)/2\rfloor$, if $f:[2p]\rightarrow[k]$ is such that there are exactly $\delta$ pair blocks of $\lambda$ on which $f$ takes two values, then necessarily  $\sharp(\gamma_f^{-1}\lambda)+k-|\im(f)|\geq p+k-2\delta$. Indeed, the case $\delta=0$ is already known. So let us describe precisely what happens in the case $\delta=1$, i.e.~when there is exactly $1$ pair block of $\lambda$ on which $f$ takes $2$ values.\\
$\bullet$ If amongst these $2$ values, at least $1$ of them is also taken on another pair block of $\lambda$, then there exist transpositions $\tau,\tau'$ and a function $g$ satisfying $g\circ\lambda=g$, such that $\gamma_f=\gamma_g\tau\tau'$ and $|\im(f)|=|\im(g)|$. Hence,
\[ \sharp(\gamma_f^{-1}\lambda) = \sharp(\tau'\tau\gamma_f^{-1}\lambda) \geq \sharp(\gamma_g^{-1}\lambda) -2 = p+|\im(g)|-2 = p+|\im(f)|-2. \]\\
$\bullet$ If none of these $2$ values is also taken on another pair block of $\lambda$, then there exist a transposition $\tau$ and a function $g$ satisfying $g\circ\lambda=g$, such that $\gamma_f=\gamma_g\tau$ and $|\im(f)|=|\im(g)|+1$. Hence,
\[ \sharp(\gamma_f^{-1}\lambda) = \sharp(\tau\gamma_f^{-1}\lambda) \geq \sharp(\gamma_g^{-1}\lambda) -1 = p+|\im(g)|-1 = p+|\im(f)|-2. \]
And this generalizes in a similar way to $\delta>1$. Yet, for a given $0\leq\delta\leq \lfloor(p+k)/2\rfloor$, there are ${p \choose \delta}k^p(k-1)^{\delta}$ functions which take $2$ values on exactly $\delta$ pair blocks of $\lambda$ (assuming of course that $p\geq k$). So we eventually get
\begin{align*} \mathbf{E}_{G_{\A\B}\sim GUE(d_\A d_\B)}\, \mathrm{Tr}\left[\left(\sum_{j=1}^k\widetilde{G}_{\A\B^k}(j)\right)^{2p}\right] & \geq \mathrm{Cat}_p\,k^p\left(\sum_{\delta=0}^{\lfloor(p+k)/2\rfloor}{p \choose \delta}(k-1)^{\delta}d_\B^{-2\delta}\right)d_\B^{p+k}d_\A^{p+1}\\
& \geq \mathrm{Cat}_p\,k^p\left(\sum_{\delta=0}^{\lfloor(p+k)/2\rfloor}{\lfloor(p+k)/2\rfloor \choose \delta}\left(\frac{k-1}{d_\B^2}\right)^{\delta}\right)d_\B^{p+k}d_\A^{p+1}\\
& = \mathrm{Cat}_p\,k^p\left(1+\frac{k-1}{d_\B^2}\right)^{\lfloor(p+k)/2\rfloor}d_\B^{p+k}d_\A^{p+1}. \end{align*}

One can then argue as in the proof of the derivation of Proposition \ref{prop:gaussian-infty} from Proposition \ref{prop:gaussian-p} that we additionally have $\mathbf{E}\left\| \sum_{j=1}^{k} \widetilde{G}_{\A\B^k}\right\|_{\infty}\sim\lim_{p\rightarrow+\infty} \left( \mathbf{E}\, \mathrm{Tr}\left[ \left( \sum_{j=1}^{k} \widetilde{G}_{\A\B^k}\right)^{2p} \right] \right)^{1/2p}$, when either $d_\B\rightarrow+\infty$ or $d_\A\rightarrow+\infty$. This automatically yields the two announced statements on the mean width of $\mathcal{E}_k(\C^{d_\A}{:}\C^{d_\B})$.
\end{proof}

\begin{remark}
In the situation where $d_\B$ is fixed, if we had an exact expression
\[ \forall\ p\in\N,\ \mathbf{E}_{G_{\A\B}\sim GUE(d_\A d_\B)}\, \mathrm{Tr}\left[\left(\sum_{j=1}^k\widetilde{G}_{\A\B^k}(j)\right)^{2p}\right] \underset{d_\A\rightarrow+\infty}{\sim} M(d_\A,p), \]
then we would be able to conclude without any further argument that
\[ \mathbf{E}\left\| \sum_{j=1}^{k} \widetilde{G}_{\A\B^k}\right\|_{\infty} \underset{d_\A\rightarrow+\infty}{\sim} \lim_{p\rightarrow+\infty} M(d_\A,p)^{1/2p}. \]
This would indeed follow from the convergence result of \cite{HT} for non-commutative polynomials in multi-variables with matrix coefficients (in our case, $d_\B^2$ variables and $d_\B^k\times d_\B^k$ coefficients).
\end{remark}

The asymmetry in the definition of $k$-extendibility appears more strikingly in this unbalanced setting. Indeed, for a finite $k\in\N$, a given state on $\mathrm{A}\otimes\mathrm{B}$ may be $k$-extendible with respect to $\mathrm{B}$ but not $k$-extendible with respect to $\mathrm{A}$. It is only in the limit $k\rightarrow +\infty$ that there is equivalence between the two notions: a state on $\mathrm{A}\otimes\mathrm{B}$ is $k$-extendible with respect to $\mathrm{B}$ for all $k\in\N$ if and only if it is $k$-extendible with respect to $\mathrm{A}$ for all $k\in\N$ (and if and only if it is separable).

However, what Theorem \ref{th:unbalanced} stipulates is that, even for a finite $k\in\N$, when both subsystems grow, being $k$-extendible with respect to either one or the other are two constraints which are, on average, equivalently restricting. On the contrary, what Theorem \ref{th:unbalanced'} shows is that when only one subsystem grows, and the other remains of fixed size, being $k$-extendible with respect to the bigger one is, on average, a tougher constraint than being $k$-extendible with respect to the smaller one (as one would have probably expected).

This is to be put in perspective with some of the original observations made in \cite{DPS}. It was indeed noticed that checking whether a state on $\C^{d}\otimes\C^{d'}$ is $k$-extendible with respect to $\C^{d'}$ requires space resources which scale as $\left[{d'+k-1 \choose k}d\right]^2$ when implemented. It was therefore advised that in the unbalanced situation of $d$ ``big'' and $d'$ ``small'', one should check $k$-extendibility with respect to $\C^{d'}$ rather than $\C^d$, the former being much more economical. On the other hand, it comes out from our study that, in this case, an entangled state is likely to fail passing the $k$-extendibility test for a smaller $k$ when the extension is searched with respect to $\C^d$ than when it is searched with respect to $\C^{d'}$. But understanding the precise trade-off seems out of reach at the moment.

\section{Miscellaneous questions}
\label{section:miscellaneous}

\subsection{What about the mean width of the set of $k$-extendible states for ``big'' $k$?}

All the statements proven sofar, regarding either the $k$-extendibility of random-induced states or the mean width of the set of $k$-extendible states, converge towards the same (expected) conclusion: for any given $k\in\N$, the $k$-extendibility criterion becomes a very weak necessary condition for separability when the dimension of the considered bipartite system increases. So the natural question at that point is: what can be said about the $k$-extendibility criterion on $\C^d\otimes\C^d$ when $k\equiv k(d)$ is allowed to grow in some way with $d$? Unfortunately, most of the results we established rely at some point on the assumption that $k$ is a fixed parameter, and therefore do not seem to be directly generalizable to the case where $k$ depends on $d$.

There is at least one estimate though that remains valid in this setting, which is the lower bound on the mean width of $k$-extendible states.

\begin{theorem} \label{th:w(k(d)-ext)-lower}
There exist positive constants $c_d\rightarrow_{d\rightarrow+\infty}1$ such that, for any $k(d)\in\N$, the mean width of the set of $k(d)$-extendible states on $\C^d\otimes\C^d$ satisfies
\[ w\left(\mathcal{E}_{k(d)}(\C^d{:}\C^d)\right) \geq c_d\, \frac{2}{\sqrt{k(d)}}\frac{1}{d}. \]
\end{theorem}

\begin{proof}
Let $d,k(d)\in\N$. For any $p\in\N$, the exact expression for the $2p$-order moment established in Proposition \ref{prop:gaussian-p-preliminary} of course remains true. So by the same arguments as in the proof of Proposition \ref{prop:gaussian-p}, we still have in that case at least the lower bound
\begin{align*}
\mathbf{E}_{G_{\A\B}\sim GUE(d^2)} \mathrm{Tr}\left[\left(\sum_{j=1}^{k(d)}\widetilde{G}_{\A\B^{k(d)}}(j)\right)^{2p}\right] = & \, \underset{f:[2p]\rightarrow[k(d)]}{\sum} \underset{\lambda\in\mathfrak{P}^{(2)}(2p)}{\sum} d^{\sharp(\gamma^{-1}\lambda)+\sharp(\gamma_f^{-1}\lambda)+k(d)-|\im(f)|}\\
\geq & \, \mathrm{Cat}_p\,k(d)^pd^{2p+k(d)+1}.
\end{align*}
This lower bound on moments in turn guarantees, as explained in the derivation of Proposition \ref{prop:gaussian-infty} from Proposition \ref{prop:gaussian-p}, that there exist positive constants $c_d\rightarrow_{d\rightarrow+\infty}1$ such that we have the inequality
\[ \mathbf{E}_{G_{\A\B}\sim GUE(d^2)} \left\| \underset{j=1}{\overset{k(d)}{\sum}}\widetilde{G}_{\A\B^{k(d)}}(j) \right\|_{\infty} 
\geq c_d\, 2\sqrt{k(d)}d ,\]
which yields the announced lower bound for the mean width of $\mathcal{E}_{k(d)}(\C^d{:}\C^d)$.
\end{proof}

Theorem \ref{th:w(k(d)-ext)-lower} only provides a lower bound on the asymptotic mean width of $\mathcal{E}_{k}(\C^d{:}\C^d)$ when $k$ is allowed to depend on $d$. It is nevertheless already an interesting piece of information. Indeed, as mentioned in Section \ref{section:w-comparison}, we know from \cite{AS1} that the mean width of the set of separable states on $\C^d\otimes\C^d$ is of order $1/d^{3/2}$. Theorem \ref{th:w(k(d)-ext)-lower} therefore asserts that, on $\C^d\otimes\C^d$, one has to go at least to $k$ of order $d$ to obtain a set of $k$-extendible states whose mean width scales as the one of the set of separable states.

Furthermore, it may be worth mentioning that the proof of Proposition \ref{prop:gaussian-infty} actually provides additional information, namely an upper bound on the mean width of $k$-extendible states which remains valid for a quite wide range of $k$.

\begin{theorem} \label{th:w(k(d)-ext)-upper}
For any $d,k(d)\in\N$, provided that $k(d)<d^{2/7}$ and $d$ is big enough, the mean width of the set of $k(d)$-extendible states on $\C^d\otimes\C^d$ satisfies
\[ w\big(\mathcal{E}_{k(d)}(\C^d{:}\C^d)\big) \leq \frac{2}{\sqrt{k(d)}}\frac{1}{d}\exp\left(\frac{k(d)^{7/5}\ln d}{d^{2/5}}\right). \]
\end{theorem}

\begin{proof}
Let $d,k(d)\in\N$ with $k(d)<d^{2/7}$. Taking $p_d=(d/k(d))^{2/5}$ in equation \eqref{eq:upper-bound-gaussian} (which is indeed, as required, bigger than $k(d)$ and smaller than $(2d/k(d))^{1/2}$ for $d$ big enough) we get
\[ \mathbf{E}_{G_{\A\B}\sim GUE(d^2)} \left\|\sum_{j=1}^{k(d)}\widetilde{G}_{\A\B^k}(j)\right\|_{\infty} \leq 2\sqrt{k(d)}\,d\, \exp\left(\frac{k(d)^{7/5}}{2d^{2/5}}\ln d +\frac{k(d)^{2/5}}{2d^{2/5}}\left[\ln d + \ln \left(\frac{5}{4}\right)\right] \right). \]
The latter quantity is smaller than $2\sqrt{k(d)}\,d\,\exp\left(k(d)^{7/5}\ln d /d^{2/5}\right)$ for $d$ big enough, which yields the advertised upper bound for the mean width of $\mathcal{E}_{k(d)}(\C^d{:}\C^d)$.
\end{proof}

Of course, the upper bound provided by Theorem \ref{th:w(k(d)-ext)-upper} is interesting only for $k(d)<k_0(d)=d^{2/7}/(\ln d)^{5/7}$. Nevertheless, since the set of $k$-extendible states contains the set of $k'$-extendible states for all $k'\geq k$, we also have as a (potentially weak) consequence of Theorem \ref{th:w(k(d)-ext)-upper} that for $k(d)\geq k_0(d)$ and $d$ big enough,
\[ w\big(\mathcal{E}_{k(d)}(\C^d{:}\C^d)\big) \leq w\big(\mathcal{E}_{k_0(d)}(\C^d{:}\C^d)\big) \leq \frac{2e\,(\ln d)^{5/14}}{d^{8/7}} . \]

Theorems \ref{th:w(k(d)-ext)-lower} and \ref{th:w(k(d)-ext)-upper} together imply in particular the following: in the regime where $k(d)$ grows with $d$ slower than $d$ itself, both the ratio $w\big(\mathcal{S}(\C^d{:}\C^d)\big)/w\big(\mathcal{E}_{k(d)}(\C^d{:}\C^d)\big)$ and the ratio $w\big(\mathcal{E}_{k(d)}(\C^d{:}\C^d)\big)/w\big(\mathcal{D}(\C^d\otimes\C^d)\big)$ are unbounded. To rephrase it, for $k(d)$ having this growth rate, the set of $k(d)$-extendible states lies ``strictly in between'' the set of separable states and the set of all states from an asymptotic size point of view.

\subsection{When is a random-induced state with high probability $k$-extendible?}

The result provided by Theorem \ref{th:not-kext} is only one-sided: it tells us that if $s<s(k,d)=d^2(k-1)^2/4k$, then a random mixed state on $\C^d\otimes\C^d$ obtained by partial tracing on $\C^s$ a uniformly distributed pure state on $\C^d\otimes\C^d\otimes\C^s$ is with high probability not $k$-extendible. But what can be said about the case $s>s(k,d)$? Or more generally, can one find a reasonable $s'(k,d)$ such that if $s>s'(k,d)$, then a random mixed state on $\C^d\otimes\C^d$ obtained by partial tracing on $\C^s$ a uniformly distributed pure state on $\C^d\otimes\C^d\otimes\C^s$ is with high probability $k$-extendible?

By the arguments discussed in extensive depth in \cite{ASY}, one can assert at least that there exists a universal constant $c>0$ such that $ckd^2\log^2d$ is a possible value for such $s'(k,d)$. We will not repeat the whole reasoning here, but let us still give the key ideas underlying it.

Define $\overline{\mathcal{E}}_k(\C^d{:}\C^d)$ the translation of $\mathcal{E}_k(\C^d{:}\C^d)$ by its center of mass, the maximally mixed state $\Id/d^2$, i.e.~\[ \overline{\mathcal{E}}_k(\C^d{:}\C^d) =\left\{ \rho-\frac{\Id}{d^2} \st \rho\in\mathcal{E}_k(\C^d{:}\C^d) \right\}. \]
Define also $\overline{\mathcal{E}}_k(\C^d{:}\C^d)^{\circ}$ the convex body polar to $\overline{\mathcal{E}}_k(\C^d{:}\C^d)$, i.e.~\[ \overline{\mathcal{E}}_k(\C^d{:}\C^d)^{\circ} = \left\{ \Delta\in\mathcal{H}(d^2) \st \forall\ X\in\overline{\mathcal{E}}_k(\C^d{:}\C^d),\ \tr\left(\Delta X\right)\leq 1 \right\}. \]
What then has to be specifically determined is (see \cite{ASY}, Section 2, for further comments)
\[ \left\{ s\in\N \st \E_{\rho\sim\mu_{d^2,s}}\sup_{\Delta\in\overline{\mathcal{E}}_k(\C^d{:}\C^d)^{\circ}}\tr\left(\left(\rho-\frac{\Id}{d^2}\right)\Delta\right)<1 \right\}, \]
One can first of all use the fact that, roughly speaking, when $d,s\rightarrow+\infty$, the random matrix $\rho-\Id/d^2$ for $\rho\sim\mu_{d^2,s}$ ``looks the same as'' the random matrix $G/d^2\sqrt{s}$ for $G\sim GUE(d^2)$ (see \cite{ASY}, Proposition 3.1 and Remark 3.2 as well as Appendices A and B, for precise majorization statements and proofs). In particular, there exists a constant $C>0$ such that, for all $d,s\in\N$ with (say) $d^2\leq s\leq d^3$, we have the upper bound
\[ \E_{\rho\sim\mu_{d^2,s}}\sup_{\Delta\in\overline{\mathcal{E}}_k(\C^d{:}\C^d)^{\circ}}\tr\left(\left(\rho-\frac{\Id}{d^2}\right)\Delta\right)\leq \frac{C}{d^2\sqrt{s}} \E_{G\sim GUE(d^2)}\sup_{\Delta\in\overline{\mathcal{E}}_k(\C^d{:}\C^d)^{\circ}}\tr\left(G\Delta\right). \]
Next, due to the fact that, again putting it vaguely, the convex body $\overline{\mathcal{E}}_k(\C^d{:}\C^d)$ is ``sufficiently well-balanced'' (see \cite{ASY}, Section 4 as well as Appendices C and D, for a complete exposition of the $\ell$-position argument), we know that there exists a constant $C'>0$ such that, for all $d\in\N$, we have the upper bound
\[ \left(\E_{G\sim GUE(d^2)}\sup_{\Delta\in\overline{\mathcal{E}}_k(\C^d{:}\C^d)^{\circ}}\tr\left(G\Delta\right)\right) \left(\E_{G\sim GUE(d^2)}\sup_{\Delta\in\overline{\mathcal{E}}_k(\C^d{:}\C^d)}\tr\left(G\Delta\right)\right) \leq C'd^4\log d. \]
Now, $\E_{G\sim GUE(d^2)}\sup_{\Delta\in\overline{\mathcal{E}}_k(\C^d{:}\C^d)}\tr\left(G\Delta\right)$ is nothing else than the Gaussian mean width of $\overline{\mathcal{E}}_k(\C^d{:}\C^d)$, which is the same as the Gaussian mean width of $\mathcal{E}_k(\C^d{:}\C^d)$, so for which we have an estimate thanks to Theorem \ref{th:w(k-ext)}, namely $w_G\left(\overline{\mathcal{E}}_k(\C^d{:}\C^d)\right) \sim_{d\rightarrow+\infty} 2d/\sqrt{k}$.

Putting everything together, we see that
\[ \E_{\rho\sim\mu_{d^2,s}}\sup_{\Delta\in\overline{\mathcal{E}}_k(\C^d{:}\C^d)^{\circ}}\tr\left(\left(\rho-\frac{\Id}{d^2}\right)\Delta\right)\leq \widetilde{C}\frac{\sqrt{k}d\log d}{\sqrt{s}}, \]
for some constant $\widetilde{C}>0$ independent of $d,s,k\in\N$, which implies as claimed that if $s> \widetilde{C}^2kd^2\log^2d$, then $\E_{\rho\sim\mu_{d^2,s}}\sup_{\Delta\in\overline{\mathcal{E}}_k(\C^d{:}\C^d)^{\circ}}\tr\left(\left(\rho-\Id/d^2\right)\Delta\right)<1$.

\begin{remark}
Let us briefly comment on a notable difference, from a convex geometry point of view, between the $k$-extendibility criterion and other common separability criteria. In the case of $k$-extendibility, computing the support function of $\overline{\mathcal{E}}_k$ is easier than computing the support function of its polar $\overline{\mathcal{E}}_k^{\circ}$, while for other separability relaxations it is usually the opposite. Indeed, for a given traceless unit Hilbert-Schmidt norm Hermitian $\Delta$ on $\C^d\otimes\C^d$, we have for instance the closed formulas
\[ h_{\overline{\mathcal{E}}_k(\C^d{:}\C^d)}(\Delta) = \sup \left\{ \tr(\Delta X) \st X\in\overline{\mathcal{E}}_k(\C^d{:}\C^d) \right\} = \left\|\widetilde{\Delta}\right\|_{\infty}, \]
\[ h_{\overline{\mathcal{P}}(\C^d{:}\C^d)^{\circ}}(\Delta) = \sup \left\{ \tr(\Delta X) \st X\in\overline{\mathcal{P}}(\C^d{:}\C^d)^{\circ} \right\} = d^2\left\|\Delta^{\Gamma}\right\|_{\infty}, \]
whereas the dual quantities $h_{\overline{\mathcal{E}}_k(\C^d{:}\C^d)^{\circ}}(\Delta)$ and $h_{\overline{\mathcal{P}}(\C^d{:}\C^d)}(\Delta)$ cannot be written in such a simple way.

This explains why in the case of $\mathcal{E}_k$ it is the mean width that can be exactly computed, contrary to the threshold value which can only be approximated, while for other approximations of $\mathcal{S}$ the reverse generally happens.
\end{remark}

\section{Appendix A: Combinatorics of permutations and partitions: short summary of standard facts}
\label{appendix:combinatorics}

Let $p\in\N$. We denote by $\mathfrak{S}(p)$ the set of permutations on $\{1,\ldots,p\}$. For any $\pi\in\mathfrak{S}(p)$, we denote by $\sharp(\pi)$ the number of cycles in the decomposition of $\pi$ into a product of disjoint cycles, and by $|\pi|$ the minimal number of transpositions in the decomposition of $\pi$ into a product of transpositions. We also define $\gamma\in\mathfrak{S}(p)$ as the canonical full cycle $(p\,\ldots\,1)$. More generally, we shall say that $c$ is the canonical full cycle on a set $\{i_1,\ldots,i_p\}$ with $i_1<\cdots<i_p$ if $c=(i_p\,\ldots\,i_1)$.

Some standard results related to $\mathfrak{S}(p)$ are gathered below (see e.g.~\cite{NS}, Lectures 9 and 23, for more details).

\begin{lemma}
\label{lemma:cycles-transpositions}
For any $\pi\in\mathfrak{S}(p)$, $\sharp(\pi)+|\pi|=p$.
\end{lemma}

\begin{lemma}
\label{lemma:distance-general}
$d:(\pi,\varsigma)\in\mathfrak{S}(p)\times\mathfrak{S}(p)\mapsto|\pi^{-1}\varsigma|$ defines a distance on $\mathfrak{S}(p)$, so that for any $\pi,\varsigma\in\mathfrak{S}(p)$,
\begin{equation}
\label{eq:geodesic}
|\varsigma^{-1}\pi|+|\pi|=d(\varsigma,\pi)+d(\id,\pi)\geq d(\id,\varsigma) =|\varsigma|,
\end{equation}
with equality in \eqref{eq:geodesic} if and only if $\pi$ lies on the geodesic between $\id$ and $\varsigma$. And whenever this is not the case, there exists $\delta\in\{1,\ldots, p-1\}$ such that $|\varsigma^{-1}\pi|+|\pi|= |\varsigma|+2\delta$.
\end{lemma}

\begin{definition}
A partition of $\{1,\ldots,p\}$ is a family $\lambda=\{I_1,\ldots,I_L\}$ of disjoint non-empty subsets of $\{1,\ldots,p\}$ whose union is $\{1,\ldots,p\}$. The sets $I_1,\ldots,I_L$ are called the blocks of $\lambda$. If each of them contains exactly $2$ elements, $\lambda$ is said to be a pair partition of $\{1,\ldots,p\}$. We shall denote by $\mathfrak{P}(p)$ the set of partitions of $\{1,\ldots,p\}$, and by $\mathfrak{P}^{(2)}(p)$ the set of pair partitions of $\{1,\ldots,p\}$. Note that $\mathfrak{P}^{(2)}(p)=\emptyset$ if $p$ is odd. Remark also that, whenever $p$ is even, the set of pair partitions of $\{1,\ldots,p\}$ is in bijection with the set of pairings on $\{1,\ldots,p\}$ (i.e.~ the set of permutations on $\{1,\ldots,p\}$ which are a product of $p/2$ disjoint transpositions). We shall therefore make no distinction between both.

A partition of $\{1,\ldots,p\}$ is said to be non-crossing if there does not exist $i<j<k<l$ in $\{1,\ldots,p\}$ such that $i,k$ belong to the same block, $j,l$ belong to the same block, and $i,j$ belong to different blocks. We shall denote by $NC(p)$ the set of non-crossing partitions of $\{1,\ldots,p\}$, and by $NC^{(2)}(p)$ the set of pair non-crossing partitions of $\{1,\ldots,p\}$. Note that $NC^{(2)}(p)=\emptyset$ if $p$ is odd.
\end{definition}

A well-known combinatorial result regarding non-crossing partitions is the following.

\begin{lemma} \label{lemma:catalan}
The number of non-crossing partitions of $\{1,\ldots,p\}$ and the number of pair non-crossing partitions of $\{1,\ldots,2p\}$ are both equal to the $p^{th}$ Catalan number
\[ \mathrm{Cat}_p=\frac{1}{p+1}{2p \choose p}. \]
More precisely, for any $1\leq m\leq p$, the number of non-crossing partitions of $\{1,\ldots,p\}$ which are composed of exactly $m$ blocks is equal to the $(p,m)^{th}$ Narayana number
\[ \mathrm{Nar}_p^m=\frac{1}{p+1}{p+1 \choose m}{p-1 \choose m-1}. \]
Obviously, these numbers are such that $\sum_{m=1}^p\mathrm{Nar}_p^m=\mathrm{Cat}_p$.
\end{lemma}

With these definitions in mind, we can now state a special case of particular interest of Lemma \ref{lemma:distance-general}.

\begin{lemma}
\label{lemma:distance}
Denote by $\gamma$ the canonical full cycle on $\{1,\ldots,p\}$. Then, for any $\pi\in\mathfrak{S}(p)$,
\begin{equation}
\label{eq:geodesic'}
|\gamma^{-1}\pi|+|\pi|\geq |\gamma|= p-1,
\end{equation}
with equality in \eqref{eq:geodesic'} if and only if $\pi$ lies on the geodesic between $\id$ and $\gamma$. The latter subset of $\mathfrak{S}(p)$ is in bijection with the set of non-crossing partitions of $\{1,\ldots,p\}$ (by the mapping which associates to a given partition the product of the canonical full cycles on each of its blocks). We shall thus write $\pi\in NC(p)$ in such case, not distinguishing a geodesic permutation from its corresponding non-crossing partition.

More generally, let $\{I_1,\ldots,I_L\}$ be a partition of $\{1,\ldots,p\}$ and denote by $\gamma_1,\ldots,\gamma_L$ the canonical full cycles on $I_1,\ldots,I_L$. Then, for any $\pi\in\mathfrak{S}(p)$,
\begin{equation}
\label{eq:geodesic''}
|(\gamma_1\cdots\gamma_L)^{-1}\pi|+|\pi|\geq |\gamma_1\cdots\gamma_L|= p-L,
\end{equation}
with equality in \eqref{eq:geodesic''} if and only if $\pi$ lies on the geodesic between $\id$ and $\gamma_1\cdots\gamma_L$. The latter subset of $\mathfrak{S}(p)$ is in bijection with the set of non-crossing partitions of $\{1,\ldots,p\}$ which are finer than $I_1\sqcup\cdots\sqcup I_L$, which itself is in bijection with $NC(|I_1|)\times\cdots\times NC(|I_L|)$.
\end{lemma}

Combining Lemma \ref{lemma:distance} with Lemma \ref{lemma:catalan}, we can in fact say the following: Let $\varsigma\in\mathfrak{S}(p)$ and assume that its decomposition into disjoint cycles is $\varsigma=c_1\cdots c_L$ where, for each $1\leq i\leq L$, $c_i$ is of length $p_i$ (hence with $p_1+\cdots+p_L=p$). Then, for any $\pi\in\mathfrak{S}(p)$, $|\varsigma^{-1}\pi|+|\pi|\geq p-L$, and
\[ \left|\left\{ \pi\in\mathfrak{S}(p) \st |\varsigma^{-1}\pi|+|\pi| = p-L \right\}\right| = \mathrm{Cat}_{p_1}\times\cdots\times\mathrm{Cat}_{p_L}. \]
Having this easy observation in mind might be useful later on.



\section{Appendix B: Computing moments of Gaussian matrices: Wick formula and genus expansion}
\label{appendix:wick}

When computing expectations of Gaussian random variables, a useful tool is the Wick formula (see e.g.~\cite{Zvonkin} or \cite{NS}, Lecture 22, for a proof).

\begin{lemma}[Gaussian Wick formula] Let $X_1,\ldots,X_q$ be jointly Gaussian centered random variables (real or complex).
\begin{align*}
& \text{If}\ q=2p+1\ \text{is odd},\ \text{then}\ \E\left[X_1\cdots X_q\right] = 0.\\
& \text{If}\ q=2p\ \text{is even},\ \text{then}\ \E\left[X_1\cdots X_q\right] = \underset{\{\{i_1,j_1\},\ldots,\{i_p,j_p\}\}\in\mathfrak{P}^{(2)}(2p)}{\sum} \,\underset{m=1}{\overset{p}{\prod}}\E\left[X_{i_m}X_{j_m}\right].
\end{align*}
\label{lemma:wick}
\end{lemma}

\subsection{Moments of GUE matrices}
\label{appendix:wick-gaussian}

A first important application of Lemma \ref{lemma:wick} is to the computation of the moments of matrices from the Gaussian Unitary Ensemble. Indeed, for any $q\in\N$, we have
\[ \E_{G\sim GUE(n)} \tr\left(G^q\right) = \underset{1\leq l_1,\ldots,l_q\leq n}{\sum} \E\left[G_{l_1,l_2}\cdots G_{l_q,l_1}\right], \]
where the $G_{i,j}$, $1\leq i,j\leq n$, are centered Gaussian random variables satisfying $\E[G_{i,j}G_{i',j'}]=\delta_{i=j',j=i'}$. So what we get applying the Wick formula is that, for any $p\in\N$,
\begin{align*}
& \E_{G\sim GUE(n)} \tr\left(G^{2p+1}\right) = 0 \\
& \E_{G\sim GUE(n)} \tr\left(G^{2p}\right) = \underset{\lambda\in\mathfrak{P}^{(2)}(2p)}{\sum} n^{\flat(\lambda)},
\end{align*}
where for each pair partition $\lambda=\{\{i_1,j_1\},\ldots,\{i_p,j_p\}\}$ of $\{1,\ldots,2p\}$, $\flat(\lambda)$ is the number of free parameters $l_1,\ldots,l_{2p}\in\{1,\ldots,n\}$ when imposing that $\forall\ 1\leq m\leq p,\ l_{i_m+1}=l_{j_m},\ l_{j_m+1}=l_{i_m}$. Identifying the pair partition $\{\{i_1,j_1\},\ldots,\{i_p,j_p\}\}$ with the pairing $(i_1\,j_1)\ldots(i_p\,j_p)$ and denoting by $\gamma$ the canonical full cycle $(2p\,\ldots\,1)$, the latter condition can be written as $\forall\ 1\leq i\leq 2p,\ l_{\gamma^{-1}\lambda(i)}=l_{i}$. So in fact, $\flat(\lambda)= \sharp(\gamma^{-1}\lambda)$ and the expression above becomes
\[ \E_{G\sim GUE(n)} \tr\left(G^{2p}\right) = \underset{\lambda\in\mathfrak{P}^{(2)}(2p)}{\sum} n^{\sharp(\gamma^{-1}\lambda)}. \]
We thus have the so-called \textit{genus expansion} (see e.g.~\cite{NS}, Lecture 22)
\[ \E_{G\sim GUE(n)} \tr\left(G^{2p}\right) = \sum_{\delta=0}^{\lfloor p/2\rfloor} P(2p,\delta)n^{p+1-2\delta}, \]
where for each $0\leq\delta\leq\lfloor p/2\rfloor$, we defined $P(2p,\delta)$ as the number of pairings of $\{1,\ldots,2p\}$ having \textit{genus} $\delta$, i.e.~\[ P(2p,\delta)=\left|\{\lambda\in\mathfrak{P}^{(2)}(2p) \st \sharp(\gamma^{-1}\lambda)=p+1-2\delta\}\right|= \left|\{\lambda\in\mathfrak{P}^{(2)}(2p) \st \sharp(\gamma^{-1}\lambda)+\sharp(\lambda)=2p+1-2\delta\}\right|. \]
Equivalently, $P(2p,\delta)$ is the number of pairings of $\{1,\ldots,2p\}$ having a defect $2\delta$ of being on the geodesics between $\id$ and $\gamma$. Hence, $P(2p,0)$ is the number of pairings of $\{1,\ldots,2p\}$ lying exactly on the geodesics between $\id$ and $\gamma$, i.e.~ the number of non-crossing pair partitions of $\{1,\ldots,2p\}$. So $P(2p,0)=\left|NC^{(2)}(2p)\right|=\mathrm{Cat}_p$, and we recover the well-known asymptotic estimate
\[ \E_{G\sim GUE(n)} \tr\left(G^{2p}\right) \underset{n\rightarrow+\infty}{\sim} \mathrm{Cat}_p\, n^{p+1} .\]


\subsection{Moments of Wishart matrices}
\label{appendix:wick-wishart}

A second important application of Lemma \ref{lemma:wick} is to the computation of the moments of matrices from the Wishart Ensemble. In such case, a graphical way of visualising the Wick formula has been developed in \cite{CN}, to which the reader is referred for further details and proofs, a brief summary only being provided here.

In the graphical formalism, a matrix $X:\C^m\rightarrow\C^n$ is represented by a ``box'' with two ``gates'', one specifying the size $m$ at its entrance and the other specifying the size $n$ at its exit. For $X:\C^m\rightarrow\C^n$  and $Y:\C^n\rightarrow\C^m$, the product $XY:\C^n\rightarrow\C^n$ is represented by a wire connecting the exit of $Y$ to the entrance of $X$. For $Z:\C^m\rightarrow\C^m$, the trace $\tr(Z)$ is represented by a wire connecting the exit and the entrance of $Z$.

Let $W$ be a $(n,s)$-Wishart matrix, i.e.~$W=GG^{\dagger}$ with $G$ a $n\times s$ matrix with independent complex normal entries. Representing by $\blacklozenge$ a $n$-dimensional gate and by $\blacktriangledown$ a $s$-dimensional gate, the quantity $\tr(W^p)$ is then graphically represented by $p$ boxes $G$ and $p$ boxes $G^{\dagger}$ connected by wires in the following way.

\begin{center}
\begin{tikzpicture} [scale=0.8]
\node[draw=lightgray, minimum height=0.8cm, minimum width=0.8cm, fill=lightgray, =white] (G1) at (0,0) {$G$};
\node[draw=lightgray, minimum height=0.8cm, minimum width=0.8cm, fill=lightgray] (G1*) at (1.5,0) {$G^{\dagger}$};
\node[draw=lightgray, minimum height=0.8cm, minimum width=0.8cm, fill=lightgray] (Gp) at (5.5,0) {$G$};
\node[draw=lightgray, minimum height=0.8cm, minimum width=0.8cm, fill=lightgray] (Gp*) at (7,0) {$G^{\dagger}$};
\draw(G1.west) node{$\blacklozenge$}; \draw(G1.east) node{$\blacktriangledown$};
\draw(G1*.east) node{$\blacklozenge$}; \draw(G1*.west) node{$\blacktriangledown$};
\draw(Gp.west) node{$\blacklozenge$}; \draw(Gp.east) node{$\blacktriangledown$};
\draw(Gp*.east) node{$\blacklozenge$}; \draw(Gp*.west) node{$\blacktriangledown$};
\draw (G1.east) -- (G1*.west); \draw (Gp.east) -- (Gp*.west);
\draw (G1.west) to[out=155,in=25] (Gp*.east);
\draw (G1*.east) -- (2.5,0);
\draw (Gp.west) -- (4.5,0);
\draw[dotted] (3,0) -- (4,0);
\end{tikzpicture}
\end{center}

For any $\alpha\in\mathfrak{S}(p)$, we will denote by $\mathcal{G}_{\alpha}$ the diagram obtained from the one above by ``erasing'' the boxes, just keeping their gates, and then connecting, for each $1\leq i\leq p$, the entrance of the $i^{th}$ box $G$ to the exit of the $\alpha(i)^{th}$ box $G^{\dagger}$, and the exit of the $i^{th}$ box $G$ to the entrance of the $\alpha(i)^{th}$ box $G^{\dagger}$. Doing so, $\sharp(\gamma^{-1}\alpha)$ loops connecting $n$-dimensional gates and $\sharp(\alpha)$ loops connecting $s$-dimensional gates are obtained. And the graphical version of the Wick formula tells us that
\[ \E_{W\sim\mathcal{W}_{n,s}}\tr(W^p) = \sum_{\alpha\in\mathfrak{S}(p)} \mathcal{D}_{\alpha} = \sum_{\alpha\in\mathfrak{S}(p)} n^{\sharp(\gamma^{-1}\alpha)}s^{\sharp(\alpha)} .\]

In the special case where $s=n$, this can be rewritten as a so-called \textit{genus expansion} (see e.g.~\cite{CN})
\[ \E_{W\sim\mathcal{W}_{n,n}}\tr(W^p) = \sum_{\delta=0}^{\lfloor p/2\rfloor}S(p,\delta)n^{p+1-2\delta}, \]
where for each $0\leq\delta\leq\lfloor p/2\rfloor$, we defined $S(p,\delta)$ as the number of permutations on $\{1,\ldots,p\}$ having \textit{genus} $\delta$, i.e.~$S(p,\delta)= \left| \{\alpha\in\mathfrak{S}(p) \st \sharp(\gamma^{-1}\alpha)+\sharp(\alpha)=p+1-2\delta\} \right|$. Since $\{\alpha\in\mathfrak{S}(p) \st \sharp(\gamma^{-1}\alpha)+\sharp(\alpha)=p+1\}=NC(p)$, we have $S(p,0)=\mathrm{Cat}_p$ and hence recover the well-known asymptotic estimate
\[ \E_{W\sim\mathcal{W}_{n,n}} \tr\left(W^{p}\right) \underset{n\rightarrow+\infty}{\sim} \mathrm{Cat}_p\, n^{p+1} .\]


\section{Appendix C: A needed combinatorial fact: relating the number of cycles in some specific permutations on either $[p]\times[k]$ or $[p]$}
\label{appendix:technical}

Let $\alpha\in\mathfrak{S}(p)$ and $f:[p]\rightarrow[k]$. We define $\hat{\alpha}_f$ on $[p]\times[k]$ as
\[ \forall\ (i,r)\in[p]\times[k],\ \hat{\alpha}_f(i,r)=
\begin{cases} (\alpha(i),f(\alpha(i)))\ \text{if}\ r=f(i)\\
(i,r)\ \text{if}\ r\neq f(i)\end{cases}. \]
We also define $\hat{\gamma}$ on $[p]\times[k]$ as $(\gamma,\id)$, where $\gamma\in\mathfrak{S}(p)$ is the canonical full cycle $(p\,\ldots\,1)$.

We would like to understand what is the number of cycles in $\hat{\gamma}^{-1}\hat{\alpha}_f$. For that, it will be convenient to do a bit of rewriting. Let us first extend the definition of $\hat{\alpha}_f$ and $\hat{\gamma}$ to $[p]\times\left(\{0\}\cup[k]\right)$. We shall denote by $\bar{\alpha}_f$ and $\bar{\gamma}$ the respective extensions. Note that since $f$ takes values in $[k]$, we have $\bar{\alpha}_f(i,0)=(i,0)$ for all $i\in[p]$.

We will now make two easy observations.

\begin{fact}
\label{fact:alpha_f}
For any $f:[p]\rightarrow[k]$, define for each $i\in[p]$, $\bar{\tau}_f^{(i)}$ as the transposition on $[p]\times\left(\{0\}\cup[k]\right)$ which swaps $(i,0)$ and $(i,f(i))$, and set $\bar{\beta}_f=\bar{\tau}_f^{(1)}\cdots\bar{\tau}_f^{(p)}$. We then have, for any $\alpha\in\mathfrak{S}(p)$,
\[ \bar{\alpha}_f=\bar{\beta}_f^{-1}\bar{\alpha}'\bar{\beta}_f,\ \text{where}\ \forall\ (i,r)\in[p]\times\left(\{0\}\cup[k]\right),\ \bar{\alpha}'(i,r)=\begin{cases} (\alpha(i),r)\ \text{if}\ r=0\\
(i,r)\ \text{if}\ r\neq 0 \end{cases}. \]
\end{fact}

The advantage of expressing $\bar{\alpha}_f$ in this way is that $\bar{\alpha}'$ is particularly simple: it acts as $\alpha\times\id$ on $[p]\times\{0\}$ and does nothing on $[p]\times[k]$. Furthermore, due to the cyclicity of $\sharp(\cdot)$, a direct consequence of Fact \ref{fact:alpha_f} is that
$\sharp(\bar{\gamma}^{-1}\bar{\alpha}_f) = \sharp(\bar{\gamma}_f^{-1}\bar{\alpha}')$, where $\bar{\gamma}_f=\bar{\beta}_f\bar{\gamma}\bar{\beta}_f^{-1}$. It may then be easily checked that $\bar{\gamma}_f$ decomposes into $k+1$ disjoint cycles as stated in Fact \ref{fact:gamma_f} below.

\begin{fact}
\label{fact:gamma_f}
For any $f:[p]\rightarrow[k]$, we have
\[ \bar{\gamma}_f=\bar{c}_1\cdots\bar{c}_k\bar{c}, \]
with $\bar{c}=\left((p,f(p))\ldots(1,f(1))\right)$, and for each $r\in[k]$, $\bar{c}_r=\left((p,s_r(p))\ldots(1,s_r(1))\right)$, where for each $i\in[p]$, $s_r(i)=0$ if $f(i)=r$ and $s_r(i)=r$ if $f(i)\neq r$.
\end{fact}

\begin{example} For the sake of concreteness, let us have a look at a simple example. In the case where $p=4$, $k=3$, and $f$ is defined by $f(1)=f(2)=f(4)=1$, $f(3)=2$, we obtain that the cycles in $\bar{\gamma}_f$ are $\bar{c}_1=((4,0)(3,1)(2,0)(1,0))$, $\bar{c}_2=((4,2)(3,0)(2,2)(1,2))$, $\bar{c}_3=((4,3)(3,3)(2,3)(1,3))$, $\bar{c}=((4,1)(3,2)(2,1)(1,1))$. This is schematically represented in Figure \ref{fig:bargamma_f}, where the elements in $\bar{c}_i$ are marked by ``$i$'', for $i\in\{1,2,3\}$, and the elements in $\bar{c}$ are marked by ``$\bullet$''.

\begin{figure}[h]
\caption{$f:[4]\rightarrow[3]$ such that $f^{-1}(1)=\{1,2,4\}$, $f^{-1}(2)=\{3\}$, $f^{-1}(3)=\emptyset$. }
\begin{center}
\begin{tikzpicture} [scale=0.7]
\draw (6,4.3) node {$ $};
\draw [<->] (0,-0.5) -- (4,-0.5); \draw [<->] (-0.5,0) -- (-0.5,4);
\draw (2,-1) node {$i\in[4]$}; \draw (-2,2) node {$r\in\{0\}\cup[3]$};
\draw (0,0) -- (4,0); \draw (0,1) -- (4,1); \draw (0,2) -- (4,2); \draw (0,3) -- (4,3); \draw (0,4) -- (4,4);
\draw (0,0) -- (0,4); \draw (1,0) -- (1,4); \draw (2,0) -- (2,4); \draw (3,0) -- (3,4); \draw (4,0) -- (4,4);
\draw (0.5,0.5) node {$1$}; \draw (1.5,0.5) node {$1$}; \draw (2.5,0.5) node {$2$}; \draw (3.5,0.5) node {$1$};
\draw (0.5,1.5) node {$\bullet$}; \draw (1.5,1.5) node {$\bullet$}; \draw (2.5,1.5) node {$1$}; \draw (3.5,1.5) node {$\bullet$};
\draw (0.5,2.5) node {$2$}; \draw (1.5,2.5) node {$2$}; \draw (2.5,2.5) node {$\bullet$}; \draw (3.5,2.5) node {$2$};
\draw (0.5,3.5) node {$3$}; \draw (1.5,3.5) node {$3$}; \draw (2.5,3.5) node {$3$}; \draw (3.5,3.5) node {$3$};
\end{tikzpicture}
\end{center}
\label{fig:bargamma_f}
\end{figure}

\end{example}

\begin{lemma}
\label{lemma:bargamma_f-gamma_f}
Let $f:[p]\rightarrow[k]$ and define $\gamma_f\in\mathfrak{S}(p)$ as $\gamma_f=\gamma_{f=1}\cdots\gamma_{f=k}$, where for each $r\in[k]$, $\gamma_{f=r}$ is the canonical full cycle on $f^{-1}(r)$. Then, for any $\alpha\in\mathfrak{S}(p)$,
\begin{equation}\label{eq:bargamma_f-gamma_f} \sharp(\bar{\gamma}_f^{-1}\bar{\alpha}') = \sharp(\gamma_f^{-1}\alpha) + 1 + k - |\im(f)|.\end{equation}
\end{lemma}

\begin{proof} In $\bar{\gamma}_f^{-1}\bar{\alpha}'$ there are, first of all:\\
$\bullet$ $k-|\im(f)|$ cycles of the form $((1,r)\ldots(p,r))$ for $r\in[k]\setminus\im (f)$, because for any $(i,r)\in[p]\times\left([k]\setminus\im (f)\right)$, $\bar{\gamma}_f^{-1}\bar{\alpha}'(i,r)=\bar{\gamma}_f^{-1}(i,r)=(\gamma^{-1}(i),r)$.\\
$\bullet$ $1$ cycle $((1,f(1))\ldots(p,f(p)))$, because for any $i\in[p]$, $\bar{\gamma}_f^{-1}\bar{\alpha}'(i,f(i))=\bar{\gamma}_f^{-1}(i,f(i))=(\gamma^{-1}(i),f(\gamma^{-1}(i)))$.\\
For the cycles in $\bar{\gamma}_f^{-1}\bar{\alpha}'$ which belong to none of these two categories, there are two crucial observations to be made. First, for any $i,j\in[p]$, $(i,0)$ and $(j,0)$ belong to the same cycle of $\bar{\gamma}_f^{-1}\bar{\alpha}'$ if and only if $i$ and $j$ belong to the same cycle of $\gamma_f^{-1}\alpha$. And second, for each $i\in[p]$ and $r\in[k]\setminus\{f(\alpha(i))\}$, there exists $j\in[p]$ such that $(i,r)$ belongs to the same cycle of $\bar{\gamma}_f^{-1}\bar{\alpha}'$ as $(j,0)$. Indeed, for any $i\in[p]$, we have on the one hand
\[ \gamma_f^{-1}\alpha(i)= (\gamma^{-1})^{L+1}\alpha(i),\ \text{with}\ L=\inf\{ l\geq 0 \st f((\gamma^{-1})^{l+1}\alpha(i))= f(\alpha(i))\}. \]
While we have on the other hand,
\begin{align*}
f(\gamma^{-1}\alpha(i))=f(\alpha(i))\ & \Rightarrow\ \bar{\gamma}_f^{-1}\bar{\alpha}'(i,0) =(\gamma^{-1}\alpha(i),0) =(\gamma_f^{-1}\alpha(i),0), \\
f(\gamma^{-1}\alpha(i))\neq f(\alpha(i))\ & \Rightarrow\ \begin{cases}\forall\ 0\leq l\leq L-1,\ (\bar{\gamma}_f^{-1}\bar{\alpha}')^l(i,0)=((\gamma^{-1})^{l}\gamma^{-1}\alpha(i),f(\alpha(i)))\\
(\bar{\gamma}_f^{-1}\bar{\alpha}')^{L}(i,0)=((\gamma^{-1})^{L}\gamma^{-1}\alpha(i),0) =(\gamma_f^{-1}\alpha(i),0) \end{cases}.
\end{align*}
So there are in fact exactly $\sharp(\gamma_f^{-1}\alpha)$ remaining cycles in $\bar{\gamma}_f^{-1}\bar{\alpha}'$.
\end{proof}

\begin{example} Looking at the same example as before, namely $p=4$, $k=3$, and $f$ such that $f^{-1}(1)=\{1,2,4\}$, $f^{-1}(2)=\{3\}$, $f^{-1}(3)=\emptyset$, we see that, for $\alpha=(14)(23)$, the cycles in $\bar{\gamma}_f^{-1}\bar{\alpha}'$ are:\\
$\bullet$ $((1,3)(2,3)(3,3)(4,3))$, because $3\notin\im (f)$.\\
$\bullet$ $((1,1)(2,1)(3,2)(4,1))$, because $f(1)=1$, $f(2)=1$, $f(3)=2$ and $f(4)=1$.\\
$\bullet$ $(1,0)$ and $((2,0)(2,2)(1,2)(4,2)(3,0)(4,0)(3,1))$, corresponding to the cycles $(1)$ and $(2,3,4)$ in $\gamma_f^{-1}\alpha$.
\end{example}

Putting together these preliminary technical results, we straightforwardly obtain Proposition \ref{prop:geodesic-alpha_f} below.

\begin{proposition}
\label{prop:geodesic-alpha_f}
Let $f:[p]\rightarrow[k]$ and define $\gamma_f\in\mathfrak{S}(p)$ as $\gamma_f=\gamma_{f=1}\cdots\gamma_{f=k}$, where for each $r\in[k]$, $\gamma_{f=r}$ is the canonical full cycle on $f^{-1}(r)$. Then, for any $\alpha\in\mathfrak{S}(p)$,
\begin{equation}\label{eq:geodesic-alpha_f} \sharp(\hat{\gamma}^{-1}\hat{\alpha}_f) = \sharp(\gamma_f^{-1}\alpha) + k - |\im(f)|. \end{equation}
\end{proposition}

\begin{proof} This is a direct consequence of Lemma \ref{lemma:bargamma_f-gamma_f}, just noticing that $\sharp(\bar{\gamma}^{-1}\bar{\alpha}_f)=\sharp(\hat{\gamma}^{-1}\hat{\alpha}_f)+1$.
\end{proof}

\section{Appendix D: Moments of ``modified'' GUE matrices (proof)}
\label{appendix:gaussian}

The goal of this Appendix is to generalize the methodology described in Appendix \ref{appendix:wick-gaussian} in order to compute the $2p$-order moments of the matrix $\sum_{j=1}^k\widetilde{G}_{\A\B^k}(j)$. Recall that this issue arises when trying to estimate the mean width of the set of $k$-extendible states. We are thus dealing here, not with standard GUE matrices, but with $d^2$-dimensional GUE matrices which are tensorized with $d^{k-1}$-dimensional identity matrices.

For any $i_1,\ldots,i_{2p}\in[k]$, we can write
\[ \mathrm{Tr}\left[\underset{j=1}{\overset{2p}{\overrightarrow{\prod}}}\widetilde{G}_{\A\B^k}(i_j)\right] = \underset{\vec{l}_1,\ldots,\vec{l}_{2p}\in[d]^{k+1}}{\sum} \widetilde{G}_{\A\B^k}(i_1)_{\vec{l}_1,\vec{l}_2}\cdots\widetilde{G}_{\A\B^k}(i_{2p})_{\vec{l}_{2p},\vec{l}_1} ,\]
where for each $j\in[2p]$ and each $\vec{l}_j=\left(a_j,b_j^1,\ldots,b_j^k\right),\vec{l}_{j+1}=\left(a_{j+1},b_{j+1}^1,\ldots,b_{j+1}^k\right)\in[d]^{k+1}$, we have \[ \widetilde{G}_{\A\B^k}(i_j)_{\vec{l}_j,\vec{l}_{j+1}}=G_{(a_j,b_j^{i_j}),(a_{j+1},b_{j+1}^{i_j})} \delta_{\vec{b}_j\setminus b_j^{i_j}=\vec{b}_{j+1}\setminus b_{j+1}^{i_j}}. \]
Consequently, for any $f:[2p]\rightarrow[k]$, we have
\[ \mathrm{Tr}\left[\underset{i=1}{\overset{2p}{\overrightarrow{\prod}}}\widetilde{G}_{\A\B^k}(f(i))\right] = \underset{a_1,\ldots,a_{2p}\in[d]}{\sum}\ \underset{\vec{b}_1,\ldots,\vec{b}_{2p}\in I_f}{\sum} G_{(a_1,b_1^{f(1)}),(a_2,b_2^{f(1)})}\cdots G_{(a_{2p},b_{2p}^{f(2p)}),(a_1,b_1^{f(2p)})}, \]
where $I_f=\left\{\vec{b}_1,\ldots,\vec{b}_{2p}\in[d]^k \st \forall\ i\in[2p],\ \forall\ r\in[k]\setminus\{f(i)\},\ b_{i+1}^r=b_i^r\right\}$.

What we therefore get by the Wick formula for Gaussian matrices is that, for any $f:[2p]\rightarrow[k]$,
\[ \mathbf{E}_{G_{\A\B}\sim GUE(d^2)} \mathrm{Tr} \left[\underset{i=1}{\overset{2p}{\overrightarrow{\prod}}}\widetilde{G}_{\A\B^k}(f(i))\right] = \underset{\lambda\in\mathfrak{P}^{(2)}(2p)}{\sum} d^{\flat(\lambda)+\flat(\hat{\lambda}_f)}, \]
where for each pair partition $\lambda=\{\{i_1,j_1\},\ldots,\{i_p,j_p\}\}$ of $\{1,\ldots,2p\}$, $\flat(\lambda)$ is the number of free parameters $a_1,\ldots,a_{2p}\in[d]$ when imposing that $\forall\ 1\leq m\leq p,\ a_{i_m+1}=a_{j_m},\ a_{j_m+1}=a_{i_m}$, and $\flat(\hat{\lambda}_f)$ is the number of free parameters $\vec{b}_1,\ldots,\vec{b}_{2p}\in I_f$ when imposing that $\forall\ 1\leq m\leq p,\ b_{i_m+1}^{f(i_m)}=b_{j_m}^{f(j_m)},\ b_{j_m+1}^{f(j_m)}=b_{i_m}^{f(i_m)}$. As noticed before, identifying the pair partition $\{\{i_1,j_1\},\ldots,\{i_p,j_p\}\}$ with the pairing $(i_1\,j_1)\,\ldots\,(i_p\,j_p)$ and denoting by $\gamma$ the canonical full cycle $(2p\,\ldots\,1)$, the latter conditions may be written as
\[ \forall\ i\in[2p],\ a_{\gamma^{-1}\lambda(i)}=a_{i}\ \text{and}\ \forall\ r\in[k],\ \begin{cases} b_{\gamma^{-1}\lambda(i)}^{f(\lambda(i))}=b_{i}^r\ \text{if}\ r=f(i) \\ b_{\gamma^{-1}(i)}^r=b_{i}^r\ \text{if}\ r\neq f(i) \end{cases}. \]
So in fact, $\flat(\lambda)=\sharp(\gamma^{-1}\lambda)$ and $\flat(\hat{\lambda}_f)=\sharp(\hat{\gamma}^{-1}\hat{\lambda}_f)$, where
\[ \forall\ (i,r)\in[2p]\times[k],\ \hat{\gamma}(i,r)=(\gamma(i),r)\ \text{and}\ \hat{\lambda}_f(i,r)=\begin{cases}  (\lambda(i),f(\lambda(i)))\ \text{if}\ r=f(i) \\ (i,r)\ \text{if}\ r\neq f(i) \end{cases}. \]
What is more, we know by Proposition \ref{prop:geodesic-alpha_f} that $\sharp(\hat{\gamma}^{-1}\hat{\lambda}_f) = \sharp(\gamma_f^{-1}\lambda) + k - |\im(f)|$, where $\gamma_f$ is the product of the canonical full cycles on the level sets of $f$.

Let us summarize.

\begin{proposition}
\label{prop:gaussian-p-preliminary}
For any $d\in\N$ and any $p\in\N$, we have
\begin{align*} \mathbf{E}_{G_{\A\B}\sim GUE(d^2)} \mathrm{Tr}\left[\left(\sum_{j=1}^k\widetilde{G}_{\A\B^k}(j)\right)^{2p}\right] = & \sum_{f:[2p]\rightarrow[k]}\mathbf{E}_{G_{\A\B}\sim GUE(d^2)} \mathrm{Tr}\left[\underset{i=1}{\overset{2p}{\overrightarrow{\prod}}}\widetilde{G}_{\A\B^k}(f(i))\right] \\
= & \underset{f:[2p]\rightarrow[k]}{\sum} \underset{\lambda\in\mathfrak{P}^{(2)}(2p)}{\sum} d^{\sharp(\gamma^{-1}\lambda)+\sharp(\gamma_f^{-1}\lambda) + k - |\im(f)|}. \end{align*}
\end{proposition}

\section{Appendix E: Moments of partially transposed ``modified'' GUE matrices (proof)}
\label{appendix:gaussian-gamma}

The goal of this Appendix is to compute the $2p$-order moments of $\sum_{j=1}^k \widetilde{G}_{\A\B^k}(j)^{\Gamma}$, where $\Gamma$ stands here for the partial transposition over the $\lceil k/2\rceil$ last $\mathrm{B}$ subsystems. Recall that this issue arises when trying to estimate the mean width of the set of $k$-PPT-extendible states.

Using the same notation as in Appendix \ref{appendix:gaussian}, and reasoning in a completely analogous way, we have that, for any $f:[2p]\rightarrow[k]$,
\[ \mathrm{Tr}\left[\underset{i=1}{\overset{2p}{\overrightarrow{\prod}}}\widetilde{G}_{\A\B^k}(f(i))^{\Gamma}\right] = \underset{a_1,\ldots,a_{2p}\in[d]}{\sum}\ \underset{\vec{b}_1,\ldots,\vec{b}_{2p}\in I_f}{\sum} G_{(a_1,b_{x_1}^{f(1)}),(a_2,b_{\bar{x}_1}^{f(1)})}\cdots G_{(a_{2p},b_{x_{2p}}^{f(2p)}),(a_1,b_{\bar{x}_{2p}}^{f(2p)})}, \]
where for each $1\leq i\leq 2p$, $x_i=\begin{cases} i\ \text{if}\ f(i)\leq \lfloor k/2\rfloor \\ i+1\ \text{if}\ f(i)> \lfloor k/2\rfloor \end{cases}$ and $\bar{x}_i=\begin{cases} i+1\ \text{if}\ f(i)\leq \lfloor k/2\rfloor \\ i\ \text{if}\ f(i)> \lfloor k/2\rfloor \end{cases}$.

What we therefore get by the Wick formula for Gaussian matrices is that, for any $f:[2p]\rightarrow[k]$,
\[ \mathbf{E}_{G_{\A\B}\sim GUE(d^2)} \mathrm{Tr}\left[\underset{i=1}{\overset{2p}{\overrightarrow{\prod}}}\widetilde{G}_{\A\B^k}(f(i))^{\Gamma}\right] = \underset{\lambda\in\mathfrak{P}^{(2)}(2p)}{\sum} d^{\flat(\lambda)+\flat(\check{\lambda}_f)}, \]
where for each pair partition $\lambda=\{\{i_1,j_1\},\ldots,\{i_p,j_p\}\}$ of $\{1,\ldots,2p\}$,  $\flat(\check{\lambda}_f)$ is the number of free parameters $\vec{b}_1,\ldots,\vec{b}_{2p}\in [d]^k$ when imposing that for all $i\in[2p]$, first $b_{\gamma^{-1}(i)}^r=b_{i}^r$ if $r\neq f(i)$, and second the one condition $b_{\gamma^{-1}\lambda(i)}^{f(\lambda(i))}=b_{i}^{f(i)}$ if $f(i),f(\lambda(i))\leq\lfloor k/2\rfloor$ or $f(i),f(\lambda(i))>\lfloor k/2\rfloor$, while the two conditions $b_{\gamma^{-1}\lambda(i)}^{f(\lambda(i))}=b_{i}^{f(i)}$ and $b_{\lambda(i)}^{f(\lambda(i))}=b_{\gamma^{-1}(i)}^{f(i)}$ if $f(i)\leq\lfloor k/2\rfloor$, $f(\lambda(i))>\lfloor k/2\rfloor$ or $f(i)>\lfloor k/2\rfloor$, $f(\lambda(i))\leq\lfloor k/2\rfloor$.

Let us rephrase what we just established. Fix $\lambda\in\mathfrak{P}^{(2)}(2p)$. For functions $f:[2p]\rightarrow\{1,\ldots,\lfloor k/2\rfloor\}\equiv\left[\lfloor k/2\rfloor\right]$ or $f:[2p]\rightarrow\{\lfloor k/2\rfloor+1,\ldots,k\}\equiv\left[\lceil k/2\rceil\right]$, the number of free parameters associated to the pair $(\lambda,f)$ is the same as the one observed in Appendix \ref{appendix:gaussian}. On the contrary, for functions $f$ which are not of this form, extra matching conditions are imposed. So these will for sure not contribute to the dominating term in the expansion of $\E\mathrm{Tr}\left[\left(\sum_{j=1}^k\widetilde{G}_{\A\B^k}(j)^{\Gamma}\right)^{2p}\right]$ into powers of $d$. Consequently, we have the asymptotic estimate
\[ \E_{G_{\A\B}\sim GUE(d^2)}\mathrm{Tr}\left[\left(\sum_{j=1}^k\widetilde{G}_{\A\B^k}(j)^{\Gamma}\right)^{2p}\right]
 \underset{d\rightarrow+\infty}{\sim} \underset{\lambda\in\mathfrak{P}^{(2)}(2p)}{\sum}\, \underset{f:[2p]\rightarrow\left[\lfloor k/2\rfloor\right]\,\text{or}\,\left[\lceil k/2\rceil\right]}{\sum} d^{\sharp(\gamma^{-1}\lambda)+\sharp(\gamma_f^{-1}\lambda)+k-|\im(f)|}. \]

\section{Appendix F: Moments of ``modified'' Wishart matrices (proof)}
\label{appendix:wishart}

The goal of this Appendix is to generalize the methodology described in Appendix \ref{appendix:wick-wishart} in order to compute the $p$-order moments of the matrix $\sum_{j=1}^k\widetilde{W}_{\A\B^k}(j)$. Recall that this issue arises when trying to characterize $k$-extendibility of random-induced states. We are thus dealing here, not with standard Wishart matrices, but with $(d^2,s)$-Wishart matrices which are tensorized with $d^{k-1}$-dimensional identity matrices.

Representing by $\bullet$ a $d$-dimensional gate and by $\blacktriangledown$ a $s$-dimensional gate, the matrix $\widetilde{W}_{\A\B^k}(1)$, for instance, may be graphically represented as in Figure \ref{fig:W_AB^k}.

\begin{figure}[h]
\caption{$\widetilde{W}_{\A\B^k}(1)=X_{\A\B^k}(1)X_{\A\B^k}(1)^{\dagger}$, with $X_{\A\B^k}(1)=G_{\A\B_1}\otimes\Id_{\B_2\ldots \B_k}$}
\begin{center}
\begin{tikzpicture} [scale=0.9]
\node[draw=lightgray, minimum height=2.4cm, minimum width=2cm, fill=lightgray] at (0,-1.6) { };
\node[draw=lightgray, minimum height=2.4cm, minimum width=2cm, fill=lightgray] at (3,-1.6) { };
\node[draw=lightgray, minimum height=0.8cm, minimum width=2cm, fill=lightgray] (X) at (0,0) { };
\node[draw=lightgray, minimum height=0.8cm, minimum width=2cm, fill=lightgray] (X*) at (3,0) { };
\draw (0,0.5) node{} node[above]{$X_{\A\B^k}(1)$}; \draw (3,0.5) node{} node[above]{$X_{\A\B^k}(1)^{\dagger}$};
\draw(X.north west) node{$\bullet$} node[left]{$\A$}; \draw(X.south west) node{$\bullet$} node[left]{$B_1$}; \draw(X.east) node{$\blacktriangledown$};
\draw (-1.1,-1.2) node{$\bullet$} node[left]{$B_2$}; \draw (-1.1,-2.8) node{$\bullet$} node[left]{$B_k$}; \draw (1.1,-1.2) node{$\bullet$}; \draw (1.1,-2.8) node{$\bullet$};
\draw(X*.north east) node{$\bullet$} node[right]{$\A$}; \draw(X*.south east) node{$\bullet$} node[right]{$B_1$}; \draw(X*.west) node{$\blacktriangledown$};
\draw (1.9,-1.2) node{$\bullet$}; \draw (1.9,-2.8) node{$\bullet$}; \draw (4.1,-1.2) node{$\bullet$} node[right]{$B_2$}; \draw (4.1,-2.8) node{$\bullet$} node[right]{$B_k$};
\draw (X.east) -- (X*.west); \draw (-1,-1.2) -- (4,-1.2); \draw (-1,-2.8) -- (4,-2.8);
\draw[dotted] (-1.3,-1.5) -- (-1.3,-2.5);  \draw[dotted] (4.3,-1.5) -- (4.3,-2.5);
\end{tikzpicture}
\end{center}
\label{fig:W_AB^k}
\end{figure}

The products $\widetilde{W}_{\A\B^k}(1)\widetilde{W}_{\A\B^k}(1)$ and $\widetilde{W}_{\A\B^k}(1)\widetilde{W}_{\A\B^k}(2)$, for instance, are then obtained by the wirings represented in Figure \ref{fig:prod-W_AB^k}.

\begin{figure}[h]
\caption{$\widetilde{W}_{\A\B^k}(1)\widetilde{W}_{\A\B^k}(1)$ (on the left) and $\widetilde{W}_{\A\B^k}(1)\widetilde{W}_{\A\B^k}(2)$ (on the right)}
\begin{center}
\begin{tikzpicture} [scale=0.9]
\node[draw=lightgray, minimum height=2.4cm, minimum width=0.8cm, fill=lightgray] at (0,-1.6) { };
\node[draw=lightgray, minimum height=2.4cm, minimum width=0.8cm, fill=lightgray] at (1.6,-1.6) { };
\node[draw=lightgray, minimum height=0.8cm, minimum width=0.8cm, fill=lightgray] (X) at (0,0) { };
\node[draw=lightgray, minimum height=0.8cm, minimum width=0.8cm, fill=lightgray] (X*) at (1.6,0) { };
\draw(X.north west) node{$\bullet$} node[left]{$\A$}; \draw(X.south west) node{$\bullet$} node[left]{$B_1$}; \draw(X.east) node{$\blacktriangledown$};
\draw (-0.4,-1.2) node{$\bullet$} node[left]{$B_2$}; \draw (-0.4,-2.8) node{$\bullet$} node[left]{$B_k$};
\draw(X*.north east) node{$\bullet$}; \draw(X*.south east) node{$\bullet$}; \draw(X*.west) node{$\blacktriangledown$};
\draw (2,-1.2) node{$\bullet$}; \draw (2,-2.8) node{$\bullet$};
\draw (X.east) -- (X*.west);
\draw[dotted] (-0.6,-1.5) -- (-0.6,-2.5);
\node[draw=lightgray, minimum height=2.4cm, minimum width=0.8cm, fill=lightgray] at (3.2,-1.6) { };
\node[draw=lightgray, minimum height=2.4cm, minimum width=0.8cm, fill=lightgray] at (4.8,-1.6) { };
\node[draw=lightgray, minimum height=0.8cm, minimum width=0.8cm, fill=lightgray] (Y) at (3.2,0) { };
\node[draw=lightgray, minimum height=0.8cm, minimum width=0.8cm, fill=lightgray] (Y*) at (4.8,0) { };
\draw(Y.north west) node{$\bullet$}; \draw(Y.south west) node{$\bullet$}; \draw(Y.east) node{$\blacktriangledown$};
\draw (2.8,-1.2) node{$\bullet$}; \draw (2.8,-2.8) node{$\bullet$};
\draw(Y*.north east) node{$\bullet$} node[right]{$\A$}; \draw(Y*.south east) node{$\bullet$} node[right]{$B_1$}; \draw(Y*.west) node{$\blacktriangledown$};
\draw (5.2,-1.2) node{$\bullet$} node[right]{$B_2$}; \draw (5.2,-2.8) node{$\bullet$} node[right]{$B_k$};
\draw (Y.east) -- (Y*.west);
\draw (-0.4,-1.2) -- (5.2,-1.2); \draw (-0.4,-2.8) -- (5.2,-2.8);
\draw[dotted] (5.4,-1.5) -- (5.4,-2.5);
\draw (X*.north east) -- (Y.north west); \draw (X*.south east) -- (Y.south west);

\node[draw=lightgray, minimum height=2.4cm, minimum width=0.8cm, fill=lightgray] at (8,-1.6) { };
\node[draw=lightgray, minimum height=2.4cm, minimum width=0.8cm, fill=lightgray] at (9.6,-1.6) { };
\node[draw=lightgray, minimum height=0.8cm, minimum width=0.8cm, fill=lightgray] (X') at (8,0) { };
\node[draw=lightgray, minimum height=0.8cm, minimum width=0.8cm, fill=lightgray] (X'*) at (9.6,0) { };
\draw(X'.north west) node{$\bullet$} node[left]{$\A$}; \draw(X'.south west) node{$\bullet$} node[left]{$B_1$}; \draw(X'.east) node{$\blacktriangledown$};
\draw (7.6,-1.2) node{$\bullet$} node[left]{$B_2$}; \draw (7.6,-2.8) node{$\bullet$} node[left]{$B_k$};
\draw(X'*.north east) node{$\bullet$}; \draw(X'*.south east) node{$\bullet$}; \draw(X'*.west) node{$\blacktriangledown$};
\draw (10,-1.2) node{$\bullet$}; \draw (10,-2.8) node{$\bullet$};
\draw (X'.east) -- (X'*.west);
\draw (7.6,-1.2) -- (10,-1.2);
\draw[dotted] (7.4,-1.5) -- (7.4,-2.5);
\node[draw=lightgray, minimum height=2.4cm, minimum width=0.8cm, fill=lightgray] at (11.2,-1.6) { };
\node[draw=lightgray, minimum height=2.4cm, minimum width=0.8cm, fill=lightgray] at (12.8,-1.6) { };
\node[draw=lightgray, minimum height=0.8cm, minimum width=0.8cm, fill=lightgray] (Y') at (11.2,0) { };
\node[draw=lightgray, minimum height=0.8cm, minimum width=0.8cm, fill=lightgray] (Y'*) at (12.8,0) { };
\draw(Y'.north west) node{$\bullet$}; \draw(Y'.south west) node{$\bullet$}; \draw(Y'.east) node{$\blacktriangledown$};
\draw (10.8,-1.2) node{$\bullet$}; \draw (10.8,-2.8) node{$\bullet$};
\draw(Y'*.north east) node{$\bullet$} node[right]{$\A$}; \draw(Y'*.south east) node{$\bullet$} node[right]{$B_2$}; \draw(Y'*.west) node{$\blacktriangledown$};
\draw (13.2,-1.2) node{$\bullet$} node[right]{$B_1$}; \draw (13.2,-2.8) node{$\bullet$} node[right]{$B_k$};
\draw (Y'.east) -- (Y'*.west);
\draw (10.8,-1.2) -- (13.2,-1.2);
\draw[dotted] (13.4,-1.5) -- (13.4,-2.5);
\draw (7.6,-2.8) -- (13.2,-2.8);
\draw (X'*.north east) -- (Y'.north west);
\draw (X'*.south east) -- (10.8,-1.2); \draw (10,-1.2) -- (Y'.south west);
\end{tikzpicture}
\end{center}
\label{fig:prod-W_AB^k}
\end{figure}

So what we get by the graphical Wick formula for Wishart matrices is that for any $f:[p]\rightarrow[k]$,
\[ \mathbf{E}_{W_{\A\B}\sim\mathcal{W}_{d^2,s}} \mathrm{Tr}\left[\underset{i=1}{\overset{p}{\overrightarrow{\prod}}}\widetilde{W}_{\A\B^k}(f(i))\right] = \underset{\alpha\in\mathfrak{S}(p)}{\sum} \mathcal{D}_{f,\alpha} = \underset{\alpha\in\mathfrak{S}(p)}{\sum} d^{\sharp(\gamma^{-1}\alpha)}d^{\sharp(\hat{\gamma}^{-1}\hat{\alpha}_f)}s^{\sharp(\alpha)}, \]
where $\hat{\alpha}_f$ is defined by
\[ \hat{\alpha}_f:(i,r)\in [p]\times[k]\mapsto
\begin{cases}(\alpha(i),f(\alpha(i)))\ \text{if}\ r=f(i)\\
(i,r)\ \text{if}\ r\neq f(i)\end{cases}, \]
and where $\hat{\gamma}$ stands for $\gamma$ applied to the first argument. Indeed, for each $\alpha\in\mathfrak{S}(p)$, there are $\sharp(\gamma^{-1}\alpha)$ loops connecting the $d$-dimensional gates corresponding to $\A$, $\sharp(\hat{\gamma}^{-1}\hat{\alpha}_f)$ loops connecting the $d$-dimensional gates corresponding to $B_1,\ldots,B_k$, and $\sharp(\alpha)$ loops connecting $s$-dimensional gates. This is because for each $1\leq i\leq p$, on  subsystems $\A$ and $B_{f(i)}$, the entrances (respectively the exit) of the $i^{th}$ box $X_{\A\B^k}(f(i))$ are connected to the exits (respectively the entrance) of the $\alpha(i)^{th}$ box $X_{\A\B^k}(f(\alpha(i)))^{\dagger}$.

What happens in the special case $p=2$ and $k=2$ is detailed in Figures \ref{fig:p=2_k=2_1} and \ref{fig:p=2_k=2_2} below as an illustration.

\begin{figure}[h]
\caption{$f(1)=f(2)=1$. On the left, $\alpha=\id$: $\mathcal{D}_{f,\alpha}=d^3s^2$. On the right, $\alpha=(1\,2)$: $\mathcal{D}_{f,\alpha}=d^5s$.}
\begin{center}
\begin{tikzpicture} [scale=0.7]
\draw (-1,0) node {$B_2$}; \draw (-1,1) node {$B_1$}; \draw (-1,2) node {$\A$};
\draw (8,0) node {$B_2$}; \draw (8,1) node {$B_1$}; \draw (8,2) node {$\A$};
\draw (0,0) node {$\bullet$}; \draw (0,1) node {$\bullet$}; \draw (0,2) node {$\bullet$};
\draw (3,0) node {$\bullet$}; \draw (3,1) node {$\bullet$}; \draw (3,2) node {$\bullet$};
\draw (1,3/2) node {$\blacktriangledown$}; \draw (2,3/2) node {$\blacktriangledown$};
\draw (4,0) node {$\bullet$}; \draw (4,1) node {$\bullet$}; \draw (4,2) node {$\bullet$};
\draw (7,0) node {$\bullet$}; \draw (7,1) node {$\bullet$}; \draw (7,2) node {$\bullet$};
\draw (5,3/2) node {$\blacktriangledown$}; \draw (6,3/2) node {$\blacktriangledown$};
\draw (0,0) -- (7,0); \draw (3,1) -- (4,1); \draw (3,2) -- (4,2);
\draw (1,3/2) -- (2,3/2); \draw (5,3/2) -- (6,3/2);
\draw (0,0) to[out=-160,in=-20] (7,0);
\draw (0,1) to[out=-160,in=-20] (7,1); \draw (0,1) to[out=30,in=180] (3,1); \draw (4,1) to[out=0,in=150] (7,1);
\draw (0,2) to[out=160,in=20] (7,2); \draw (0,2) to[out=0,in=150] (3,2); \draw (4,2) to[out=30,in=180] (7,2);
\draw (1,3/2) to[bend left=90] (2,3/2); \draw (5,3/2) to[bend left=90] (6,3/2);

\draw (10,0) node {$B_2$}; \draw (10,1) node {$B_1$}; \draw (10,2) node {$\A$};
\draw (19,0) node {$B_2$}; \draw (19,1) node {$B_1$}; \draw (19,2) node {$\A$};
\draw (11,0) node {$\bullet$}; \draw (11,1) node {$\bullet$}; \draw (11,2) node {$\bullet$};
\draw (14,0) node {$\bullet$}; \draw (14,1) node {$\bullet$}; \draw (14,2) node {$\bullet$};
\draw (12,3/2) node {$\blacktriangledown$}; \draw (13,3/2) node {$\blacktriangledown$};
\draw (15,0) node {$\bullet$}; \draw (15,1) node {$\bullet$}; \draw (15,2) node {$\bullet$};
\draw (18,0) node {$\bullet$}; \draw (18,1) node {$\bullet$}; \draw (18,2) node {$\bullet$};
\draw (16,3/2) node {$\blacktriangledown$}; \draw (17,3/2) node {$\blacktriangledown$};
\draw (11,0) -- (18,0); \draw (14,1) -- (15,1); \draw (14,2) -- (15,2);
\draw (12,3/2) -- (13,3/2); \draw (16,3/2) -- (17,3/2);
\draw (11,0) to[out=-160,in=-20] (18,0);
\draw (11,1) to[out=-160,in=-20] (18,1); \draw (11,1) to[bend right=10] (18,1); \draw (14,1) to[bend right=45] (15,1);
\draw (11,2) to[out=160,in=20] (18,2); \draw (11,2) to[bend left=10] (18,2); \draw (14,2) to[bend left=45] (15,2);
\draw (12,3/2) to[out=-30, in=180] (16,3/2); \draw (13,3/2) to[out=0, in=150] (17,3/2);
\end{tikzpicture}
\end{center}
\label{fig:p=2_k=2_1}
\end{figure}

\begin{figure}[h]
\caption{$f(1)=1$, $f(2)=2$. On the left, $\alpha=\id$: $\mathcal{D}_{f,\alpha}=d^3s^2$. On the right, $\alpha=(1\,2)$: $\mathcal{D}_{f,\alpha}=d^3s$.}
\begin{center}
\begin{tikzpicture} [scale=0.7]
\draw (-1,0) node {$B_2$}; \draw (-1,1) node {$B_1$}; \draw (-1,2) node {$\A$};
\draw (8,0) node {$B_1$}; \draw (8,1) node {$B_2$}; \draw (8,2) node {$\A$};
\draw (0,0) node {$\bullet$}; \draw (0,1) node {$\bullet$}; \draw (0,2) node {$\bullet$};
\draw (3,0) node {$\bullet$}; \draw (3,1) node {$\bullet$}; \draw (3,2) node {$\bullet$};
\draw (1,3/2) node {$\blacktriangledown$}; \draw (2,3/2) node {$\blacktriangledown$};
\draw (4,0) node {$\bullet$}; \draw (4,1) node {$\bullet$}; \draw (4,2) node {$\bullet$};
\draw (7,0) node {$\bullet$}; \draw (7,1) node {$\bullet$}; \draw (7,2) node {$\bullet$};
\draw (5,3/2) node {$\blacktriangledown$}; \draw (6,3/2) node {$\blacktriangledown$};
\draw (0,0) -- (3,0); \draw (4,0) -- (7,0); \draw (3,2) -- (4,2);
\draw (1,3/2) -- (2,3/2); \draw (5,3/2) -- (6,3/2);
\draw (0,0) to[out=-160,in=-90] (7,1); \draw (0,1) to[out=-90,in=-20] (7,0);
\draw (3,0) to[out=0, in=180] (4,1); \draw (3,1) to[out=0, in=180] (4,0);
\draw (0,1) to[out=30, in=180] (3,1); \draw (4,1) to[out=0, in=150] (7,1);
\draw (0,2) to[out=160,in=20] (7,2); \draw (0,2) to[out=0,in=150] (3,2); \draw (4,2) to[out=30,in=180] (7,2);
\draw (1,3/2) to[bend left=90] (2,3/2); \draw (5,3/2) to[bend left=90] (6,3/2);

\draw (10,0) node {$B_2$}; \draw (10,1) node {$B_1$}; \draw (10,2) node {$\A$};
\draw (19,0) node {$B_1$}; \draw (19,1) node {$B_2$}; \draw (19,2) node {$\A$};
\draw (11,0) node {$\bullet$}; \draw (11,1) node {$\bullet$}; \draw (11,2) node {$\bullet$};
\draw (14,0) node {$\bullet$}; \draw (14,1) node {$\bullet$}; \draw (14,2) node {$\bullet$};
\draw (12,3/2) node {$\blacktriangledown$}; \draw (13,3/2) node {$\blacktriangledown$};
\draw (15,0) node {$\bullet$}; \draw (15,1) node {$\bullet$}; \draw (15,2) node {$\bullet$};
\draw (18,0) node {$\bullet$}; \draw (18,1) node {$\bullet$}; \draw (18,2) node {$\bullet$};
\draw (16,3/2) node {$\blacktriangledown$}; \draw (17,3/2) node {$\blacktriangledown$};
\draw (11,0) -- (14,0); \draw (15,0) -- (18,0); \draw (14,2) -- (15,2);
\draw (12,3/2) -- (13,3/2); \draw (16,3/2) -- (17,3/2);
\draw (11,0) to[out=-160, in=-90] (18,1); \draw (11,1) to[out=-90, in=-20] (18,0);
\draw (14,0) to[out=0, in=-135] (15,1); \draw (14,1) to[out=-45, in=180] (15,0);
\draw (11,1) to[bend right=15] (18,1); \draw (14,1) to[bend left=45] (15,1);
\draw (11,2) to[out=160,in=20] (18,2); \draw (11,2) to[bend left=10] (18,2); \draw (14,2) to[bend left=45] (15,2);
\draw (12,3/2) to[out=-30, in=180] (16,3/2); \draw (13,3/2) to[out=0, in=150] (17,3/2);
\end{tikzpicture}
\end{center}
\label{fig:p=2_k=2_2}
\end{figure}

Finally, we also know by Proposition \ref{prop:geodesic-alpha_f} that for any $\alpha\in\mathfrak{S}(p)$ and $f:[p]\rightarrow[k]$, denoting by $\gamma_f$ the product of the canonical full cycles on the level sets of $f$, we have $\sharp(\hat{\gamma}^{-1}\hat{\alpha}_f) = \sharp(\gamma_f^{-1}\alpha) + k - |\im(f)|$.

Putting everything together, we eventually come to the result summarized in Proposition \ref{prop:wishart-p-preliminary} below.

\begin{proposition} \label{prop:wishart-p-preliminary} For any $d,s\in\N$ and any $p\in\N$, we have
\begin{align*} \mathbf{E}_{W_{\A\B}\sim\mathcal{W}_{d^2,s}} \mathrm{Tr}\left[\left(\sum_{j=1}^k\widetilde{W}_{\A\B^k}(j)\right)^p\right] = & \sum_{f:[p]\rightarrow[k]}\mathbf{E}_{W_{\A\B}\sim\mathcal{W}_{d^2,s}} \mathrm{Tr}\left[\underset{i=1}{\overset{p}{\overrightarrow{\prod}}}\widetilde{W}_{\A\B^k}(f(i))\right] \\
= & \underset{f:[p]\rightarrow[k]}{\sum}\underset{\alpha\in\mathfrak{S}(p)}{\sum} d^{\sharp(\gamma^{-1}\alpha)+\sharp(\gamma_f^{-1}\alpha)+k-|\im(f)|}s^{\sharp(\alpha)}.
\end{align*}
\end{proposition}

\section{Appendix G: Counting geodesics vs non-geodesics pairings and permutations}
\label{appendix:geodesics}

Let us recall once and for all two notation that we will use repeatedly in this section, and that were introduced in Lemma \ref{lemma:catalan}. For any $p,m\in\N$ with $m\leq p$, we denote by $\mathrm{Cat}_p=\frac{1}{p+1}{2p \choose p}$ the $p^{th}$ Catalan number, and by $\mathrm{Nar}_p^m=\frac{1}{p+1}{p+1 \choose m}{p-1 \choose m-1}$ the $(p,m)^{th}$ Narayana number.

\subsection{Number of pairings of $2p$ elements which are not on the geodesics between the identity and the canonical full cycle}

\begin{lemma} \label{lemma:number-pairings-defect}
Let $p\in\N$ and denote by $\gamma$ the canonical full cycle on $\{1,\ldots,2p\}$. For any $0\leq\delta\leq\lfloor p/2\rfloor$, define the set of pairings having a defect $2\delta$ of being on the geodesics between $\id$ and $\gamma$ as
\[ \mathfrak{P}^{(2)}_{\delta}(2p) = \{\lambda\in\mathfrak{P}^{(2)}(2p) \st \sharp(\gamma^{-1}\lambda)=p+1-2\delta\}. \]
Then, the cardinality of $\mathfrak{P}^{(2)}_{\delta}(2p)$ is upper bounded by $\mathrm{Cat}_p \left(p^4/4\right)^{\delta}$.
\end{lemma}

To prove Lemma \ref{lemma:number-pairings-defect} (and later on Lemma \ref{lemma:number-functions,pairings-defect}) we will need the simple observation below. Roughly speaking, it will allow us to assume without loss of generality that, in the decomposition of an element of $\mathfrak{P}^{(2)}_{\delta}(2p)$ into $p$ disjoint transpositions, the ones ``creating'' the $2\delta$ geodesic defects are the $2\delta$ first ones.

\begin{fact} \label{fact:ordering-defects}
Let $\varsigma$ be a permutation on $\{1,\ldots,q\}$ and $\tau_1,\tau_2,\tau_3$ be three disjoint transpositions on $\{1,\ldots,q\}$, for some integer $q\geq 6$. Define $\varsigma^{(1)}=\varsigma\,\tau_1$, $\varsigma^{(2)}=\varsigma\,\tau_1\tau_2$, $\varsigma^{(3)}=\varsigma\,\tau_1\tau_2\tau_3$, and assume that
\begin{equation} \label{eq:not-canonical} \sharp(\varsigma^{(1)})=\sharp(\varsigma)+1,\ \sharp(\varsigma^{(2)})=\sharp(\varsigma)+2,\  \sharp(\varsigma^{(3)})=\sharp(\varsigma)+1. \end{equation}
Then, there exists a permutation $\pi$ of the three indices $\{1,2,3\}$ such that, defining this time $\varsigma_{\pi}^{(1)}=\varsigma\,\tau_{\pi(1)}$, $\varsigma_{\pi}^{(2)}=\varsigma\,\tau_{\pi(1)}\tau_{\pi(2)}$, $\varsigma_{\pi}^{(3)}=\varsigma\,\tau_{\pi(1)}\tau_{\pi(2)}\tau_{\pi(3)}$, we have
\[ \sharp(\varsigma_{\pi}^{(1)})=\sharp(\varsigma)+1,\ \sharp(\varsigma_{\pi}^{(2)})=\sharp(\varsigma),\  \sharp(\varsigma_{\pi}^{(3)})=\sharp(\varsigma)+1. \]
\end{fact}

\begin{proof}
Assume that $\varsigma$ and $\tau_1=(i_1\,j_1),\tau_2=(i_2\,j_2),\tau_3=(i_3\,j_3)$ satisfy equation \eqref{eq:not-canonical}. This means that $i_1,j_1$ belong to the same cycle of $\varsigma$, $i_2,j_2$ belong to the same cycle of $\varsigma^{(1)}$, and $i_3,j_3$ belong to two different cycles of $\varsigma^{(2)}$. So let us inspect all the scenarios which may occur.\\
$\bullet$ $c_1 \overset{(i_1\,j_1)}{\rightarrow} c_1^xc_1^y$ and $c_2 \overset{(i_2\,j_2)}{\rightarrow} c_2^xc_2^y$, with $c_1,c_2$ two different cycles of $\varsigma$: If $i_3\in c_1^x$ and $j_3\in c_1^y$ then the re-ordering $1,3,2$ is suitable. If $i_3\in c_2^x$ and $j_3\in c_2^y$ then the re-ordering $2,3,1$ is suitable. If $i_3\in c_1^a$ and $j_3\in c_2^b$, for $a,b\in\{x,y\}$, then both re-orderings $1,3,2$ and $2,3,1$ are suitable. And similarly when the roles of $i_3$ and $j_3$ are exchanged.\\
$\bullet$ $c \overset{(i_1\,j_1)}{\rightarrow} c'c'' \overset{(i_2\,j_2)}{\rightarrow} c^xc^yc^z$, with $c$ a cycle of $\varsigma$, while $c^z=c''$ and $c' \overset{(i_2\,j_2)}{\rightarrow} c^xc^y$: If $i_3\in c^x$ and $j_3\in c^y$ then the re-ordering $2,3,1$ is suitable. If $i_3\in c^a$, for $a\in\{x,y\}$, and $j_3\in c^z$ then the re-ordering $1,3,2$ is suitable. And similarly when the roles of $i_3$ and $j_3$ are exchanged.
\end{proof}

\begin{figure}[h]
\caption{Case $p=6$ and $\sharp(\varsigma)=1$. From left to right: $\delta=0$, $\delta=1$, $\delta=2$ and $\delta=3$.}
\begin{center}
\begin{tikzpicture} [scale=0.4]
\draw [->] (-0.5,0) -- (7,0); \draw [->] (0,-0.5) -- (0,8);
\draw (0,8.7) node {$n(q)$}; \draw (7.5,0) node {$q$};
\draw (-0.5,-0.7) node {$0$}; \draw (-0.5,1) node {$1$}; \draw (0,7) node {$-$}; \draw (-0.5,7) node {$7$}; \draw (6,0) node {$+$}; \draw (6,-0.7) node {$6$};
\draw (0,1) -- (6,7);
\draw (0,1) node {$\bullet$}; \draw (1,2) node {$\bullet$}; \draw (2,3) node {$\bullet$}; \draw (3,4) node {$\bullet$}; \draw (4,5) node {$\bullet$}; \draw (5,6) node {$\bullet$}; \draw (6,7) node {$\bullet$};

\draw [->] (9.5,0) -- (17,0); \draw [->] (10,-0.5) -- (10,6);
\draw (10,6.7) node {$n(q)$}; \draw (17.5,0) node {$q$};
\draw (9.5,-0.7) node {$0$}; \draw (9.5,1) node {$1$}; \draw (10,5) node {$-$}; \draw (9.5,5) node {$5$}; \draw (16,0) node {$+$}; \draw (16,-0.7) node {$6$};
\draw (10,1) -- (11,2); \draw (11,2) -- (12,1); \draw (12,1) -- (16,5);
\draw (10,1) node {$\bullet$}; \draw (11,2) node {$\bullet$}; \draw (12,1) node {$\bullet$}; \draw (13,2) node {$\bullet$}; \draw (14,3) node {$\bullet$}; \draw (15,4) node {$\bullet$}; \draw (16,5) node {$\bullet$};

\draw [->] (19.5,0) -- (27,0); \draw [->] (20,-0.5) -- (20,4);
\draw (20,4.7) node {$n(q)$}; \draw (27.5,0) node {$q$};
\draw (19.5,-0.7) node {$0$}; \draw (19.5,1) node {$1$}; \draw (20,3) node {$-$}; \draw (19.5,3) node {$3$}; \draw (26,0) node {$+$}; \draw (26,-0.7) node {$6$};
\draw (20,1) -- (21,2); \draw (21,2) -- (22,1); \draw (22,1) -- (23,2); \draw (23,2) -- (24,1); \draw (24,1) -- (26,3);
\draw (20,1) node {$\bullet$}; \draw (21,2) node {$\bullet$}; \draw (22,1) node {$\bullet$}; \draw (23,2) node {$\bullet$}; \draw (24,1) node {$\bullet$}; \draw (25,2) node {$\bullet$}; \draw (26,3) node {$\bullet$};

\draw [->] (29.5,0) -- (37,0); \draw [->] (30,-0.5) -- (30,3);
\draw (30,3.7) node {$n(q)$}; \draw (37.5,0) node {$q$};
\draw (29.5,-0.7) node {$0$}; \draw (29.5,1) node {$1$}; \draw (36,0) node {$+$}; \draw (36,-0.7) node {$6$};
\draw (30,1) -- (31,2); \draw (31,2) -- (32,1); \draw (32,1) -- (33,2); \draw (33,2) -- (34,1); \draw (34,1) -- (35,2); \draw (35,2) -- (36,1);
\draw (30,1) node {$\bullet$}; \draw (31,2) node {$\bullet$}; \draw (32,1) node {$\bullet$}; \draw (33,2) node {$\bullet$}; \draw (34,1) node {$\bullet$}; \draw (35,2) node {$\bullet$}; \draw (36,1) node {$\bullet$};
\end{tikzpicture}
\end{center}
\label{fig:pairings}
\end{figure}
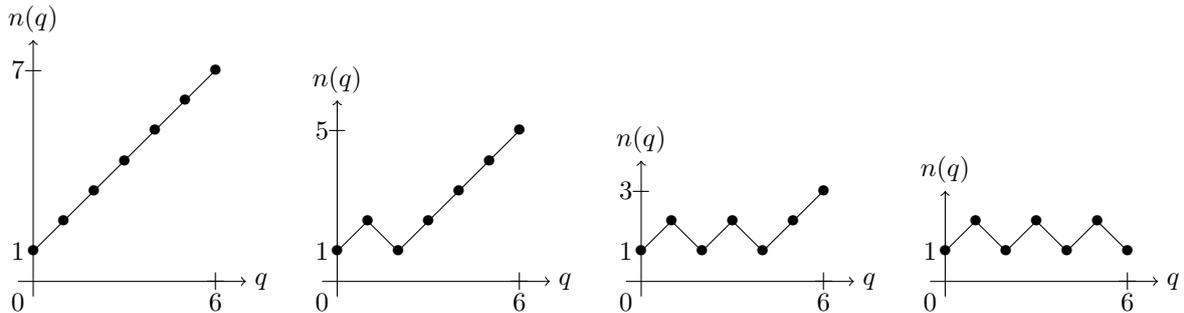

As an immediate consequence of Fact \ref{fact:ordering-defects}, we have the following: Let $\varsigma\in\mathfrak{S}(2p)$ and $\lambda=\tau_1\cdots\tau_p\in\mathfrak{P}^{(2)}(2p)$. Define for each $1\leq q\leq p$, $\varsigma^{(q)}=\varsigma\,\tau_1\cdots\tau_q$, as well as $\varsigma^{(0)}=\varsigma$. Assume next that, for some $0\leq\delta\leq\lfloor (p+\sharp(\varsigma))/2\rfloor$,
\[ \sharp(\varsigma^{(p)})=\sharp(\varsigma)+p-2\delta. \]
Then, there exists a permutation $\pi$ of the $p$ indices $\{1,\ldots,p\}$ such that, defining this time for each $1\leq q\leq p$, $\varsigma_{\pi}^{(q)}=\varsigma\,\tau_{\pi(1)}\cdots\tau_{\pi(q)}$, as well as $\varsigma_{\pi}^{(0)}=\varsigma$, we have
\begin{equation} \label{eq:canonical} \forall\ 1\leq q\leq p,\ \begin{cases} \sharp(\varsigma_{\pi}^{(q)})=\sharp(\varsigma_{\pi}^{(q-1)})-1\ \text{if}\ q\in\{2\epsilon \st 1\leq\epsilon\leq\delta\} \\  \sharp(\varsigma_{\pi}^{(q)})=\sharp(\varsigma_{\pi}^{(q)})+1\ \text{if}\ q\notin\{2\epsilon \st 1\leq\epsilon\leq\delta\} \end{cases}. \end{equation}
Since $\lambda=\tau_1\cdots\tau_p=\tau_{\pi(1)}\cdots\tau_{\pi(p)}$, we see that we may always assume without loss of generality that, given $\varsigma$, the transpositions $\tau_1,\ldots,\tau_p$ in the decomposition of $\lambda$ are ordered so that $\lambda$ is under the canonical form \eqref{eq:canonical}. The behaviour of the function $q\in[p]\mapsto \sharp(\varsigma^{(q)})$ under this hypothesis, depending on the value of $\delta$, is represented in Figure \ref{fig:pairings} (in the special case $p=6$ and $\sharp(\varsigma)=1$).

With this result in mind, let us now turn to the proof of Lemma \ref{lemma:number-pairings-defect}.

\begin{proof}[Proof of Lemma \ref{lemma:number-pairings-defect}]
Given $\lambda=(i_1\,j_1)\cdots (i_p\,j_p)\in\mathfrak{P}^{(2)}(2p)$, we will always assume from now that the transpositions $(i_1\,j_1),\ldots,(i_p\,j_p)$ in its decomposition are ordered so that $\lambda$ is under the canonical form \eqref{eq:canonical} for $\gamma^{-1}$. This means the following: defining, for each $1\leq q\leq p$, the permutation $\widetilde{\lambda}^{(q)}=\gamma^{-1}(i_1\,j_1)\cdots (i_q\,j_q)$ and the integer $n(q)=\sharp(\widetilde{\lambda}^{(q)})$, as well as $\widetilde{\lambda}^{(0)}=\gamma^{-1}$ and $n(0)=\sharp(\widetilde{\lambda}^{(0)})=1$, we have, for any $0\leq\delta\leq\lfloor p/2\rfloor$,
\[ \lambda\in\mathfrak{P}^{(2)}_{\delta}(2p)\ \Leftrightarrow\ \forall\ 1\leq q\leq p,\ \begin{cases} n(q)=n(q-1)-1\ \text{if}\ q\in\{2\epsilon \st 1\leq\epsilon\leq\delta\} \\  n(q)=n(q-1)+1\ \text{if}\ q\notin\{2\epsilon \st 1\leq\epsilon\leq\delta\} \end{cases}. \]
In particular, $\lambda\in NC^{(2)}(2p)\ \Leftrightarrow\ \forall\ 1\leq q\leq p,\ n(q)=n(q-1)+1$, and we know that there are precisely $\mathrm{Cat}_p$ possibilities to build such pairing $\lambda$.
This implies that, for each $1\leq\delta\leq\lfloor p/2\rfloor$, there are necessarily less than
${2p \choose 2}\cdots{2(p-\delta+1) \choose 2}\times\mathrm{Cat}_{p-2\delta}$ possibilities to build a $\lambda\in\mathfrak{P}^{(2)}_{\delta}(2p)$.
Indeed, for the choice of the $2\delta$ first disjoint transpositions we can use the trivial upper bound that would consist in picking them completely arbitrarily, while the $p-2\delta$ last ones have to be chosen so that they form a non-crossing pairing of the $2p-4\delta$ not yet selected indices. Now, we just have to observe that
\begin{align*} {2p \choose 2}\cdots{2(p-2\delta+1) \choose 2}\times \mathrm{Cat}_{p-2\delta} =\, & \frac{2p\cdots(2p-4\delta+1)}{2^{2\delta}}\times\frac{(2p-4\delta)!}{(p-2\delta)!(p-2\delta+1)!} \\
=\, & \frac{1}{2^{2\delta}}\times\frac{p!(p+1)!}{(p-2\delta)!(p-2\delta+1)!}\times \frac{(2p)!}{p!(p+1)!} \\
\leq\, & \frac{p^{4\delta}}{2^{2\delta}}\times\mathrm{Cat}_p, \end{align*}
which completes the proof.
\end{proof}

\subsection{One needed generalization: bounding the number of pairings of $2p$ elements which are not on the geodesic path between the identity and a product of (few) cycles}

The proof of Proposition \ref{prop:gaussian-infty} crucially relies at some point on a statement of the same kind as the one appearing in Lemma \ref{lemma:number-pairings-defect}. Nevertheless, what we actually need there is a slight generalization of the latter. More specifically, we have to bound the number of pairings which have some defect of lying on the geodesics between the identity and, not only a full cycle, but also a product of (few) cycles. So let us give the following extension of Lemma \ref{lemma:number-pairings-defect}, which is really directed towards the application that we have in mind.

\begin{lemma} \label{lemma:number-functions,pairings-defect}
Let $p\in\N$. For any $f:[2p]\rightarrow[k]$ and any $0\leq\delta\leq \left\lfloor \left(p+|\im(f)|\right)/2\right\rfloor$, define the set of pairings having a defect $2\delta$ of being on the geodesics between $\id$ and $\gamma_f$ (the product of the canonical full cycles on each of the $|\im(f)|$ level sets of $f$) as
\[ \mathfrak{P}^{(2)}_{f,\delta}(2p) = \{\lambda\in\mathfrak{P}^{(2)}(2p) \st \sharp(\gamma_f^{-1}\lambda)=p+|\im(f)|-2\delta\}. \]
Then, for any $0\leq\delta\leq\lfloor p/2\rfloor$, we have the upper bound
\[ \left|\left\{(f,\lambda) \st \lambda\in\mathfrak{P}^{(2)}_{f,\delta}(2p) \right\} \right| \leq k^{p+2\delta}\mathrm{Cat}_p \left(\frac{p^4}{4}\right)^{\delta}. \]
\end{lemma}

\begin{proof}
We will follow the same strategy and employ the same notation as in the proof of Lemma \ref{lemma:number-pairings-defect}. Given $\lambda=(i_1\,j_1)\cdots(i_p\,j_p)\in\mathfrak{P}^{(2)}(2p)$ and $f:[2p]\rightarrow[k]$, we will always assume that the transpositions $(i_1\,j_1),\ldots,(i_p\,j_p)$ in the decomposition of $\lambda$ are ordered so that $\lambda$ is under the canonical form \eqref{eq:canonical} for $\gamma_f^{-1}$. This means the following: defining, for each $1\leq q\leq p$, $\widetilde{\lambda}^{(q)}=\gamma_f^{-1}(i_1\,j_1)\cdots (i_q\,j_q)$ and $n(q)=\sharp(\widetilde{\lambda}^{(q)})$, as well as $\widetilde{\lambda}^{(0)}=\gamma_f^{-1}$ and $n(0)=\sharp(\widetilde{\lambda}^{(0)})=|\im(f)|$, we have, for any $0\leq\delta\leq\lfloor (p+|\im(f)|)/2\rfloor$,
\begin{equation} \label{eq:b} \lambda\in\mathfrak{P}^{(2)}_{f,\delta}(2p)\ \Leftrightarrow\ \forall\ 1\leq q\leq p,\ \begin{cases} n(q)=n(q-1)-1\ \text{if}\ q\in\{2\epsilon \st 1\leq\epsilon\leq\delta\} \\  n(q)=n(q-1)+1\ \text{if}\ q\notin\{2\epsilon \st 1\leq\epsilon\leq\delta\} \end{cases}. \end{equation}

In particular, $\lambda\in \mathfrak{P}^{(2)}_{f,0}(2p)\ \Leftrightarrow\ \forall\ 1\leq q\leq p,\ n(q)=n(q-1)+1$, and we know that there are precisely $k^p\,\mathrm{Cat}_p$ possibilities to build a pair $(f,\lambda)$ satisfying this condition (because the latter holds if and only if both constraints $\lambda\in NC^{(2)}(2p)$ and $f\circ\lambda=f$ are fulfilled).

In the case $1\leq\delta\leq\lfloor p/2\rfloor$, notice that after the $2\delta$ first steps, we are left with a permutation $\overline{\varsigma}$ having $|\im(f)|$ cycles, and we have to impose that the partial pairing $\overline{\lambda}=(i_{2\delta+1}\,j_{2\delta+1})\cdots(i_{2p}\,j_{2p})$ lies on the geodesics between $\id$ and $\overline{\varsigma}$. Now, the number of such partial pairings is the same as the number of partial pairings lying on the geodesics between $\id$ and $\gamma_{\overline{f}}$, for any function $\overline{f}:[2p]\rightarrow[k]$ whose level sets are the supports of the cycles of $\overline{\varsigma}$. Hence, to build a pair $(\overline{f},\lambda)$ meeting our requirements, we have at most $k^{4\delta}{2p \choose 2}\cdots{2(p-\delta+1) \choose 2} \times k^{p-2\delta}\,\mathrm{Cat}_{p-2\delta}$ possibilities. Indeed, for the $2\delta$ first disjoint transpositions we can use the trivial upper bound that would consist in picking them, as well as the values of $\overline{f}$ on them, completely arbitrarily, while for the $p-2\delta$ last ones we have to impose that they are non-crossing and that $\overline{f}$ takes only one value on a given transposition. Now, we know from the proof of Lemma \ref{lemma:number-pairings-defect} that ${2p \choose 2}\cdots{2(p-\delta+1) \choose 2}\mathrm{Cat}_{p-2\delta}\leq(p^4/4)^{\delta}\mathrm{Cat}_p$. So we get as announced that there are less than $k^{p+2\delta}\,\mathrm{Cat}_p(p^4/4)^{\delta}$ pairs $(f,\lambda)$ satisfying condition \eqref{eq:b}.
\end{proof}

\subsection{One needed adaptation: bounding the number of permutations of $p$ elements which are not on the geodesic path between the identity and a product of (few) cycles}

The proof of Proposition \ref{prop:wishart-infty} requires a statement analogous to the one appearing in Lemma \ref{lemma:number-functions,pairings-defect}, but for permutations instead of pairings. In order to derive it, we need first to explicit a bit how an element of $\mathfrak{S}(p)$ can be put in one-to-one correspondence with an element of $\mathfrak{P}^{(2)}(2p)$ whose pairs are all composed of one even integer and one odd integer.

To a full cycle $c=(i_l\,\ldots\,i_1)$ on $\{1,\ldots,l\}$ we associate the pairing $\lambda_c=(2i_1\,2i_2-1)\cdots(2i_l\,2i_1-1)$ on $\{1,\ldots,2l\}$. The reverse operation is obtained by collapsing the two elements $2i$ and $2i-1$ to a single element $i$ for each $1\leq i\leq p$. Then as expected, we associate to a general permutation $\alpha=c_1\cdots c_m\in\mathfrak{S}(p)$ the pairing $\lambda_{\alpha}=\lambda_{c_1}\cdots\lambda_{c_m}\in\mathfrak{P}^{(2)}(2p)$.

Observe that, denoting by $\gamma$ the canonical full cycle either on $\{1,\ldots,p\}$ or on $\{1,\ldots,2p\}$, we have
\begin{equation} \label{eq:permutation-pairing} \forall\ \alpha\in\mathfrak{S}(p),\ \sharp(\alpha) + \sharp(\gamma^{-1}\alpha) = \sharp(\gamma^{-1}\lambda_{\alpha}). \end{equation}
Indeed, the cycles of $\gamma^{-1}\lambda_{\alpha}$ are precisely cycles of the form $(2i_i\,\ldots\,2i_l)$ for $(i_l\,\ldots\,i_1)$ a cycle of $\alpha$ (supported on even integers) and of the form $(2i_{l'}-1\,\ldots\,2i_1-1)$ for $(i_{l'}\,\ldots\,i_1)$ a cycle of $\gamma^{-1}\alpha$ (supported on odd integers). So what equation \eqref{eq:permutation-pairing} shows is that the elements of $\mathfrak{S}(p)$ having a given geodesic defect are in bijection with the elements of $\mathfrak{P}^{(2)}(2p)$ with even-odd pairs only and having the same geodesic defect (between $\id$ and $\gamma$ in both cases). In particular, we recover the well-known bijection between $NC(p)$ and $NC^{(2)}(2p)$ (because a non-crossing pairing is necessarily composed of even-odd pairs only).

Next, for any function $g$, either from $[p]$ to $[k]$ or from $[2p]$ to $[k]$, we will denote by $\gamma_g$ the permutation, either on $\{1,\ldots,p\}$ or on $\{1,\ldots,2p\}$, which is the product of the canonical full cycles on the level sets of $g$. For any function $f:[p]\rightarrow[k]$, we define the function $\widetilde{f}:[2p]\rightarrow[k]$ by $\widetilde{f}(2i)=\widetilde{f}(2i-1)=f(i)$ for each $1\leq i\leq p$. It is then easy to see that we have more generally
\[ \forall\ f:[p]\rightarrow[k],\ \forall\ \alpha\in\mathfrak{S}(p),\ \sharp(\alpha) + \sharp(\gamma_f^{-1}\alpha) = \sharp(\gamma_{\widetilde{f}}^{-1}\lambda_{\alpha}). \]

This simple observation will allow us to derive, as a slight adaptation of Lemma \ref{lemma:number-functions,pairings-defect}, a corresponding estimate for permutations instead of pairings.

\begin{lemma} \label{lemma:number-functions,permutations-defect}
Let $p\in\N$.
For any $f:[p]\rightarrow[k]$, any $0\leq\delta\leq \left\lfloor \left(p+|\im(f)|\right)/2\right\rfloor$, and any $1\leq m\leq p-2\delta$, define the set of permutations which are composed of $m$ disjoint cycles and which have a defect $2\delta$ of being on the geodesics between $\id$ and $\gamma_f$ (the product of the canonical full cycles on each of the $|\im(f)|$ level sets of $f$) as
\[ \mathfrak{S}_{f,\delta,m}(p) = \{\alpha\in\mathfrak{S}(p) \st \sharp(\alpha)=m\ \text{and}\ \sharp(\gamma_f^{-1}\alpha)+\sharp(\alpha)=p+|\im(f)|-2\delta\}. \]
Then, for any $0\leq\delta\leq\lfloor p/2\rfloor$ and any $1\leq m\leq p-2\delta$, we have the upper bound
\[ \left|\big\{(f,\alpha) \st \alpha\in\mathfrak{S}_{f,\delta,m}(p) \big\} \right| \leq \left(4k^4p^4\right)^{\delta}\sum_{\epsilon=0}^{2\delta}k^{m-\epsilon}\mathrm{Nar}_p^{m-\epsilon}. \]
\end{lemma}

\begin{proof}
We just observed that, for any $0\leq\delta\leq\lfloor p/2\rfloor$ and $1\leq m\leq p-2\delta$, the following equivalence holds
\begin{equation} \label{eq:c} \alpha\in\mathfrak{S}_{f,\delta,m}(p)\ \Leftrightarrow\ \sharp(\alpha)=m\ \text{and}\ \lambda_{\alpha}\in\mathfrak{P}^{(2)}_{\widetilde{f},\delta}, \end{equation}
where $\mathfrak{P}^{(2)}_{\widetilde{f},\delta}$ denotes the set of pairings having a defect $2\delta$ of lying on the geodesics between $\id$ and $\gamma_{\widetilde{f}}$, as defined in Lemma \ref{lemma:number-functions,pairings-defect}.

In particular, $\alpha\in\mathfrak{S}_{f,0,m}(p)\ \Leftrightarrow\ \sharp(\alpha)=m\ \text{and}\ \lambda_{\alpha}\in\mathfrak{P}^{(2)}_{\widetilde{f},0}$, and we know that there are precisely $k^m\mathrm{Nar}_p^m$ possibilities to build a pair $(f,\alpha)$ satisfying this condition (because the latter holds if and only if the three constraints $\sharp(\alpha)=m$, $\alpha\in NC(p)$ and $f\circ\alpha=f$ are fulfilled).

For the case $1\leq\delta\leq\lfloor p/2\rfloor$, we will mimic the proof of Lemma \ref{lemma:number-functions,pairings-defect}. So let $(f,\alpha)$ be such that $\alpha\in\mathfrak{S}_{f,\delta,m}(p)$ and assume without loss of generality that the transpositions $(i_1\,j_1),\ldots,(i_p\,j_p)$ in $\lambda_{\alpha}$ are ordered so that $\lambda_{\alpha}$ is under the canonical form \eqref{eq:canonical} for $\gamma_{\widetilde{f}}$. This means that the partial pairing $(i_{2\delta+1}\,j_{2\delta+1})\cdots(i_p\,j_p)$ is on the geodesics between $\id$ and some $\overline{\varsigma}$ with $|\im(f)|$ cycles, and the number of such partial pairings is the same as the number of partial pairings being on the geodesics between $\id$ and some $\gamma_{\overline{f}}$ with $|\im(\overline{f})|=|\im(f)|$. Hence, to count how many ways there are of constructing what happens on $\{i_1,j_1,\ldots,i_{2\delta},j_{2\delta}\}$, we have the trivial upper bound that would arise if picking the $2\delta$ first transpositions in $\lambda_{\alpha}$, as well as the values of $\overline{f}$ on them, completely arbitrarily. This yields a number of possibilities of at most $k^{4\delta}{2p \choose 2}\cdots{2(p-\delta+1) \choose 2}$. While on $\{i_{2\delta+1},j_{2\delta+1},\ldots,i_{2p},j_{2p}\}$, we have to impose that the $p-2\delta$ last transpositions in $\lambda_{\alpha}$ are non-crossing, and that, when collapsed into a permutation of $p-2\delta$ elements, the latter has between $m-2\delta$ and $m$ cycles and the function $\overline{f}$ takes only one value on each of them. This leaves us with a number of possibilities of at most $\sum_{\epsilon=0}^{2\delta}k^{m-\epsilon}\mathrm{Nar}_{p-2\delta}^{m-\epsilon}$. Putting everything together, we see that the number of pairs $(f,\alpha)$ satisfying condition \eqref{eq:c} is less than
\[ k^{4\delta}{2p \choose 2}\cdots{2(p-\delta+1) \choose 2}\sum_{\epsilon=0}^{2\delta}k^{m-\epsilon}\mathrm{Nar}_{p-2\delta}^{m-\epsilon} \leq k^{4\delta}\left(2p^2\right)^{2\delta}\sum_{\epsilon=0}^{2\delta}k^{m-\epsilon}\mathrm{Nar}_p^{m-\epsilon}, \]
which is exactly what we wanted to show.
\end{proof}

\begin{remark}
The upper bound we established in Lemma \ref{lemma:number-functions,permutations-defect} is probably far from optimal (e.g.~it is likely that the exponent $4\delta$ in the polynomial pre-factor in $k$ and $p$ can be improved). But this does not really matter for our specific goal. Nonetheless, in the special case of non-geodesic permutations between $\id$ and $\gamma$ on $\{1,\ldots,p\}$, it is in fact quite easy to obtain an upper bound which scales as $p^{3\delta}$ for the ratio between the number of $2\delta$ non-geodesic permutations with a given number of cycles and the number of geodesic permutations with the same number of cycles. We present the result in Lemma \ref{lemma:number-permutations-defect} below, the problem being that the proof method does not seem to generalize so straightforwardly to the case that we truly need, that is the one of non-geodesic permutations between $\id$ and $\gamma_f$.

Note also that very similar looking upper bounds had previously been derived regarding the cardinality of the set $\mathfrak{S}_{\delta}(p) = \{\alpha\in\mathfrak{S}(p) \st \sharp(\gamma^{-1}\alpha)+\sharp(\alpha)=p+1-2\delta\}$, which is the union of the sets $\mathfrak{S}_{\delta,m}(p)$ defined in Lemma \ref{lemma:number-permutations-defect}, for $1\leq m\leq p$. In particular, it was established in \cite{Montanaro}, Lemma 12, that for any $0\leq\delta\leq\lfloor p/2\rfloor$,
\[ \big|\mathfrak{S}_{\delta}(p)\big| \leq \big|\mathfrak{S}_{0}(p)\big|\,p^{3\delta} = \mathrm{Cat}_p \,p^{3\delta}.\]
However, this is definitely even less enough for our purpose: the latter really requires an upper bound on the number of permutations which have a given defect and a given number of cycles in terms of the number of permutations which have no defect and the same (or a related) number of cycles.

On the other hand, one may have hoped for a stronger result than these simply counting ones. For instance something like
\[ d(\id,\alpha)+d(\alpha,\gamma)=d(\id,\gamma)+2\delta\ \Rightarrow\ \exists\ \alpha' \st d(\alpha,\alpha')=2\delta'\ \text{and}\ d(\id,\alpha')+d(\alpha',\gamma)=d(\id,\gamma), \]
with $\delta'\leq \theta\delta$ and with the mapping $\phi:\alpha\mapsto\alpha'$ satisfying $\left|\phi^{-1}(\alpha')\right|\leq p^{\kappa\delta}$, for some coefficients $\theta,\kappa$. However, determining whether this kind of statement holds or not seems to remain an open question.
\end{remark}

\begin{lemma} \label{lemma:number-permutations-defect}
Let $p\in\N$ and denote by $\gamma$ the canonical full cycle on $\{1,\ldots,p\}$. For any $0\leq\delta\leq\lfloor p/2\rfloor$ and $1\leq m\leq p-2\delta$, define the set of permutations which are composed of $m$ disjoint cycles and which are $2\delta$-away from the geodesics between $\id$ and $\gamma$ as
\[ \mathfrak{S}_{\delta,m}(p) = \{\alpha\in\mathfrak{S}(p) \st \sharp(\alpha)=m\ \text{and}\ \sharp(\gamma^{-1}\alpha)+\sharp(\alpha)=p+1-2\delta\}. \]
Then, the cardinality of $\mathfrak{S}_{\delta,m}(p)$ is upper bounded in terms of the cardinality of $\mathfrak{S}_{0,m}(p)$ as
\[ \big|\mathfrak{S}_{\delta,m}(p)\big| \leq \big|\mathfrak{S}_{0,m}(p)\big|\left(\frac{p^3}{2}\right)^{\delta}. \]
\end{lemma}

\begin{proof} Let $p\in\N$ and $1\leq m\leq p$. We know from \cite{GS}, Theorems 4.1 and 4.2, that there exist polynomials $P_{q}$ of degree $q$, for $0\leq q\leq \lfloor p/2\rfloor$, such that for any $0\leq\delta\leq \lfloor(p-m)/2\rfloor$,
\begin{equation} \label{eq:S_delta,m} \big|\mathfrak{S}_{\delta,m}(p)\big| = \frac{p!}{2^{2\delta}(2\delta)!} {p+1-2\delta \choose m} \sum_{\epsilon=0}^{\delta} {p-1 \choose m-1+2\epsilon}P_{\epsilon}(m)P_{\delta-\epsilon}(p+1-m-2\delta) .\end{equation}
What is more, one can check from the explicit expression provided there for the polynomials $P_q$ that, for any $x\geq 0$, $P_q(x)\leq (2x)^q$. As a particular instance of equation \eqref{eq:S_delta,m}, we have
\[ \big|\mathfrak{S}_{0,m}(p)\big| = p!\, {p+1 \choose m}{p-1 \choose m-1}. \]
And as a consequence, we get by a brutal upper bounding that, for any $1\leq\delta\leq \lfloor(p-m)/2\rfloor$,
\begin{align*}
\frac{\big|\mathfrak{S}_{\delta,m}(p)\big|}{\big|\mathfrak{S}_{0,m}(p)\big|} = & \,\frac{1}{2^{2\delta}(2\delta)!}\prod_{i=0}^{m-1}\frac{p+1-2\delta-i}{p+1-i} \sum_{\epsilon=0}^{\delta}\prod_{j=0}{2\epsilon-1}\frac{p-m-j}{m+j} P_{\epsilon}(m)P_{\delta-\epsilon}(p+1-m-2\delta) \\
\leq  & \,\frac{1}{2^{2\delta}(2\delta)!} \sum_{\epsilon=0}^{\delta}\left(\frac{p-m}{m}\right)^{2\epsilon}\, 2^{\epsilon}\,m^{\epsilon}\, 2^{\delta-\epsilon}\,(p+1-m-2\delta)^{\delta-\epsilon} \\
\leq  & \,\frac{1}{2^{2\delta}(2\delta)!} \times (\delta+1)\,2^{\delta}\,p^{3\delta}\\
\leq & \,\left(\frac{p^3}{2}\right)^{\delta}.
\end{align*}
And this is precisely the claimed upper bound.
\end{proof}

\begin{remark} There is a close link between the problem we are concerned with and the one of finding tractable expressions for the so-called \textit{connection coefficients} of the symmetric group (the reader is referred e.g.~to \cite{GJ} for more on that topic). Closed formulas are actually known for the connection coefficients of $\mathfrak{S}(p)$, involving the characters of its irreducible representations. But unfortunately, they are not really handleable in there full generality. And it seems it is only in some specific cases that more manageable forms can been obtained (i.e.~in the first place as a sum of positive terms, so that one can see more easily what its order of magnitude is). The two situations which are well-understood are, on the one hand when the function $f$ is constant (which corresponds to the case where $\gamma_f$ is the canonical full cycle, and hence has a particularly simple cycle type, that is treated e.g.~in \cite{GS}), and on the other hand when the defect $2\delta$ is $0$ (which corresponds to the case of so-called \textit{top connection coefficients}).
\end{remark}

\section{Appendix H: Extra remarks on the convergence of the studied random matrix ensembles}
\label{appendix:convergence}

For any Hermitian $M$ on $\C^n$, we shall denote by $\lambda_1(M),\ldots,\lambda_n(M)\in\R$ its eigenvalues, and by $N_M$ its eigenvalue distribution, i.e.~the probability measure on $\R$ defined by
\[ N_M = \frac{1}{n}\sum_{i=1}^n\delta_{\lambda_i(M)}. \]
In words, for any $I\subset\R$, $N_M(I)$ is the proportion of eigenvalues of $M$ which belong to $I$.

\subsection{``Modified'' Wishart ensemble}

Fix $k\in\N$ and $c>0$. Then, for each $d\in\N$, let $W\sim\mathcal{W}_{d^2,cd^2}$ and define the random positive semidefinite matrix $W_d$ on $(\C^d)^{\otimes k+1}$ by
\begin{equation} \label{eq:W_d} W_d=\frac{1}{d^2}\sum_{j=1}^k\widetilde{W}(j). \end{equation}
Proposition \ref{prop:wishart-p} establishes that when $d\rightarrow+\infty$, the eigenvalue distribution of $W_d$ converges in moments towards a Mar\v{c}enko-Pastur distribution of parameter $ck$. But a stronger result actually holds, namely that there is convergence in probability of $N_{W_d}$ towards $\mu_{MP(ck)}$. What is meant is made precise in Theorem \ref{th:convergenceW} below.

\begin{theorem} \label{th:convergenceW}
For any $I\in\R$ and any $\e>0$,
\[ \underset{d\rightarrow+\infty}{\lim} \P_{W\sim\mathcal{W}_{d^2,cd^2}}\left(\left|N_{W_d}(I)-\mu_{MP(ck)}(I)\right|>\e\right) =0 ,\]
where the matrix $W_d$ is as defined in equation \eqref{eq:W_d}.
\end{theorem}

Theorem \ref{th:convergenceW} is a direct consequence of the estimate on the $p$-order moments $\mathbf{E}\, \mathrm{Tr}\left[\left(\sum_{j=1}^k\widetilde{W}_{\A\B^k}(j)\right)^p\right]$ from Proposition \ref{prop:wishart-p}, combined with the estimate on the $p$-order variances $\mathbf{Var}\, \mathrm{Tr}\left[\left(\sum_{j=1}^k\widetilde{W}_{\A\B^k}(j)\right)^p\right]$ from Proposition \ref{prop:wishart-p-var} below. The proof, which follows a quite standard procedure, may be found detailed in \cite{AGZ} and sketched in \cite{Aubrun1}.

\begin{proposition}\label{prop:wishart-p-var} Let $p\in\N$. For any constant $c>0$,
\[ \mathbf{Var}_{W_{\A\B}\sim\mathcal{W}_{d^2,cd^2}}\, \mathrm{Tr}\left[\left(\underset{j=1}{\overset{k}{\sum}}\widetilde{W}_{\A\B^k}(j)\right)^p\right] \underset{d\rightarrow+\infty}{=} o\left(d^{2p+k+1}\right). \]
\end{proposition}

\begin{proof}
Let $p\in\N$. We already know that $\left(\mathbf{E}\, \mathrm{Tr}\left[\left(\sum_{j=1}^k\widetilde{W}_{\A\B^k}(j)\right)^p\right]\right)^2 \sim_{d\rightarrow +\infty} \left(\mathrm{M}_{MP(ck)}^{(p)}d^{2p+k+1} \right)^2$ thanks to Proposition \ref{prop:wishart-p}. Consequently, the only thing that remains to be shown in order to establish Proposition \ref{prop:wishart-p-var} is that we also have $\mathbf{E}\,\left( \mathrm{Tr}\left[\left(\sum_{j=1}^k\widetilde{W}_{\A\B^k}(j)\right)^p\right]\right)^2 \sim_{d\rightarrow +\infty} \left(\mathrm{M}_{MP(ck)}^{(p)}d^{2p+k+1} \right)^2$. The combinatorics involved in the proof of the latter estimate is very similar to the one already appearing in the proof of the former. We will therefore skip some of the details here.

To begin with, let us fix a few additional notation. We define $\gamma_1=(p\,\ldots\,1)$ and $\gamma_2=(2p\,\ldots\,p+1)$ as the canonical full cycles on $\{1,\ldots,p\}$ and $\{p+1,\ldots,2p\}$ respectively. Also, for each functions $f_1:\{1,\ldots,p\}\rightarrow[k]$, $f_2:\{p+1,\ldots,2p\}\rightarrow[k]$, we define the function $f_{1,2}:[2p]\rightarrow[k]$ by $f_{1,2}=f_1$ on $\{1,\ldots,p\}$ and $f_{1,2}=f_2$ on $\{p+1,\ldots,2p\}$. Then, by a slight generalization of Proposition \ref{prop:geodesic-alpha_f} we have that, for any $\alpha\in\mathfrak{S}(2p)$,
\[ \sharp((\hat{\gamma}_1\hat{\gamma}_2)^{-1}\hat{\alpha}_{f_{1,2}})=\sharp((\gamma_{1\,f_1}\gamma_{2\,f_2})^{-1}\alpha)+ 2k -|\im(f_1)|-|\im(f_2)|. \]
We can thus derive from the graphical calculus for Wishart matrices (in complete analogy to the way formula \eqref{eq:p-moment} was obtained) that, for any $d\in\N$,
\begin{equation} \label{eq:varp} \mathbf{E}_{W_{\A\B}\sim\mathcal{W}_{d^2,cd^2}} \left(\mathrm{Tr}\left[\left(\sum_{j=1}^k\widetilde{W}_{\A\B^k}(j)\right)^p\right]\right)^2 = \sum_{\substack{f_1:\{1,\ldots,p\}\rightarrow[k]\\f_2:\{p+1,\ldots,2p\}\rightarrow[k]}}\sum_{\alpha\in\mathfrak{S}(2p)} c^{\sharp(\alpha)} d^{n(\alpha,f_1,f_2)}, \end{equation}
where for each $\alpha\in\mathfrak{S}(2p)$ and $f_1:\{1,\ldots,p\}\rightarrow[k]$, $f_2:\{p+1,\ldots,2p\}\rightarrow[k]$,
\[ n(\alpha,f_1,f_2)= 2\sharp(\alpha)+ \sharp((\gamma_1\gamma_2)^{-1}\alpha)+ \sharp((\gamma_{1\,f_1}\gamma_{2\,f_2})^{-1}\alpha)+2k-|\im(f_1)|-|\im(f_2)|. \]
Yet, by Lemma \ref{lemma:cycles-transpositions} and equation \eqref{eq:geodesic''} in Lemma \ref{lemma:distance}, we get: First, for any $\alpha\in\mathfrak{S}(2p)$,
\begin{equation}
\label{eq:alpha-var}
\sharp(\alpha)+\sharp((\gamma_1\gamma_2)^{-1}\alpha)= 4p-\left(|\alpha|+|(\gamma_1\gamma_2)^{-1}\alpha|\right)\leq 4p-|\gamma_1\gamma_2|= 2p+\sharp(\gamma_1\gamma_2) =2p+2,
\end{equation}
with equality if and only if $\alpha=\alpha_1\alpha_2$ where $\alpha_1\in NC(\{1,\ldots,p\})$, $\alpha_2\in NC(\{p+1,\ldots,2p\})$. And second, for any $\alpha\in\mathfrak{S}(2p)$ and any $f_1:\{1,\ldots,p\}\rightarrow[k]$, $f_2:\{p+1,\ldots,2p\}\rightarrow[k]$,
\begin{equation}
\label{eq:alpha_f'-var}
\sharp(\alpha)+\sharp((\gamma_{1\,f_1}\gamma_{2\,f_2})^{-1}\alpha)\leq 2p+\sharp(\gamma_{1\,f_1}\gamma_{2\,f_2}) =2p+|\im(f_1)|+|\im(f_2)|,
\end{equation}
with equality if and only if $\alpha=\alpha_1\alpha_2$ where $\alpha_1\in NC(\{1,\ldots,p\})$ and $f_1\circ\alpha_1=f_1$, $\alpha_2\in NC(\{p+1,\ldots,2p\})$ and $f_2\circ\alpha_2=f_2$. So putting equations \eqref{eq:alpha-var} and \eqref{eq:alpha_f'-var} together, we get in the end that for any $\alpha\in\mathfrak{S}(2p)$ and $f_1:\{1,\ldots,p\}\rightarrow[k]$, $f_2:\{p+1,\ldots,2p\}\rightarrow[k]$,
\[ n(\alpha,f_1,f_2) \leq 4p+2k+2, \]
with equality if and only if $\alpha=\alpha_1\alpha_2$ where $\alpha_1\in NC(\{1,\ldots,p\})$ and $f_1\circ\alpha_1=f_1$, $\alpha_2\in NC(\{p+1,\ldots,2p\})$ and $f_2\circ\alpha_2=f_2$.

We thus get that, asymptotically, the dominant term in formula \eqref{eq:varp} factorizes as
\begin{align*}
\mathbf{E}\, \left(\mathrm{Tr}\left[\left(\sum_{j=1}^k\widetilde{W}_{\A\B^k}(j)\right)^p\right]\right)^2 & \underset{d\rightarrow+\infty}{\sim} d^{4p+2k+2} \sum_{\substack{\alpha_1\in NC(\{1,\ldots,p\})\\ \alpha_2\in NC(\{p+1,\ldots,2p\})}} \sum_{\substack{f_1:\{1,\ldots,p\}\rightarrow[k],\,f_1\circ\alpha_1=f_1 \\f_2:\{p+1,\ldots,2p\}\rightarrow[k],\,f_2\circ\alpha_2=f_2}}  c^{\sharp(\alpha_1\alpha_2)}\\
& \underset{d\rightarrow+\infty}{\sim} \left( d^{2p+k+1}\sum_{\alpha\in NC(p)}\sum_{\underset{f\circ\alpha=f} {f:[p]\rightarrow[k]}}  c^{\sharp(\alpha)} \right)^2,
\end{align*}
where the last equality is simply because $\sharp(\alpha_1\alpha_2)=\sharp(\alpha_1)+\sharp(\alpha_2)$. And hence,
\[ \mathbf{E}_{W_{\A\B}\sim\mathcal{W}_{d^2,cd^2}} \left(\mathrm{Tr}\left[\left(\sum_{j=1}^k\widetilde{W}_{\A\B^k}(j)\right)^p\right]\right)^2 \underset{d\rightarrow+\infty}{\sim} \left(d^{2p+k+1}\sum_{\alpha\in NC(p)}(ck)^{\sharp(\alpha)}\right)^2 = \left(d^{2p+k+1}\mathrm{M}_{MP(ck)}^{(p)}\right)^2, \]
which is exactly what we needed to conclude the proof.
\end{proof}

\begin{figure}[h] \caption{Spectral distribution of $\left(W_{\A\B_1}\otimes\Id_{\B_2}+W_{\A\B_2}\otimes\Id_{\B_1}\right)/d^2$, for $W_{\A\B}\sim\mathcal{W}_{d^2,d^2}$ vs Mar\v{c}enko-Pastur distribution of parameter $2$.}
\label{fig:wishart}
\begin{center}
\includegraphics[width=12cm]{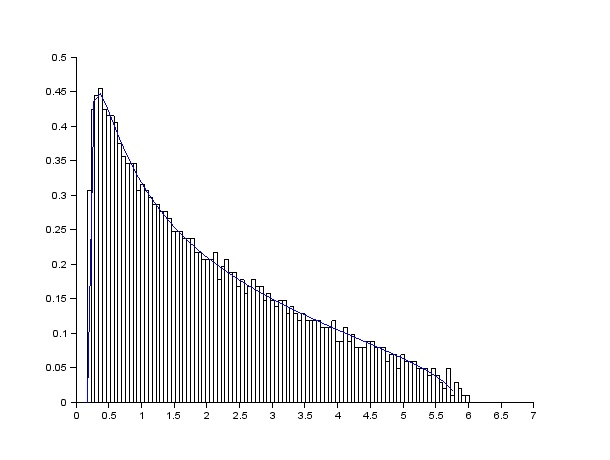}
\end{center}
\end{figure}

Let us illustrate the result stated in Theorem \ref{th:convergenceW} in the simplest case of $2$-extendibility and uniformly distributed mixed states. In Figure \ref{fig:wishart}, the spectral distribution of $W_d=\left(W_{\A\B_1}\otimes\Id_{\B_2}+W_{\A\B_2}\otimes\Id_{\B_1}\right)/d^2$, for $W_{\A\B}\sim\mathcal{W}_{d^2,d^2}$, and a Mar\v{c}enko-Pastur distribution of parameter $2$ are plotted together. The empirical eigenvalue histogram is done in dimension $d=12$, from $100$ repetitions.

\subsection{``Modified'' GUE ensemble}

Fix $k\in\N$. Then, for each $d\in\N$, let $G\sim GUE(d^2)$ and define the random Hermitian matrix $G_d$ on $(\C^d)^{\otimes k+1}$ by
\begin{equation} \label{eq:G_d} G_d=\frac{1}{d}\sum_{j=1}^k\widetilde{G}(j). \end{equation}
In complete analogy to what was explained in the case of Wishart matrices, Proposition \ref{prop:gaussian-p} establishes that when $d\rightarrow+\infty$, the eigenvalue distribution of $G_d$ converges in moments towards a centered semicircular distribution of parameter $k$. But here again, there is in fact convergence in probability of $N_{G_d}$ towards $\mu_{SC(k)}$, which is made precise in Theorem \ref{th:convergenceG} below.

\begin{theorem} \label{th:convergenceG}
For any $I\in\R$ and any $\e>0$,
\[ \underset{d\rightarrow+\infty}{\lim} \P_{G\sim GUE(d^2)}\left(\left|N_{G_d}(I)-\mu_{SC(k)}(I)\right|>\e\right) =0, \]
where the matrix $G_d$ is as defined in equation \eqref{eq:G_d}.
\end{theorem}

As already explained in the Wishart case, this follows directly from the moment's estimate in Proposition \ref{prop:gaussian-p}, together with the variance's estimate, for all $p\in\N$,
\begin{equation} \label{eq:gaussian-p-var} \mathbf{Var}_{G_{\A\B}\sim GUE(d^2)}\, \mathrm{Tr}\left[ \left( \underset{j=1}{\overset{k}{\sum}}\widetilde{G}_{\A\B^k}(j) \right)^{2p} \right] \underset{d\rightarrow+\infty}{=} o\left(d^{2p+k+1}\right) .\end{equation}
The proof follows the exact same lines as the one of Proposition \ref{prop:wishart-p-var} and is not repeated here.

\begin{figure}[h] \caption{Spectral distribution of $\left(G_{\A\B_1}\otimes\Id_{\B_2}+G_{\A\B_2}\otimes\Id_{\B_1}\right)/d^2$, for $G_{\A\B}\sim GUE(d^2)$ vs Centered semicircular distribution of parameter $2$.}
\label{fig:gaussian}
\begin{center}
\includegraphics[width=12cm]{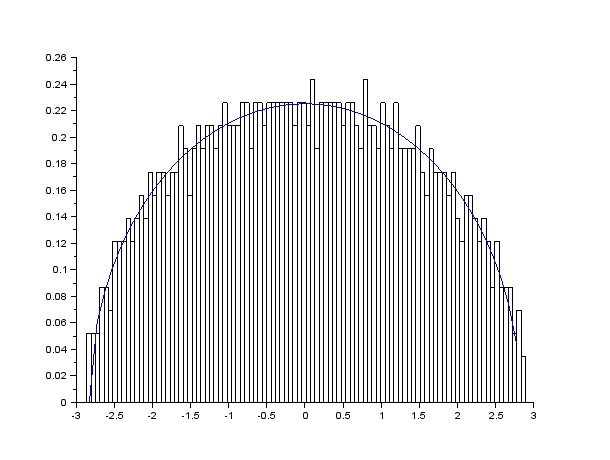}
\end{center}
\end{figure}

Let us illustrate the result stated in Theorem \ref{th:convergenceG} in the simplest case of $2$-extendibility. In Figure \ref{fig:gaussian}, the spectral distribution of $G_d=\left(G_{\A\B_1}\otimes\Id_{\B_2}+G_{\A\B_2}\otimes\Id_{\B_1}\right)/d^2$, for $G_{\A\B}\sim GUE(d^2)$, and a centered semicircular distribution of parameter $2$ are plotted together. The empirical eigenvalue histogram is done in dimension $d=10$, from $100$ repetitions.

\begin{remark}
We may in fact say even more on the convergence of the random matrix sequences $(W_d)_{d\in\N}$ and $(G_d)_{d\in\N}$ defined by equations \eqref{eq:W_d} and \eqref{eq:G_d} respectively. Namely,
\[ N_{W_d}\overset{a.s.}{\underset{d\rightarrow+\infty}{\rightarrow}} \mu_{MP(ck)}\ \text{and}\  N_{G_d}\overset{a.s.}{\underset{d\rightarrow+\infty}{\rightarrow}} \mu_{SC(k)}.\]
To establish this almost sure convergence result, the only thing that has to be verified is that, for any $p\in\N$, the series of variances
\begin{equation} \label{eq:summable} \sum_{d=1}^{+\infty} \mathbf{Var}\left[\frac{1}{d^{k+1}}\mathrm{Tr}\left(W_d^p\right)\right]\ \text{and} \ \sum_{d=1}^{+\infty} \mathbf{Var}\left[\frac{1}{d^{k+1}}\mathrm{Tr}\left(G_d^{2p}\right)\right] \end{equation}
are summable. Indeed, almost sure convergence will then automatically follow from a standard application of the Chebyshev inequality and the Borel--Cantelli lemma. And condition \eqref{eq:summable} actually holds, as a consequence of the fact that, for any $p\in\N$,
\[ \mathbf{Var}\left[\frac{1}{d^{k+1}}\mathrm{Tr}\left(W_d^p\right)\right]= O\left(d^{-2}\right)\ \text{and}\ \mathbf{Var}\left[\frac{1}{d^{k+1}}\mathrm{Tr}\left(G_d^{2p}\right)\right]= O\left(d^{-2}\right). \]
\end{remark}

\subsection{Asymptotic freeness of certain Gaussian matrices}

Let us fix a few definitions and notation. Given $n\in\N$ and $[\Omega,P]$ a classical probability space, we define the free probability space $\left[\mathcal{M}_n(L^{\infty}[\Omega,P]),\varphi_n\right]$, where $\mathcal{M}_n(L^{\infty}[\Omega,P])$ is the set of $n\times n$ matrices with entries in $L^{\infty}[\Omega,P]$ and $\varphi_n(\cdot)=\E\tr(\cdot)/n$ is the normalized trace function on $\mathcal{M}_n(L^{\infty}[\Omega,P])$. The two particular examples we shall focus on in the sequel are the ones we have already been extensively dealing with, namely GUE and Wishart matrices.


\begin{lemma} \label{lem:phi-G}
Given two finite-dimensional Hilbert spaces $\A\equiv\C^{d_\A}$, $\B\equiv\C^{d_\B}$, and $G$ a random GUE matrix on $\A\otimes\B$, we define the following random matrices on $\A\otimes\B_1\otimes\B_2$:
\[ \widetilde{G}_{1}=\frac{1}{\sqrt{d_\A d_\B}}\,G_{\A\B_1}\otimes\Id_{\B_2}\ \text{and}\ \widetilde{G}_{2}=\frac{1}{\sqrt{d_\A d_\B}}\,G_{\A\B_2}\otimes\Id_{\B_1}. \]
Then, for any $p\in\N$ and any function $f:[2p]\rightarrow[2]$,
\begin{equation} \label{eq:phi-G} \lim_{d_\A\leq d_\B\rightarrow+\infty} \varphi_{d_\A d_\B^2}\left(\widetilde{G}_{f(1)}\ldots\widetilde{G}_{f(2p)}\right) = \left|\left\{\lambda\in NC^{(2)}(2p) \st f\circ\lambda =f \right\} \right|. \end{equation}
\end{lemma}

\begin{proof}
We know from the proof of Proposition 2.3 (and using the same notation as those employed there) that
\[ \varphi_{d_\A d_\B^2}\left(\widetilde{G}_{f(1)}\ldots\widetilde{G}_{f(2p)}\right) = \sum_{\lambda\in\mathfrak{P}^{(2)}(2p)} d_\A^{\sharp(\gamma^{-1}\lambda)-p-1}d_B^{\sharp(\gamma_f^{-1}\lambda)-p-|\im(f)|}. \]
Now, as explained there as well, for any $\lambda\in\mathfrak{P}^{(2)}(2p)$, on the one hand $\sharp(\gamma^{-1}\lambda)\leq p+1$ with equality iff $\lambda\in NC^{(2)}(2p)$, and on the other hand $\sharp(\gamma_f^{-1}\lambda)\leq p+ |\im(f)|$ with equality iff $\lambda\in NC^{(2)}(2p)$ and $f\circ\lambda=f$. The asymptotic estimate \eqref{eq:phi-G} therefore immediately follows.
\end{proof}

\begin{theorem} \label{th:free-G}
Let $G_{\A\B}$ be a random GUE matrix on $\A\otimes\B$. Then, the random matrices $G_{\A\B_1}\otimes\Id_{\B_2}$ and $G_{\A\B_2}\otimes\Id_{\B_1}$ on $\A\otimes\B_1\otimes\B_2$ are asymptotically free.
\end{theorem}

\begin{proof}
Theorem \ref{th:free-G} is a direct consequence of Lemma \ref{lem:phi-G} (see e.g.~\cite{NS}, proof of Proposition 22.22, for an entirely analogous argument). Indeed, as $d_\A,d_\B\rightarrow+\infty$, the two empirical spectral distributions $\mu_{\widetilde{G}_{1}}$ and $\mu_{\widetilde{G}_{2}}$ both converge to the semicircular distribution with mean $0$ and variance $1$. And equation \eqref{eq:phi-G} is exactly the rule for computing mixed moments in two free such semicircular distributions (see e.g.~\cite{NS}, Lecture 12).
\end{proof}

\begin{lemma} \label{lem:phi-W}
Given two finite-dimensional Hilbert spaces $\A\equiv\C^{d_\A}$, $\B\equiv\C^{d_\B}$, and $W$ a random Wishart matrix on $\A\otimes\B$ with parameter $c d_\A d_\B\in\N$, we define the following random matrices on $\A\otimes\B_1\otimes\B_2$:
\[ \widetilde{W}_{1}=\frac{1}{d_\A d_\B}\,W_{\A\B_1}\otimes\Id_{\B_2}\ \text{and}\ \widetilde{W}_{2}=\frac{1}{d_\A d_\B}\,W_{\A\B_2}\otimes\Id_{\B_1}. \]
Then, for any $p\in\N$ and any function $f:[p]\rightarrow[2]$,
\begin{equation} \label{eq:phi-W} \lim_{d_\A\leq d_\B\rightarrow+\infty} \varphi_{d_\A d_\B^2} \left(\widetilde{W}_{f(1)}\ldots\widetilde{W}_{f(p)}\right) = \sum_{\underset{f\circ\alpha=f}{\alpha\in NC(p)}}c^{\sharp(\alpha)}. \end{equation}
\end{lemma}

\begin{proof}
We know from the proof of Proposition 6.2 (and using the same notation as those employed there) that
\[ \varphi_{d_\A d_\B^2} \left(\widetilde{W}_{f(1)}\ldots\widetilde{W}_{f(p)}\right) = \sum_{\alpha\in\mathfrak{S}(p)} c^{\sharp(\alpha)}d_\A^{\sharp(\alpha)+\sharp(\gamma^{-1}\alpha)-p-1}d_\B^{\sharp(\alpha)+\sharp(\gamma_f^{-1}\lambda)-p-|\im(f)|}. \]
Now, as explained there as well, for any $\alpha\in\mathfrak{S}(p)$, on the one hand $\sharp(\alpha)+\sharp(\gamma^{-1}\alpha)\leq p+1$ with equality iff $\alpha\in NC(p)$, and on the other hand $\sharp(\alpha)+\sharp(\gamma_f^{-1}\alpha)\leq p+ |\im(f)|$ with equality iff $\alpha\in NC(p)$ and $f\circ\alpha=f$. The asymptotic estimate \eqref{eq:phi-W} therefore immediately follows.
\end{proof}

\begin{theorem} \label{th:free-W}
Let $W_{\A\B}$ be a random Wishart matrix on $\A\otimes\B$ with parameter $c d_\A d_\B\in\N$. Then, the random matrices $W_{\A\B_1}\otimes\Id_{\B_2}$ and $W_{\A\B_2}\otimes\Id_{\B_1}$ on $\A\otimes\B_1\otimes\B_2$ are asymptotically free.
\end{theorem}

\begin{proof}
Theorem \ref{th:free-W} is a direct consequence of Lemma \ref{lem:phi-W} (see e.g.~\cite{NS}, proof of Proposition 22.22, for an entirely analogous argument). Indeed, as $d_\A,d_\B\rightarrow+\infty$, the two empirical spectral distributions $\mu_{\widetilde{W}_{1}}$ and $\mu_{\widetilde{W}_{2}}$ both converge to the Mar\v{c}enko-Pastur distribution with parameter $c$. And equation \eqref{eq:phi-W} is exactly the rule for computing mixed moments in two free such Mar\v{c}enko-Pastur distributions (see e.g.~\cite{NS}, Lectures 12 and 13).
\end{proof}


\part{Making use of permutation-symmetry to tackle multiplicativity issues}
\label{part:symmetry}

Chapter \ref{chap:k-extendibility} in the previous part already made it quite clear that symmetries play an important role in quantum information theory. This part stands up much more ardently for such claim.

\smallskip

It is an ubiquitous issue in quantum information theory (but in fact in many other fields as well) to determine whether certain quantities have a multiplicative/additive behaviour. Indeed, a situation where perfect multiplicativity/additivity holds can usually be interpreted as follows: given a device that enables accomplishing a task with a certain performance, there is no way of combining two copies of this device which would allow performing better than when using them independently. In most cases though, this is not what happens: clever use of correlations does help. This assertion is actually even more true in the quantum than in the classical world: many quantities which are additive in classical information theory (and whose quantum analogues were therefore initially conjectured to be additive as well) have in fact been proved to fail additivity in quantum information theory. This is precisely what makes asymptotic performances much harder to quantify than single-copy ones. However, whenever the study of a multi-copy scenario can be reduced, in some manner or another, to that of an i.i.d.~one, then the analysis becomes easy again. This is exactly the motivation behind de Finetti type statements, all of them having in common to make the most of the permutation-symmetry of the problem under consideration.

\smallskip

In Chapter \ref{chap:deFinetti} we first of all draw the outline of a very flexible de Finetti reduction. What we mean is that the latter has a broad scope of application (e.g.~both quantum states and classical probability distributions) and that it allows keeping track of additional information that one may have on the considered object, apart from its permutation-symmetry. It is especially well-suited to the case where this extra knowledge consists of linear constraints that the object satisfies, one instance of which is extensively studied in Chapter \ref{chap:SNOS}. It may however still be useful in the case of convex constraints. In particular, it permits to derive the following connection: given any convex set of states $\mathcal{K}$, the exponential decay under tensoring of the maximum fidelity function $F(\cdot,\mathcal{K})$ implies that of the support function $h_{\mathcal{K}}(\cdot)$, and vice versa. We show-case this link on the example of the set of separable states. The latter is of prime importance because of its implications in a multitude of fields, most notably quantum computing and quantum Shannon theory. In this specific case, weak multiplicativity under tensoring does hold, but with a dependence on the ambient dimensions, and we do not know whether or not it could be removed. Finally, we examine how our techniques can be extended to the infinite-dimensional setting, in order to prove e.g.~security of continuous variable quantum key distribution against general attacks, under assumptions as minimal as possible.

\smallskip

Chapter \ref{chap:SNOS} entirely consists of one major application of the de Finetti reduction established in Chapter \ref{chap:deFinetti}, namely to the study of the parallel repetition of multi-player non-local games. More precisely, we are interested in the following problem: if players sharing certain correlations are not able to win a given game with probability $1$, does there probability of winning $n$ instances of that game played in parallel decrease exponentially to $0$ (with $n$), and if so at which rate? We solve this question in (almost) full generality in the case where the only limitation which is assumed on the physical power of the players is that they cannot signal information instantaneously from one another. For that, we introduce the notion of sub-no-signalling correlations between players. We then show that, if players sharing such correlations have a probability below $1-\delta$ of winning one instance of a game, then they have an exponentially decaying probability of winning a fraction above $1-\delta$ of $n$ instances of this game played in parallel. This result translates into the analogous one for players sharing no-signalling correlations (a more commonly studied set of allowed strategies) in two cases: when the game only involves two players, or when the game is such that all potential queries to the players have a non-zero probability of being actually asked.

\smallskip

One major question which remains open from that point is whether the flexible constrained de Finetti reduction of Chapter \ref{chap:deFinetti} could be used to study the parallel repetition problem, not only for no-signalling players as in Chapter \ref{chap:SNOS}, but also for quantum players (i.e.~players whose strategy is dictated by the outcomes of local measurements which they perform on a shared entangled state). Let us enlarge a bit on this issue. The standard proof technique to tackle parallel repetition, in the quantum case but also, originally, in the classical and no-signalling cases, consists in iteratively assuming that the players have won a given instance of the game and then studying how this affects their winning probability in the others. Hence, if you can show that, conditioned on the event ``the players have already won $k$ instances of the game'', the probability is high that they lose in at least $1$, resp.~most, of the $n-k$ remaining instances, you get exponential decay of the probability of winning all, resp.~a fraction above the game value of, $n$ instances of the game played in parallel. The main drawback of this approach is probably its ``locality'', which makes it not so straightforward and not so easily generalizable to more than $2$ players as the more ``global'' de Finetti approach. That is why finding a way of attacking the quantum parallel repetition problem via de Finetti reductions would be desirable. One obstacle seems however to be that there are certain steps from the standard route which remain unavoidable. One of them (which is actually crucial as well in Chapter \ref{chap:deFinetti}, when studying exponential decay under tensoring of support functions of quantum state sets) is some kind of reconstruction step. Phrased informally: you have to be able to say that, if your strategy (or quantum state) almost satisfies the constraints defining your set of interest, then there must exist a strategy (or quantum state) which exactly satisfies them and which is not too far away from it. The hope that, more often than not, this quite natural expectancy will turn out to be true is at the heart of our whole constrained de Finetti reduction philosophy. Nonetheless, it is not always possible (or at least so easy) to get nice quantitative versions of this intuition.

\smallskip

Let us bring this discussion to a close by connecting it with more convex geometry orientated considerations. What we are always doing in this part is asking how a sequence of convex sets under examination $\{\mathcal{K}_n,\ n\in\N\}$ compares to the sequence of projective tensor power sets $\{\mathcal{K}_1^{\hat{\otimes} n},\ n\in\N\}$. Specifically, what we usually want to know is how much bigger than $\mathcal{K}_1^{\hat{\otimes} n}$ is our considered $\mathcal{K}_n$ in a given tensor power direction $u^{\otimes n}$. And in the cases where giving an answer valid for any such $u^{\otimes n}$ looks out of reach, attempting to still say what happens for most of them (in a way to be defined) could be something interesting to explore.

\newpage
\textbf{\LARGE{Part IV -- Table of contents}}
\parttoc

\chapter{Flexible constrained de Finetti reductions and applications}
\chaptermark{Flexible constrained de Finetti reductions and applications}
\label{chap:deFinetti}

\textsf{Based on ``Flexible constrained de Finetti reductions and applications'', in collaboration with A. Winter \cite{L-W}.}

\bigskip

De Finetti theorems show how sufficiently exchangeable states are well-approximated by convex combinations of i.i.d.~states. Recently,
it was shown that in many quantum information applications a more relaxed \emph{de Finetti reduction} (i.e.~only a matrix inequality
between the symmetric state and one of de Finetti form) is enough, and that it leads to more concise and elegant arguments.

Here we show several uses and general flexible applicability of a \emph{constrained de Finetti reduction} in quantum information theory, which was recently discovered by Duan, Severini and Winter. In particular we show that the technique can accommodate other symmetries commuting with the permutation action, and permutation-invariant linear constraints.
We then demonstrate that, in some cases, it is also fruitful with convex constraints, in particular separability in a bipartite setting. This is a constraint particularly interesting in the context of the complexity class $\mathrm{QMA}(2)$ of interactive quantum Merlin-Arthur games with unentangled provers, and our results relate to the soundness gap amplification of $\mathrm{QMA}(2)$ protocols by parallel repetition. It is also relevant for grasping the regularization of certain entropic channel parameters.
Finally, we explore an extension to infinite-dimensional systems, which usually pose inherent problems to de Finetti techniques in the quantum case.

\section{Introduction}
\label{sec:intro}
The main motivation behind all de Finetti type theorems is to reduce
the study of permutation-invariant scenarios to that of i.i.d.~ones,
which are often much easier to understand (see Chapter \ref{chap:symmetries}, Section \ref{sec:deFinetti}, for a broader exposition).
In many information theoretic situations, the problem is posed in such
a way that one almost directly sees that the solution is (or is without
loss of generality) permutation-invariant. Furthermore, in many
scenarios one needs only to upper bound (and not to accurately approximate)
a permutation-invariant object by i.i.d.~ones. The seminal \textit{de Finetti
reduction} (aka \textit{post-selection lemma}) of Christandl, K\"{o}nig and Renner \cite{C-K-R}
was precisely designed for that: for any permutation-invariant
state $\rho$ on $\mathrm{H}^{\otimes n}$, with $d=|\mathrm{H}|$ the
``local'' Hilbert space dimension,
\begin{equation}
  \label{eq:CKR}
  \rho \leq (n+1)^{d^2} \int_{\sigma\in\mathcal{D}(\mathrm{H})} \sigma^{\otimes n} \,\mathrm{d}\sigma,
\end{equation}
where $\mathrm{d}\sigma$ denotes the uniform probability measure over the set of mixed states $\mathcal{D}(\mathrm{H})$ on $\mathrm{H}$,
and the inequality refers to the matrix ordering (for Hermitians $A,B$, $A\leq B$ means
that $B-A$ is positive semidefinite).
The beauty of this statement is that on the right hand side we have a
universal object: one and the same convex combination provides the
upper bound to all permutation-invariant states.
At the same time, though, its very universality can be a drawback:
every permutation-invariant state (quantum or classical) is upper bounded
by the same convex combination of tensor power states, so that any
other a priori information (apart from its permutation-invariance),
that one may have on it, is lost. In~\cite{D-S-W}, Appendix B, it was shown that
at the sole cost of slightly increasing the polynomial pre-factor in
front of the upper bounding de Finetti operator, it is actually possible
to make it depend on the state of interest, or on some property
that this state has, including in the integral on the right hand side
of equation~(\ref{eq:CKR}) a fidelity term between $\rho$ and the
i.i.d.~state $\sigma^{\otimes n}$.
In~\cite{D-S-W}, this \emph{constrained de Finetti reduction}
was applied to prove a coding
theorem in a setting with adversarially chosen channel.
In Chapter \ref{chap:SNOS} another application to parallel repetition
of no-signalling games is given.

In Section~\ref{sec:finite-de-finetti}, we first review the constrained de
Finetti reduction of~\cite{D-S-W} and give an alternative proof of it
(Subsection~\ref{subsec:finite-general}). We then show that certain linear constraints lead to very simple
and at the same time useful forms of the de Finetti reduction, such
that certain ``unwanted'' contributions in the integral on the right hand
side of equation~(\ref{eq:CKR}) are either completely absent or exponentially
suppressed (Subsection~\ref{subsec:linear-constraints}).
Next, in Sections~\ref{sec:sep1} and \ref{sec:sep2} we study in depth
the case of separability, a convex constraint. In particular we show
that there are several essentially equivalent ways of thinking about
the exponential decay of the fidelity term.
Inspired by separability, in Section~\ref{sec:general} we present an
axiomatic treatment of a wider class of convex constraints.
Finally, in Section~\ref{sec:infinite} we move to de Finetti reductions
in the infinite-dimensional case.

\section{Flexible de Finetti reductions for finite-dimensional symmetric quantum systems}
\label{sec:finite-de-finetti}

\subsection{A general constrained de Finetti reduction}
\label{subsec:finite-general}
Before getting into more specific statements, let us fix once and for
all some notation that we shall use throughout the
whole chapter (in addition to the general ones specified in Chapter \ref{chap:motivations}, Section \ref{sec:notation}). Most of them, and many more related ones, are also introduced in Chapter \ref{chap:symmetries}, Section \ref{sec:sym}, to which the reader is referred for all the basics about the symmetric subspace. Given a Hilbert space $\mathrm{H}$, we denote by $\Sym^n(\mathrm{H})$ the $n$-symmetric subspace of $\mathrm{H}^{\otimes n}$. In the case where $d=|\mathrm{H}|<+\infty$, $\dim\left(\Sym^n(\mathrm{H})\right)={n+d-1 \choose n}$ and the orthogonal projector onto $\Sym^n(\mathrm{H})$ may be written as
\[ P_{\Sym^n(\mathrm{H})} = {n+d-1 \choose n} \int_{\ket{\psi}\in S_{\mathrm{H}}} \proj{\psi}^{\otimes n}\mathrm{d}\psi, \]
where $\mathrm{d}\psi$ stands for the uniform probability measure
on the unit sphere $S_{\mathrm{H}}$ of $\mathrm{H}$. A state $\rho$ on $\mathrm{H}^{\otimes n}$ is then called permutation-invariant (or simply symmetric) if for any permutation $\pi\in\mathfrak{S}(n)$, it is invariant under the action of the associated permutation unitary $U(\pi)\in\mathcal{U}(\mathrm{H})$, i.e.~$U(\pi)\rho U(\pi)^{\dagger} = \rho$.
This can be expressed equivalently by saying that there exists a
unit vector $\ket{\psi}\in\Sym^n(\mathrm{H}\otimes\mathrm{H}')$, where $|\mathrm{H}'|=d$, such that
$\rho = \tr_{\mathrm{H}'^{\otimes n}} \proj{\psi}$.

Going from rigid to more flexible de Finetti reductions relies essentially on
the so-called ``pinching trick'', which we state formally as Lemma \ref{lemma:pinching} below. This is a generalization of results appearing in \cite{Hayashi} and \cite{HO}.

\begin{lemma}
\label{lemma:pinching}
Let $\mathrm{H}$ be a Hilbert space and $M_1,\ldots,M_r$ be operators on $\mathrm{H}$.
Then, for any state $\rho$ on $\mathrm{H}$,
\[
  \sum_{i,j=1}^r M_i\rho M_j^{\dagger} \leq r\sum_{i=1}^r M_i\rho M_i^{\dagger}.
\]
\end{lemma}

\begin{proof}
To prove that Lemma \ref{lemma:pinching} holds for any state on $\mathrm{H}$, it is
sufficient to prove that it holds for any pure state on $\mathrm{H}$. Let therefore
$\ket{\psi}$ be a unit vector in $\mathrm{H}$. Then, for any unit vector $\ket{\varphi}$ in $\mathrm{H}$,
we have by the Cauchy-Schwarz inequality
\[\begin{split}
  \bra{\varphi} \left(\sum_{i,j=1}^r M_i\proj{\psi} M_j^{\dagger}\right) \ket{\varphi}
     &=    \left| \sum_{i=1}^r \bra{\varphi}M_i\ket{\psi} \right|^2 \\
     &\leq r \sum_{i=1}^r \left|\bra{\varphi}M_i\ket{\psi} \right|^2 \\
     &=    \bra{\varphi} \left( r\sum_{i=1}^r M_i\proj{\psi} M_i^{\dagger}\right) \ket{\varphi},
\end{split}\]
which concludes the proof.
\end{proof}

With this tool at hand, we are ready to get, first of all,
the pure state version of the flexible de Finetti reduction.

\begin{proposition}
  \label{prop:ps-pure}
  Any unit vector $\ket{\theta}\in\Sym^n\left(\mathrm{H}\right)$ satisfies
  \[
     \proj{\theta} \leq {n+d-1 \choose n}^3 \int_{\ket{\psi}\in S_{\mathrm{H}}}
             \left|\langle\theta|\psi^{\otimes n}\rangle\right|^2 \proj{\psi}^{\otimes n}\mathrm{d}\psi.
  \]
\end{proposition}
\begin{proof}
Let $\ket{\theta}\in\Sym^n\left(\mathrm{H}\right)$ be a unit vector. Then,
\[
  \proj{\theta} = P_{\Sym^n(\mathrm{H})} \proj{\theta} P_{\Sym^n(\mathrm{H})}^{\dagger}
                = {n+d-1 \choose n}^2 \int_{\ket{\psi},\ket{\varphi}\in S_{\mathrm{H}}}
                                \proj{\psi}^{\otimes n} \proj{\theta}
                                \proj{\varphi}^{\otimes n} \,\mathrm{d}\psi\,\mathrm{d}\varphi.
\]
Now observe, setting $r={n+d-1 \choose n}^2$, that the span of $\left\{ \proj{\psi}^{\otimes n},\ \ket{\psi}\in S_{\mathrm{H}} \right\}$, subject to the condition of having trace $1$, has dimension $r-1$. So by Caratheodory's theorem, we know that there exist $\{p_1,\ldots,p_r\}$, a convex combination, and $\{\psi_1,\ldots,\psi_r\}$, a set of unit vectors in $\mathrm{H}$, such that
\begin{equation} \label{eq:cara} \int_{\ket{\psi}\in S_{\mathrm{H}}} \proj{\psi}^{\otimes n}\mathrm{d}\psi = \sum_{i=1}^r p_i \proj{\psi_i}^{\otimes n}. \end{equation}
We can therefore rewrite
\begin{align*}
\proj{\theta} & = r \sum_{i,j=1}^r p_ip_j \proj{\psi_i}^{\otimes n}\proj{\theta}\proj{\psi_j}^{\otimes n} \\
& \leq r^2 \sum_{i=1}^r p_i^2\left|\braket{\theta}{\psi_i^{\otimes n}}\right|^2 \proj{\psi_i}^{\otimes n} \\
& \leq r^{3/2} \sum_{i=1}^r p_i\left|\braket{\theta}{\psi_i^{\otimes n}}\right|^2 \proj{\psi_i}^{\otimes n},
\end{align*}
where the next to last inequality is by Lemma \ref{lemma:pinching}, and the last inequality is because, for each $1\leq i\leq r$, $p_i\leq 1/\sqrt{r}$ (which can be seen by contracting both sides of equation \eqref{eq:cara} with $\bra{\psi_i^{\otimes n}}\cdot\ket{\psi_i^{\otimes n}}$). And consequently, since this holds for any ensemble $\{p_i,\,\psi_i\}_{1\leq i\leq r}$ satisfying equation \eqref{eq:cara}, we have by convex combination
\[ \proj{\theta} \leq r^{3/2} \int_{\ket{\psi}\in S_{\mathrm{H}}} \left|\langle\theta|\psi^{\otimes n}\rangle\right|^2 \proj{\psi}^{\otimes n} \,\mathrm{d}\psi, \]
which is precisely the advertised result.
\end{proof}

From Proposition \ref{prop:ps-pure}, we can now easily derive the general
mixed state version of our flexible de Finetti reduction, which was also obtained in \cite{D-S-W} by a slightly different route. In Theorem \ref{th:ps-mixed} below, as well as in the remainder of this chapter, $F(\rho,\sigma)$ stands for the fidelity between quantum states (or classical probability distributions) $\rho$ and $\sigma$, defined as $F(\rho,\sigma)=\|\sqrt{\rho}\sqrt{\sigma}\|_1$.

\begin{theorem}[cf.~\cite{D-S-W}, Lemma 18]
  \label{th:ps-mixed}
  Any symmetric state $\rho$ on $\mathrm{H}^{\otimes n}$ satisfies
  \[
     \rho \leq {n+d^2-1 \choose n}^3 \int_{\ket{\psi}\in S_{\mathrm{H}\otimes\mathrm{H}'}}
        F\left(\rho,\sigma(\psi)^{\otimes n}\right)^2 \sigma(\psi)^{\otimes n} \,\mathrm{d}\psi,
  \]
  where for a unit vector $\ket{\psi}\in\mathrm{H}\otimes\mathrm{H}'$,
  $\sigma(\psi) = \tr_{\mathrm{H}'}\proj{\psi}$ is the reduced state of $\proj{\psi}$ on $\mathrm{H}$.
\end{theorem}
\begin{proof}
As noted before, there exists a unit vector $\ket{\theta}\in\Sym^n\left(\mathrm{H}\otimes\mathrm{H}'\right)$
such that $\rho=\tr_{\mathrm{H}'^{\otimes n}}\proj{\theta}$.
By Proposition \ref{prop:ps-pure}, we have
\[
  \proj{\theta} \leq {n+d^2-1 \choose n}^3 \int_{\ket{\psi}\in S_{\mathrm{H}\otimes\mathrm{H}'}}
       \left|\langle\theta|\psi^{\otimes n}\rangle\right|^2
       \proj{\psi}^{\otimes n} \,\mathrm{d}\psi.
\]
Thus, after partial tracing over $\mathrm{H}'^{\otimes n}$, we obtain
\[
  \rho\leq {n+d^2-1 \choose n}^3 \int_{\ket{\psi}\in S_{\mathrm{H}\otimes\mathrm{H}'}}
       \left|\langle\theta|\psi^{\otimes n}\rangle\right|^2
       \sigma(\psi)^{\otimes n} \,\mathrm{d}\psi.
\]
To get the announced result, we then just have to notice that,
by monotonicity of the fidelity under the CPTP map $\tr_{\mathrm{H}'^{\otimes n}}$,
we have for each $\ket{\psi}\in\mathrm{H}\otimes\mathrm{H}'$,
\[
  \left|\langle\theta|\psi^{\otimes n}\rangle\right|
       =    F\left(\proj{\theta}, \proj{\psi}^{\otimes n} \right)
       \leq F\left(\rho,\sigma(\psi)^{\otimes n}\right). \qedhere
\]
\end{proof}

\subsection{Linear constraints}
\label{subsec:linear-constraints}

Let $\rho$ be a symmetric state on $\mathrm{H}^{\otimes n}$.
What Theorem \ref{th:ps-mixed} tells us is that there exists a
probability measure $\mu$ over the set of states on $\mathrm{H}$ such that
\begin{equation}
  \label{eq:ps-fidelity}
   \rho \leq {n+d^2-1 \choose n}^3 \int_{\sigma\in\in\mathcal{D}(\mathrm{H})}
                            F\left(\rho,\sigma^{\otimes n}\right)^2
                            \sigma^{\otimes n} \,\mathrm{d}\mu(\sigma) \leq (n+1)^{3d^2} \int_{\sigma\in\in\mathcal{D}(\mathrm{H})}
                            F\left(\rho,\sigma^{\otimes n}\right)^2
                            \sigma^{\otimes n} \,\mathrm{d}\mu(\sigma).
\end{equation}
It may be worth pointing out that $\mu$ is in fact the uniform probability measure over the set of mixed states on $\mathrm{H}$ (with respect to the Hilbert--Schmidt distance), since the latter is equivalently characterized as the partial trace over an environment $\mathrm{H}'$ having same dimension as $\mathrm{H}$ of uniformly distributed pure states on $\mathrm{H}\otimes\mathrm{H}'$ (see e.g.~\cite{ZS} for a proof).

Observe that, contrary to the original de Finetti reduction, where the upper bound is the same for every symmetric state, we here have a highly
state-dependent upper bound, where only states which have a high
fidelity with the state of interest $\rho$ are given an important weight.
This is especially useful when one knows that $\rho$ satisfies some
additional property. Indeed, one would then expect that, amongst states
of the form $\sigma^{\otimes n}$, only those approximately satisfying
this same property should have a non-negligible fidelity weight. There
are at least two archetypical cases where this intuition can easily be seen to be true.

\begin{corollary}
  \label{cor:product-image}
  Let $\mathcal{N}:\mathcal{L}(\mathrm{H})\rightarrow\mathcal{L}(\mathrm{K})$ be a quantum channel, with $d=|\mathrm{H}|<+\infty$.
  Assume that $\rho$ is a symmetric state on $\mathrm{H}^{\otimes n}$, which
  is additionally satisfying $\mathcal{N}^{\otimes n}(\rho)=\tau_0^{\otimes n}$,
  for some given state $\tau_0$ on $\mathrm{K}$. Then,
  \[
     \rho \leq (n+1)^{3d^2} \int_{\sigma\in\mathcal{D}(\mathrm{H})}
             F\left(\tau_0,\mathcal{N}(\sigma)\right)^{2n} \sigma^{\otimes n}\,\mathrm{d}\mu(\sigma).
  \]
\end{corollary}
\begin{proof}
This follows directly from inequality \eqref{eq:ps-fidelity}
by monotonicity of the fidelity under the CPTP map $\mathcal{N}$,
and by multiplicativity of the fidelity on tensor products.
\end{proof}

This especially implies that, under the hypotheses of
Corollary \ref{cor:product-image}, we have: for any $0<\delta<1$,
setting
$\mathcal{K}_{\delta}=\left\{ \sigma\in\mathcal{D}(\mathrm{H}) \st F\left(\tau_0,\mathcal{N}(\sigma)\right)\geq 1-\delta\right\}$,
\[
   \rho \leq (n+1)^{3d^2} \left( \int_{\sigma\in\mathcal{K}_{\delta}} \sigma^{\otimes n}\mathrm{d}\mu(\sigma)
        + (1-\delta)^{2n}\int_{\sigma\notin\mathcal{K}_{\delta}} \sigma^{\otimes n}\mathrm{d}\mu(\sigma) \right).
\]
Such flexible de Finetti reduction, for states which satisfy the constraint of
being sent to a certain tensor power state by a certain tensor power
CPTP map, has already been fruitfully applied, for instance in the context of
zero-error communication via quantum channels \cite{D-S-W}.

\smallskip

Another linear constraint is that of a fixed point equation.
\begin{corollary}
  \label{cor:fixed-point}
  Let $\mathcal{N}:\mathcal{L}(\mathrm{H})\rightarrow\mathcal{L}(\mathrm{H})$ be a quantum channel, with $d=|\mathrm{H}|<+\infty$.
  Assume that $\rho$ is a symmetric state on $\mathrm{H}^{\otimes n}$, which
  is additionally satisfying $\mathcal{N}^{\otimes n}(\rho)=\rho$. Then,
  \[
     \rho \leq (n+1)^{3d^2} \int_{\sigma\in\mathcal{D}(\mathrm{H})}
         F\left(\rho,\mathcal{N}(\sigma)^{\otimes n}\right)^2 \mathcal{N}(\sigma)^{\otimes n}\mathrm{d}\mu(\sigma).
  \]
\end{corollary}

\begin{proof}
Apply $\mathcal{N}^{\otimes n}$ on both sides of inequality \eqref{eq:ps-fidelity},
and use once more the monotonicity of the fidelity under the CPTP map $\mathcal{N}$.
\end{proof}

This means that, under the assumptions of Corollary \ref{cor:fixed-point}, there actually exists a probability measure $\widetilde{\mu}$ over the set of states on $\mathrm{H}$ which belong to the range of $\mathcal{N}$ such that
\begin{equation}
  \label{eq:fixed-point}
  \rho \leq (n+1)^{3d^2} \int_{\sigma\in\mathrm{Range}(\mathcal{N})}
      F\left(\rho,\sigma^{\otimes n}\right)^2 \sigma^{\otimes n}\mathrm{d}\widetilde{\mu}(\sigma).
\end{equation}

A case of particular interest for equation \eqref{eq:fixed-point} is the following. Let $G$ be a subgroup of the unitary group on $\mathrm{H}$, equipped with its Haar measure $\mu_G$ (unique normalised left and right invariant measure over $G$). Its associated twirl is the quantum channel $\mathcal{T}_G:\mathcal{L}(\mathrm{H})\rightarrow\mathcal{L}(\mathrm{H})$ defined by
\[ \mathcal{T}_G:\sigma\mapsto\int_{U\in G}U\sigma U^{\dagger}\mathrm{d}\mu_G(U). \]
The range of $\mathcal{T}_G$ is then precisely the set of states on $\mathrm{H}$ in the commutant of $G$, i.e.~\[ \mathcal{K}_G=\left\{\sigma\in\mathcal{D}(\mathrm{H}) \st \forall\ U\in G,\ [\sigma,U]=0 \right\}. \]
Hence, there exists a probability measure $\widetilde{\mu}$ over $\mathcal{K}_G$ such that, if $\rho$ is a symmetric state on $\mathrm{H}^{\otimes n}$ satisfying $\mathcal{T}_G^{\otimes n}(\rho)=\rho$, then
\[
  \rho \leq (n+1)^{3d^2} \int_{\sigma\in\mathcal{K}_G}
         F\left(\rho,\sigma^{\otimes n}\right)^2 \sigma^{\otimes n}\mathrm{d}\widetilde{\mu}(\sigma).
\]

Another situation where equation \eqref{eq:fixed-point} might be especially useful is when $\mathcal{N}$ is a quantum-classical channel, so that its range can be identified with the set of classical probability distributions. We get in that case the corollary below.
\begin{corollary}
  \label{cor:probability}
  Let $\mathcal{X}$ be a finite alphabet and let $P_{\mathcal{X}^n}$ be a
  symmetric probability distribution on $\mathcal{X}^n$. There exists
  a universal probability measure $\mathrm{d}Q_{\mathcal{X}}$ over the set of
  probability distributions on $\mathcal{X}$ such that
  \[
    P_{\mathcal{X}^n} \leq (n+1)^{3|\mathcal{X}|^2} \int_{Q_{\mathcal{X}}}
              F\left(P_{\mathcal{X}^n},Q_{\mathcal{X}}^{\otimes n}\right)^2
              Q_{\mathcal{X}}^{\otimes n}\mathrm{d}Q_{\mathcal{X}},
  \]
  where the inequality sign signifies point-wise inequality between probability
  distributions on $\mathcal{X}^n$.
\end{corollary}

\begin{proof}
This is a special case of Corollary \ref{cor:fixed-point}.
Indeed, we can make the identification $\mathcal{X}\equiv\{1,\ldots,d\}$,
where $d=|\mathcal{X}|$. So let $\mathrm{H}$ be a $d$-dimensional Hilbert space, and denote by $\{\ket{1},\ldots,\ket{d}\}$ an orthonormal basis of $\mathrm{H}$. We can then define the ``classical'' state $\rho$ on $\mathrm{H}^{\otimes n}$ by
\[
  \rho= \sum_{1\leq x_1,\ldots,x_n\leq d} P(x_1,\ldots,x_n)\proj{x_1\otimes\cdots\otimes x_n},
\]
and the quantum-classical channel $\mathcal{N}:\mathcal{L}(\mathrm{H})\rightarrow\mathcal{L}(\mathrm{H})$ by
\[
  \mathcal{N}:\sigma \mapsto \sum_{1\leq x\leq d} Q_{\sigma}(x)\proj{x}
                             = \sum_{1\leq x\leq d} \proj{x}\sigma\proj{x}.
\]
By assumption on $P$, $\rho$ is a symmetric state on $\mathrm{H}^{\otimes n}$, which is additionally, by construction, a fixed point of $\mathcal{N}^{\otimes n}$. Hence, by Corollary \ref{cor:fixed-point},
\[ \rho \leq (n+1)^{3d^2} \int_{\sigma\in\mathcal{D}(\mathrm{H})} F\left(\rho,\mathcal{N}(\sigma)^{\otimes n}\right)^2 \mathcal{N}(\sigma)^{\otimes n}\mathrm{d}\mu(\sigma). \]
By the way $\rho$ and $\mathcal{N}$ have been designed, this actually translates into the point-wise inequality
\[ \forall\ 1\leq x_1,\ldots,x_n\leq d,\ P(x_1,\ldots,x_n) \leq (n+1)^{3d^2} \int_{\sigma\in\mathcal{D}(\mathrm{H})} F\left(P,Q_{\sigma}^{\otimes n}\right)^2 Q_{\sigma}(x_1)\cdots Q_{\sigma}(x_n)\mathrm{d}\mu\left(Q_{\sigma}\right), \]
which is exactly the announced result.
\end{proof}

This flexible de Finetti reduction for probability distributions turns out
to be especially useful when studying the parallel repetition of multi-player
non-local games, as exemplified in Chapter \ref{chap:SNOS}.

\begin{remark}
Note that all these results generalize to non-normalized permutation
invariant positive semidefinite operators on finite-dimensional spaces (or positive
distributions on finite alphabets). One just has to extend the usual
definition of the fidelity by setting $F(M,N)=\|\sqrt{M}\sqrt{N}\|_1$
for any positive semidefinite operators (or positive distributions) $M,N$.
\end{remark}

\subsection{On to convex constraints?}
\label{subsec:convex-constraints}
We just saw that, in the case where the symmetric state $\rho$ under
consideration is additionally known to satisfy certain linear constraints,
it is possible to upper bound it by a de Finetti operator where
either no or exponentially small weight is given to tensor power
states which do not satisfy this same constraint. But what about
the case where the a priori information on $\rho$ is that it
belongs or not to a certain convex subset of states? This is the
question we investigate in the sequel, focussing first in
Sections \ref{sec:sep1} and \ref{sec:sep2} on the paradigmatic example
of the set of separable states, and then describing in
Section \ref{sec:general} the general setting in which similar conclusions hold.

\section{Exponential decay and concentration of $h_{\cS}$ via de Finetti reduction approach}
\label{sec:sep1}

As we just mentioned, we will now be interested for a while in the case where the underlying Hilbert space is a tensor product Hilbert space $\mathrm{H} = \mathrm{A}\otimes \mathrm{B}$, and the kind of symmetric states on $\mathrm{H}^{\otimes n}$ that we will look at are those which additionally satisfy the (convex but non-linear) constraint of being separable across the bipartite cut $\A^{\otimes n}{:}\B^{\otimes n}$. For such a state $\rho$, one can of course still write down a de Finetti reduction of the form
\[ \rho \leq (n+1)^{3|\A|^2|\B|^2} \int_{\sigma\in\mathcal{D}(\mathrm{A}\otimes\mathrm{B})}
                     F\left(\rho,\sigma^{\otimes n}\right)^2 \sigma^{\otimes n}\,\mathrm{d}\mu(\sigma). \]
And what we would like to understand is whether it is possible to argue that only the states $\sigma^{\otimes n}$ which are such that $\sigma$ is separable across the bipartite cut $\A{:}\B$ are given a non exponentially small weight in this integral representation. As we shall see, this question is especially relevant when analysing the multiplicative behaviour of the support function of the set of biseparable states.

So let us specify a bit what we have in mind. Given a positive operator $M$ on $\mathrm{A}\otimes \mathrm{B}$, its maximum overlap with states which are separable across the bipartite cut $\mathrm{A}{:}\mathrm{B}$, which we denote by $\mathcal{S}(\mathrm{A}{:}\mathrm{B})$, is defined as
\[
  h_{\cS(\mathrm{A}{:}\mathrm{B})}(M) = \sup_{\sigma\in\mathcal{S}(\mathrm{A}{:}\mathrm{B})} \tr(M\sigma).
\]
Our interest is in understanding how this quantity behaves
under tensoring. Concretely, this means that we want to know,
for any $n\in\N$, how $h_{\cS(\mathrm{A}^n{:}\mathrm{B}^n)}(M^{\otimes n})$ relates to $h_{\cS(\mathrm{A}{:}\mathrm{B})}(M)$
(where the former quantity is defined as the maximum overlap of $M^{\otimes n}$ with states which are separable across the bipartite cut $\mathrm{A}^{\otimes n}{:}\mathrm{B}^{\otimes n}$). Note that here, and in the remainder of this chapter, we use the shorthand notation $\mathrm{A}^n,\mathrm{B}^n$ for $\mathrm{A}^{\otimes n},\mathrm{B}^{\otimes n}$ whenever they appear subscript. Because $h_{\cS(\mathrm{A}{:}\mathrm{B})}$ is linear homogeneous in its argument, we can always rescale $M$ by a positive constant such that $0\leq M \leq \Id$, meaning that $M$ can be interpreted as a POVM element of the binary test with operators $(M,\Id-M)$. We shall make this assumption throughout from now on.
Then, it is easy to see that, for any $n\in\N$, we have the inequalities
\begin{equation}
  \label{eq:h_SEP-h_SEPn}
  \left(h_{\cS(\mathrm{A}{:}\mathrm{B})}(M)\right)^n \leq h_{\cS(\mathrm{A}^n{:}\mathrm{B}^n)}(M^{\otimes n}) \leq h_{\cS(\mathrm{A}{:}\mathrm{B})}(M) \leq 1.
\end{equation}
But in the case where $h_{\cS(\mathrm{A}{:}\mathrm{B})}(M)<1$, the gap between the lower and upper bounds in
equation \eqref{eq:h_SEP-h_SEPn} grows exponentially with $n$, making these inequalities very little
informative.

This problem is interesting in itself, but also because it connects to plethora of others, some of them even outside the purely quantum information range of applications. The reader is referred to \cite{HM} for a full list of problems which are exactly or approximately equivalent to estimating $h_{\cS}$. Two notable applications of $h_{\cS}$ arise in quantum computing and in quantum Shannon theory.

The first is to $\mathrm{QMA}(2)$, the class of quantum Merlin-Arthur interactive proof systems with two unentangled provers. The setting is that a verifier requires states $\alpha$ and $\beta$ from separate provers which are assumed to be computationally unlimited, and then performs a binary test with POVM $(M,\Id-M)$ on the separable state $\alpha\otimes\beta$. The maximum probability of passing the test that the provers can achieve evidently equals precisely $h_{\cS}(M)$. For complexity theoretic considerations (in particular the so-called soundness gap amplification) it is important to understand how well many instances of the same test, performed in parallel, can be passed, either all $n$, leading to $h_{\cS}(M^{\otimes n})$, or $t$ out of $n$, where $t > n\, h_{\cS}(M)$.

The second application appears in the problem of minimum output entropies of quantum channels, and their asymptotic behaviour. Namely, a quantum channel $\mathcal{N} : \mathcal{L}(\A) \rightarrow\mathcal{L}(\B)$ can be represented in Stinespring form by $\mathcal{N}(X) = \tr_{\mathrm{E}} (V X V^{\dagger})$, with an isometry $V:\mathrm{A}\hookrightarrow \mathrm{B}\otimes\mathrm{E}$. Its minimum output R\'{e}nyi $p$-entropy is given by
\[ S_p^{\min}(\mathcal{N}) = \min_{\rho\in\mathcal{D}(\mathrm{A})} S_p(\mathcal{N}(\rho)),\ \text{where}\ \forall\ \sigma\in\mathcal{D}(\mathrm{B}),\ S_p(\sigma) = -\frac{p}{p-1}\log \|\sigma\|_p. \]
Taking the limit, we recover the von Neumann entropy $S(\sigma)$ for $p=1$ and $-\log\lambda_{\max}(\sigma)$ for $p=\infty$ (see Chapter \ref{chap:channel-compression}, Sections \ref{sec:channel-intro} and \ref{sec:applications}, for further details). From this, it is not hard to see that, with $M = VV^{\dagger}$ the projector onto the range of $V$, i.e.~the subspace $V(\mathrm{A}) \subset \mathrm{B}\otimes \mathrm{E}$, we have $S_{\infty}^{\min}(\mathcal{N}) = -\log h_{\cS(\B{:}\rE)}(M)$. In quantum Shannon theory, the behaviour of $S_p^{\min}(\mathcal{N}^{\otimes n})$, and in particular of $S_{\infty}^{\min}(\mathcal{N}^{\otimes n})=-\log h_{\cS(\B^n{:}\rE^n)}(M^{\otimes n})$ as $n$ grows is of great interest.

\subsection{Some general facts about ``filtered by measurements'' distance measures}

We need to introduce first a few definitions and properties regarding ``filtered by measurements'' distance measures. The reader is e.g.~referred to Chapter \ref{chap:zonoids}, Section \ref{section:POVM}, for more precise definitions.

Let $\mathrm{H}$ be a Hilbert space and let $\mathbf{M}$ be a set of POVMs on $\mathrm{H}$. For any states $\rho,\sigma$ on $\mathrm{H}$, we define their measured by $\mathbf{M}$ trace-norm distance as
\[ d_{\mathbf{M}}(\rho,\sigma)=\sup_{\mathcal{M}\in\mathbf{M}}\frac{1}{2}\left\|\mathcal{M}(\rho)-\mathcal{M}(\sigma)\right\|_1, \]
and their measured by $\mathbf{M}$ fidelity distance as
\[ F_{\mathbf{M}}(\rho,\sigma)=\inf_{\mathcal{M}\in\mathbf{M}}F\left(\mathcal{M}(\rho),\mathcal{M}(\sigma)\right). \]
We have the well-known relations between these two distances (see e.g.~\cite{ChNi}, Chapter 9)
\begin{equation} \label{eq:trace-fidelity}
1-F_{\mathbf{M}}\leq d_{\mathbf{M}} \leq \left(1-F_{\mathbf{M}}^2\right)^{1/2}.
\end{equation}
We further define, for any set of states $\mathcal{K}$ on $\mathrm{H}$, the measured by $\mathbf{M}$ trace-norm distance of $\rho$ to $\mathcal{K}$ as
\[ d_{\mathbf{M}}\left(\rho,\mathcal{K}\right)=\inf_{\sigma\in\mathcal{K}}d_{\mathbf{M}}(\rho,\sigma), \]
and the measured by $\mathbf{M}$ fidelity distance of $\rho$ to $\mathcal{K}$ as
\[ F_{\mathbf{M}}\left(\rho,\mathcal{K}\right)=\sup_{\sigma\in\mathcal{K}}F_{\mathbf{M}}(\rho,\sigma), \]

In the sequel, we shall consider the case where $\mathrm{H}=\mathrm{A}\otimes \mathrm{B}$ is a bipartite Hilbert space, with $|\mathrm{A}|,|\mathrm{B}|<+\infty$. In this setting, we will be especially interested in the set of separable states $\mathcal{S}$ and in the set of separable POVMs $\mathbf{SEP}$ on $\mathrm{H}$ (in the bipartite cut $\mathrm{A}{:}\mathrm{B}$).

\begin{lemma} \label{lemma:F_SEP}
Let $\mathrm{A}_1,\mathrm{B}_1,\mathrm{A}_2,\mathrm{B}_2$ be Hilbert spaces. Then, for any states $\rho_1$ on $\mathrm{A}_1\otimes \mathrm{B}_1$ and $\rho_2$ on $\mathrm{A}_2\otimes \mathrm{B}_2$, we have
\[ F\big(\rho_1\otimes\rho_2,\mathcal{S}(\mathrm{A}_1\mathrm{A}_2{:}\mathrm{B}_1\mathrm{B}_2)\big) \leq F_{\mathbf{SEP}}\big(\rho_1,\mathcal{S}(\mathrm{A}_1{:}\mathrm{B}_1)\big) F\big(\rho_2,\mathcal{S}(\mathrm{A}_2{:}\mathrm{B}_2)\big). \]
\end{lemma}

\begin{proof}
The proof is directly inspired from \cite{Piani}, adapted here to the case of fidelities rather than relative entropies.

Let $\mathcal{M}_1\equiv \big(M_1^{(i)}\big)_{i\in I}\in\mathbf{SEP}(\mathrm{A}_1{:}\mathrm{B}_1)$. Then, by monotonicity of the fidelity under the CPTP map $\mathcal{M}_1\otimes\mathcal{I\!d}_2$, we have
\begin{align*} \sup_{\sigma_{12}\in\mathcal{S}(\mathrm{A}_1\mathrm{A}_2{:}\mathrm{B}_1\mathrm{B}_2)} F\left(\rho_1\otimes\rho_2,\sigma_{12}\right) \leq & \, \sup_{\sigma_{12}\in\mathcal{S}(\mathrm{A}_1\mathrm{A}_2{:}\mathrm{B}_1\mathrm{B}_2)} F\left(\mathcal{M}_1\otimes\mathcal{I\!d}_2(\rho_1\otimes\rho_2),\mathcal{M}_1\otimes\mathcal{I\!d}_2(\sigma_{12})\right)\\
= & \, F\left(\mathcal{M}_1\otimes\mathcal{I\!d}_2(\rho_1\otimes\rho_2),\mathcal{M}_1\otimes\mathcal{I\!d}_2(\widetilde{\sigma}_{12})\right), \end{align*}
for some $\widetilde{\sigma}_{12}\in\mathcal{S}(\mathrm{A}_1\mathrm{A}_2{:}\mathrm{B}_1\mathrm{B}_2)$. Next,
\[ F\left(\mathcal{M}_1\otimes\mathcal{I\!d}_2(\rho_1\otimes\rho_2), \mathcal{M}_1\otimes\mathcal{I\!d}_2(\widetilde{\sigma}_{12})\right) = \sum_{i\in I}\sqrt{\tr\left(M_1^{(i)}\rho_1\right)}\sqrt{\tr\left(M_1^{(i)}\widetilde{\sigma}_1\right)} F\left(\rho_2,\widetilde{\sigma}_2^{(i)}\right), \]
where $\widetilde{\sigma}_1=\tr_{\mathrm{A}_2\mathrm{B}_2}\left(\widetilde{\sigma}_{12}\right)$ and for each $i\in I$, $\widetilde{\sigma}_2^{(i)}= \tr_{\mathrm{A}_1\mathrm{B}_1}\big(M_1^{(i)}\otimes\Id_2\widetilde{\sigma}_{12}\big)/ \tr_{\mathrm{A}_1\mathrm{B}_1}\big(M_1^{(i)}\widetilde{\sigma}_1\big)$, so $\widetilde{\sigma}_1\in\mathcal{S}(\mathrm{A}_1{:}\mathrm{B}_1)$ and for each $i\in I$, $\widetilde{\sigma}_2^{(i)}\in\mathcal{S}(\mathrm{A}_2{:}\mathrm{B}_2)$. Hence, for each $i\in I$, $F\big(\rho_2,\widetilde{\sigma}_2^{(i)}\big)\leq \sup_{\sigma_2\in\mathcal{S}(\mathrm{A}_2{:}\mathrm{B}_2)} F\left(\rho_2,\sigma_2\right)$, and subsequently
\begin{align*}
\sum_{i\in I}\sqrt{\tr\big(M_1^{(i)}\rho_1\big)}\sqrt{\tr\big(M_1^{(i)}\widetilde{\sigma}_1\big)} F\big(\rho_2,\widetilde{\sigma}_2^{(i)}\big) \leq & \left(\sup_{\sigma_2\in\mathcal{S}_{\mathrm{A}_2:\mathrm{B}_2}} F\left(\rho_2,\sigma_2\right) \right) F\left(\mathcal{M}_1(\rho_1),\mathcal{M}_1(\widetilde{\sigma}_1)\right)\\
\leq & \left(\sup_{\sigma_2\in\mathcal{S}(\mathrm{A}_2{:}\mathrm{B}_2)} F\left(\rho_2,\sigma_2\right)\right) \left(\sup_{\sigma_1\in\mathcal{S}(\mathrm{A}_1{:}\mathrm{B}_1)} F\left(\mathcal{M}_1(\rho_1),\mathcal{M}_1(\sigma_1)\right)\right).
\end{align*}
We thus have shown that, for any $\mathcal{M}_1\in\mathbf{SEP}(\mathrm{A}_1{:}\mathrm{B}_1)$,
\[ F\big(\rho_1\otimes\rho_2,\mathcal{S}(\mathrm{A}_1\mathrm{A}_2{:}\mathrm{B}_1\mathrm{B}_2)\big) \leq \left(\sup_{\sigma_1\in\mathcal{S}(\mathrm{A}_1{:}\mathrm{B}_1)} F\left(\mathcal{M}_1(\rho_1),\mathcal{M}_1(\sigma_1)\right) \right) F\left(\rho_2,\mathcal{S}(\mathrm{A}_2{:}\mathrm{B}_2)\right). \]
Taking the infimum over $\mathcal{M}_1\in\mathbf{SEP}(\mathrm{A}_1{:}\mathrm{B}_1)$, we get precisely the statement in Lemma \ref{lemma:F_SEP}.
\end{proof}

\begin{theorem}
  \label{th:F_SEP}
  Let $\mathrm{A},\mathrm{B}$ be Hilbert spaces, and let $\rho$ be a state on $\mathrm{A}\otimes\mathrm{B}$.
  Then, for any $n\in\N$,
  \[
    F\big(\rho^{\otimes n},\mathcal{S}(\mathrm{A}^n{:}\mathrm{B}^n)\big)
               \leq F_{\mathbf{SEP}(\mathrm{A}{:}\mathrm{B})}\big(\rho,\mathcal{S}(\mathrm{A}{:}\mathrm{B})\big)^n.
  \]
\end{theorem}

\begin{proof}
Theorem \ref{th:F_SEP} is a direct corollary of Lemma \ref{lemma:F_SEP}, obtained by iterating the latter.
\end{proof}

\subsection{Weak multiplicativity of $h_{\cS}$}
With these facts prepared, we can now derive our main theorem.

\begin{theorem}
  \label{th:h_sep-n}
  Let $M$ be an operator on the tensor product Hilbert space $\mathrm{A}\otimes \mathrm{B}$, satisfying $0 \leq M \leq \Id$, and set $r=\|M\|_2$.
  If $h_{\cS(\mathrm{A}{:}\mathrm{B})}(M)\leq 1-\delta$, for some $0<\delta<1$,
  then for any $n\in\N$,
  \[
    h_{\cS(\mathrm{A}^n{:}\mathrm{B}^n)}(M^{\otimes n}) \leq \left(1-\frac{\delta^2}{5r^2}\right)^n.
  \]
\end{theorem}

\begin{proof}
Let $\rho\in\mathcal{S}(\mathrm{A}^{n}{:}\mathrm{B}^{n})$. Our goal will be first of all to show that $\tr\left(M^{\otimes n}\rho\right)\leq 2(n+1)^{3|\mathrm{A}|^2|\mathrm{B}|^2}\left(1-\delta^2/2r^2\right)^n$.
Now observe that, denoting, for each permutation $\pi\in\mathfrak{S}(n)$, by $U(\pi)\in\mathcal{U}\big((\A\otimes\B)^{\otimes n}\big)$ the associated permutation unitary,
\[ \tr\left(M^{\otimes n}\rho\right) = \tr\left( \left(\frac{1}{n!}\sum_{\pi\in\mathfrak{S}(n)}U(\pi)M^{\otimes n}U(\pi)^{\dagger}\right) \rho\right) = \tr\left(M^{\otimes n} \left(\frac{1}{n!}\sum_{\pi\in\mathfrak{S}(n)}U(\pi)^{\dagger}\rho U(\pi)\right) \right). \]
The first equality is by $n$-symmetry of $M^{\otimes n}$ and the second one is by cyclicity
of the trace. Hence, for our purposes, we may actually assume without loss of generality
that $\rho\in\mathcal{S}(\mathrm{A}^{n}{:}\mathrm{B}^{n})$ is $n$-symmetric.

Yet, if $\rho$ is an $n$-symmetric state on $\left(\mathrm{A}\otimes\mathrm{B}\right)^{\otimes n}$,
we know by Theorem \ref{th:ps-mixed} that there exists a probability measure
$\mu$ on the set of states on $\mathrm{A}\otimes\mathrm{B}$ such that
\[
  \rho \leq (n+1)^{3|\mathrm{A}|^2|\mathrm{B}|^2} \int_{\sigma\in\mathcal{D}(\mathrm{A}\otimes\mathrm{B})}
                     F\left(\rho,\sigma^{\otimes n}\right)^2 \sigma^{\otimes n}\,\mathrm{d}\mu(\sigma).
\]
So, by multiplicativity of the trace on tensor products, we get in that case
\[
  \tr\left(M^{\otimes n}\rho\right)
      \leq (n+1)^{3|\mathrm{A}|^2|\mathrm{B}|^2} \int_{\sigma\in\mathcal{D}(\mathrm{A}\otimes\mathrm{B})}
                     F\left(\rho,\sigma^{\otimes n}\right)^2 \tr\left(M\sigma\right)^n\,\mathrm{d}\mu(\sigma).
\]
Consequently, for any $0<\epsilon<1$, setting $\mathcal{K}_{\epsilon}=\left\{\sigma\in\mathcal{D}(\mathrm{A}\otimes\mathrm{B}) \st \left\|\sigma-\mathcal{S}(\mathrm{A}{:}\mathrm{B})\right\|_2\leq \epsilon/r\right\}$, we have, upper bounding either $F\left(\rho,\sigma^{\otimes n}\right)$ or $\tr\left(M\sigma\right)$ by $1$,
\[ \tr\left(M^{\otimes n}\rho\right) \leq\, (n+1)^{3|\mathrm{A}|^2|\mathrm{B}|^2} \left( \int_{\sigma\in\mathcal{K}_{\epsilon}} \tr\left(M\sigma\right)^n \mathrm{d}\mu(\sigma) + \int_{\sigma\notin\mathcal{K}_{\epsilon}} F\left(\rho,\sigma^{\otimes n}\right)^2 \mathrm{d}\mu(\sigma)\right). \]
Now, if $\sigma\in\mathcal{K}_{\epsilon}$, this means that there exists $\tau\in\mathcal{S}(\mathrm{A}{:}\mathrm{B})$ such that $\|\sigma-\tau\|_2\leq\epsilon/r$, so that
\[
  \tr(M\sigma) =    \tr(M\tau)+\tr(M(\sigma-\tau))
               \leq \tr(M\tau)+ \|M\|_2\|\sigma-\tau\|_2
               \leq 1-\delta+\epsilon.
\]
The next to last inequality is simply by the Cauchy--Schwarz inequality,
while the last one is by assumption on $M$, $\tau$, $\sigma$. And if
$\sigma\notin\mathcal{K}_{\epsilon}$, then
\[ F\left(\rho,\sigma^{\otimes n}\right) \leq F\left(\sigma^{\otimes n}, \mathcal{S}(\mathrm{A}^{n}{:}\mathrm{B}^{n})\right) \leq F_{\mathbf{SEP}(\mathrm{A}{:}\mathrm{B})}\left(\sigma, \mathcal{S}(\mathrm{A}{:}\mathrm{B})\right)^n \leq \left(1-\frac{\epsilon^2}{4r^2}\right)^{n/2}. \]
The first inequality is because $\rho\in\mathcal{S}(\mathrm{A}^{n}{:}\mathrm{B}^{n})$, the second one is by Theorem \ref{th:F_SEP}, and the third one is obtained by combining equation \eqref{eq:trace-fidelity} with the known lower bound $\left\|\sigma-\mathcal{S}(\mathrm{A}{:}\mathrm{B})\right\|_{\mathbf{SEP}(\A{:}\B)}\geq \left\|\sigma-\mathcal{S}(\mathrm{A}{:}\mathrm{B})\right\|_2$ (see e.g.~\cite{LW1}).

Putting everything together, we obtain in the end that for any $0<\epsilon<1$,
\begin{equation} \label{eq:epsilon} \tr\left(M^{\otimes n}\rho\right) \leq\, (n+1)^{3|\mathrm{A}|^2|\mathrm{B}|^2} \left( (1-\delta+\epsilon)^n + \left(1-\frac{\epsilon^2}{4r^2}\right)^n \right). \end{equation}
In particular, choosing $\epsilon=2r^2\left((1+\delta/r^2)^{1/2}-1\right)$ in equation \eqref{eq:epsilon}, so that $\epsilon^2/4r^2=\delta-\epsilon\geq \delta^2/5r^2$, we get $\tr\left(M^{\otimes n}\rho\right) \leq 2(n+1)^{3|\mathrm{A}|^2|\mathrm{B}|^2} \left(1-\delta^2/5r^2\right)^n$. And consequently
\begin{equation}
  \label{eq:h_SEPn}
  h_{\cS(\mathrm{A}^n{:}\mathrm{B}^n)}(M^{\otimes n}) \leq 2(n+1)^{3|\mathrm{A}|^2|\mathrm{B}|^2}\left(1-\frac{\delta^2}{5r^2}\right)^n.
\end{equation}

In order to conclude, we just need to remove the polynomial pre-factor in equation \eqref{eq:h_SEPn}.
Assume that there exists a constant $C>0$ such that
$h_{\cS(\mathrm{A}^N{:}\mathrm{B}^N)}(M^{\otimes N})\geq C\left(1-\delta^2/5r^2\right)^N$ for some $N\in\N$.
Then, we would have for any $n\in\N$,
\[
  h_{\cS(\mathrm{A}^{Nn}{:}\mathrm{B}^{Nn})}\left(M^{\otimes Nn}\right) \geq \left(h_{\cS(\mathrm{A}^N{:}\mathrm{B}^N)}\left(M^{\otimes N}\right)\right)^n
                                     \geq C^n\left(1-\frac{\delta^2}{5r^2}\right)^{Nn}.
\]
On the other hand, equation \eqref{eq:h_SEPn} says that we also have
\[
  h_{\cS(\mathrm{A}^{Nn}{:}\mathrm{B}^{Nn})}\left(M^{\otimes Nn}\right) \leq 2(Nn+1)^{3|\mathrm{A}|^2|\mathrm{B}|^2} \left(1-\frac{\delta^2}{5r^2}\right)^{Nn}.
\]
Letting $n$ grow, we see that the only option to make these two inequalities
compatible is to have $C\leq 1$, which is precisely what we wanted to show.
\end{proof}

The conclusion of Theorem \ref{th:h_sep-n} had already been obtained via completely
different techniques than the one presented here (and even with slightly better constants).
However, the good thing about the de Finetti reduction approach is that it
gives, almost for free, not only this exponential decay result for the behaviour
of $h_{\cS}$ under tensoring, but also some kind of concentration statement.
To be precise, assume that $M$ is an operator on $\A\otimes\B$, satisfying $0 \leq M \leq \Id$
and $h_{\cS(\mathrm{A}{:}\mathrm{B})}(M)\leq 1-\delta$ for some $0<\delta<1$.
Then, $M$ can be identified with a binary test that a separable state is
guaranteed to pass only with probability $h_{\cS(\mathrm{A}{:}\mathrm{B})}(M)\leq 1-\delta$,
while there exists some (entangled) state that would pass it with probability $h_{\mathcal{D}(\mathrm{A}\otimes\mathrm{B})}(M)=\|M\|_{\infty}$,
which may be $1$.
Hence, a natural question would be: performing this test $n$ times in parallel,
what is the probability that a separable state passes a certain fraction $t/n$ of them?
Such maximum probability is nothing else than $h_{\cS(\mathrm{A}^n{:}\mathrm{B}^n)}\left(M^{(t/n)}\right)$, where the operator $M^{(t/n)}$ on $(\A\otimes\B)^{\otimes n}$ is defined as
\[ M^{(t/n)} := \sum_{I\subset[n],\,|I|\geq t}M^{\otimes I}\otimes\Id^{\otimes I^c}. \]
Above (and several times later on as well) we use the following shorthand notation: for any $1\leq k\leq n$ and any bipartition $\{i_1,\ldots,i_k\}\sqcup\{j_1,\ldots,j_{n-k}\}=\{1,\ldots,n\}$,
\[ M_{\A\B}^{\otimes \{i_1,\ldots,i_k\} }\otimes\Id_{\A\B}^{\otimes \{j_1,\ldots,j_{n-k}\} } := M_{\A_{i_1}\B_{i_1}}\otimes\cdots\otimes M_{\A_{i_k}\B_{i_k}}\otimes\Id_{\A_{j_1}\B_{j_1}}\otimes\cdots\otimes \Id_{\A_{j_{n-k}}\B_{j_{n-k}}}. \]
Obviously, if $t < (1-\delta)n$ then the answer is asymptotically $1$,
whereas for $t=n$ the answer is $h_{\cS(\mathrm{A}^n{:}\mathrm{B}^n)}(M^{\otimes n})$, which decays
exponentially fast with $n$ as established in Theorem \ref{th:h_sep-n}.
But is such exponential amplification of the failing probability already
true for $t$ just slightly above $(1-\delta)n$? Theorem \ref{th:h_sep-t/n}
answers this question affirmatively.

\begin{theorem}
  \label{th:h_sep-t/n}
  Let $M$ be an operator on the tensor product Hilbert space $\mathrm{A}\otimes \mathrm{B}$,
  satisfying $0 \leq M \leq \Id$, and set $r=\|M\|_2$.
  If $h_{\cS(\mathrm{A}{:}\mathrm{B})}(M)\leq 1-\delta$ for some $0<\delta<1$, then for any
  $n,t\in\N$ with $t\geq(1-\delta+\alpha)n$ for some $0<\alpha\leq\delta$, we have
  \[
    h_{\cS(\mathrm{A}^n{:}\mathrm{B}^n)}\left(M^{(t/n)}\right)\leq\exp\left(-n\frac{\alpha^2}{5r^2}\right).
  \]
\end{theorem}

\begin{proof}
Following the exact same lines as in the proof of Theorem \ref{th:h_sep-n}, we now have in place of equation \eqref{eq:epsilon}
\begin{equation} \label{eq:epsilon'} h_{\cS(\mathrm{A}^n{:}\mathrm{B}^n)}\left(M^{(t/n)}\right) \leq\, (n+1)^{3|\mathrm{A}|^2|\mathrm{B}|^2} \left( \exp\left[-2n(\alpha-\epsilon)^2\right] + \exp\left[-n\frac{\epsilon^2}{4r^2}\right] \right). \end{equation}
This is indeed a consequence of Hoeffding's inequality (and of the fact that $e^{-x}\geq 1-x$ for any $x>0$). So in particular, choosing $\epsilon=\alpha\left(1-(\sqrt{2}-1)/(8r^2-1)\right)$ in equation \eqref{eq:epsilon'}, so that $\epsilon^2/4r^2=2(\alpha-\epsilon)^2\geq\alpha^2/5r^2$, and removing the polynomial pre-factor by the same trick as in the proof of Theorem \ref{th:h_sep-n}, we get as announced
\[ h_{\cS(\mathrm{A^n}{:}\mathrm{B}^n)}\left(M^{(t/n)}\right)\leq\exp\left(-n\frac{\alpha^2}{5r^2}\right). \qedhere \]
\end{proof}

\section{Exponential decay and concentration of $h_{\cS}$ via entanglement measure approach}
\label{sec:sep2}

\subsection{Quantifying the disturbance induced by measurements}

Let us state first a few technical lemmas that we will need later on to establish our main result. The reader is referred to Chapter \ref{chap:QIT}, Section \ref{sec:Shannon}, for all quantum Shannon theory definitions which are used in this section (entropy and relative entropy, mutual information etc.).

\begin{lemma}
\label{lemma:post-POVM}
Let $\rho$ be a state on $\mathrm{U}\otimes\mathrm{V}$ and $T$ be an operator on $\mathrm{U}$, satisfying $0\leq T\leq \Id$. Define next $p=\tr_{\mathrm{U}\mathrm{V}}\left[\left(T_{\mathrm{U}}\otimes\Id_{\mathrm{V}}\right)\rho_{\mathrm{U}\mathrm{V}}\right]$ as the probability of obtaining the first outcome when the two-outcome POVM $\left(T_{\mathrm{U}}\otimes\Id_{\mathrm{V}}, (\Id_{\mathrm{U}}-T_{\mathrm{U}})\otimes\Id_{\mathrm{V}}\right)$ is performed on $\rho_{\mathrm{U}\mathrm{V}}$, and $\tau_{\mathrm{V}}=\tr_{\mathrm{U}}\left[\left(T_{\mathrm{U}}\otimes\Id_{\mathrm{V}}\right)\rho_{\mathrm{U}\mathrm{V}}\right]/p$ as the corresponding post-measurement state on $\mathrm{V}$. Also, denote by $\rho_{\mathrm{V}}=\tr_{\mathrm{U}}\left[\rho_{\mathrm{U}\mathrm{V}}\right]$ the reduced state of $\rho_{\mathrm{U}\mathrm{V}}$ on $\mathrm{V}$. Then,
\[ D\left(\tau_{\mathrm{V}}\big\|\rho_{\mathrm{V}}\right) \leq - \log p.\]
\end{lemma}

\begin{proof}
Note that $\rho_{\mathrm{V}}= p\tau_{\mathrm{V}} +(1-p)\sigma_{\mathrm{V}}$, where $\sigma_{\mathrm{V}}=\tr_{\mathrm{U}}\left[\left((\Id_{\mathrm{U}}-T_{\mathrm{U}})\otimes\Id_{\mathrm{V}}\right)\rho_{\mathrm{U}\mathrm{V}}\right]/(1-p)$. We therefore have the operator inequality $p\tau_{\mathrm{V}}\leq\rho_{\mathrm{V}}$. And hence,
\[ D\left(\tau_{\mathrm{V}}\big\|\rho_{\mathrm{V}}\right) = \tr\left[\tau_{\mathrm{V}}\left(\log\tau_{\mathrm{V}}-\log\rho_{\mathrm{V}}\right)\right] \leq \tr\left[\tau_{\mathrm{V}}\left(\log\tau_{\mathrm{V}}-\log (p\tau_{\mathrm{V}})\right)\right] = -\log p, \]
the next to last inequality being because $\log$ is an operator monotone function.
\end{proof}

Let us recall the definition of the squashed entanglement $E_{sq}$, introduced in \cite{CW}:
\[ E_{sq}\left(\rho_{\mathrm{A}\mathrm{B}}\right) = \inf\left\{ \frac{1}{2} I(\mathrm{A}{:}\mathrm{B}|\mathrm{E})\left(\rho_{\mathrm{A}\mathrm{B}\mathrm{E}}\right) \st \tr_{\mathrm{E}}\left(\rho_{\mathrm{A}\mathrm{B}\mathrm{E}}\right)=\rho_{\mathrm{A}\mathrm{B}} \right\}. \]

\begin{lemma} \label{lemma:E_sq}
Let $M_{\mathrm{A}\mathrm{B}}$ be an operator on the tensor product Hilbert space $\mathrm{A}\otimes\mathrm{B}$, satisfying $0\leq M_{\mathrm{A}\mathrm{B}}\leq \Id$, and let $\alpha_{\mathrm{A}^n},\beta_{\mathrm{B}^n}$ be states on $\mathrm{A}^{\otimes n},\mathrm{B}^{\otimes n}$ respectively. Next, fix $1\leq k\leq n-1$, and define
\begin{align*}
& p_k=\tr_{\mathrm{A}^n\mathrm{B}^n}\left[\left(M_{\mathrm{A}\mathrm{B}}^{\otimes k}\otimes\Id_{\mathrm{A}\mathrm{B}}^{\otimes n-k}\right)\alpha_{\mathrm{A}^n}\otimes\beta_{\mathrm{B}^n}\right]\\
& \tau(k)_{\mathrm{A}^{n-k}\mathrm{B}^{n-k}}=\frac{1}{p_k}\tr_{\mathrm{A}^k\mathrm{B}^k}\left[\left(M_{\mathrm{A}\mathrm{B}}^{\otimes k}\otimes\Id_{\mathrm{A}\mathrm{B}}^{\otimes n-k}\right)\alpha_{\mathrm{A}^n}\otimes\beta_{\mathrm{B}^n}\right].
\end{align*}
Then, it holds that
\[ \sum_{j=k+1}^n E_{sq}\left(\tau(k)_{\mathrm{A}_j\mathrm{B}_j}\right) \leq \frac{1}{2}\log \frac{1}{p_k} .\]
\end{lemma}

\begin{proof}
By Lemma \ref{lemma:post-POVM}, with $\mathrm{U}=\mathrm{A}^{\otimes k}\otimes\mathrm{B}^{\otimes k}$, $\mathrm{V}=\mathrm{A}^{\otimes n-k}\otimes\mathrm{B}^{\otimes n-k}$, $T_{\mathrm{U}}=M_{\mathrm{A}\mathrm{B}}^{\otimes k}$ and $\rho_{\mathrm{U}\mathrm{V}}=\alpha_{\mathrm{A}^n}\otimes\beta_{\mathrm{B}^n}$, we know that
\[ D\left(\tau(k)_{\mathrm{A}^{n-k}\mathrm{B}^{n-k}}\big\|\alpha_{\mathrm{A}^{n-k}}\otimes\beta_{\mathrm{B}^{n-k}}\right) \leq \log \frac{1}{p_k}. \]
Now, observe that
\begin{align*}
D\left(\tau(k)_{\mathrm{A}^{n-k}\mathrm{B}^{n-k}}\big\|\alpha_{\mathrm{A}^{n-k}}\otimes\beta_{\mathrm{B}^{n-k}}\right) \geq & \, D\left(\tau(k)_{\mathrm{A}^{n-k}\mathrm{B}^{n-k}}\big\|\tau(k)_{\mathrm{A}^{n-k}}\otimes\tau(k)_{\mathrm{B}^{n-k}}\right) \\
= & \, I\left(\mathrm{A}_{k+1}\ldots \mathrm{A}_n{:}\mathrm{B}_{k+1}\ldots \mathrm{B}_n\right)\left(\tau(k)_{\mathrm{A}^{n-k}\mathrm{B}^{n-k}}\right)\\
= & \sum_{j=k+1}^n I\left(\mathrm{A}_j{:}\mathrm{B}_{k+1}\ldots \mathrm{B}_n|\mathrm{A}_{k+1}\ldots \mathrm{A}_{j-1}\right)\left(\tau(k)_{\mathrm{A}^{n-k}\mathrm{B}^{n-k}}\right)\\
\geq & \sum_{j=k+1}^n I\left(\mathrm{A}_j{:}\mathrm{B}_j|\mathrm{A}_{k+1}\ldots \mathrm{A}_{j-1}\right)\left(\tau(k)_{\mathrm{A}^{n-k}\mathrm{B}^{n-k}}\right) \\
\geq & \sum_{j=k+1}^n 2\,E_{sq}\left(\tau(k)_{\mathrm{A}_j\mathrm{B}_j}\right).
\end{align*}
The first inequality is due to the fact that, given a bipartite state $\tau_{\mathrm{U}\mathrm{V}}$ on $\mathrm{U}\otimes\mathrm{V}$, for any states $\rho_{\mathrm{U}}$, $\rho_{\mathrm{V}}$ on $\mathrm{U}$, $\mathrm{V}$ respectively, $D(\tau_{\mathrm{U}\mathrm{V}}\|\rho_{\mathrm{U}}\otimes\rho_{\mathrm{V}})\geq D(\tau_{\mathrm{U}\mathrm{V}}\|\tau_{\mathrm{U}}\otimes\tau_{\mathrm{V}})$.
The third equality and the fourth inequality follow from the chain rule and the monotonicity under discarding of subsystems, respectively, for the quantum mutual information.
And the last inequality is by definition of the squashed entanglement.
\end{proof}

\begin{remark}
\label{remark:E_I}
Observe that under the assumptions of Lemma \ref{lemma:E_sq},
we actually have the stronger conclusion
\[
  \sum_{j=k+1}^n E_{I}\left(\tau(k)_{\mathrm{A}_j\mathrm{B}_j}\right) \leq \frac{1}{2}\log \frac{1}{p_k},
\]
where $E_I$ is the \emph{conditional entanglement of mutual information (CEMI)}
introduced in \cite{HWY}:
\[
  E_I\left(\rho_{\mathrm{A}\mathrm{B}}\right) = \inf \left\{ \frac{1}{2}\big[ I(\mathrm{A}\mathrm{A}'{:}\mathrm{B}\mathrm{B}')\left(\rho_{\mathrm{A}\mathrm{A}'\mathrm{B}\mathrm{B}'}\right)
                                    - I(\mathrm{A}'{:}\mathrm{B}')\left(\rho_{\mathrm{A}\mathrm{A}'\mathrm{B}\mathrm{B}'}\right)\big]
                              \st \tr_{\mathrm{A}'\mathrm{B}'}\left(\rho_{\mathrm{A}\mathrm{A}'\mathrm{B}\mathrm{B}'}\right) =\rho_{\mathrm{A}\mathrm{B}} \right\}.
\]
CEMI is always at least as large as squashed entanglement: for any state $\rho_{\mathrm{A}\mathrm{B}}$, $E_I(\rho_{\mathrm{A}\mathrm{B}}) \geq E_{sq}(\rho_{\mathrm{A}\mathrm{B}})$. But the precise relation between these two entanglement measures is unknown.
In~\cite{HWY} it was furthermore shown that, like squashed entanglement, CEMI is additive,
and more generally super-additive in the sense that
\[
  E_I(\rho_{\mathrm{A}_1\mathrm{A}_2{:}\mathrm{B}_1\mathrm{B}_2}) \geq E_I(\rho_{\mathrm{A}_1{:}\mathrm{B}_1}) + E_I(\rho_{\mathrm{A}_2{:}\mathrm{B}_2}).
\]
However, unlike squashed entanglement, there is no simple proof of
monogamy of CEMI, and it may well not hold in general.
\end{remark}

\begin{lemma}[cf.~\cite{Hol}, Lemma 8.6]
\label{lemma:recursivity}
Let $0<\nu<1$ and $c>0$. Let also $n\in\N$ and assume that $(p_k)_{1\leq k\leq n}$ is a sequence of numbers satisfying $1> p_1\geq\cdots\geq p_n >0$ and
\[ \forall\ 1\leq k\leq n-1,\ p_{k+1} \leq p_k \left( \sqrt{\frac{c}{n-k}\log\frac{1}{p_k}} + \nu \right). \]
Then, for any $0<\gamma<1-\nu$ such that $p_1\leq \nu+\gamma$, we have
\[ \forall\ 1\leq k\leq n,\ p_k\leq (\nu+\gamma)^{\min(k,k_0)}\ \text{for}\ k_0=\frac{\gamma^2}{c\log[1/(\nu+\gamma)]+\gamma^2}(n+1). \]
\end{lemma}

\begin{proof}
To prove Lemma \ref{lemma:recursivity}, we only have to show that
\begin{equation} \label{eq:rec} \forall\ 1\leq k\leq k_0,\ p_k\leq (\nu+\gamma)^k. \end{equation}
Indeed, the case $k>k_0$ then directly follows from the assumption that the sequence $(p_k)_{1\leq k\leq n}$ is non-increasing, so that $p_k\leq p_{k_0} \leq (\nu+\gamma)^{k_0}$.

Let us establish \eqref{eq:rec} by recursivity. The statement obviously holds for $k=1$ since $p_1\leq \nu+\gamma$ by hypothesis. So assume next that it holds for some $k\leq k_0-1$. If $p_k\leq(\nu+\gamma)^{k+1}$, then clearly $p_{k+1}\leq p_k\leq(\nu+\gamma)^{k+1}$. Otherwise, by the way $p_{k+1}$ is related to $p_k$, we then have
\[ p_{k+1}\leq (\nu+\gamma)^k\left(\sqrt{\frac{c(k+1)\log[1/(\nu+\gamma)]}{(n-k)}}+\nu\right),\]
and the latter quantity is smaller than $(\nu+\gamma)^{k+1}$ if
\[ \frac{k+1}{n-k} \leq \frac{\gamma^2}{c\log[1/(\nu+\gamma)]}, \]
which can be checked to be equivalent to $k+1\leq k_0$. Hence in both cases, the statement holds for $k+1$.
\end{proof}

\begin{corollary}[cf.~\cite{Hol}, Lemma 8.6]
\label{cor:recursivity}
Let $(p_k)_{1\leq k\leq n}$ be a sequence of numbers satisfying the assumptions of Lemma \ref{lemma:recursivity}, and the additional condition $p_1\leq 1-(1-\nu)/2$. Then,
\[ p_n\leq \left(1-\frac{(1-\nu)^2}{8 c}\right)^n. \]
\end{corollary}

\begin{proof}
Corollary \ref{cor:recursivity} follows from applying Lemma \ref{lemma:recursivity} in the particular case $\gamma=(1-\nu)/2$. Indeed, we then have $\nu+\gamma=1-(1-\nu)/2=(1+\nu)/2$, so that
\[ k_0= \frac{(1-\nu)^2}{4c\log[2/(1+\nu)]+(1-\nu)^2}(n+1) \geq \frac{(1-\nu)^2}{4c\log[2^{1-\nu}]+1}(n+1) \geq \frac{1-\nu}{4c}n, \]
And consequently,
\[ p_n \leq \left(1-\frac{1-\nu}{2}\right)^{n(1-\nu)/(4c)} \leq \left(1-\frac{(1-\nu)^2}{8 c}\right)^n, \]
which is exactly the announced upper bound for $p_n$.
\end{proof}

\subsection{Weak multiplicativity of $h_{\cS}$}

Our approach in this section, to prove the multiplicative behaviour of $h_{\cS}$, is directly inspired from the seminal angle of attack to the parallel repetition problem for classical non-local games: our Theorem \ref{th:h_sep-h_qext} is an analogue of the exponential decay results by Raz \cite{Raz} and Holenstein \cite{Hol}, while our Theorem \ref{th:w_sep-h_qext} is an analogue of the concentration bound result by Rao \cite{Rao}. Indeed, here in the same spirit as theirs, we want to make precise the following intuition: if the initial state of a system on $(\A\otimes\B)^{\otimes n}$ is product across the cut $\A^n{:}\B^n$, then performing a measurement $(M,\Id-M)$ on only a few subsystems $\A\otimes\B$ should not create too much correlations in the post-measurement state on the remaining subsystems.

Before we prove the main result of this section,
we need to recall one last definition: for any $q\in\N$,
a state $\rho_{\mathrm{A}\mathrm{B}}$ on a bipartite Hilbert space $\mathrm{A}\otimes \mathrm{B}$
is said to be $q$-extendible with respect to $\mathrm{B}$
if there exists a state $\rho_{\mathrm{A}\mathrm{B}^q}$ on $\mathrm{A}\otimes \mathrm{B}^{\otimes q}$
that is invariant under any permutation of the $\mathrm{B}$-subsystems and such
that $\rho_{\mathrm{A}\mathrm{B}}=\tr_{\mathrm{B}^{q-1}}\rho_{\mathrm{A}\mathrm{B}^q}$ (cf.~Chapter \ref{chap:k-extendibility}). We shall denote by $\mathcal{E}_q(\mathrm{A}{:}\mathrm{B})$ the set of
$q$-extendible states with respect to $\mathrm{B}$ on $\mathrm{A}\otimes \mathrm{B}$, and by $h_{\mathcal{E}_q(\mathrm{A}{:}\mathrm{B})}$ its associated support function.

\begin{theorem}
\label{th:h_sep-h_qext}
Let $M$ be an operator on the tensor product Hilbert space $\mathrm{A} \otimes \mathrm{B}$, satisfying $0 \leq M \leq \Id$. Then, for any $q\in\N$,
\begin{equation}
  \label{eq:qext-bound}
  h_{\cS(\mathrm{A}^n{:}\mathrm{B}^n)}\left(M^{\otimes n}\right) \leq \left( 1-\frac{\left(1-h_{\mathcal{E}_q(\mathrm{A}{:}\mathrm{B})}(M)\right)^2}{8\ln 2\,q^2}\right)^n . \end{equation}
And consequently, if $h_{\cS(\mathrm{A}{:}\mathrm{B})}(M)\leq 1-\delta$ for some $0<\delta<1$, then
\begin{equation}
  \label{eq:sep-bound}
  h_{\cS(\mathrm{A}^n{:}\mathrm{B}^n)}\left(M^{\otimes n}\right) \leq \left(1 -\frac{\delta^4}{512\ln 2\,d^4}\right)^n,
\end{equation}
assuming $|\mathrm{A}|=|\mathrm{B}|=d$.
\end{theorem}

\begin{proof}
To establish the first statement \eqref{eq:qext-bound}, we have to show that,
\[
  \forall\ \rho_{\mathrm{A}^n\mathrm{B}^n}\in\mathcal{S}(\mathrm{A}^{n}{:}\mathrm{B}^{n}),\ \tr\left(M_{\mathrm{A}\mathrm{B}}^{\otimes n}\rho_{\mathrm{A}^n\mathrm{B}^n} \right) \leq \left( 1-\frac{\left(1-h_{\mathcal{E}_q(\mathrm{A}{:}\mathrm{B})}(M_{\mathrm{A}\mathrm{B}})\right)^2}{8\ln 2\,q^2}\right)^n .
\]
Note that, with this aim in view, we can without loss of generality focuss only on states which are extremal in $\mathcal{S}(\mathrm{A}^{n}{:}\mathrm{B}^{n})$, namely on states which are product across the cut $\mathrm{A}^{\otimes n}{:}\mathrm{B}^{\otimes n}$. So let $\alpha_{\mathrm{A}^n}\otimes\beta_{\mathrm{B}^n}$ be such a state, and set $p_0=1$, $\tau(0)_{\mathrm{A}^n\mathrm{B}^n}=\alpha_{\mathrm{A}^n}\otimes\beta_{\mathrm{B}^n}$. In the sequel, we will use the following notation: given $I_k\subset[n]$ with $|I_k|=k$, define $M^{(I_k)}_{\mathrm{A}^n\mathrm{B}^n}$ as
\[ M^{(I_k)}_{\mathrm{A}^n\mathrm{B}^n} = M_{\mathrm{A}\mathrm{B}}^{\otimes I_k}\otimes\Id_{\mathrm{A}\mathrm{B}}^{\otimes I_k^c}. \]
Then, for each $1\leq k\leq n$, construct recursively
\begin{align*}
& p_{k}= \tr_{\mathrm{A}^n\mathrm{B}^n} \left[M^{(I_k)}_{\mathrm{A}^n\mathrm{B}^n}\alpha_{\mathrm{A}^n}\otimes\beta_{\mathrm{B}^n}\right], \\
& \tau(k)_{\mathrm{A}_{I_k^c}\mathrm{B}_{I_k^c}}= \frac{1}{p_k} \tr_{\mathrm{A}_{I_k}\mathrm{B}_{I_k}} \left[M^{(I_k)}_{\mathrm{A}^n\mathrm{B}^n}\alpha_{A^n}\otimes\beta_{B^n}\right],
\end{align*}
with $i_{k}$ chosen in $I_{k-1}^c$ such that
\[ E_{sq}\left(\tau(k-1)_{\mathrm{A}_{i_k}\mathrm{B}_{i_k}}\right)\leq \frac{1}{n-k+1}\,\frac{1}{2}\log \frac{1}{p_{k-1}}. \]
We know that this is possible. Indeed, assuming that $p_{k-1}$, $\tau(k-1)_{\mathrm{A}_{I_{k-1}^c}\mathrm{B}_{I_{k-1}^c}}$ have been constructed, Lemma \ref{lemma:E_sq} guarantees that
\[ \frac{1}{n-k+1}\sum_{j=1}^{n-k+1}E_{sq}\left(\tau(k-1)_{\mathrm{A}_{I_{k-1}^c}\mathrm{B}_{I_{k-1}^c}}\right)\leq \frac{1}{n-k+1}\,\frac{1}{2}\log \frac{1}{p_{k-1}}, \]
so that there necessarily exists an index $i\in I_{k-1}^c$ such that $E_{sq}\big(\tau(k-1)_{\mathrm{A}_{i}\mathrm{B}_{i}}\big)$ is smaller than the quantity on the right-hand-side of the average upper bound above.

Now, notice that the $p_k$, $0\leq k\leq n$, are related by the recursion formula
\[ \forall\ 0\leq k\leq n-1,\ p_{k+1} = p_k\tr_{\mathrm{A}_{i_{k+1}}\mathrm{B}_{i_{k+1}}}\left(M_{\mathrm{A}_{i_{k+1}}\mathrm{B}_{i_{k+1}}} \tau(k)_{\mathrm{A}_{i_{k+1}}\mathrm{B}_{i_{k+1}}}\right), \]
where, by the way the $\tau(k)$, $0\leq k\leq n$, are built
\[ E_{sq}\left(\tau(k)_{\mathrm{A}_{i_{k+1}}\mathrm{B}_{i_{k+1}}}\right) \leq \frac{1}{n-k}\frac{1}{2}\log \frac{1}{p_k}. \]
Yet, we know from \cite{LiW} that this implies that
\begin{equation} \label{eq:E_sq-q-ext} \exists\ \sigma_{\mathrm{A}\mathrm{B}}\in\mathcal{E}_q(\mathrm{A}{:}\mathrm{B}):\ \left\| \tau(k)_{\mathrm{A}\mathrm{B}} - \sigma_{\mathrm{A}\mathrm{B}} \right\|_1 \leq \sqrt{2\ln 2}(q-1)\sqrt{\frac{1}{n-k}\frac{1}{2}\log \frac{1}{p_k}} \leq \sqrt{\frac{\ln 2\, q^2}{n-k}\log \frac{1}{p_k}}. \end{equation}
And therefore,
\[ p_{k+1} \leq p_k \left(\left\|M_{\mathrm{A}\mathrm{B}}\right\|_{\infty}\left\| \tau(k)_{\mathrm{A}\mathrm{B}} - \sigma_{\mathrm{A}\mathrm{B}} \right\|_1 + \tr\left(M_{\mathrm{A}\mathrm{B}}\sigma_{\mathrm{A}\mathrm{B}}\right) \right) \leq p_k \left(\sqrt{\frac{\ln 2\, q^2}{n-k}\log \frac{1}{p_k}} + h_{\mathcal{E}_q(\mathrm{A}{:}\mathrm{B})}\left(M_{\mathrm{A}\mathrm{B}}\right) \right). \]
With this upper bound, and because we also clearly have $1> p_1\geq\cdots\geq p_n> 0$ as well as the requirement $p_1\leq h_{\cS(\mathrm{A}{:}\mathrm{B})}(M)\leq h_{\mathcal{E}_q(\mathrm{A}{:}\mathrm{B})}(M)\leq 1-(1-h_{\mathcal{E}_q(\mathrm{A}{:}\mathrm{B})})/2$, it follows from Corollary \ref{cor:recursivity} that
\[ \tr\left(M_{\mathrm{A}\mathrm{B}}^{\otimes n}\alpha_{\A^n}\otimes\beta_{\B^n}\right)=p_n \leq \left( 1-\frac{\left(1-h_{\mathcal{E}_q(\mathrm{A}{:}\mathrm{B})}(M_{\mathrm{A}\mathrm{B}})\right)^2}{8\ln 2\,q^2}\right)^n, \]
which is precisely what we wanted to prove.

From there, the second statement \eqref{eq:sep-bound} easily follows.
Indeed, in the case where $|\mathrm{A}|=|\mathrm{B}|=d$, we know from \cite{CKMR} that,
for any $q\in\N$, if $\rho_{\mathrm{A}\mathrm{B}}\in\mathcal{E}_q(\mathrm{A}{:}\mathrm{B})$ then there exists $\sigma_{\mathrm{A}\mathrm{B}}\in\mathcal{S}(\mathrm{A}{:}\mathrm{B})$ such that $\|\rho_{\mathrm{A}\mathrm{B}}-\sigma_{\mathrm{A}\mathrm{B}}\|_1\leq 2d^2/q$,
so that $h_{\mathcal{E}_q(\mathrm{A}{:}\mathrm{B})}\leq h_{\cS(\mathrm{A}{:}\mathrm{B})}+2d^2/q$. Hence, if $h_{\cS(\mathrm{A}{:}\mathrm{B})}(M_{\mathrm{A}\mathrm{B}})\leq 1-\delta$,
making the choice $q=4d^2/\delta$, in order to have $h_{\mathcal{E}_q(\mathrm{A}{:}\mathrm{B})}(M_{\mathrm{A}\mathrm{B}})\leq 1-\delta/2$,
yields, after a straightforward computation, exactly the announced exponential decay result.
\end{proof}

The scaling as $(\delta/d)^4$ in the upper bound provided by equation \eqref{eq:sep-bound} of Theorem \ref{th:h_sep-h_qext} is much worse than the scaling as $(\delta/d)^2$ in the upper bound provided by Theorem \ref{th:h_sep-n}. However, equation \eqref{eq:qext-bound} of Theorem \ref{th:h_sep-h_qext}, which relates $h_{\cS(\mathrm{A}^n{:}\mathrm{B}^n)}(M^{\otimes n})$ to $h_{\mathcal{E}_q(\mathrm{A}{:}\mathrm{B})}(M)$, may be of interest in some specific cases, namely when $M$ has a maximum overlap with $q$-extendible states which is already of the same order as its maximum overlap with separable states for $q \ll d^2$.

\begin{remark}
By Remark \ref{remark:E_I}, we see that we could also have done the recursive construction described in the proof of Theorem \ref{th:h_sep-h_qext} by imposing instead that, for each $0\leq k\leq n-1$,
\[ p_{k+1} = p_k\tr_{\mathrm{A}_{i_{k+1}}\mathrm{B}_{i_{k+1}}}\left(M_{\mathrm{A}_{i_{k+1}}\mathrm{B}_{i_{k+1}}} \tau(k)_{\mathrm{A}_{i_{k+1}}\mathrm{B}_{i_{k+1}}}\right),\ \text{with}\ E_I\left(\tau(k)_{\mathrm{A}_{i_{k+1}}\mathrm{B}_{i_{k+1}}}\right) \leq \frac{1}{n-k}\frac{1}{2}\log \frac{1}{p_k}. \]
Now, it is an open question to determine whether there exists a dimension-independent constant $C>0$ such that an implication of the following form holds true:
\begin{equation} \label{eq:E_I-conjecture} E_I\left(\rho_{\mathrm{A}\mathrm{B}}\right)\leq \epsilon\ \Rightarrow\ \exists\ \sigma_{\mathrm{A}\mathrm{B}}\in\mathcal{S}(\mathrm{A}{:}\mathrm{B}):\ \left\| \rho_{\mathrm{A}\mathrm{B}} - \sigma_{\mathrm{A}\mathrm{B}} \right\|_1 \leq C\sqrt{\epsilon}. \end{equation}
If Conjecture \eqref{eq:E_I-conjecture} indeed held, this would imply that the $(p_k)_{1\leq k\leq n}$ satisfy
\[
  \forall\ 0\leq k\leq n-1,\
  p_{k+1} \leq p_k \left(\sqrt{\frac{C^2/2}{n-k}\log \frac{1}{p_k}} + h_{\cS(\mathrm{A}{:}\mathrm{B})}\left(M_{\mathrm{A}\mathrm{B}}\right) \right).
\]
And hence eventually, the following dimension-free exponential decay result for $h_{\cS(\mathrm{A}{:}\mathrm{B})}$:
\[
  h_{\cS(\mathrm{A}{:}\mathrm{B})}(M)\leq 1-\delta\ \Rightarrow\ h_{\cS(\mathrm{A}^n{:}\mathrm{B}^n)}\left(M^{\otimes n}\right) \leq \left( 1-\frac{\delta^2}{4C^2}\right)^n . \]
In fact, if a more general variant of Conjecture \eqref{eq:E_I-conjecture} held, with $C\sqrt{\epsilon}$ replaced by $\varphi(\epsilon)$ for $\varphi$ a (universal) non-decreasing function such that $\varphi(0)=0$, then one could prove analogously that
\[ h_{\cS(\mathrm{A}{:}\mathrm{B})}(M)\leq 1-\delta\ \Rightarrow\ h_{\cS(\mathrm{A}^n{:}\mathrm{B}^n)}\left(M^{\otimes n}\right) \leq \left(1-\frac{\varphi^{-1}(\delta)}{4}\right)^n. \]
The way property \eqref{eq:E_sq-q-ext} of strong faithfulness of squashed entanglement with respect to $q$-extendible states, is proved in \cite{LiW} is relying on the breakthrough result by Fawzi and Renner \cite{FR} that small conditional mutual information does imply approximate recoverability. Now, in an even stronger manner than $E_{sq}(\rho)$ being small means that the conditional mutual information of any extension of $\rho$ is small, $E_I(\rho)$ being small is a condition that is expressible as a bunch of conditional mutual information of extensions of $\rho$ being simultaneously small. So it could be that recoverability results (in particular the best one up-to-date \cite{JRSWW}, which carries the advantage over the original one \cite{FR} of being universal and explicit) would help in an attempt to prove a strong faithfulness property of CEMI with respect to separable states such as \eqref{eq:E_I-conjecture}.
\end{remark}

\begin{theorem}
\label{th:w_sep-h_qext}
Let $M$ be an operator on the tensor product Hilbert space $\mathrm{A}\otimes \mathrm{B}$, satisfying $0 \leq M \leq \Id$. If $h_{\cS(\mathrm{A}{:}\mathrm{B})}(M)\leq 1-\delta$ for some $0<\delta<1$, then for any $n,t\in\N$ with $t\geq(1-\delta+\alpha)n$ for some $0<\alpha\leq\delta$, we have
\[ h_{\cS(\mathrm{A}^n{:}\mathrm{B}^n)}\left(M^{(t/n)}\right)\leq\left(1-\frac{\alpha^5}{2048\ln 2\, d^4\,(2\delta-\alpha)}\right)^n, \]
assuming $|\mathrm{A}|=|\mathrm{B}|=d$.
\end{theorem}

\begin{proof}
The proof of this theorem follows a very similar route to that of
Theorem \ref{th:h_sep-h_qext}: for any given state $\alpha_{\mathrm{A}^n}\otimes\beta_{\mathrm{B}^n}$ which is product across the cut $\mathrm{A}^{\otimes n}{:}\mathrm{B}^{\otimes n}$, we want to show that the probability that it passes at least $t$ amongst $n$ tests defined by $M_{\mathrm{A}\mathrm{B}}$ is upper bounded as
\[
  P_t(\alpha_{\mathrm{A}^n}\otimes\beta_{\mathrm{B}^n}) \leq \left(1-\frac{\alpha^5}{2048\ln 2\, d^4\,(2\delta-\alpha)}\right)^n.
\]

In that aim, we start by defining the following deterministic set, number and state: $I_0=\emptyset$, $p_{I_0}=1$ and $\tau(I_0)_{\mathrm{A}^n\mathrm{B}^n}=\alpha_{\mathrm{A}^n}\otimes\beta_{\mathrm{B}^n}$. Then, for each $1\leq k\leq n$, we construct recursively the following random set, number and state: pick $i_k$ uniformly at random in $I_{k-1}^c$, and define
\begin{align*}
& I_k=I_{k-1}\cup\{i_k\}, \\
& p_{I_k}= \tr_{\mathrm{A}^n\mathrm{B}^n} \left[M^{(I_k)}_{\mathrm{A}^n\mathrm{B}^n} \alpha_{\mathrm{A}^n}\otimes\beta_{\mathrm{B}^n}\right],\\
& \tau(I_k)_{\mathrm{A}_{I_k^c}\mathrm{B}_{I_k^c}}= \frac{1}{p_{I_k}}\tr_{\mathrm{A}_{I_k}\mathrm{B}_{I_k}} \left[M^{(I_k)}_{\mathrm{A}^n\mathrm{B}^n} \alpha_{\mathrm{A}^n}\otimes\beta_{\mathrm{B}^n}\right].
\end{align*}
Lemma \ref{lemma:E_sq} guarantees that, on average, for each $0\leq k\leq n-1$,
\[
  E_{sq}\left(\overline{\tau}(I_k)_{\mathrm{A}_{i_{k+1}}\mathrm{B}_{i_{k+1}}}\right)
        \leq \frac{1}{n-k}\frac{1}{2}\log \frac{1}{\overline{p}_{I_k}},
\]
so that, on average, for any $q\in\N$,
\[ \overline{p}_{I_{k+1}} \leq \overline{p}_{I_k} \left(\sqrt{\frac{\ln 2\, q^2}{n-k}\log \frac{1}{\overline{p}_{I_k}}} + h_{\mathcal{E}_q(\mathrm{A}{:}\mathrm{B})}\left(M_{\mathrm{A}\mathrm{B}}\right) \right). \]
In particular, we can make the choice $q=8d^2/\alpha$, in order to have $h_{\mathcal{E}_q(\mathrm{A}{:}\mathrm{B})}(M_{\mathrm{A}\mathrm{B}})\leq 1-\delta+\alpha/4$. And we thus get from Lemma \ref{lemma:recursivity}, after computation, that on average,
\[ \overline{p}_{I_{k_0}} \leq \left(1-\delta+\frac{\alpha}{2}\right)^{k_0}\ \text{for}\ k_0= \frac{\alpha^4}{1024\ln 2\,d^4\,\log[1/(1-\delta+\alpha/2)]+\alpha^4}(n+1)\geq \frac{\alpha^4}{1024\ln 2\,d^4\,(2\delta-\alpha)}\,n. \]

To finish off the proof, we just have to observe (cf.~\cite{Rao}, Section 8) that
\begin{align*} P_t(\alpha_{\mathrm{A}^n}\otimes\beta_{\mathrm{B}^n}) \leq & \sum_{I_{k_0}\subset[n],\,|I_{k_0}|=k_0}\frac{1}{{(1-\delta+\alpha)n \choose k_0}}p_{I_{k_0}}\\
\leq & \frac{{n \choose k_0}}{{(1-\delta+\alpha)n \choose k_0}} \overline{p}_{I_{k_0}}\\
\leq & \left(\frac{n-k_0+1}{(1-\delta+\alpha)n-k_0+1}\right)^{k_0}\left(1-\delta+\frac{\alpha}{2}\right)^{k_0},
\end{align*}
where the last inequality follows from the fact that $\prod_{i=0}^{l-1}(a+i)/(b+i)\leq (a/b)^l$, combined with the upper bound on $\overline{p}_{I_{k_0}}$. In the end, we can therefore conclude that
\[ P_t(\alpha_{\mathrm{A}^n}\otimes\beta_{\mathrm{B}^n}) \leq \left(1-\frac{\alpha}{2}\right)^{k_0}\leq \left(1-\frac{\alpha^5}{2048\ln 2\, d^4\,(2\delta-\alpha)}\right)^n, \]
where the first inequality follows from the fact that $(1-\delta+\alpha/2)/(1-\delta+\alpha)\leq 1-\alpha/2$, while the second inequality is a consequence of the lower bound on $k_0$.
\end{proof}

\begin{remark}
Here again, we see by Remark \ref{remark:E_I} that, if
Conjecture \eqref{eq:E_I-conjecture} held, then we could have
obtained in the proof of Theorem \ref{th:w_sep-h_qext} that, on average
\[
  \overline{p}_{I_{k_0}} \leq \left(1-\delta+\frac{\alpha}{2}\right)^{k_0}\
  \text{for}\
  k_0= \frac{\alpha^2}{8\log[1/(1-\delta+\alpha/2)]+\alpha^2}(n+1)\geq \frac{\alpha^2}{8(2\delta-\alpha)}\,n.
\]
And hence eventually, the following dimension-free concentration result for $h_{\cS(\mathrm{A}{:}\mathrm{B})}$:
\[
  h_{\cS(\mathrm{A}{:}\mathrm{B})}(M)\leq 1-\delta\ \Rightarrow\ \forall\ 0<\alpha<\delta,\ \forall\ t\geq(1-\delta+\alpha)n,\
  h_{\cS(\mathrm{A}^n{:}\mathrm{B^n})}\left(M^{(t/n)}\right) \leq \left( 1-\frac{\alpha^3}{16C^2(2\delta-\alpha)}\right)^n .
\]
\end{remark}

\section{Equivalence between weak multiplicativity of support functions and of maximum fidelities}
\label{sec:general}

In the previous Sections \ref{sec:sep1} and \ref{sec:sep2}, we studied in great depth one particular example of convex constraint on quantum states, namely the separability one. We showed in this specific case that there is a strong connection between the (weakly) multiplicative behaviour under tensoring of either the support function $h_{\cS}$ or the maximum fidelity $F(\cdot,\mathcal{S})$. We would now like to describe, more generally, which kind of convex sets of states exhibit a similar feature.

So let us fix $d\in\N$, $\mathrm{H}$ a $d$-dimensional Hilbert space, and assume that we have a sequence of convex sets
of states $\cK^{(n)}$ on $\mathrm{H}^{\otimes n}$, $n\in\N$, with
the following stability properties (under permutation and partial trace):
\begin{equation}
  \label{eq:sym-stability}
  \rho\in\cK^{(n)}\ \Rightarrow\ \forall\ \pi\in\mathfrak{S}(n),\ U(\pi)\rho U(\pi)^{\dagger}\in\cK^{(n)}\ \text{and}\
  \tr_{\mathrm{H}}\rho\in\cK^{(n-1)}.
\end{equation}
Note that requirement \eqref{eq:sym-stability} implies in particular that,
if $\rho^{\otimes n}\in\cK^{(n)}$, then $\rho\in\cK^{(1)}$. In view of our subsequent discussion, it would be meaningless not to impose that the opposite holds as well, i.e.~that, if $\rho\in\cK^{(1)}$, then $\rho^{\otimes n}\in\cK^{(n)}$. This means in other words that, for each $n\in\N$, $\cK^{(n)}$ is assumed to contain the so-called $n^{\text{th}}$ projective tensor power of $\cK^{(1)}$, which is defined as
\[ \left(\cK^{(1)}\right)^{\hat{\otimes} n} := \mathrm{conv}\left\{\rho_1\otimes\cdots\otimes\rho_n,\ \rho_1,\ldots,\rho_n\in\cK^{(1)} \right\}. \]

\subsection{Exponential decay and concentration of $h_{\mathcal{K}}$ from multiplicativity of $F(\cdot,\mathcal{K})$}

Given an operator $M$ on $\mathrm{H}$, satisfying $0\leq M\leq\Id$, define the support function of $\cK^{(n)}$ at $M^{\otimes n}$ as
\[ h_{\cK^{(n)}}\left(M^{\otimes n}\right) = \underset{\sigma\in\cK^{(n)}}{\sup}\tr\left(M^{\otimes n}\sigma\right). \]
Define also more generally, for any $0\leq t\leq n$, $h_{\cK^{(n)}}\left(M^{(t/n)}\right)$ as the maximum probability for a state in $\cK^{(n)}$ to pass a fraction $t/n$ of $n$ binary tests $(M,\Id-M)$ performed in parallel. The question we are next interested in is to understand how $h_{\cK^{(n)}}(M^{\otimes n})$ and $h_{\cK^{(n)}}\left(M^{(t/n)}\right)$ relate to $h_{\cK^{(1)}}(M)$.

Hence, assume also that these sets $\cK^{(n)}$ satisfy the following condition: there exists a non-decreasing function $f:\epsilon\in]0,1[\mapsto f(\epsilon)\in]0,1[$ such that, for any state $\rho$ on $\C^d$ and any $0<\epsilon<1$,
\begin{equation} \label{eq:exp-decay-F} \left\|\rho-\cK^{(1)}\right\|_2\geq\epsilon\ \Rightarrow\ F\left(\rho^{\otimes n},\cK^{(n)} \right)^2\leq\left(1-f(\epsilon)\right)^n. \end{equation}

Then, under assumption \eqref{eq:exp-decay-F} for the sets $\cK^{(n)}$, the following holds: for any operator $M$ on $\mathrm{H}$, satisfying $0\leq M\leq\Id$, and $\|M\|_2=r$ for some $0\leq r\leq \sqrt{d}$, and any $0<\alpha\leq\delta<1$,
\begin{align*} & h_{\cK^{(1)}}(M)\leq 1-\delta\ \Rightarrow\ h_{\cK^{(n)}}(M^{\otimes n})\leq \left(1-g(\delta,r)\right)^n,\\
& \forall\ t\geq (1-\delta+\alpha)n,\ h_{\cK^{(n)}}\left(M^{(t/n)}\right)\leq e^{-ng'(\alpha,r)}, \end{align*}
where $g(\delta,r)=f(\epsilon(\delta,r))$ for $0<\epsilon(\delta,r)<1$ the solution of the equation $f(\epsilon)=\delta-r\epsilon$ and $g'(\alpha,r)=f(\epsilon(\alpha,r))$ for $0<\epsilon(\alpha,r)<1$ the solution of the equation $f(\epsilon)=2(\alpha-r\epsilon)^2$.

To come to these statements, the strategy is entirely analogous to the one adopted in the proofs of Theorems \ref{th:h_sep-n} and \ref{th:h_sep-t/n}. It is therefore only sketched below. First of all, when looking for a state $\rho\in\cK^{(n)}$ maximizing $\tr\left(M^{\otimes n}\rho\right)$, one can in fact assume without loss of generality that $\rho$ is $n$-symmetric. And for such state $\rho$, reasoning as in the proof of Theorem \ref{th:h_sep-n}, we know that we have
\[ \tr\left(M^{\otimes n}\rho\right) \leq (n+1)^{3d^2} \int_{\sigma\in\mathcal{D}(\mathrm{H})} F\left(\rho,\sigma^{\otimes n}\right)^2 \tr\left(M\sigma\right)^n \mathrm{d}\mu(\sigma). \]
Hence, we get as a consequence of hypothesis \eqref{eq:exp-decay-F} that, for any $0<\epsilon<1$,
\[ \tr\left(M^{\otimes n}\rho\right) \leq (n+1)^{3d^2} \left( \left(1-\delta+r\epsilon\right)^n + \left(1-f(\epsilon)\right)^n\right). \]
So choosing $\epsilon$ such that $f(\epsilon)=\delta-r\epsilon$ and $\rho$ such that $\tr\left(M^{\otimes n}\rho\right)= h_{\cK^{(n)}}(M^{\otimes n})$ yields in particular
\[ h_{\cK^{(n)}}(M^{\otimes n}) \leq 2(n+1)^{3d^2}\left(1-g(\delta,r)\right)^n. \]
Similarly, it follows from hypothesis \eqref{eq:exp-decay-F} as well that, for any $0<\epsilon<1$,
\[ h_{\cK^{(n)}}\left(M^{(t/n)}\right)\leq (n+1)^{3d^2}\left( \exp\left[-2n(\alpha-r\epsilon)^2\right] + \exp\left[-nf(\epsilon)\right] \right), \]
so that choosing $\epsilon$ such that $f(\epsilon)=2(\alpha-r\epsilon)^2$ gives
\[ h_{\cK^{(n)}}\left(M^{(t/n)}\right)\leq 2(n+1)^{3d^2} e^{-ng'(\alpha,r)}. \]
In both cases the polynomial pre-factor $2(n+1)^{3d^2}$ can then be removed by the exact same argument as in the proof of Theorem \ref{th:h_sep-n}.

\subsection{Weak multiplicativity of $F(\cdot,\mathcal{K})$ from exponential decay and concentration of $h_{\mathcal{K}}$}

We would now like to go in the other direction. Namely, let us assume this time that these sets $\cK^{(n)}$ satisfy the following condition: there exists a function $f:(\alpha,d)\in]0,1[\times\N\mapsto f(\alpha,d)\in]0,1[$, non-decreasing in $\alpha$ and non-increasing in $d$, such that, for any operator $M$ on $\mathrm{H}$, satisfying $0\leq M\leq\Id$, and any $0<\alpha\leq\delta<1$,
\begin{equation} \label{eq:exp-decay-h} h_{\cK^{(1)}}(M)\leq 1-\delta\ \Rightarrow\ \forall\ t\geq (1-\delta+\alpha)n,\ h_{\cK^{(n)}}\left(M^{(t/n)}\right)\leq e^{-nf(\alpha,d)}. \end{equation}

Then, under assumption \eqref{eq:exp-decay-h} for the sets $\cK^{(n)}$, the following holds: for any state $\rho$ on $\mathrm{H}$ and any $0<\epsilon<1$,
\[ \frac{1}{2}\left\|\rho-\cK^{(1)}\right\|_1 \geq \epsilon\ \Rightarrow\ F\left(\rho^{\otimes n},\cK^{(n)}\right) \leq 2e^{-ng(\epsilon,d)}, \]
where $g(\epsilon,d)=f(\alpha(\epsilon,d),d)/2$ for $0<\alpha(\epsilon,d)<1$ the solution of the equation $f(\alpha,d)/2=(\alpha-\epsilon)^2$.

Here is the strategy to derive such result: Imagine you are given a state on $\mathrm{H}^{\otimes n}$, which you know is either $\rho^{\otimes n}$ or in $\cK^{(n)}$, and you want to decide between these two hypotheses. For that, you can design a binary test $(T_+,T_-)$ such that outcome $+$ is obtained with a high probability $p$ if the state was $\rho^{\otimes n}$ and outcome $-$ is obtained with a high probability $q$ if the state was in $\cK^{(n)}$. Then, clearly
\[ F\left(\rho^{\otimes n},\cK^{(n)}\right) \leq F\left((p,1-p),(q,1-q)\right) \leq \sqrt{1-p}+\sqrt{1-q}. \]
Therefore, if both error probabilities $1-p$ and $1-q$ are exponentially small, the conclusion follows.

In the present case, the fact that $\left\|\rho-\cK^{(1)}\right\|_1=2\epsilon$, implies that there exist $0\leq M\leq \Id$ and $\epsilon<\eta<1$ such that $\tr(M\rho)=1-\eta+\epsilon$ whereas $h_{\cK^{(1)}}(M)= 1-\eta$. So consider the binary POVM $(M_0,M_1)=(M,\Id-M)$ performed $n$ times in parallel, and the corresponding binary test $(T_+,T_-)$ with $+$ being the event ``outcome $0$ is obtained more than $(1-\eta+\alpha)n$ times'' and $-$ being the event ``outcome $0$ is obtained less than $(1-\eta+\alpha)n$ times'', for some $0<\alpha<\epsilon$ to be chosen later. Define next, for each $1\leq i\leq n$, the random variable $X_i$, respectively $Y_i$, as the outcome of measurement number $i$ given that the state was $\rho^{\otimes n}$, respectively in $\cK^{(n)}$. We then have
\begin{align*}
& 1-p = \P\left(-\big|\rho^{\otimes n}\right) = \P\left(\sum_{i=1}^n X_i < (1-\eta+\alpha)n\right),\\
& 1-q = \P\left(+\big|\cK^{(n)}\right) = \P\left(\sum_{i=1}^n Y_i > (1-\eta+\alpha)n\right).
\end{align*}
Yet on the one hand, $X_1,\ldots,X_n$ are independent Bernoulli random variables with expectation $1-\eta+\epsilon$, so by Hoeffding's inequality
\[ \P\left(\sum_{i=1}^n X_i < (1-\eta+\alpha)n\right) \leq e^{-2n(\epsilon-\alpha)^2}. \]
While on the other hand, for any $0\leq t\leq n$, $\P\left(\sum_{i=1}^n Y_i > t\right) = h_{\cK^{(n)}}\left(M^{(t/n)}\right)$, so assumption \eqref{eq:exp-decay-h} guarantees that
\[ \P\left(\sum_{i=1}^n Y_i > (1-\eta+\alpha)n\right) \leq e^{-nf(\alpha,d)}. \]
Hence, putting everything together, we eventually obtain that, for any $0<\alpha<\epsilon$,
\[ F\left(\rho^{\otimes n},\cK^{(n)}\right) \leq e^{-n(\epsilon-\alpha)^2} + e^{-nf(\alpha,d)/2}, \]
which yields the wanted result after choosing $\alpha$ such that $f(\alpha,d)/2=(\epsilon-\alpha)^2$.

\begin{remark}
Note that requirement \eqref{eq:sym-stability} is clearly fulfilled by the sets $\mathcal{S}_{\mathrm{A}^n:\mathrm{B}^n}$ of biseparable states on $\left(\mathrm{A}\otimes\mathrm{B}\right)^{\otimes n}$. Furthermore, they satisfy requirements \eqref{eq:exp-decay-F} and \eqref{eq:exp-decay-h} as well, with $f(\epsilon)=\epsilon^2/4$ and $f(\alpha,d^2)=\alpha^2/5d^2$.

It may also be worth emphasizing that conditions \eqref{eq:exp-decay-F} and \eqref{eq:exp-decay-h} are just strengthened and quantitative versions of the following stability property for the sets $\cK^{(n)}$: $\rho\notin\cK^{(1)}\ \Rightarrow\ \rho^{\otimes n}\notin\cK^{(n)}$, i.e.~ equivalently $\rho^{\otimes n}\in\cK^{(n)}\ \Rightarrow\ \rho\in\cK^{(1)}$.
\end{remark}

\subsection{One simple example}

Let us look at what the previous discussion becomes in the case of the simplest possible sequence $\{\mathcal{K}^{(n)},\ n\in\N\}$ satisfying requirement \eqref{eq:sym-stability}, namely when there exists a set of states $\mathcal{K}$ on $\mathrm{H}$ such that, for each $n\in\N$, $\mathcal{K}^{(n)}$ is exactly the $n^{\text{th}}$ projective tensor power of $\mathcal{K}$, i.e.~\[ \mathcal{K}^{(n)}= \mathcal{K}^{\hat{\otimes} n} := \mathrm{conv}\left\{\rho_1\otimes\cdots\otimes\rho_n,\ \rho_1,\ldots,\rho_n\in\mathcal{K} \right\}. \]
Then, assumption \eqref{eq:exp-decay-h} is clearly satisfied, in the following way: for any operator $M$ on $\mathrm{H}$, satisfying $0\leq M\leq\Id$, and any $0<\alpha\leq\delta<1$,
\[ h_{\cK^{(1)}}(M)\leq 1-\delta\ \Rightarrow\ \forall\ t\geq (1-\delta+\alpha)n,\ h_{\cK^{(n)}}\left(M^{(t/n)}\right)\leq e^{-n2\alpha^2}. \]
This is a consequence of Hoeffding's inequality, following an argument similar to the one detailed in the previous subsection. And by the result established in the latter, this implies that: for any state $\rho$ on $\mathrm{H}$ and any $0<\epsilon<1$,
\[ F\left(\rho,\cK^{(1)}\right)\leq e^{-\epsilon}\ \Rightarrow\ F\left(\rho^{\otimes n},\cK^{(n)}\right) \leq 2e^{-n\epsilon^2/8}. \]
This is because $F(\rho,\cK^{(1)})\leq e^{-\epsilon}\ \Rightarrow\ \|\rho-\cK^{(1)}\|_1/2 \geq 1-e^{-\epsilon}\geq \epsilon/2$.

In connection with the discussion developed in Sections \ref{sec:sep1} and \ref{sec:sep2}, we see that we are actually facing the following interesting open question: how differently do $\mathcal{S}(\mathrm{A}^n{:}\mathrm{B}^n)$ and $\mathcal{S}(\mathrm{A}{:}\mathrm{B})^{\hat{\otimes} n}$ behave, from the (more or less equivalent) points of view of support functions and maximum fidelity functions?

\section{De Finetti reductions for infinite-dimensional symmetric quantum systems}
\label{sec:infinite}

All quantum de Finetti theorems and reductions require a bound on the dimension of the involved
Hilbert spaces. So what can be said about symmetric states on $\mathrm{H}^{\otimes n}$ when $\mathrm{H}$ is an infinite-dimensional Hilbert space? What extra assumptions do we need on them in order to be able to reduce their study to that of states in some de Finetti form? The original de Finetti reduction of \cite{C-K-R} was especially designed to prove the security of QKD protocols against general attacks. Yet, showing security of continuous variable QKD is also a major issue. This was the motivation behind the infinite-dimensional de Finetti type theorem of \cite{CR}. Our ultimate goal here is the same, which we rather try to achieve via a de Finetti reduction under constraints.

\subsection{Infinite-dimensional post-selection lemma}

Let $\mathrm{H}$ be an infinite-dimensional Hilbert space, and let
$\bar{\mathrm{H}}\subset\mathrm{H}$ be a finite $d$-dimensional subspace of $\mathrm{H}$.
Denote by $\{\ket{j}\}_{j\in\N}$ an orthonormal basis of $\mathrm{H}$, chosen such
that $\{\ket{j}\}_{1\leq j\leq d}$ is an orthonormal basis of $\bar{\mathrm{H}}$.
Then, for any $n,k\in\N$, the $(n+k)$-symmetric subspace of
$\bar{\mathrm{H}}^{\otimes n}\otimes\mathrm{H}^{\otimes k}$ is defined as
\[ \Sym^{n+k}\left(\bar{\mathrm{H}},\mathrm{H}\right) := \Span\left\{ \sum_{\pi\in\mathfrak{S}(n+k)} \ket{j_{\pi(1)}}\otimes\cdots\otimes\ket{j_{\pi(n+k)}} \st j_1\leq\cdots\leq j_{n+k},\ \forall\ 1\leq q\leq n,\ j_q\leq d\right\}. \]
Note that, denoting by $\bar{\mathrm{H}}_{\perp}\subset\mathrm{H}$ the orthogonal complement of $\bar{\mathrm{H}}$, i.e.~ $\mathrm{H}=\bar{\mathrm{H}}\oplus\bar{\mathrm{H}}_{\perp}$, we have
\[ \Sym^{n+k}\left(\bar{\mathrm{H}},\mathrm{H}\right) \subset \mathrm{V}^{n+k}\left(\bar{\mathrm{H}},\mathrm{H}\right) := \bigoplus_{\underset{|I|\geq n}{I\subset[n+k]}} \bar{\mathrm{H}}_{\perp}^{\otimes I^c}\otimes\Sym\left(\bar{\mathrm{H}}^{\otimes I}\right). \]

\begin{lemma}
\label{lemma:ps-infinite}
Let $\mathrm{H}$ be an infinite-dimensional Hilbert space, and let $\bar{\mathrm{H}}\subset\mathrm{H}$ be a finite $d$-dimensional subspace of $\mathrm{H}$. Let also $n,k\in\N$. Then, any unit vector $\ket{\theta}\in\Sym^{n+k}\left(\bar{\mathrm{H}},\mathrm{H}\right)$ satisfies
\[ \ketbra{\theta}{\theta} \leq \left[\sum_{q=0}^k{n+k \choose q} {n+d-1 \choose n}^3 \right] \sum_{\underset{|I|\geq n}{I\subset[n+k]}} \int_{\ket{x}\in S_{\bar{\mathrm{H}}}} \epsilon(\theta_x)_{\bar{\mathrm{H}}_{\perp}^{I^c}}\otimes\ketbra{x}{x}_{\bar{\mathrm{H}}}^{\otimes I} \mathrm{d}x , \]
where for all $0\leq q\leq k$ and all unit vector $\ket{x}\in \bar{\mathrm{H}}$, $\epsilon(\theta_x)_{\bar{\mathrm{H}}_{\perp}^{k-q}}=\tr_{\bar{\mathrm{H}}^{ n+q}}\big[\big(\Id_{\bar{\mathrm{H}}_{\perp}}^{\otimes k-q}\otimes\ketbra{x}{x}_{\bar{\mathrm{H}}}^{\otimes n+q}\big)\ketbra{\theta}{\theta}\big]$ is a sub-normalized state on $\bar{\mathrm{H}}_{\perp}^{\otimes k-q}$. 
\end{lemma}

\begin{proof}
Since $\Sym^{n+k}\left(\bar{\mathrm{H}},\mathrm{H}\right)\subset \mathrm{V}^{n+k}\left(\bar{\mathrm{H}},\mathrm{H}\right)$, any unit vector $\ket{\theta}\in\Sym^{n+k}\left(\bar{\mathrm{H}},\mathrm{H}\right)$ satisfies
\begin{align*}
\ketbra{\theta}{\theta} = & P_{\mathrm{V}^{n+k}\left(\bar{\mathrm{H}},\mathrm{H}\right)} \ketbra{\theta}{\theta} P_{\mathrm{V}^{n+k}\left(\bar{\mathrm{H}},\mathrm{H}\right)}^{\dagger}\\
= & {n+d-1 \choose n}^2 \sum_{\underset{|I|,|J|\geq n}{I,J\subset[n+k]}} \int_{\ket{x},\ket{y}\in S_{\bar{\mathrm{H}}}} \left(\Id_{\bar{\mathrm{H}}_{\perp}}^{\otimes I^c}\otimes\ketbra{x}{x}_{\bar{\mathrm{H}}}^{\otimes I}\right) \ketbra{\theta}{\theta} \left(\Id_{\bar{\mathrm{H}}_{\perp}}^{\otimes J^c}\otimes\ketbra{y}{y}_{\bar{\mathrm{H}}}^{\otimes J}\right)^{\dagger} \mathrm{d}x\mathrm{d}y.
\end{align*}
Now, by Lemma \ref{lemma:pinching} (and using the same Caratheodory argument as in the proof of Proposition \ref{prop:ps-pure}), we have
\begin{align*}
& \sum_{\underset{|I|,|J|\geq n}{I,J\subset[n+k]}} \int_{\ket{x},\ket{y}\in S_{\bar{\mathrm{H}}}} \left(\Id_{\bar{\mathrm{H}}_{\perp}}^{\otimes I^c}\otimes\ketbra{x}{x}_{\bar{\mathrm{H}}}^{\otimes I}\right) \ketbra{\theta}{\theta} \left(\Id_{\bar{\mathrm{H}}_{\perp}}^{\otimes J^c}\otimes\ketbra{y}{y}_{\bar{\mathrm{H}}}^{\otimes J}\right)^{\dagger} \mathrm{d}x\mathrm{d}y\\
& \ \ \ \leq \left[\sum_{q=0}^k{n+k \choose q} {n+d-1 \choose n}\right] \sum_{\underset{|I|\geq n}{I\subset[n+k]}} \int_{\ket{x}\in S_{\bar{\mathrm{H}}}} \left(\Id_{\bar{\mathrm{H}}_{\perp}}^{\otimes I^c}\otimes\ketbra{x}{x}_{\bar{\mathrm{H}}}^{\otimes I}\right) \ketbra{\theta}{\theta} \left(\Id_{\bar{\mathrm{H}}_{\perp}}^{\otimes I^c}\otimes\ketbra{x}{x}_{\bar{\mathrm{H}}}^{\otimes I}\right)^{\dagger} \mathrm{d}x.\\
\end{align*}
We then just have to notice that, for any $0\leq q\leq k$ and any unit vector $\ket{x}\in \bar{\mathrm{H}}$,
\[ \big(\Id_{\bar{\mathrm{H}}_{\perp}}^{\otimes k-q}\otimes\ketbra{x}{x}_{\bar{\mathrm{H}}}^{\otimes n+q}\big) \ketbra{\theta}{\theta} \big(\Id_{\bar{\mathrm{H}}_{\perp}}^{\otimes k-q}\otimes\ketbra{x}{x}_{\bar{\mathrm{H}}}^{\otimes n+q}\big)^{\dagger} = \tr_{\bar{\mathrm{H}}^{ n+q}}\big[\big(\Id_{\bar{\mathrm{H}}_{\perp}}^{\otimes k-q}\otimes\ketbra{x}{x}_{\bar{\mathrm{H}}}^{\otimes n+q}\big)\ketbra{\theta}{\theta}\big]\otimes\ketbra{x}{x}_{\bar{\mathrm{H}}}^{\otimes n+q}, \]
in order to actually get the advertised result.
\end{proof}

\subsection{An application}

One application of Lemma \ref{lemma:ps-infinite} is to the security analysis of quantum cryptographic schemes, when there is no a priori bound on the dimension of the information carriers. This problem was originally investigated in \cite{CR} via a de Finetti theorem specifically designed for it. It was shown there that, under experimentally verifiable conditions, it is possible to ensure the security of quantum key distribution (QKD) protocols with continuous variables against general attacks. We show here that similar conclusions can be reached using the de Finetti reduction of Lemma \ref{lemma:ps-infinite}.

We look at things from the exact same point of view as the one adopted in \cite{CR}. Let $\mathrm{H}$ be an infinite-dimensional Hilbert space and let $X,Y$ be two canonical operators on $\mathrm{H}$. Then, denote by $\Lambda=X^2+Y^2$ the corresponding Hamiltonian, fix $\lambda_0>0$, and define
\[ \bar{\mathrm{H}}:=\left\{ \ket{\theta}\in\mathrm{H} \st \Lambda\ket{\theta}=\lambda\ket{\theta},\ \lambda\leq\lambda_0\right\}, \]
finite-dimensional subspace of $\mathrm{H}$ spanned by the eigenvectors of $\Lambda$ with associated eigenvalue at most $\lambda_0$.

Let $n,k\in\N$, with $n\geq 2k$, and let $\rho^{(n+2k)}$ be a $(n+2k)$-symmetric state on $\mathrm{H}^{\otimes n+2k}$. Next, define the two events $\mathcal{A}$ and $\mathcal{B}$ as
\begin{align*}
\mathcal{A}\ = & \ \text{``}\ \forall\ 1\leq q\leq k,\ \tr\left(\Lambda\rho_q^{(1)}\right) \leq \lambda_0\ \text{''}\\
\mathcal{B}\ = & \ \text{``}\ \exists\ \ket{\theta^{(n+k)}}\in\Sym^{n+k}\left(\bar{\mathrm{H}}\otimes\bar{\mathrm{H}}',\mathrm{H}\otimes\mathrm{H}'\right):\ \rho^{(n+k)}=\tr_{\mathrm{H}'^{n+k}} \ketbra{\theta^{(n+k)}}{\theta^{(n+k)}}\ \text{''},
\end{align*}
where for all $1\leq q\leq k$, $\rho_q^{(1)}=\tr_{\mathrm{H}^{n+2k}\setminus\mathrm{H}_q}\rho^{(n+2k)}$, and $\rho^{(n+k)}=\tr_{\mathrm{H}_{k+1}\cdots\mathrm{H}_{2k}}\rho^{(n+2k)}$. We know by Lemma III.3 in \cite{CR} that there exist universal constants $C_0,c>0$ such that, whenever $\lambda_0\geq C_0\log (n/k)$, we have
\[ \P\left(\mathcal{A}\wedge\neg\mathcal{B}\right) \leq e^{-ck^3/n^2} . \]
In words, this means the following. Fix a threshold $\lambda_0\geq C_0\log (n/k)$, and assume that when measuring the energy $\Lambda$ on the $k$ first subsystems of $\rho^{(n+2k)}$, only values below $\lambda_0$ are obtained. Then, with probability greater than $1-e^{-ck^3/n^2}$, the remaining $n+k$ subsystems of $\rho^{(n+2k)}$ have a purification which is the symmetrization of a state with more than $n$ subsystems supported in $\bar{\mathrm{H}}\otimes\bar{\mathrm{H}}'$.

Now, let $\widetilde{\rho}$ be a state on $\mathrm{H}^{\otimes n+k}$ such that $\widetilde{\rho}=\tr_{\mathrm{H}'^{n+k}} \ketbra{\widetilde{\theta}}{\widetilde{\theta}}$ for some $\ket{\widetilde{\theta}} \in \Sym^{n+k}\left(\bar{\mathrm{H}}\otimes\bar{\mathrm{H}}',\mathrm{H}\otimes\mathrm{H}'\right)$. 
Denoting by $d$ the dimension of $\bar{\mathrm{H}}$, we have by Lemma \ref{lemma:ps-infinite} that $\ket{\widetilde{\theta}}$ satisfies
\[ \ketbra{\widetilde{\theta}}{\widetilde{\theta}} \leq \left[\sum_{q=0}^k {n+k \choose q}{n+d^2-1 \choose n}^3\right] \sum_{\underset{|I|\geq n}{I\subset[n+k]}} \int_{\ket{x}\in S_{\bar{\mathrm{H}}\otimes\bar{\mathrm{H}}'}} \epsilon(\widetilde{\theta}_x)_{(\mathrm{H}\mathrm{H}')^{ I^c}}\otimes\proj{x}_{\bar{\mathrm{H}}\bar{\mathrm{H}}'}^{\otimes I} \mathrm{d}x. \]
And hence, after partial tracing over $\mathrm{H}'^{\otimes n+k}$, we finally get
\[ \widetilde{\rho} \leq \left[\sum_{q=0}^k {n+k \choose q}{n+d^2-1 \choose n}^3\right] \sum_{\underset{|I|\geq n}{I\subset[n+k]}} \int_{\ket{x}\in S_{\bar{\mathrm{H}}\otimes\bar{\mathrm{H}}'}} \varepsilon(\widetilde{\theta}_x)_{\mathrm{H}^{I^c}}\otimes\sigma(x)_{\bar{\mathrm{H}}}^{\otimes I} \mathrm{d}x, \]
where for all $0\leq q\leq k$ and all unit vector $\ket{x}\in \bar{\mathrm{H}}\otimes\bar{\mathrm{H}}'$, $\sigma(x)_{\bar{\mathrm{H}}}=\tr_{\bar{\mathrm{H}}'}\proj{x}_{\bar{\mathrm{H}}\bar{\mathrm{H}}'}$ is the reduced state of $\proj{x}$ on $\bar{\mathrm{H}}$, and
$\varepsilon(\widetilde{\theta}_x)_{\mathrm{H}^{k-q}}=\tr_{\mathrm{H}'^{k-q}}\epsilon(\widetilde{\theta}_x)_{(\mathrm{H}\mathrm{H}')^{k-q}}$ is the reduced sub-normalized state of $\epsilon(\widetilde{\theta}_x)$ on $\mathrm{H}^{\otimes k-q}$.


Putting everything together, we can eventually get the following: Let $n\in\N$ and $k=\lfloor n^{\alpha} \rfloor$ for a given $\alpha$ fulfilling $2/3<\alpha<1$. Let also $\lambda_0$ be a threshold such that on the one hand $\lambda_0\geq C_0\log n$, where $C_0>0$ is a universal constant, and on the other hand $d\leq n^{\beta}$ for a given $\beta$ fulfilling $0<\beta<1/2$. Suppose next that $\rho^{(n+2k)}$ is a $(n+2k)$-symmetric state on $\mathrm{H}^{\otimes n+2k}$ such that event $\mathcal{A}$ holds. Then, with probability greater than $1-e^{-cn^{3\alpha-2}}$, where $c>0$ is a universal constant, the reduced state $\rho^{(n+k)}$ of $\rho^{(n+2k)}$ on $\mathrm{H}^{\otimes n+k}$ satisfies
\[ \rho^{(n+k)} \leq (Cn)^{n^{\alpha}+n^{2\beta}} \sum_{\underset{|I|\geq n}{I\subset[n+k]}} \int_{\sigma_{\bar{\mathrm{H}}}\in\mathcal{D}(\bar{\mathrm{H}})} \varepsilon(\rho,\sigma)_{\mathrm{H}^{I^c}}\otimes\sigma_{\bar{\mathrm{H}}}^{\otimes I} \mathrm{d}\mu(\sigma_{\bar{\mathrm{H}}}), \]
where $C>0$ is a universal constant, $\mu$ is a probability measure on the set of states on $\bar{\mathrm{H}}$, and for each $0\leq q\leq k$ and each state $\sigma$ on $\bar{\mathrm{H}}$, $\varepsilon(\rho,\sigma)_{\mathrm{H}^{k-q}}$ is a sub-normalized state on $\mathrm{H}^{\otimes k-q}$.

Now, let $\mathcal{N}:\mathcal{L}(\mathrm{H})\rightarrow\mathcal{L}(\mathrm{K})$ be a quantum channel, and assume that there exists some $0<\delta<1$ such that
\[ \sup\left\{ \left\|\mathcal{N}(\sigma)\right\|_1 \st \sigma\in\mathcal{D}(\bar{\mathrm{H}})\right\} \leq \delta. \]
This implies in particular that, for any $0\leq q\leq k$, and any states $\varepsilon$ on $\mathrm{H}^{\otimes k-q}$, $\sigma$ on $\bar{\mathrm{H}}$, we have
\[ \left\|\mathcal{N}^{\otimes n+k}\left(\varepsilon\otimes\sigma^{\otimes n+q}\right)\right\|_1 = \left\|\mathcal{N}^{\otimes k-q}\left(\varepsilon\right)\right\|_1\left\| \mathcal{N}(\sigma)\right\|_1^{n+q} \leq \delta^{n+q}. \]
And subsequently, by what precedes, for any state $\rho^{(n+2k)}$ on $\mathrm{H}^{\otimes n+2k}$ such that event $\mathcal{A}$ holds, denoting by $\rho^{(n+k)}$ its reduced state on $\mathrm{H}^{\otimes n+k}$, it holds with probability greater than $1-e^{-cn^{3\alpha-2}}$ that
\begin{align*} \left\|\mathcal{N}^{\otimes n+k}\big(\rho^{(n+k)}\big)\right\|_1 \leq & \, (Cn)^{n^{\alpha}+n^{2\beta}} \sum_{q=0}^{k}{n+k \choose q} \sup\left\{ \left\|\mathcal{N}^{\otimes n+k}\left(\varepsilon\otimes\sigma^{\otimes n+ q}\right)\right\|_1 \st \varepsilon\in\mathcal{D}(\mathrm{H}^{\otimes k-q}),\ \sigma\in\mathcal{D}(\bar{\mathrm{H}}) \right\}\\
\leq & \, (Cn)^{n^{\alpha}+n^{2\beta}}\sum_{q=0}^{k}{n+k \choose q}\delta^{n+q}\\
\leq & \, (C'n)^{n^{\alpha}+n^{2\beta}}\, \delta^n, \end{align*}
where $C'>0$ is a universal constant. By the way $\alpha,\beta$ have been chosen, this means that, for any $\tilde{\delta}>\delta$ and $n\geq n_{\tilde{\delta}}$, we have with high probability
\[
  \sup\left\{ \left\|\mathcal{N}^{\otimes n+k}\big(\rho^{(n+k)}\big)\right\|_1 \st \rho^{(n+k)}=\tr_{\mathrm{H}^{\otimes k}}\rho^{(n+2k)}\ \text{with}\ \rho^{(n+2k)}\in\mathcal{D}(\mathrm{H}^{\otimes n+2k})\ \text{s.t.}\ \mathcal{A}\ \text{holds} \right\}
  \leq \tilde{\delta}^n.
\]

\section{Conclusion and outlook}

We have reviewed (and given a new proof of) the constrained de Finetti
reduction of \cite{D-S-W}. We have demonstrated its adaptability to various situations
where one would like to impart a known constraint satisfied by a
permutation symmetric state onto the i.i.d.~states occurring in the
operator with which to compare it. We have seen that our technique works especially well in the case of linear constraints (see \cite{D-S-W} and Chapter \ref{chap:SNOS} for two developed such applications).

We have then spent considerable effort on a particularly interesting
convex constraint, separability. Apart from the obvious relevance
to entanglement theory, the constrained de Finetti reduction provides a
very natural framework in which to derive bounds on the success
probability of parallel repetitions of tests, and has immediate
applications in the parallel repetition of $\mathrm{QMA}(2)$, quantum
Merlin-Arthur interactive proof systems with two unentangled provers (see \cite{HM} for further details).
Conversely, we showed that certain progress in entanglement theory
(on the conjectured faithfulness properties of the CEMI entanglement
measure for instance) would imply even stronger, dimension-independent bounds, which
would show in particular that the soundness gap of $\mathrm{QMA}(2)$ can
be amplified exponentially by parallel repetition, without any other
devices. It is curious to see that the progress on questions like this
can depend on the properties of a simple, but little-understood
entanglement measure such as CEMI, and we would like to recommend
its study to the reader's attention. Indeed, it seems to be the best candidate so far for a \textit{magical}, or even \textit{supercalifragilistic} entanglement measure. The latter is defined as one which has the post-selection property with respect to an initial product state and measurement on a separate subsystem (cf.~Lemma \ref{lemma:post-POVM}), is super-additive, and satisfies a universal faithfulness bound with respect to the trace-norm distance (cf.~Conjecture \eqref{eq:E_I-conjecture}).

We have also presented a more abstract framework of convex
constraints, that allows us to demonstrate in greater generality
the interplay between the multiplicative behaviour
of $(i)$ the support function and $(ii)$ the maximum fidelity function. The way $(i)$ is derived from $(ii)$ is via our de Finetti reduction with fidelity weight in the upper bounding operator. And $(ii)$ is obtained from $(i)$ by constructing a test whose failure probability decays exponentially under parallel repetition.

Finally, seeing that the de Finetti reductions had been so far always
limited by the finite dimensionality of the system
involved, we have made first steps towards an extension of
the main technical tool to infinite-dimensional systems under
suitable constraints. It remains to be seen how widely it or a variation can be applied to quantum cryptography in continuous variable systems \cite{CR,L-GP-R-C}, or similar problems.

\chapter{Parallel repetition and concentration for (sub-)no-signalling games}
\chaptermark{Parallel repetition and concentration for (sub-)no-signalling games}
\label{chap:SNOS}

\textsf{Based on ``Parallel repetition and concentration for (sub-)no-signalling games via a flexible constrained de Finetti reduction'', in collaboration with A. Winter \cite{LW2}.}

\bigskip

We use a recently discovered constrained de Finetti reduction (aka ``Post-Selection Lemma'') to study the parallel repetition of multi-player non-local games under no-signalling strategies. Since the technique allows us to reduce general strategies to independent plays, we obtain parallel repetition (corresponding to winning all rounds) in the same way as exponential concentration of the probability to win a fraction larger than the value of the game.

Our proof technique leads us naturally to a relaxation of no-signalling (NS) strategies, which we dub \emph{sub-no-signalling (SNOS)}. While for two players the two concepts coincide, they differ for three or more players. Our results are most complete and satisfying for arbitrary number of sub-no-signalling players, where we get universal parallel repetition and concentration for any game, while the no-signalling case is obtained as a corollary, but only for games with ``full support''.

\section{Non-local games and no-signalling strategies}
\label{sec:games}
A multi-player non-local game is played between cooperating but non-communicating
players. Each player receives an input from some input alphabet and has to
produce an output in some output alphabet. The common goal of the players
is to satisfy some pre-defined predicate on their inputs and outputs. For that,
they may agree on a strategy before the game starts, but are then not allowed to
communicate anymore. Such games are especially relevant in theoretical physics in
the context of the foundations of quantum mechanics and quantum information,
and in computer science where they arise in multi-prover interactive proof
systems.
Indeed, they may provide an intuitive and quantitative understanding of the role
played by various degrees of correlations in global systems which are composed of
several local subsystems. These games also arise in complexity theory, under the
formulation of multi-provers with some shared resources producing a protocol that
should convince a referee, or in cryptography as attacks from malicious parties
having a more or less restricted physical power.

The \emph{value} of a game is the maximum winning probability of the players,
over all allowed joint strategies, using possibly some prescribed correlation
resource such as shared randomness, quantum entanglement or no-signalling
correlations. It has been a subject of considerable study how the availability of different
resources affects the values of certain games \cite{Bell,C-H-S-H,Tsi,P-R,B-L-M-P-P-R}.

In this context, a natural question is how the value of a game behaves when $n$ independent
instances of the game are played simultaneously, i.e.~each player gets
$n$ independent inputs and has to provide $n$ outputs such that each
game instance is won (or a large fraction of them). This is the parallel
repetition problem (which obviously has the exact same flavor as the question of how support functions behave under tensoring, that was at the heart of Chapter \ref{chap:deFinetti}). Playing independently the optimal single-game strategy
on all $n$ game instances will result in an exponentially decreasing
winning probability. But although that was found paradoxical at first, this
is in general not optimal \cite{Feige,F-V}.
For classical two-player games, Raz~\cite{Raz}, later simplified and
improved by Holenstein~\cite{Hol}, established the first general parallel
repetition theorem, showing that the value of $n$ repetitions decreases
exponentially for every game. Holenstein~\cite{Hol} also proved an analogous
parallel repetition theorem for the no-signalling value of general
two-player games. Only recently, parallel repetition theorems
were proved for the entangled value of two-player games: for general games, nothing better than a polynomial decay result is known up to now~\cite{K-V}, while exponential decay results have been established in several special cases (perfect parallel repetition for XOR games~\cite{C-S-U-U}, exponential decrease under parallel repetition for unique games~\cite{K-R-T}, projection games~\cite{D-S-V}, free games~\cite{C-S,J-P-Y}).

Multi-player games have received little attention until recently. And apart from the result in \cite{C-W-Y} (containing both classical and quantum statements), only in the no-signalling setting~\cite{B-F-S,AF-R-V}.
The present work has the same focus on multiple no-signalling players,
albeit we will find that the theory becomes much more satisfying
for \emph{sub-no-signalling} players.

Specifically, we will consider here $\ell$-player games $G$ with input alphabets
$\cX_1,\ldots,\cX_\ell$ and output alphabets $\cA_1,\ldots,\cA_\ell$.
By way of notation,
\[ \underline{\cX} := \bigtimes_{i=1}^\ell \cX_i\ \text{and}\ \underline{\cA} := \bigtimes_{i=1}^\ell \cA_i. \]
Furthermore,
for any subset $I\subset [\ell]$ of indices,
\[ \cX_I := \bigtimes_{i\in I}\cX_i\ \text{and}\ \cA_I := \bigtimes_{i\in I} \cA_i. \]
For any $I,J\subset [\ell]$, given $T$ a probability distribution on $\cX_I$, resp.~$P$ a conditional probability distribution on $\cA_J|\cX_I$, we may denote it by $T_{\cX_I}$, resp.~$P_{\cA_J|\cX_I}$, when confusion on the considered alphabets is at risk. We may also sometimes use the abbreviation ``p.d.'' for ``probability distribution''.

From now on, we will be interested in making minimal a priori assumptions on how powerful
the $\ell$ players may be. This will naturally lead us to considering that
their common strategy to win the game $G$ could be any no-signalling (or even
sub-no-signalling) strategy, which we define now.

\begin{definition} \label{def:NS-SNOS}
  The sets of \emph{no-signalling} and \emph{sub-no-signalling}
  correlations, denoted respectively $\NS(\underline{\cA}|\underline{\cX})$
  and $\SNOS(\underline{\cA}|\underline{\cX})$,
  consist of non-negative densities $P(\underline{a}|\underline{x}) \geq 0$
  defined as follows:
  \begin{equation}
    \label{eq:NS}
    P \in \NS(\underline{\cA}|\underline{\cX})
      :\Leftrightarrow
      \forall\ I\subsetneq[\ell],\ \exists\ Q(\cdot|x_I)\text{ p.d.'s on }\cA_I\text{ s.t.~}
      \forall\ \underline{x},a_I,\ P(a_I|\underline{x}) = Q(a_I|x_I),
  \end{equation}
  \begin{equation}
    \label{eq:SNOS}
    P \in \SNOS(\underline{\cA}|\underline{\cX})
      :\Leftrightarrow
      \forall\ I\subsetneq[\ell],\ \exists\ Q(\cdot|x_I)\text{ p.d.'s on }\cA_I\text{ s.t.~}
      \forall\ \underline{x},a_I,\ P(a_I|\underline{x}) \leq Q(a_I|x_I).
  \end{equation}
  Here, $P(a_I|\underline{x})$ denotes the marginal density,
  \[
    P(a_I|\underline{x}) = \sum_{a_{I^c} \in \cA_{I^c}} P(\underline{a}=a_I a_{I^c}|\underline{x}).
  \]
\end{definition}

\begin{remark}
Note that under this definition,
$\NS(\underline{\cA}|\underline{\cX}) \subset \SNOS(\underline{\cA}|\underline{\cX})$,
but the latter is a strictly larger set (e.g.~it always contains the
all-zero density). Furthermore,
$P \in \NS(\underline{\cA}|\underline{\cX})$
iff $P \in \SNOS(\underline{\cA}|\underline{\cX})$ and
$P$ is \emph{normalized} in the sense that for all
$\underline{x} \in \underline{\cX}$,
$\sum_{\underline{a}} P(\underline{a}|\underline{x}) = 1$.
Indeed, $\NS$ consists of conditional probability distributions, while $\SNOS$
allows, given each input, a total ``probability'' of less than or equal to $1$.

Also, it can be shown that in equation~(\ref{eq:NS}), only sets of the
form $I=[\ell]\setminus i$ need to be considered. This is because
the no-signalling conditions take the form of equations and
this subset spans the set of all equations required (cf.~\cite{Han}, Lemma 2.7). The analogous
statement for sub-no-signalling is not known and likely false.
Nevertheless, one might in other contexts consider to relax
the conditions of equation~(\ref{eq:SNOS}) to hold only for a selected family
of subsets $I \subset [\ell]$.
\end{remark}

\begin{figure}[h]
\caption{An $\ell$-player non local game}
\begin{center}
\begin{tikzpicture} [scale=1]
\node[draw=lightgray, minimum height=1cm, minimum width=1cm, fill=lightgray] (X) at (0,0) {Player $1$};
\node[draw=white, minimum height=0.5cm, minimum width=1cm, fill=white] (X') at (0,1.5) {$x_1\in\mathcal{X}_1$};
\node[draw=white, minimum height=0.5cm, minimum width=1cm, fill=white] (X'') at (0,-1.5) {$a_1\in\mathcal{A}_1$};
\node[draw=lightgray, minimum height=1cm, minimum width=1cm, fill=lightgray] (Y) at (3,0) {Player $\ell$};
\node[draw=white, minimum height=0.5cm, minimum width=1cm, fill=white] (Y') at (3,1.5) {$x_\ell\in\mathcal{X}_\ell$};
\node[draw=white, minimum height=0.5cm, minimum width=1cm, fill=white] (Y'') at (3,-1.5) {$a_\ell\in\mathcal{A}_\ell$};
\node[draw=white, minimum height=0.5cm, minimum width=1cm, fill=white] at (6.5,1.5) {w.p. $T(x_1\ldots x_\ell)$};
\node[draw=white, minimum height=0.5cm, minimum width=1cm, fill=white] at (6.5,-1.5) {w.p. $P(a_1\ldots a_\ell|x_1\ldots x_\ell)$};
\node[draw=white, minimum height=0.5cm, minimum width=6cm, fill=white] at (4,-2.5) {The players win iff $V(a_1\ldots a_\ell,x_1\ldots x_\ell)=1$};
\draw [->] (X'.south) -- (X.north); \draw [->] (X.south) -- (X''.north); \draw [->] (Y'.south) -- (Y.north); \draw [->] (Y.south) -- (Y''.north);
\draw [dotted] (X.east) -- (Y.west); \draw [dotted] (X'.east) -- (Y'.west); \draw [dotted] (X''.east) -- (Y''.west);
\end{tikzpicture}
\end{center}
\label{fig:game}
\end{figure}
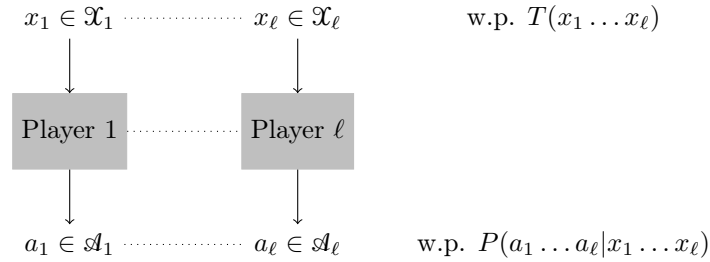

An $\ell$-player game $G$ is characterized by a probability distribution
$T(\underline{x})$ on the queries $\underline{\cX}$, and a binary
predicate $V(\underline{a},\underline{x}) \in \{0,1\}$
on the answers and queries $\underline{\cA}\times\underline{\cX}$, as illustrated in Figure \ref{fig:game}. The no-signalling, resp.~ sub-no-signalling, value of the game, denoted $\omega_{\NS}(G)$, resp.~$\omega_{\SNOS}(G)$, is the maximum of the winning probability
\[ \P\left(\text{win}\right) = \E V(\underline{A},\underline{X})
                    = \sum_{\underline{a},\underline{x}} T(\underline{x}) V(\underline{a},\underline{x})
                                                                          P(\underline{a}|\underline{x}) \]
over all $P\in\NS(\underline{\cA}|\underline{\cX})$, resp.~$P\in \SNOS(\underline{\cA}|\underline{\cX})$, where the distribution of
$\underline{X}=X_1\ldots X_\ell$ and $\underline{A}=A_1\ldots A_\ell$
is as expected,
\[ \forall\ \underline{x},\underline{a},\ \P\left(\underline{X}=\underline{x},\, \underline{A}=\underline{a}\right)
                           = T(\underline{x}) P(\underline{a}|\underline{x}). \]

In words, the (sub-)no-signalling value of a game is the maximal probability of winning it when no limitation is assumed on the power of the players, apart from the fact that they cannot signal information instantaneously from one another. In the sub-no-signalling case, constraints are relaxed even more: players are not forced to always produce an output, and it is only required that their strategy ``looks as if it were no-signalling'' (even though they may have ``hidden'' in their abstentions the fact that it is signalling). In Section \ref{sec:end} we briefly discuss other kinds of restrictions that one may put on the players' physical power, such as shared randomness or shared quantum entanglement only.

\subsection{Two-player SNOS $\mathbf{\equiv}$ NS}

Not surprisingly, the no-signalling and sub-no-signalling values of
games are related. We start by showing that for any two-player game $G$,
they are identical, i.e.~$\omega_{\NS}(G) = \omega_{\SNOS}(G)$.
As $\NS \subset \SNOS$, the
inequality ``$\leq$'' is evident, and we only need to prove the opposite
inequality ``$\geq$''. This follows from the following structural lemma.

\begin{lemma}[cf.~\cite{Ito}]
  \label{lemma:2-player-SNOS}
  Let $P \in \SNOS(\cA\times\cB|\cX\times\cY)$ be a two-player
  sub-no-signalling correlation. Then there exists a no-signalling
  correlation $P' \in \NS(\cA\times\cB|\cX\times\cY)$ with $P\leq P'$
  pointwise, i.e.~$P(ab|xy) \leq P'(ab|xy)$ for all $a,b,x,y$.
\end{lemma}


\begin{proof}
  If $P$ is normalized, i.e.~if for all $x,y$,
  $\sum_{ab} P(ab|xy)=1$, there is nothing to prove because
  $P$ is already no-signalling.

  Otherwise, there exist $x,y$ with weight
  $\sum_{ab} P(ab|xy) =: w < 1$. By sub-no-signalling assumption, we have
  distributions $Q(a|x)$ and $Q(b|y)$ dominating the marginals:
  \[
    \forall\ a,b,\ P(a|xy) \leq Q(a|x),\ P(b|xy) \leq Q(b|y).
  \]
  As the total weight of both marginals of $P(\cdot|xy)$ is $w < 1$, we can find
  $a$ and $b$ such that
  \[
    P(a|xy) < Q(a|x), \quad P(b|xy) < Q(b|y),
  \]
  so we can increase $P(ab|xy)$ by some $\epsilon > 0$
  to $P'(ab|xy) = P(ab|xy) + \epsilon$ and still satisfy the
  sub-no-signalling conditions. By choosing $\epsilon$ maximally
  so, we can reduce the total number of strict inequality signs
  in the SNOS conditions. Iterating this procedure we arrive
  at a sub-no-signalling correlation $P'$ with all inequalities
  met with equality, i.e.~a no-signalling correlation.

  Another presentation of this argument appeals to compactness.
  Consider the following set of correlations:
  \[ \mathrm{X}_{P,Q} :=\left\{ P'\ :\ \forall\ a,b,x,y,\ P'(ab|xy)\geq P(ab|xy),\ P'(a|x)\leq Q(a|x),\ P'(b|y)\leq Q(b|y) \right\}. \]
  $\mathrm{X}_{P,Q}$ being compact and $P'\mapsto \sum_{xy} \sum_{ab} P'(ab|xy)$ being continuous,
  \[ \sup \left\{ \sum_{xy} \sum_{ab} P'(ab|xy)\ :\ P'\in\mathrm{X}_{P,Q} \right\} \]
  is actually attained. If it were less than $|\cX\times\cY|$, we could use the procedure
  above to increase the objective function, contradicting that it
  is a maximum.
\end{proof}

Note that the ``bumping up'' procedure described above, in order to transform any two-player sub-no-signalling strategy into a no-signalling one dominating it pointwise, may fail for more players. The two-player case is indeed special, due to non-overlapping of the two SNOS or NS constraints. However, already in the case of three players, even just the three inequalities $P_{\cA_i\cA_j|\cX_i\cX_j\cX_k}\leq Q_{\cA_i\cA_j|\cX_i\cX_j}$ may be impossible to bring simultaneously to equalities by pointwise increment.

\subsection{Multi-player SNOS vs NS}

Clearly, $\omega_{\NS}(G) \leq \omega_{\SNOS}(G)$ for every game,
and there are examples of games (with game distribution $T$ having
strictly smaller than full support) where $\omega_{\NS}(G) < 1$
but $\omega_{\SNOS}(G)=1$, for instance the \textit{anticorrelation
game} (and likewise a number of games where frustration prohibits
extension of a winning sub-no-signalling strategy to a normalized,
no-signalling one).

\begin{example}
\normalfont
{\textbf (cf.~\cite{AF-R-V}, Appendix A)}
Consider the three-player \emph{anti-correlation game} $A_3$, which
has binary input and output for all players and game
distribution $T$ supported on $\{0,1\}^3\setminus\{111\}$,
i.e.~$111$ does not occur as a triple of questions.
The winning predicate is that if any two inputs are $1$,
say $x_i=x_j=1$, then the corresponding outputs must be
different, $a_i \neq a_j$. While if there are zero or only a single
$1$ amongst the inputs, outputs may be arbitrary.

It is straightforward to verify that the following correlation
is in $\SNOS\left(\{0,1\}^3|\{0,1\}^3\right)$ and wins the game with certainty:
\[
  P(a_1a_2a_3|x_1x_2x_3) = \begin{cases}
                             0                         & \text{ if } x_1x_2x_3=111, \\
                             1/8                   & \text{ if } \exists\ 1\leq i\neq j\leq 3:\ x_i=x_j=0, \\
                             \delta_{a_i,1-a_j}/4 & \text{ if } \exists\ 1\leq i\neq j\leq 3:\ x_i=x_j=1 \text{ and } x_1x_2x_3 \neq 111.
                           \end{cases}
\]
So $\omega_{\SNOS}(A_3)=1$. On the other hand, for, say, $T$ uniform on $\{011,101,110\}$,
one can check by elementary means that $\omega_{\NS}(A_3) = 2/3$.
\end{example}

However, for a game distribution $T$ having full support, a simple reasoning shows that
$\omega_{\NS}(G) < 1$ implies $\omega_{\SNOS}(G) < 1$. Indeed,
we show the contrapositive, assuming that $\omega_{\SNOS}(G) = 1$.
Because of the full support of $T$, this implies that for the
optimal sub-no-signalling strategy $P$ and every $\underline{x}$,
\[
  1 =    \sum_{\underline{a}} V(\underline{a},\underline{x}) P(\underline{a}|\underline{x})
    \leq \sum_{\underline{a}} P(\underline{a}|\underline{x})
    \leq 1,
\]
hence equality (i.e.~normalization) holds for all $\underline{x}$.
Thus, $P$ is really a no-signalling correlation and so
$\omega_{\NS}(G)=1$.
In fact, we can show something stronger, namely the following quantitative relationship.

\begin{lemma}
  \label{lemma:SNOS-NS-multiplayers}
  Consider a game distribution $T$ with full support on $\underline{\cX}$.
  Then there exists $\Gamma=\Gamma(T)\geq 0$, which only depends
  on $T$, such that for every game $G$ with query distribution
  $T$,
  \[ \omega_{\SNOS}(G) \geq 1-\epsilon \ \Rightarrow\ \omega_{\NS}(G) \geq 1-(\Gamma+1)\epsilon. \]
  The definition of $\Gamma$ can be taken from~\cite{B-F-S} or~\cite{AF-R-V}.
\end{lemma}

\begin{proof}
Take an optimal strategy $P\in\SNOS(\underline{\cA}|\underline{\cX})$,
so that $P(a_I|\underline{x}) \leq Q(a_I|x_I)$ for all $I$, $a_I$, $\underline{x}$. Then,
\[
  \sum_{\underline{a},\underline{x}} T(\underline{x}) P(\underline{a}|\underline{x})
                                                \geq \sum_{\underline{a},\underline{x}} T(\underline{x})V(\underline{a},\underline{x}) P(\underline{a}|\underline{x}) = \omega_{\SNOS}(G) \geq 1-\epsilon.
\]
And so we get, for all $I$,
\[
  \bigl\| T_{\underline{\cX}}P_{\cA_I|\underline{\cX}} - T_{\underline{\cX}}Q_{\cA_I|\cX_I} \bigr\|_1
    =    \sum_{a_I,\underline{x}}  T(\underline{x}) \bigl( Q(a_I|x_I) - P(a_I|\underline{x}) \bigr)
    \leq \epsilon,
\]
because the difference term in the sum is non-negative.

Now simply ``bump up'' the sub-normalized probability distribution
$P_{\underline{\cA}|\underline{\cX}}$ to a
properly normalized conditional probability distribution
$P_{\underline{\cA}|\underline{\cX}}'$, adding at most an averaged weight over $T_{\underline{\cX}}$ of $\epsilon$, and hence, for all $I$,
\[
  \frac12 \bigl\|  T_{\underline{\cX}}P_{\cA_I|\underline{\cX}}' -  T_{\underline{\cX}}Q_{\cA_I|\cX_I} \bigr\|_1 \leq \epsilon.
\]

At this point we can invoke the stability of linear programmes,
used in~\cite{B-F-S} and~\cite{AF-R-V} to conclude that there
is $\Gamma = \Gamma(T)\geq 0$ such that there is a no-signalling
correlation $P_{\underline{\cA}|\underline{\cX}}'' \in \NS(\underline{\cA}|\underline{\cX})$ with
\[
  \frac12 \bigl\| T_{\underline{\cX}}P_{\underline{\cA}|\underline{\cX}}''
                   - T_{\underline{\cX}}P_{\underline{\cA}|\underline{\cX}}' \bigr\|_1 \leq \Gamma\epsilon.
\]
This gives
\begin{align*} \omega_{\NS}(G) \geq & \sum_{\underline{a},\underline{x}} T(\underline{x})V(\underline{a},\underline{x}) P''(\underline{a}|\underline{x})\\
\geq & \sum_{\underline{a},\underline{x}} T(\underline{x})V(\underline{a},\underline{x}) P'(\underline{a}|\underline{x}) -\Gamma\epsilon \\
\geq & \sum_{\underline{a},\underline{x}} T(\underline{x})V(\underline{a},\underline{x}) P(\underline{a}|\underline{x}) -\Gamma\epsilon \\
\geq &\, 1-(\Gamma+1)\epsilon,
\end{align*}
where we have used the total variational bound on $P''-P'$, the fact that
$P'$ dominates $P$ and the assumption on the probability of winning $G$ when played $P$.
\end{proof}

The rest of the chapter is structured as follows: In Section~\ref{sec:main}
we introduce parallel repetition of games, and state our main
results, which improve upon, and partly clarify, earlier findings
by Holenstein~\cite{Hol}, Buhrman \emph{et al.}~\cite{B-F-S}
and Arnon-Friedman \emph{et al.}~\cite{AF-R-V}.
In Section~\ref{sec:constrained-postselection}, we present the
main technical tool, one of the constrained de Finetti reductions from Chapter \ref{chap:deFinetti},
adapted to our present needs, followed by the proofs of the
main theorems and corollaries in Section~\ref{sec:proofs}. We
conclude in Section~\ref{sec:end}.

\section{Parallel repetition: definitions and main results}
\label{sec:main}

Given an $\ell$-player game $G$, with probability distribution
$T(\underline{x})$ on $\underline{\cX}$ and binary
predicate $V(\underline{a},\underline{x}) \in \{0,1\}$
on $\underline{\cA}\times\underline{\cX}$, we are interested
in playing the same game $n$ times independently in parallel,
and in looking at the probability of winning all $n$ or a subset of $t$ of them. The reader is referred to Chapter \ref{chap:deFinetti} for entirely analogous considerations, but in the setting of support functions of convex sets of states instead of multi-player non-local games.

Formally, the \emph{$n$-fold parallel repetition of $G$}
is the $\ell$-player game $G^n$ having the product probability distribution on $\underline{\cX}^n$
\[ T^{\otimes n}(\underline{x}^n)=T\bigl(\underline{x}^{(1)})\cdots T(\underline{x}^{(n)}\bigr), \]
and the product binary predicate on $\underline{\cA}^n\times\underline{\cX}^n$
\[ V^{\otimes n}(\underline{a}^n,\underline{x}^n)
  =V\bigl(\underline{a}^{(1)},\underline{x}^{(1)})\cdots V(\underline{a}^{(n)},\underline{x}^{(n)}\bigr) \in \{0,1\}. \]
The no-signalling, resp.~sub-no-signalling, value of this $n$-fold parallel repetition game,
denoted $\omega_{\NS}(G^n)$, resp.~$\omega_{\SNOS}(G^n)$, is thus the maximum
of the winning probability
\[ \P\left(\text{win}\right) = \sum_{\underline{a}^n,\underline{x}^n} T^{\otimes n}(\underline{x}^n) V^{\otimes n}(\underline{a}^n,\underline{x}^n)
                                                                          P(\underline{a}^n|\underline{x}^n) \]
over all $P\in\NS(\underline{\cA}^n|\underline{\cX}^n)$, resp.~$P\in \SNOS(\underline{\cA}^n|\underline{\cX}^n)$.

In words, the players win $G^n$ if they win all $n$ instances of $G$ played
in parallel. So we obviously always have (for the allowed set of strategies
being $X \in \{\NS,\SNOS\}$)
\begin{equation}
  \label{eq:G-G^n}
  \big(\omega_X(G)\big)^n \leq \omega_X(G^n) \leq \omega_X(G).
\end{equation}
However, in the case where $\omega_X(G)<1$, the gap between the lower and
upper bounds in equation~\eqref{eq:G-G^n} grows exponentially with $n$, making
equation \eqref{eq:G-G^n} very little informative. The parallel repetition problem is thus the following:
If none of the players' allowed strategies can make them win $1$ instance
of $G$ with probability $1$, does it necessarily imply that they have an
exponentially decaying probability of winning $n$ of them at the same time?
And if so at which rate?

More generally, we can study the game $G^{t/n}$, whose winning predicate is
defined as winning any $t$ (or more) out of $n$ repetitions \cite{B-F-S}, i.e.~\[ V^{t/n}(\underline{a}^n,\underline{x}^n) := \left\{ \sum_{i=1}^n V\bigl(\underline{a}^{(i)},\underline{x}^{(i)}\bigr) \geq t \right\} = \begin{cases} 1 & \text{ if }\sum_{i=1}^n V\bigl(\underline{a}^{(i)},\underline{x}^{(i)}\bigr) \geq t, \\
0 & \text{ otherwise}. \end{cases} \]
Note that, with our notation, $G^n = G^{n/n}$.

\medskip
The main results of the present chapter are gathered below, where we set $C_{\ell}:=2^{\ell+1}-3$.

\begin{theorem}[Parallel repetition for the sub-no-signalling value of $\ell$-player games] \label{th:repetitionSNOS}
Let $G$ be an $\ell$-player game such that
$\omega_{\SNOS}(G)\leq 1-\delta$ for some $0<\delta<1$.
Then, for any $n\in\N$, and any $t \geq (1-\delta+\alpha)n$ for some $0<\alpha\leq\delta$, we have
\begin{align*}
& \omega_{\SNOS}(G^n) \leq \left(1-\frac{\delta^2}{5C_{\ell}^2}\right)^n,\\
& \omega_{\SNOS}(G^{t/n}) \leq \exp\left( -n \frac{\alpha^2}{5C_{\ell}^2} \right).
\end{align*}
\end{theorem}

As immediate consequences or refinements of Theorem \ref{th:repetitionSNOS}, we can get
parallel repetition results for the no-signalling value of multiplayer games
in some particular instances.

\begin{corollary}[Parallel repetition for the no-signalling value of full support $\ell$-player games]
\label{cor:repetitionSNOS}
Let $G$ be an $\ell$-player game whose distribution $T$ has full support,
and such that $\omega_{\NS}(G)\leq 1-\delta$ for some $0<\delta<1$.
Then, for any $n\in\N$, and any $t \geq (1-\delta+\alpha)n$ for some $0<\alpha\leq\delta$, we have
\begin{align*}
& \omega_{\NS}(G^n) \leq \left(1-\frac{\delta^2}{5C_{\ell}^2(\Gamma+1)^2}\right)^n,    \\
& \omega_{\NS}(G^{t/n}) \leq \exp\left( -n \frac{\alpha^2}{5C_{\ell}^2(\Gamma+1)^2} \right),
\end{align*}
where $\Gamma=\Gamma(T)\geq 0$ is the constant from Lemma \ref{lemma:SNOS-NS-multiplayers},
which only depends on $T$.
\end{corollary}

Note that the constant $\Gamma$ in this corollary depends on the game,
and in the worst case carries a heavy dependence on the players' alphabet
sizes. This is in contrast to Holenstein's two-player result for no-signalling games,
which has no alphabet dependence at all \cite{Hol}. This is generalized
in our Theorem~\ref{th:repetitionSNOS}, since for two players we know by
Lemma \ref{lemma:2-player-SNOS} that $\NS \equiv \SNOS$,
and we could directly read off bounds with constants already improving on
Holenstein's. Looking a little into the proof allows us to optimize the
constants even more, which we record as follows.

\begin{theorem}[Parallel repetition for the no-signalling value of $2$-player games, cf.~Holenstein~\cite{Hol}]
\label{thm:2-player}
Let $G$ be a $2$-player game with
$\omega_{\NS}(G)\leq 1-\delta$ for some $0<\delta<1$.
Then, for any $n\in\N$, and any $t \geq (1-\delta+\alpha)n$ for some $0<\alpha\leq\delta$, we have
\begin{align*}
& \omega_{\NS}(G^n) \leq \left(1-\frac{\delta^2}{27}\right)^n,    \\
& \omega_{\NS}(G^{t/n}) \leq \exp\left( -n \frac{\alpha^2}{33} \right).
\end{align*}
\end{theorem}

\section{Constrained de Finetti reduction}
\label{sec:constrained-postselection}

De Finetti reductions are a useful tool when trying to understand any permutation-invariant information processing task. Indeed, these enable to restrict the analysis to that of i.i.d.~scenarios, which are usually trivially understood, as it was exemplified in Chapter \ref{chap:deFinetti}. In the context of multi-player games played $n$ times in parallel, one would like to use the fact that the numbering of the $n$ instances of the repeated game is irrelevant to reduce the study of strategies for the latter to the study of so-called
\emph{de Finetti strategies} (i.e.~convex combinations of $n$ i.i.d.~strategies).

The seminal de Finetti reduction (aka post-selection) lemma was stated in \cite{C-K-R}, later finding applications in many areas of quantum information theory, from quantum cryptography \cite{L-GP-R-C} to quantum Shannon theory \cite{B-C-R}. Our proofs though, will rely on two more recently established de Finetti reduction results, which are stated below. Just to fix some definitions: we will say that a (sub-)probability distribution $P_{\cZ^n}$, resp.~a conditional (sub-)probability distribution $P_{\cB^n|\cY^n}$, is $n$-symmetric if for any permutation $\pi$ of $n$ elements, $\forall\ z^n,\ P(\pi(z^n))=P(z^n)$, resp.~$\forall\ b^n,y^n,\ P(\pi(b^n)|\pi(y^n))=P(b^n|y^n)$.

\begin{lemma}[de Finetti reduction for conditional p.d.'s, \cite{AF-R}]
\label{lemma:dF-conditional}
Let $\cB,\cY$ be finite alphabets. There exists a probability measure $dR_{\cB|\cY}$ on the set
of conditional probability distributions $R_{\cB|\cY}$ such that, for any $n$-symmetric
conditional probability distribution $P_{\cB^n|\cY^n}$,
\[
  P_{\cB^n|\cY^n} \leq \mathrm{poly}(n) \int_{R_{\cB|\cY}} R_{\cB|\cY}^{\otimes n}\,\mathrm{d}R_{\cB|\cY},
\]
where the polynomial pre-factor may be upper bounded as $\mathrm{poly}(n)\leq (n+1)^{|\cB||\cY|}$.
\end{lemma}

\begin{lemma}[Constrained de Finetti reduction for (sub-)p.d.'s, {[Chapter \ref{chap:deFinetti}, Section \ref{sec:finite-de-finetti}, Corollary \ref{cor:probability}]}]
\label{lemma:dF-constrained}
Let $\cZ$ be a finite alphabet. There exists a probability measure $dQ_{\cZ}$ on the
set of probability distributions $Q_{\cZ}$ on $\cZ$ such that, for any $n$-symmetric
(sub-)probability distribution $P_{\cZ^n}$ on $\cZ^n$,
\[
  P_{\cZ^n} \leq \mathrm{poly}(n) \int_{Q_{\cZ}} F\left(P_{\cZ^n},Q_{\cZ}^{\otimes n}\right)^2Q_{\cZ}^{\otimes n}\,\mathrm{d}Q_{\cZ},\]
where the polynomial pre-factor may be upper bounded as  $\mathrm{poly}(n)\leq (n+1)^{3|\cZ|^2}$.
\end{lemma}

In Lemma \ref{lemma:dF-constrained} above, as well as in the remainder of this chapter, $F(P,Q)$ stands for the fidelity between probability distributions $P$ and $Q$, defined as $F(P,Q)=\|\sqrt{P}\sqrt{Q}\|_1$.

We are now ready to present the technical lemma that will allow us in Section \ref{sec:proofs} to reduce
the study of strategies for repeated games to the study of so-called
\emph{de Finetti strategies}, and hence prove our main results.

\begin{lemma}[de Finetti reduction for sub-no-signalling correlations]
\label{lemma:deFinetti}
There exists a probability measure $dQ$ on the set of probability distributions $Q$ on $\underline{\cA}\times\underline{\cX}$ such that for any probability distribution $T$ on $\underline{\cX}$ and
any $P\in \SNOS(\underline{\cA}^n|\underline{\cX}^n)$ an $n$-symmetric sub-no-signalling correlation,
it holds that
\begin{equation}
  \label{eq:definetti}
  T_{\underline{\cX}}^{\otimes n} P_{\underline{\cA}^n|\underline{\cX}^n}
       \leq \mathrm{poly}(n)
         \int_{Q_{\underline{\cA}\underline{\cX}}} \widetilde{F}\left(Q_{\underline{\cA}\underline{\cX}}\right)^{2n}
               Q_{\underline{\cA}\underline{\cX}}^{\otimes n}\,\mathrm{d}Q_{\underline{\cA}\underline{\cX}},
\end{equation}
where we defined
\[
  \widetilde{F}\left(Q_{\underline{\cA}\underline{\cX}}\right)
     := \underset{\emptyset \neq I\varsubsetneq[\ell]}{\min}\, \underset{R_{\cA_I|\cX_I}}{\max}\,
                 F\left(T_{\underline{\cX}}R_{\cA_I|\cX_I},Q_{\cA_I\underline{\cX}}\right).
\]
We mention for the sake of completeness that the $\mathrm{poly}(n)$
pre-factor in equation~\eqref{eq:definetti} may be upper bounded by
$(n+1)^{3|\underline{\cA}|^2|\underline{\cX}|^2+2|\underline{\cA}||\underline{\cX}|}$.
\end{lemma}

\begin{proof}
Since $T_{\underline{\cX}}^{\otimes n} P_{\underline{\cA}^n|\underline{\cX}^n}$ is an $n$-symmetric sub-probability distribution on $(\underline{\cA}\underline{\cX})^n$, we first of all have by Lemma~\ref{lemma:dF-constrained} that
\[ T_{\underline{\cX}}^{\otimes n} P_{\underline{\cA}^n|\underline{\cX}^n} \leq \mathrm{poly}(n) \int_{Q_{\underline{\cA}\underline{\cX}}} F\left(T_{\underline{\cX}}^{\otimes n}P_{\underline{\cA}^n|\underline{\cX}^n},Q_{\underline{\cA}\underline{\cX}}^{\otimes n}\right)^2Q_{\underline{\cA}\underline{\cX}}^{\otimes n}\,\mathrm{d}Q_{\underline{\cA}\underline{\cX}}. \]
Notice next that, for any $\emptyset \neq I\varsubsetneq[\ell]$,
\[
  F\left(T_{\underline{\cX}}^{\otimes n}P_{\underline{\cA}^n|\underline{\cX}^n},Q_{\underline{\cA}\underline{\cX}}^{\otimes n}\right) \leq F\left(T_{\underline{\cX}}^{\otimes n}P_{\cA_I^n|\underline{\cX}^n},Q_{\cA_I\underline{\cX}}^{\otimes n}\right) \leq F\left(T_{\underline{\cX}}^{\otimes n}P'_{\cA_I^n|\cX_I^n},Q_{\cA_I\underline{\cX}}^{\otimes n}\right).
\]
The first inequality is by monotonicity of the fidelity under stochastic
maps (in particular taking marginals). While the second inequality is
because $P\in \SNOS(\underline{\cA}^n|\underline{\cX}^n)$, so that
$P_{\cA_I^n|\underline{\cX}^n}\leq P'_{\cA_I^n|\cX_I^n}$ for some
conditional p.d.~$P'_{\cA_I^n|\cX_I^n}$, and because the fidelity is order-preserving.

What is more, for any $\emptyset \neq I\varsubsetneq[\ell]$, $P'_{\cA_I^n|\cX_I^n}$ can be chosen to be an $n$-symmetric conditional probability distribution. Indeed, if it were not, its $n$-symmetrization would still upper bound $P_{\cA_I^n|\underline{\cX}^n}$ (since the latter is by assumption $n$-symmetric). We then have by Lemma~\ref{lemma:dF-conditional} that
\[
  P'_{\cA_I^n|\cX_I^n} \leq \mathrm{poly}(n) \int_{R_{\cA_I|\cX_I}} R_{\cA_I|\cX_I}^{\otimes n}\,\mathrm{d}R_{\cA_I|\cX_I}, \]
and subsequently, using first, once more, that the fidelity is order-preserving, and second that it is multiplicative on tensor products,
\[ F\left(T_{\underline{\cX}}^{\otimes n}P'_{\cA_I^n|\cX_I^n},Q_{\cA_I\underline{\cX}}^{\otimes n}\right) \leq \mathrm{poly}(n) \underset{R_{\cA_I|\cX_I}}{\max}\, F\left(T_{\underline{\cX}}^{\otimes n} R_{\cA_I|\cX_I}^{\otimes n},Q_{\cA_I\underline{\cX}}^{\otimes n}\right) = \mathrm{poly}(n) \underset{R_{\cA_I|\cX_I}}{\max}\, F\left(T_{\underline{\cX}} R_{\cA_I|\cX_I},Q_{\cA_I\underline{\cX}}\right)^n. \]

Recapitulating, we get
\[
  T_{\underline{\cX}}^{\otimes n} P_{\underline{\cA}^n|\underline{\cX}^n}
    \leq \mathrm{poly}(n) \int_{Q_{\underline{\cA}\underline{\cX}}} \left(\underset{\emptyset \neq I\varsubsetneq[\ell]}{\min}\, \underset{R_{\cA_I|\cX_I}}{\max}\, F\left(T_{\underline{\cX}}R_{\cA_I|\cX_I},Q_{\cA_I\underline{\cX}}\right)\right)^{2n}Q_{\underline{\cA}\underline{\cX}}^{\otimes n}\,\mathrm{d}Q_{\underline{\cA}\underline{\cX}},
\]
as announced.
\end{proof}

\section{Proofs of the main Theorems}
\label{sec:proofs}

In this section we prove Theorem \ref{th:repetitionSNOS}, Corollary~\ref{cor:repetitionSNOS} and Theorem \ref{thm:2-player}.

We need first of all the following extension of Lemma 9.5 in \cite{Hol}:
\begin{lemma}
  \label{lemma:Hol-Lemma-9.5}
  For $\underline{\cZ} = \bigtimes_{\!\!j=1}^m\, \cZ_j$ and
  $\underline{\cB} = \bigtimes_{\!\!j=1}^m\, \cB_j$,
  consider probability distributions $T$ on $\underline{\cZ}$
  and $P$ on $\underline{\cB}\times\underline{\cZ}$ satisfying
  \[
    \frac12 \bigl\| P_{\underline{\cZ}} - T_{\underline{\cZ}} \bigr\|_1 \leq \epsilon_0.
  \]
  If for each $j\in[m]$ there exists a conditional probability distribution
  $Q(b_j|z_j)$ such that
  \[
    \bigl\| P_{\cB_j\underline{\cZ}} - T_{\underline{\cZ}}Q_{\cB_j|\cZ_j} \bigr\|_1 \leq \epsilon_j,
  \]
  then there exists a conditional probability distribution
  $P'(\underline{b}|\underline{z})$ such that, for each $j\in[m]$, $P'(b_j|\underline{z}) = P'(b_j|z_j)$ for all $b_j,\underline{z}$, and
  \[ \frac12 \bigl\| T_{\underline{\cZ}} P'_{\underline{\cB}|\underline{\cZ}} - P_{\underline{\cB}\underline{\cZ}} \bigr\|_1 \leq \epsilon_0 + \sum_{j=1}^m 2\epsilon_j. \]
\end{lemma}

\begin{proof}
This works exactly as the proof of Lemma 9.5 in \cite{Hol},
which is a successive application ($m$ times) of Lemma 9.4 in the same
paper. The latter relies on the fact that the statistical distance $\|P_1-P_2\|_1/2$ between two probability distributions $P_1,P_2$ can be equivalently characterized as the minimum probability that $X_1$ differs from $X_2$ over pairs of random variables $(X_1,X_2)$ sampled from $P$ having $(P_1,P_2)$ as marginals.
\end{proof}

Note that the conditions enforced in Lemma \ref{lemma:Hol-Lemma-9.5}
are not enough to ensure no-signalling of $P'$ for three or more players. They would be sufficient though to guarantee that $P'$ satisfies the relaxed no-signalling constraints considered in \cite{Ros}, namely that any group of $\ell-1$ players together cannot signal to the remaining player. Nevertheless, we can leverage this result to approximate the given
no-signalling correlation by a sub-no-signalling correlation.

\begin{lemma}
  \label{lemma:Hol-SNOS}
  Let $P$ be a probability distribution on $\underline{\cA}\times\underline{\cX}$
  and $T$ be a probability distribution on $\underline{\cX}$. If the no-signalling conditions
  (\ref{eq:NS}) hold approximately, namely
  \[
    \forall\ I\subsetneq[\ell],\ \exists\ Q(\cdot|x_I)\text{ p.d.'s on }\cA_I\text{ s.t.~}
      \frac12 \bigl\| P_{\cA_I\underline{\cX}}
                      - T_{\underline{\cX}} Q_{\cA_I|\cX_I} \bigr\|_1 \leq \epsilon_I,
  \]
  then there exists a sub-no-signalling correlation $P' \in \SNOS(\underline{\cA}|\underline{\cX})$
  that approximates $P$, in the sense that
  \[
    \frac12 \bigl\| T_{\underline{\cX}}P'_{\underline{\cA}|\underline{\cX}}
                                         - P_{\underline{\cA}\underline{\cX}} \bigr\|_1
                 \leq \epsilon_{\emptyset} + \sum_{\emptyset\neq I\subsetneq[\ell]} 2\epsilon_I.
  \]
  In the two-player case $\ell=2$, $P'$ can be chosen to be
  no-signalling itself, $P'\in \NS(\underline{\cA}|\underline{\cX})$.
\end{lemma}

\begin{proof}
We will apply Lemma~\ref{lemma:Hol-Lemma-9.5}, with $m=2^\ell-2$,
the index $j$ identifying a non-empty and non-full set
$\emptyset \neq I \subsetneq [\ell]$ (for instance via the
expansion of $j$ into $\ell$ binary digits). The local input and
output alphabets are
\[
  \cZ_j = \prod_{i\in I} \cX_i,\quad \cB_j = \prod_{i\in I} \cA_i,
\]
and the distribution we apply it to is
\[
  \widehat{P}(\underline{b}\underline{z})
   = \begin{cases}
       P(\underline{a}\underline{x}) & \text{ if } \forall j,\ b_j = (a_i:i\in I),\ z_j = (x_i:i\in I), \\
       0                             & \text{ otherwise}.
     \end{cases}
\]
Likewise, the prior distribution on $\underline{\cZ}$ is given by
\[
  \widehat{T}(\underline{z})
   = \begin{cases}
       T(\underline{x}) & \text{ if } \forall j,\ z_j = (x_i:i\in I), \\
       0                & \text{ otherwise},
     \end{cases}
\]
and we use the conditional distributions $Q(b_j|z_j) = Q(a_I|x_I)$.

Now, the prerequisites of Lemma~\ref{lemma:Hol-Lemma-9.5} are given,
with $\epsilon_j=\epsilon_I$, and thus we get a conditional probability
distribution $\widehat{P}'$ with $\widehat{P}'(b_j|\underline{z}) = \widehat{P}'(b_j|z_j)$
for all $j$, and
\[
  \frac12 \bigl\| \widehat{T}_{\underline{\cZ}}\widehat{P}'_{\underline{\cB}|\underline{\cZ}}
                                       - \widehat{P}_{\underline{\cB}\underline{\cZ}} \bigr\|_1
                                               \leq \epsilon_0 + \sum_{j=1}^n 2\epsilon_j =: \epsilon.
\]
We would like to conclude here by ``pulling back'' this conditional
distribution to a (it would seem: no-signalling) correlation on
$\underline{\cA}\times\underline{\cX}$, except that $P'$ has support
outside the image of the diagonal embedding
\begin{align*}
  \Delta : \underline{\cA} &\longrightarrow \underline{\cB} \\
             \underline{a} &\longmapsto     \underline{b} \text{ s.t.~} \forall j,\ b_j = (a_i:i\in I),
\end{align*}
and likewise for $\Delta:\underline{\cX} \longrightarrow \underline{\cZ}$.

To resolve this issue, we simply remove this part of the distribution,
and define the desired sub-normalized conditional densities by letting
\[
  P'(\underline{a}|\underline{x})
       := \widehat{P}'\bigl(\Delta(\underline{a})|\Delta(\underline{x})\bigr).
\]
From this we see directly that
\[
  \frac12 \bigl\| T_{\underline{\cX}} P'_{\underline{\cA}|\underline{\cX}}
                                       - P_{\underline{\cA}\underline{\cX}} \bigr\|_1
                                                                            \leq \epsilon,
\]
because $\widehat{P}(\underline{b},\underline{z}) = P(\underline{a},\underline{x})$
for $\underline{b} = \Delta(\underline{a})$ and $\underline{z} = \Delta(\underline{x})$,
and it is $0$ outside the image of $\Delta$.

It remains to check that $P'$ is sub-no-signalling. Let
$\emptyset \neq I \subsetneq [\ell]$ be a subset with corresponding index
$1\leq j \leq 2^\ell-2$. Let also $\underline{x}\in\underline{\cX}$,
$a_I\in \cA_I$ be tuples, and set $\underline{z} = \Delta(\underline{x})$,
$\underline{b} = \Delta(\underline{a})$ (so that $z_j=x_I\in \cX_I = \cZ_j$, $b_j = a_I \in \cA_I = \cB_j$).
Then,
\begin{align*}
P'(a_I|\underline{x}) = & \sum_{a_{I^c}\in\cA_{I^c}} P'(\underline{a}|\underline{x}) \\
= & \sum_{a_{I^c}\in\cA_{I^c}} \widehat{P}'\bigl(\Delta(\underline{a})|\Delta(\underline{x})\bigr) \\
\leq & \sum_{b_k \in \cB_k,\ k\neq j} \widehat{P}'\bigl(\underline{b}|\underline{z}\bigr)\\
= & \,\widehat{P}'(b_j|\underline{z}\bigr) \\
= & \,\widehat{P}'(b_j|z_j) =:   Q'(a_I|x_I).
\end{align*}
Here, we have used the definition of the marginal and of $P'$.
The inequality in the third line is because we enlarge the domain
of the summation, and the equality in the last line is by the marginal property
of $\widehat{P}'$.

The last claim, regarding $\ell=2$ players, is the original
Lemma 9.5 in \cite{Hol}.
\end{proof}

We are now ready to prove our main theorem, namely the parallel
repetition and concentration results for the sub-no-signalling value of multi-player games.

\begin{proof}[Proof of Theorem~\ref{th:repetitionSNOS}]
Let $P_{\underline{\cA}^n|\underline{\cX}^n}$ be a sub-no-signalling correlation which is optimal to win the game $G^n$. The distribution $T_{\underline{\cX}}^{\otimes n}$ and the predicate $V_{\underline{\cA}\underline{\cX}}^{\otimes n}$ of $G^n$ being $n$-symmetric, we can assume without loss of generality that $P_{\underline{\cA}^n|\underline{\cX}^n}$ is also $n$-symmetric. Indeed, since for any permutation $\pi$ of $n$ elements, $T\circ\pi=T$ and $V\circ\pi=V$, playing $G^n$ with $P$ or with $P\circ\pi$ yields the same winning probability. And therefore, if $P$ is an optimal strategy then so is its symmetrization over all permutations of $n$ elements. Hence, by Lemma \ref{lemma:deFinetti},
\[
  T_{\underline{\cX}}^{\otimes n} P_{\underline{\cA}^n|\underline{\cX}^n}
      \leq \mathrm{poly}(n) \int_{Q_{\underline{\cA}\underline{\cX}}} \widetilde{F}\left(Q_{\underline{\cA}\underline{\cX}}\right)^{2n}Q_{\underline{\cA}\underline{\cX}}^{\otimes n}\,\mathrm{d}Q_{\underline{\cA}\underline{\cX}}.
\]
Now, fix $0<\epsilon<1$ and define
\[
  \mathcal{P}_{\epsilon}
    :=\left\{ Q_{\underline{\cA}\underline{\cX}}\ :\ \max_{\emptyset \neq I\varsubsetneq[\ell]} \min_{R_{\cA_I|\cX_I}}\, \frac{1}{2}\|T_{\underline{\cX}}R_{\cA_I|\cX_I}-Q_{\cA_I\underline{\cX}}\|_1 \leq \epsilon \right\}.
\]
Observe that, by well-known relations between fidelity and trace-distance (see e.g.~\cite{F-vdG}), if $Q_{\underline{\cA}\underline{\cX}}\notin\mathcal{P}_{\epsilon}$,
then automatically $\widetilde{F}\left(Q_{\underline{\cA}\underline{\cX}}\right)^2\leq 1-\epsilon^2$.
Hence,
\[
  T_{\underline{\cX}}^{\otimes n} P_{\underline{\cA}^n|\underline{\cX}^n}
     \leq \mathrm{poly}(n)\left( \int_{Q_{\underline{\cA}\underline{\cX}}\in\mathcal{P}_{\epsilon}} Q_{\underline{\cA}\underline{\cX}}^{\otimes n}\,\mathrm{d}Q_{\underline{\cA}\underline{\cX}} + (1-\epsilon^2)^{n}\int_{Q_{\underline{\cA}\underline{\cX}}\notin\mathcal{P}_{\epsilon}} Q_{\underline{\cA}\underline{\cX}}^{\otimes n}\,\mathrm{d}Q_{\underline{\cA}\underline{\cX}} \right).
\]
On the other hand, if $Q_{\underline{\cA}\underline{\cX}}\in\mathcal{P}_{\epsilon}$,
then by definition
\[
  \forall\ \emptyset\neq I\varsubsetneq[\ell],\ \exists\ R_{\cA_I|\cX_I}:\
  \frac{1}{2}\|T_{\underline{\cX}}R_{\cA_I|\cX_I}-Q_{\cA_I\underline{\cX}}\|_1 \leq \epsilon.
\]
By Lemma \ref{lemma:Hol-SNOS}, the latter condition implies that there exists
a sub-no-signalling correlation $R'_{\underline{\cA}|\underline{\cX}}$ such that
\[ \frac{1}{2}\|T_{\underline{\cX}}R'_{\underline{\cA}|\underline{\cX}}-Q_{\underline{\cA}\underline{\cX}}\|_1 \leq C_{\ell}\epsilon,\ \text{where}\ C_{\ell}=1+2(2^{\ell}-2)=2^{\ell+1}-3. \]
Yet, the winning probability when playing $G$ with a strategy $R'_{\underline{\cA}|\underline{\cX}}\in \SNOS(\underline{\cA}|\underline{\cX})$ is, by assumption on $G$, at most $1-\delta$. So the average of the predicate of $G$ over $Q_{\underline{\cA}\underline{\cX}}\in\mathcal{P}_{\epsilon}$ is at most $1-\delta+2C_{\ell}\epsilon$.
Putting everything together, we eventually get that the winning probability when playing $G^n$ with strategy $P_{\underline{\cA}^n|\underline{\cX}^n}$ is upper bounded as
\begin{equation}
  \label{eq:split}
  \P(\text{win}) \leq \mathrm{poly}(n)\left( (1-\delta+2C_{\ell}\epsilon)^n + (1-\epsilon^2)^n \right).
\end{equation}
Choosing in equation~\eqref{eq:split}
\[ \epsilon=C_{\ell}\left(\left(1+\frac{\delta}{C_{\ell}^2}\right)^{1/2}-1\right)\geq \frac{99\delta}{200C_{\ell}},\ \text{so that}\ \epsilon^2\geq\frac{\delta^2}{5C_{\ell}^2}, \]
and recalling that $P_{\underline{\cA}^n|\underline{\cX}^n}$ is, by hypothesis, an optimal sub-no-signalling strategy, we obtain
\begin{equation}
  \label{eq:poly-exp}
  \omega_{\SNOS}(G^n) \leq \mathrm{poly}(n) \left(1-\frac{\delta^2}{5C_{\ell}^2}\right)^n.
\end{equation}

In order to conclude, we have to remove the polynomial pre-factor. So assume that there exists a constant
$C>0$ such that for some $N\in\N$, $\omega_{\SNOS}(G^N)\geq C\left(1-\delta^2/5C_{\ell}^2\right)^N$.
Then, for any $n\in\N$, we would have
\[ \omega_{\SNOS}(G^{Nn}) \geq \left(\omega_{\SNOS}(G^{N})\right)^{n}
                          \geq C^{n}\left(1-\frac{\delta^2}{5C_{\ell}^2}\right)^{Nn}. \]
On the other hand, however, we still have by equation~\eqref{eq:poly-exp}
\[ \omega_{\SNOS}(G^{Nn}) \leq \mathrm{poly}(Nn) \left(1-\frac{\delta^2}{5C_{\ell}^2}\right)^{Nn}. \]
Letting $n$ grow, we see that the only option to make these two conditions compatible is to have $C\leq 1$, which is precisely what we wanted to show.

Following the exact same lines as above, we also get the
concentration bound.
Indeed, we now have in place of equation~\eqref{eq:split} that, for any $0<\epsilon<1$,
\begin{equation} \label{eq:split-t}
  \omega_{\SNOS}(G^{t/n})
    \leq \mathrm{poly}(n)\left( \exp\left[-2n(\alpha-2C_{\ell}\epsilon)^2\right]
                                          + \exp\left[-n\epsilon^2\right] \right),
\end{equation}
by Hoeffding's inequality (and because $e^{-x}\geq 1-x$ for any $x>0$).

The announced upper bound follows from choosing in equation~\eqref{eq:split-t}
\[ \epsilon = \frac{(4C_{\ell}-\sqrt{2})\alpha}{8C_{\ell}^2-1} \geq \frac{5(20-\sqrt{2})\alpha}{199C_{\ell}},\ \text{so that}\ \epsilon^2\geq\frac{\alpha^2}{5C_{\ell}^2}, \]
and removing the polynomial pre-factor by the same trick as before.
\end{proof}

\begin{proof}[Proof of Corollary~\ref{cor:repetitionSNOS}]
By Lemma \ref{lemma:SNOS-NS-multiplayers}, we know that if $G$ is
an $\ell$-player game with full support satisfying
$\omega_{\NS}(G)\leq 1-\delta$,
then $\omega_{\SNOS}(G)\leq 1-\delta/(\Gamma+1)$.
And thus by Theorem \ref{th:repetitionSNOS},
\[
  \omega_{\NS}(G^n) \leq \omega_{\SNOS}(G^n) \leq \left(1-\frac{\delta^2}{5C_{\ell}^2(\Gamma+1)^2}\right)^n.
\]
The concentration bound for $\omega_{\NS}(G^{t/n})$ follows analogously.
\end{proof}

\begin{proof}[Proof of Theorem~\ref{thm:2-player}]
We follow the exact same reasoning as in the proof of Theorem \ref{th:repetitionSNOS}, and keep the same notation. In the case $\ell=2$, we have by Lemma \ref{lemma:Hol-SNOS} that, for any $0<\epsilon<1$,
\[ Q_{\underline{\cA}\underline{\cX}}\in\mathcal{P}_{\epsilon}\ \Rightarrow\ \exists\ R'_{\underline{\cA}\underline{\cX}}\in\NS(\underline{\cA}|\underline{\cX}):\  \frac{1}{2}\|T_{\underline{\cX}}R'_{\underline{\cA}|\underline{\cX}}-Q_{\underline{\cA}\underline{\cX}}\|_1 \leq 5\epsilon. \]
Yet, if the winning probability when playing $G$ with a strategy $R'_{\underline{\cA}|\underline{\cX}}\in \NS(\underline{\cA}|\underline{\cX})$ is, by assumption on $G$, at most $1-\delta$, then the average of the predicate of $G$ over $Q_{\underline{\cA}\underline{\cX}}\in\mathcal{P}_{\epsilon}$ is at most $1-\delta+5\epsilon$. This is because we are here dealing with normalised probability distributions. Hence, for any $0<\epsilon<1$,
\begin{align*}
& \omega_{\NS}(G^n) \leq \mathrm{poly}(n)\left( (1-\delta+5\epsilon)^n + (1-\epsilon^2)^n \right),\\
& \omega_{\SNOS}(G^{t/n}) \leq \mathrm{poly}(n)\left( \exp\left[-2n(\alpha-5\epsilon)^2\right] + \exp\left[-n\epsilon^2\right] \right).
\end{align*}
We can now choose $\epsilon=(\sqrt{29}-5)\delta/2$ in the parallel repetition
estimate and $\epsilon=(10-\sqrt{2})\alpha/49$ in the concentration bound one,
and argue as in the proof of Theorem \ref{th:repetitionSNOS} to remove the
polynomial pre-factor, which yields the two advertised results.
\end{proof}

\section{Discussion}
\label{sec:end}

Our main contribution in the present chapter is a concentration result for the
sub-no-signalling value of multi-player games under parallel repetition. In
fact, we believe that our work is the first to recognize the intrinsic interest
of the class of sub-no-signalling correlations, which appears naturally as a
relaxation of the no-signalling ones.

Specifically, if an $\ell$-player game $G$ has $\SNOS$ value $1-\delta$, then the
probability for $\SNOS$ players to win a fraction at least $1-\delta+\alpha$ of
$n$ instances of $G$ played in parallel is at most $\exp(-nC_{\ell}\alpha^2)$,
where $C_{\ell}>0$ is a constant which only depends on the number $\ell$ of
players. As mentioned in \cite{B-F-S}, such result, valid for games involving strictly more than $2$ players and where not all queries are asked, could potentially find applications in position-based cryptography \cite{B-C-F-G-G-O-S,F-K-T-W}. In the case $\ell=2$, this is actually equivalent to the analogous
concentration result for the no-signalling value of $G$, thus with a universal
constant $c=C_2$ in the exponential bound. And we know we cannot hope for a
better dependence in $\alpha$ than the obtained one, even in the special case
$\alpha=\delta$. Indeed, as explained in \cite{K-R}, strong parallel repetition in general does not hold for
no-signalling players. In the case $\ell>2$, our result implies a concentration
bound for the no-signalling value of $G$, but only if its input distribution has
full support. Besides, the constant in the exponential bound is this time highly
game-dependent (dependence on the sizes of the input and output alphabets, and
on the smallest weight occurring in the input distribution). This is fully
comparable to previous work in this direction due to Buhrman, Fehr and
Schaffner~\cite{B-F-S}, and Arnon-Friedman, Renner and Vidick~\cite{AF-R-V}.

Hence, the most immediate open problem at that point is regarding games with
non-full support in the case of three or more players (e.g.~the anti-correlation game): does a parallel repetition
result hold for the no-signalling value of such multi-player games?
Answering this question probably requires to understand first whether in
Corollary~\ref{cor:repetitionSNOS}, the presence of the game parameter $\Gamma$
is really necessary or is just an artifact of the proof technique. In other words,
does the rate at which the no-signalling value of a game decays under parallel
repetition truly depends on the game distribution?

Another issue that would be worth investigating is whether constrained de Finetti
reductions could also be used to establish parallel repetition results for the
classical or quantum value of multi-player games. Formally, the sets of classical
correlations $\text{C}(\underline{\cA}|\underline{\cX})$ and quantum correlations
$\text{Q}(\underline{\cA}|\underline{\cX})$ are defined as follows:
\[
  P \in \text{C}(\underline{\cA}|\underline{\cX})
      :\Leftrightarrow
      \forall\ \underline{x},\underline{a},\
        P(\underline{a}|\underline{x})
           =\sum_{m\in\mathcal{M}}Q(m)P_1(a_1|x_1\,m)\cdots P_{\ell}(a_{\ell}|x_{\ell}\,m),
\]
for some p.d.~$Q$ on some alphabet $\mathcal{M}$ and some p.d.'s $P_i(\cdot|x_i\,m)$ on $\cA_i$.
\[
  P \in \text{Q}(\underline{\cA}|\underline{\cX})
      :\Leftrightarrow
      \forall\ \underline{x},\underline{a},\
         P(\underline{a}|\underline{x})
            =\bra{\psi}M(x_1)_{a_1}\otimes\cdots\otimes M(x_{\ell})_{a_{\ell}}\ket{\psi},
\]
for some pure state $\ketbra{\psi}{\psi}$ on $\mathrm{H}_1\otimes\cdots\otimes\mathrm{H}_{\ell}$ and some POVMs $M(x_i)$ on $\mathrm{H}_i$.\\
And the classical, resp.~quantum, value of an $\ell$-player game $G$ with distribution $T$ and predicate $V$, denoted $\omega_{\text{C}}(G)$, resp.~$\omega_{\text{Q}}(G)$, is then naturally defined as the maximum, resp.~supremum, of the winning probability
\[ \label{eq:NS-SNOS-value}
  \P\left(\text{win}\right) = \sum_{\underline{a},\underline{x}} T(\underline{x}) V(\underline{a},\underline{x}) P(\underline{a}|\underline{x}) \]
over all $P\in\text{C}(\underline{\cA}|\underline{\cX})$, resp.~$P\in\text{Q}(\underline{\cA}|\underline{\cX})$.

In the classical case, the first parallel repetition result for two-player games
was established by Raz~\cite{Raz}, and later improved by Holenstein~\cite{Hol},
while Rao~\cite{Rao} gave a concentration bound. However, the proof techniques are
arguably not as straightforward as via de Finetti reductions, and do not
generalise directly to any number $\ell$ of players. In the quantum case, even
less is known. The best parallel repetition result up to now is the one established by
Chailloux and Scarpa~\cite{C-S} (subsequently improved by Chung, Wu and Yuen~\cite{C-W-Y}),
which applies to two-player ($\ell$-player) free games, and from there to games
with full support.
That is why being able to export
ideas from the de Finetti approach to these two cases would be of great interest.
Roughly speaking, the problem we are facing is the following: Given an $n$-symmetric
correlation $P_{\underline{\cA}^n|\underline{\cX}^n}$, we can always write the
first step in the proof of Lemma~\ref{lemma:deFinetti}, i.e.~\begin{equation} \label{eq:dFstrategy}
  T_{\underline{\cX}}^{\otimes n} P_{\underline{\cA}^n|\underline{\cX}^n}
       \leq \mathrm{poly}(n)
            \int_{Q_{\underline{\cA}\underline{\cX}}}
                    F\left(T_{\underline{\cX}}^{\otimes n}P_{\underline{\cA}^n|\underline{\cX}^n},
                           Q_{\underline{\cA}\underline{\cX}}^{\otimes n}\right)^2
                    Q_{\underline{\cA}\underline{\cX}}^{\otimes n}\,\mathrm{d}Q_{\underline{\cA}\underline{\cX}}.
\end{equation}
Now, we would like to argue that if $P_{\underline{\cA}^n|\underline{\cX}^n}$ is a
classical, resp.~quantum, correlation, then the p.d.'s $Q_{\underline{\cA}\underline{\cX}}$
for which the fidelity weight in the r.h.s.~of equation~\eqref{eq:dFstrategy} is not exponentially small are
necessarily close to being of the form $T_{\underline{\cX}}R_{\underline{\cA}|\underline{\cX}}$
for some classical, resp.~quantum, correlation $R_{\underline{\cA}|\underline{\cX}}$.
This was precisely our proof philosophy in the no-signalling case. However, the
fact that the classical and quantum conditions are not properties that one can
read off on the marginals, contrary to the no-signalling one, seems to be a first
obstacle to surmount.

One related legitimate question would be the following: is it possible to make an even stronger statement than the one that, as explained above, we either are looking for (in the classical and quantum cases) or already have (in the no-signalling case)? Namely, could we upper bound $T_{\underline{\cX}}^{\otimes n}P_{\underline{\cA}^n|\underline{\cX}^n}$ by a de Finetti distribution analogous to that in the r.h.s.~of equation~\eqref{eq:dFstrategy}, but with weight strictly $0$ on p.d.'s $Q_{\underline{\cA}\underline{\cX}}$
which are not of the form $T_{\underline{\cX}}R_{\underline{\cA}|\underline{\cX}}$, for $R_{\underline{\cA}|\underline{\cX}}$ belonging to the same class as $P_{\underline{\cA}^n|\underline{\cX}^n}$? The answer to this question is no. Indeed, such improved de Finetti reduction would imply a strong parallel repetition result, which we know does not hold (see \cite{AF-R-V} for a similar discussion). So the best we can hope for is really to show that the fidelity weight in our upper bounding de Finetti distribution is exponentially small on the  p.d.'s which are too far from being of the desired form.

Finally, let us briefly comment on the main spirit difference between the present work and the one by Arnon-Friedman \emph{et al.}~\cite{AF-R-V}. Our approach consists in using a more ``flexible'' de Finetti reduction, in which the information on the correlation $P_{\underline{\cA}^n|\underline{\cX}^n}$ and the p.d.~$T_{\underline{\cX}}^{\otimes n}$ of interest are kept in the upper bounding de Finetti distribution, through the fidelity weight $F(T_{\underline{\cX}}^{\otimes n}P_{\underline{\cA}^n|\underline{\cX}^n},Q_{\underline{\cA}\underline{\cX}}^{\otimes n})^2$. Whereas in \cite{AF-R-V}, any initial correlation is first upper bounded by the same universal de Finetti correlation, on which a test (specifically tailored to the considered game distribution) is performed in a second step, that has the property of letting pass, resp.~rejecting, with high probability the strategies which are no-signalling, resp.~too signalling. So it seems in the end that both approaches are quite closely related: in our case, the ``signalling test'' which is applied to a given p.d.~$Q_{\underline{\cA}\underline{\cX}}$ is nothing else than the maximal fidelity of $Q_{\underline{\cA}\underline{\cX}}$ to the set of p.d.'s of the form $T_{\underline{\cX}}R_{\underline{\cA}|\underline{\cX}}$, with $R_{\underline{\cA}|\underline{\cX}}$ no-signalling, being above or below a certain threshold value. Also, it would be interesting (and potentially fruitful) to investigate whether one could combine in some way the techniques yielding Lemmas \ref{lemma:dF-conditional} and \ref{lemma:dF-constrained}, to get a de Finetti reduction result that would have the advantages of both: namely, that is designed for conditional p.d.'s while at the same carrying the relevant information on the conditional p.d.~it is applied to.


\part{Outlook and perspectives}

Let us wrap up with a (highly subjective) selection of further prospects, bringing together several of the different topics that were evoked over these pages (and therefore expected to be of less specific nature than the ones already put forward at the end of each chapter). When we allude to notions that were broached in a previous chapter, it is tacitely understood that we refer the reader to it for all needed definitions and notation.

\section*{Other models of random states}
\label{sec:random-states}

In many places of this manuscript, we were interested in studying whether random states satisfied a given property or not (e.g.~being locally almost indistinguishable in Chapter \ref{chap:data-hiding}, being separable or satisfying a separability criterion in Chapters \ref{chap:SDrelaxations} and \ref{chap:k-extendibility} etc.). Depending on the context, the random state model that we chose slightly differed, but in the end all of them were qualitatively comparable. Let us be a bit more specific. Let $d,s\in\N$ and consider the following random state models on $\C^d$ (which were all used at some point or another):
\begin{itemize}[topsep=0cm,itemsep=-1em,parsep=0cm,leftmargin=*]
\item \textit{Random induction}: $\rho=\tr_{\C^s}\ketbra{\psi}{\psi}$ with $\psi$ a uniformly distributed unit vector in $\C^d\otimes\C^s$, i.e.~equivalently $\rho=W/\tr W$ with $W$ a Wishart matrix of parameter $s$ on $\C^d$. \\
\item \textit{Random mixture}: $\rho=\big(\sum_{i=1}^s\ketbra{\psi_i}{\psi_i}\big)/s$ with $\psi_1,\ldots,\psi_s$ independent and uniformly distributed unit vectors in $\C^d$. \\
\item \textit{Maximally mixed + Gaussian noise}: In the case where $s>d$, $\rho=\big(\Id+G/2\sqrt{s}\big)/d$ with $G$ a traceless GUE matrix on $\C^d$ (which is a state only with high probability). \\
\item \textit{Random renormalized projection}: In the case where $s<d$, $\rho=P_E/s$ with $P_E$ the projector onto $E$ a uniformly distributed $s$-dimensional subspace of $\C^d$. \\
\end{itemize}
When we talk about ``random states'', without any further specification, we usually refer to the first mentioned model. However, it may happen that one is only interested, for instance, in computing the average of a given convex function $f:\mathcal{H}(\C^d)\rightarrow\R$ over $\Delta=\rho-\Id/d$ with $\rho$ a random state distributed in some way. This was more or less our case throughout this whole manuscript, where we always needed to estimate the average of the support function of a convex body in $\cH(\C^d)$ (a set of either measurements or states). Then, in this situation, the nice thing is that the values obtained for $\E f$ over each of these four random state ensembles will all be of the same order (or even sometimes equivalent) as $d,s\rightarrow+\infty$. Let us briefly try to explain why (see Appendices A and B in \cite{ASY} for more details, and Section \ref{sec:local-POVM-proofs} in Chapter \ref{chap:data-hiding} of this manuscript for a similar reasoning put in practice). First of all, define a function $\widetilde{f}:\R^d\rightarrow\R$, associated to $f$, by
\[ \widetilde{f}(x)=\int_{\mathcal{U}(\C^d)}f\left(U\diag(x)U^{\dagger}\right)\mathrm{d}U. \]
Since all four mentioned shifted random state models are obviously unitarily invariant, $\widetilde{f}$ is such that, for $\Delta$ distributed according to one of these, $\E f(\Delta)=\E \widetilde{f}(\spec(\Delta))$. Next, for each of the considered shifted random state models, the limiting spectral distribution is well-known (Mar\v{c}enko-Pastur, semi-circular, Bernoulli, with certain parameters). And these distributions can be compared to one another, in the sense that: for each two of them $\mu,\mu'$, there exist constants $c,C>0$ such that, for some random variables $X,X'$ sampled from $\mu,\mu'$, $cX\leq X'\leq CX$. Now, these asymptotic comparisons can be turned into non-asymptotic ones, because each of the empirical spectral distributions converges both weakly and in terms of extreme eigenvalues. Specifically, there exist constants $c_{d,s},C_{d,s}>0$ (going to $c,C$ as $d,s\rightarrow+\infty$) such that, for some $\Delta,\Delta'$ sampled from the two ensembles having $\mu,\mu'$ as limiting spectra,
\[ c_{d,s}\spec(\Delta)\preceq\spec(\Delta')\preceq C_{d,s}\spec(\Delta). \]
By elementary properties of majorization, this in turn implies (due to the convexity and the invariance under coordinate permutation of $\widetilde{f}$) that
\[ c_{d,s}\widetilde{f}(\spec(\Delta))\leq\widetilde{f}(\spec(\Delta'))\leq C_{d,s}\widetilde{f}(\spec(\Delta)). \]
And eventually, simply taking expectations, what we wanted to show, namely
\[ c_{d,s}\E f(\Delta)\leq\E f(\Delta')\leq C_{d,s}\E f(\Delta). \]
Hence in a nutshell, the conclusion of this digression: depending on the problem we are looking at (and especially on the extra conditions we want to impose), working with one type or the other of these random states may be more convenient, but in the end the picture is usually more or less the same for all of them.

Yet, there are many settings where we would need to understand completely different kinds of random states. One example which might be worth mentioning is that of random tensor product states, i.e.~states of the form $\rho_1\otimes\cdots\otimes\rho_n$ with $\rho_1,\ldots,\rho_n$ random states (either independent or correlated in some way). Indeed, whenever one wants to study the multiplicative/additive behaviour of a certain function, this is precisely the type of states which shows up. And a quite natural wonder that one may have is: given a function which is known to violate multiplicativity/additivity in general, is it nevertheless (weakly) multiplicative/additive in typical scenarios? Such investigations were for instance carried on in \cite{Montanaro} or \cite{FN} for the output entropy of quantum channels, yielding partial results. But there would be much more to explore. As a sample, closely related to preoccupations in this manuscript:
\begin{itemize}[topsep=0cm,itemsep=0cm,parsep=0cm,leftmargin=*]
\item For $\mathbf{M}$ being a set of locally restricted POVMs, such as $\mathbf{SEP}$ or $\mathbf{PPT}$, does the distinguishability norm $\|\cdot\|_{\mathbf{M}}$ (cf.~Chapter \ref{chap:data-hiding}) generically exhibit a multiplicative behaviour? That is concretely, we ask if it is true that for random bipartite states $\rho_{\A\B},\sigma_{\A\B}$, we have with high probability
\[ \big\|\left(\rho_{\A\B}-\sigma_{\A\B}\right)^{\otimes n}\big\|_{\mathbf{M(\A^n{:}\B^n)}} \simeq \left\|\rho_{\A\B}-\sigma_{\A\B}\right\|_{\mathbf{M(\A{:}\B)}}^n. \]
\item For $\mathcal{K}$ being the set of separable states or one of its relaxations, does the maximum fidelity function $F(\cdot,\mathcal{K})$ (cf.~Chapter \ref{chap:deFinetti}) generically exhibit a multiplicative behaviour? That is concretely, we ask if it is true that for a random bipartite state $\rho_{\A\B}$, we have with high probability
\[ F\big(\rho_{\A\B}^{\otimes n},\mathcal{K}(\A^n{:}\B^n)\big) \simeq F\big(\rho_{\A\B},\mathcal{K}(\A{:}\B)\big)^n. \]
\end{itemize}
At first sight, one of the main difficulties seems to lie in the fact that there is much less invariance in a random tensor product state than in a random state: e.g.~the distribution of $\rho^{\otimes n}$, for $\rho$ a random state on $\C^d$, is not invariant under conjugation by any unitary on $(\C^d)^{\otimes n}$, but only under conjugation by those of the form $U^{\otimes n}$ with $U$ a unitary on $\C^d$. Another major obstacle is that, even once the average behaviour is identified, asserting concentration around it is not so easy. Indeed, tiny fluctuations for $\rho$ may have huge repercussions for $\rho^{\otimes n}$.

Note that the related question of studying random separable states has already been considered in \cite{AHH}, whose main result summarizes as follows: If $s\simeq d^n$, then the distribution of $\big( \sum_{i=1}^s \ketbra{\psi_i^1}{\psi_i^1} \otimes\cdots\otimes \ketbra{\psi_i^n}{\psi_i^n} \big)/s$, with the $\psi_i^1,\ldots,\psi_i^n$ independent and uniformly distributed unit vectors in $\C^d$, is roughly the same as the distribution of $\big(\sum_{i=1}^s\ketbra{\psi_i}{\psi_i}\big)/s$, with the $\psi_i$ independent and uniformly distributed unit vectors in $(\C^d)^{\otimes n}$.

\section*{Quantifying the generic amount of entanglement in multipartite quantum systems}
\label{sec:generic-entanglement}

In Part \ref{part:entanglement} we looked at separability vs entanglement properties in generic high-dimensional multipartite systems from various angles. One common trait of the functions we were trying to quantify though was that they were almost always expressible as the support function of a certain convex set (of either measurements or states). Nevertheless, most more operational measures of entanglement are not of that kind, but rather of entropic form. And loosely speaking, the presence of $\log$'s in the formulas defining them makes the problem of estimating their expected value much less attackable by our techniques. This said, in the seminal exploration of typical entanglement, launched in \cite{HLW}, bounds were however put on the expected value of several entropic measures of correlations. In short, the main conclusion of this study was that ``bound-entangled-like'' features, namely both high entanglement of formation and low distillable entanglement, are in fact quite generic in large bipartite quantum systems. But being able to derive more precise estimates would be highly desirable. One possible application would be in disproving that a given entanglement measure obeys a given property (e.g.~of monogamy-type, faithfulness-type etc.). Indeed, counter-examples to the long-standing additivity of minimum output entropy conjecture were precisely provided by appropriately chosen random channels \cite{Hastings,ASW}. So it is reasonable to hope that, similarly, random states could be good candidates to rule out certain behaviours for such or such measure of entanglement.

There are nonetheless a few entanglement measures for which getting quite tight bounds (or even the exact asymptotics) for their typical value is actually possible (see \cite{ADHLPW}, in collaboration with G.~Adesso, S.~Di Martino, M.~Huber, M.~Piani and A.~Winter). Examples include the entanglement of formation or the relative entropy of entanglement. The paradigmatic idea could be summarized as follows: whenever a minimization/maximization has to be performed over a certain set of states, the ``obvious'' candidate states usually already work pretty well in high dimension. On the other hand, archetypical examples of entanglement measures which cannot be easily tackled are all the regularized ones. Indeed, for a non-additive entanglement measure $E$, even if the generic behaviour of $E(\rho)$ is well understood, this is generally not the case of its regularized version $E^{\infty}(\rho)=\lim_{n\rightarrow+\infty}E(\rho^{\otimes n})/n$. This is for the reason evoked before that properties of tensor power random states remain ill homed in on.

Another category of entanglement measures, which was briefly evoked in the concluding section of Chapter \ref{chap:data-hiding}, is that of ``filtered through measurements'' entanglement measures (initially introduced and studied in \cite{Piani}). For the sake of concreteness, let us focus here on one specific example, based on the relative entropy distance measure $D(\cdot\|\cdot)$ between two quantum states (or two classical probability distributions), as defined in Chapter \ref{chap:QIT}, Section \ref{sec:Shannon}. So now, given a state $\rho$ on some bipartite system $\A\otimes\B$, its relative entropy of entanglement is defined as
\[ D\left(\rho\|\mathcal{S}\right) = \inf_{\sigma\in\mathcal{S}} D(\rho\|\sigma), \]
and for any class of POVMs $\mathbf{M}$ on $\A\otimes\B$, its relative entropy of entanglement filtered through $\mathbf{M}$ is defined as
\[ D_{\mathbf{M}}\left(\rho\|\mathcal{S}\right) = \inf_{\sigma\in\mathcal{S}} \sup_{\mathcal{M}\in\mathbf{M}} D\left(\mathcal{M}(\rho)\|\mathcal{M}(\sigma)\right). \]
One may then ask: for $\mathbf{M}$ being a class of locally restricted POVMs on $\A\otimes\B$, such as $\mathbf{SEP}$ or $\mathbf{PPT}$, what is the generic value of $D_{\mathbf{M}}\left(\rho\|\mathcal{S}\right)$ for $\rho$ a uniformly distributed random state on $\A\otimes\B$? Of course, we know by Pinsker's inequality that $D_{\mathbf{M}}\left(\rho\|\mathcal{S}\right)\geq \|\rho-\mathcal{S}\|_{\mathbf{M}}^2/(2\ln 2)$. So a lower bound on the typical $\|\rho-\mathcal{S}\|_{\mathbf{M}}$, which is something we establish in Chapter \ref{chap:data-hiding}, immediately implies a lower bound on the typical $D_{\mathbf{M}}\left(\rho\|\mathcal{S}\right)$. But what about an upper bound? And more generally, is it possible to directly estimate the correct order of magnitude for $D_{\mathbf{M}}\left(\rho\|\mathcal{S}\right)$? In order to do so, it seems that a useful notion could be that of distinguishability ordering rather than distinguishability norm. Let us explicit a bit what we have in mind. Given a class of POVMs $\mathbf{M}$ and two states $\rho,\sigma$ on $\A\otimes\B$, one can define the minimal dominating constant between $\rho$ and $\sigma$, filtered through $\mathbf{M}$, as
\[ \alpha_{\mathbf{M}}(\rho,\sigma) = \inf\left\{\alpha \st \forall\ \mathcal{M}\in\mathbf{M},\  \mathcal{M}(\rho)\leq \alpha\,\mathcal{M}(\sigma) \right\}. \]
The reason why we claim that this is an interesting figure of merit is simply because, for two probability distributions $p,q$, knowing that $p\leq Cq$, for some constant $C>1$, implies that $D(p\|q)\leq \log C$. So an upper bound on $\alpha_{\mathbf{M}}(\rho,\mathcal{S})$ straightforwardly translates into an upper bound on $D_{\mathbf{M}}\left(\rho\|\mathcal{S}\right)$. Understanding the quantity $\alpha_{\mathbf{M}}(\rho,\sigma)$ requires a bit more work than understanding the quantity $\|\rho-\sigma\|_{\mathbf{M}}$, but can nevertheless be done in some cases (see a forthcoming work in collaboration with M.~Christandl).

\section*{In search of a supercalifragilistic measure of entanglement}
\label{sec:entanglement-measure}

One of our main preoccupations in Chapter \ref{chap:deFinetti} was to see under which conditions the support function of a set of quantum states exhibits a multiplicative behaviour. As a case study, we scrutinized what happens for the set of separable states. Let us start with commenting a bit more on the relevance of this example, linking it to the topic of Chapter \ref{chap:channel-compression}. Let $P$ be the projector onto some subspace $\mathrm{C}\subset\A\otimes\B$, and write $P=VV^{\dagger}$, with $V:\mathrm{C}\hookrightarrow\A\otimes\B$ an isometry. Define next the quantum channel $\mathcal{N}:\mathcal{L}(\mathrm{C})\rightarrow\mathcal{L}(\A)$ by $\mathcal{N}(X)=\tr_{\B}(VXV^{\dagger})$. It is then easy to see that
\[ \underset{\sigma\in\mathcal{S}(\A{:}\B)}{\sup} \tr(P\sigma) =: h_{\mathcal{S}(\A{:}\B)}(P) = \|\mathcal{N}\|_{1{\rightarrow}\infty} := \underset{\rho\in\mathcal{D}(\mathrm{C})}{\sup} \|\mathcal{N}(\rho)\|_{\infty}. \]
Hence, asking if $h_{\mathcal{S}(\A^n{:}\B^n)}(P^{\otimes n})\simeq \big(h_{\mathcal{S}(\A{:}\B)}(P)\big)^n$ is nothing else than asking if $\|\mathcal{N}^{\otimes n}\|_{1{\rightarrow}\infty}\simeq\big(\|\mathcal{N}\|_{1{\rightarrow}\infty}\big)^n$, i.e.~equivalently if $S_{\infty}^{\min}(\mathcal{N}^{\otimes n})\simeq n\, S_{\infty}^{\min}(\mathcal{N})$. It is now a well-known fact that the so-called additivity conjecture for $S_{\infty}^{\min}$ is false: inequality can be strict in general between $S_{\infty}^{\min}(\mathcal{N}^{\otimes 2})$ and $2\, S_{\infty}^{\min}(\mathcal{N})$, with even drastic examples where $S_{\infty}^{\min}(\mathcal{N}^{\otimes 2})\sim S_{\infty}^{\min}(\mathcal{N})$. The two archetypical cases for which this happens are when the associated projector $P$ is either the projector onto the anti-symmetric subspace (see \cite{GHP}) or a random projector (see \cite{HW}). However, it has also been proved more recently that, precisely in these two situations, the extreme $2$-copy behaviour does not reflect the $n$-copy behaviour for large $n$, where weak additivity of $S_{\infty}^{\min}(\mathcal{N})$ actually holds (see \cite{CSW} and \cite{Montanaro}). Hence the natural question of whether this could not in fact be what occurs for any quantum channel $\mathcal{N}$.

Chapter \ref{chap:deFinetti} put forward two ways of showing a weakly multiplicative behaviour of $h_{\mathcal{S}(\A{:}\B)}(P)$. Indeed, let $\rho^{(n)}$ be a permutation-invariant state on $(\A\otimes\B)^{\otimes n}$ which is separable across the cut $\A^{\otimes n}{:}\B^{\otimes n}$, and assume that you want to upper bound $\tr(P^{\otimes n} \rho^{(n)})$. Then, you can follow either the ``global'' strategy of using a de Finetti reduction for $\rho^{(n)}$ (cf.~Section \ref{sec:sep1}), or the more ``local'' strategy of iteratively projecting $\rho^{(n)}$ on $\A\otimes\B$ and working with the conditional state on the remaining subsystems (cf.~Section \ref{sec:sep2}). The two approaches can be seen through, but with the annoying feature of eventually yielding a dimensional dependence in the exponentially decaying bound. Concretely, setting $d=\max(|\A|,|\B|)$, one can obtain a statement of the form:
\begin{equation} \label{eq:h_sep-dim} h_{\mathcal{S}(\A{:}\B)}(P)\leq 1-\varepsilon\ \Rightarrow\ h_{\mathcal{S}(\A^n{:}\B^n)}(P^{\otimes n})\leq (1-\delta(\varepsilon,d))^n, \end{equation}
with $\delta$ a function which is non-decreasing in $\varepsilon$ and non-increasing in $d$, vanishing for $\varepsilon=0$ or $d\rightarrow+\infty$. Now, it could well be that equation \eqref{eq:h_sep-dim} actually holds with a dimensionally independent function $\delta(\varepsilon)$.

In order to prove something is that direction via the ``local'' technique, it would be enough to find a function $E$, quantifying the amount of entanglement in a bipartite quantum state, which would satisfy the two following conditions:
\begin{itemize}[topsep=0cm,itemsep=-1em,parsep=0cm,leftmargin=*]
\item Monogamy-like property: $I(\mathrm{A}\mathrm{A}'{:}\mathrm{B}\mathrm{B}')(\rho_{\A\A'\B\B'}) \geq E(\rho_{\A\B}) + E(\rho_{\A'\B'})$ (see Chapter \ref{chap:QIT}, Section \ref{sec:Shannon}, for the definition of the mutual information $I$). \\
\item Strong faithfulness property: $E(\rho_{\A\B})\leq\varepsilon\ \Rightarrow\ \left\|\rho_{\A\B}-\mathcal{S}(\A{:}\B)\right\|_1\leq \delta(\varepsilon)$, with $\delta$ a universal function (i.e.~in particular not depending on $|\A|$ and $|\B|$).
\end{itemize}
The reason why this is not so easy is because monogamy and faithfulness are two properties of entanglement measures which usually exclude one another (for a more precise formulation, see \cite{ADHLPW}, in collaboration with G.~Adesso, S.~Di Martino, M.~Huber, M.~Piani and A.~Winter). One candidate though for an entanglement measure which would combine the two above mentioned features is the conditional entanglement of mutual information introduced in \cite{HWY} (cf.~Chapter \ref{chap:deFinetti}, Section \ref{sec:sep2}).

We already mentioned the one-to-one correspondence between estimating $h_{\mathcal{S}}$ and estimating $S_{\infty}^{\min}$. There is another important one, namely with estimating the acceptance probability in a $\mathrm{QMA}(2)$ protocol with unentangled provers. Also out of purely quantum information scope, there is a plethora of problems which are equivalent to pulling apart the two options $h_{\mathcal{S}(\A{:}\B)}(P)=1$ and $h_{\mathcal{S}(\A{:}\B)}(P)<1$: distinguishing $\|\psi\|_{\text{inj}}=1$ from $\|\psi\|_{\text{inj}}<1$ for $\psi\in\A\otimes\B\otimes\mathrm{C}$ such that $\|\psi\|\leq 1$, distinguishing $\|M\|_{2{\rightarrow}4}=1$ from $\|M\|_{2{\rightarrow}4}<1$ for $M:\mathrm{C}\rightarrow\A\otimes\B$ such that $\|M\|_{\infty}\leq 1$ etc. The reader is referred to \cite{HM} for the precise definitions of a $\mathrm{QMA}(2)$ protocol, the injective tensor norm, the $(2{\rightarrow}4)$-norm, and for a full list of other (roughly) equivalent discrimination problems. Anyway, all this to say that having an efficient way of amplifying an $\varepsilon$-gap in $h_{\mathcal{S}}$ by parallel repetition would have significant implications!

\section*{Relaxations of separability or absolute separability and related complexity issues}
\label{sec:sep-relaxtations}

As was highlighted in many different ways over these pages, any simple necessary condition for separability (meaning that there exists an efficient way of checking it) is doomed to become very rough as the dimensions of the local Hilbert spaces grow. Let us focus here on two such relaxations of separability, which are of very distinct nature, even though deciding membership can be cast as an SDP for both of them: positivity under partial transposition and $k$-extendibility. The PPT criterion is believed to be, in several senses, the best amongst separability criteria based on positive maps. Nonetheless, it is not even known whether the set of PPT states has a mean width and a volume radius which asymptotically differ from those of the set of all states. On the contrary, concerning the set of $k$-extendible states, Theorem \ref{th:w(k-ext)} in Chapter \ref{chap:k-extendibility}, Section \ref{section:w(k-ext)}, tells us that
\[ \forall\ k\in\N,\ \mathrm{vrad}\left(\mathcal{E}_k(\C^d{:}\C^d)\right) \leq w\left(\mathcal{E}_k(\C^d{:}\C^d)\right) \underset{d\rightarrow+\infty}{\sim}\frac{2}{\sqrt{k}}\frac{1}{d}. \]
Now, we just have to recall that for the set of all states we have
\[ w\left(\mathcal{D}(\C^d\otimes\C^d)\right) \underset{d\rightarrow+\infty}{\sim} \frac{2}{d}\ \text{and}\ \mathrm{vrad}\left(\mathcal{D}(\C^d\otimes\C^d)\right) \underset{d\rightarrow+\infty}{\sim} \frac{e^{-1/4}}{d}. \]
And we thus get the existence of a constant $0<c_0<1$ and a dimension $d_0\in\N$ such that, for any $d\geq d_0$,
\[ \forall\ k\geq 2,\ w\left(\mathcal{E}_k(\C^d{:}\C^d)\right)\leq c_0\, w\left(\mathcal{D}(\C^d\otimes\C^d)\right)\ \text{and}\ \forall\ k\geq 7,\ \mathrm{vrad}\left(\mathcal{E}_k(\C^d{:}\C^d)\right)\leq c_0\, \mathrm{vrad}\left(\mathcal{D}(\C^d\otimes\C^d)\right). \]
Hence to summarize: for a fixed $k\geq 2$, the size of $\mathcal{E}_k(\C^d{:}\C^d)$ differs from that of $\mathcal{D}(\C^d\otimes\C^d)$ only by an absolute factor, but one can at least certify that the latter is strictly smaller than $1$ (for all $k\geq 2$ if size is measured in terms of mean width, and for sure as soon as $k\geq 7$ if size is measured in terms of volume radius). Whereas oppositely, it could be that the ratio between the size of $\mathcal{P}(\C^d{:}\C^d)$ and that of $\mathcal{D}(\C^d\otimes\C^d)$ approaches $1$ as $d$ increases. It may be worth pointing out that, when volume radius is used as size parameter, this problem is equivalent to: does there exist an absolute constant $c>0$ such that
\[ \forall\ d\in\N,\ \P_{\rho\sim\mu_{d^2,d^2}}\left(\rho\in\mathcal{P}(\C^d{:}\C^d)\right)\leq e^{-cd^4}\ ? \]
As a side technical comment, let us mention that, for the case of $k$-extendibility, we already have as a corollary of Theorem \ref{th:not-kext} in Chapter \ref{chap:k-extendibility}, Section \ref{section:non-k-ext-witness}, that
\[ \forall\ k\geq 6,\ \exists\ c_k>0:\ \forall\ d\in\N,\ \P_{\rho\sim\mu_{d^2,d^2}}\left(\rho\in\mathcal{E}_k(\C^d{:}\C^d)\right)\leq e^{-c_kd^2}. \]
But by all the remarks above, we see that (at least for $k\geq 7$) this probability estimate is far from optimal: the dimensional dependence in the concentration bound can actually be improved from $d^2$ to $d^4$.

However complicated to characterize the set of separable states is, it could be that its so-called \textit{absolute} version is much easier to grasp. The latter, initially introduced in \cite{KZ}, is defined as the set of states which remain separable under conjugation by any unitary, i.e.~it encompasses a notion of separability for states which only depends on their eigenvalues, and not on their eigenvectors. Similarly, one can define, for each separability criterion, the set of states which absolutely satisfy it. Quite surprisingly, there are strong evidences towards the fact that, for instance, being absolutely PPT could already be equivalent to being absolutely separable. This conjecture is known to hold in the special case of $\C^2\otimes\C^d$, for any $d\in\N$ \cite{John}. If it were true in general, it would imply an extremely simple description of absolutely separable states, as a finite number of semidefinite conditions \cite{Hild}. With this kind of motivations in mind, trying to understand the hierarchy of absolutely $k$-extendible states looks like an interesting project: does it collapse to absolutely separable states for a finite $k$? and if so a dimension dependent or independent $k$? etc.

\section*{Typical degradability or anti-degradability of quantum channels}
\label{sec:degradability}

Let $\cN:\cL(\A)\rightarrow\cL(\B)$ be a quantum channel, such that $\cN(X)=\tr_{\rE}(VXV^{\dagger})$ for some isometry $V:\A\hookrightarrow\B\otimes\rE$. Its complementary channel is then the quantum channel $\cN^c:\cL(\A)\rightarrow\cL(\rE)$ defined by $\cN^c(X)=\tr_{\B}(VXV^{\dagger})$. $\cN$ is called \textit{degradable} if there exists a quantum channel $\mathcal{F}:\cL(\B)\rightarrow\cL(\rE)$ such that $\cN^c=\mathcal{F}\circ\cN$, and it is called \textit{anti-degradable} if $\cN^c$ is degradable (i.e.~if there exists a quantum channel $\mathcal{F}:\cL(\rE)\rightarrow\cL(\B)$ such that $\cN=\mathcal{F}\circ\cN^c$). These channels are interesting from many respects. One main feature of theirs is that an anti-degradable channel does not have any quantum capacity, while a degradable channel has an additive coherent information, and therefore a quantum capacity admitting a single-letter formula. In \cite{RSSW}, an approximate notion of (anti-)degradability was more generally defined. And it was shown there that, with this definition, the smallest $\varepsilon$ such that a channel $\cN$ is $\varepsilon$-anti-degradable can be used to give an upper bound on the quantum capacity of $\cN$. This is an important result because this minimal $\varepsilon$ can be efficiently determined (by solving an SDP), hence providing an easily computable upper bound on the quantum capacity, a quantity for which no method to even just estimate it is known in general.

Besides, another nice fact about (anti-)degradability is that it is a property which can be easily characterized at the level of the Choi states of either $\cN$ or $\cN^c$. Specifically, $\cN$ is anti-degradable if and only if $\tau(\cN)$ is a $2$-extendible state, and thus conversely, $\cN$ is degradable if and only if $\tau(\cN^c)$ is a $2$-extendible state (cf.~Chapter \ref{chap:k-extendibility}). So now, assume that $\cN:\cL(\A)\rightarrow\cL(\B)$ is a random quantum channel, defined by $\cN(X)=\tr_{\rE}(VXV^{\dagger})$ for $V:\A\hookrightarrow\B\otimes\rE$ a uniformly distributed isometry. Then, as $|\A|,|\B|,|\rE|$ grow, $\tau(\cN)$ is more or less distributed as a random state on $\A\otimes\B$ induced by some environment $\rE$ (this hand-waving claim can be made mathematically precise). Hence, understanding when such random channels are typically (anti-)degradable or not roughly boils down to understanding when random-induced states are typically $2$-extendible or not. And this is now something that we know quite well how to do thanks to the techniques developed in Chapter \ref{chap:k-extendibility}. An interesting question at that point would therefore be: can this approach be extended in order to estimate, given a random channel, its minimal $\varepsilon$-anti-degradability (and subsequently an upper bound on its quantum capacity)?

There are several reasons why this does not look like a completely straightforward affair. Characterizing when a random channel is with high probability exactly (anti-)degradable is probably not the hardest part: from the results of Chapter \ref{chap:k-extendibility}, one can reasonably expect that there should be a threshold for anti-degradability vs non-anti-degradability at $|\rE|=c|\A||\B|$, for some universal constant $c>0$ (and correspondingly, one for degradability vs non-degradability at $|\B|=c|\A||\rE|$). The main difficulty rather comes from the fact that the notion of $\varepsilon$-(anti-)degradability introduced in \cite{RSSW} is not simply being $\varepsilon$-close to an exactly (anti-)degradable channel, but instead being $\varepsilon$-close to satisfying the (anti-)degradability condition. It is a problem we already stumbled upon in Chapters \ref{chap:deFinetti} and \ref{chap:SNOS} (in a totally different context) that, even though sounding like a sensible hope, it is not always true that the latter (for some $\varepsilon$) implies the former (for some $\delta(\varepsilon)$).


\bibliographystyle{plain}
\bibliography{references}
\addcontentsline{toc}{part}{Bibliography}

\addtocontents{toc}{\protect\mbox{}\protect\hrulefill\par}

\end{document}